\documentclass[dvipdfm,11pt,a4paper]{article}
\usepackage[dvips]{graphicx,color}
\usepackage{url}
\usepackage{amssymb}
\usepackage[centertags]{amsmath}
\usepackage{multirow}
\usepackage{subfigure}
\usepackage[affil-it]{authblk}
\usepackage{cite}
\usepackage{lineno}
\usepackage[bookmarks=false]{hyperref}

\graphicspath{
{Atmospheric/} {ExecutiveSummary/} {Geoneutrino/} {IndirectDarkMatter/} {Introduction/}
{MassHierarchy/} {MixingParameters/} {NucleonDecay/} {OtherExotic/} {Solar/}
{Sterile/} {Summary/} {SupernovaBurst/} {SupernovaDiffuse/} {Appendix/}
}

\makeatletter
\def\blfootnote{\xdef\@thefnmark{}\@footnotetext}
\renewcommand{\thefigure}{\ifnum \c@section>\z@ \thesection-\fi \@arabic\c@figure}
\renewcommand{\thetable}{\ifnum \c@section>\z@ \thesection-\fi \@arabic\c@table}
\@addtoreset{equation}{section}
\@addtoreset{table}{section}
\@addtoreset{figure}{section}
\@addtoreset{footnote}{section}
\makeatother

\begin{document}

\title{\LARGE\bf Neutrino Physics with JUNO}

\newcommand{\ECUST} {1}
\newcommand{\IHEP} {2}
\newcommand{\USTC} {3}
\newcommand{\INFNMI} {4}
\newcommand{\IPHC} {5}
\newcommand{\OhioSU} {6}
\newcommand{\INRRAS} {7}
\newcommand{\NTU} {8}
\newcommand{\UNIPD} {9}
\newcommand{\LLR} {10}
\newcommand{\CPPM} {11}
\newcommand{\APC} {12}
\newcommand{\WHU} {13}
\newcommand{\INFNMB} {14}
\newcommand{\PTMB} {15}
\newcommand{\NUU} {16}
\newcommand{\Tsinghua} {17}
\newcommand{\NJU} {18}
\newcommand{\NCEPU} {19}
\newcommand{\UNIMB} {20}
\newcommand{\ULB} {21}
\newcommand{\MIT} {22}
\newcommand{\ANL} {23}
\newcommand{\INFNPD} {24}
\newcommand{\Hawaii} {25}
\newcommand{\UOULU} {26}
\newcommand{\WuYi} {27}
\newcommand{\TUM} {28}
\newcommand{\IHEGCAGS} {29}
\newcommand{\JINR} {30}
\newcommand{\FZJ} {31}
\newcommand{\SJTU} {32}
\newcommand{\BNU} {33}
\newcommand{\Hamburg} {34}
\newcommand{\CIAE} {35}
\newcommand{\SDU} {36}
\newcommand{\HIT} {37}
\newcommand{\YerPhI} {38}
\newcommand{\HEPHY} {39}
\newcommand{\UMD} {40}
\newcommand{\DGUT} {41}
\newcommand{\EKUT} {42}
\newcommand{\CUPrague} {43}
\newcommand{\CQU} {44}
\newcommand{\SYSU} {45}
\newcommand{\NCTU} {46}
\newcommand{\UIUC} {47}
\newcommand{\UH} {48}
\newcommand{\GXU} {49}
\newcommand{\UCAS} {50}
\newcommand{\JLU} {51}
\newcommand{\XMU} {52}
\newcommand{\INFNFerrara} {53}
\newcommand{\PKU} {54}
\newcommand{\INFNRoma} {55}
\newcommand{\UMDG} {56}
\newcommand{\UCChile} {57}
\newcommand{\INFNPG} {58}
\newcommand{\INFNLNF} {59}
\newcommand{\BNL} {60}
\newcommand{\Minnesota} {61}
\newcommand{\MPP} {62}
\newcommand{\Columbia} {63}
\newcommand{\Aachen} {64}
\newcommand{\JYU} {65}
\newcommand{\Mainz} {66}
\newcommand{\SichuanU} {67}
\newcommand{\SUBATECH} {68}
\newcommand{\Nankai} {69}
\newcommand{\INFNGenova} {70}
\newcommand{\XJTU} {71}
\newcommand{\ITP} {72}

\author[\ECUST]{Fengpeng~An}
\author[\IHEP]{Guangpeng~An}
\author[\USTC]{Qi~An}
\author[\INFNMI]{Vito~Antonelli}
\author[\IPHC]{Eric~Baussan}
\author[\OhioSU]{John~Beacom}
\author[\INRRAS]{Leonid~Bezrukov}
\author[\NTU]{Simon~Blyth}
\author[\UNIPD]{Riccardo~Brugnera}
\author[\LLR]{Margherita~Buizza~Avanzini}
% ...

\date{}
\maketitle
\clearpage

\begin{abstract}

The Jiangmen Underground Neutrino Observatory (JUNO), a 20 kton multi-purpose
underground liquid scintillator detector, was proposed with the
determination of the neutrino mass hierarchy as a primary physics goal.
The excellent energy resolution and the large fiducial volume anticipated for
the JUNO detector offer exciting opportunities for addressing many important
topics in neutrino and astro-particle physics. In this document, we present
the physics motivations and the anticipated performance of the JUNO detector
for various proposed measurements.

Following an introduction summarizing the current status and open
issues in neutrino physics, we discuss how the detection of
antineutrinos generated by a cluster of nuclear power plants allows
the determination of the neutrino mass hierarchy at a 3-4$\sigma$
significance with six years of running of JUNO. The measurement of
antineutrino spectrum with excellent energy resolution will also
lead to the precise determination of the neutrino oscillation parameters
$\sin^2\theta_{12}$, $\Delta m^2_{21}$, and $|\Delta m^2_{ee}|$ to an
accuracy of better than 1\%, which will play a crucial role in the
future unitarity test of the MNSP matrix.

The JUNO detector is capable of observing not only antineutrinos from the power plants, but also neutrinos/antineutrinos from
terrestrial and extra-terrestrial sources, including supernova burst neutrinos, diffuse supernova neutrino background,
geoneutrinos, atmospheric neutrinos, and solar neutrinos. As a result of JUNO's large size, excellent energy resolution,
and vertex reconstruction capability, interesting new data
on these topics can be collected. For example, a neutrino burst from a typical core-collapse supernova at a distance of
10~kpc would lead to $\sim5000$ inverse-beta-decay events and $\sim2000$ all-flavor neutrino-proton elastic scattering events in JUNO,
which are of crucial importance for understanding the mechanism of supernova explosion and for exploring novel phenomena such as collective
neutrino oscillations. Detection of neutrinos from all past core-collapse supernova explosions in the visible universe with JUNO
would further provide valuable information on the cosmic star-formation rate and the average core-collapse neutrino energy spectrum.
Antineutrinos originating from the radioactive decay of Uranium and Thorium in the Earth can be detected in JUNO with a rate of $\sim 400$ events per year,
significantly improving the statistics of existing geoneutrino event samples. Atmospheric neutrino events collected in JUNO can
provide independent inputs for determining the mass hierarchy and the octant of the $\theta_{23}$ mixing angle. Detection of the
$^7$Be and $^8$B solar neutrino events at JUNO would shed new light on the solar metallicity problem and examine the transition region
between the vacuum and matter dominated neutrino oscillations.

Regarding light sterile neutrino topics, sterile neutrinos with $10^{-5}~{\rm eV}^2 < \Delta m^2_{41} < 10^{-2}~{\rm eV}^2$ and a
sufficiently large mixing angle $\theta^{}_{14}$ could be identified through a precise measurement of the reactor antineutrino energy spectrum.
Meanwhile, JUNO can also provide us excellent opportunities to test the eV-scale sterile neutrino hypothesis, using either the
radioactive neutrino sources or a cyclotron-produced neutrino beam.

The JUNO detector is also sensitive to several other beyond-the-standard-model physics.
Examples include the search for proton decay via the $p \to K^+ + \bar \nu$ decay channel,
search for neutrinos resulting from dark-matter annihilation in the Sun,
search for violation of Lorentz invariance via the sidereal
modulation of the reactor neutrino event rate, and search for the effects of non-standard interactions.

The proposed construction of the JUNO detector will provide a unique
facility to address many outstanding crucial questions in particle
and astrophysics in a timely and cost-effective fashion. It holds the
great potential for further advancing our quest to understanding the
fundamental properties of neutrinos, one of the building blocks of our
Universe.

\end{abstract}

\clearpage

\setcounter{section}{0}
\tableofcontents
\clearpage
\listoffigures
\clearpage
\listoftables
\clearpage

%\linenumbers

\section{Introduction}
\label{sec:intro}

\blfootnote{Editors: Jun Cao (caoj@ihep.ac.cn) and Zhi-zhong Xing (xingzz@ihep.ac.cn)}

\subsection{Neutrino Oscillations in a Nutshell}
\label{subsec:nutshell}

The standard electroweak model is a successful theory which not only
unifies the electromagnetic and weak interactions but also
explains almost all the phenomena of this nature observed at or below
the electroweak scale. When this theory was first formulated by Weinberg
in 1967 \cite{Weinberg:1967tq}, its particle content was so economical that
the neutrinos were assumed to be massless and hence there was no lepton
flavor mixing. But just one year later the solar neutrinos were observed
by Davis {\it et al} \cite{Davis:1968cp}, and a deficit of their flux as compared
with the prediction from the standard solar model was also established
by Bahcall {\it et al} \cite{Bahcall:1968hc,Bahcall:1981zh}.
Such an anomaly turned out to be solid evidence for new physics
beyond the standard model, because it was found to be attributed
to the neutrino oscillation --- a spontaneous and periodic
change from one neutrino flavor to another, which does not take place
unless neutrinos have finite masses and lepton flavors are mixed.
Flavor oscillations can therefore serve as a powerful tool to study
the intrinsic properties of massive neutrinos and probe other kinds of
new physics.

\subsubsection{Flavor mixing and neutrino oscillation probabilities}
\label{subsubsec:mixing}

In the standard model the fact that the quark fields interact
with both scalar and gauge fields leads to a nontrivial mismatch between
their mass and flavor eigenstates, which is just the dynamical reason
for quark flavor mixing and CP violation. Although a standard
theory for the origin of tiny neutrino masses has not been established,
one may expect a straightforward extension of the standard model in which
the phenomena of lepton flavor mixing and CP violation emerge for a
similar reason. In this case the weak charged-current interactions of leptons
and quarks can be written as
\begin{eqnarray}
-{\cal L}^{}_{\rm cc} = \frac{g}{\sqrt{2}} \left[
\overline{\left(e \hspace{0.3cm} \mu \hspace{0.3cm} \tau
\right)^{}_{\rm L}} \ \gamma^\mu \
U \left(\begin{matrix} \nu^{}_1 \cr \nu^{}_2 \cr \nu^{}_3
\cr \end{matrix}\right)^{}_{\rm L} W^-_\mu ~ + ~
\overline{\left(u \hspace{0.3cm} c \hspace{0.3cm} t \right)^{}_{\rm L}}
\ \gamma^\mu \ V \left(\begin{matrix} d \cr
s \cr b \cr \end{matrix}\right)^{}_{\rm L} W^+_\mu \right] +
{\rm h.c.} \; ,
%     (1.1)
\end{eqnarray}
where all the fermion fields are the mass eigenstates, $U$ is the
$3\times 3$ Maki-Nakagawa-Sakata-Pontecorvo (MNSP) matrix \cite{Maki:1962mu,Pontecorvo:1967fh},
and $V$ denotes the $3\times 3$ Cabibbo-Kobayashi-Maskawa (CKM) matrix
\cite{Cabibbo:1963yz,Kobayashi:1973fv}. Note that unitarity is the only but powerful
constraint, imposed by the standard model itself, on $V$. This property,
together with the freedom of redefining the phases of six quark fields,
allows one to parametrize $V$ in terms of only four independent parameters,
such as three mixing angles and one CP-violating phase. In contrast,
whether the MNSP matrix $U$ is unitary or not depends on the mechanism
of neutrino mass generation~\cite{Xing:2011zza}
%%%%%%%%%%%%%%%%%%%%%%%%%%%%%%%%%%%%%%%
\footnote{For example, $U$ is exactly unitary in the type-II seesaw
mechanism~\cite{Konetschny:1977bn,Magg:1980ut,Cheng:1980qt,Lazarides:1980nt,Mohapatra:1980yp},
but it is non-unitary in the type-I~\cite{Minkowski:1977sc,Yanagida:1979as,GellMann:1980vs,Glashow:1980unknown1,Mohapatra:1979ia,Schechter:1980gr}
and type-III~\cite{Foot:1988aq} seesaw mechanisms due to the mixing of three
light neutrinos with some heavy degrees of freedom.}.
%%%%%%%%%%%%%%%%%%%%%%%%%%%%%%%%%%%%%%%
The bottom line is that any possible deviation of $U$ from unitarity
must be small, at most at the percent level, as constrained by
the available experimental data~\cite{Antusch:2006vwa,Antusch:2014woa}. That is
why $U$ is simply assumed to be unitary in dealing with current
neutrino oscillation data.

Given the basis in which the flavor eigenstates of three charged leptons
are identified with their mass eigenstates, the flavor eigenstates
of three active neutrinos and $n$ sterile neutrinos read as
\begin{eqnarray}
\left(\begin{matrix} \nu^{}_e \cr \nu^{}_\mu \cr \nu^{}_\tau \cr \vdots \cr
\end{matrix}
\right) = \left(\begin{matrix} U^{}_{e1} & U^{}_{e2} & U^{}_{e3} & \cdots \cr
U^{}_{\mu 1} & U^{}_{\mu 2} & U^{}_{\mu 3} & \cdots \cr
U^{}_{\tau 1} & U^{}_{\tau 2} & U^{}_{\tau 3} & \cdots \cr
\vdots & \vdots & \vdots & \ddots \cr \end{matrix} \right)
\left(\begin{matrix} \nu^{}_1 \cr \nu^{}_2 \cr \nu^{}_3 \cr \vdots \cr
\end{matrix}
\right) \; .\label{eq:intro:UPMNS}
%     (1.2)
\end{eqnarray}
where $\nu^{}_i$ is a neutrino mass eigenstate with the physical
mass $m^{}_i$ (for $i=1,2, \cdots, 3+n$). Eq. (1.1) tells us that a $\nu^{}_\alpha$
neutrino can be produced from the $W^+ + \ell^-_\alpha \to \nu^{}_\alpha$
interaction, and a $\nu^{}_\beta$ neutrino can be detected through the
$\nu^{}_\beta \to W^+ + \ell^-_\beta$ interaction (for
$\alpha, \beta = e, \mu, \tau$). So the $\nu^{}_\alpha \to \nu^{}_\beta$
oscillation may happen if the $\nu^{}_i$ beam with energy
$E \gg m^{}_i$ travels a proper distance $L$ in vacuum.
The amplitude of the $\nu^{}_\alpha \to \nu^{}_\beta$ oscillation
turns out to be
\begin{eqnarray}
A(\nu^{}_\alpha \to \nu^{}_\beta) & \hspace{-0.2cm} =
\hspace{-0.2cm} & \sum_i \left[A(W^+ + \ell^-_\alpha \to \nu^{}_i) \cdot
{\rm Prop}(\nu^{}_i) \cdot A(\nu^{}_i \to W^+ + \ell^-_\beta) \right]
\nonumber \\
& \hspace{-0.2cm} = \hspace{-0.2cm} &
\frac{1}{\sqrt{(U U^\dagger)^{}_{\alpha\alpha}
(U U^\dagger)^{}_{\beta\beta}}} \sum_i \left[U^*_{\alpha i}
\exp\left(\displaystyle -{\rm i} \frac{m^2_i L}{2E}\right) U^{}_{\beta
i}\right] \;
%     (1.3)
\end{eqnarray}
in the plane-wave expansion approximation \cite{Xing:2012kh,Li:2015oal}, where
$A(W^+ + \ell^-_\alpha \to \nu^{}_i) = U^*_{\alpha
i}/\sqrt{(U U^\dagger)_{\alpha\alpha}}~$, ${\rm Prop}(\nu^{}_i)$ and
$A(\nu^{}_i \to W^+ + \ell^-_\beta) = U^{}_{\beta
i}/\sqrt{(U U^\dagger)_{\beta\beta}}$ describe the production of
$\nu^{}_\alpha$ at the source, the propagation of free $\nu^{}_i$ over
a distance $L$ and the detection of $\nu^{}_\beta$ at the detector,
respectively. It is then straightforward to obtain the probability of the
$\nu^{}_\alpha \to \nu^{}_\beta$ oscillation
$P(\nu^{}_\alpha \to \nu^{}_\beta) \equiv
| A(\nu^{}_\alpha \to \nu^{}_\beta)|^2$; i.e.,
\begin{eqnarray}
P(\nu^{}_\alpha \to \nu^{}_\beta) = \frac{\displaystyle
\sum^{}_i |U^{}_{\alpha i}|^2
|U^{}_{\beta i}|^2 + 2 \sum^{}_{i<j} \left[ {\rm Re} \left(
U^{}_{\alpha i} U^{}_{\beta j} U^*_{\alpha j} U^*_{\beta i} \right)
\cos \Delta^{}_{ij} - {\rm Im} \left( U^{}_{\alpha i} U^{}_{\beta j}
U^*_{\alpha j} U^*_{\beta i} \right) \sin\Delta^{}_{ij} \right]}
{\displaystyle (UU^\dagger)^{}_{\alpha\alpha}
(UU^\dagger)^{}_{\beta\beta}} \; ,
%     (1.4)
\end{eqnarray}
where $\Delta^{}_{ij} \equiv \Delta m^2_{ij} L/(2E)$ with $\Delta
m^2_{ij} \equiv m^2_i - m^2_j$. The probability of the
$\overline{\nu}^{}_\alpha \to \overline{\nu}^{}_\beta$ oscillation
can easily be read off from Eq. (1.4) by making the replacement
$U \to U^*$.

As for the ``disappearance" reactor antineutrino oscillations to
be studied in this Yellow Book, we may simply take $\alpha = \beta = e$
for Eq. (1.4) and then arrive at
\begin{eqnarray}
P(\overline{\nu}^{}_e \to \overline{\nu}^{}_e) = 1 -
\frac{4}{\displaystyle \left(\sum_i |U^{}_{e i}|^2 \right)^2}
\sum^{}_{i<j} \left(|U^{}_{e i}|^2 |U^{}_{e j}|^2 \sin^2
\frac{\Delta m^{2}_{ij} L}{4 E} \right) \; . \label{eq:intro:peegeneral}
%     (1.5)
\end{eqnarray}
Note that the denominator on the right-hand side of Eq. (1.5) is not
equal to unity if there are heavy sterile antineutrinos which mix
with the active antineutrinos but do not take part in the flavor
oscillations (forbidden by kinematics). Note also that the
terrestrial matter effects on $P(\overline{\nu}^{}_e \to
\overline{\nu}^{}_e)$ are negligibly small in most cases, because
the typical value of $E$ is only a few MeV and that
of $L$ is usually less than 100 km for a realistic
reactor-based $\overline{\nu}^{}_e \to \overline{\nu}^{}_e$
oscillation experiment.

If the $3\times 3$ MNSP matrix $U$ is exactly unitary, it can be
parametrized in terms of three flavor mixing angles and three
CP-violating phases in the following ``standard" way:
\begin{eqnarray}
U & \hspace{-0.2cm} = \hspace{-0.2cm} &
\left( \begin{matrix} 1 & 0 & 0 \cr 0 & c^{}_{23} & s^{}_{23} \cr
0 & -s^{}_{23} & c^{}_{23} \cr \end{matrix} \right)
\left( \begin{matrix} c^{}_{13} & 0 & s^{}_{13} e^{-{\rm i}\delta} \cr
0 & 1 & 0 \cr -s^{}_{13} e^{{\rm i}\delta} & 0 & c^{}_{13} \cr
\end{matrix} \right)
\left( \begin{matrix} c^{}_{12} & s^{}_{12} & 0 \cr
-s^{}_{12} & c^{}_{12} & 0 \cr 0 & 0 & 1 \cr \end{matrix} \right) P^{}_\nu
\nonumber \\
& \hspace{-0.2cm} = \hspace{-0.2cm} &
\left( \begin{matrix} c^{}_{12} c^{}_{13} & s^{}_{12}
c^{}_{13} & s^{}_{13} e^{-{\rm i} \delta} \cr -s^{}_{12} c^{}_{23} -
c^{}_{12} s^{}_{13} s^{}_{23} e^{{\rm i} \delta} & c^{}_{12}
c^{}_{23} - s^{}_{12} s^{}_{13} s^{}_{23} e^{{\rm i} \delta} &
c^{}_{13} s^{}_{23} \cr s^{}_{12} s^{}_{23} - c^{}_{12} s^{}_{13}
c^{}_{23} e^{{\rm i} \delta} & -c^{}_{12} s^{}_{23} - s^{}_{12}
s^{}_{13} c^{}_{23} e^{{\rm i} \delta} & c^{}_{13} c^{}_{23} \cr
\end{matrix} \right) P^{}_\nu \; ,
%     (1.6)
\end{eqnarray}
where $c^{}_{ij} \equiv \cos\theta^{}_{ij}$ and $s^{}_{ij} \equiv
\sin\theta^{}_{ij}$ (for $ij = 12, 13, 23$) are defined, and
$P^{}_\nu = {\rm Diag}\{e^{{\rm i}\rho}, e^{{\rm i}\sigma}, 1\}$ denotes
the diagonal Majorana phase matrix which has nothing to do with
neutrino oscillations. In this case,
\begin{eqnarray}
P(\overline{\nu}^{}_e \to \overline{\nu}^{}_e) = 1 -
\sin^2 2\theta^{}_{12} c^4_{13} \sin^2
\frac{\Delta m^{2}_{21} L}{4 E} - \sin^2 2\theta^{}_{13}
\left[ c^2_{12} \sin^2 \frac{\Delta m^{2}_{31} L}{4 E}
+ s^2_{12} \sin^2 \frac{\Delta m^{2}_{32} L}{4 E}
\right] \; ,
%     (1.7)
\end{eqnarray}
in which $\Delta m^2_{32} = \Delta m^2_{31} - \Delta m^2_{21}$.
The oscillation terms driven by $\Delta m^2_{21}$ and
$\Delta m^2_{31} \simeq \Delta m^2_{32}$ can therefore be used
to determine $\theta^{}_{12}$ and $\theta^{}_{13}$, respectively.
The JUNO experiment aims to measure the flux rate and energy spectrum
of $\overline{\nu}^{}_e \to \overline{\nu}^{}_e$ oscillations
to an unprecedentedly good degree of accuracy, especially to pin down the
sign of $\Delta m^2_{31}$ or equivalently the neutrino mass ordering.

To test unitarity of the $3\times 3$ MNSP matrix $U$ or to probe
possible sterile neutrino effects in the foreseeable future,
one should better make use of Eq. (1.5) instead of Eq. (1.7) to
analyze the relevant experimental data. This point will be made much
clearer later in the following sections.

\subsubsection{Known and unknown neutrino oscillation parameters}

In the standard three-flavor framework without any extra neutrino
species, there are six independent parameters which govern the behaviors of
neutrino oscillations: the neutrino mass-squared differences
$\Delta m^2_{21}$ and $\Delta m^2_{31}$, the flavor mixing angles
$\theta^{}_{12}$, $\theta^{}_{13}$ and $\theta^{}_{23}$, and the
Dirac CP-violating phase $\delta$.
Since 1998, a number of atmospheric, solar,
accelerator and reactor experiments \cite{Agashe:2014kda} have provided us with very
compelling evidence for neutrino (or antineutrino) oscillations,
from which $\Delta m^2_{21}$, $|\Delta m^2_{31}|$, $\theta^{}_{12}$,
$\theta^{}_{13}$ and $\theta^{}_{23}$ have well been determined.
The ongoing and future neutrino oscillation experiments are expected
to fix the sign of $\Delta m^2_{31}$ and probe the value of $\delta$.

A global three-flavor analysis of the currently available experimental data on
solar (SNO, Super-Kamiokande, Borexino), atmospheric (Super-Kamiokande),
accelerator (MINOS, T2K) and reactor (KamLAND, Daya Bay, RENO) neutrino
(or antineutrino) oscillations has recently been done by several
groups~\cite{Capozzi:2013csa,Forero:2014bxa,Gonzalez-Garcia:2014bfa}. For simplicity, here we only quote
the main results of Ref.~\cite{Capozzi:2013csa} as summarized in Table~1.1
%%%%%%%%%%%%%%%%%%%%%%%%%%%%%%%%%%%%%%%%%%%%%%%%%%%%%%%%%
\footnote{Note that the notations $\delta m^2 \equiv m^2_2 - m^2_1$ and
$\Delta m^2 \equiv m^2_3 - (m^2_1 + m^2_2)/2$ have been used in
Ref. \cite{Capozzi:2013csa}.
Their relations with $\Delta m^2_{21}$ and $\Delta m^2_{31}$ are rather
simple: $\Delta m^2_{21} = \delta m^2$ and
$\Delta m^2_{31} = \Delta m^2 + \delta m^2/2$.},
%%%%%%%%%%%%%%%%%%%%%%%%%%%%%%%%%%%%%%%%%%%%%%%%%%%%%%%%%
in which the normal neutrino mass ordering ($m^{}_1 < m^{}_2 < m^{}_3$)
and the inverted neutrino mass ordering ($m^{}_3 < m^{}_1 < m^{}_2$)
are separately considered. Some comments are in order.
\begin{itemize}
\item     The output values of $\theta^{}_{13}$, $\theta^{}_{23}$ and
$\delta$ in such a global fit are sensitive to the sign of
$\Delta m^2_{31}$. That is why
it is crucial to determine whether $\Delta m^2_{31} >0$ or
$\Delta m^2_{31} <0$ (i.e., whether $m^{}_1$ or $m^{}_3$ is
the smallest neutrino mass) in JUNO and a few other experiments.

\item     The hint $\delta \neq 0^\circ$ (or $180^\circ$) at the
$1\sigma$ level is preliminary but
encouraging, because it implies a potential effect of leptonic CP
violation which is likely to show up in some long-baseline
neutrino oscillation experiments in the foreseeable future.

\item     The possibility $\theta^{}_{23} = 45^\circ$ cannot be
excluded at the $2\sigma$ level, and thus a more precise
determination of $\theta^{}_{23}$ is desirable in order to resolve
its octant. The latter is important since
it can help fix the geometrical structure of the MNSP matrix $U$.
\end{itemize}
Note also that $|U^{}_{\mu i}| = |U^{}_{\tau i}|$ (for $i=1,2,3$),
the so-called $\mu$-$\tau$ permutation
symmetry of $U$ itself, holds if either the
conditions $\theta^{}_{13} = 0^\circ$ and $\theta^{}_{23} = 45^\circ$
or the conditions $\delta = 90^\circ$ (or $270^\circ$) and
$\theta^{}_{23} = 45^\circ$ are satisfied. Now that
$\theta^{}_{13} = 0^\circ$ has definitely been ruled out by the
Daya Bay experiment \cite{An:2012eh,An:2012bu,An:2013zwz}, it is imperative
to know the values of $\theta^{}_{23}$ and $\delta$ as accurately
as possible, so as to fix the strength of $\mu$-$\tau$ symmetry
breaking associated with the structure of $U$.
%%%%%%%%%%%%%%%%%%%%%%%%%% Table 1.1 %%%%%%%%%%%%%%%%%%%%%%%%%%%%
%%%%%%%%%%%%%%%%%%%%%%%%%%%%%%%%%%%%%%%%%%%%%%%%%%%%%%%%%%%%%%%%%
\begin{table}[t]
\vspace{-0.25cm}
\begin{center}
\caption{The best-fit values, together with the 1$\sigma$, 2$\sigma$
and 3$\sigma$ intervals, for the six three-flavor neutrino oscillation
parameters from a global analysis of current experimental data
\cite{Capozzi:2013csa}.}
\vspace{0.5cm}
\begin{tabular}{c|c|c|c|c}
\hline
\hline
Parameter & Best fit & 1$\sigma$ range & 2$\sigma$ range & 3$\sigma$ range \\
\hline
\multicolumn{5}{c}{Normal neutrino mass ordering
$(m^{}_1 < m^{}_2 < m^{}_3$)} \\ \hline
%------------------------------------------------------------
$\Delta m^2_{21}/10^{-5} ~{\rm eV}^2$
& $7.54$  & 7.32 --- 7.80 & 7.15 --- 8.00 & 6.99 --- 8.18 \\
%------------------------------------------------------------
$\Delta m^2_{31}/10^{-3} ~ {\rm eV}^2$~ & $2.47$
& 2.41 --- 2.53 & 2.34 --- 2.59 & 2.26 --- 2.65 \\
%------------------------------------------------------------
$\sin^2\theta_{12}/10^{-1}$
& $3.08$ & 2.91 --- 3.25 & 2.75 --- 3.42 & 2.59 --- 3.59 \\
%------------------------------------------------------------
$\sin^2\theta_{13}/10^{-2}$
& $2.34$ & 2.15 --- 2.54 & 1.95 --- 2.74 & 1.76 --- 2.95 \\
%------------------------------------------------------------
$\sin^2\theta_{23}/10^{-1}$
& $4.37$  & 4.14 --- 4.70 & 3.93 --- 5.52 & 3.74 --- 6.26 \\
%------------------------------------------------------------
$\delta/180^\circ$ &  $1.39$ & 1.12 --- 1.77 & 0.00 --- 0.16
$\oplus$ 0.86 --- 2.00 & 0.00 --- 2.00 \\ \hline
%%%%%%%%%%%%%%%%%%%%%%%%%%%%%%%%%%%%%%%%%%%%%%%%%%%%%%%%%%%%%
\multicolumn{5}{c}{Inverted neutrino mass ordering
$(m^{}_3 < m^{}_1 < m^{}_2$)} \\ \hline
%------------------------------------------------------------
$\Delta m^2_{21}/10^{-5} ~{\rm eV}^2$
& $7.54$  & 7.32 --- 7.80 & 7.15 --- 8.00 & 6.99 --- 8.18 \\
%------------------------------------------------------------
$\Delta m^2_{13}/10^{-3} ~ {\rm eV}^2$~ & $2.42$
& 2.36 --- 2.48 & 2.29 --- 2.54 & 2.22 --- 2.60 \\
%------------------------------------------------------------
$\sin^2\theta_{12}/10^{-1}$
& $3.08$ & 2.91 --- 3.25 & 2.75 --- 3.42 & 2.59 --- 3.59 \\
%------------------------------------------------------------
$\sin^2\theta_{13}/10^{-2}$
& $2.40$ & 2.18 --- 2.59 & 1.98 --- 2.79 & 1.78 --- 2.98 \\
%------------------------------------------------------------
$\sin^2\theta_{23}/10^{-1}$
& $4.55$  & 4.24 --- 5.94 & 4.00 --- 6.20 & 3.80 --- 6.41 \\
%------------------------------------------------------------
$\delta/180^\circ$ &  $1.31$ & 0.98 --- 1.60 & 0.00 --- 0.02
$\oplus$ 0.70 --- 2.00 & 0.00 --- 2.00 \\ \hline\hline
%%%%%%%%%%%%%%%%%%%%%%%%%%%%%%%%%%%%%%%%%%%%%%%%%%%%%%%%%%%%%
\end{tabular}
\end{center}
\end{table}
%%%%%%%%%%%%%%%%%%%%%%%%%%%%%%%%%%%%%%%%%%%%%%%%%%%%%%%%%%%%%%%%%
%%%%%%%%%%%%%%%%%%%%%%%%%%%%%%%%%%%%%%%%%%%%%%%%%%%%%%%%%%%%%%%%%

\subsection{Open Issues of Massive Neutrinos}
\label{subsec:openissue}

There are certainly many open questions in neutrino physics, but
here we concentrate on some intrinsic flavor issues of massive
neutrinos which may more or less be addressed in the JUNO
experiment.

\subsubsection{The nature of neutrinos and their mass spectrum}

{\it Question (1): Dirac or Majorana?} ---
If a massive neutrino is the Dirac particle, just like the
electron, then it must be distinguishable from its antiparticle
because they possess the opposite lepton numbers. By definition,
a massive Majorana neutrino is its own antiparticle \cite{Majorana:1937vz},
leading to lepton number violation as a direct consequence.
The tiny masses of three known neutrinos make it extremely
difficult to identify their nature, i.e., whether they are
the Dirac or Majorana particles. At present the only experimentally
feasible way to probe the Majorana nature of massive neutrinos is to
observe the neutrinoless double-beta ($0\nu\beta\beta$) decays
of some even-even nuclei, $N(A,Z) \to N(A,Z+2) + 2 e^-$,
which occur via an exchange of the virtual Majorana neutrinos
between two associated beta decays \cite{Furry:1939qr}. The effective
neutrino mass term in the $0\nu\beta\beta$ decay is defined as
\begin{eqnarray}
\langle m\rangle^{}_{ee} \equiv \sum^{}_i \left(
m^{}_i U^2_{e i} \right) \; ,
%     (1.8)
\end{eqnarray}
which is in general sensitive to all the three neutrino masses
(including the sign of $\Delta m^2_{31}$),
two of the three flavor mixing angles and two of the three
CP-violating phases.

So far no convincing evidence for an occurrence of the
$0\nu\beta\beta$ decay has been established, although a lot of
experimental efforts have been made in the past few decades.
Within the standard three-flavor scheme, the inverted neutrino
mass ordering (i.e., $\Delta m^2_{31} <0$) or a near neutrino
mass degeneracy will allow $|\langle m\rangle^{}_{ee}| \geq 0.01$ eV
\cite{Rodejohann:2011mu,Dell'Oro:2014yca,Bilenky:2014uka}, perhaps accessible in the next-generation
$0\nu\beta\beta$-decay experiments.

{\it Question (2): Normal or inverted mass ordering?} ---
Now that $\Delta m^2_{21} > 0$ has been fixed but
the sign of $\Delta m^2_{31}$ remains unknown, we are
left with two possibilities for the mass ordering of three
active neutrinos whose family indices are specified by
the weak charged-current interactions in Eq. (1.1):
the normal case $m^{}_1 < m^{}_2 < m^{}_3$ and the inverted
case $m^{}_3 < m^{}_1 < m^{}_2$. The former is ``normal" in
the sense that it is parallel to the mass ordering of three
charged leptons or three quarks of the same electric charge (i.e.,
$m^{}_e \ll m^{}_\mu \ll m^{}_\tau$, $m^{}_u \ll m^{}_c \ll m^{}_t$
and $m^{}_d \ll m^{}_s \ll m^{}_b$, as shown in Fig. 1.1).
However, a good (model-independent) theoretical reason for
either $\Delta m^2_{31} > 0$ or $\Delta m^2_{31} < 0$ has been
lacking
%%%%%%%%%%%%%%%%%%%%%%%%%%%%%%%%%
\footnote{If the neutrino mass ordering is finally found to be
inverted, one may always ``renormalize" it to
$m^\prime_1 < m^\prime_2 < m^\prime_3$ by setting
$m^\prime_1 = m^{}_3$, $m^\prime_2 = m^{}_1$
and $m^\prime_3 = m^{}_2$, equivalent to a
transformation $(\nu^{}_1, \nu^{}_2, \nu^{}_3) \to
(\nu^\prime_2, \nu^\prime_3, \nu^\prime_1)$. In this case
the matrix elements of $U$ must be reordered in a self-consistent
way: $U \to U^\prime$, in which $U^\prime_{\alpha 1} =
U^{}_{\alpha 3}$, $U^\prime_{\alpha 2} = U^{}_{\alpha 1}$ and
$U^\prime_{\alpha 3} = U^{}_{\alpha 2}$ (for $\alpha = e, \mu, \tau$).
Therefore, such a treatment does not change any physical content
of massive neutrinos.}.
%%%%%%%%%%%%%%%%%%%%%%%%%%%%%%%%%%%%%%%%%%%%%%%%%%%%%%%%%%%%%%%%%
%%%%%%%%%%%%%%%%%% Fig. 1.1 %%%%%%%%%%%%%%%
\begin{figure*}
\centering
\includegraphics[width=.8\textwidth]{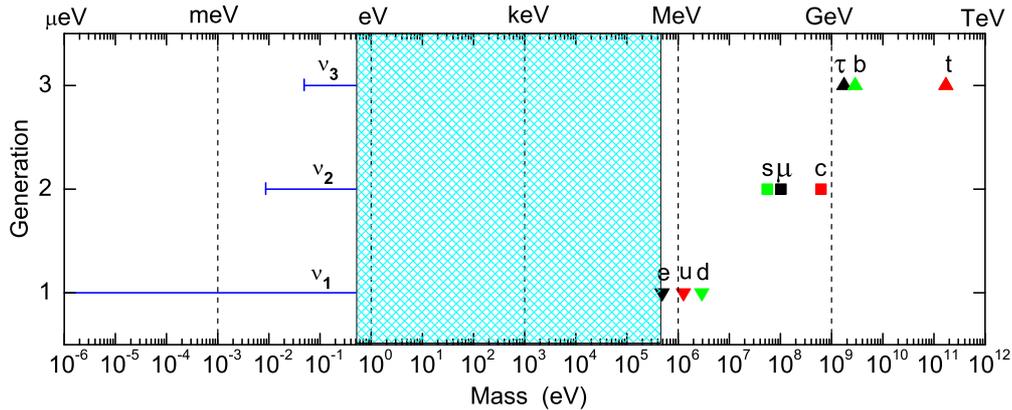}
\vspace{-0.23cm}
\caption{A schematic illustration of the ``flavor hierarchy'' and
``flavor desert'' in the fermion mass spectrum at the
electroweak scale \cite{Li:2010vy}. Here the neutrino masses are assumed
to have a normal ordering.}
\end{figure*}
%%%%%%%%%%%%%%%%%%%%%%%%%%%%%%%%%%%%%%%%%

The sign of $\Delta m^2_{31}$ is of fundamental importance because
it can impact on many important processes in particle physics,
astrophysics and cosmology. The $0\nu\beta\beta$ decay is just a
typical example of this kind. In the atmospheric and long-baseline
accelerator neutrino experiments, the behaviors
of flavor oscillations are sensitive to a combination of $\Delta m^2_{31}$
and the terrestrial matter term $2\sqrt{2} \ G^{}_{\rm F} N^{}_e E$,
where $N^{}_e$ denotes the number density of electrons in the matter
background and $E$ is the neutrino beam energy~\cite{Bernabeu:2003yp}.
The supernova neutrino oscillations are also affected by the sign of
$\Delta m^2_{31}$ for a similar reason~\cite{Dighe:1999bi}.
Meanwhile, the interference effects of two atmospheric mass-squared
differences in reactor neutrino vacuum oscillations are sensitive to the
sign of $\Delta m^2_{31}$~\cite{Petcov:2001sy}.
It is therefore possible to probe the neutrino
mass ordering via the atmospheric, long-baseline accelerator, reactor
and supernova neutrino oscillations~\cite{Qian:2015waa,Patterson:2015xja}.
On the other hand, the ``right" sign of $\Delta m^2_{31}$ may help the seesaw and
leptogenesis mechanisms \cite{Fukugita:1986hr} work well to simultaneously
interpret the origin of tiny neutrino masses and the observed baryon
number asymmetry of the Universe \cite{Buchmuller:2004nz,Davidson:2008bu}.

{\it Question (3): The absolute mass scale?} --- Since the flavor
oscillations of massive neutrinos are only sensitive to the neutrino
mass-squared differences, a determination of the absolute neutrino
mass scale has to rely on some non-oscillation experiments.
Searching for the $0\nu\beta\beta$ decay is one of the feasible ways
for this purpose if massive neutrinos are the Majorana particles,
because the magnitude of its effective mass $\langle m\rangle^{}_{ee}$
is governed by $m^{}_i$ as shown in Eq. (1.8). The upper bound of
$|\langle m\rangle^{}_{ee}|$ has been set to be about $0.2$ eV
by the present $0\nu\beta\beta$-decay experiments~\cite{Agostini:2013mzu},
and the sensitivities of the future experiments are likely to reach
the meV level~\cite{Rodejohann:2011mu,Dell'Oro:2014yca,Bilenky:2014uka}. Another way is to detect the effective
neutrino mass
\begin{eqnarray}
\langle m\rangle^{}_e \equiv \sqrt{\sum_i \left(m^2_i |U^{}_{e i}|^2
\right)} \;
%     (1.9)
\end{eqnarray}
in the beta decays, such as $^3_1 {\rm H} \to ~
^3_2 {\rm He} + e^- + \overline{\nu}^{}_e$. The most promising
experiment of this kind is the KATRIN experiment, which may
hopefully probe $\langle m\rangle^{}_e$ with a sensitivity of
about $0.2$ eV~\cite{Bornschein:2003xi}.

Furthermore, one may get useful information on the mass scale of
light neutrinos from cosmology and astrophysics. A global analysis
of current cosmological data (especially those on the cosmic
microwave background and large-scale structures) has actually
provided us with the most powerful sensitivity to the sum of
light neutrino masses,
\begin{eqnarray}
\Sigma^{}_\nu  \equiv \sum_i m^{}_i \; .
%     (1.10)
\end{eqnarray}
For example, $\Sigma^{}_\nu < 0.23$ eV has recently been reported
by the Planck Collaboration at the $95\%$ confidence level~\cite{Ade:2013zuv}.
Given the values of $\Delta m^2_{21}$ and $|\Delta m^2_{31}|$
extracted from current neutrino oscillation data, the results of
$|\langle m\rangle^{}_{ee}|$, $\langle m\rangle^{}_e$ and $\Sigma^{}_\nu$
are all sensitive to the sign of $\Delta m^2_{31}$.

Let us restrict ourselves to the standard three-flavor case to
illustrate the correlation among the above three effective mass terms
in Fig. 1.2, where the $3\sigma$ ranges of $\Delta m^2_{21}$,
$|\Delta m^2_{31}|$, $\theta^{}_{12}$ and $\theta^{}_{13}$ as given
in Table 1.1 are input, and the CP-violating phases of $U$ are
all allowed to vary in the $[0^\circ, 360^\circ)$ interval.
%%%%%%%%%%%%%%%%%% Fig. 1.2 %%%%%%%%%%%%%%%
\begin{figure*}
\centering
\includegraphics[width=.9\textwidth]{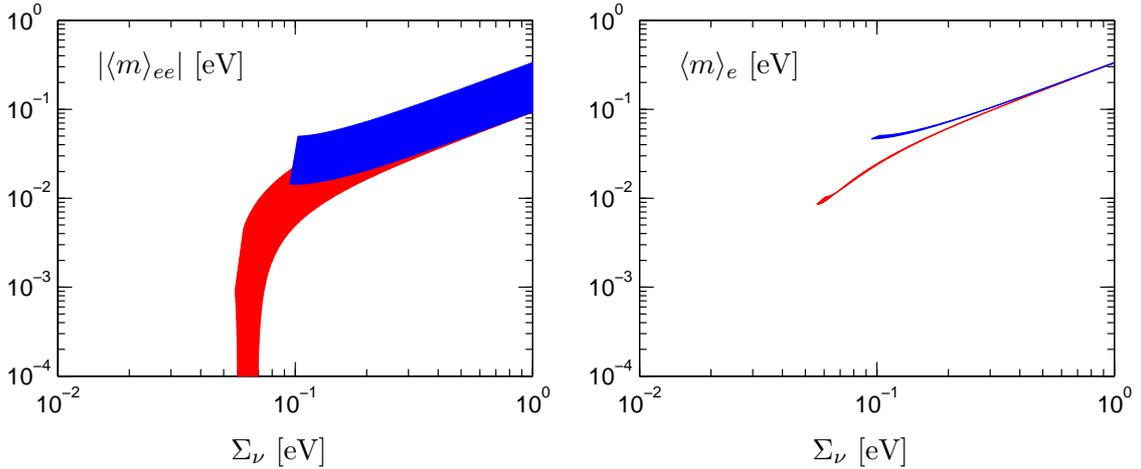}
\vspace{-0.15cm}
\caption{An illustration of the correlation between $\Sigma^{}_\nu$ and
$|\langle m\rangle^{}_{ee}|$ (left panel) and that between $\Sigma^{}_\nu$
and $\langle m\rangle^{}_e$ (right panel) by using the $3\sigma$
inputs as given in Table 1.1. Here the red (blue) region corresponds to
the normal (inverted) neutrino mass ordering.}
\end{figure*}
%%%%%%%%%%%%%%%%%%%%%%%%%%%%%%%%%%%%%%%%%

\subsubsection{Lepton flavor mixing pattern and CP violation}

{\it Question (4): The octant of $\theta^{}_{23}$?} --- Although the
smallest neutrino mixing angle $\theta^{}_{13}$ has already been
determined in the Daya Bay experiment~\cite{An:2012eh,An:2012bu,An:2013zwz}, the geometric
structure of the $3\times 3$ MNSP matrix $U$ cannot be fully fixed until the
octant of $\theta^{}_{23}$ and the value of $\delta$ are both known.
Current experimental data strongly support
$|U^{}_{\mu i}| \simeq |U^{}_{\tau i}|$
(for $i=1,2,3$), i.e., an approximate $\mu$-$\tau$ permutation symmetry
of $U$ itself \cite{Xing:2014zka}. In particular, $\theta^{}_{23} = 45^\circ$
is favored in many neutrino mass models as a consequence of the exact
$\mu$-$\tau$ symmetry or some other kinds of flavor symmetries.
In this sense the deviation of $\theta^{}_{23}$ from $45^\circ$
serves for a useful model discriminator \cite{Luo:2014upa,Zhou:2014sya} and deserves a precise
measurement in the upcoming atmospheric and long-baseline neutrino
oscillation experiments.

Taking account of $\sin\theta^{}_{12} < 1/\sqrt{2}$ and
$\sin\theta^{}_{13} \ll 1$, one may roughly expect
$|U^{}_{e1}| > |U^{}_{\mu 3}| \sim |U^{}_{\tau 3}| \sim |U^{}_{\mu 2}|
\sim |U^{}_{\tau 2}| \sim |U^{}_{e 2}| > |U^{}_{\mu 1}| \sim
|U^{}_{\tau 1}| > |U^{}_{e 3}|$, where ``$\sim$" means that the two
MNSP matrix elements are comparable in magnitude. Hence the pattern
of $U$ seems to be partly anarchical and partly hierarchical.
In comparison, the CKM quark flavor mixing
matrix $V$ is found to possess a clearly hierarchical structure:
$|V^{}_{tb}| > |V^{}_{ud}| > |V^{}_{cs}| \gg
|V^{}_{us}| > |V^{}_{cd}| \gg |V^{}_{cb}| > |V^{}_{ts}| \gg
|V^{}_{td}| > |V^{}_{ub}|$ since its three mixing angles satisfy
$\vartheta^{}_{12} \gg \vartheta^{}_{23} \gg \vartheta^{}_{13}$,
which should have something to do with the strong quark mass hierarchies.

{\it Question (5): The Dirac CP-violating phase $\delta$?} --- In the
standard three-flavor scheme the phase parameter $\delta$ of $U$ is
fundamentally important because it uniquely controls the strength of
leptonic CP and T violation in neutrino oscillations. Under CPT
invariance, the CP- and T-violating asymmetries
${\cal A}^{}_{\alpha\beta} \equiv P(\nu^{}_\alpha \to \nu^{}_\beta) -
P(\overline{\nu}^{}_\alpha \to \overline{\nu}^{}_\beta)
= P(\nu^{}_\alpha \to \nu^{}_\beta) - P(\nu^{}_\beta \to \nu^{}_\alpha)$
in vacuum are explicitly given by
\begin{eqnarray}
{\cal A}^{}_{\alpha\beta} =  2 \sin 2\theta^{}_{12}
\cos\theta^{}_{13} \sin 2\theta^{}_{13} \sin 2\theta^{}_{23}
\sin\delta \sum_\gamma \epsilon^{}_{\alpha\beta\gamma}
\sin\frac{\Delta m^{2}_{21} L}{4E} \sin\frac{\Delta m^{2}_{31} L} {4E}
\sin\frac{\Delta m^{2}_{32} L}{4E} \; ,
%     (1.11)
\end{eqnarray}
in which the Greek subscripts run over $e$, $\mu$ and $\tau$.
It becomes obvious that CP or T violation is a three-flavor
``appearance" effect, and a measurement of this effect will allow
us to determine the value of $\delta$. Given the best-fit values
$\theta^{}_{12} \simeq 33.7^\circ$, $\theta^{}_{13} \simeq 8.8^\circ$
and $\theta^{}_{23} \simeq 40.7^\circ$ in the normal neutrino mass ordering
case, for example, the coefficient in front of the oscillating term of
${\cal A}^{}_{\alpha\beta}$ is about $0.5 \times \sin\delta$.
Current neutrino oscillation data seem to hint at
$\delta \sim 270^\circ$ (see Table 1.1) --- an encouraging implication
of large CP violation in the lepton sector.

In a realistic medium- or long-baseline
neutrino oscillation experiment, however, the terrestrial matter
effects may modify the oscillation behaviors and thus affect the
determination of $\delta$. This kind of contamination is negligible
for a variety of experiments provided the neutrino beam energy $E$
and the baseline length $L$ satisfy the condition
$10^{-7} \left(L/{\rm km}\right)^2 \left({\rm GeV}/E\right) \ll 1$.
If the unitarity of the $3\times 3$
MNSP matrix $U$ is slightly violated due to the existence of extra
species of massive neutrinos, it is also possible for
new CP-violating effects to show up in neutrino oscillations
\cite{Xing:2011zza}.

{\it Question (6): The Majorana CP-violating phases $\rho$ and $\sigma$?}
--- If the Majorana nature of massive neutrinos is finally established
through a convincing measurement of the $0\nu\beta\beta$
decay, one will be left with a question which is probably most challenging
in neutrino physics --- how to determine the CP-violating phases
$\rho$ and $\sigma$ in the standard three-flavor scheme? Because the
$0\nu\beta\beta$ decay is a CP-conserving process, its effective mass term
$\langle m\rangle^{}_{ee}$ can only provide some indirect information
on the combinations of $\delta$, $\rho$ and $\sigma$. Hence a direct
determination of $\rho$ and $\sigma$ depends on the observation of
those processes which are both lepton-number-violating and CP-violating.
Although the measurement of neutrino-antineutrino oscillations can in
principle allow us to probe all the three CP-violating phases and
even the absolute neutrino mass scale \cite{Xing:2013woa}, it is in practice
impossible to do such an experiment since the corresponding oscillation
probabilities are suppressed by the factors
$m^2_i/E^2 \lesssim 10^{-12}$.

The $3\times 3$ Majorana neutrino mass matrix $M^{}_\nu$ can be
reconstructed in terms of three neutrino masses, three flavor mixing
angles and three CP-violating phases in the basis where the flavor
eigenstates of three charged leptons are identified with their mass
eigenstates, and its six independent elements are
\begin{eqnarray}
\langle m\rangle^{}_{\alpha \beta} \equiv \sum_i \left(m^{}_i
U^{}_{\alpha i} U^{}_{\beta i} \right) \; ,
%     (1.12)
\end{eqnarray}
where $\alpha$ and $\beta$ run over $e$, $\mu$ and $\tau$.
Current experimental constraints on the magnitudes of
$\langle m\rangle^{}_{\alpha\beta}$ can be found in Ref. \cite{Xing:2015zha}.
While a theoretical model is always possible to predict the moduli and
phases of $\langle m\rangle^{}_{\alpha \beta}$, its correctness or
wrongness will not be testable until sufficient information about
the CP-violating phases of $U$ is experimentally obtained.

\subsubsection{Extra neutrino species and unitarity tests}

{\it Question (7): Extra light or heavy sterile neutrinos?} ---
One of the fundamental questions in neutrino physics and cosmology
is whether there exist extra species of neutrinos which do not
directly participate in the standard weak interactions.
Such {\it sterile} neutrinos are certainly hypothetical,
but their possible existence is either theoretically motivated or
experimentally implied \cite{Abazajian:2012ys}. For example, the canonical (type-I)
seesaw mechanism \cite{Minkowski:1977sc,Yanagida:1979as,GellMann:1980vs,Glashow:1980unknown1,Mohapatra:1979ia,Schechter:1980gr}
provides an elegant interpretation of the
small masses of $\nu^{}_i$ (for $i=1,2,3$) with the help of two or
three heavy sterile neutrinos, and the latter can even help account
for the observed matter-antimatter asymmetry of the Universe via the
leptogenesis mechanism~\cite{Fukugita:1986hr}. On the experimental side, the LSND~\cite{Aguilar:2001ty}, MiniBooNE~\cite{AguilarArevalo:2010wv}
and reactor antineutrino~\cite{Mention:2011rk}
anomalies can all be explained as the active-sterile antineutrino
oscillations in the assumption of one or two species of sterile
antineutrinos whose masses are below 1 eV~\cite{Kopp:2013vaa,Giunti:2013aea}.
Furthermore, a careful analysis of the existing data on the
cosmic microwave background
anisotropy, galaxy clustering and supernovae Ia seems
to favor one species of sterile neutrinos at the sub-eV
mass scale~\cite{Hamann:2010bk,Hamann:2011ge,Giusarma:2011ex}.
On the other hand, sufficiently long-lived sterile
neutrinos in the keV mass range might serve as a good candidate for
warm dark matter if they were present in the early Universe \cite{Bode:2000gq}.
That is why a lot of attention has been paid to sterile neutrinos.
No matter how small or how large the mass scale of sterile neutrinos
is, they are undetectable unless they mix with three active neutrinos
to some extent. The active-sterile neutrino mixing can slightly
modify the behaviors of the standard three-flavor neutrino
oscillations, as shown in Eqs. (1.2)---(1.4).

As for the $\overline{\nu}^{}_e \to \overline{\nu}^{}_e$ oscillation
in a reactor antineutrino experiment, its probability
$P(\overline{\nu}^{}_e \to \overline{\nu}^{}_e)$ is governed by
Eq. (1.5). The heavy sterile antineutrinos do not participate in
any flavor oscillations, but they may violate the unitarity of
the $3\times 3$ MNSP matrix. In comparison, the light sterile
antineutrinos can contribute extra oscillation terms to
$P(\overline{\nu}^{}_e \to \overline{\nu}^{}_e)$. The JUNO
experiment will therefore be a good playground to probe or
constrain the effects of sterile antineutrinos.

{\it Question (8): Direct and indirect non-unitary effects?} ---
In the presence of small mixing between 3 active and
$n$ sterile neutrinos, the $3\times 3$ MNSP matrix becomes
a submatrix of the $(3+n) \times (3+n)$ unitary matrix which
describes the overall flavor mixing effects,
as shown in Eq. (1.2). Hence the $3\times 3$ MNSP matrix itself
must be non-unitary. From the point of view of neutrino oscillations,
one may classify its possible non-unitary effects into three
categories \cite{Xing:2012kh,Li:2015oal}:
\begin{itemize}
\item     the {\it indirect} non-unitary effect arising from the heavy
sterile neutrinos which are kinematically forbidden to take part in
neutrino oscillations;

\item     the {\it direct} non-unitary effect caused by the light sterile
neutrinos which are able to participate in neutrino oscillations;

\item     the {\it interplay} of the direct and indirect non-unitary
effects in a flavor mixing scenario including both light and heavy
sterile neutrinos.
\end{itemize}
An experimental test of the unitarity of the $3\times 3$ MNSP matrix
is therefore important to probe or constrain the flavor mixing parameters
of possible new physics associated with sterile neutrinos, and it can
theoretically shed light on the underlying dynamics responsible for
neutrino mass generation and lepton flavor mixing (e.g., the
$3\times 3$ MNSP matrix is exactly
unitary in the type-II~\cite{Konetschny:1977bn,Magg:1980ut,Cheng:1980qt,Lazarides:1980nt,Mohapatra:1980yp}
seesaw mechanism but non-unitary in the
type-I~\cite{Minkowski:1977sc,Yanagida:1979as,GellMann:1980vs,Glashow:1980unknown1,Mohapatra:1979ia,Schechter:1980gr}
and type-III~\cite{Foot:1988aq} seesaw mechanisms).

In the quark sector, the unitarity of the CKM matrix $V$ has been
tested to an impressive degree of accuracy. For instance,
$|V^{}_{ud}|^2 + |V^{}_{us}|^2 + |V^{}_{ub}|^2 = 0.9999 \pm 0.0006$
and $|V^{}_{ud}|^2 + |V^{}_{cd}|^2 + |V^{}_{td}|^2 = 1.000 \pm 0.004$
for the first row and column of $V$, respectively \cite{Agashe:2014kda}.
Hence the room for possible new physics which may violate the
unitarity of $V$ must be extremely small. In comparison, a preliminary
constraint on the sum of $|U^{}_{e1}|^2$, $|U^{}_{e2}|^2$ and
$|U^{}_{e3}|^2$ in the lepton sector is
\begin{eqnarray}
|U^{}_{e1}|^2 + |U^{}_{e2}|^2 + |U^{}_{e3}|^2 = 0.9979 \cdots 0.9998 \;
%     (1.13)
\end{eqnarray}
at the $90\%$ confidence level \cite{Antusch:2006vwa,Antusch:2014woa},
implying that the $3\times 3$ MNSP matrix $U$
is allowed to be non-unitary only at the ${\cal O}(10^{-3})$ level.
But it should be noted that such a stringent constraint is obtained
in the assumption of {\it minimal unitarity violation}, and there
might be the effects of unitarity violation at the percent level
in the lepton sector.
The JUNO experiment will allow us to determine $|U^{}_{e1}|$,
$|U^{}_{e2}|$ and $|U^{}_{e3}|$ to a much better degree of
accuracy via a precision measurement of the
$\overline{\nu}^{}_e \to \overline{\nu}^{}_e$ oscillation, and then
examine whether the sum
$|U^{}_{e1}|^2 + |U^{}_{e2}|^2 + |U^{}_{e3}|^2$
deviates from one or not.

Of course, there are many more open questions, which are more or
less associated with massive neutrinos and their various consequences,
in particle physics, astrophysics and cosmology. Typical examples
of this kind include how to detect the cosmic neutrino (or antineutrino)
background, how to detect the supernova neutrino burst and (or) the
supernova relic neutrino background, how to detect the ultrahigh-energy
cosmic neutrinos, etc. In any case the JUNO experiment is expected to
clarify a part of the flavor issues in the lepton sector and help
resolve some of the fundamental problems about the origin and evolution
of the Universe.

\subsection{JUNO Experiment}
\label{subsec:juno}

The Jiangmen Underground Neutrino Observatory (JUNO) is a multi-purpose neutrino experiment. It was proposed in 2008 to determine the neutrino mass hierarchy by detecting reactor antineutrinos from the Daya Bay nuclear power plant (NPP)~\cite{Zhan:2008id,Zhan:2009rs,yfwang2008,caoj2009}, thus formerly known as ``Daya Bay II experiment". The mass hierarchy determination requires equal baselines from the detector to all reactor cores to avoid cancellation of the oscillation dephasing effect. Due to the complex and unclear layout of the future nuclear power plants in the neighborhood, the experiment was moved to Jiangmen city in Guangdong province in August 2012, and named as JUNO in 2013.

The site location is optimized to have the best sensitivity for the mass hierarchy determination, which is at 53 km from both the Yangjiang and Taishan NPPs~\cite{Li:2013zyd}.
The neutrino detector is a liquid scintillator (LS) detector with a 20 kton fiducial mass, deployed in a laboratory 700 meter underground.
The experimental site and the detector will be described in this section.

The JUNO project was approved by Chinese Academy of Sciences in February 2013. Data taking is expected in 2020.

\subsubsection{Experimental site}
\label{subsubsec:site}

\begin{figure}[htb!]
\centering
\includegraphics[width=0.7\textwidth]{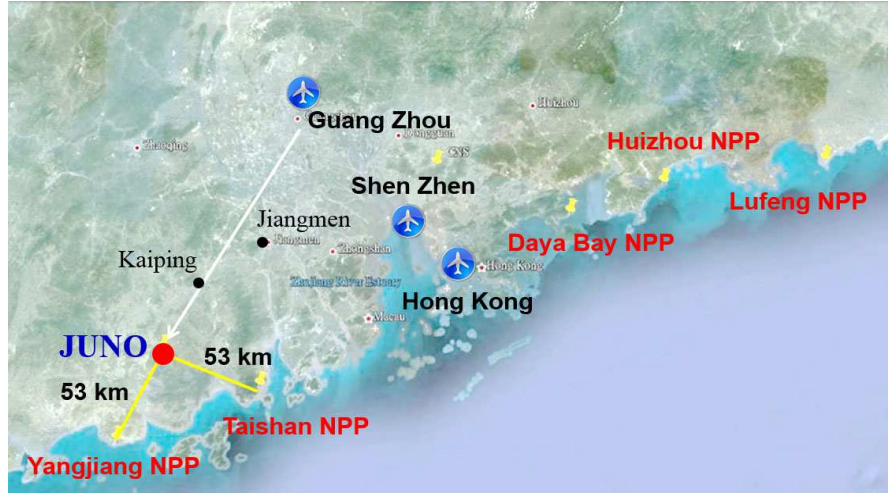}
\caption{Location of the JUNO site. The distances to the nearby Yangjiang NPP and Taishan NPP are both 53 km. Daya Bay NPP is 215 km away. Huizhou and Lufeng NPPs have not been approved yet. Three metropolises, Hong Kong, Shenzhen, and Guangzhou, are also shown.
\label{fig:intro:location}
}
\end{figure}

The JUNO experiment locates in Jinji town, Kaiping city, Jiangmen city, Guangdong province. The geographic location is east longitude 112$^\circ$31'05" and north latitude 22$^\circ$07'05". The experimental site is 43 km to the southwest of the Kaiping city, a county-level city in the prefecture-level city Jiangmen in Guangdong province. There are five big cities, Guangzhou, Hong Kong, Macau, Shenzhen, and Zhuhai,  all in $\sim$ 200~km drive distance, as shown in Fig.~\ref{fig:intro:location}.

The experimental site is at $\sim$ 53~km from the Yangjiang NPP and Taishan NPP. Yangjiang NPP has six reactor cores of 2.9~GW$_{\rm th}$ each (themal power). All cores are the 2nd generation pressurized water reactors CPR1000, which is a derivative of Framatone M310, with improvements on safety, refueling, and conventional island design. They are very similar in terms of nuclear core design. The distances between any two cores of Yangjiang NPP are between 88~m and 736~m. The first core started construction on Dec.~16, 2008 and began commercial operations on Mar.~26, 2014. The 6th core started construction on Dec.~23, 2013. All six cores will be running when JUNO starts data taking in 2020. Taishan NPP has planned four cores of 4.59~GW$_{\rm th}$ each. All cores are the 3rd generation pressurized water reactors EPR. The distances between any two cores are between 252~m and 1110~m. The first two cores started construction on Sep. 1, 2009 and Apr. 15, 2010, respectively. The first core is expected to begin commercial operation in 2015. The construction of the 3rd and 4th cores have not started yet. The total thermal power of the Yangjiang and Taishan NPPs would be 35.73~GW$_{\rm th}$. It is possible that the last two cores in Taishan will not be available by 2020, in which case the total power will be 26.55~GW$_{\rm th}$ when JUNO will start data taking.

Daya Bay complex includes Daya Bay NPP, Ling Ao NPP, and Ling Ao-II NPP in a spread of 1.1~km, each with 2 cores of 2.9~GW$_{\rm th}$. The Daya Bay and Ling Ao cores are Framatone M310 and the Ling Ao-II cores are CPR1000. The Daya Bay complex is 215~km away from the JUNO detector, and will contribute about 2.8\% of the reactor antineutrino events. There are proposals for new NPPs in Huizhou and Lufeng, which is unclear now. The Huizhou site is 265~km from the JUNO detector and the Lufeng site is more than 300~km. There is no other NPP or planned NPP in 500~km around the JUNO experimental site. The thermal power of all cores and the baselines are listed in Table~\ref{tab:intro:NPP}. The distances from the detector site to the Yangjiang and Taishan cores are surveyed with a Global Positioning System (GPS) to a precision of 1~meter. All these NPPs are constructed and operated by the China General Nuclear Power Group (CGNPG).

\begin{table}[htb]
\centering
\begin{tabular}{|c|c|c|c|c|c|c|}\hline\hline
Cores & YJ-C1 & YJ-C2 & YJ-C3 & YJ-C4 & YJ-C5  & YJ-C6 \\
\hline Power (GW) & 2.9 & 2.9 & 2.9 & 2.9 & 2.9 & 2.9 \\ \hline
Baseline(km) & 52.75 & 52.84 & 52.42 & 52.51 & 52.12 & 52.21 \\
\hline\hline
Cores & TS-C1 & TS-C2 & TS-C3 & TS-C4 & DYB  & HZ \\
\hline Power (GW) & 4.6 & 4.6 & 4.6 & 4.6 & 17.4 & 17.4 \\ \hline
Baseline(km) & 52.76 & 52.63 & 52.32 & 52.20 & 215 & 265 \\
\hline
\end{tabular}
\caption{Summary of the thermal power and baseline to the JUNO detector for the
Yangjiang (YJ) and Taishan (TS) reactor cores, as well as the remote reactors of Daya Bay (DYB) and
Huizhou (HZ).\label{tab:intro:NPP}}
\end{table}

In absence of high mountains in the allowed area where the sensitivity to the mass hierarchy is optimized, the detector will be deployed in an underground laboratory under the Dashi hill. The elevation of the hill above the detector is 268~m, and that of the dome and the floor of the underground experimental hall is -433~m and -460~m, respectively. The detector is located in a cylindrical pit. The elevation of the detector center is -481.25~m. Therefore, the vertical overburden for the detector is more than 700~m. The experimental hall is designed to have two accesses. One is a 616~m-deep vertical shaft, and the other is a 1340~m long tunnel with a slope of 42.5\%. The rock is granite. The average rock density along a 650~m borehole is measured to be 2.61 g/cm$^3$. The activities of the $^{238}$U, $^{232}$Th, and $^{40}$K in the rock around the experimental hall are measured to be 130, 113, and 1062 Bq/kg, respectively. The muon rate and average energy in the JUNO detector are expected to be 0.0030 Hz/m$^2$ and 215 GeV estimated by simulation with the surveyed mountain profile taken into account.

\subsubsection{JUNO Detector}
\label{subsubsec:detector}

\begin{figure}[htb!]
\centering
\includegraphics[width=0.7\textwidth]{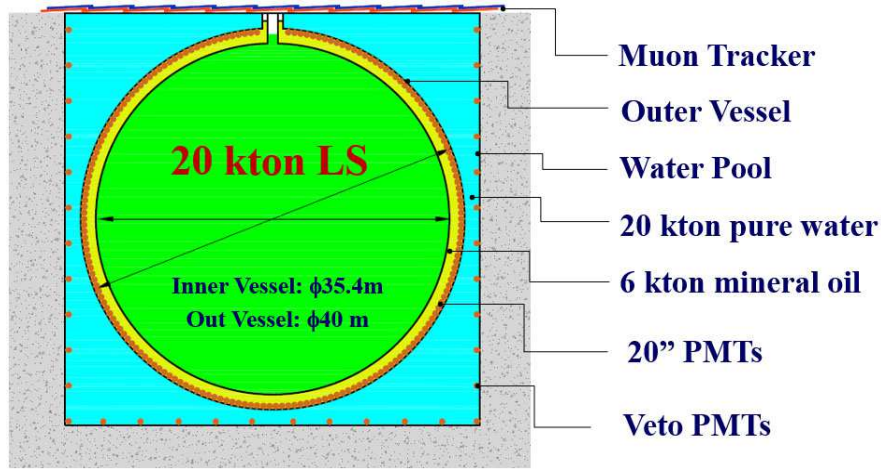}
\caption{A schematic view of the JUNO detector.
\label{fig:intro:det}}
\end{figure}

The JUNO detector consists of a central detector, a water Cherenkov detector and a muon tracker. The central detector is a liquid scintillator (LS) detector of 20 kton fiducial mass with an designed energy resolution of $3\%/\sqrt{E{\rm (MeV)}}$. The central detector is submerged in a water pool to be shielded from natural radioactivity from the surrounding rock and air. The water pool is equipped with Photomultiplier Tubes (PMTs) to detect the Cherenkov light from cosmic muons, acting as a veto detector. On top of the water pool, there is another muon detector to accurately measure the muon tracks. A schematic view of the JUNO detector is shown in Fig.~\ref{fig:intro:det}. The detector design is still developing in the carrying on of R\&D.

To achieve a $3\%/\sqrt{E{\rm (MeV)}}$ energy resolution is very challenging. A Monte Carlo simulation has been developed based on the Monte Carlo of the Daya Bay experiment, as described in Sec.~\ref{subsec:intro:sim}. The photoelectron yield has been tuned according to the Daya Bay data. To reach the required energy resolution, the following improvements from Daya Bay have to be accomplished.
\begin{itemize}
\item The PMT photocathode covergage $\geq 75$\%.
\item The PMT photocathode quantum efficiency $\geq 35$\%.
\item The attenuation length of the liquid scintillator $\geq 20$ m at 430 nm, which corresponds to an absorption length of 60 m with a Rayleigh scattering length of 30 m~\footnote{The Rayleigh scattering length of Linear alkybenzene was measured to be $28.2\pm1.0$~m at 430 nm recently~\cite{Liu:2015hwa}}.
\end{itemize}

The liquid scintillator has similar recipe as the Daya Bay LS without gadolinium loading. Linear alkylbenzene (LAB), a straight alkyl chain of 10-13 carbons attached to a benzene ring [10], is used as the detection medium due to its excellent transparency, high flash point, low chemical reactivity, and good light yield. The liquid scintillator also consists of 3 g/L 2,5-diphenyloxazole (PPO) as the fluor and 15 mg/L p-bis-(o-methylstyryl)-benzene (bis-MSB) as the wavelength shifter.

The density of the LS is 0.859 g/ml. Twenty thousand ton LS is contained in a spherical container of radius of 17.7 m. The light emitted by the LS is watched by about 17,000 20-inch PMTs. PMTs are installed on a spherical structure of a radius of 19.5 m, and submerged in a buffer liquid to protect the LS from the radioactivity of the PMT glass.

The mechanics of the central detector is also very challenging. Two options are under R\&D. The ``Acrylic Sphere" option uses an acrylic vessel to contain the LS. The buffer liquid is water, which is connected with the outer water Cherenkov detector but being optically separated. The PMTs are installed on the inner surface of the truss structure, which also supports the acrylic sphere. The ``Balloon" option uses nylon bag instead of acrylic to contain the LS. The buffer liquid is non-scintillation LAB or mineral oil contained in a stainless steel sphere. The PMTs are installed on the inner surface of the stainless steel vessel. For both options, the PMTs have special protection in case of implosion. Taking into account the implosion container and mechanical clearance, the photocathode coverage can reach 75\% to 78\% for various options.

The central detector is submerged in a cylindrical water pool. At least 2 m water from any direction protects the central detector from the surrounding rock radioactivity. About 1600 20-inch PMTs are installed in the water pool. The muon detection efficiency is expected to be similar as that of the Daya Bay water Cherenkov detector, which is 99.8\%.

The earth magnetic field intensity is about 0.5 gauss at the experimental site. It could have significant negative impact on the photoelectron collection efficiency of the large size PMTs. Both compensation coils surrounding the water pool and high-$\mu$ metal shielding for individual PMTs will be installed.

On the top of the water pool, muon tracker will be installed to accurately measure the muon direction. Plastic scintillator strips decommissioned from the target tracker of the OPERA experiment~\cite{Adam:2007ex} will be reused as the JUNO top tracker. The OPERA target tracker is composed of 62 walls with a sensitive area of 6.7$\times$6.7 m$^2$ each. Each wall consists of four vertical (x) and four horizontal (y) modules. A target tracker module is composed of 64 scintillating strips, 6.7 m long and 26.4 mm wide. Each strip is read out on both sides by a Hamamatsu 64-channel multi-anode PMT. The total surface which could be covered by the 62 x-y walls is 2783 m$^2$. Radioactivity from the surrounding rock of the experimental hall will induce extremely high noise rate in the plastic scintillator strips. Multi-layer design, at least 3 x-y layers, is needed to suppress the radioactivity background. Distance between two adjacent super-layers will be between 1~m and 1.5~m. The muon tracker will cover more than 25\% of the area of the top surface of the water pool.

A chimney for calibration operation will connect the central detector to outside from the top. Special radioactivity shielding and muon detector will be designed for the chimney.

\clearpage

\newcommand{\nua}[1]{\ensuremath{\rlap
           {\kern-2.5pt\ensuremath
           {\overset{\scriptscriptstyle(-)}{\phantom{\nu}}}}
           {\ensuremath{{\nu}_{#1}}}}}

\section{Identifying the Neutrino Mass Hierarchy}
\label{sec:mh}

\blfootnote{Editors: Yufeng Li (liyufeng@ihep.ac.cn) and Liang Zhan (zhanl@ihep.ac.cn)}
\blfootnote{Major contributor: Xin Qian}

\subsection{Introduction and motivation}
\label{subsec:mh:intro}

After the discovery of non-zero $\theta_{13}$ in
recent reactor~\cite{An:2012eh,An:2012bu,Abe:2011fz,Ahn:2012nd} and accelerator~\cite{Abe:2011sj,Adamson:2011qu}
neutrino experiments, the present status of the
standard three-flavor neutrino oscillation~\cite{Capozzi:2013csa,Forero:2014bxa,Gonzalez-Garcia:2014bfa,Fogli:2012ua,Tortola:2012te,GonzalezGarcia:2012sz}
can be summarized as follows:
\begin{itemize}
\item three non-zero mixing angles~\cite{Agashe:2014kda} $\theta_{12}$, $\theta_{23}$, and
$\theta_{13}$ in the MNSP~\cite{Maki:1962mu,Pontecorvo:1967fh} lepton mixing matrix have been
measured with the precision from $4\%$ to $10\%$
\footnote{precision in terms of $\sin^2\theta_{12}$, $\sin^2\theta_{23}$, and $\sin^2\theta_{13}$.},
\item two independent mass-squared differences $|\Delta m^2_{31}|=|m^2_3-m^2_1|$ (or $|\Delta
m^2_{32}|=|m^2_3-m^2_2|$) and $\Delta m^2_{21}=m^2_2 - m^2_1$ have
been measured with the precision better than $4\%$~\cite{Agashe:2014kda},
\item the neutrino mass hierarchy (i.e., sign of the mass-squared difference $\Delta m^2_{31}$) is unknown,
\item the octant of the mixing angle $\theta_{23}$ (i.e., $\theta_{23}<\pi/4$ or $\theta_{23}>\pi/4$) is unknown,
\item the leptonic CP-violating phase $\delta$ in the MNSP matrix is unknown.
\end{itemize}
Therefore, the determination of the neutrino mass hierarchy and octant of the mixing angle $\theta_{23}$, as well as the
measurement of the leptonic CP-violating phase constitutes the main
focus of future neutrino oscillation experiments.

\begin{figure}
\begin{centering}
\includegraphics[angle=-90, width=0.55\textwidth]{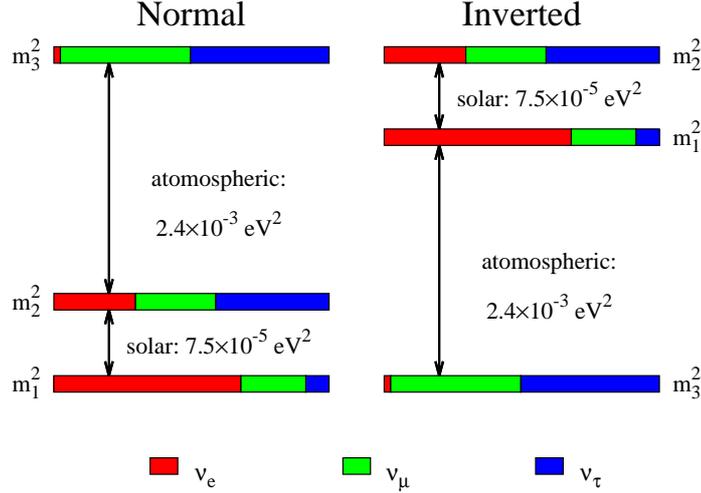}
\par\end{centering}
\caption{\label{fig:mh:pattern} Illustration for the patterns of
normal and inverted neutrino mass hierarchies.}
\end{figure}

The neutrino mass hierarchy (MH) answers the question whether the
third generation ($\nu_3$ mass eigenstate) is heavier or lighter
than the first two generations ($\nu_1$ and $\nu_2$). As shown in
Fig.~\ref{fig:mh:pattern}, the normal mass hierarchy (NH) refers to
$m_3> m_1$ and the inverted mass hierarchy (IH) refers to $m_3<m_1$.

The relatively large value of $\theta_{13}$ has provided excellent
opportunities to resolve the MH in different neutrino oscillation
configurations, which include
\begin{itemize}
\item the medium baseline ($\sim$50 km) reactor antineutrino $\bar\nu_{e}\to\bar\nu_{e}$ oscillation experiments
(JUNO~\cite{Zhan:2008id,Zhan:2009rs,Li:2013zyd,Qian:2012xh,Kettell:2013eos} and RENO-50~\cite{Kim:2014rfa}),
\item the long-baseline accelerator (anti-)neutrino $\nua{\mu}\to\nua{e}$ oscillation experiments
(NO$\nu$A~\cite{Ayres:2004js} and DUNE~\cite{Adams:2013qkq}),
\item  the atmospheric (anti-)neutrino $\nua{\mu}\to\nua{\mu}$ oscillation experiments (INO~\cite{Ahmed:2015jtv}, PINGU~\cite{Aartsen:2014oha}, ORCA~\cite{VanElewyck:2015una},
DUNE~\cite{Adams:2013qkq} and Hyper-K~\cite{Abe:2011ts,Abe:2015zbg}).
\end{itemize}
While the last two methods depend on the matter effect in neutrino
oscillations (the charge-current interaction between (anti-)$\nu_e$
and electrons in the matter), the first method with reactor
antineutrinos at a medium baseline only relies on the oscillation
interference between $\Delta m^2_{31}$ and $\Delta m^2_{32}$ with
$\Delta m^2_{ij} = m^2_i - m^2_j$~\cite{Zhan:2008id,Zhan:2009rs,Li:2013zyd,Qian:2012xh,Kettell:2013eos}.

Besides the neutrino oscillation experiments of determining the MH, the octant of $\theta_{23}$ and the lepton CP-violating phase,
the absolute neutrino mass scale and nature of the massive neutrinos (i.e., the
Majorana or Dirac type) are questions of fundamental importance to be answered in future neutrino non-oscillation probes,
including beta decays, neutrinoless double beta decays and cosmological observations.

The determination of the MH has profound impacts on our understanding of the neutrino physics,
neutrino astronomy and neutrino cosmology.
\begin{itemize}

\item First, as illustrated in Fig.~\ref{fig:mh:dbd}~\cite{Bilenky:2012qi}, MH
helps to define the goal of neutrinoless double beta
decay ($0\nu\beta\beta$) search experiments, which aim to reveal whether neutrinos are
Dirac or Majorana particles. In particular, the chance to observe
$0\nu\beta\beta$ in the next-generation double beta decay
experiments is greatly enhanced for an inverted MH and the Majorana
nature of massive neutrinos. New techniques beyond the next
generation are needed to explore the region covered by a normal MH.
\begin{figure}
\begin{centering}
\includegraphics[width=0.5\textwidth]{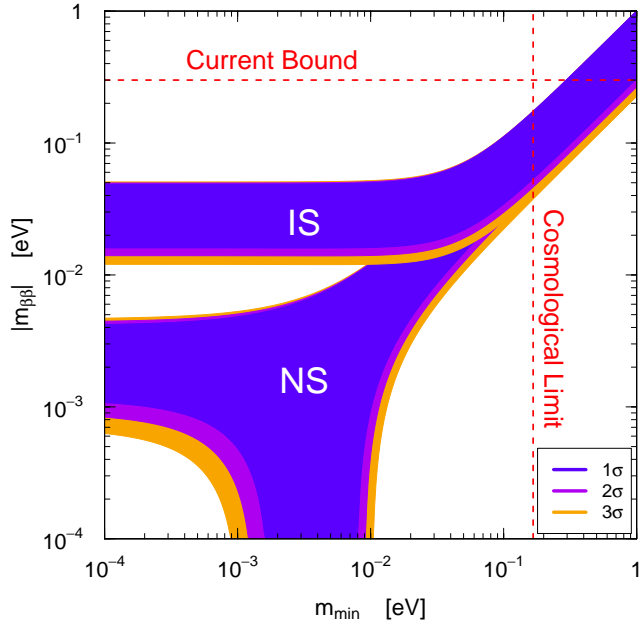}
\par\end{centering}
\caption{\label{fig:mh:dbd} Values of the effective Majorana mass
$|m_{\beta\beta}|$ as a function of the lightest neutrino
mass in the normal (NS, with $m_{\mathrm{min}}=m_{1}$) and inverted
(IS, with $m_{\mathrm{min}}=m_{3}$) neutrino mass spectra after the
measurement of non-zero $\theta_{13}$. The plot is taken from~\cite{Bilenky:2012qi}.}
\end{figure}

\item Second, MH is a crucial factor for measuring the
lepton CP-violating phase. In the long-baseline accelerator
(anti-)neutrino oscillation experiments, %$\nua{\mu}\to\nua{e}$
degenerate solutions for the MH and CP phase emerge, and
the wrong MH would give a fake local minimum for the CP phase, thus
reduce the significance of the CP measurement. This effect is even
more important for accelerator neutrino experiments with a
shorter baseline such as Hyper-K~\cite{Abe:2011ts,Abe:2015zbg} and MOMENT~\cite{Cao:2014bea}.
Therefore, a determination of the MH independent of the CP phase is
important for the future prospect of neutrino physics.

\item Third, MH is a key parameter of the neutrino astronomy and neutrino
cosmology. On one hand, the spectral splits~\cite{Raffelt:2007cb} in supernova neutrino
fluxes would provide a smoking gun for collective
neutrino oscillations induced by the neutrino self-interaction in
the dense environment. The split patterns are significantly
different for the normal and inverted MHs. MH is also important for
the supernova nucleosynthesis, where the prediction of the $^7{\rm
Li}/^{11}{\rm B}$ ratio is also distinct for different
MHs~\cite{Kajino:2014bra}. On the other hand, MH may have important implications on the
cosmological probe of the neutrino mass scale (i.e., $\sum m_\nu$).
As shown in Fig.~\ref{fig:mh:cosmo}, in the case of an inverted MH,
future combined cosmological constraints would have a very
high-precision detection, with $1\sigma$ error shown as a blue
band.  In the case of a normal MH, future cosmology would detect the
lowest $\sum m_\nu$ at a level of $\sim 4 \sigma$.
\begin{figure}
\begin{centering}
\includegraphics[width=0.5\textwidth]{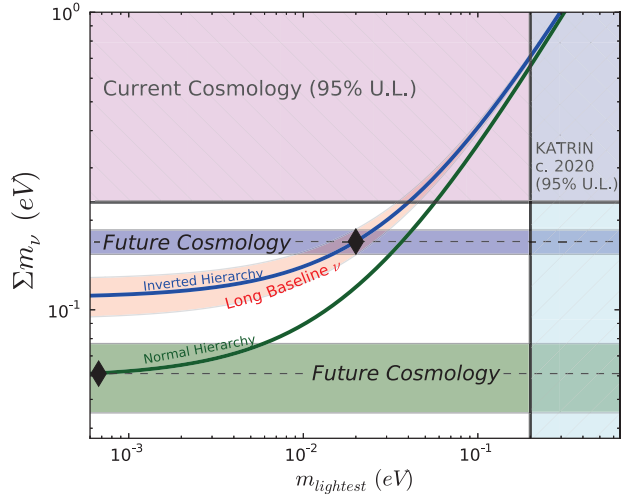}
\par\end{centering}
\caption{\label{fig:mh:cosmo} The current constraints and forecast
sensitivity of cosmology to the neutrino mass in relation to MH~\cite{Abazajian:2013oma}.
In the case of an inverted MH in the upper curve,
future combined cosmological constraints would have a very
high-precision detection, with $1\sigma$ error shown as the blue
band. In the case of a normal MH in the lower curve, future
cosmology would detect the lowest $\sum m_\nu$ at a level of $\sim 4
\sigma$.}
\end{figure}

\item Fourth, MH is one of the most important discriminators for model building
of the neutrino masses and flavor mixing. To understand the origin of neutrino mass generation, the MH information is crucial.
Due to the similar and complementary aspects of quarks and leptons,
the normal MH could be related to the quark mass spectrum and
attributed to the relations of Grand Unified Theories (GUTs). On the other hand, the inverted MH predicts
a nearly-degenerate spectrum between the first and second mass eigenstates, which could be explained
in the models with the discrete or $U(1)$ flavor symmetries. Therefore, MH is a critical parameter to understand the origin of neutrino masses and mixing.

\end{itemize}

JUNO is designed to
resolve the neutrino MH using precision spectral measurements of
reactor antineutrino oscillations. Before giving the quantitative
calculation of the MH sensitivity, we shall briefly review the
principle of this method. The electron antineutrino survival
probability in vacuum can be written as \cite{Li:2013zyd,Qian:2012xh,Minakata:2007tn}:
\begin{eqnarray}\label{eq:mh:osc}
P_{\bar{\nu}_{e}\rightarrow\bar{\nu}_{e}} &=&
1-\sin^{2}2\theta_{13}(\cos^{2}\theta_{12}\sin^{2}\Delta_{31}+\sin^{2}\theta_{12}\sin^{2}{\Delta_{32}})-
\cos^{4}\theta_{13}\sin^{2}2\theta_{12}\sin^{2}\Delta_{21}\\
&=& 1-\frac{1}{2}\sin^{2}2\theta_{13}
\left[1-\sqrt{1-\sin^{2}2\theta_{12}\sin^{2}\Delta_{21}}\cos(2|\Delta_{ee}|\pm\phi)\right]
-\cos^{4}\theta_{13}\sin^{2}2\theta_{12}\sin^{2}\Delta_{21},\nonumber
\end{eqnarray}
where $\Delta_{ij}\equiv\Delta m_{ij}^{2}L/4E$, in which $L$ is the
baseline, $E$ is the antineutrino energy,
\[
\sin\phi=\frac{c_{12}^{2}\sin(2s_{12}^{2}\Delta_{21})-s_{12}^{2}\sin(2c_{12}^{2}\Delta_{21})}{\sqrt{1-\sin^{2}2\theta_{12}\sin^{2}\Delta_{21}}}\,,\,
\cos\phi=\frac{c_{12}^{2}\cos(2s_{12}^{2}\Delta_{21})+s_{12}^{2}\cos(2c_{12}^{2}\Delta_{21})}{\sqrt{1-\sin^{2}2\theta_{12}\sin^{2}\Delta_{21}}}\,,
\]
and~\cite{Nunokawa:2005nx,deGouvea:2005hk}
\begin{eqnarray}
\Delta m^2_{ee} = \cos^2\theta_{12}\Delta m^2_{31} +
\sin^2\theta_{12}\Delta m^2_{32}\,. \label{eq:mh:dmee}
\end{eqnarray}
The $\pm$ sign in the last term of Eq.~\eqref{eq:mh:osc} is decided
by the MH with plus sign for the normal MH and minus sign for the
inverted MH.

\begin{figure}%[p!]
\begin{center}
\begin{tabular}{cc}
\includegraphics[width=0.45\textwidth]{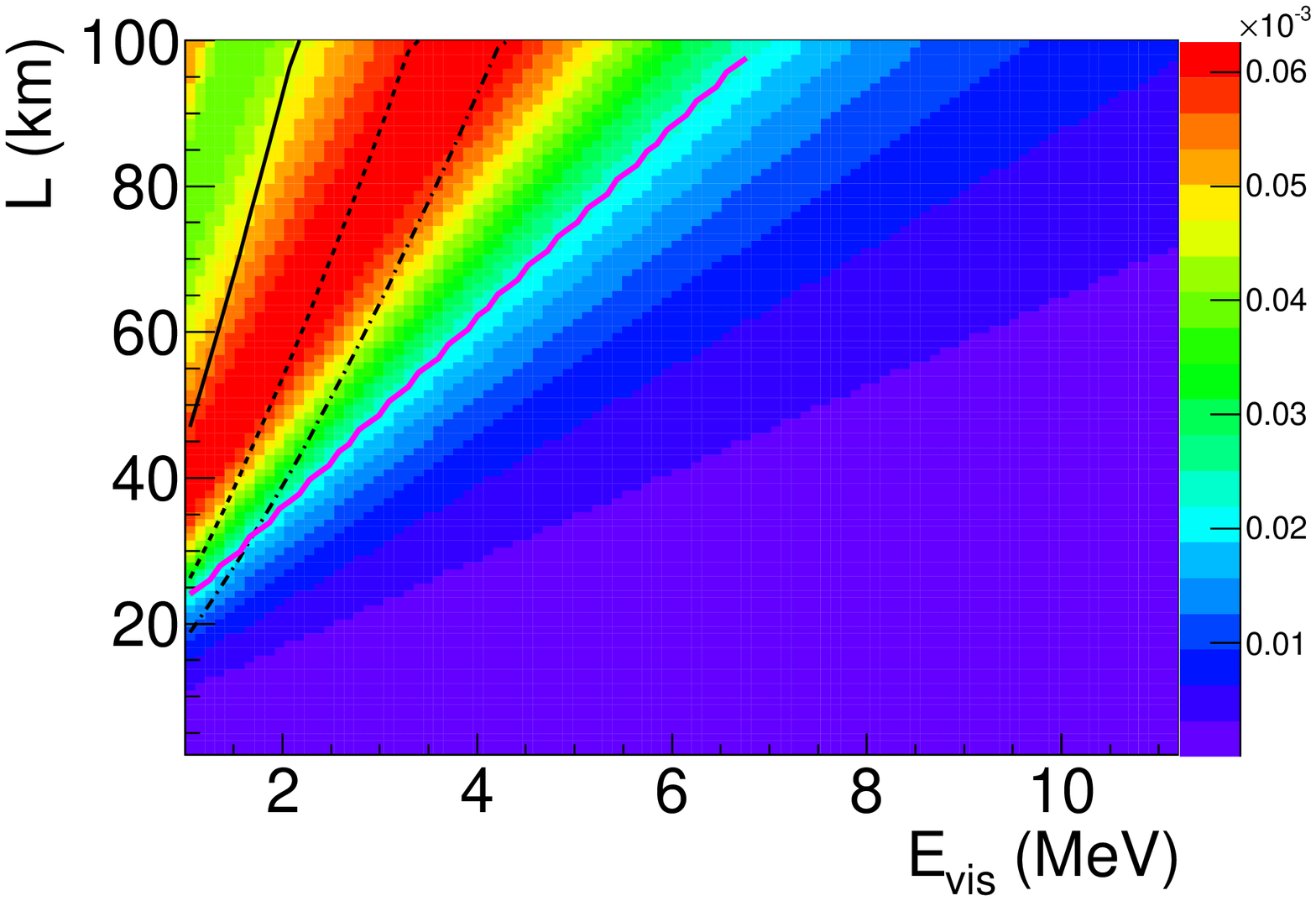}
&
\includegraphics[width=0.45\textwidth]{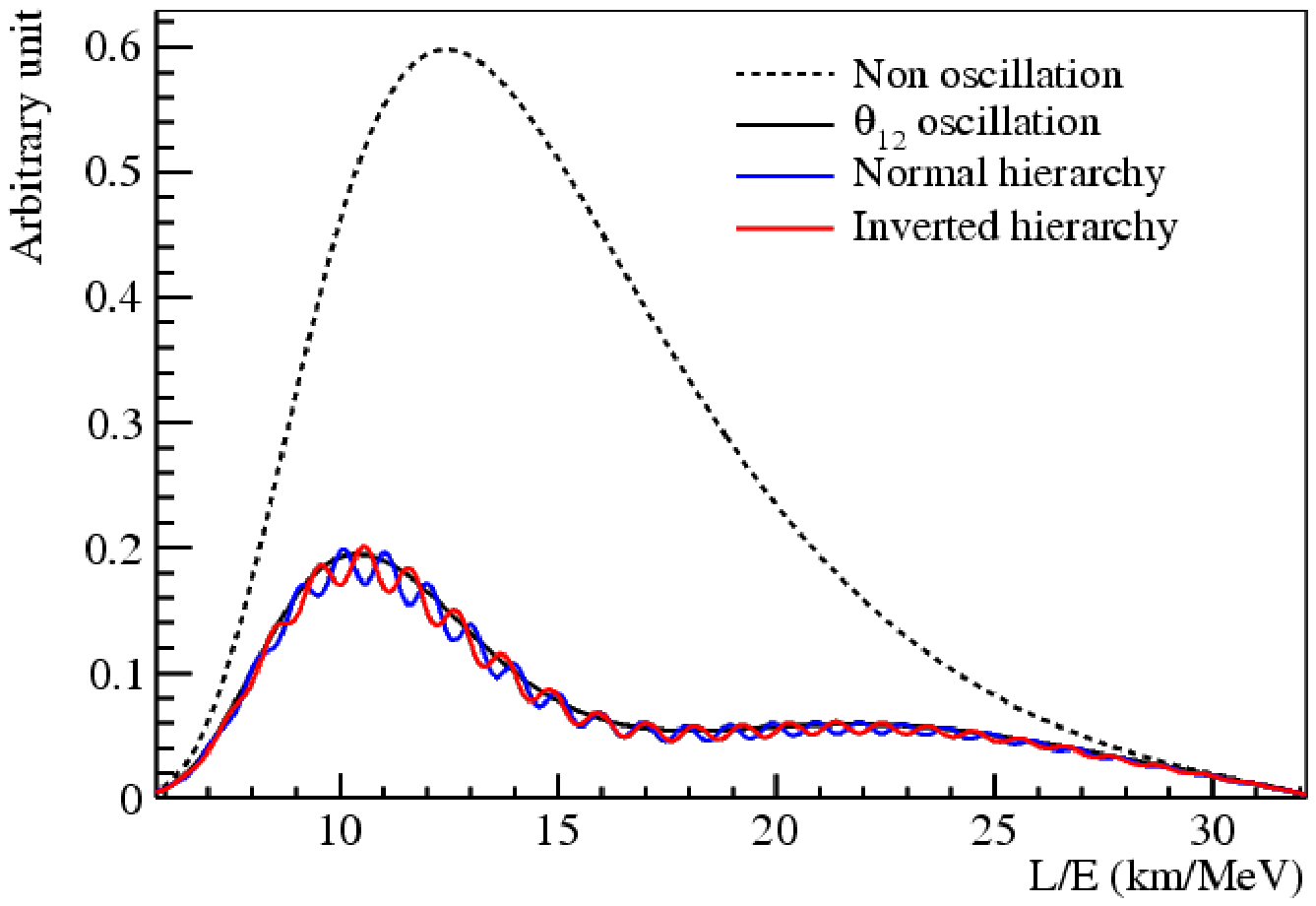}
\end{tabular}
\end{center}
\caption{\label{fig:mh:contour} (left panel) The effective
mass-squared difference shift $\Delta m^2_{\phi}$~\cite{Qian:2012xh} as a
function of baseline (y-axis) and visible prompt energy $E_{\rm vis} \simeq
E_{\nu}-0.8\,{\rm MeV}$ (x-axis). The legend of color code
is shown in the right bar, which represents the size of $\Delta m^2_{\phi}$
in eV$^2$. The solid, dashed, and dotted lines
represent three choices of detector energy resolution with
2.8\%, 5.0\%, and 7.0\% at 1 MeV, respectively. The purple
solid line represents the approximate boundary of degenerate mass-squared
difference. (right panel) The relative shape difference~\cite{Zhan:2008id,Zhan:2009rs}
of the reactor antineutrino flux for different neutrino MHs.}
\end{figure}
In a medium-baseline reactor antineutrino experiment (e.g., JUNO),
oscillation of the atmospheric mass-squared difference manifests
itself in the energy spectrum as the multiple cycles. The spectral
distortion contains the MH information, and can be understood with the
left panel of Fig.~\ref{fig:mh:contour} which shows the energy and
baseline dependence of the extra effective mass-squared difference,
\begin{eqnarray}
\Delta m^2_{\phi}=4E\phi/L\,,
\end{eqnarray}
At baseline of $\sim50\,\rm km$, $\Delta m^2_{\phi}$ at the low
energy ($\sim$ 3 MeV) is larger than the $\Delta m^2_{\phi}$ at the
high energy ($\sim$ 6 MeV). For NH, the effective mass-squared
difference $2|\Delta m^2_{ee}|+\Delta m^2_{\phi}$ in
Eq.~(\ref{eq:mh:osc}) at low energies will be larger than that at
high energies and vice versa for IH, in which the effective
mass-squared difference is $2|\Delta m^2_{ee}|-\Delta m^2_{\phi}\,$.
Therefore, the advancement or retardance of the oscillation phase
illustrated in the right panel of Fig.~\ref{fig:mh:contour} contains
the useful MH information. However, in order to extract the MH information
from the spectral distortion, an excellent energy resolution
($3\%/\sqrt{E}$), a good understanding of the energy response
(better than $1\%$), and a large statistics (${\cal O}(100\rm k)$
inverse beta decay events) are required.
\begin{figure}%[p!]
\begin{center}
\begin{tabular}{cc}
\includegraphics*[width=0.45\textwidth]{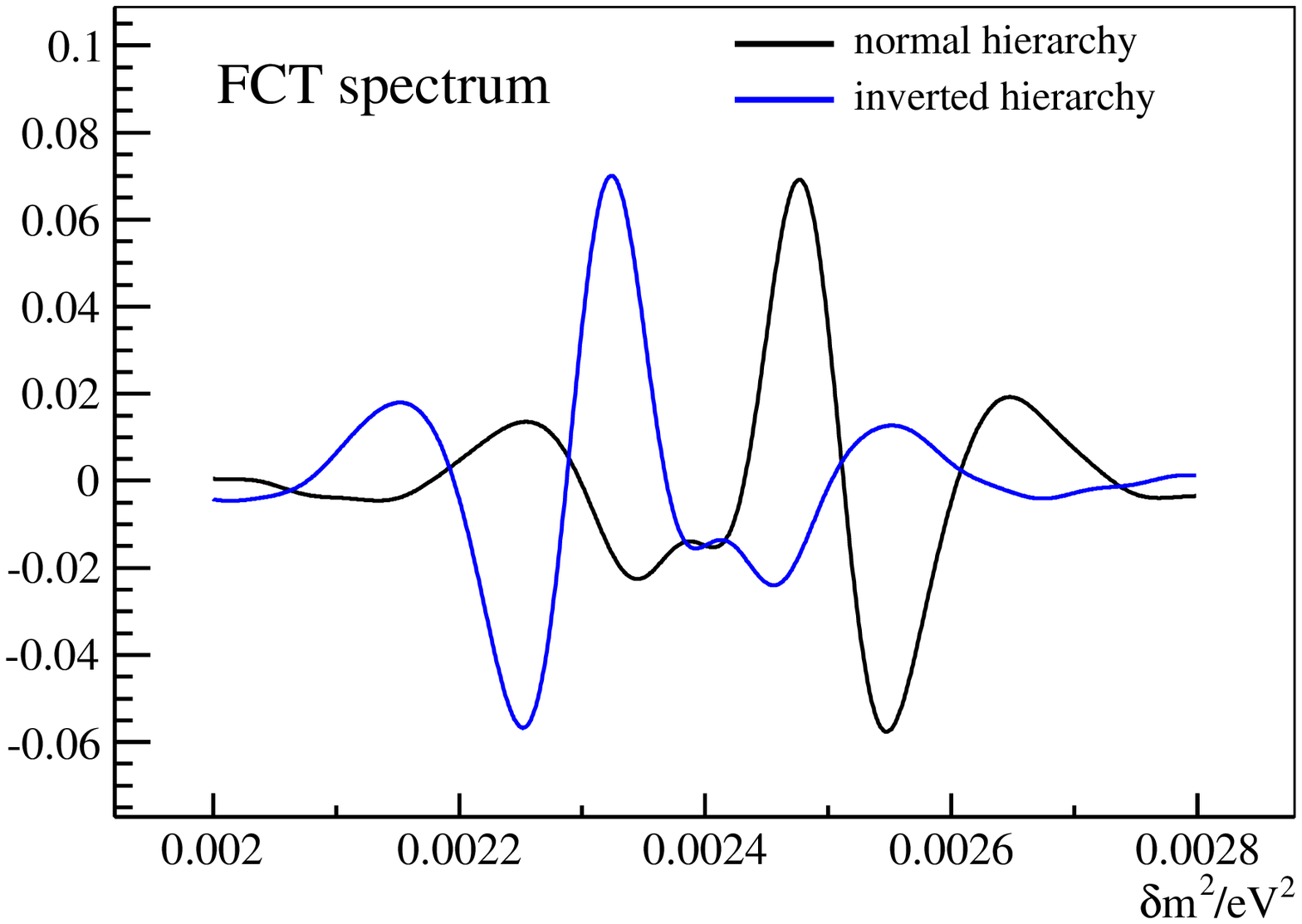}
&
\includegraphics*[width=0.45\textwidth]{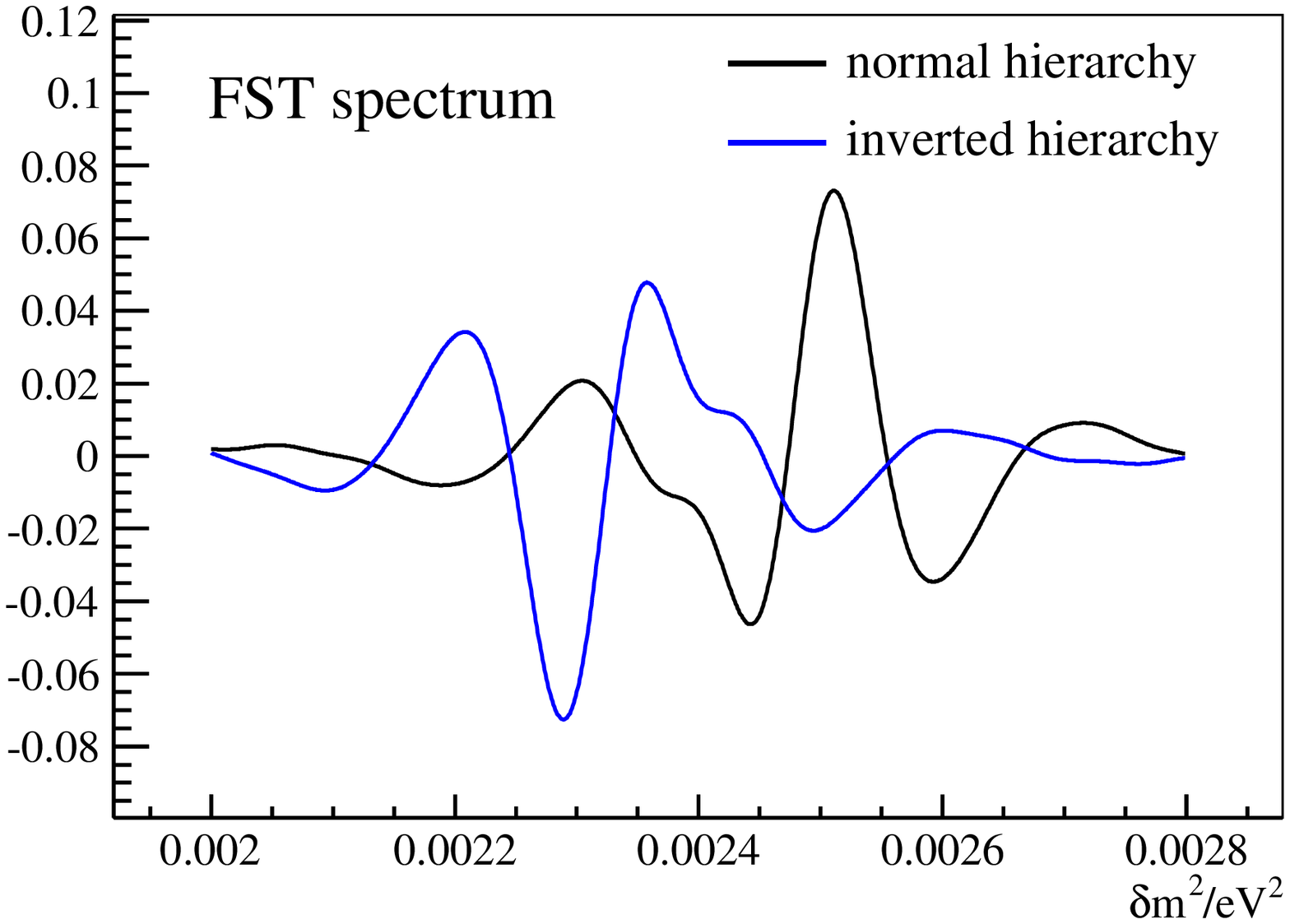}
\end{tabular}
\end{center}
 \caption{The Fourier cosine transform (FCT) (left panel) and Fourier sine transform (FST) (right panel) of the reactor antineutrino energy spectrum.
  The solid and dashed lines are for the normal MH and the inverted MH, respectively.}
\label{fig:mh:FT}
\end{figure}

The oscillation interference effect is more evident in the frequency domain after a Fourier transform of the $L/E$ spectrum of reactor antineutrinos~\cite{Zhan:2008id,Zhan:2009rs,Learned:2006wy}.
In Fig~\ref{fig:mh:FT} we illustrate the Fourier sine transform (FST) and Fourier
cosine transform (FCT) of the reactor antineutrino energy spectrum, where the FST and FCT frequency spectra are defined as
\begin{eqnarray}
\label{eq:FT}
{\rm FST}(\omega) = \int^{t_{\rm max}}_{t_{\rm min}}F(t) \sin (\omega t)\mathrm{d}t\,, \;\quad\quad\;
{\rm FCT}(\omega) = \int^{t_{\rm max}}_{t_{\rm min}}F(t) \cos(\omega
t)\mathrm{d}t\,,
\end{eqnarray}
with $\omega = 2.54\times\Delta m^{2}_{ij}$ being the frequency, and $t=L/E$ being the variable in the $L/E$ space, varying from $t_{\rm min}
= {L}/{E_{\rm max}}$ to $t_{\rm max} = {L}/{E_{\rm min}}$. $F(L/E)$ is written as
\begin{eqnarray}
F(L/E) = \phi(E)\sigma(E) P_{\bar{\nu}_{e}\rightarrow\bar{\nu}_{e}}(L/E)\,,
\end{eqnarray}
where $\phi(E)$, $\sigma(E)$ and $P_{\bar{\nu}_{e}\rightarrow\bar{\nu}_{e}}(L/E)$ are the reactor antineutrino spectrum,
the interaction cross section and the oscillation probability, respectively.
$\phi(E)$ and $\sigma(E)$  will be discussed in the next section, and $P_{\bar{\nu}_{e}\rightarrow\bar{\nu}_{e}}(L/E)$ is defined in Eq.~(\ref{eq:mh:osc}).
Distinctive features of the FST and FCT spectra for normal and inverted MHs can be observed in Fig~\ref{fig:mh:FT}.
On the FCT spectrum (left panel), a valley appears at the left of the prominent peak for the IH, and a peak appears at the left of the valley for the NH.
On the FST spectrum (right panel), there is a clear valley for the IH, and a clear peak for NH. Therefore, we can distinguish the MH from the Fourier transform spectra
without any prior information on the neutrino mass-squared differences. More details on properties of the FCT and FST spectra in the MH determination can
be found in Refs.~\cite{Zhan:2008id,Zhan:2009rs}.

Beside the aforementioned interference between $|\Delta m^2_{31}|$
and $|\Delta m^2_{32}|$, the precision measurement of $|\Delta
m^2_{ee}|$ in a medium-baseline reactor experiment can reveal
additional information regarding the MH, when combined with the
precision $|\Delta m^2_{\mu\mu}|$ measurements from the future muon
(anti-)neutrino disappearance~\cite{Nunokawa:2005nx,deGouvea:2005hk}. Using the
convention of Refs.~\cite{Li:2013zyd,Nunokawa:2005nx}, we have
\begin{eqnarray}
|\Delta m^2_{ee}|-|\Delta m^2_{\mu\mu}|&=& \pm \Delta
m^2_{21}(\cos2\theta_{12}-\sin2\theta_{12}\sin\theta_{13}\tan\theta_{23}\cos\delta)
\label{eq:mh:dmemu}\,,
\end{eqnarray}
where the positive and negative signs correspond to normal and
inverted MHs, respectively. The precision measurements of both
$|\Delta m^2_{\mu\mu}|$ and $|\Delta m^2_{ee}|$ would provide new
information regarding the neutrino MH. Therefore, by combining these
two types of information (interference and precision $|\Delta
m^2_{ee}|$ measurement), JUNO will have a robust path to resolve the
neutrino MH~\cite{Li:2013zyd}.

\subsection{Signal and Background}

\subsubsection{Reactor neutrino signal}
\label{subsec:mh:sigbkg:neusel}
Reactor neutrinos are electron antineutrinos emitted from subsequent $\beta$-decays of instable fission fragments. All reactors close to JUNO are pressurized water reactors (PWR), the same type as the Daya Bay reactors. In these reactors, fissions of four fuel isotopes, $^{235}\rm U$, $^{238}\rm U$, $^{239}\rm Pu$, and $^{241}\rm Pu$, generate more than 99.7\% of the thermal power and reactor antineutrinos. Reactor neutrino fluxes per fission of each isotope are determined by inversion of the measured $\beta$ spectra of fission products~\cite{VonFeilitzsch:1982jw,Schreckenbach:1985ep,Hahn:1989zr,Huber:2011wv,Mueller:2011nm}
or by calculation with the nuclear database~\cite{Vogel:1980bk,Dwyer:2014eka}.
Their fission rates in a reactor can be estimated with the core simulation and thermal power measurements. The reactor neutrino flux can be predicted as
\begin{equation}
\label{eq:mnuflux}
\Phi(E_{\nu})=\frac{W_{\rm th}}{\sum_i f_i e_i}\cdot\sum_i f_i \cdot S_{i}(E_{\nu}),
\end{equation}
Where $W_{\rm th}$ is the thermal power of the reactor, $f_i$, $e_i$, and $S_i(E_{\nu})$ are fission fraction, the thermal energy released in each fission, and the neutrino flux per fission for the $i$-th isotope, respectively. Such a prediction is expected to carry an uncertainty of 2-3\%~\cite{An:2012eh}. Recently, reactor neutrino experiments
(Daya Bay~\cite{DYBbump}, RENO~\cite{Seon-HeeSeofortheRENO:2014jza} Double Chooz~\cite{Abe:2014bwa})
found a large discrepancy between the predicted and measured spectra in the 4-6 MeV region.
Model independent prediction based on the new precision measurements could avoid this bias,
and might be able to improve the precision to 1\%. Detailed description on the reactor neutrino flux can be found in the Appendix of Sec.~\ref{sec:app}.

\par
\begin{figure}
\begin{center}
\includegraphics[width=0.5\textwidth]{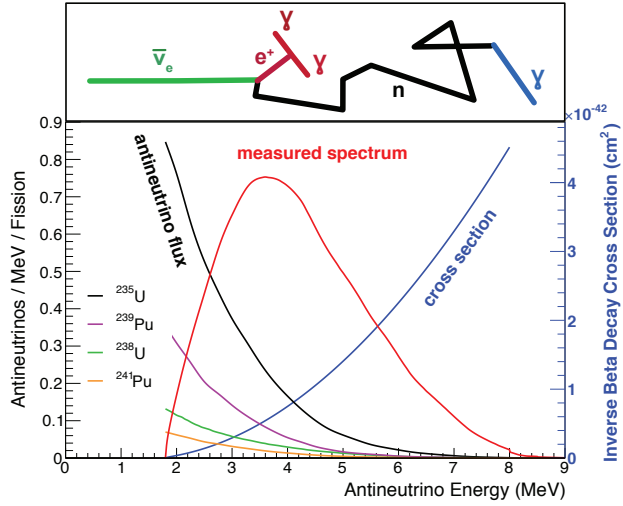}
\caption{The observable $\overline{\nu}_{e}$ spectrum (red line) is a product of the
antineutrino flux from reactor and the cross section of inverse beta decay (blue line). The contributions of four fission isotopes to the antineutrino flux are shown for a typical pressurized water reactor. The steps involved in the detection are schematically drawn on the top of the figure~\cite{Vogel:2015wua}.
\label{fig:nuspec}}
\end{center}
\end{figure}

JUNO measures the reactor neutrino signal via the inverse beta decay (IBD) reaction
\begin{equation}
\bar{\nu}_e +p \rightarrow e^+ + n \,.
\end{equation}
The reactor antineutrino $\bar{\nu}_e$ interacts with a proton, creating a positron ($e^+$) and a neutron. The positron quickly deposits its energy and annihilates into two 511-keV $\gamma$-rays, which gives a prompt signal. The neutron scatters in the detector until being thermalized. It is then captured by a proton $\sim\,200\ \mu$s later and releases a 2.2-MeV $\gamma$-ray. The coincidence of the prompt-delayed signal pair in such a short time significantly reduces backgrounds. Positron carries almost all energy of the neutrino in this reaction. Therefore, the observable neutrino spectrum shown in Fig.~\ref{fig:nuspec} can be obtained from the prompt signal with a $\sim\,$0.8 MeV shift. With reactors of 36 GW thermal power at 53 km, a 20-kton LS detector will have 83 IBD events per day.

The accidental background, $^8$He/$^9$Li, fast neutron and ($\alpha$, n) background are the major backgrounds for the reactor neutrino oscillation analysis. Fiducial volume cut can significantly reduce the accidental background and the ($\alpha$, n) background. Energy selection, time coincidence, and vertex correlation of the prompt and delayed signals are required for the reactor antineutrino selection to further suppress the accidental background. To reject the cosmogenic backgrounds such as $^{9}$Li/$^{8}$He and fast neutron, muon veto cuts are necessary and need be optimized to reduce the loss of detector live time and dead volume. Detailed discussion on backgrounds will be presented later. A set of preliminary antineutrino selection criteria is listed below:
\begin{itemize}
  \item fiducial volume cut $r<17$ m;
  \item the prompt energy cut 0.7 MeV $<E_p<$ 12 MeV;
  \item the delayed energy cut 1.9 MeV $<E_d<$ 2.5 MeV;
  \item time interval between the prompt and delayed signal $\Delta T<$1.0 ms;
  \item the prompt-delayed distance cut $R_{p-d}<$1.5 m;
  \item Muon veto criteria:
  \begin{itemize}
    \item for muons tagged by Water Pool, veto the whole LS volume for 1.5 ms
    \item for good tracked muons in central detector and water Cerenkov detector, veto the detector volume within $R_{d2\mu}<3$ m and $T_{d2\mu}<1.2$ s
    \item for the tagged, non-trackable muons in central detector, veto the whole LS volume for 1.2 s
  \end{itemize}
\end{itemize}

The antineutrino selection efficiency due to the fiducial volume is 91.8\%. The energy cut, time cut, and vertex cut have efficiencies of 97.8\%, 99.1\%, and 98.7\%, respectively, using Geant4-based Monte Carlo (MC) studies described in Sec.~\ref{sec:app}. Assuming 99\% muons have good reconstructed track, the efficiency of above muon veto cut is estimated to be 83\% by using the toy MC method. Tab.~\ref{tab:mh:sigbkg} summarizes the efficiencies of antineutrino selection cuts and the corresponding reduction to various backgrounds, which will be discussed in the next subsection. JUNO will observe 60 IBD events per day, with about 6\% backgrounds.
\begin{table}[htb]
\centering
\begin{tabular}{|c|c|c|c|c|c|c|c|}\hline\hline
Selection & IBD efficiency & IBD & Geo-$\nu$s & Accidental & $^9$Li/$^8$He & Fast $n$ & $(\alpha, n)$ \\ \hline
- & - & 83 & 1.5 & $\sim5.7\times10^4$ & 84 & - & - \\ \hline
Fiducial volume & 91.8\% & 76 & 1.4 &  & 77 & 0.1 & 0.05 \\ \cline{1-4}\cline{6-6}
Energy cut & 97.8\% & & & 410 &  &  &  \\ \cline{1-2}
Time cut & 99.1\% & 73 & 1.3 &  & 71 &  &  \\ \cline{1-2}\cline{5-5}
Vertex cut & 98.7\% & & & 1.1 &  &  &  \\ \cline{1-6}
Muon veto & 83\% & 60 & 1.1 & 0.9  & 1.6 &  &  \\ \hline
Combined & 73\% & 60  & \multicolumn{5}{c|}{3.8} \\ \hline
\hline
\end{tabular}
\caption{The efficiencies of antineutrino selection cuts, signal and backgrounds rates.
\label{tab:mh:sigbkg}}
\end{table}

\subsubsection{Background estimation}
\label{subsec:mh:sigbkg:bkgest}

\paragraph{Accidental background}
The rate of accidental backgrounds can be calculated as $R_{\rm acc} = R_p\cdot R_d \cdot \Delta T$, where $R_p$ and $R_d$ are the rate of prompt and delayed signals, respectively, and $\Delta T$ is the time coincidence window. A fiducial volume cut is essential to significantly suppress such background. The accidental background consists of mainly three types of random coincidence: (radioactivity, radioactivity), (radioactivity, cosmogenic isotope) and (radioactivity, spallation neutrons):
\begin{itemize}
  \item (radioactivity, radioactivity): The singles rate obtained from MC simulation is about 7.6 Hz after fiducial volume cut (see Sec.~\ref{subsec:intro:bkg:natural}),%~\ref{sec:app},
  in which the faction of neutron-like signals is $\sim$8\%. Thus the rate of prompt-delayed coincidence within 1.0 ms is $\sim\,$410/day.
  In addition, a toy MC study gives a factor of 380 suppression by requiring $R_{p-d}<$1.5 m, where $R_{p-d}$ is the distance between the prompt-delayed pair,
  thus the accidental background rate is reduced to 1.1/day.
  \item (radioactivity, cosmogenic isotope): based on the rates of cosmogenic isotopes in Sec.~\ref{subsec:intro:bkg:cosmogenic},
  the neutron-like singles from cosmogenic isotopes is estimated to be $\sim$340/day.
  The rate of accidental coincidence between radioactivity and those isotopes is $<$0.01/day after $\Delta T<1.0$ ms and $R_{p-d}<$1.5 m cut.
  \item (radioactivity, spallation neutrons): Though the total rate of spallation neutrons is 1.8 Hz,
  after 1.5 ms muon veto the rate is reduced to $\sim$45/day.
  The coincidence between radioactivity and the residual spallation neutrons is negligible after the time and spatial cut.
\end{itemize}
Thus the total rate of accidental backgrounds is estimated to be 0.9/day, after taking into account the efficiency of muon veto.
During data taking, the rate of radioactivity can be precisely monitored, so can the neutron-like events from muon spallation.
So the uncertainty of accidental background rate can be controlled within 1\% and the uncertainty of spectrum shape
is negligible due to the large statistics of prompt-like singles.

\paragraph{$^{9}$Li/$^{8}$He}
As noted in Sec.~\ref{subsec:intro:bkg:cosmogenic}, the $\beta$-n decays from cosmogenic $^8$He and $^9$Li can mimic IBD interactions, thus are the most serious correlated background to reactor antineutrinos. The $^9$Li and $^8$He production cross section is often modelled empirically as being proportional to $E_{\mu}^{0.74}$, where $E_{\mu}$ is the average energy of the muon at the detector.
Considering the cross section measured in the KamLAND detector~\cite{Abe:2009aa}, $2.2\times$10$^{-7}\mu^{-1}$g$^{-1}$cm$^2$ for $^9$Li and $0.7\times$10$^{-7}\mu^{-1}$g$^{-1}$cm$^2$ for $^8$He, the predicted $^9$Li and $^8$He production rate at JUNO is 150 and 50 per day, respectively.
The branching ratio of the $\beta$-n decay is 51\% for $^9$Li and 16\% for $^8$He, thus the total rate of $\beta$-n decays is 84/day. Taking into account the fiducial volume cut, the rate is reduced to 77/day. The delayed energy cut and time cut efficiencies for $^9$Li are the same as those for IBDs,
as shown in Tab.~\ref{tab:mh:sigbkg}, while the prompt energy cut efficiency for $^9$Li is $\sim$97\%. Thus the background rate is reduced to $\sim$71/day. The FLUKA simulation results in Tab.~\ref{tab:intro:isotopes} are smaller than this empirical extrapolation. It is pointed out that scaling with the muon average energy is not as accurate as the scaling with the energy loss of individual muons, which may explain the differences between the FLUKA-estimated yields and the empirical extrapolation~\cite{Li:2015kpa}.

In practice, the rate of $^9$Li/$^8$He can be measured from the distribution of the time since the last muon using the known decay times for these isotopes~\cite{Wen:2006hx}.
A toy MC based on the simulated muon data has been performed, and it's expected that $<$3\% rate uncertainty can be achieved with 6 years data.

The $^9$Li/$^8$He background is correlated with the parent muon in time and space. The lateral distance between the muon-induced isotopes and the parent muon trajectory
is roughly exponential. The most effective approach to reject $^{9}$Li/$^{8}$He background is to veto a sufficient detector volume
along the muon trajectory for a relative long time, e.g, a few times of the isotope's lifetime. Muons that are accompanied by electromagnetic or hadronic showers,
usually named as showering muons, are the dominant producers ($>$85\%) of the radioactive isotopes.
With a muon simulation for JUNO, it is found that the showering muon rate is $\sim$0.5 Hz.
The simulation also suggests that after producing a shower, the muon still survives and its direction changes negligibly.
Thus the critical issue is how well the muon track reconstruction is, for both non-showering muons and showering muons.

For single non-showering muon, the track can be reconstructed using the first hit-time on PMTs.
The methods to reconstruct the track of showering muons and the tracks in muon bundles (see Sec.~\ref{subsec:intro:bkg:muon}) are under development.
Initial track can be guessed by using the PMTs near the injecting point and outgoing point.
The locations of spallation neutrons, if there is any, can be used to constrain their parent muon tracks.
Particularly, the high multiplicity of spallation neutrons from the showering processes can give good estimation of where the showering happens,
then it's possible to veto a spherical volume around the showering point to reject $^9$Li/$^8$He.
If showering muons are poorly reconstructed, a whole volume cut is essential to reject the $^9$Li/$^8$He background.

To estimate the residual $^9$Li/$^8$He after the muon veto cuts listed in Sec.~\ref{subsec:mh:sigbkg:neusel},
we assume the efficiency of good track reconstruction for non-showering muon is more than 99\%, based on the experience from KamLAND.
For showering muons, we expect the track reconstruction can also reach 99\% efficiency.
Then the efficiency of the muon veto cuts is estimated to be 83\% using toy MC, and 2.3\% of $^9$Li/$^8$He will survive,
thus the final residual $^9$Li and $^8$He background is 1.6/day.

The uncertainties on the residual $^9$Li/$^8$He background rate are mainly from the uncertainty from muon track reconstruction and
the position reconstruction uncertainty of IBD candidates. We assume 20\% relative uncertainty on the residual background rate.
With the full statistics of $^9$Li/$^8$He events, the energy spectrum shape can be measured quite well. We can assign 10\% bin-to-bin shape uncertainty.

\paragraph{Fast neutron}
The cosmic muons that only passing the surrounding rock of the water pool, as well as the corner clipping muons with
very short track length in water, are not able to be tagged. The energetic neutrons produced by those muons can form
a fast neutron background by scattering off a proton and then being captured in the LS detector. Based on a full simulation without
optical processes, the rate of fast neutrons is estimated to be $\sim$0.1/day. The signals produced by the energetic
neutrons are found to be concentrated at the top of detector and near the equator where the water shielding is minimum.
For the MH analysis, we assume the relative rate uncertainty is 100\%. The prompt energy spectrum is consistent with a flat distribution.
The tagged fast-neutrons can actually provide good information about the energy spectrum. In this analysis, we assume the shape uncertainty is 20\%.

\paragraph{$^{13}$C$(\alpha, n)^{16}$O background}
The alpha particles from the U, Th radioactivities can react with the $^{13}$C in LS. The $^{13}$C$(\alpha, n)^{16}$O reaction could
lead to a correlated background if the neutron is fast enough or there is a gamma from the de-excitation of the $^{16}$O excited states.
Based on the estimated natural radioactivity concentrations, the $(\alpha, n)$ background rate is estimated to be 0.05/day for the ``Acrylic Sphere" option,
and 0.01/day for the ``Balloon" option due to the lower U/Th concentration. The highest energy of alpha's from U/Th is about 9 MeV,
and the cross section of $^{13}$C$(\alpha, n)^{16}$O reaction is known with a $\sim$20\% uncertainty for an alpha with energy $<$10 MeV.
Thus if the rate of alpha particles is well measured, the $(\alpha, n)$ background can be predicted precisely.
In this analysis, a 50\% relative uncertainty for both the background rate and the energy spectrum shape is conservatively assumed.

\paragraph{Geo-neutrino background}
Antineutrinos produced from radioactive decays of Th and U inside the Earth constitute the geo-neutrino flux,
which will also contribute to the background of reactor antineutrinos.
The total event rate of geo-neutrinos at the JUNO site is 1.5/day, where contributions from Th and U are
23\% and 77\% respectively. After the IBD efficiency cut, the remaining
geo-neutrino background is 1.1/day. The relative rate uncertainty of geo-neutrinos is estimated as 30\%,
where the crust uncertainty is 18\% and the uncertainty of mantle prediction is assumed to be 100\%.
Experimentally the rate of geo-neutrinos can be measured with much better precision using the JUNO detector itself.
Finally, we assume the relative shape uncertainty of geo-neutrinos to be 5\%,
because the $\beta$-decay spectra of Th and U are well known from the nuclear physics.
More details on the geo-neutrino prediction and measurement are discussed in Sec.~\ref{sec:geo}.

The background rates and the reduction with the antineutrino selection cuts are summarized in
Tab.~\ref{tab:mh:sigbkg}.

\subsection{The MH Sensitivity}
 \label{subsec:mh:sensitivity}

\subsubsection{Basic Setup and Definition}
 \label{subsec:mh:sensitivity:setup}

In JUNO simulation, we assume a 20 kt LS
detector, and the total thermal power of the two reactor complexes as 36
GW$_{th}$. We use the nominal running time of six years (i.e., 2000 effective days) and a detector energy
resolution of $3\%/\sqrt{E{\rm (MeV)}}$ as a benchmark. The IBD detection efficiency is estimated as 73\% as show in
Tab.~\ref{tab:mh:sigbkg}. The energy $E$
is referred as the visible energy of the IBD events [$E{\rm
(MeV)}\simeq E_{\nu}{\rm (MeV)} -0.8$]. A normal MH is assumed to be
the true one (unless mentioned explicitly) while the conclusion
is the same for the other assumption. The relevant oscillation
parameters are taken from the latest global analysis~\cite{Capozzi:2013csa,Fogli:2012ua} as
$\Delta m^2_{21}=7.54\times10^{-5}\,{\rm eV}^{-2}$, $(\Delta
m^2_{31}+\Delta m^2_{32})/2=2.43\times10^{-3}\,{\rm eV}^{-2}$,
$\sin^2\,\theta_{13}=0.024$ and $\sin^2\theta_{12}=0.307$. The
CP-violating phase will be specified when needed. Corrections to
$\Delta m^2_{21}$ and $\sin^2\theta_{12}$ from terrestrial
matter effects are around 0.5\%-1\% (see Sec.~\ref{sec:prec}) and the
induced uncertainties are negligibly small ($<0.1\%$).
Finally, the reactor antineutrino flux model from ILL and Vogel {\it
et al.}~\cite{VonFeilitzsch:1982jw,Schreckenbach:1985ep,Hahn:1989zr,Vogel:1980bk} is adopted in our simulation\footnote{We
have tried both the old~\cite{VonFeilitzsch:1982jw,Schreckenbach:1985ep,Hahn:1989zr,Vogel:1980bk} and new
evaluations~\cite{Huber:2011wv,Mueller:2011nm} of the reactor antineutrino fluxes.
Both evaluations give consistent results on the MH determination.}.
Because only two of the three mass-squared differences ($\Delta
m^2_{21}$, $\Delta m^2_{31}$ and $\Delta m^2_{32}$) are independent,
we choose $\Delta m^2_{21}$ and $\Delta m^2_{ee}$ [see
Eq.~(\ref{eq:mh:dmee})] as our working parameters.

To obtain the MH sensitivity, we employ the least-squares method and
construct a $\chi^2$ function as \footnote{A
different definition with the Poisson $\chi^2$ function yields the
consistent MH sensitivity \cite{Qian:2012xh,Kettell:2013eos}.},
\begin{equation}
\chi^2_{\text{REA}}=\sum^{N_{\text{bin}}}_{i=1}\frac{[M_{i} -
T_{i}(1+\sum_k \alpha_{ik}\epsilon_{k})]^2}{M_{i}} +
\sum_k\frac{\epsilon^2_{k}}{\sigma^2_k}\,,\label{eq:mh:chiREA}
\end{equation}
where $M_{i}$ is the measured neutrino events in the
$i$-th energy bin, %$B^n_d$ is the corresponding background,
$T_{i}$ is the predicted neutrino events with
oscillations, $\sigma_k$ is the systematic uncertainty,
$\epsilon_{k}$ is the corresponding pull parameter, and
$\alpha_{ik}$ is the fraction of neutrino event contribution of the
$k$-th pull parameter to the $i$-th energy bin. The considered
systematic uncertainties include the correlated (absolute) reactor
uncertainty ($2\%$), the uncorrelated (relative) reactor uncertainty
($0.8\%$), the spectrum shape uncertainty ($1\%$) and the
detector-related uncertainty ($1\%$). We use 200 equal-size bins for
the incoming neutrino energy between 1.8 MeV and 8.0 MeV.

We fit the spectrum assuming the normal MH or inverted MH with the chisquare
method and take the difference of the minima as a measure of the
MH sensitivity. The discriminator of the MH can be defined as
\begin{equation}
\Delta \chi^2_{\text{MH}}=|\chi^2_{\rm min}(\rm N)-\chi^2_{\rm
min}(\rm I)|, \label{eq:mh:chisquare}
\end{equation}
where the minimization process is implemented for all the relevant
oscillation parameters. Note that two local minima for each MH
[$\chi^2_{\rm min}(\rm N)$ and $\chi^2_{\rm min}(\rm I)$] can be
located at different positions of $|\Delta m^2_{ee}|$.

\subsubsection{Baseline Optimization}
\begin{figure}%[p!]
\begin{center}
\begin{tabular}{cc}
\includegraphics*[bb=20 20 290 232, width=0.43\textwidth]{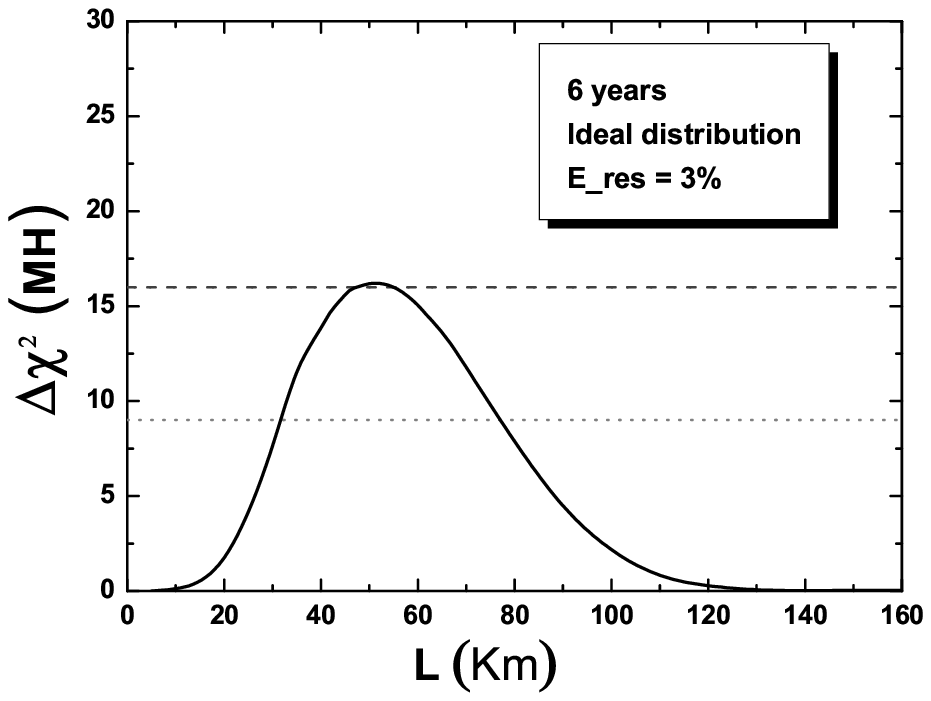}
&
\includegraphics*[bb=20 16 284 214, width=0.42\textwidth]{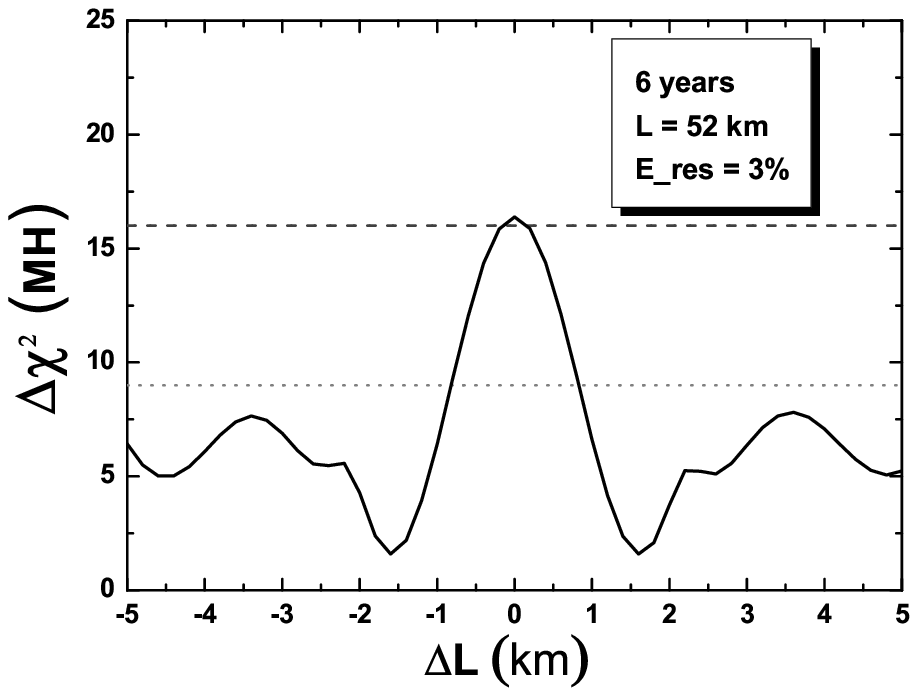}
\end{tabular}
\end{center}
\caption{The MH discrimination ability as the function of the
baseline (left panel) and function of the baseline difference of two
reactors (right panel).} \label{fig:mh:baseline}
\end{figure}
%%%%%%%%%%%%%%%%%%%% Fig. x %%%%%%%%%%%%%%%%%%%%%%%%%%%%%%
\begin{figure}%[p!]
\begin{center}
\begin{tabular}{c}
\includegraphics*[bb=26 22 292 222, width=0.5\textwidth]{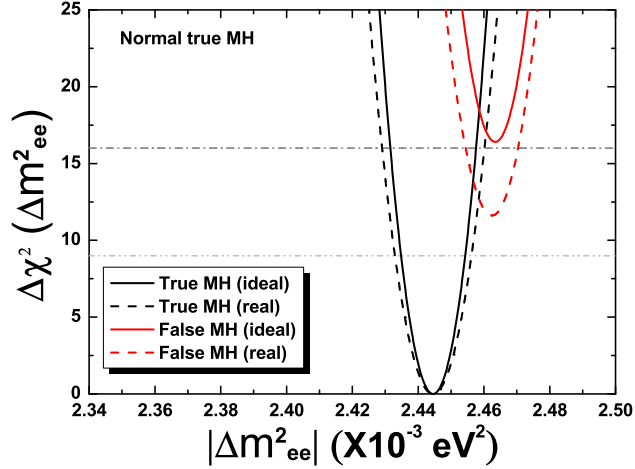}
\end{tabular}
\end{center}
\caption{The comparison of the MH sensitivity for the ideal and
actual distributions of the reactor cores. The real distribution
gives a degradation of $\Delta \chi^2_{\text{MH}}\simeq5$.}
\label{fig:mh:ideal}
\end{figure}
%%%%%%%%%%%%%%%%%%%% Fig. x %%%%%%%%%%%%%%%%%%%%%%%%%%%%%%
The discriminator defined in Eq.~(\ref{eq:mh:chisquare}) can be used
to obtain the optimal baseline, which are shown in the left panel of
Fig.~\ref{fig:mh:baseline}. A sensitivity of $\Delta
\chi^2_{\text{MH}}\simeq16$ is obtained for the ideal case with
identical baselines at around 50 km. The impact of the baseline
difference due to multiple reactor cores is shown in the right panel
of Fig.~\ref{fig:mh:baseline}, by keeping the baseline of one
reactor unchanged and varying that of another. A rapid oscillatory
behavior is observed and demonstrates the importance of reducing the baseline
differences of reactor cores. The worst case is at $\Delta L
\sim 1.7$~km, where the $|\Delta m^2_{ee}|$ related oscillation is
cancelled between two reactors.

Considering the baseline optimization and impact of the baseline
difference, we select of the experimental site. A
candidate site was identified by taking account of the physical
performance and detailed geological survey. With the spatial coordinates of
the experimental site and reactor cores, the actual power and baseline distributions
for the reactor cores of Yangjiang (YJ) and Taishan (TS) NPPs are
shown in Tab.~\ref{tab:intro:NPP}. The remote reactors in the
Daya Bay (DYB) and the possible Huizhou (HZ) NPP are also included.
The reduction of sensitivity due to the actual distribution of
reactor cores is shown in Fig.~\ref{fig:mh:ideal}, which gives a
degradation of $\Delta \chi^2_{\text{MH}}\simeq5$. The
degradation includes $\Delta \chi^2_{\text{MH}}\simeq3$ due to the baseline differences
of Taishan and Yangjiang NPP, and
$\Delta \chi^2_{\text{MH}}\simeq1.7$ with the inclusion
of Daya Bay and Huizhou NPPs. Other NPPs in operation and construction are much further away from
the experimental site (larger than 400 km) and can be neglected in the MH studies (reduction of $\Delta \chi^2_{\text{MH}}<0.2$).
In all the following, the actual spacial distribution of reactor cores
as shown in Tab.~\ref{tab:intro:NPP} is taken into account. %\ref{tab:intro:NPP}
%%
%\begin{table}%[p!]
%\centering
%\begin{tabular}{|c|c|c|c|c|c|c|}\hline\hline
%Cores & YJ-C1 & YJ-C2 & YJ-C3 & YJ-C4 & YJ-C5  & YJ-C6 \\
%\hline Power (GW) & 2.9 & 2.9 & 2.9 & 2.9 & 2.9 & 2.9 \\ \hline
%Baseline(km) & 52.75 & 52.84 & 52.42 & 52.51 & 52.12 & 52.21 \\
%\hline\hline
%Cores & TS-C1 & TS-C2 & TS-C3 & TS-C4 & DYB  & HZ \\
%\hline Power (GW) & 4.6 & 4.6 & 4.6 & 4.6 & 17.4 & 17.4 \\ \hline
%Baseline(km) & 52.76 & 52.63 & 52.32 & 52.20 & 215 & 265 \\
%\hline
%\end{tabular}
%\caption{Summary of the power and baseline distribution for the
%Yangjiang (YJ) and Taishan (TS) reactor complexes, as well as the
%remote reactors of Daya Bay (DYB) and Huizhou (HZ).
%\label{tab:mh:distribution}}
%\end{table}

\subsubsection{Requirement on the Energy Resolution}
\begin{figure}%[p!]
\begin{center}
\begin{tabular}{c}
\includegraphics*[width=0.6\textwidth]{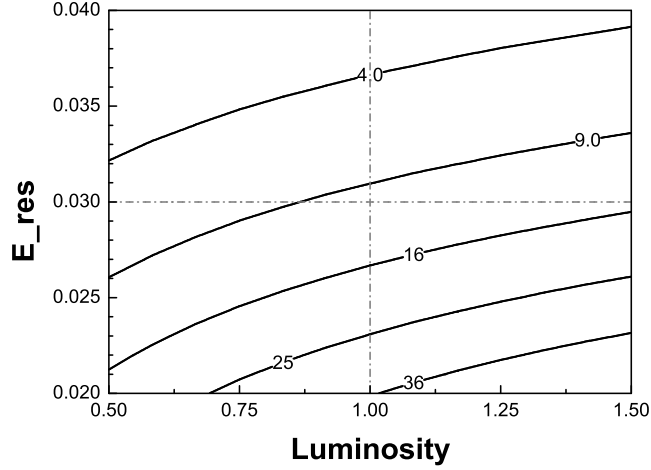}
%&
%\includegraphics*[width=0.5\textwidth]{Figs/MH_bterm.eps}
\end{tabular}
\end{center}
\caption{The iso-$\Delta \chi^2_{\text{MH}}$ contour plot as the
function of the event statistics (luminosity) and the energy resolution, where the vertical dash-dotted line stands
for the nominal running of six years with 80\% signal efficiency.} \label{fig:MH:resolution}
\end{figure}

The energy resolution as
or better than the size of $\Delta m^2_{21}/|\Delta m^2_{31}|$ is
required in order to precisely measure both the fast oscillations
(driven by $\Delta m^2_{31}$ and $\Delta m^2_{32}$) and slow
oscillation (driven by $\Delta m^2_{21}$) at a medium baseline. In
our nominal setup, the detector energy resolution $3\%/\sqrt{E{\rm
(MeV)}}$ is defined from the photon-electron statistics (1200
p.e./MeV), with the energy $E$ defined as the visible energy of the
positron. To show the effects of the energy resolution and event
statistics, we illustrate the iso-$\Delta \chi^2_{\text{MH}}$
contour plot as a function of the two key factors in Fig.~\ref{fig:MH:resolution}.
The nominal luminosity is defined as
the IBD event statistics in Sec.~\ref{subsec:mh:sensitivity:setup}.
From the figure, we can observe that the energy resolution of
$2.6\%/\sqrt{E{\rm (MeV)}}$ and $2.3\%/\sqrt{E{\rm (MeV)}}$ is
required to achieve the sensitivity of $\Delta
\chi^2_{\text{MH}}\simeq16$ or $\Delta \chi^2_{\text{MH}}\simeq25$
with the nominal statistics, respectively. With an increase of the
statistics by $50\%$, the energy resolution of $2.9\%/\sqrt{E{\rm
(MeV)}}$ and $2.6\%/\sqrt{E{\rm (MeV)}}$ are required to achieve the
same sensitivity of $\Delta \chi^2_{\text{MH}}\simeq16$ or $\Delta
\chi^2_{\text{MH}}\simeq25$.

For a real experimental environment, there are other important factors beyond the photon-electron statistics
that affect the energy resolution, such as the dark noise from PMT and electronics,
the detector non-uniformity and vertex resolution, as well as the PMT charge resolution. A generic parametrization
for the detector energy resolution is defined as
\begin{equation}
\frac{\sigma_{E}}{E} = \sqrt{\left(\frac{a}{\sqrt{E}}\right)^2+b^2+\left(\frac{c}{E}\right)^2}\,\;,\label{eq:mh:abcterms}
\end{equation}
where the visible energy $E$ is in the unit of MeV.

Based on our numerical calculation of the MH sensitivity in terms of $\Delta \chi^2_{\text{MH}}\,$,
we find an approximate relation for effects of non-stochastic terms (i.e., $b$, $c$)
using the equivalent $a$ term,
\begin{equation}
%\frac{\sigma_{E}}{E} =
\sqrt{\left(\frac{a}{\sqrt{E}}\right)^2+b^2+\left(\frac{c}{E}\right)^2}
\simeq \sqrt{\left(\frac{a}{\sqrt{E}}\right)^2+\left(\frac{1.6\;b}{\sqrt{E}}\right)^2+\left(\frac{c}{1.6\;\sqrt{E}}\right)^2}
\,\;,\label{eq:mh:aterm}
\end{equation}
which indicates that the influence of $b$ is $1.6$ times larger than the $a$ term, and
$c$ is less significant than $a$ by a factor of $1.6$. Therefore, a requirement for the resolution of
${a}/{\sqrt{E}}$ better than $3\%$ is equivalent to the following requirement,
\begin{equation}
%\frac{\sigma_{E}}{E} =
\sqrt{\left({a}\right)^2+\left({1.6\times b}\right)^2+\left(\frac{c}{1.6}\right)^2}\leq 3\%\;.\label{eq:mh:abc}
\end{equation}
Using Fig.~\ref{fig:MH:resolution} and the approximation in Eq.~(\ref{eq:mh:aterm}), we
can study different effects of detector design parameters and optimize the corresponding requirements.

The energy resolution of the JUNO detector is projected in Appendix~\ref{subsubsec:reconstruction} with a full MC simulation.
Toy MC is also used to study the degradation due to the PMT charge resolution, dark noise, quantum efficiency variation, and smearing from the vertex reconstruction, as shown in Tab.~\ref{tab:intro:energyres}.
Besides the detector response and reconstruction, the variation of the neutron recoil energy also degrades the resolution of the reconstructed neutrino energy, which introduces a degradation of $\Delta \chi^2_{\rm MH} \simeq 0.1$ on the MH sensitivity.

\subsubsection{Statistical Interpretation}

In this section, we shall present a brief summary of the MH
statistics and relation to the sensitivity. The following discussion
is crucial to properly understand the sensitivity results shown in
Fig.~\ref{fig:mh:ideal}. The determination of MH is equivalent to
resolving the sign of $\Delta m^2_{31}$. From the
statistics point of view, the determination of MH is a test to
distinguish two discrete hypotheses (NH vs. IH).

First let us employ the commonly used approach in the Frequentist
statistics. Given a null hypothesis $H_0$ and the alternative
hypothesis $H_1$, we can choose a test statistic $T$ in order to
test whether data can reject the null hypothesis $H_0$. The
confidence level (CL) $(1-\alpha)$ to reject $H_0$ is related to the
type-I error rate $\alpha$, where,
\begin{itemize}
\item {\it type-I error rate $\alpha$} is defined as the probability of
rejecting the null hypothesis $H_0$, if $H_0$ is true.
\end{itemize}
From the definition, one can define the relation between a critical
value of the observation $T^{\alpha}_{c}$ and the the type-I error
rate $\alpha$ as
\begin{equation}
 \int_{T_c^\alpha}^\infty p(T|H_{0}) dT = \alpha \,,
\end{equation}
with $p(T|H_{0})$ being the probability distribution function of $T$
given that $H_{0}$ is true. Moreover, we can further define the
conversion between the double-sided Gaussian $n\sigma$ and the value
of $\alpha$ as
\begin{equation}
\label{eq:MH:sigma-alpha} \alpha(n) = \frac{2}{\sqrt{2\pi}}
\int_{n}^{\infty} dx \, e^{-x^2/2} = {\rm erfc}\left(\frac n{\sqrt
2}\right)\,,
\end{equation}
where ${\rm erfc}(x)$ is the complementary error
function. This definition implies that we identify standard
deviations of $1\sigma,2\sigma,3\sigma$ with a CL $(1-\alpha)$ of
68.27\%, 95.45\%, 99.73\%, respectively.

On the other hand, the power of a given test $T$ is related to the
type-II error rate $\beta$, where
\begin{itemize}
\item {\it type-II error rate $\beta$} is defined as the probability
of accepting the null hypothesis H$_0$, if H$_1$ is true.
\end{itemize}
According to the definition, $\beta$ is calculated as
\begin{equation}
 \beta = P(T < T_c^\alpha|H_{1}) = \int_{-\infty}^{T_c^\alpha} p(T|H_{1}) dT \,,
\end{equation}
where $p(T|H_{1})$ is the probability distribution function of
$T$ assuming the alternative hypothesis $H_{1}$ is true. $\beta$
depends on the CL $(1-\alpha)$ at which we want to reject $H$. A
small value of $\beta$ means that the type-II error rate is small,
the power of the test (which is defined as $1-\beta$) is large.

Based on different choices of $\beta$, one can have different
sensitivities, such as the median sensitivity~\cite{Schwetz:2006md,Blennow:2013oma}
($\beta=50\%$) and the crossing sensitivity~\cite{Blennow:2013oma,Qian:2012zn,Ge:2012wj,Ciuffoli:2013rza}
($\beta=\alpha$). The former is defined
for the expected sensitivity of an averaged experiment, and the
latter corresponds to the CL at which exactly one of the two
hypotheses can be rejected. By definition, the crossing sensitivity
gives smaller confidence levels than the median sensitivity and is
not necessarily connected to what would be expected from an designed
experiment \cite{Blennow:2013oma}.
\begin{figure}
\begin{center}
\begin{tabular}{c}
\includegraphics*[width=0.6\textwidth]{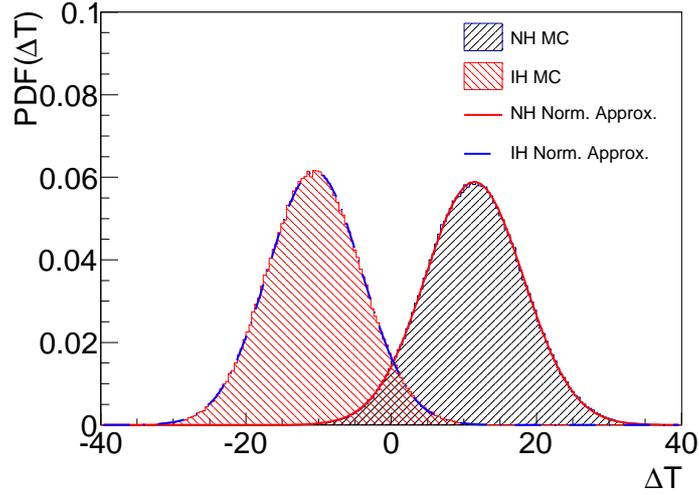}
\end{tabular}
\end{center}
\caption{\label{fig:mh:stat}  The $T$
distribution function for the JUNO nominal setup of six year running.}
\end{figure}

For the case of the MH determination at JUNO, we first define
our working test statistics similar to the discriminator in
Eq.~(\ref{eq:mh:chisquare}),
 \begin{equation}
T=\chi^2_{\rm min}(\rm I)-\chi^2_{\rm min}(\rm N).
\label{eq:mh:workingtest}
\end{equation}
According to the derivation in Refs.~\cite{Blennow:2013oma,Qian:2012zn,Ge:2012wj,Ciuffoli:2013rza},
the static $T$ follows the Gaussian distribution with
\begin{equation}
 T = \mathcal{N}\left( \pm \Delta \chi^2_{\text{MH}}, 2\sqrt{\Delta \chi^2_{\text{MH}}}\right) \,,
\label{eq:mh:Tgauss1}
\end{equation}
where the plus (minus) sign holds for the normal (inverted) MH. This
distribution is validated with an explicit MC
simulation in Fig.~{\ref{fig:mh:stat}}, showing
the excellent agreement between the numerical and analytical
calculations.

In Tab~\ref{tab:mh:sens}, we show the median sensitivity, standard
sensitivity (defined as $\sqrt{\Delta\chi^2_{\text{MH}}}\sigma$),
and crossing sensitivity respectively for the JUNO nominal setup,
where we can see that the commonly defined standard sensitivity is
very close to the median sensitivity and can be regard as the
expected sensitivity of a future experiment.
\begin{table}%[!htb]
\begin{center}
\begin{tabular}[c]{l|l|l|l} \hline\hline
 &  Median sens.  & Standard sens. & Crossing  sens.
\\ \hline
Normal MH & 3.4 $\sigma$  & 3.3 $\sigma$ & 1.9 $\sigma$  \\ \hline Inverted MH & 3.5 $\sigma$  & 3.4 $\sigma$  & 1.9 $\sigma$  \\
\hline\hline
\end{tabular}
\caption{\label{tab:mh:sens} The MH sensitivity with the JUNO
nominal setup of six year running.}
\end{center}
\end{table}

Before finishing the discussion on this approach, we want to stress
that the Frequentist approach does not directly address the question
how much one MH hypothesis is favored than the other MH hypothesis
given the experimental data. In the MH determination one can choose
the null hypothesis to be the NH, and the alternative hypothesis would
be the IH. The result of this hypothesis testing will tell us whether
the NH would be rejected or not, given the pre-defined rule.
Similarly, one should also perform a second hypothesis testing by
choosing null hypothesis to be the IH with the alternative hypothesis
being the NH. The result of the second test will tell us whether the IH
would be rejected or not.

On the other hand, the method of using the Bayesian statistics can
tell us the comparison of two MH hypotheses. The Bayes' theorem
gives the relationship of the posterior probability distribution
function $p({\rm MH}|{\rm D}, {\rm I})$, the prior probability
distribution function $p({\rm MH}|{\rm I})$ and the likelihood
function $L({\rm D}|{\rm MH}, {\rm I})$ as
\begin{equation}
p({\rm MH}|{\rm D}, {\rm I}) = \frac{L({\rm D}|{\rm MH}, {\rm
I})\cdot p({\rm MH}|{\rm I})}{\int L({\rm D}|{\rm MH}, {\rm I})\cdot
p({\rm MH}|{\rm I})}\,, \label{eq:mh:bayes}
\end{equation}
where ${\rm D}$ represents the data of a measurement, ${\rm I}$ is
the prior information on the MH hypotheses and the integration are
carried out for all the oscillation parameters. Given data from the
measurement, one can calculate $T_{\rm NH}=-2\,{\rm Log}[p({\rm
NH}|{\rm D}, {\rm I})]$ and $T_{\rm IH}=-2\,{\rm Log}[p({\rm
IH}|{\rm D}, {\rm I})]$. $T_{\rm NH}$ and $T_{\rm IH}$ are
marginalized over all nuisance parameters including unknown
parameters and systematic uncertainties of the experiment to obtain
$T_{\rm NH}^{\rm mag}$ and $T_{\rm IH}^{\rm mag}$,
respectively~\footnote{In the marginalization process, one
integrates the likelihood function over the entire phase space of
nuisance parameters. }.

Assuming that the prior information of NH
vs. IH is 50\% vs. 50\%, the probability ratio of the IH vs. NH,
which is $p(\rm IH|{\rm D},{\rm I})$ vs. $p(\rm NH|{\rm D},{\rm
I})$, can be calculated as $e^{-\Delta \tau/2}$ vs. 1, with $\Delta
\tau = T_{\rm IH}^{\rm mag} - T_{\rm NH}^{\rm mag}$. %~\cite{mXQstat}
As illustrated in Ref.~\cite{Qian:2012zn}, an approximation $\Delta \tau
\approx \Delta\chi^2_{\text{MH}}$ can be made in practice. More
generally, $\Delta \tau$ can be explicitly calculated through MC
simulations or other advanced integration techniques.
Tab.~\ref{table:MH_bayesstat} lists a few values of
$\Delta\chi^2_{\text{MH}}$ and their corresponding probability
ratios. The final results are then presented in terms of probability
ratio which is a natural and simple way to present results for the
MH determination.
\begin{table}[htp]
\centering
\begin{tabular}{|c|c|c|c|c|c|} \hline\hline
$\Delta \tau \approx \Delta\chi^2_{\text{MH}}$ & 1 & 4 & 9 & 16 & 25
\\\hline
$p(\rm IH|{\rm D},{\rm I})$ vs. $p(\rm NH|{\rm D},{\rm I})$ & 38\%
vs.  & 12\% vs.
& 1.1\% vs.  & 0.034\% vs.  & 3.7 $\times10^{-6}$ vs.  \\
& 62\% & 88\% & 98.9\% & 99.966\% & 100\% \\\hline $p(\rm IH|{\rm
D},{\rm I})/p(\rm NH|{\rm D},{\rm I})$ & 0.61 & 0.136 & 0.011 &
3.4$\times10^{-4}$ & $3.7\times10^{-6}$
\\\hline \hline
\end{tabular}
\caption{\label{table:MH_bayesstat} Probability ratios with respect
to several typical $\Delta\chi^2_{\text{MH}}$ values.}% ~\cite{Qian:2012zn}.}
\end{table}

\subsection{Systematics}
In this section, we shall discuss the effects of systematics in
the MH measurement in reactor antineutrino oscillations, which
includes the reactor related uncertainties, detector related
uncertainties, background related uncertainties and the energy
related uncertainties.

\subsubsection{Reactor related uncertainties}
As discussed before, the MH information is encoded in the spectral shape of reactor antineutrino oscillations.
The absolute normalization uncertainty from the reactor flux at the current level
has negligible impact on the MH determination. Therefore only the reactor-related shape uncertainty is
considered in the following.

We incorporate the shape uncertainty to each bin by modifying the $\chi^2$ definition in Eq.~(\ref{eq:mh:chiREA}) as follows:
\begin{equation}
\chi^2_{\text{REA}}=\sum^{N_{\text{bin}}}_{i=1}\frac{[M_{i} -
T_{i}(1+\sum_k \alpha_{ik}\epsilon_{k})]^2}{M_{i}+(\sigma M_i)^2} +
\sum_k\frac{\epsilon^2_{k}}{\sigma^2_k}\,,\label{eq:mh:chiREAb2b}
\end{equation}
where $\sigma$ denotes the relative (uncorrelated) shape
uncertainty. The $\Delta \chi^2_{\text{REA}}$ degrades less than 1 when we set
the 1\% shape uncertainty ($\sigma = 1\%$), while the degradation is
about 2 in $\Delta \chi^2_{\text{REA}}$ when we increase $\sigma$ to $2\%$.
Therefore, careful estimation and reduction of the energy uncorrelated shape uncertainties are mandatory to achieve the required sensitivity.

Because the model predictions for the reactor antineutrino spectrum are inconsistent
with the measurement from ongoing reactor experiments (i.e., the bump between $4\sim6$ MeV)~\cite{DYBbump,Seon-HeeSeofortheRENO:2014jza,Abe:2014bwa}.
Moreover, a recent theoretical calculation trying to understand the above inconsistency observes additional high-frequency fine structures~\cite{Dwyer:2014eka}
in the reactor antineutrino spectrum. Both of the mentioned spectral structures may induce additional systematics of the shape uncertainty. MC studies
of the MH sensitivity on the effects of these spectral structures are carried out, which show that the changes in $\Delta \chi^2_{\text{REA}}$
can be controlled to be less than 1. In summary, it is crucial to control the effect of reactor spectral structures to reduce systematics of the shape uncertainty.

\subsubsection{Detector related uncertainties}
Similar to the reactor normalization uncertainty,
the uncertainty in the detection absolute efficiency also has negligible impact on the MH determination.
Therefore it is desirable to study the energy related uncertainties.

%%%%%%%%%%%%%%%%%%%% Fig. 13 %%%%%%%%%%%%%%%%%%%%%%%%%%%%%%
\begin{figure}%[p!]
\begin{center}
\begin{tabular}{c}
\includegraphics*[width=0.6\textwidth]{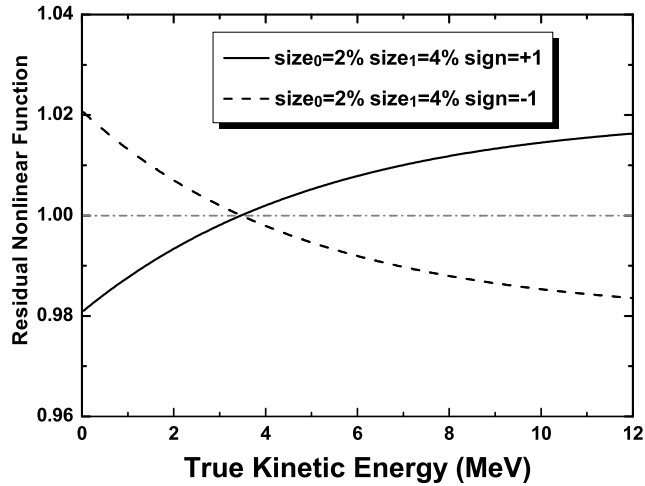}
%\\
%\includegraphics*[width=0.6\textwidth]{./Figs/nonlinear.eps}
\end{tabular}
\end{center}
\caption{Two classes of typical examples for the residual non-linear
functions in our simulation.\label{fig:mh:nlcurve}}
\end{figure}
%%%%%%%%%%%%%%%%%%%%%%%%%%%%%%%%%%%%%%%%%%%%%%%%%%%%%%%%%%
%%%%%%%%%%%%%%%%%%%% Fig. 14 %%%%%%%%%%%%%%%%%%%%%%%%%%%%%%
\begin{figure}%[p!]
\begin{center}
\begin{tabular}{cc}
\includegraphics*[bb=25 20 295 228, width=0.46\textwidth]{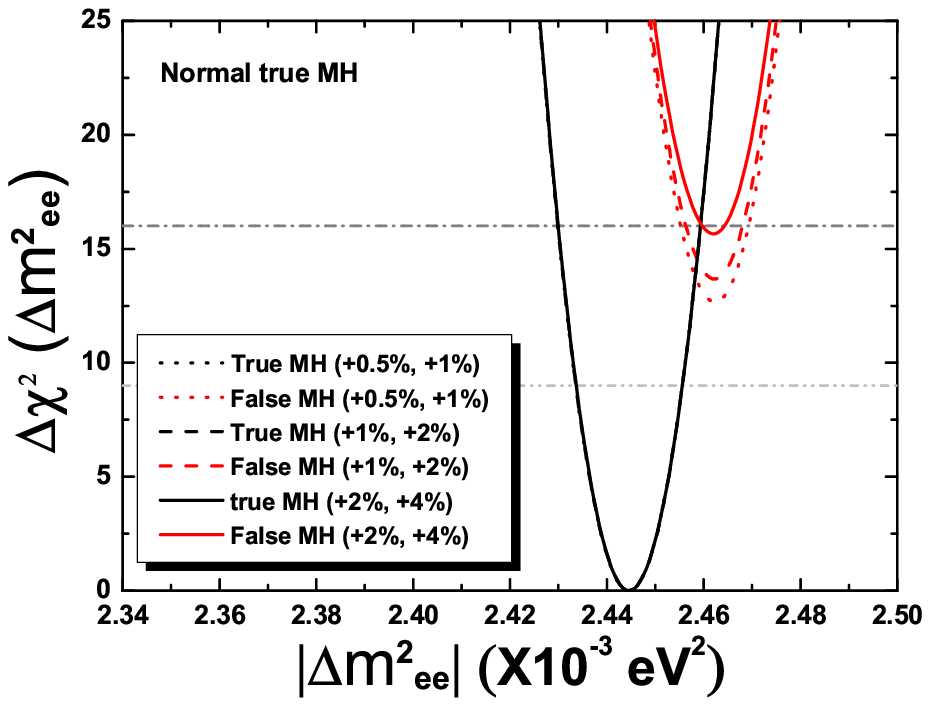}
&
\includegraphics*[bb=25 20 295 240, width=0.46\textwidth]{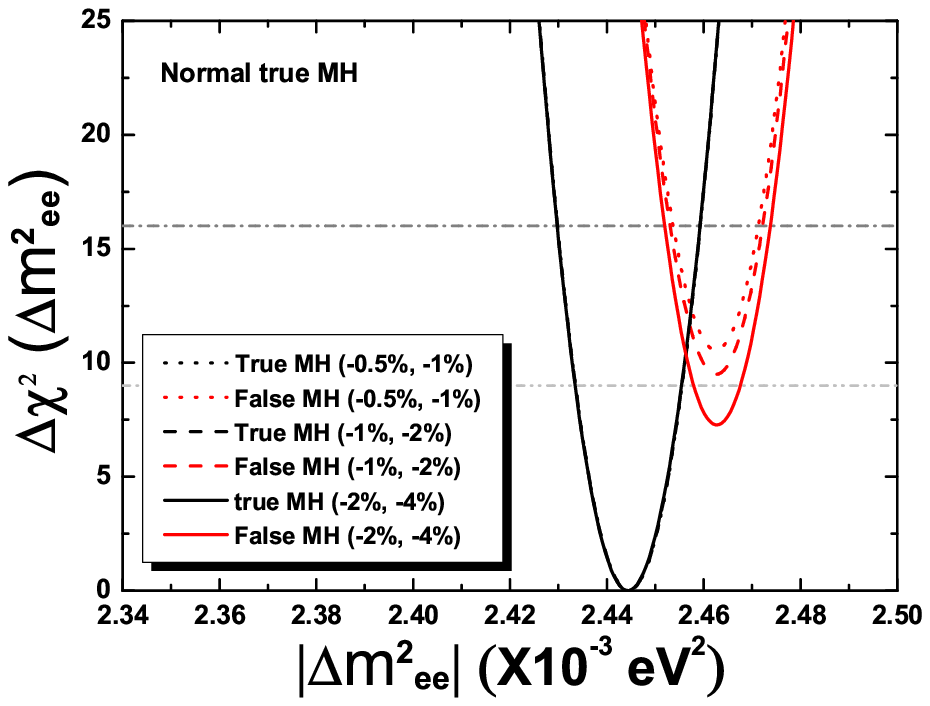}
\\
%\includegraphics*[bb=25 20 295 228, width=0.46\textwidth]{./Figs/Fixed_Inv_POS.eps}
%&
%\includegraphics*[bb=25 20 295 240, width=0.46\textwidth]{./Figs/Fixed_Inv_NEG.eps}
\end{tabular}
\end{center}
\caption{$\Delta \chi^2$ distribution as functions of the free parameters $|\Delta m^{2}_{ee}|$,
where the normal MH is assumed, and the plus (left) and minus (right) signs of the non-linearity
curves in Fig~\ref{fig:mh:nlcurve} are implemented, respectively.
The dashed and dotted lines are for the cases with reduced sizes of the non-linearity. \label{fig:mh:nlfixed}}
\end{figure}
%%%%%%%%%%%%%%%%%%%%%%%%%%%%%%%%%%%%%%%%%%%%%%%%%%%%%%%%%%
%%%%%%%%%%%%%%%%%%%% Fig. 13 %%%%%%%%%%%%%%%%%%%%%%%%%%%%%%
\begin{figure}%[p!]
\begin{center}
\begin{tabular}{c}
\includegraphics*[width=0.6\textwidth]{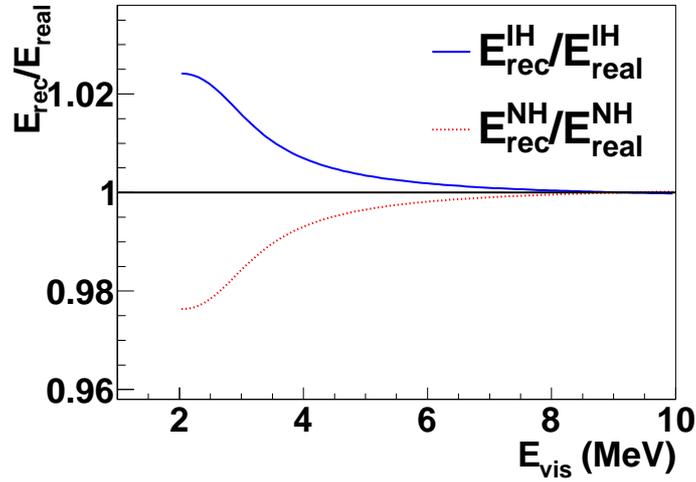}
\end{tabular}
\end{center}
\caption{The non-linearity models with the largest effects of mimicking between the normal MH and inverted MH.
\label{fig:mh:xinmodel}}
\end{figure}
%%%%%%%%%%%%%%%%%%%%%%%%%%%%%%%%%%%%%%%%%%%%%%%%%%%%%%%%%%
A dedicated calibration of the detector energy non-linearity response is another critical factor
to obtain reliable sensitivity of the MH determination.
\begin{figure}%[p!]
\begin{center}
\begin{tabular}{cc}
\includegraphics*[bb=25 20 295 228, width=0.46\textwidth]{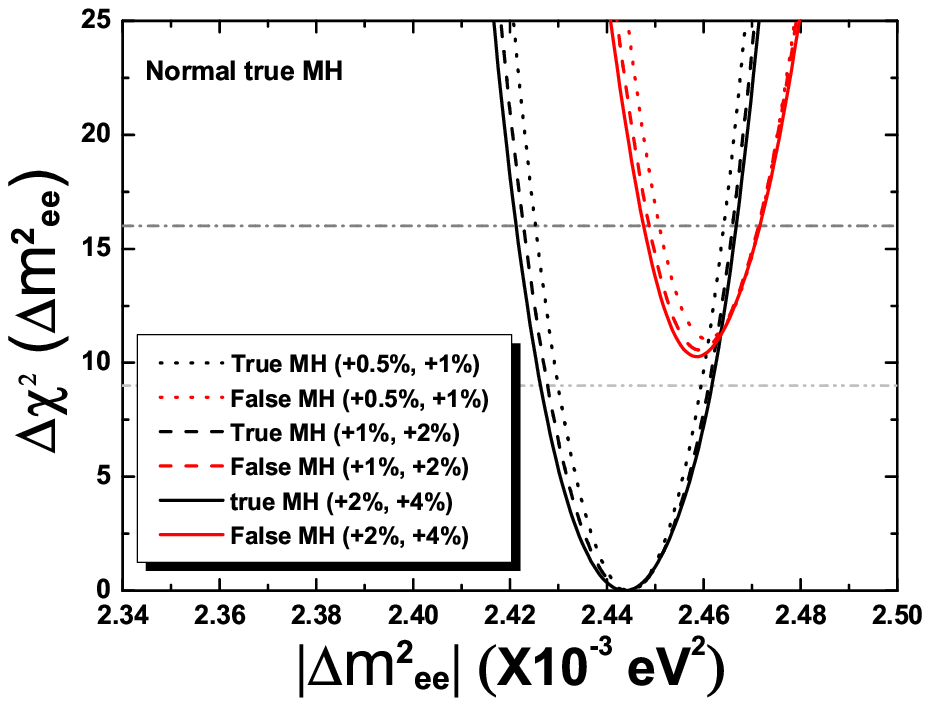}
&
\includegraphics*[bb=25 20 295 240, width=0.46\textwidth]{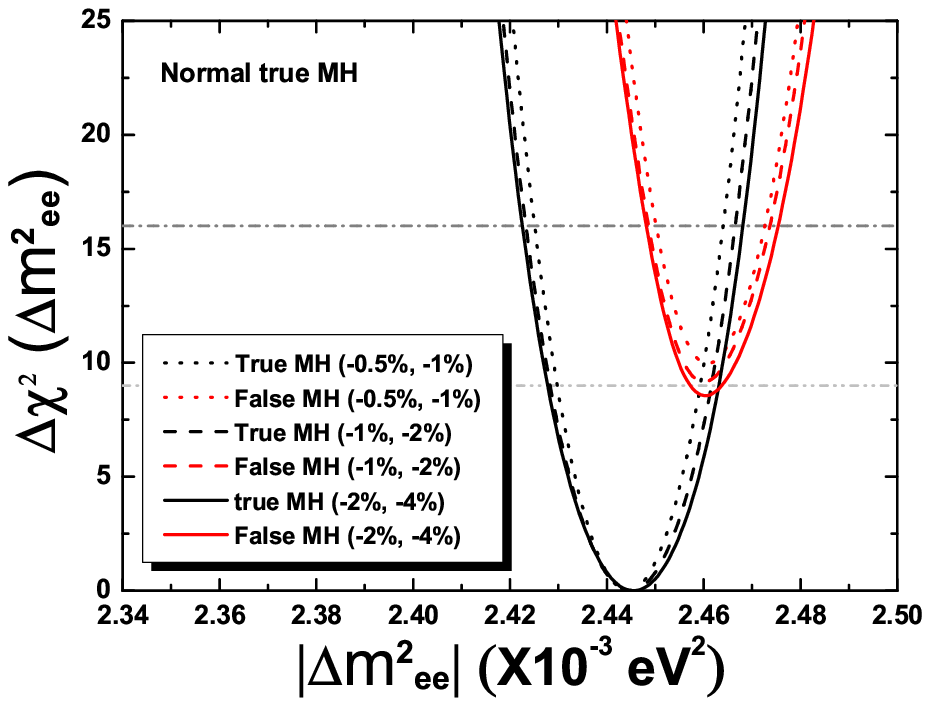}
%\\
%\includegraphics*[bb=25 20 295 228, width=0.46\textwidth]{./Figs/Free_Inv_POS.eps}
%&
%\includegraphics*[bb=25 20 295 240, width=0.46\textwidth]{./Figs/Free_Inv_NEG.eps}
\end{tabular}
\end{center}
\caption{Effects of two classes of energy non-linearity models [with plus (left) and minus (right) signs] in determination
of the MH with the self-calibration effect,
where the normal MH is assumed. \label{fig:mh:nlfree}}
\end{figure}
%%%%%%%%%%%%%%%%%%%%%%%%%%%%%%%%%%%%%%%%%%%%%%%%%%%%%%%%%%
The uncertainty from the detector nonlinearity response can distort the antineutrino
spectrum and is crucial for JUNO, since a precise energy spectrum
of reactor antineutrinos is required to resolve the MH.
Assuming the energy non-linearity correction is imperfect, we study the impact on the sensitivity by
including in our simulation a residual non-linearity between the
measured and expected neutrino spectra. By including the residual
non-linearity with the assumed form shown in Fig~\ref{fig:mh:nlcurve},
we obtain the $\Delta \chi^2$
distribution as functions of the free parameter $|\Delta m^{2}_{ee}|$ in Fig.~\ref{fig:mh:nlfixed},
where the normal MH is assumed and the plus (left) and minus (right) signs of the non-linearity curves are implemented, respectively.
The solid, dashed and dotted lines are for the cases with different sizes of the non-linearity.

From Fig.~\ref{fig:mh:nlfixed}, we observe that non-linearity with the minus sign would significantly reduce the sensitivity
of MH determination for the true normal MH. In principle, there is the worst case of non-linearity that the wrong MH may perfectly
mimic the true one, which defines as
\begin{equation}
\frac{E_{\rm rec}}{E_{\rm true}} = \frac{2|\Delta' m^2_{ee}| + \Delta m^2_{\phi}}{2|\Delta m^2_{ee}|  - \Delta m^2_{\phi}}\;,
\label{eq:escale}
\end{equation}
where $\Delta m^2_{ee}$ and $\Delta' m^2_{ee}$ are the effective mass-squared differences
in Eq.~(\ref{eq:mh:dmee}) for the true and false MHs, respectively. Thanks to the current measurements of
the neutrino oscillation parameters, we can illustrate the specific non-linearity curves for the normal MH
and inverted MH in Fig.~\ref{fig:mh:xinmodel}. With this residual non-linearity in the measurement of Eq.~(\ref{eq:mh:chiREA}),
we can obtain the degradation of $\Delta \chi^2\simeq2.5$ in agreement with the effect in the right panel of
Fig.~\ref{fig:mh:nlfixed}.

For the reactor antineutrino experiment at a medium baseline, we can observe multiple peaks of the
$\Delta m^2_{ee}$ induced oscillation. Each of the peak position carries the
information of $\Delta m^2_{ee}$. This redundancy can be used to
evaluate the energy scale at different energies, providing a self-calibration way of measuring the energy non-linearity~\cite{Li:2013zyd}.
To illustrate, we consider a test quadratic non-linear function in the prediction of Eq.
(\ref{eq:mh:chiREA}),
where the coefficients of the function are arbitrary and will be determined in the fitting.
Therefore, in the simplest way, we illustrate the self-calibration effect in Fig.~\ref{fig:mh:nlfree},
where the normal MH is assumed. We observe that the increase and reduction of the MH sensitivity
due to unknown energy non-linearity can be resolved to some extent, and consistent sensitivity of the MH
determination are obtained. Notice that the width of the $\Delta \chi^2$ functions in Fig.~\ref{fig:mh:nlfree} is broadened,
because additional uncertainties from the parameters of test quadratic non-linearity are introduced.

\subsubsection{Background related uncertainties}

\begin{figure}
\begin{center}
\includegraphics[width=0.5\textwidth]{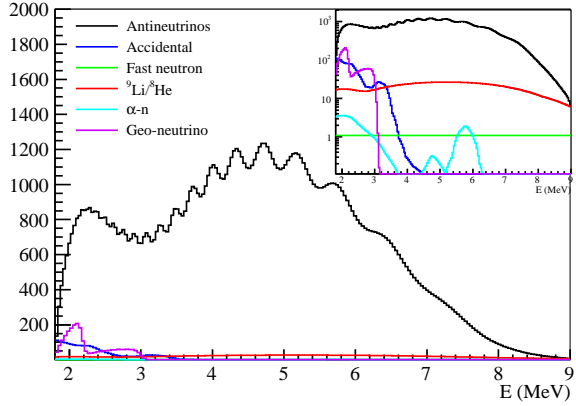}
\caption{Spectra for the antineutrino signal and five kinds of main backgrounds, including the accidental, $^8$He/$^9$Li, fast neutron, and
$^{13}$C$(\alpha, n)^{16}$O and geo-neutrinos.} \label{fig:backgrounds}
\end{center}
\end{figure}
\begin{table}%[!htb]
\centering
\begin{tabular}{|c|c|c|c|}\hline\hline
Event type & Rate (per day) & Rate uncertainty (relative) & Shape uncertainty \\
IBD candidates & 60 & - & - \\
Geo-$\nu$s & 1.1 & 30\% & 5\%\\
Accidental signals & 0.9 & 1\% & negligible \\
Fast-$n$ & 0.1 & 100\% & 20\%\\
$^9$Li-$^8$He & 1.6 & 20\% & 10\%\\
$^{13}$C$(\alpha, n)^{16}$O  & 0.05 & 50\% & 50\% \\
\hline\hline
\end{tabular}
\caption{The background summary table for the analysis of reactor antineutrinos.
\label{tab:MH:bkg}}
\end{table}
We further study the effects of background related uncertainties.
From Tab.~\ref{tab:MH:bkg}, the total background to signal (B/S) ratio is 6.3$\%$, which contributes to a reduction of
$\Delta \chi^2_{\text{MH}}\simeq0.6$. Second, the rate uncertainties of backgrounds are negligible for the MH determination
since they are nicely constrained in the precision spectral measurements. Finally, the expected energy spectra for five kinds of main backgrounds
are shown in Fig.~\ref{fig:backgrounds}. The total background shape uncertainties
contribute to a $0.4\%$ bin-to-bin uncertainty, which can further reduce the MH sensitivity by $\Delta \chi^2_{\text{MH}}\simeq0.1$.

\subsubsection{Systematics summary}
To conclude, we summarize the decomposition of experimental systematics in the MH determination in Tab.~\ref{eq:mh:decompos}.

\begin{itemize}
\item Ideal distribution of reactor cores with the equal baseline of $52.5\,{\rm km}$ gives the
MH sensitivity of $\Delta \chi^2_{\text{MH}}\simeq16$.
\item In reality, the real baseline distribution of reactor cores in Taishan and Yangjiang NPPs from Tab.~\ref{tab:intro:NPP}
induces a degradation of $\Delta \chi^2_{\text{MH}}\simeq3$.
\item An additional reduction of $\Delta \chi^2_{\text{MH}}\simeq1.7$ is obtained due to inclusion of Daya Bay and Huizhou NPPs.
\item The reactor shape uncertainty of $1\%$ will further degrade the $\Delta \chi^2_{\text{MH}}$ by 1.
\item The statistical and shape uncertainties of backgrounds with the estimation of Tab.~\ref{tab:MH:bkg} contribute
to $\Delta \chi^2_{\text{MH}}\simeq-0.6$ and $\Delta \chi^2_{\text{MH}}\simeq-0.1$, respectively.
\item As will be discussed in the next subsection, an increase of $\Delta \chi^2_{\text{MH}}\simeq+8$
can be obtained by including a measurement of $|\Delta m^{2}_{\mu\mu}|$ at the 1\% precision level.
\end{itemize}

\begin{table}[!htb]
\begin{center}
\begin{tabular}[c]{l|l|l|l|l|l|l|l} \hline\hline
  & Stat. & Core dist. & DYB \& HZ & Shape & B/S (stat.) & B/S (shape) & $|\Delta m^{2}_{\mu\mu}|$ \\
  \hline
  Size & $52.5\,{\rm km}$  & Tab.~\ref{tab:intro:NPP} & Tab.~\ref{tab:intro:NPP}& 1\% & 6.3\% & 0.4\% & 1\%  \\
  \hline
 $\Delta \chi^2_{\text{MH}}$ & $+16$ & $-3$ & $-1.7$ & $-1$ & $-0.6$ & $-0.1$ & $+(4-12)$ \\
\hline\hline
\end{tabular}
\caption{\label{eq:mh:decompos} Different contributions for the MH determination.
The first column is the statistical-only scenario with the equal baseline of $52.5\,{\rm km}$, the second column considers the real distribution (dist.) of reactor cores,
the third column defines the contribution of remote DYB and HZ NPPs, the fourth column stands for the reduction of the reactor shape uncertainty,
the fifth and sixth columns are the contributions of the background statistical and shape uncertainties, the seventh column is the enhanced sensitivity
from additional information of $|\Delta m^{2}_{\mu\mu}|$. }
\end{center}
\end{table}

\subsection{MH Sensitivity with Precision \texorpdfstring{$|\Delta m^{2}_{ee}|$}{Dm2ee}
and \texorpdfstring{$|\Delta m^{2}_{\mu\mu}|$}{Dm2mumu} Measurements}
\label{subsec:mh:meemmumu}

Due to the intrinsic difference between $|\Delta m^2_{ee}|$ and $|\Delta
m^2_{\mu\mu}|$, precise measurements of these two mass-squared differences
can provide additional sensitivity to MH, besides the sensitivity
from the interference effects.
To incorporate the contribution from the $|\Delta m^2_{\mu\mu}|$
measurement in long-baseline muon-neutrino oscillation
experiments, we define the following the extra pull function
\begin{equation}
\chi^2_{\text{pull}}(|\Delta m^2_{\mu\mu}|)= \frac{(|\Delta
m^2_{\mu\mu}|-|\overline{\Delta m^2_{\mu\mu}}|)^2}{\sigma^2(\Delta
m^2_{\mu\mu})}\,,\label{eq:mh:chipull}
\end{equation}
where $\overline{|\Delta m^2_{\mu\mu}|}$ and $\sigma(\Delta
m^2_{\mu\mu})$ are the central value and $1\sigma$ uncertainty of
the measurement. The combined $\chi^2$ function is
defined as
\begin{equation}
\chi^2_{\text{ALL}}=\chi^2_{\text{REA}}+\chi^2_{\text{pull}}(|\Delta
m^2_{\mu\mu}|)\,.\label{eq:mh:chiALL}
\end{equation}
Because two of the three mass-squared differences ($\Delta
m^2_{21}$, $\Delta m^2_{31}$ and $\Delta m^2_{32}$) are independent,
we choose $\Delta m^2_{21}$ and $\Delta m^2_{ee}$ defined in
Eq.~(\ref{eq:mh:dmee}) as the free parameters. Proper
values of $\Delta m^2_{\mu\mu}$ can be calculated by the relations
in Eq.~(\ref{eq:mh:dmemu}).

%%%%%%%%%%%%%%%%%%%% Fig. 16 %%%%%%%%%%%%%%%%%%%%%%%%%%%%%%
\begin{figure}%[p!]
\begin{center}
\begin{tabular}{cc}
\includegraphics*[bb=25 20 295 228, width=0.46\textwidth]{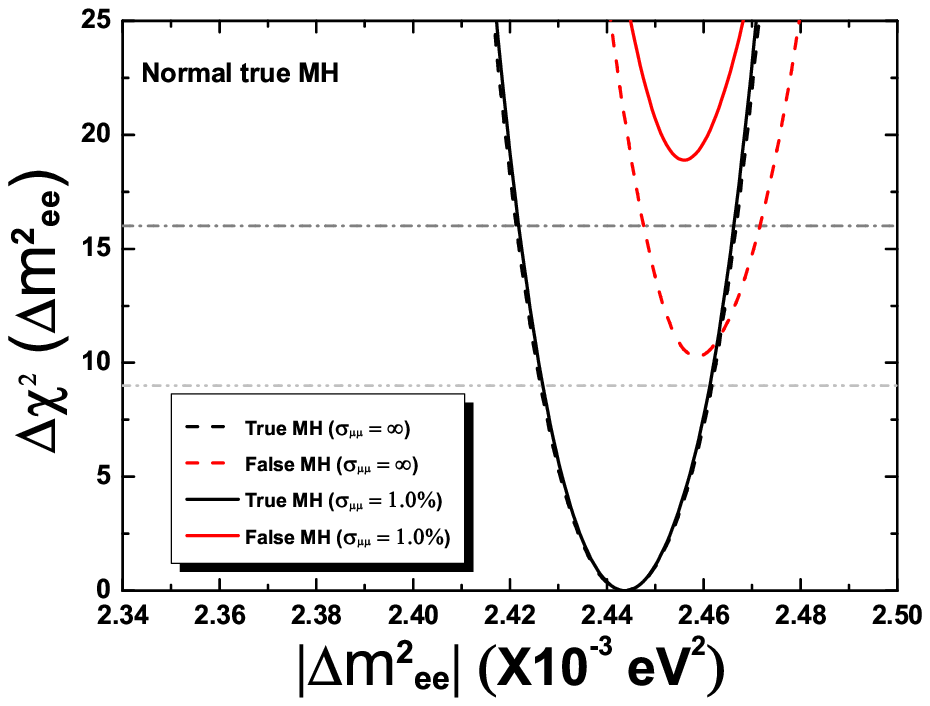}
&
\includegraphics*[bb=25 20 295 228, width=0.46\textwidth]{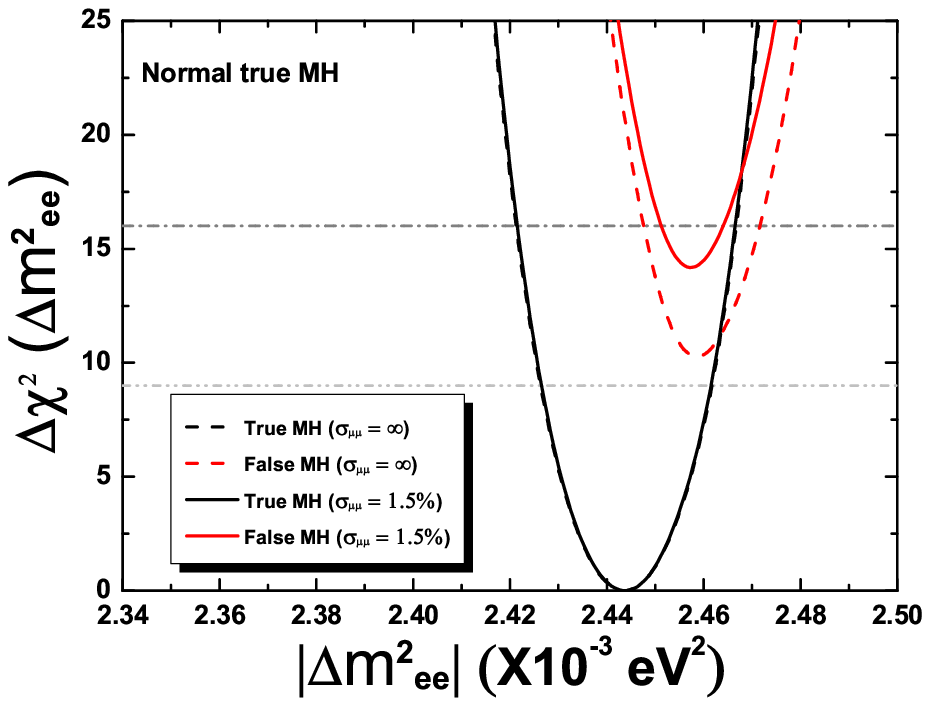}
\end{tabular}
\end{center}
\caption{the reactor-only (dashed) and combined (solid)
distributions of the $\Delta\chi^2$ function in Eq.
(\ref{eq:mh:chiREA}) and Eq. (\ref{eq:mh:chiALL}), where a $1\%$
(left panel) or $1.5\%$ (right panel) relative error of $\Delta
m^2_{\mu\mu}$ is assumed and the CP-violating phase ($\delta$) is
assigned to be $90^\circ/270^\circ$ ($\cos\delta=0$) for
illustration. The black and red lines are for the true (normal) and
false (inverted) neutrino MH, respectively.
\label{fig:mh:chi2eemumu}}
\end{figure}
%%%%%%%%%%%%%%%%%%%%%%%%%%%%%%%%%%%%%%%%%%%%%%%%%%%%%%%%%%

To illustrate the effect of the external $|\Delta m^2_{\mu\mu}|$
measurement, we calculate the separated and
combined $\chi^2$ functions in Eqs.
(\ref{eq:mh:chiREA}) and (\ref{eq:mh:chiALL}) in
Fig.~\ref{fig:mh:chi2eemumu}, where a $1\%$ (left panel) or $1.5\%$
(right panel) relative error of $\Delta m^2_{\mu\mu}$ is assumed.
The black and red lines are for the true (normal) and false
(inverted) MHs, respectively. The dashed and solid lines are for the
reactor-only [in Eq.~(\ref{eq:mh:chiREA})] and combined
distributions. Here a fixed CP-violating phase ($\cos \delta =0$) is assumed
for illustration. We can get a value of $\Delta
\chi^2_{\text{MH}}\simeq10$ for the reactor-only analysis in
the $\chi^2$ method. As for the contribution from the external
$|\Delta m^2_{\mu\mu}|$ measurement, it is almost negligible if we
choose the true (normal) MH in the fitting program. However, if the fitting
MH is the false (inverted) one, the central value of $\Delta
m^2_{ee}$ in the $\chi^2_{\text{pull}}$ function will change by two
times the difference in Eq.~(\ref{eq:mh:dmemu}), which
accordingly results in a significant contribution to the combined
$\chi^2$ function. Finally we can achieve $\Delta
\chi^2_{\text{MH}}\simeq19$ and $\Delta \chi^2_{\text{MH}}\simeq14$
for the $1\%$ and $1.5\%$ relative errors of the $|\Delta
m^2_{\mu\mu}|$ measurement, respectively. Considering the whole parameter space of $\delta^2_{\text{MH}}$
from 0 to $2\pi$, $\Delta \chi^2$ can range from 14 to 22 for the $1\%$ relative precision of $|\Delta
m^2_{\mu\mu}|$~\cite{Li:2013zyd}.

\subsection{Conclusions}
\label{subsec:mh:summary}
The determination of the neutrino mass hierarchy is of great importance in neutrino physics, since the MH provides a crucial input for future searches of neutrinoless double beta decays, observation of supernoca neutrino bursts, cosmological probe of neutrino properties, and model building of the neutrino masses and flavor mixing.

Thanks to the relatively large $\theta_{13}$ discovered in recent reactor and accelerator neutrino experiments,
precise measurements of the reactor antineutrino spectrum at a medium baseline of about 50 km can
probe the interference effect of two fast oscillation modes (i.e., oscillations induced by $\Delta
m^2_{31}$ and $\Delta m^2_{32}$) and sensitive to the neutrino MH.
The corresponding sensitivity depends strongly on the energy resolution,
the baseline differences and
energy response functions. Moreover, the MH sensitivity can be
improved by including a measurement of the effective mass-squared
difference in the long-baseline muon-neutrino disappearance
experiment due to flavor dependence of the effective mass-squared
differences.

We have calculated the MH sensitivity at JUNO taking into account the real
spatial distribution of reactor complexes, reactor related uncertainties,
detector related uncertainties and background related uncertainties.
We demonstrated that a median sensitivity of $\sim3\sigma$ can be achieved
with the reasonable assumption of the systematics and six years of running.
We emphasized that the reactor shape uncertainty and detector non-linearity response,
are the important factors to be dealt with.
In addition, we have studied the additional sensitivity by including precision measurements
of $|\Delta m^2_{\mu\mu}|$ from long baseline muon (anti)neutrino disappearance.
A confidence level of $\Delta \chi^2_{\text{MH}}\sim14$ ($3.7\,\sigma$) or $\Delta
\chi^2_{\text{MH}}\sim19$ ($4.4\,\sigma$) can be obtained, for the $|\Delta m^2_{\mu\mu}|$ uncertainty of 1.5\% or 1\%.

Besides the spectral measurement of reactor antineutrino oscillations, there are
other methods to resolve the MH using the matter-induced oscillation of accelerator or atmospheric neutrinos.
Worldwide, there are many ongoing and planed experiments designed in this respect. These include the
long baseline accelerator neutrino experiments (i.e. NO$\nu$A and DUNE) and atmospheric neutrino experiments
(i.e., INO, PINGU, Hyper-K). Using different oscillation patterns, different neutrino sources and different detector techniques,
they are complementary in systematics and contain a great amount of synergies.
Therefore, the mass hierarchy, being one of the most important undetermined
fundamental parameters in neutrino physics, clearly
deserves multiple experiments with preferably different experimental techniques.
A consistent resolution of the MH from all these
experiments will greatly increase our confidence in the MH
determination.

\clearpage

\newcommand{\obb}{0\mbox{$\nu\beta\beta$}}
\newcommand{\onbb}{neutrino-less double beta decay}
\newcommand{\ba}{\begin{array}{c}}
\newcommand{\baz}{\begin{array}{cc}}
\newcommand{\bad}{\begin{array}{ccc}}
\newcommand{\bea}{\begin{equation} \begin{array}{c}}
\newcommand{\eea}{ \end{array} \end{equation}}
\newcommand{\ea}{\end{array}}
\newcommand{\D}{\displaystyle}
\newcommand{\dms}{\mbox{$\Delta m^2_{\odot}$}}
\newcommand{\dma}{\mbox{$\Delta m^2_{\rm A}$}}
\newcommand{\meff}{\mbox{$\left| m_{ee} \right|$}}
\newcommand{\eV}{\mbox{ eV}}
\newcommand{\ppp}{\mbox{$(+++)$ }}
\newcommand{\pmm}{\mbox{$(+--)$ }}
\newcommand{\mpm}{\mbox{$(-+-)$ }}
\newcommand{\mmp}{\mbox{$(--+)$ }}

\newcommand{\be}{\begin{eqnarray}}
\newcommand{\ee}{\end{eqnarray}}
\newcommand{\etal}{{\it et al.}}
\def\nue{{\nu_e}}
\def\anue{{\bar\nu_e}}
\def\numu{{\nu_{\mu}}}
\def\anumu{{\bar\nu_{\mu}}}
\def\nutau{{\nu_{\tau}}}
\def\anutau{{\bar\nu_{\tau}}}
\def\lsim{\:\raisebox{-0.5ex}{$\stackrel{\textstyle<}{\sim}$}\:}
\def\gsim{\:\raisebox{-0.5ex}{$\stackrel{\textstyle>}{\sim}$}\:}
\newcommand{\ms}{\Delta m^2_{21}}
\newcommand{\ma}{\Delta m^2_{31}}
\newcommand{\sstwos}{\sin^2 2\theta_{12}}
\newcommand{\sss}{\sin^2 \theta_{12}}
\newcommand{\sch}{\sin^2 \theta_{13}}
\newcommand{\stch}{\sin^2 2\theta_{13}}
\newcommand{\sa}{\sin^2 \theta_{23}}
\newcommand{\css}{\cos^2 \theta_{12}}
\newcommand{\csh}{\cos^2 \theta_{13}}
\def\ltap{\ \raisebox{-.4ex}{\rlap{$\sim$}} \raisebox{.4ex}{$<$}\ }
\def\gtap{\ \raisebox{-.4ex}{\rlap{$\sim$}} \raisebox{.4ex}{$>$}\ }

\newcommand{\meffnh}{\mbox{$\left| m_{ee} \right|$}^{\rm NH}}
\newcommand{\meffih}{\mbox{$\left| m_{ee} \right|$}^{\rm IH}}
\newcommand{\meffqd}{\mbox{$\left| m_{ee} \right|$}^{\rm QD}}
\newcommand{\meffnhmin}{\mbox{$\left| m_{ee} \right|$}^{\rm NH}_{\rm min}}
\newcommand{\meffihmin}{\mbox{$\left| m_{ee} \right|$}^{\rm IH}_{\rm min}}
\newcommand{\meffqdmin}{\mbox{$\left| m_{ee} \right|$}^{\rm QD}_{\rm min}}
\newcommand{\meffnhmax}{\mbox{$\left| m_{ee} \right|$}^{\rm NH}_{\rm max}}
\newcommand{\meffihmax}{\mbox{$\left| m_{ee} \right|$}^{\rm IH}_{\rm max}}
\newcommand{\meffqdmax}{\mbox{$\left| m_{ee} \right|$}^{\rm QD}_{\rm max}}

\newcommand{\nme}{\mbox{nuclear matrix elements}}

\section{Precision Measurements of Neutrinos}
\label{sec:prec}

\blfootnote{Editors: Yufeng Li (liyufeng@ihep.ac.cn)}
\blfootnote{Major contributor: Xin Qian}

\subsection{Introduction and motivation}
\label{subsec:prec:intro}

JUNO is designed to collect a large number of reactor antineutrino events with excellent energy resolution
[3\%/$\sqrt E (\rm MeV)$] and accurate energy determination [better than 1\%].
Therefore, besides the determination of the neutrino
mass hierarchy (MH)~\cite{Li:2013zyd,Qian:2012xh}, in 6 years JUNO will also allow a detailed study of various
aspects of neutrino oscillations, including the extractions of the mixing parameters
$\theta_{12}$, $\Delta m^2_{21}$, and $|\Delta m^2_{ee}|$, and probe the fundamental
properties of neutrino oscillations. JUNO will be:
\begin{itemize}
\item the first experiment to simultaneously observe the
neutrino oscillation driven by both atmospheric and solar neutrino
mass-squared differences (Fig.~\ref{fig:prec:juno_spec}).  The pronounced
dip around 3 MeV corresponds to the solar $\Delta m^2$, and the rapid
oscillations correspond to the atmospheric $\Delta m^2$.
\item the first experiment to observe multiple
 oscillation cycles of the atmospheric $\Delta m^2$
  (Fig.~\ref{fig:prec:juno_spec}).
\item the well place to have an unprecedented precision measurement of $\sin^2\theta_{12}$, $\Delta m^2_{21}$
and $|\Delta m^2_{ee}|$ to better than 1\%.
\end{itemize}
Furthermore, together with long-baseline neutrino experiments (DUNE~\cite{Adams:2013qkq},
Hyper-K~\cite{Abe:2011ts,Abe:2015zbg}), JUNO will usher in the new era
of precision neutrino oscillation experiments.

It has to be stressed that on top of measuring the oscillation parameters
precision tests of the oscillation pattern in a model-independent way is also very important
to probe new physics beyond the Standard Model.
The precision measurement of neutrino oscillation parameters is a
very powerful tool to test the standard 3-flavor neutrino model
($\nu \rm SM$). In particular, precision measurement of the
fundamental parameter $\theta_{12}$ will:
\begin{itemize}
\item play a crucial role in the future unitarity test of the MNSP matrix $U$.
The combination of short-baseline reactor antineutrino experiments
(e.g., Daya Bay~\cite{An:2012eh}), medium-baseline reactor antineutrino
experiments as JUNO and solar neutrino experiments
(e.g., SNO~\cite{Aharmim:2011vm}) will enable the first direct unitarity test of the
MNSP matrix~\cite{Antusch:2006vwa,Antusch:2014woa,Xing:2012kh,Qian:2013ora}:
%%%%%%%%%%%%%%%%%%%% Eq. 1 %%%%%%%%%%%%%%%%%%%%%%%%%%%%%%
\begin{eqnarray}
|U_{e1}|^2 + |U_{e2}|^2 + |U_{e3}|^2 \stackrel{?}{=} 1\,.
\label{eq:prec:unitarity}
\end{eqnarray}
%%%%%%%%%%%%%%%%%%%% Eq. 1 %%%%%%%%%%%%%%%%%%%%%%%%%%%%%%
With the combination of Daya Bay, JUNO and SNO, the
above unitarity condition will be tested with the precision of 2.5\%
\cite{Qian:2013ora}. The precision is limited by the solar neutrino
measurements and could be improved to better than 1\% with future
precision solar neutrino measurements.
\item narrow down the parameter space of the effective mass (i.e., $|m_{ee}|$) of the
neutrinoless double beta decay allowing a conclusive test for the
 scenario of inverted mass hierarchy (IH). In the case
of $m_3 \leq 0.05\,{\rm eV}$, the minimal value of the effective
neutrino mass for IH can be written as~\cite{Lindner:2005kr}
%%%%%%%%%%%%%%%%%%%% Fig. 1 %%%%%%%%%%%%%%%%%%%%%%%%%%%%%%
\begin{figure}%[H]
\begin{centering}
\includegraphics[width=0.6\textwidth]{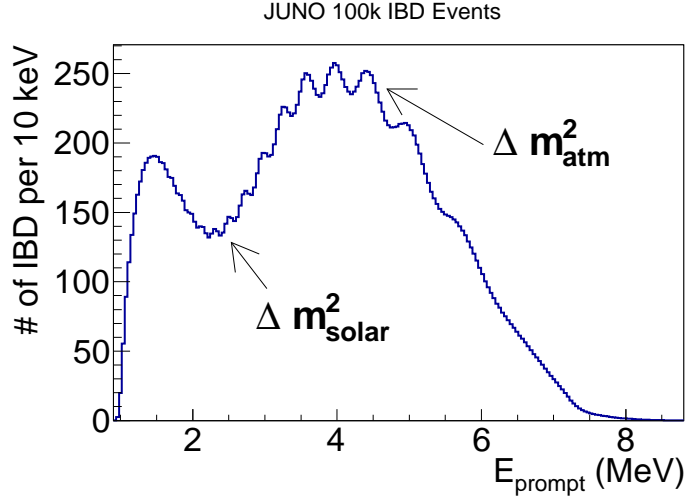}
\par\end{centering}
\caption{\label{fig:prec:juno_spec} The expected prompt energy
spectrum of JUNO with a nominal luminosity for six years of data
taking with a 20 kt detector and 36 GW$_{th}$ reactor power (a total
of 100k IBD events). A $3\%/\sqrt{E}$ energy resolution is assumed.}
\end{figure}
%%%%%%%%%%%%%%%%%%%% Fig. 1 %%%%%%%%%%%%%%%%%%%%%%%%%%%%%
%%%%%%%%%%%%%%%%%%%% Eq. 2 %%%%%%%%%%%%%%%%%%%%%%%%%%%%%%
\begin{eqnarray}
\label{eq:prec:mee_min}
|m_{ee}|_{\rm min}({\rm IH})\simeq
\cos^2\theta_{13}\sqrt{|\Delta m^2_{31}|}\cos2\theta_{12}\,.
\end{eqnarray}
%%%%%%%%%%%%%%%%%%%% Eq. 2 %%%%%%%%%%%%%%%%%%%%%%%%%%%%%%
As the uncertainty of $\sin^2\theta_{12}$ shrinks from the current level
to better than $1\%$, the lower limit of the effective mass for IH
can increase by a factor of two (see Fig.~\ref{fig:prec:dbd}). In a
background-dominated neutrinoless double beta
decay experiment, a factor of two improvement in the effective mass sensitivity corresponds to a
combined factor of 16 improvement for the experimental parameters of
the running time, detector mass, background level and energy
resolution~\cite{Dueck:2011hu,Ge:2015bfa}. Therefore, the precision measurement
of $\sin^2\theta_{12}$ is crucial for the next generation of the
neutrinoless double beta decay experiments, which aim to
cover the whole parameter space corresponding to IH.
%%%%%%%%%%%%%%%%%%%% Fig. 2 %%%%%%%%%%%%%%%%%%%%%%%%%%%%%%
\begin{figure}[tb]
\begin{center}
\begin{picture}(0,0)%
\includegraphics{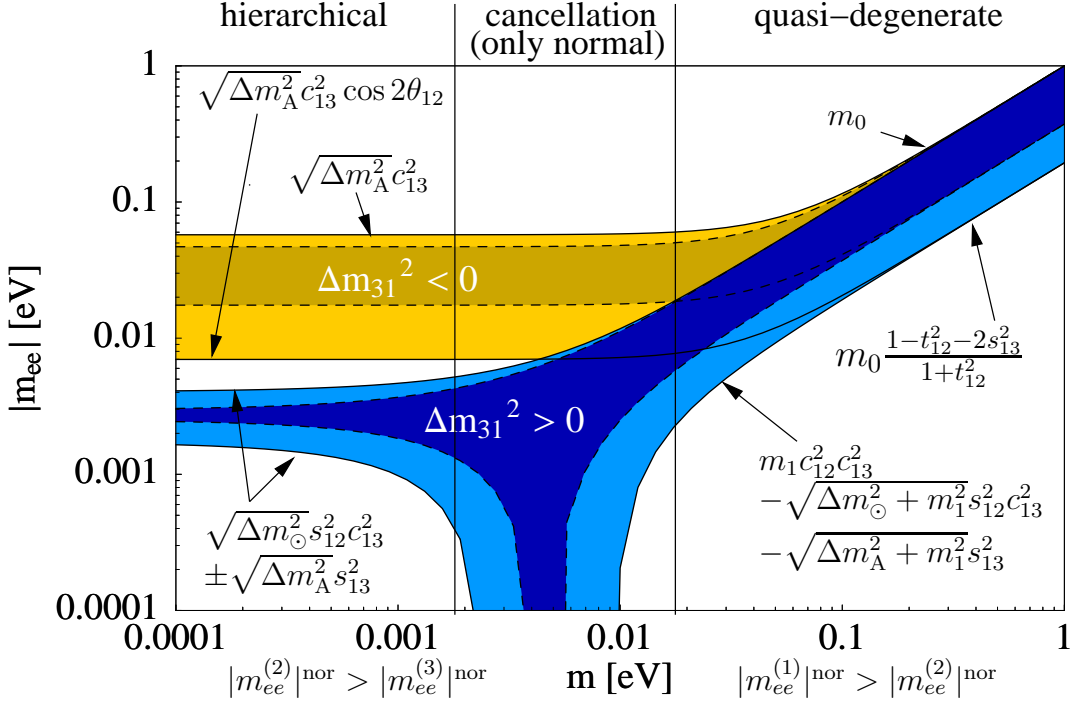}%
\end{picture}%
\setlength{\unitlength}{4144sp}%
\begingroup\makeatletter\ifx\SetFigFont\undefined%
\gdef\SetFigFont#1#2#3#4#5{%
  \reset@font\fontsize{#1}{#2pt}%
  \fontfamily{#3}\fontseries{#4}\fontshape{#5}%
  \selectfont}%
\fi\endgroup%
\begin{picture}(6931,4174)(3646,-4493)
\put(4816,-3526){\makebox(0,0)[lb]{\smash{{\SetFigFont{12}{14.4}{\rmdefault}{\mddefault}{\updefault}{\color[rgb]{0,0,0}$\sqrt{\dms} s_{12}^{2} c_{13}^{2}$}%
}}}}
\put(4771,-906){\makebox(0,0)[lb]{\smash{{\SetFigFont{12}{14.4}{\rmdefault}{\mddefault}{\updefault}{\color[rgb]{0,0,0}$\sqrt{\dma} c_{13}^{2} \cos 2\theta_{12}$}%
}}}}
\put(8101,-3111){\makebox(0,0)[lb]{\smash{{\SetFigFont{12}{14.4}{\rmdefault}{\mddefault}{\updefault}{\color[rgb]{0,0,0}$m_1 c_{12}^{2} c_{13}^{2}$}%
}}}}
\put(5311,-1406){\makebox(0,0)[lb]{\smash{{\SetFigFont{12}{14.4}{\rmdefault}{\mddefault}{\updefault}{\color[rgb]{0,0,0}$\sqrt{\dma} c_{13}^{2}$}%
}}}}
\put(8506,-1051){\makebox(0,0)[lb]{\smash{{\SetFigFont{12}{14.4}{\rmdefault}{\mddefault}{\updefault}{\color[rgb]{0,0,0}$m_0$}%
}}}}
\put(4951,-4426){\makebox(0,0)[lb]{\smash{{\SetFigFont{12}{14.4}{\rmdefault}{\mddefault}{\updefault}{\color[rgb]{0,0,0}$|m_{ee}^{(2)}|^{\rm nor}>|m_{ee}^{(3)}|^{\rm nor}$}%
}}}}
\put(7966,-4426){\makebox(0,0)[lb]{\smash{{\SetFigFont{12}{14.4}{\rmdefault}{\mddefault}{\updefault}{\color[rgb]{0,0,0}$|m_{ee}^{(1)}|^{\rm nor}>|m_{ee}^{(2)}|^{\rm nor}$}%
}}}}
\put(8101,-3346){\makebox(0,0)[lb]{\smash{{\SetFigFont{12}{14.4}{\rmdefault}{\mddefault}{\updefault}{\color[rgb]{0,0,0}$- \sqrt{\dms +m_1^{2} }s_{12}^{2}c_{13}^{2}$}%
}}}}
\put(8551,-2491){\makebox(0,0)[lb]{\smash{{\SetFigFont{14}{16.8}{\rmdefault}{\mddefault}{\updefault}{\color[rgb]{0,0,0}$m_0 \frac{1-t_{12}^{2}-2 s_{13}^2}{1+t_{12}^{2}}$}%
}}}}
\put(8101,-3661){\makebox(0,0)[lb]{\smash{{\SetFigFont{12}{14.4}{\rmdefault}{\mddefault}{\updefault}{\color[rgb]{0,0,0}$-\sqrt{\dma+m_1^{2}}s_{13}^{2}$}%
}}}}
\put(4816,-3796){\makebox(0,0)[lb]{\smash{{\SetFigFont{12}{14.4}{\rmdefault}{\mddefault}{\updefault}{\color[rgb]{0,0,0}$\pm \sqrt{\dma} s_{13}^{2}$}%
}}}}
\end{picture}%
\vspace*{8pt}
\caption{\label{fig:prec:dbd}
The main properties of the effective mass $|m_{ee}|$ as a function of
the smallest neutrino mass~\cite{Rodejohann:2012xd}.
Here $m_{}$ denotes the common mass for the quasi-degenerate region and $t_{ij}=\tan\theta_{ij}$, $s_{ij}
=\sin\theta_{ij}$,  $c_{ij} =\cos\theta_{ij}$. Furthermore, $\Delta
m^2_A$ and $\Delta m^2_{\odot}$ stands for the atmospheric
mass-squared difference and the solar mass-squared difference,
respectively.}
\end{center}
\end{figure}
\item be a powerful discriminator for models of the neutrino masses and mixing.
First, $\theta_{12}$ is more sensitive than other mixing angles to
the quantum corrections since $\Delta
m^2_{21} \ll |\Delta m^2_{31}|$. Therefore,
neutrino-mixing models will be better constrained when the accuracy
of $\theta_{12}$ is improved. Second, taking the prediction of the
tri-bimaximal mixing (TBM)~\cite{Harrison:2002er,Xing:2002sw,He:2003rm} for $\theta_{12}$ as an
example:
%%%%%%%%%%%%%%%%%%%% Eq. 3 %%%%%%%%%%%%%%%%%%%%%%%%%%%%%%
\begin{eqnarray}
\label{eq:prec:t12TBM}
\theta_{12}^{\rm{TBM}}=\arcsin{\frac{1}{\sqrt{3}}} \simeq
35.3^{\circ}\,,
\end{eqnarray}
%%%%%%%%%%%%%%%%%%%% Eq. 3 %%%%%%%%%%%%%%%%%%%%%%%%%%%%%%
the value of non-zero $\theta_{13}$ may induce further
corrections for $\theta_{12}$. Depending on the sign of $\theta_{12}$
corrections, two categories of mixing models, TM1 and TM2, are defined respectively~\cite{Albright:2008rp}.
TM1 and TM2 correspond to the mixing matrix keeping the
first or second column of TBM unchanged. A measurement of
$\sin^2\theta_{12}$ better than $1\%$ is a powerful tool to discriminate
between TM1 and TM2 and may shed light on the mechanism of the
neutrino masses and mixing.
\end{itemize}

The muon (anti)neutrino and electron antineutrino disappearance
effectively measure $\Delta
m^2_{\mu\mu}$ and $\Delta m^2_{ee}$~\cite{Nunokawa:2005nx,deGouvea:2005hk} (two
different combinations of $\Delta m^2_{31}$ and $\Delta m^2_{32}$),
respectively. When combined with the precision $|\Delta
m^2_{\mu\mu}|$ measurements from muon (anti)neutrino disappearance,
the precision measurement of $|\Delta m^2_{ee}|$ will:
\begin{itemize}
\item test the mass sum rule:
%%%%%%%%%%%%%%%%%%%% Eq. 4 %%%%%%%%%%%%%%%%%%%%%%%%%%%%%%
\begin{eqnarray}
\Delta m^2_{13} + \Delta m^2_{21} + \Delta m^2_{32} \stackrel{?}{=}
0\,,\label{eq:prec:masssumrule}
\end{eqnarray}
%%%%%%%%%%%%%%%%%%%% Eq. 4 %%%%%%%%%%%%%%%%%%%%%%%%%%%%%%
which is an important prediction of the $\nu \rm SM$. New physics
like the light sterile neutrinos or non-standard interactions may induce non-trivial corrections
to the effective oscillation frequencies $\Delta m^2_{}$ . Therefore, a
precision test of the sum rule is an important probe of new
physics beyond the Standard Model.

\item reveal additional information regarding the neutrino mass hierarchy.
As discussed in Refs.~\cite{Li:2013zyd,Nunokawa:2005nx}, precision measurements
of both $|\Delta m^2_{ee}|$ and $|\Delta m^2_{\mu\mu}|$ would
provide new information for the neutrino mass hierarchy. A
quantitative calculation is performed in Ref.~\cite{Li:2013zyd},
illustrating a significant improvement for the sensitivity of the
neutrino mass hierarchy. The median sensitivity is increased from
$3\div3.5\sigma$ to $4\div4.5\sigma$~\cite{Li:2013zyd} by considering a
precision of $1\%$ for $|\Delta m^2_{\mu\mu}|$.
An individual $1.5\%$ precision is estimated for both
T2K~\cite{Itow:2001ee} and NO$\nu$A~\cite{Ayres:2004js}. Therefore, a combined
precision level~\cite{Agarwalla:2013qfa} of $1\%$ would be achievable
by the moment when JUNO will start taking data.
\end{itemize}
In the following sections, the sensitivity of JUNO for precision measurements will be illustrated.
Precision measurements of the oscillation
parameters will be presented in Sec.~\ref{subsec:prec:parameters}.
Moreover, the strategy for testing the unitarity of
the lepton mixing matrix and the contribution of JUNO in
this respect will be presented in Sec.~\ref{subsec:prec:unitarity}.
Finally, a summary will be given in Sec.~\ref{subsec:prec:summary}.

\subsection{Precision measurements of oscillation parameters}
\label{subsec:prec:parameters}

In the standard
three-neutrino mixing framework, the relevant mass and mixing
parameters for JUNO are $\theta_{12}$, $\theta_{13}$, $\Delta m^2_{21}$ and
$\Delta m^2_{ee}$ (as the linear combination of $\Delta m^2_{31}$
and $\Delta m^2_{32}$, see the definition in Eq.~(\ref{eq:mh:dmee})).
In the era of precision measurements, matter
effects may not be negligible and deserve careful evaluations. In
the presence of terrestrial matter effects, the survival probability
of reactor antineutrinos is written as
%%%%%%%%%%%%%%%%%%%% Eq. 5 %%%%%%%%%%%%%%%%%%%%%%%%%%%%%%
\begin{eqnarray}
  P(\bar\nu_{\rm e} \to \bar\nu_{\rm e}) =
  1&-&  \sin^22\theta^{\rm m}_{13}\left[\cos^2\theta^{\rm m}_{12} \sin^2{\Delta^{\rm m}_{31}}
  + \sin^2\theta^{\rm m}_{12} \sin^2{\Delta^{\rm m}_{32}}\right] \nonumber \\ %\sin^2 \frac{\Delta m^{2\,\rm M}_{32}}{4E}
  &-&  \cos^4\theta^{\rm m}_{13}\sin^22\theta^{\rm m}_{12} \sin^2{\Delta^{\rm m}_{21}}\,, %\sin^2 \frac{\Delta m^{2\,\rm M}_{21}}{4E}
  \label{eq:prec:peeM}
\end{eqnarray}
%%%%%%%%%%%%%%%%%%%% Eq. 5 %%%%%%%%%%%%%%%%%%%%%%%%%%%%%%
with $\Delta^{\rm m}_{ij}\equiv(\lambda^2_{i}-\lambda^2_{j})L/4E$, and
%%%%%%%%%%%%%%%%%%%% Eq. 6 %%%%%%%%%%%%%%%%%%%%%%%%%%%%%%
\begin{eqnarray}
\sin^22\theta^{\rm m}_{12}&\simeq&
\sin^22\theta_{12}\left(1+2\frac{A_{\rm
CC}}{\Delta m^{2}_{21}}\cos2\theta_{12}\right)\,,\nonumber \\
\lambda^2_{2}-\lambda^2_{1}&\simeq& \Delta
m^{2}_{21}\left(1-\frac{A_{\rm CC}}{\Delta
m^{2}_{21}}\cos2\theta_{12}\right)\,, \label{eq:prec:paraM}
\end{eqnarray}
%%%%%%%%%%%%%%%%%%%% Eq. 6 %%%%%%%%%%%%%%%%%%%%%%%%%%%%%%
and
%%%%%%%%%%%%%%%%%%%% Eq. 7 %%%%%%%%%%%%%%%%%%%%%%%%%%%%%%
\begin{eqnarray}
\sin^22\theta^{\rm m}_{13}&\sim&
\sin^22\theta_{13}\left[1+{\cal O}(\frac{A_{\rm CC}}
{\Delta m^{2}_{31}}\sin^2\theta_{13})\right]\,,\nonumber \\
\lambda^2_{3}-\lambda^2_{1}&\sim& \Delta m^{2}_{31}\left[1+{\cal
O}(\frac{A_{\rm CC}}{\Delta m^{2}_{31}}\sin^2\theta_{13})\right]\,.
\label{eq:prec:parashift}
\end{eqnarray}
%%%%%%%%%%%%%%%%%%%% Eq. 7 %%%%%%%%%%%%%%%%%%%%%%%%%%%%%%
Note that $\lambda^2_{i}$ is the $i$th energy eigenvalue of the Hamiltonian in matter,
and $A_{\rm CC}=2\sqrt{2}E G_{\rm F}N_{\rm e}$ with $N_{\rm e}$ being the electron number density in matter and $G_{\rm F}$ the Fermi constant..
For the typical neutrino energy $E$ and matter density $\rho(x)$, we obtain
%%%%%%%%%%%%%%%%%%%% Eq.8 %%%%%%%%%%%%%%%%%%%%%%%%%%%%%%
\begin{eqnarray}
\frac{A_{\rm CC}}{\Delta m^{2}_{21}}\cos2\theta_{12}&\simeq& 0.5\%\times
\frac{E}{4\,{\rm MeV}}\times\frac{\rho(x)}{3\,{\rm g/cm}}\,,\nonumber \\
\frac{A_{\rm CC}}{\Delta m^{2}_{31}}\sin^2\theta_{13}&\simeq&
1.5\times10^{-5}\times\frac{E}{4\,{\rm
MeV}}\times\frac{\rho(x)}{3\,{\rm g/cm}}\,.
\label{eq:prec:shiftsize}
\end{eqnarray}
%%%%%%%%%%%%%%%%%%%% Eq. 8 %%%%%%%%%%%%%%%%%%%%%%%%%%%%%%
For measurements with precision better than $1\%$,
matter effects are negligible for the atmospheric neutrino
oscillation parameters (i.e., $\theta_{13}$ and $\Delta m^2_{31}$),
but they are sizable (0.5\%-0.1\%) for the solar neutrino oscillation parameters (i.e.,
$\theta_{12}$ and $\Delta m^2_{21}$).
Therefore, it is important to consider matter effects
when analyzing the real data. However, matter effects can be neglected
for sensitivity studies as the corrections only affect
the central values of oscillation parameters and not the sensitivity of
MH and oscillation parameters.

In the history of neutrino oscillation observations, three different
oscillation modes are observed in terms of effective two-flavor oscillations.
The Super-Kamiokande Collaboration~\cite{Ashie:2004mr} observed the oscillation
driven by ($\Delta m^2_{32}$, $\sin^22\theta_{23}$) in
the atmospheric muon neutrino disappearance channel (shown in the
upper panel of Fig.~\ref{fig:prec:loe_2nu}). Later on, two reactor
antineutrino experiments also presented the electron antineutrino survival
probability as a function of $L/E$. The middle panel of
Fig.~\ref{fig:prec:loe_2nu} illustrates the oscillation determined by ($\Delta m^2_{21}$,
$\sin^22\theta_{12}$) at the KamLAND detector~\cite{Gando:2010aa} with
almost two complete cycles. The third case is shown in the lower panel of
Fig.~\ref{fig:prec:loe_2nu}, which is characterized by ($\Delta m^2_{31}$,
$\sin^22\theta_{13}$) and observed in the Daya Bay spectral
observation~\cite{An:2015rpe}.
%%%%%%%%%%%%%%%%%%%% Fig. 3 %%%%%%%%%%%%%%%%%%%%%%%%%%%%%%
\begin{figure}
\begin{center}
\begin{tabular}{c}
\includegraphics[bb=55 225 482 566,width=0.52\textwidth]{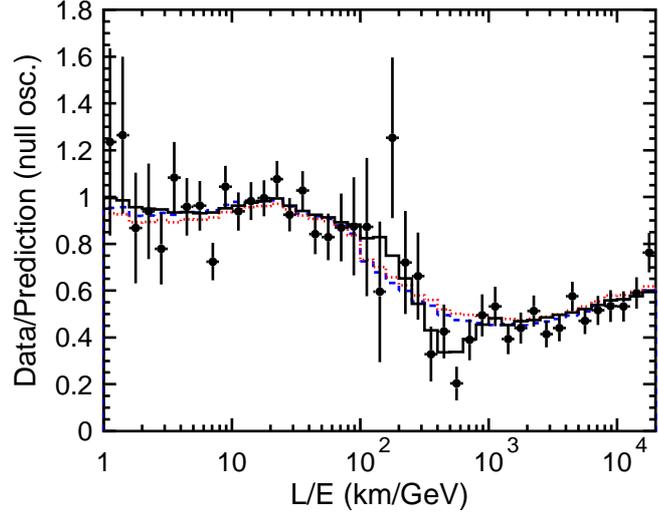}
\\
\includegraphics[angle=270,width=0.52\textwidth]{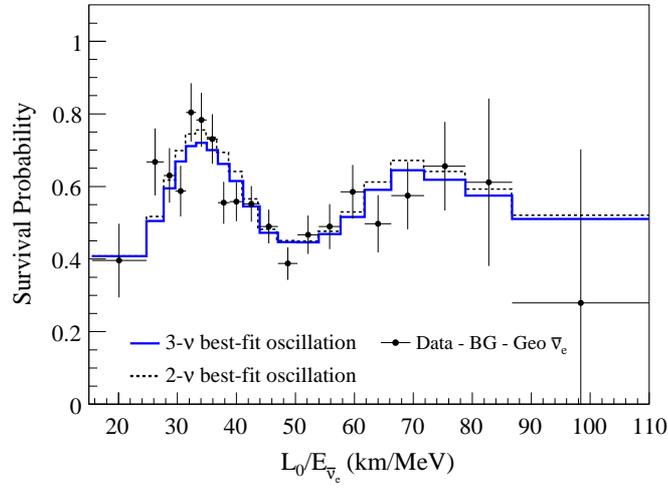}
\\
%\vspace{+1.0cm}
\includegraphics[bb=10 4 550 350,width=0.52\textwidth]{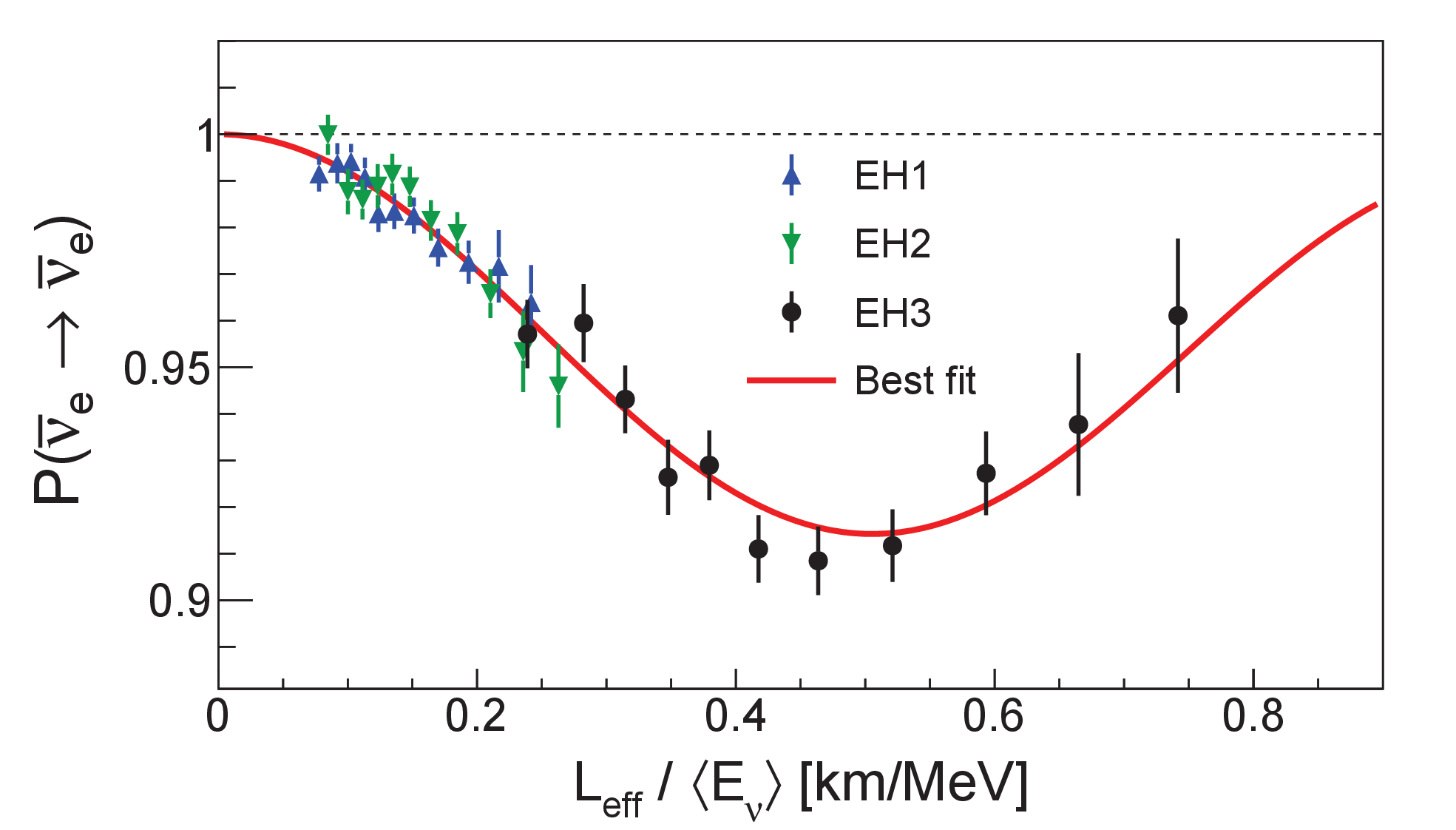}
\end{tabular}
\end{center}
\caption{\label{fig:prec:loe_2nu} The observed neutrino events over
the non-oscillation predictions as functions of $L_{\rm eff}$ (the effective baseline) over $E$
(the neutrino energy) for Super-K~\cite{Ashie:2004mr} atmospheric neutrino
oscillations (top), KamLAND~\cite{Gando:2010aa} reactor antineutrino oscillations (middle)
and Daya Bay~\cite{An:2015rpe} reactor antineutrino oscillations (bottom). They
correspond to three different modes of effective two-flavor oscillations
controlled by ($\Delta m^2_{32}$, $\sin^22\theta_{23}$),
($\Delta m^2_{21}$, $\sin^22\theta_{12}$) and ($\Delta m^2_{31}$,
$\sin^22\theta_{13}$), respectively.}
\end{figure}
%%%%%%%%%%%%%%%%%%%% Fig. 3 %%%%%%%%%%%%%%%%%%%%%%%%%%%%%%
JUNO could be the first to observe an oscillation pattern
containing two independent oscillation frequencies and multiple oscillation cycles.
Because the reactor-detector distances
are almost identical as required by the MH measurement,
antineutrinos from different reactors generate
nearly identical energy spectra without smearing the oscillation
patterns. This represents an important advantage for extracting
the oscillation parameters with high precision. Fig.~\ref{fig:prec:juno_spec} shows the
predicted prompt energy spectrum for the IBD events. Multiple
oscillation patterns corresponding to the solar and atmospheric
$\Delta m^2$ scales are clearly visible.

%%%%%%%%%%%%%%%%%%%%%%%%%%%%%%%%%%%%%%%%%%%%%%%%%%%%%%%%%
\begin{table}%[!htb]
\begin{center}
\begin{tabular}[c]{l|l|l|l|l|l} \hline\hline
  & $\Delta m^2_{21}$ & $|\Delta m^2_{31}|$ & $\sin^2\theta_{12}$ &
  $\sin^2\theta_{13}$ & $\sin^2\theta_{23}$ \\ \hline
 Dominant Exps. & KamLAND  & MINOS & SNO & Daya Bay & SK/T2K  \\  \hline
 Individual 1$\sigma$ & 2.7\%~\cite{Gando:2010aa} & 4.1\%~\cite{Adamson:2013whj} & 6.7\%~\cite{Aharmim:2011vm}
 & 6\%~\cite{An:2015rpe} & 14\%~\cite{Wendell:2014dka,Abe:2014ugx} \\  \hline
 Global 1$\sigma$ & 2.6\% & 2.7\% & 4.1\% & 5.0\% & 11\%  \\
\hline\hline
\end{tabular}
\caption{\label{tab:prec:current} Current precision for the five
known oscillation parameters from the dominant experiments
and the latest global analysis~\cite{Gonzalez-Garcia:2014bfa}.}
\end{center}
\end{table}
%%%%%%%%%%%%%%%%%%%%%%%%%%%%%%%%%%%%%%%%%%%%%%%%%%%%%%%%%

Current precision for five known oscillation parameters are
summarized in Table \ref{tab:prec:current}, where both the results
from individual experiments and from the latest global analysis~\cite{Gonzalez-Garcia:2014bfa} are
presented. Most of the oscillation parameters have
been measured with an accuracy better than $10\%$. The least accurate case is for
$\theta_{23}$, where the octant ambiguity hinders a precision
determination.
Among the four oscillation parameters accessible by JUNO,
$\theta_{13}$ can not be measured with a precision better
than the Daya Bay one, which is expected to reach a 4\%
precision for this smallest mixing angle after 5 years
of running. Therefore, we only discuss the prospect for
precision measurements of $\theta_{12}, \Delta m^2_{21}$,
and $|\Delta m^2_{ee}|$\footnote{There will be two degenerated solutions for
$|\Delta m^2_{ee}|$ in case of undetermined MH.}.

With the nominal setup~\cite{Li:2013zyd} described in the MH
measurement, the expected accuracy for the three relevant
parameters is shown in Fig.~\ref{fig:prec:threepara}, where the solid lines
show the accuracy with all the other oscillation parameters
fixed and the dashed lines show the accuracy with free oscillation
parameters. The precision (dashed lines) of $0.54\%$, $0.24\%$ and
$0.27\%$ can be obtained for $\sin^2\theta_{12}$, $\Delta m^2_{21}$
and $\Delta m^2_{ee}$, respectively, after 6 years of running.

%%%%%%%%%%%%%%%%%%%% Fig. 4 %%%%%%%%%%%%%%%%%%%%%%%%%%%%%%
\begin{figure}
\begin{center}
\begin{tabular}{c}
\includegraphics[width=0.5\textwidth]{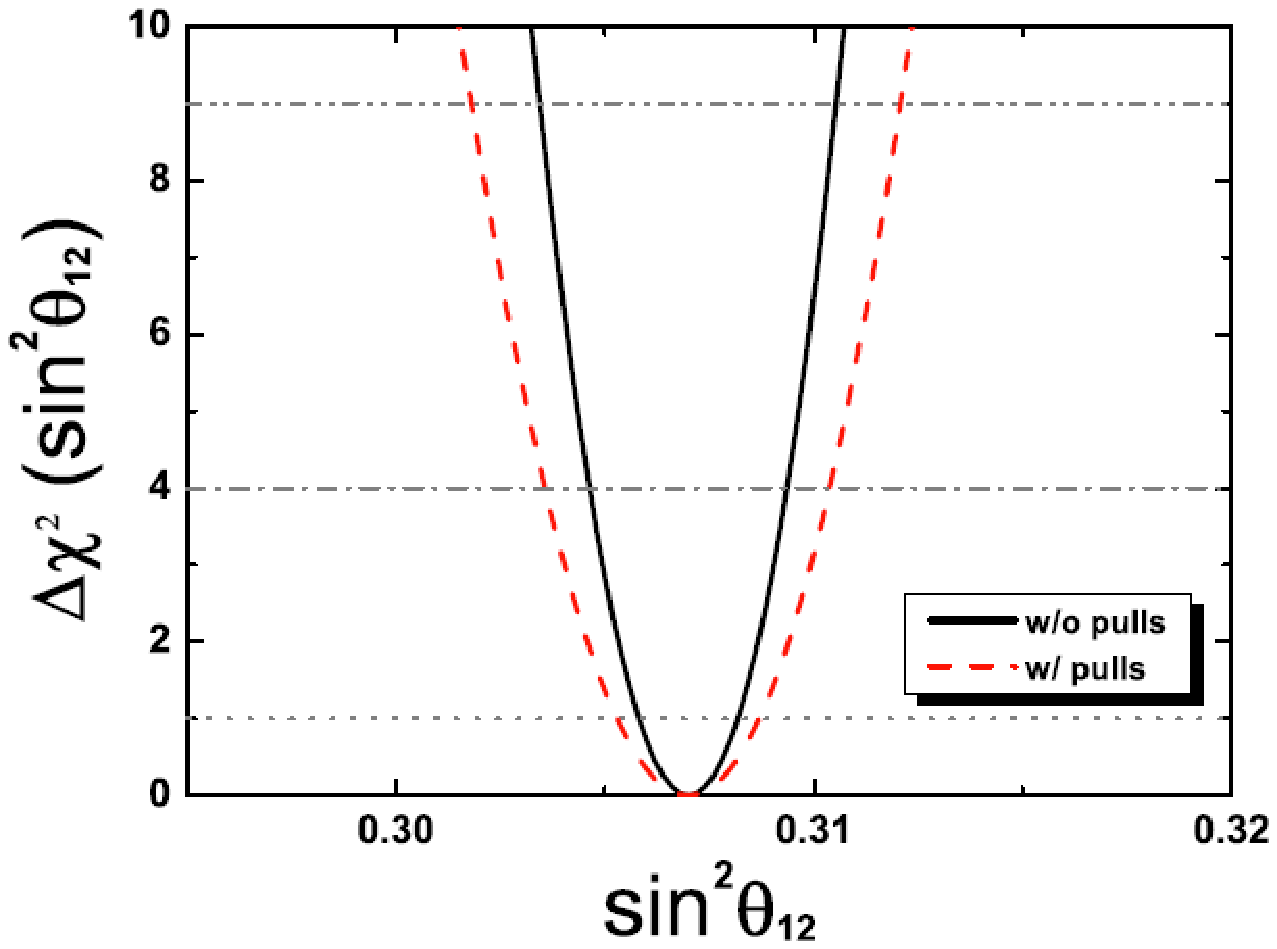}
\\
\includegraphics[width=0.5\textwidth]{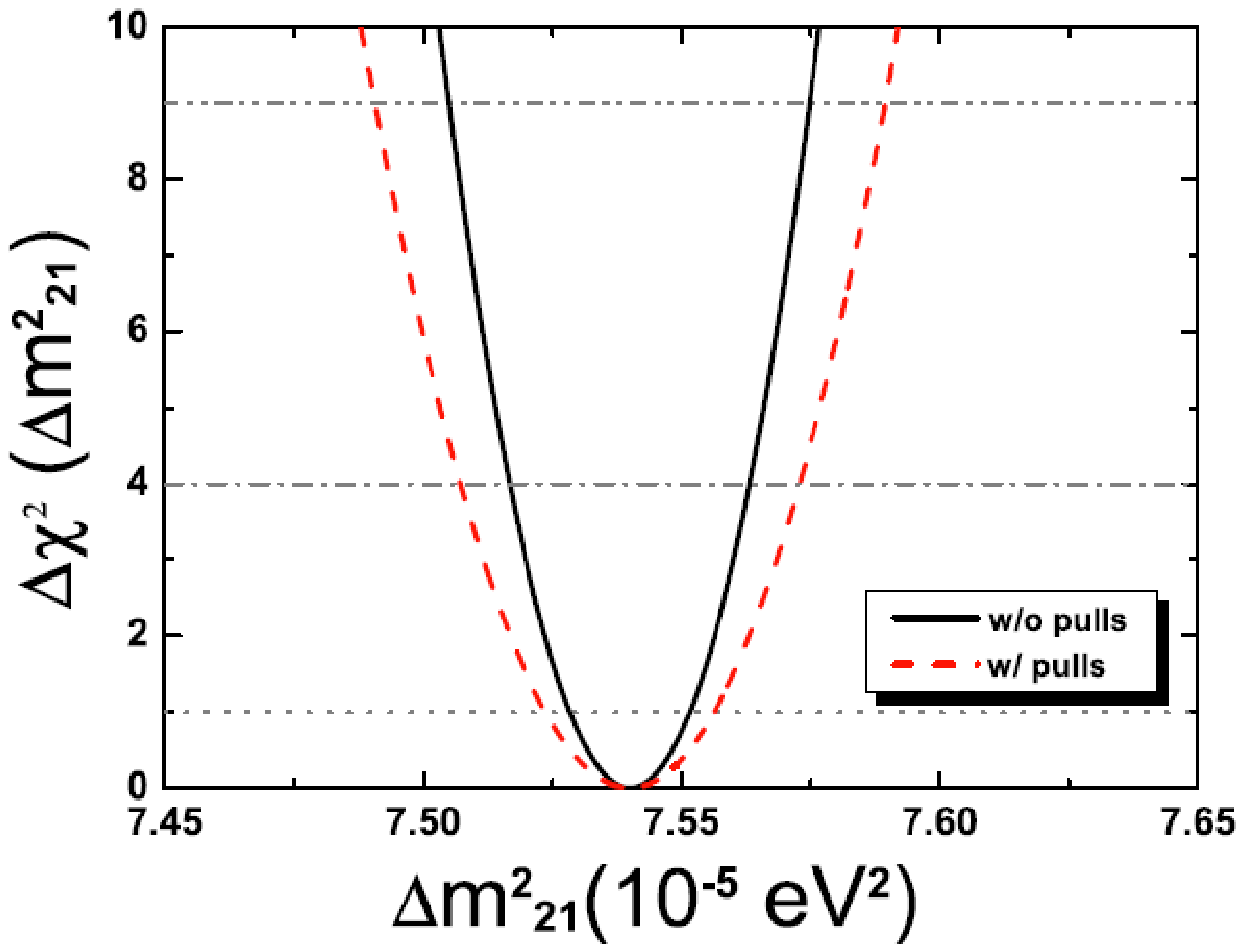}
\\
\includegraphics[width=0.5\textwidth]{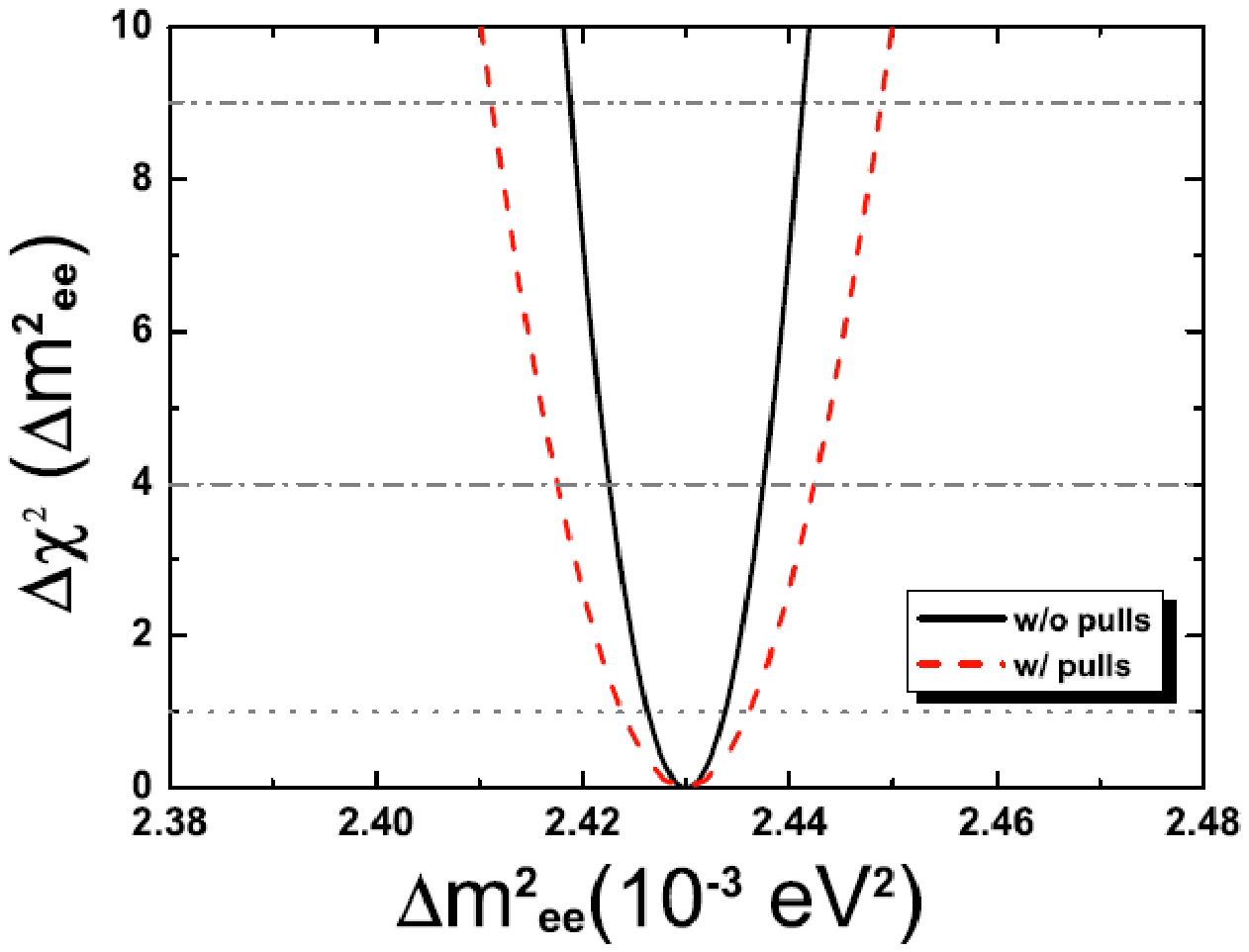}
\end{tabular}
\end{center}
\caption{ \label{fig:prec:threepara} Expected accuracy for
 $\sin^2\theta_{12}$, $\Delta m^2_{21}$ and
$\Delta m^2_{ee}$ after 6 years of running at JUNO
(i.e., ${\cal O}(100\,\rm k$) events).
The solid curves are obtained with all
other oscillation parameters fixed, while the parameters are
set free for the dashed curves.}
\end{figure}
%%%%%%%%%%%%%%%%%%%%%%%%%%%%%%%%%%%%%%%%%%%%%%%%%%%%%%%%%
%%%%%%%%%%%%%%%%%%%% Fig. 5 %%%%%%%%%%%%%%%%%%%%%%%%%%%%%%
\begin{figure}
\begin{center}
\begin{tabular}{c}
\includegraphics[width=0.6\textwidth]{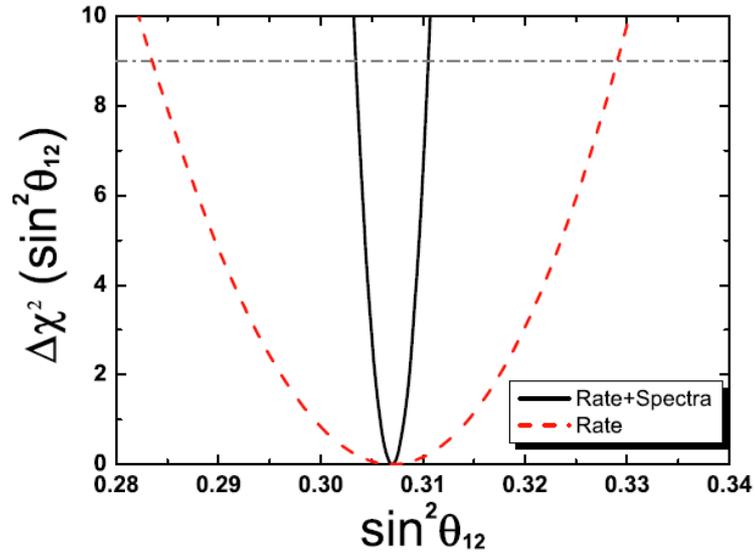}
\end{tabular}
\end{center}
\caption{ \label{fig:prec:th12shape} The precision of
$\sin^2\theta_{12}$ with the rate plus shape information
(solid curve) and rate-only information (dashed curve).}
\end{figure}
%%%%%%%%%%%%%%%%%%%%%%%%%%%%%%%%%%%%%%%%%%%%%%%%%%%%%%%%%
%%%%%%%%%%%%%%%%%%%% Fig. 6 %%%%%%%%%%%%%%%%%%%%%%%%%%%%%%
\begin{figure}
\begin{center}
\begin{tabular}{c}
\includegraphics[width=0.6\textwidth]{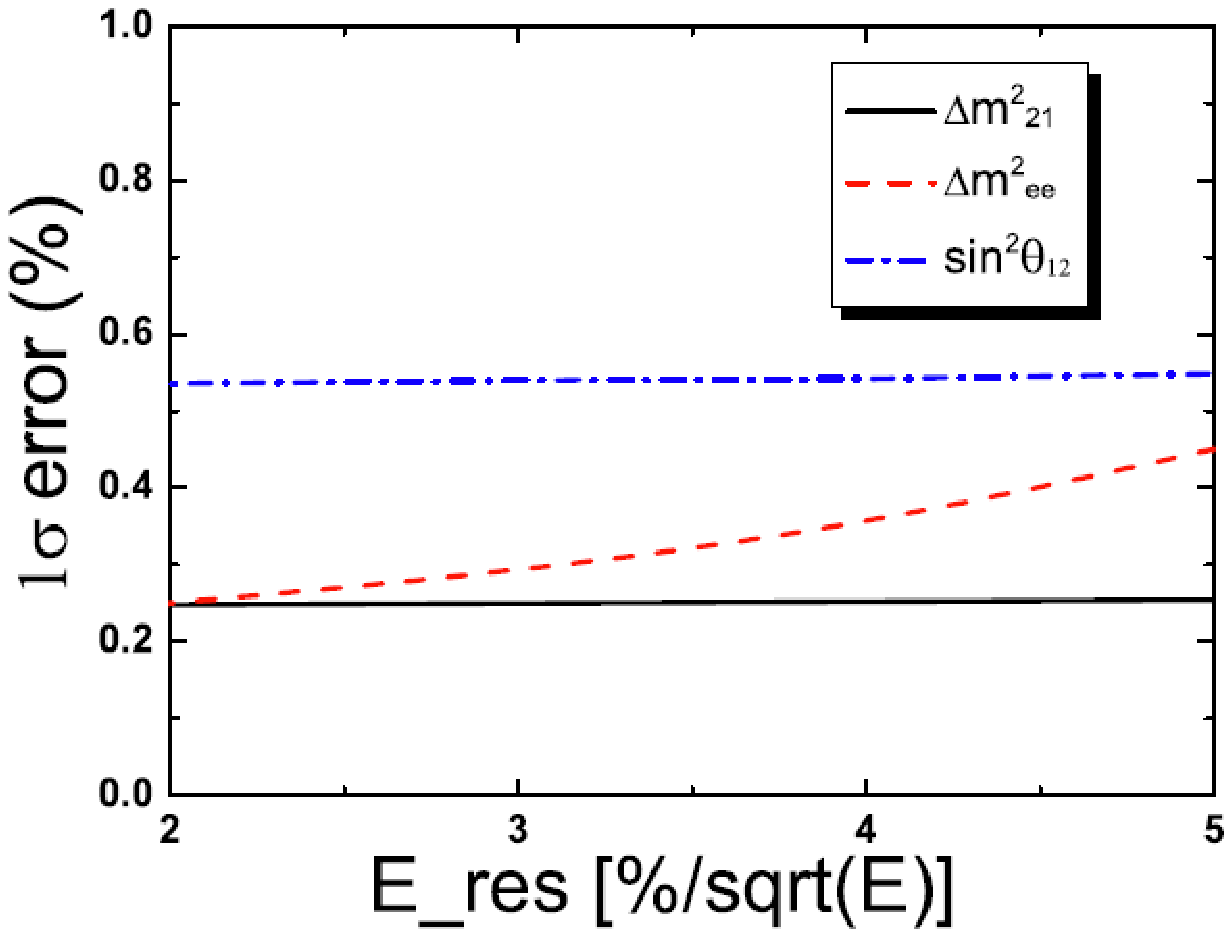}
\end{tabular}
\end{center}
\caption{ \label{fig:prec:erec} Dependence of the precision of $\sin^2\theta_{12}$, $\Delta
m^2_{21}$ and $\Delta m^2_{ee}$ with the neutrino energy resolution. }
\end{figure}
%%%%%%%%%%%%%%%%%%%%%%%%%%%%%%%%%%%%%%%%%%%%%%%%%%%%%%%%

Several comments are listed as follows:
\begin{itemize}
\item Although only one single detector is considered, the precision on $\theta_{12}$
at the sub-percent level is achievable because most of the sensitivity is from the spectral information.
This property is illustrated in Fig.~\ref{fig:prec:th12shape},
showing the $\theta_{12}$ accuracy with both the rate and shape
information and with only the rate information.%, respectively.

\item  A precision of $|\Delta m^2_{ee}|$ similar
to $\Delta m^2_{21}$ is obtained because each fast oscillation cycle
gives a statistically independent measurement of $|\Delta m^2_{ee}|$. The
combined result from the whole spectrum has a high statistical accuracy.

\item The baseline differences may affect significantly the precision of
$\theta_{12}$ because different baselines can smear the oscillation pattern. For
comparison, the precision of $\theta_{12}$ could be improved from $0.54\%$ to $0.35\%$
if the baselines were identical for JUNO.

\item The energy resolution impacts mainly $|\Delta m^2_{ee}|$ because the
relevant information is contained in the fine structure of fast
oscillations. A quantitative dependence on the energy resolution for
all the three oscillation parameters is shown in
Fig.~\ref{fig:prec:erec} with energy resolution ranging from
$2\%$ to $5\%$.
\end{itemize}
\begin{table}%[!htb]
\begin{center}
\begin{tabular}[c]{l|l|l|l|l|l} \hline\hline
  & Nominal &  + B2B (1\%)  & + BG &
  + EL (1\%) &  + NL (1\%) \\ \hline
 $\sin^2\theta_{12}$ & 0.54\%  & 0.60\% & 0.62\% & 0.64\% & 0.67\%  \\  \hline
 $\Delta m^2_{21}$ & 0.24\% & 0.27\%  & 0.29\% & 0.44\% & 0.59\% \\  \hline
 $|\Delta m^2_{ee}|$ & 0.27\% & 0.31\% & 0.31\% & 0.35\% & 0.44\%  \\
\hline\hline
\end{tabular}
\caption{\label{tab:prec:syst} Precision of $\sin^2\theta_{12}$,
$\Delta m^2_{21}$ and $|\Delta m^2_{ee}|$ from the nominal setup to
those including additional systematic uncertainties. The systematics are added
one by one from left to right.}
\end{center}
\end{table}
In the following a study of the effects of important systematic errors,
including the bin-to-bin (B2B) energy uncorrelated uncertainty,
the energy linear scale (EL) uncertainty and the energy non-linear
(NL) uncertainty, will be discussed and the influence of background
(BG) will be presented. As a benchmark, $1\%$ precision for all the
considered systematic errors is assumed. The background level and
uncertainties are the same as in the previous chapter for the MH
determination. In Table~\ref{tab:prec:syst}, we show the precision
of $\sin^2\theta_{12}$, $\Delta m^2_{21}$ and $|\Delta m^2_{ee}|$
from the nominal setup to those including additional systematic uncertainties.
The systematics are added one by one. Note the energy-related uncertainties are more
important because the sensitivity is mostly from the spectrum
distortion due to neutrino oscillations.

In summary, for the precision measurements of oscillation parameters,
we can achieve the precision level of $0.5\%$$-$$0.7\%$ for
the three oscillation parameters $\sin^2\theta_{12}$, $\Delta m^2_{21}$
and $|\Delta m^2_{ee}|$. Therefore, precision tests of the unitarity
of the lepton mixing matrix in Eq.~(\ref{eq:prec:unitarity}), and the
mass sum rule in Eq.~(\ref{eq:prec:masssumrule}) are feasible at
unprecedented precision levels.

\subsection{Tests of the standard three-neutrino paradigm}
\label{subsec:prec:unitarity}

In this section, the strategy for testing the standard three-neutrino paradigm
including the unitarity of the lepton mixing matrix and the sum rule of the mass-squared differences will be discussed.
As only the lepton mixing elements of the electron flavor are accessible in reactor
antineutrino oscillations, we here focus on testing the
normalization condition in the first row of $U$ as shown in
Eq.~(\ref{eq:prec:unitarity}). It should be noted that the $\theta_{12}$ measurement in JUNO is mainly
from the energy spectrum measurement, and $\theta_{13}$ in Daya
Bay is from the relative rate measurement. Therefore, an absolute rate
measurement from either reactor antineutrino experiments or solar
neutrino experiments is required to anchor the total normalization
for the first row of $U$. For the test of the mass sum rule,
an additional independent mass-squared difference is needed, where the most
promising one is that from the long-baseline accelerator
muon-neutrino disappearance channel, i.e., $\Delta m^2_{\mu\mu}$.

To explain non-zero neutrino masses in new physics beyond the Standard Model (SM),
a large class of models introduces additional fermion singlets to mix with the SM neutrinos.
Thus the full neutrino mixing matrix will be enlarged, and an effective $3\times3$ \emph{non-unitary} mixing
matrix emerges when one integrates out all those heavy fermion singlets (i.e., sterile neutrinos).
The distinct effects within this class of SM extensions are well described by an effective field extension
of the SM, called the Minimal Unitarity Violation (MUV) scheme.
The MUV extension of the SM, characterized by two non-renormalizable effective operators, is defined as
\begin{eqnarray}
\label{eq:prec:MUVmodel}
{\mathcal L}_{\rm MUV} &=& {\mathcal L}_{\rm SM} +
\delta{\mathcal L}^{d=5} + \delta{\cal L}^{d=6} \;\nonumber\\
&=&{\mathcal L}_{\rm SM} + \frac{1}{2}\, c_{\alpha \beta}^{d=5} \,\left( \overline{L^c}_{\alpha} \tilde \phi^* \right) \left( \tilde \phi^\dagger \, L_{ \beta} \right) +
c^{d=6}_{\alpha \beta} \, \left( \overline{L}_{\alpha} \tilde \phi \right) i \not{\partial} \left(\tilde \phi^\dagger L_{ \beta} \right) +
{\rm H.c.}\,,
\end{eqnarray}
where $\phi$ denotes the SM Higgs field, which breaks the electroweak (EW) symmetry spontaneously
after acquiring the vacuum expectation value (vev) $v_{\rm EW} \,\simeq\, 246\,{\rm GeV}$, and $L_{\alpha}$ represents the lepton doublets.
In addition, we use the notation $\tilde{\phi} = i \tau_{2} \phi^*$.
The dimension-5 operator in Eq.~(\ref{eq:prec:MUVmodel}) generates nonzero neutrino masses after EW symmetry
breaking. On the other hand, the dimension-6 operator contributes to the kinetic terms of neutrinos, which leads to %in Eq.~(\ref{eq:prec:MUVmodel})
the \emph{non-unitary} neutrino mixing matrix $N$ after we canonically normalize the kinetic terms with a \emph{non-unitary} transformation
of the neutrino fields. Therefore, we can obtain the effective low-energy Lagrangian in the neutrino mass basis as
\begin{eqnarray}
\label{eq:prec:eff-lagr}
{\mathcal L}_{\rm MUV} &=&
\frac{1}{2}\,\left( \bar{\nu}_{i}\,i\,{\partial\hspace{-6pt}\slash}\,\nu_{i}
-\,\overline{{\nu}^{\rm c}}_{i}\,m_i\,\nu_i  + {\rm H.c.}\right)\,
-\,\frac{g}{2\sqrt{2}}\,
 \left( W^+_\mu\,\bar{l}_\alpha\,\gamma_\mu\,(1-\gamma_5)\,N_{\alpha i}\,\nu_i
   + {\rm H.c.}\right)\, \nonumber\\
&&-\,\frac{g}{2 \cos\theta_W}\,
 \left(Z_\mu\,\bar{\nu}_i\,\gamma^\mu\,(1-\gamma_5)\,
     (N^{\dagger}N)_{ij}\,\nu_j\,+ {\rm H.c.}\right)\,
+\,\dots \,
\end{eqnarray}
where ${\nu}_i$ denotes the four-component left-handed field for the $i$th neutrino mass eigenstate, and $g$ is the coupling constant of
the EW interactions. It should be noted that observable consequences of the \emph{non-unitary} neutrino mixing matrix are encoded in
the modifications of charged current (CC) and neutral current (NC) interactions of neutrinos in Eq.~(\ref{eq:prec:eff-lagr}).

To test the unitarity violation, we first need a parametrization of the \emph{non-unitary} neutrino mixing matrix $N$.
Without loss of generality, one can write $N$ as the
product of a Hermitian matrix $H$ and a unitary matrix $U$:
\begin{eqnarray}
\label{eq:prec:paraN}
N=HU\equiv (1+\eta)U\,,
\end{eqnarray}
where the elements of the $\eta$ matrix are assumed to be $\ll 1$ due to the smallness of the unitarity violation.
Alternatively, one can use the parametrization for the Hermitian combination $N N^\dagger$ as
\begin{eqnarray}
\label{eq:prec:paraNNd}
(N N^{\dagger})_{\alpha\beta} = (1_{\alpha\beta} + \varepsilon_{\alpha\beta})\,,
\end{eqnarray}
where elements of the $\varepsilon$ matrix are again assumed to be $\ll 1$, and can be related to the $\eta$ matrix as
\begin{eqnarray}
\label{eq:prec:eps-eta}
\varepsilon_{\alpha\beta} \simeq 2 \,\eta_{\alpha \beta}  \,,
\end{eqnarray}
up to higher orders of $\eta_{\alpha \beta}$ and $\varepsilon_{\alpha\beta}$.

Unitarity violation can be tested in both the neutrino oscillation and EW interaction processes. There exist significant distinctions
for the unitarity tests in the neutrino oscillation and EW interaction processes. For the neutrino oscillations, the neutrino flavors are tagged with the
corresponding charged leptons in the production or detection processes,
and the indices for neutrino mass eigenstates are distinguished using the interference effects. Thus we can determine the individual elements of
the mixing matrix from neutrino oscillations~\cite{Antusch:2006vwa,Xing:2012kh,Qian:2013ora}. On the other hand, in the EW interaction processes, neutrino mass
eighstates in the final states are not detected separately. The experiment rates correspond to sums over all possible mass eigenstates.
Therefore, only the sums of products of the matrix elements (i.e., $N N^{\dagger}$) are measurable~\cite{Antusch:2006vwa,Antusch:2014woa}.

In the following we shall briefly summarize the constraints of non-unitarity from the EW interaction processes~\cite{Antusch:2006vwa,Antusch:2014woa}, which include both high and low
energy observables:
\begin{itemize}
\item electroweak precision observables, including the weak mixing angle $\sin^2\theta_{\rm W}$, $Z$ decay parameters, the $W$ boson mass and decay widths;
\item leptonic universality tests;
\item rare charged lepton decays (e.g., $\ell_\rho \to \ell_\sigma \gamma$);
\item ${\rm CKM}$ unitarity;
\item NuTeV tests of weak interactions;
\item low energy measurements of $\sin^2\theta_{\rm W}$, including parity non-conservation in Cesium, weak charge of the proton, and M\"oller scattering.
\end{itemize}
A global analysis of all above observables is performed to obtain the constraints on the elements of $N N^{\dagger}$. At 90\% C.L., the constraints are~\cite{Antusch:2014woa}:
\begin{eqnarray}
\label{eq:prec:NNdcons}
\left| NN^\dagger \right| = \left( \begin{array}{ccc} 0.9979 - 0.9998 & < 10^{-5} & < 0.0021 \\
< 10^{-5} & 0.9996 - 1.0 & < 0.0008 \\
< 0.0021 & < 0.0008 & 0.9947 - 1.0 \end{array}\right)\,,
\end{eqnarray}
where the best-fit points for the off-diagonal $\varepsilon_{\alpha\beta}$ and for $\varepsilon_{\mu\mu}$ are at zero, and the preference
of flavor-conserving non-unitarity is observed at 90\% C.L. for $\varepsilon_{ee}$ and below 90\% CL for $\varepsilon_{\tau\tau}$. The off-diagonal
non-unitarity parameters are mainly constrained from charged lepton flavor-changing decays. Particularly, the limit for $\varepsilon_{e\mu}$
is dominated by the measurements of $\mu \to e\,\gamma$. The slight departure from zero for $\varepsilon_{ee}$ is stemmed from
the similar discrepancy in the invisible
width of the $Z$ boson, and that for $\varepsilon_{\tau\tau}$ is from the internal correlation among $\varepsilon_{\alpha\beta}$ in the MUV scheme.
The EW interaction processes are only sensitive to the elements of $N N^{\dagger}$. The unitarity test in the form of $N^{\dagger}N$ can only
be inferred indirectly from Eq.~(\ref{eq:prec:NNdcons}) in the MUV scheme, where a precision level of around 3\% can be obtained~\cite{Antusch:2006vwa}. In this respect,
the neutrino oscillations provide us an excellent opportunity to test the unitarity using the direct measurements of mixing matrix elements.% (i.e., $N_{\alpha j}$).

In the neutrino oscillation with a single transition channel, only one combination of the mixing matrix elements can be extracted from the oscillation amplitude.
In addition, an absolute measurement of the zero-distance effect gives a direct test of unitarity conditions in $N N^{\dagger}$ or $N^{\dagger}N$.
For the former case, the global analysis of different oscillation channels is needed to test unitary conditions. As for the latter case,
the level of unitarity tests from absolute rate measurements requires better understanding of the uncertainties in the neutrino flux normalization and detector
efficiency. In the following part, we shall discuss the unique role of JUNO in the global picture of unitarity tests in neutrino oscillations.

For reactor antineutrino oscillations, only the $\bar{\nu}_{e}$ survival probability is detectable, which can be expressed in the MUV scheme as
\begin{eqnarray}
P_{\bar{\nu}_{e}\rightarrow\bar{\nu}_{e}}
&=&\left(|N_{e 1}|^2 + |N_{e 2}|^2 + |N_{e 3}|^2\right)^2\nonumber\\
&-&4|N_{e 1}|^2|N_{e 2}|^2 \sin^2 \Delta_{21}
-4|N_{e3}|^2(|N_{e1}|^2 \sin^2 \Delta_{31}+|N_{e 2}|^2 \sin^2 \Delta_{32})\,,
\label{eq:prec:peenu}
\end{eqnarray}
where $\Delta_{ij}=(m_{i}^{2}-m_{j}^{2})L/4E$. It has to be noted that the mixing element $N_{\alpha i}$ is
only sensitive to the absolute measurements, which is directly compared to model predictions of the antineutrino production.
To illustrate this effect, we define the effective mixing angles as
\begin{eqnarray}
\sin^2{\tilde{\theta}_{13}}\equiv\frac{|N_{e 3}|^2}{|N_{e 1}|^2 + |N_{e 2}|^2 + |N_{e 3}|^2}\,,\quad\,\quad
\tan^2{\tilde{\theta}_{12}}\equiv\frac{|N_{e 2}|^2}{|N_{e 1}|^2}\,.
\label{eq:prec:effmix}
\end{eqnarray}
Therefore, we can rewrite Eq.~(\ref{eq:prec:peenu}) as
\begin{eqnarray}
P_{\bar{\nu}_{e}\rightarrow\bar{\nu}_{e}}
&=&\left(|N_{e 1}|^2 + |N_{e 2}|^2 + |N_{e 3}|^2\right)\times\left[1-
\cos^{4}\tilde{\theta}_{13}\sin^{2}2\tilde{\theta}_{12}\sin^{2}\Delta_{21}\right.\nonumber\\
&-&\left.\sin^{2}2\tilde{\theta}_{13}(\cos^{2}\tilde{\theta}_{12}\sin^{2}\Delta_{31}+\sin^{2}\tilde{\theta}_{12}\sin^{2}{\Delta_{32}})\right]\,.
\label{eq:prec:peenu2}
\end{eqnarray}
Comparing to the probability in Eq.~(\ref{eq:mh:osc}), it can be shown that only the rate normalization
factor contributes to the unitarity test of Eq.~(\ref{eq:prec:unitarity}),
which mainly includes the uncertainties from the reactor flux normalization and detector efficiency.
We refer this scenario as the absolute (Abs) measurement.
On the other hand, we can define a relative (Rel) measurement, which combines the reactor spectral measurements of effective mixing parameters in Eq.~(\ref{eq:prec:effmix})
and the solar neutrino measurements to test the unitarity violation in an indirect way~\cite{Qian:2013ora}.
In Fig.~\ref{fig:prec:uv}, we present the two different scenarios of the unitarity test, which show obvious advantages for the combined analysis,
for several considered uncertainties of reactor flux normalization. An additional 1\% detector efficiency uncertainty is assumed.
In the combined analysis, we assume a 4\% final projected sensitivity for $\sin^{2}2\tilde{\theta}_{13}$ at Daya Bay, and a direct measurement
of $|N_{e 2}|^2$ of the 4\% level at a future SNO-like solar neutrino experiment. Finally, we may achieve the precision level of 1.2\% for
unitarity tests in the assumption of a 10\% reactor normalization uncertainty, which could present a independent test of the reactor antineutrino anomaly.

Regarding the test of the mass sum rule in Eq.~(\ref{eq:prec:masssumrule}), JUNO is a powerful precision tool in measuring the
two independent mass-squared differences $\Delta m^2_{21}$ and $|\Delta m^2_{ee}|$ as seen in Table~\ref{tab:prec:syst},
and can test the mass sum rule by inclusion of a third independent mass-squared difference in Eq.~(\ref{eq:mh:dmemu}).
Assuming a 1\% precision level of $|\Delta m^2_{\mu\mu}|$, we can obtain the test of the mass sum rule in Eq.~(\ref{eq:prec:masssumrule}) at better
than 1.8\%, which represents an important test for the standard three-neutrino paradigm.

%%%%%%%%%%%%%%%%%%%% Fig. 7 %%%%%%%%%%%%%%%%%%%%%%%%%%%%%%
\begin{figure}
\begin{center}
\begin{tabular}{c}
\includegraphics[width=0.6\textwidth]{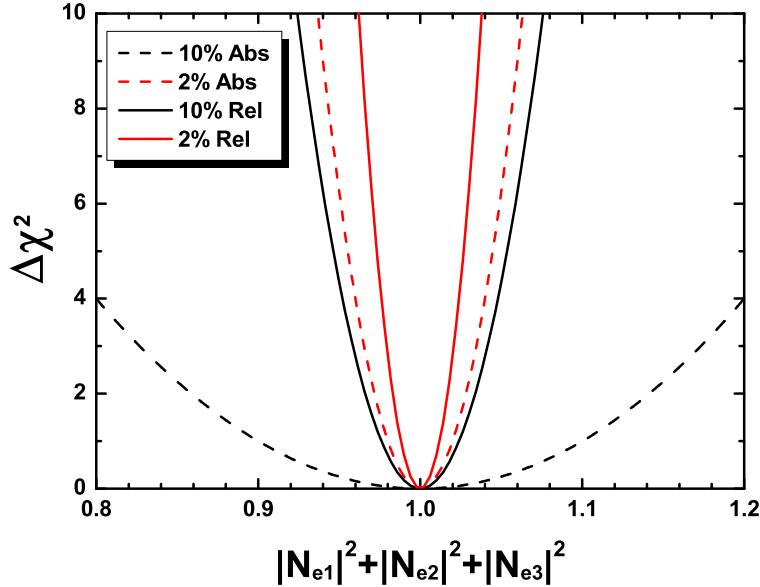}
\end{tabular}
\end{center}
\caption{\label{fig:prec:uv} Different realizations of the unitarity
violation test of the equation~\ref{eq:prec:unitarity} in reactor
antineutrino oscillations. See the text for details.}
\end{figure}
%%%%%%%%%%%%%%%%%%%%%%%%%%%%%%%%%%%%%%%%%%%%%%%%%%%%%%%%

\subsection{Conclusions}
\label{subsec:prec:summary}
Precision measurement of the oscillation parameters and test of the standard three-neutrino
framework constitute another important goal of the JUNO experiment. Among all the six oscillation parameters of neutrino oscillations, JUNO can measure
$\sin^2\theta_{12}$, $\Delta m^2_{21}$ and $|\Delta m^2_{ee}|$ to the world-leading levels of 0.7\%, 0.6\%, and 0.5\%, respectively.
As a powerful detector with huge statistics, unprecedent energy resolution,
the precision measurements of oscillation parameters are important for discrimination of the neutrino mixing patterns, the search of neutrinoless double beta decay,
and test of the standard three-neutrino paradigm.
Utilizing these above measurements, JUNO can help in testing the unitarity relation of Eq.~(\ref{eq:prec:unitarity}) and the mass sum rule of Eq.~(\ref{eq:prec:masssumrule})
to the levels of around 1.2\% and 1.8\%, respectively. Moreover,
JUNO would be the first experiment to simultaneously observe neutrino oscillations from two different frequencies,
and be the first experiment to observe more than two oscillation cycles.

\clearpage

\section{Supernova Burst Neutrinos}
\label{sec:sn}

\blfootnote{Editors: Georg Raffelt (raffelt@mpp.mpg.de) and Shun Zhou (zhoush@ihep.ac.cn)}
\blfootnote{Major contributors: Gang Guo, Yufeng Li, Jiashu Lu, and Hao Wang}

Measuring the neutrino burst from the next nearby supernova (SN) is a premier target of low-energy neutrino physics and astrophysics. For a typical galactic distance of 10~kpc and typical SN parameters, JUNO will register about 5000 events from inverse beta decay (IBD), \hbox{$\bar\nu_e+p\to n+e^+$}, comparable to Super-Kamiokande, and many events from complementary channels, notably 2000 events from all-flavor elastic neutrino-proton scattering. With more than 300 events from neutrino-electron scattering, JUNO will also be the best detector for SN $\nu_e$. Such a high-statistics signal can determine a detailed neutrino ``light curve'' spectrum, and complete flavor information. In combination with other neutrino detectors, gravitational-wave detectors, and observations in various electromagnetic channels, a detailed astrophysical multi-messenger picture will emerge. In this way the standard paradigm of stellar core collapse will be confirmed, refuted or extended. The unique particle-physics lessons pioneered by the sparse SN~1987A data for the first time will be made precise and will extend to areas that depend on high statistics, good energy resolution, or flavor information, notably in the area of neutrino
oscillations.

\subsection{Core-collapse supernovae: What, where and when?}
\label{subsec:sn:core-collapse}

The baryonic matter in a spiral galaxy like our own Milky Way participates in an on-going cycle of star formation, nuclear processing, and ejection into interstellar space. Only low-mass stars with
$M\lesssim0.85\,M_\odot$\footnote{In stellar astrophysics, masses are usually measured in units of the solar mass, 1 ${\it M}_\odot$ = 1.989 $\times 10^{33}~{\rm g}$,
corresponding to 1.20 $\times 10^{57}$ nucleons.} live longer than the age of the universe of 14 billion years, whereas the progenitors of core-collapse SNe ($M\gtrsim 6\sim 8\,M_\odot$, depending on other parameters such as metallicity~\cite{Langer:2012jz}) finish after less than a hundred million years. Intermediate-mass stars lose most of their mass during the red-giant phase in the form of a ``stellar wind'', i.e., they mostly ``evaporate'', leaving behind a white dwarf, a sub-$M_\odot$ compact star supported by electron degeneracy pressure. A more massive star, on the other hand, completes the nuclear reaction chains all the way to iron, forming a degenerate core. It ultimately collapses to nuclear density, ejects the remaining mass in a spectacular SN explosion, and emits 99\% of
the gravitational binding energy of the newly formed neutron star in the form of neutrinos and antineutrinos\footnote{In the following we will generically use the term ``neutrinos'' to include both neutrinos and antineutrinos unless we explicitly distinguish between them.} of all flavors. The standard paradigm of how a core-collapse implosion reverses to a SN explosion is that a shock wave forms at core bounce, propagates outward, stalls at a radius of 100--200~km, and is finally revived by neutrino energy deposition~\cite{Burrows:2000mk, Janka:2012wk}. If this neutrino-driven explosion mechanism indeed captures the crucial physics remains open and could be tested with a high-statistics SN neutrino observation.

Core-collapse SNe encompass the spectral types Ib, Ic, and II, whereas SNe of Type~Ia represent a different physical phenomenon~\cite{Burrows:2000mk,
Cappellaro:2000ez, Mazzali:2007et}. The progenitor consists of a
carbon-oxygen white dwarf, accreting matter from a binary companion until
collapse (single degenerate scenario) or of two merging white dwarfs (double
degenerate scenario), either way igniting explosive nuclear burning that
disrupts the entire star---no stellar remnant survives. This thermonuclear
explosion mechanism liberates a comparable amount of visible energy as a
core-collapse SN which is driven by gravitational binding energy, liberates
about 100 times more energy, but emits 99\% of it in the form of neutrinos.
The light curves of SNe~Ia are very reproducible and they have been
extensively used as cosmological standard candles, whereas core-collapse SNe
are dimmer and show more diverse light curves. Core-collapse SNe occur in
regions of active star formation, i.e., primarily the gaseous and dust-filled disks of spiral galaxies, where around 2/3 of all SNe are of core-collapse type. Little star formation takes place in elliptical galaxies and they host only SNe~Ia~\cite{Cappellaro:2000ez}.

Core collapse may sometimes produce a black hole, but usually the compact
remnant is a neutron star with a gravitational mass of up to $2\,M_\odot$,
often showing up as a pulsar. The collapse of this amount of matter to
nuclear density, i.e., to a radius of 12--15~km, liberates 10--20\% of its
rest mass as gravitational binding energy, corresponding to some
$3\times10^{53}~{\rm erg}$ or $2\times10^{59}~{\rm MeV}$, exact numbers
depending on the final neutron-star mass and the nuclear equation of state.
Around 99\% of this enormous amount of energy emerges as neutrinos, about 1\%
as kinetic energy of the explosion, and about 0.01\% as light, still
outshining the host galaxy. Star formation strongly favors low-mass stars, so
core-collapse SNe are rare, yet the integrated cosmic energy density of SN
neutrinos (the diffuse SN neutrino background or DSNB, see Sec.~5) is
comparable to the integrated photon emission from all stars~\cite{Beacom:2010kk}.

While every second a few core-collapse events happen in the visible universe, a JUNO-class detector covers only our own galaxy and its satellites, such as the Large Magellanic Cloud (LMC) at a distance of 50~kpc, the site of SN~1987A which provided the first and only observed SN neutrino signal~\cite{Koshiba:1992yb}. A SN in Andromeda, at 750~kpc our
nearest-neighbor big galaxy, would produce about one neutrino event in JUNO.
Another SN in the LMC would produce around 200~events, a significant
improvement on the SN~1987A statistics of a total of about two dozen events
in three detectors. The most likely place of the next nearby SN, however, is
the spiral disk of our own galaxy. The solar system is located close to the
mid-plane of the disk at a distance of 8.7~kpc from the galactic center. The
expected SN distance distribution has an average of around 10~kpc, usually
taken as the fiducial distance, where JUNO would register around 5000 events.

However, the distance distribution is broad~\cite{Mirizzi:2006xx,
Ahlers:2009ae, Adams:2013ana} and one cannot expect the next galactic SN to
be ``average'' or ``typical''. The distribution drops very quickly at or
around 20~kpc, a distance that would still provide a whopping $10^3$ events.
On the other extreme, the distribution becomes very small at around 2~kpc,
where the number of events would be around $10^5$. The nearest possible SN
progenitor is the red supergiant Betelgeuse (Alpha Orionis) at a distance of
around 0.2~kpc \cite{Harper:2008} producing around $10^7$ events. Such a high event rate ($\sim\,$20 MHz at peak) is a big challenge for a huge liquid scintillator detector like JUNO. The data acquisition system of JUNO will be designed to accommodate as high event rates as possible to maintain the event by event detection.

Observing the next nearby SN in neutrinos is a once-in-a-lifetime opportunity
that must not be missed. The galactic core-collapse SN rate is only one every
few decades. It can be estimated from SN statistics in external
galaxies~\cite{Cappellaro:2000ez, vandenBergh:1994, Li:2010kd}, the galactic
birth rate of massive stars~\cite{Reed:2005en}, the pulsar birth
rate~\cite{FaucherGiguere:2005ny, Keane:2008jj}, the measured galactic
abundance of the radioactive isotope $^{26}$Al which is produced in
core-collapse SNe~\cite{Diehl:2006cf}, and the historical SN rate over the
past millennium~\cite{Adams:2013ana, Strom:1994, Tammann:1994ev}. A weak
upper limit is obtained from the non-observation of a neutrino burst other
than SN~1987A since 30~June 1980 when the Baksan Scintillator Telescope (BST)
took up operation~\cite{Alekseev:1993dy}. Together with subsequent neutrino
detectors, no galactic SN neutrino burst would have been missed since that
time. Assuming a galactic core-collapse rate of one every 30--40 years, the
chances of observing a galactic SN burst over ten years of operation is
around 30\%, a great opportunity for a world-class fundamental observation. Coordination on detector maintenance with other big neutrino observatories, such as Super-Kamiokande and IceCube, will help to avoid missing a SN in all neutrino observatories simultaneously.

Our location in the mid-plane of the dust-filled galactic disk implies that
visual SN observations are strongly impeded by obscuration. Therefore, only
five historical SNe have been reported in the second millennium where the
record may be reasonably complete~\cite{Green:2003ir}. Moreover, three out of
these events were of the brighter type~Ia (SNe~1006, 1572 and 1604). The two
historical core-collapse SNe were the ``Chinese SN'' of 1054 that has
produced the Crab Nebula and Crab Pulsar and SN~1181. The Cas~A remnant and
non-pulsar hot neutron star may correspond to the uncertain SN observation of
1680 that would bring the historical second-millennium SNe up to six. These
numbers are consistent with 2/3 or more of all galactic SNe being of
core-collapse type and with a few galactic core-collapse events per century.
Today, a galactic SN almost certainly would be observed in some
electromagnetic band, notably in the near infra-red (IR)
\cite{Adams:2013ana}. Actually, an IR record of the progenitor star almost
certainly exists in the Two Micron All Sky Survey. The neutrino burst occurs
several hours before the explosion and optical outburst, leaving time to
issue a neutrino alert to the astronomical community. JUNO will join the
Supernova Early Warning System (SNEWS)~\cite{Antonioli:2004zb}, a network of
neutrino detectors, to participate in this important task.

\subsection{Neutrino signature of core collapse}
\label{subsec:sn:neutrino-signature}

\subsubsection{Neutrino-driven explosion}
\label{subsubsec:sn:neutrino-driven-explosion}

The neutrino signal from the next nearby, probably galactic, core-collapse SN
will be picked up by many detectors that will measure tens to hundreds of
events if the SN is at the fiducial distance of 10~kpc
\cite{Scholberg:2012id}. A completely different level of statistics will be
provided by the large detectors Super-Kamiokande~\cite{Abe:2013gga,
Mori:2013wua} with around $10^4$ reconstructed events and IceCube \cite{Abbasi:2011ss,Demiroers:2011am, Aartsen:2013nla} with around $10^6$ events of correlated noise increase. JUNO will play in this forefront league of large detectors that will provide detailed information about the neutrino signal and will have a number of unique capabilities, similar to the earlier LENA (Low-Energy Neutrino Astronomy) concept \cite{Wurm:2011zn}. However, what is to be expected and what will it tell us?

The standard scenario of core-collapse SN explosion is the delayed,
neutrino-driven explosion paradigm of Bethe and Wilson~\cite{Bethe:1984ux,
Bethe:1990mw}. The degenerate core of an evolved massive star becomes
unstable by pressure loss due to electron absorption and photon dissociation
on heavy nuclei. Usually this happens after completing all nuclear burning
stages when the core consists of iron, the most tightly bound nucleus---the
class of iron-core SNe. For the smallest progenitor masses of perhaps
6--$8\,M_\odot$, the dissociation begins before igniting the final burning
stage when the oxygen-neon-magnesium core has become very degenerate---the
class of electron-capture or O-Ne-Mg-core SNe. In both cases, the subsequent
implosion on a near free-fall time is halted when nuclear density of around
$3\times10^{14}~{\rm g}~{\rm cm}^{-3}$ is reached and the equation of state
stiffens. This sudden ``core bounce'' forms a shock wave that propagates
outward, ramming into the high-$Z$ material that keeps falling in at
supersonic speed. Its dissociation absorbs energy and weakens the shock wave
until it stagnates after reaching a radius of 100--200~km.

The absorption of electrons during infall produces a $\nu_e$ flux until the
core reaches densities of around $10^{12}~{\rm g}~{\rm cm}^{-3}$. Afterwards
neutrinos are trapped and the lepton number stored in the electron gas can no longer escape. This trapping at relatively low density is caused by the
coherence of neutrino scattering on large nuclei. This effect means that the
electron fraction per baryon, $Y_e$, after collapse is around 0.32, not much
smaller than that of the pre-collapse core. Most of this trapped
electron-lepton number will eventually escape by diffusion in the form of
$\nu_e$. However,  a ``prompt $\nu_e$ burst'' or ``deleptonization burst''
lasting around 10~ms (Fig.~\ref{fig:sn:SNburst}), is released when the shock
wave passes through the edge of the iron core, dissociates iron, and allows
the outer layers to deleptonize by $e^-+p\to n+\nu_e$.

During the subsequent stalled-shock phase, or standing accretion shock phase, matter keeps falling in, heating the outer layers of the forming neutron star, and powering strong neutrino emission. Neutrinos and antineutrinos of all flavors are thermally produced and in addition a $\nu_e$ excess flux emerges as a result of charged-current electron conversion fed by the electrons of the infalling material. The neutrino emission region with a thickness of a few tens of km above the proto-neutron star preferentially emits $\nu_e$ and $\bar\nu_e$. Heavy-flavor neutrino emission emerges from a somewhat deeper and hotter region. Typical neutrino energies of tens of MeV are too low to produce muon or tau leptons. Therefore, $\nu_\mu$, $\bar\nu_\mu$, $\nu_\tau$ and $\bar\nu_\tau$, often collectively denoted as $\nu_x$, are not subject to charged-current reactions but rather are produced or absorbed in pairs. During the accretion phase $L_{\nu_e}$ and $L_{\bar\nu_e}$ are similar, but perhaps twice as large as each $L_{\nu_x}$ (Fig.~\ref{fig:sn:SNburst}).

Neutrinos stream almost freely from the decoupling region or ``neutrino
sphere'', which however should be pictured as a broad region and depends on
energy. Still, $\nu_e$ and $\bar\nu_e$ occasionally interact by
charged-current reactions on nucleons on their way out, producing electrons
and positrons. The balance of neutrino energy gain and loss is such that,
midway between the neutrino sphere and the stalling shock wave, a ``gain
radius'' develops such that there is a net gain of energy behind the shock
wave. It is this effect that pumps energy back into this region, builds up
pressure, and eventually revives the shock wave: The explosion takes off and
the accretion flow is terminated. The proto-neutron star settles and the
subsequent evolution is cooling and deleptonization, a detectable signal in
JUNO lasting of order 10~s.

Numerical simulations, however, do not reliably produce explosions or
sufficiently energetic explosions: neutrinos do not deposit quite enough
energy. Spherically symmetric (one-dimensional, 1D) simulations produce
explosions for the low-mass class of electron-capture SNe, but not for the
higher-mass iron-core SNe. Early exploding models by the Livermore group had
enhanced neutrino fluxes by the assumed effect of neutron-finger convection
which today is no longer deemed realistic.

However, core-collapse SNe are generically 3D phenomena. The explosion of
SN~1987A was strongly asymmetric, as seen in electromagnetic observations. Neutron stars often have large ``kick velocities'' of several hundreds to \mbox{thousands} of km/s, pointing to strong asymmetries in their birth process. 2D and 3D \mbox{numerical} simulations show the evolution of large-scale convective overturn between neutron star and standing shock wave after some 100~ms post bounce. Other dynamical
instabilities have been discovered over the years, notably the standing
accretion shock instability (SASI)~\cite{Blondin:2002sm}, producing a large-scale sloshing or
spiral motion and strong deformations of the shock front, and very recently
the LESA phenomenon (lepton-emission self-sustained asymmetry)~\cite{Tamborra:2014aua}. Multi-D
effects may help with the explosion, for example by allowing the material to
absorb more energy from the neutrino flux, yet explosions have not been
systematically successful. A suite of axisymmetric (2D) models shows
explosions, but in 3D the same models do not explode in current simulations~\cite{Janka:2012sb}.

Systematic 3D simulations with sophisticated multi-flavor Boltzmann neutrino
transport are only beginning and still depend on significant approximations.
Therefore, the question remains if numerical improvements, notably
increased spatial resolution, will actually produce systematic explosions or
if entirely new physical ingredients are required as some authors have
argued. In principle, these could be of the astrophysical type (rotation,
magnetic fields, new hydrodynamical effects) or of the particle physics type
(e.g., energy transfer to the shock wave by new particles). It is also not
known how strongly the explosion physics depends on details of the progenitor
model and if some or many cases also fail in nature, leading to black-hole
formation. These failed SNe would produce a strong neutrino signal because
the black-hole collapse would occur only after a period of shock-wave
stagnation. The non-observation of a galactic neutrino burst since 1980
provides only a weak upper limit on the galactic rate of failed SNe.

Core-collapse SNe probably create half of the chemical elements heavier than
iron. The conditions depend on the proton-to-neutron ratio as well as the
entropy in the hot SN outflows which are determined by neutrinos and
particularly by the fluxes and spectra of $\nu_e$ and $\bar\nu_e$ through
beta processes. Observations of nucleosynthesis yields in SNe provide some
information about the conditions in this region, and conversely measuring the neutrino flavor-dependent neutrino fluxes provides crucial information on these conditions. Nucleosynthesis in the SN environment, like the explosion mechanism itself, is a major subject of numerical study that would benefit from high-statistics neutrino data.

\begin{figure}
\centering
\includegraphics[width=1.4\textwidth]{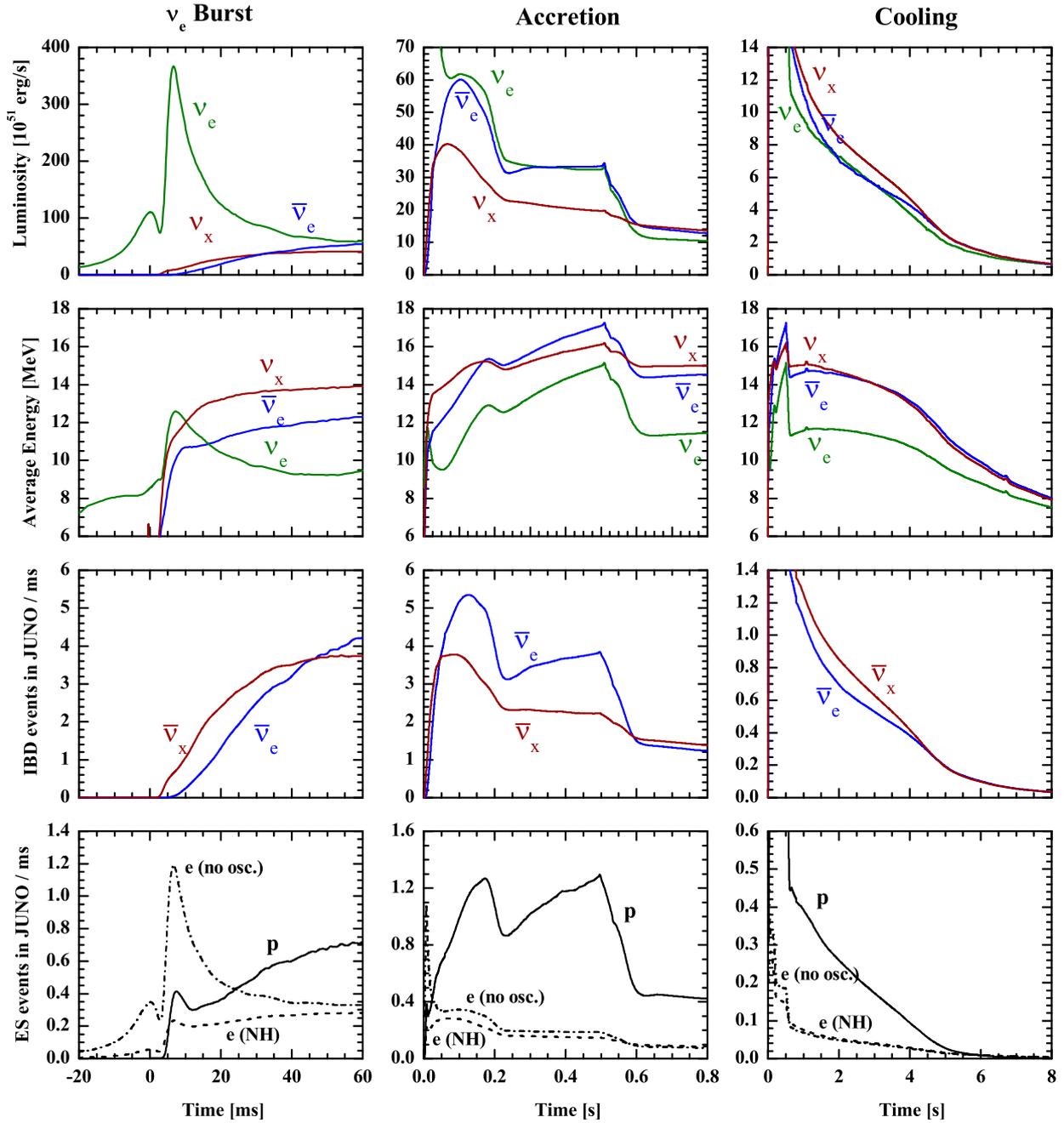}
\vspace{-1cm}
\caption{Three phases of neutrino emission from a core-collapse SN, from left to right: (1)~Infall, bounce and initial shock-wave propagation, including prompt $\nu_e$ burst. (2)~Accretion phase with significant flavor differences of fluxes and spectra and time variations of the signal. (3)~Cooling of the newly formed neutron star, only small flavor differences between fluxes and spectra. (Based on a spherically symmetric Garching model with explosion triggered by hand during 0.5--0.6~ms~\cite{GarchingModel,Mirizzi:2015eza}. See text for details.)
We show the flavor-dependent luminosities and average energies as well as the IBD rate in JUNO assuming either no flavor conversion (curves~$\bar\nu_e$) or complete flavor swap (curves~$\bar\nu_x$).
The elastic proton (electron) scattering rate uses all six species and assumes a detection threshold of 0.2~MeV of visible proton (electron) recoil energy. For the electron scattering, two extreme cases of no flavor conversion (curves no osc.) and flavor conversion with a normal neutrino mass ordering (curves NH) are presented.
\label{fig:sn:SNburst}}
\end{figure}

\subsubsection{Three-phase neutrino signal}
\label{subsubsec:sn:three-phase-signal}

If the core-collapse scenario indeed roughly follows the stages described in
the previous section, one expects a neutrino signal with three characteristic phases as shown in Fig.~\ref{fig:sn:SNburst}. In a high-statistics observation one should consider these essentially as three different experiments, each holding different and characteristic lessons for particle- and astrophysics. The shown example is a spherically symmetric model from the Garching group, based on a $27\,M_\odot$ model~\cite{GarchingModel,Mirizzi:2015eza}. The explosion is triggered ``by hand'' by quenching the accretion flow between 500--600~ms. The final neutron-star baryonic mass is $1.762\,M_\odot$ and it emits a total of $3.24\times10^{53}~{\rm erg}$ of energy in neutrinos, corresponding to a mass equivalent of $0.180\,M_\odot$. Its final gravitational mass is $1.582\,M_\odot$ and thus a typical neutron star. This model is comparable to the example from the Basel group shown in the LENA White Paper~\cite{Wurm:2011zn}, except that the Garching models include nucleon recoils in the treatment of neutrino transport, leading to a much smaller flavor-dependent spread in average neutrino energies.

For each of the three phases ($\nu_e$ Burst, Accretion, Cooling) we show the
luminosities in $\nu_e$, $\bar\nu_e$, and $\nu_x$, where the latter stands
for any of $\nu_\mu$, $\bar\nu_\mu$, $\nu_\tau$ and $\bar\nu_\tau$, and the
average energies. We further show the IBD rate in JUNO for two cases: No
flavor conversion at all (curves marked $\bar\nu_e$) and assuming complete
flavor conversion (curves marked $\bar\nu_x$). The actual neutrino flavor conversions lead to an IBD rate lying between the curves of two extreme cases. Finally we show the rate of elastic proton scattering caused by all six species, assuming a detection threshold for proton recoils of 0.2~MeV visible energy. More discussions about the detection of SN neutrinos can be found in the following section. A few comments are in order on the three phases of neutrino signals:

\vspace{0.2cm}

\noindent {\bf 1.~Infall, Bounce and Shock Propagation.}---During the core collapse, the electron capture on protons and heavy nuclei already starts to produce $\nu_e$. The first tens of ms after bounce show the
characteristic prompt $\nu_e$ burst, the emission of $\bar\nu_e$ is at first
suppressed, and emission of other flavors only begins. The flux and spectral
characteristics are rather reproducible, independently of specific
assumptions, e.g., concerning the progenitor mass and nuclear equation of
state. The IBD rate shows a characteristic rise-time difference between
$\bar\nu_e$ and $\bar\nu_x$ that can be used to diagnose neutrino flavor
conversion.

\vspace{0.2cm}

\noindent {\bf 2.~Accretion Phase (Shock Stagnation)}.---Few tens to few
hundreds of ms, depending on progenitor properties and other parameters.
Neutrino emission is powered by accretion flow. Luminosities in $\nu_e$ and
$\bar\nu_e$ perhaps as much as a factor of two larger than each of the
$\nu_x$ fluxes. Pronounced hierarchy of average energies $\langle
E_{\nu_e}\rangle<\langle E_{\bar\nu_e}\rangle<\langle E_{\nu_x}\rangle$,
energies increasing until explosion. Large-scale convection, SASI and LESA
build up, implying fast time variations and directional dependence of the
neutrino signal. The luminosity drop at around 200~ms represents the infall
of the Si/O interface---the accretion rate and therefore luminosity clearly
drops afterwards.

\vspace{0.2cm}

\noindent {\bf 3.~Cooling}.---When the explosion has taken off (here
triggered between 500--600~ms by numerically quenching the accretion flow)
the luminosity drops and is subsequently powered by cooling of the
proto-neutron star. Approximate luminosity equipartition between species and
$\langle E_{\nu_e}\rangle<\langle E_{\bar\nu_e}\rangle\sim\langle
E_{\nu_x}\rangle$. Number flux of $\nu_e$ enhanced because of
de-leptonization.

\vspace{0.2cm}

Finally, it is worth mentioning that there exist alternatives to the delayed, neutrino-driven explosion paradigm, such as the acoustic mechanism~\cite{Burrows:2005dv} and magnetohydrodynamically driven explosion~\cite{Akiyama:2002xn}. A successful explosion could also be triggered by a QCD phase transition~\cite{Gentile:1993ma,Dai:1995uj,Sagert:2008ka}. The neutrino signals in all these cases may be quite different from those shown in Fig.~\ref{fig:sn:SNburst}, and thus a high-statistics measurement of time and energy spectra of neutrino events can be used to distinguish one explosion mechanism from another.

\subsection{Detection channels in JUNO}

\subsubsection{Time-integrated event rates}

In order to estimate the expected neutrino rates in JUNO, we assume a linear alkylbenzene (LAB) based liquid scintillator and a fiducial mass of 20 kiloton, implying about $1.5\times 10^{33}$ target protons. For a typical galactic SN at 10 kpc, there will be more than 5000 neutrino events solely from the IBD channel. Such a high-statistics observation definitely allows us to probe the time-dependent features of SN neutrinos. However, the time-integrated event rates will give a first impression on the capability of JUNO detector in SN neutrino detection. For simplicity, instead of taking the simulated results in Fig.~\ref{fig:sn:SNburst} as input,
we model the time-integrated neutrino spectra as $f_\nu(E_\nu) \propto E^\alpha_\nu \exp[-(1+\alpha)E_\nu/\langle E_\nu \rangle]$ with a nominal index $\alpha = 3$ and $\langle E_\nu \rangle$ being the average neutrino energy~\cite{Keil:2002in}. Furthermore, a total energy of $3 \times 10^{53}~{\rm erg}$ is assumed to be equally distributed in neutrinos and antineutrinos of three flavors. As the average neutrino energies are both flavor- and time-dependent, we calculate the event rates for three representative values $\langle E_\nu \rangle = 12~{\rm MeV}$, $14~{\rm MeV}$ and $16~{\rm MeV}$, and in each case the average energy is taken to be equal for all flavors so that one can easily observe the impact of average neutrino energies on the event rates. The total numbers of neutrino events for the main channels in JUNO are summarized in Table~\ref{table:events}, where no neutrino flavor conversions are considered. For the numerical model introduced in the previous section, the time-dependent event rates for IBD  are displayed in the third row of Fig.~\ref{fig:sn:SNburst}, while those for elastic proton and electron scattering are given in the fourth row. The neutrino event spectra with respect to the visible energy in six main reaction channels are shown in Fig.~\ref{fig:spectra}. Some comments on the different signal channels are in order.

(1) The IBD is the dominant channel for SN neutrino detection in both scintillator and water-Cherenkov detectors, in which a large number of free protons are available. In the IBD reaction
\begin{equation}
\overline{\nu}_e + p \to e^+ + n \; ,
\label{eq: IBD}
\end{equation}
the neutrino energy threshold is $E^{\rm th}_\nu = \Delta + m_e \approx 1.806~{\rm MeV}$, where $\Delta \equiv m_n - m_p \approx 1.293~{\rm MeV}$ is the neutron-proton mass difference. The energy of the incident neutrino can be reconstructed from the positron energy via $E_\nu \approx E_e + \Delta$. The energy deposition and the annihilation of the positron with an ambient electron into 0.511-MeV $\gamma$'s give rise to a prompt signal. In addition, the neutron is captured on a free proton with an average lifetime of about $200~{\rm \mu s}$, producing a $2.2$-MeV $\gamma$. Hence the time coincidence of the prompt and delayed signals increases greatly the tagging power. A precise calculation of the IBD cross section has been performed in Ref.~\cite{Strumia:2003zx,Vogel:1999zy}. In general, the angular distribution of the positron is nearly isotropic, so it is difficult to extract the directional information of neutrinos. However, the forward shift of the neutron capture vertex may be used to further reduce backgrounds and locate the neutrino source~\cite{Vogel:1999zy,Apollonio:1999jg}.
%%%%%%%%%%%%%%%%%%%%%%%%%%%%%%%%% Table 4-1 %%%%%%%%%%%%%%%%%%%%%%%%%%%%%%%%%%
\begin{table}[!t]
\centering
\begin{tabular}{ccccccccc}
\hline
\multicolumn{1}{c}{\multirow {2}{*}{Channel}} & \multicolumn{1}{c}{} & \multicolumn{1}{c}{\multirow {2}{*}{Type}} & \multicolumn{1}{c}{} & \multicolumn{5}{c}{Events for different $\langle E_\nu \rangle$ values} \\
\cline{5-9} \multicolumn{1}{c}{} & \multicolumn{1}{c}{} & \multicolumn{1}{c}{} & \multicolumn{1}{c}{} & \multicolumn{1}{c}{$12~{\rm MeV}$} & \multicolumn{1}{c}{} & \multicolumn{1}{c}{$14~{\rm MeV}$} & \multicolumn{1}{c}{} & \multicolumn{1}{c}{$16~{\rm MeV}$} \\
\hline
\multicolumn{1}{l}{$\overline{\nu}_e + p \to e^+ + n$} & \multicolumn{1}{c}{} & \multicolumn{1}{c}{CC} & \multicolumn{1}{c}{} & \multicolumn{1}{c}{$4.3\times 10^3$} & \multicolumn{1}{c}{} & \multicolumn{1}{c}{$5.0\times 10^3$} & \multicolumn{1}{c}{} & \multicolumn{1}{c}{$5.7\times 10^3$} \\
\multicolumn{1}{l}{$\nu + p \to \nu + p$} & \multicolumn{1}{c}{} & \multicolumn{1}{c}{NC} & \multicolumn{1}{c}{} & \multicolumn{1}{c}{$0.6\times 10^3$} & \multicolumn{1}{c}{} & \multicolumn{1}{c}{$1.2\times 10^3$} & \multicolumn{1}{c}{} & \multicolumn{1}{c}{$2.0\times 10^3$} \\
\multicolumn{1}{l}{$\nu + e \to \nu + e$} & \multicolumn{1}{c}{} & \multicolumn{1}{c}{ES} & \multicolumn{1}{c}{} & \multicolumn{1}{c}{$3.6\times 10^2$} & \multicolumn{1}{c}{} & \multicolumn{1}{c}{$3.6\times 10^2$} & \multicolumn{1}{c}{} & \multicolumn{1}{c}{$3.6\times 10^2$} \\
\multicolumn{1}{l}{$\nu +~^{12}{\rm C} \to \nu +~^{12}{\rm C}^*$} & \multicolumn{1}{c}{} & \multicolumn{1}{c}{NC} & \multicolumn{1}{c}{} & \multicolumn{1}{c}{$1.7\times 10^2$} & \multicolumn{1}{c}{} & \multicolumn{1}{c}{$3.2\times 10^2$} & \multicolumn{1}{c}{} & \multicolumn{1}{c}{$5.2\times 10^2$} \\
\multicolumn{1}{l}{$\nu_e +~^{12}{\rm C} \to e^- +~^{12}{\rm N}$} & \multicolumn{1}{c}{} & \multicolumn{1}{c}{CC} & \multicolumn{1}{c}{} & \multicolumn{1}{c}{$0.5\times 10^2$} & \multicolumn{1}{c}{} & \multicolumn{1}{c}{$0.9\times 10^2$} & \multicolumn{1}{c}{} & \multicolumn{1}{c}{$1.6\times 10^2$} \\
\multicolumn{1}{l}{$\overline{\nu}_e +~^{12}{\rm C} \to e^+ +~^{12}{\rm B}$} & \multicolumn{1}{c}{} & \multicolumn{1}{c}{CC} & \multicolumn{1}{c}{} & \multicolumn{1}{c}{$0.6\times 10^2$} & \multicolumn{1}{c}{} & \multicolumn{1}{c}{$1.1\times 10^2$} & \multicolumn{1}{c}{} & \multicolumn{1}{c}{$1.6\times 10^2$} \\
\hline
\end{tabular}
\caption{Numbers of neutrino events in JUNO for a SN at a typical distance of 10 kpc, where $\nu$ collectively stands for neutrinos and antineutrinos of all three flavors and their contributions are summed over. Three representative values of the average neutrino energy $\langle E_\nu \rangle = 12~{\rm MeV}$, $14~{\rm MeV}$ and $16~{\rm MeV}$ are taken for illustration, where in each case the same average energy is assumed for all flavors and neutrino flavor conversions are not considered. For the elastic neutrino-proton scattering, a threshold of $0.2~{\rm MeV}$ for the proton recoil energy is chosen.}
	\label{table:events}
\end{table}
%%%%%%%%%%%%%%%%%%%%%%%%%%%%%%%%%%%%%%%%%%%%%%%%%%%%%%%%%%%%%%%%%%%%%%%%%%%%%%
%%%%%%%%%%%%%%%%%%%%%%%%%%%%%%%%%%%%%%%%%%%%%%%%%%%%%%%%%%%%%%%%%%%%%%%%%%%%%%
\begin{figure}[!t]
    \centering
    \includegraphics[width=0.7\textwidth]{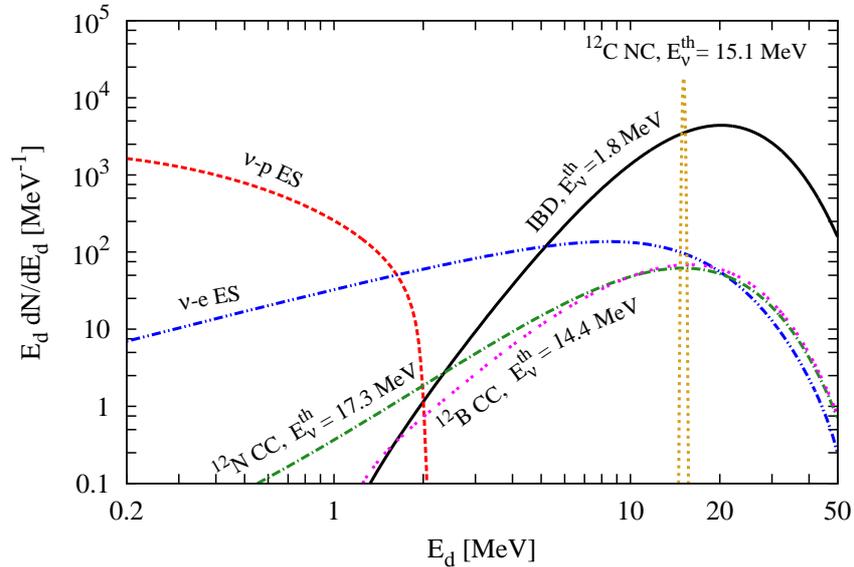}
    \vspace{-0.4cm}
    \caption{The neutrino event spectra with respect to the visible energy $E^{}_{\rm d}$ in the JUNO detector for a SN at 10 kpc, where no neutrino flavor conversions are assumed for illustration and the average neutrino energies are $\langle E^{}_{\nu_e}\rangle = 12~{\rm MeV}$, $\langle E^{}_{\overline{\nu}_e}\rangle = 14~{\rm MeV}$ and $\langle E^{}_{\nu_x}\rangle = 16~{\rm MeV}$. The main reaction channels are shown together with the threshold of neutrino energies: (1) IBD (black and solid curve), $E_{\rm d} = E^{}_\nu - 0.8~{\rm MeV}$; (2) Elastic $\nu$-$p$ scattering (red and dashed curve), $E_{\rm d}$ stands for the recoil energy of proton; (3) Elastic $\nu$-$e$ scattering (blue and double-dotted-dashed curve), $E_{\rm d}$ denotes the recoil energy of electron; (4) Neutral-current reaction ${^{12}{\rm C}}(\nu, \nu^\prime){^{12}{\rm C}^*}$ (orange and dotted curve), $E_{\rm d} \approx 15.1~{\rm MeV}$; (5) Charged-current reaction ${^{12}{\rm C}}(\nu_e, e^-){^{12}{\rm N}}$ (green and dotted-dashed curve), $E_{\rm d} = E_\nu - 17.3~{\rm MeV}$; (6) Charged-current reaction ${^{12}{\rm C}}(\overline{\nu}_e, e^+){^{12}{\rm B}}$ (magenta and double-dotted curve), $E_{\rm d} = E_\nu - 13.9~{\rm MeV}$.}
\label{fig:spectra}
\end{figure}
%%%%%%%%%%%%%%%%%%%%%%%%%%%%%%%%%%%%%%%%%%%%%%%%%%%%%%%%%%%%%%%%%%%%%%%%%%%%%%
%%%%%%%%%%%%%%%%%%%%%%%%%%%%%%%%%%%%%%%%%%%%%%%%%%%%%%%%%%%%%%%%%%%%%%%%%%%%%%

(2) As an advantage of the scintillator detector, the charged-current (CC) interaction on $^{12}{\rm C}$ takes place for both $\nu_e$ and $\overline{\nu}_e$ via
\begin{eqnarray}
&& \nu_e +{^{12}}{\rm C} \to e^- +{^{12}}{\rm N} \; , \label{eq: CCnue}\\
&& \overline{\nu}_e +{^{12}}{\rm C} \to e^+ +{^{12}}{\rm B} \; . \label{eq: CCnueb}
\end{eqnarray}
The energy threshold for $\nu_e$ is $17.34~{\rm MeV}$, while that for $\overline{\nu}_e$ is $14.39~{\rm MeV}$. The subsequent beta decays of $^{12}{\rm B}$ and $^{12}{\rm N}$ with a $20.2~{\rm ms}$ and $11~{\rm ms}$ half-life, respectively, lead to a prompt-delayed coincident signal. Hence the charged-current reactions in Eqs.~(\ref{eq: CCnue}) and (\ref{eq: CCnueb}) provide a possibility to detect separately $\nu_e$ and $\overline{\nu}_e$~\cite{Laha:2014yua}. The cross section of neutrino interaction on $^{12}{\rm C}$ has been calculated in Ref.~\cite{Fukugita:1988hg} by using a direct evaluation of nuclear matrix elements from experimental data at that time. Recent calculations based on the nuclear shell model and the random-phase approximation can be found in Ref.~\cite{Volpe:2000zn}. The cross section has been measured in the LSND experiment, and the result is well compatible with theoretical calculations~\cite{Auerbach:2001hz}.

(3) The neutral-current (NC) interaction on $^{12}{\rm C}$ is of crucial importance to probe neutrinos of non-electron flavors, i.e.,
\begin{equation}
\nu +{^{12}}{\rm C} \to \nu +{^{12}}{\rm C}^* \; ,
\label{eq: NCnux}
\end{equation}
where $\nu$ collectively denotes neutrinos and antineutrinos of all three flavors. A $15.11$-MeV $\gamma$ from the deexcitation of $^{12}{\rm C}^*$ to its ground state is a clear signal of SN neutrinos. The cross section can be found in Refs.~\cite{Fukugita:1988hg,Volpe:2000zn}, and has also been measured in the LSND experiment~\cite{Auerbach:2001hz}. Since the neutrinos of non-electron flavors $\nu_x$ have higher average energies, the neutral-current interaction is most sensitive to $\nu_x$, offering a possibility to pin down the flavor content of SN neutrinos. However, the recoil energy of $^{12}{\rm C}$ is negligible and the final-state neutrino is invisible to current detectors, implying the impossibility to reconstruct neutrino energy event-by-event in this channel. Note that the neutral-current processes are not affected by the neutrino flavor oscillations.

(4) In the elastic scattering (ES) of neutrinos on electrons, the scattered electrons carry the directional information of incident neutrinos, and thus can be used to locate the SN. This is extremely important if a SN is hidden in the galactic gas and dust clouds and the optical signal is obscured. The elastic scattering
\begin{equation}
\nu + e^- \to \nu + e^-
\label{eq: ESe}
\end{equation}
is most sensitive to $\nu_e$ because of its largest cross section, which is particularly useful in detecting the prompt $\nu^{}_e$ burst. The cross sections of neutrino- and antineutrino-electron elastic scattering have been computed and summarized in Ref.~\cite{Marciano:2003eq}, where the electroweak radiative corrections are also included. Unlike water Cherenkov detectors such as Super-Kamiokande, it is challenging for a liquid scintillator detector to determine the SN direction by reconstructing the direction of the scattered electron, unless the PMT time response is quick enough and the detector is precisely understood. For high-energy electrons, it might be possible to make use of Cherenkov light to obtain some directional information.

(5) The elastic scattering of neutrinos on protons has been proposed as a promising channel to measure SN neutrinos of non-electron flavors~\cite{Beacom:2002hs,Dasgupta:2011wg}:
\begin{equation}
\nu + p \to \nu + p \; .
\label{eq: ESp}
\end{equation}
Although the total cross section is about four times smaller than that of the IBD reaction, the contributions from all the neutrinos and antineutrinos of three flavors will compensate for the reduction of cross section, especially if the average energy of the SN neutrinos is large. In this channel, the proton recoil energy $T_p \leq 2 E^2_\nu/m_p$ is highly suppressed by the nucleon mass, so the precise determination of the proton quenching factor and a low energy threshold are required to reconstruct neutrino energy and accumulate sufficient statistics. The cross section of elastic neutrino-proton scattering was calculated a long time ago~\cite{Weinberg:1972tu}, and has been recently simplified for low-energy neutrinos~\cite{Beacom:2002hs}. However, the cross section receives a dominant contribution from the proton axial form factor, which at present is only known with a $30\%$ uncertainty if the strange-quark contribution to the nucleon spin is taken into account.

\subsubsection{Elastic neutrino-proton scattering}
%%%%%%%%%%%%%%%%%%%%%%%%%%%%%%%%%%%%%%%%%%%%%%%%%%%%%%%%%%%%%%%%%%%%%%%%%%%%%%
\begin{figure}[!t]
    \centering
    \includegraphics[width=0.7\textwidth]{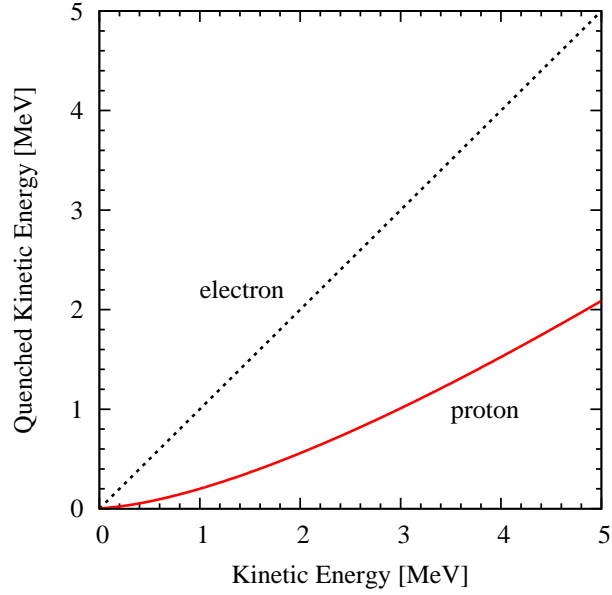}
    \caption{The light output of a recoiled proton in a LAB-based scintillator detector, where the Birk's constant is taken to be $k_{\rm B} = 0.0098~{\rm cm}~{\rm MeV}^{-1}$ according to the measurement in Ref.~\cite{vonKrosigk:2013sa}. The energy deposition rates of protons in hydrogen and carbon targets are taken from the PSTAR database~\cite{PSTAR}, and combined to give the deposition rate in a LAB-based scintillator according to the ratio of their weights in the detector.}
\label{fig:quenching}
\end{figure}
%%%%%%%%%%%%%%%%%%%%%%%%%%%%%%%%%%%%%%%%%%%%%%%%%%%%%%%%%%%%%%%%%%%%%%%%%%%%%%
%%%%%%%%%%%%%%%%%%%%%%%%%%%%%%%%%%%%%%%%%%%%%%%%%%%%%%%%%%%%%%%%%%%%%%%%%%%%%%
\begin{figure}[!t]
    \centering
    \includegraphics[width=0.75\textwidth]{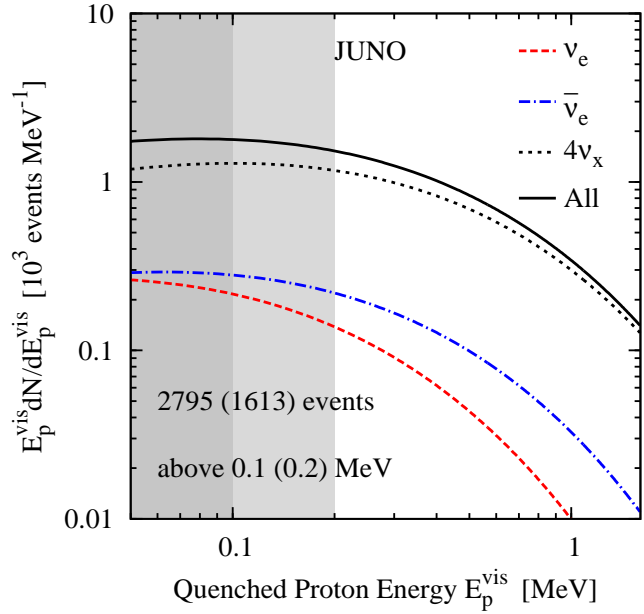}
    \vspace{-0.4cm}
    \caption{Event spectrum of the elastic neutrino-proton scattering in JUNO. The average neutrino energies are $\langle E_{\nu_e} \rangle = 12~{\rm MeV}$, $\langle E_{\overline{\nu}_e} \rangle = 14~{\rm MeV}$ and $\langle E_{\nu_x} \rangle = 16~{\rm MeV}$. The total numbers of events have been calculated by imposing an energy threshold of $0.1$ or $0.2$ MeV.}
\label{fig:junoep}
\end{figure}
%%%%%%%%%%%%%%%%%%%%%%%%%%%%%%%%%%%%%%%%%%%%%%%%%%%%%%%%%%%%%%%%%%%%%%%%%%%%%%
%%%%%%%%%%%%%%%%%%%%%%%%%%%%%%%%%%%%%%%%%%%%%%%%%%%%%%%%%%%%%%%%%%%%%%%%%%%%%%
\begin{figure}[!t]
    \centering
    \includegraphics[width=0.7\textwidth]{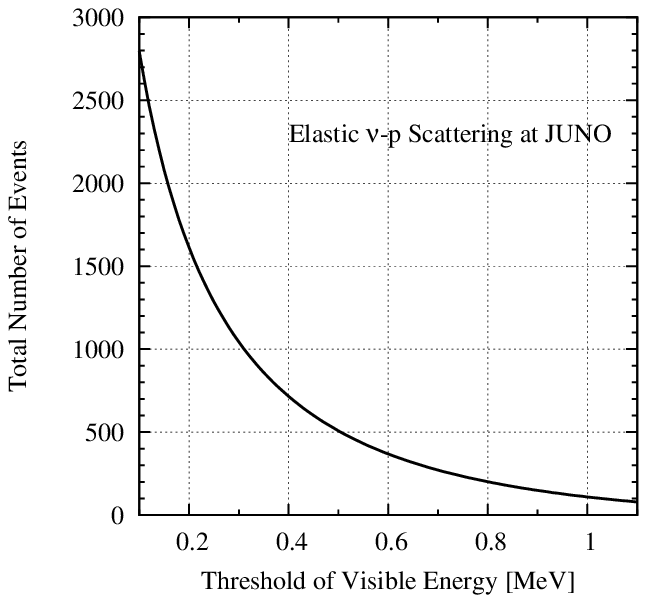}
    \vspace{-0.1cm}
    \caption{Impact of energy threshold on the elastic $\nu$-$p$ scattering events in JUNO, where the input parameters are the same as in Fig.~\ref{fig:junoep}.}
\label{fig:threshold}
\end{figure}
%%%%%%%%%%%%%%%%%%%%%%%%%%%%%%%%%%%%%%%%%%%%%%%%%%%%%%%%%%%%%%%%%%%%%%%%%%%%%%

As shown in Refs.~\cite{Beacom:2002hs,Dasgupta:2011wg}, it is possible to reconstruct the energy spectrum of $\nu_x$ at a large scintillator detector, which is very important to establish the flavor conversions and the total energy of SN neutrinos. Therefore, it is worthwhile to study in more detail the potential of JUNO to detect SN neutrinos in this channel.

For a realistic measurement of the neutrino energy spectrum, a low-energy threshold and a satisfactory reconstruction of proton recoil energy are required. The reason is simply that low-energy protons are highly ionizing particles, implying that their energy-loss rate is much higher compared to electrons of the same energy. From Fig.~\ref{fig:quenching}, one can observe that the light output of a low-energy proton is significantly quenched relative to an electron losing the same amount of energy. The measurement of the proton quenching factor for various liquid scintillators has been performed in Ref.~\cite{vonKrosigk:2013sa}, and further applied to the SN neutrino detection. For JUNO, the proton quenching factor will be measured for the ultimately implemented scintillator.

In Fig.~\ref{fig:junoep}, the event spectrum of elastic neutrino-proton scattering in JUNO has been given, where a hierarchical spectrum of average neutrino energies is assumed. It is evident that the events are dominated by the heavy-lepton flavors, for which the average energy is higher than that for the electron flavor. However, as indicated by the sophisticated numerical simulations, the average energies of $\overline{\nu}^{}_e$ and $\nu^{}_x$ are quite similar and larger than that of $\nu^{}_e$ in the late phases (see Fig.~\ref{fig:sn:SNburst}). Whereas the $\overline{\nu}^{}_e$ spectrum can be well measured in the IBD channel, the $\nu^{}_x$ spectrum can only be determined in the neutrino-proton channel. The reconstruction of neutrino energy spectrum allows us to figure out the total energy emitted in neutrinos and to extract the information of flavor conversions. Although the neutrino-proton scattering is most sensitive to the high-energy tail of the neutrino spectrum, one can fit the observed data by using a thermal spectrum with an unknown average energy~\cite{Dasgupta:2011wg}. In JUNO, the total number of events is about 1600 above a threshold of $0.2~{\rm MeV}$ and 2800 above $0.1~{\rm MeV}$. Singles from radioactivity will be controlled to tens of Hz in the JUNO detector. The major impediment is the coincidence of PMT dark noise, limiting the energy threshold at $\sim\,$0.3 MeV. A sophisticated trigger approach that rejects low energy events with vertices at the detector center is under study, which could lower the energy threshold down to $\sim\,$0.1 MeV with negligible inefficiency. In Fig.~\ref{fig:threshold}, the dependence of total events on the energy threshold is illustrated, where one can see that the events decrease dramatically for a higher energy threshold. Even at the worst case, the threshold $\sim0.7~{\rm MeV}$ as required by reactor neutrino physics, the neutrino-proton scattering is still the best channel to measure the SN neutrinos of heavy-lepton flavors. The time-dependent event rate for the Garching SN model has been shown in Fig.~\ref{fig:sn:SNburst}, where one can see that a few tens of proton scattering events can be observed for the prompt $\nu^{}_e$ burst, independently of the flavor oscillations. The reason is simply that the elastic proton scattering is governed by the NC interaction, which is flavor blind.

\subsubsection{Backgrounds}

There exist various backgrounds for SN neutrino detection, and the backgrounds may vary by the detector location and type, as well as the signal channel. In general, background is not a serious concern for a SN neutrino burst since it lasts only for about 10 seconds. Possible background sources for a given signal channel are natural radioactivities (less than 10 Hz at $> 0.7$ MeV), cosmogenic backgrounds ($\sim\,3$ Hz muon rate in the JUNO liquid scintillator detector), the SN neutrinos themselves, and other neutrinos. Detailed background estimation can be found in the Appendix.

Inverse beta decay is the major signal channel of the SN burst detection in JUNO. Reactor neutrinos and geo-neutrinos contribute as backgrounds. In a 10-second interval, there will be 0.01 IBD from reactors and 0.0002 IBD from geo-neutrinos, which is totally negligible for any detectable SN. Other backgrounds, such as that from low-energy atmospheric antineutrinos, natural radioactivities, and cosmogenic isotopes are even smaller.

The other charged-current channels that produce ${^{12}}{\rm N}$ and $^{12}{\rm B}$ as delayed signals will not suffer from the reactor neutrinos and geo-neutrinos, either. Coincidence of IBD prompt signals from the SN burst is a major obstacle to identify these events, since the delayed signal from the decay of $^{12}{\rm N}$ or $^{12}{\rm B}$ has a long lifetime of tens of ms. Vertex correlation between the prompt $e^-$ ($e^+$) and the delayed $^{12}{\rm N}$ ($^{12}{\rm B}$) signals can reject the coincidence backgrounds to a negligible level, unless the SN neutrino rate is extremely high.

The neutral-current channel that produces a $^{12}{\rm C}^*$ can be identified with the 15.11 MeV $\gamma$, which is much higher than the natural radioactivities. There are 0.1 $^8{\rm Li}$ and 0.1 $^{12}{\rm B}/^{12}{\rm N}$ in the cosmogenic isotopes in 10 seconds. With the energy selection around 15.11 MeV, they are negligible. The $^{12}{\rm C}^*$ signal locates in the energy region of the IBD prompt signal and the $^{12}{\rm N}$ and $^{12}{\rm B}$ decays of the SN neutrinos. Again, rejection can be done with vertex correlation and time correlation of the other two types of SN neutrinos.

Backgrounds should be estimated with more care for the elastic neutrino-proton scattering since they are singles of low visible energy. In the cosmogenic isotopes, there are 1.9 $^{11}{\rm C}$, 0.6 $^{7}{\rm B}$, and 0.6 other longlived isotopes in the 10 seconds interval. Radioactive decays of scintillator materials and surroundings lead to the dominant backgrounds. In the energy region from 0.2 to 1\,MeV, it will arise from the $\beta$-decay of the noble gas $^{85}$Kr and the lead-daughter $^{210}$Bi, which is fed by the long-lived isotope $^{210}$Pb ($\tau=32$\,yrs). The JUNO baseline design without distillation foresees an activity of 50~$\mu$Bq/m$^3$ for $^{85}$Kr and a concentration of $1.4\times10^{-22}$\,g/g for $^{210}$Pb. The corresponding background rates in the 10 second interval are 10 events of $^{85}$Kr and $\sim$ 70 events of $^{210}$Bi. Even assuming low spectral energies for the SN neutrinos, the proton recoil signal would be dominating by at least one order of magnitude.

The neutrino-electron scattering has similar backgrounds as the neutrino-proton scattering. It has higher visible energy since the quenching effect for the electron is small. If we select only high energy events, it has much less background than the neutrino-proton scattering.

Faked events created by random coincidences of the PMT dark noise can be rejected by vertex selection. Since the PMT hits are uniformly distributed, the vertex of the faked event is at the detector center. Fig.~\ref{fig:sn:rdarkevent} shows the vertex distribution due to statistical fluctuation versus the number of photoelectrons, while every 120 p.e. corresponds $\sim\,$0.1 MeV. Without online selection on vertex, the energy threshold will be limited at 0.3--0.4 MeV due to extremely high event rate. With an online trigger rejecting the events at the detector center and of an energy below 0.4 MeV, the threshold can be lowered to 0.1 MeV, with only a couple of percent inefficiency.

The rate of fake dark-noise triggers above a given energy threshold (defined by a coincident number of PMT hits, e.g. 240 hits for 0.2\,MeV) depends on the dark rate of the individual PMTs, $R_{\rm dn}$, and the length of the trigger gate $t_{\rm gate}$ for coincidences between hits and can be computed from Poisson statistics.
%%%%%%%%%%%%%%%%%%%%%%%%%%%%%%%%%%%%% Fig. 4.6 %%%%%%%%%%%%%%%%%%%%%%%%%%%%%
\begin{figure}[t]
\centering
\includegraphics[width=0.6\textwidth]{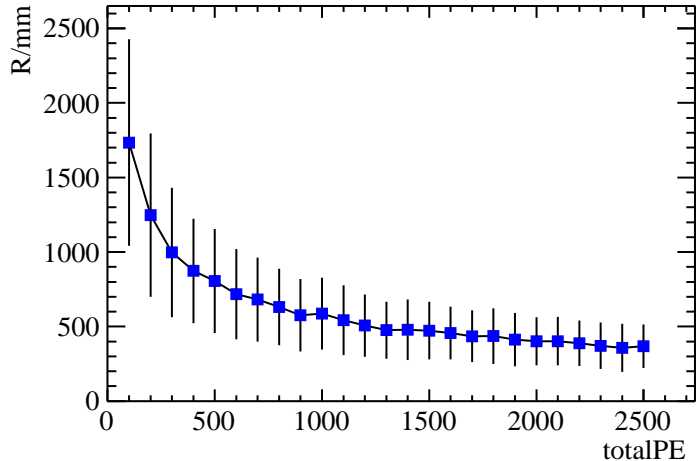}
\caption{Vertex distribution of the faked events created by coincidence of the PMT dark noise versus total photoelectrons. The data point and the error bar shows the mean value and 1 $\sigma$ uncertainty for the Maxwell-Boltzmann distribution of the radius of the faked events.
\label{fig:sn:rdarkevent}}
\end{figure}
%%%%%%%%%%%%%%%%%%%%%%%%%%%%%%%%%%%%%%%%%%%%%%%%%%%%%%%%%%%%%%%%%%%%%%%%%%%%%%

\subsubsection{Data acquisition}
%%%%%%%%%%%%%%%%%%%%%%%%%%%%%%%%% Fig. 4.7 %%%%%%%%%%%%%%%%%%%%%%%%%%%%%%%%%
\begin{figure}[!t]
    \centering
    \includegraphics[width=0.8\textwidth]{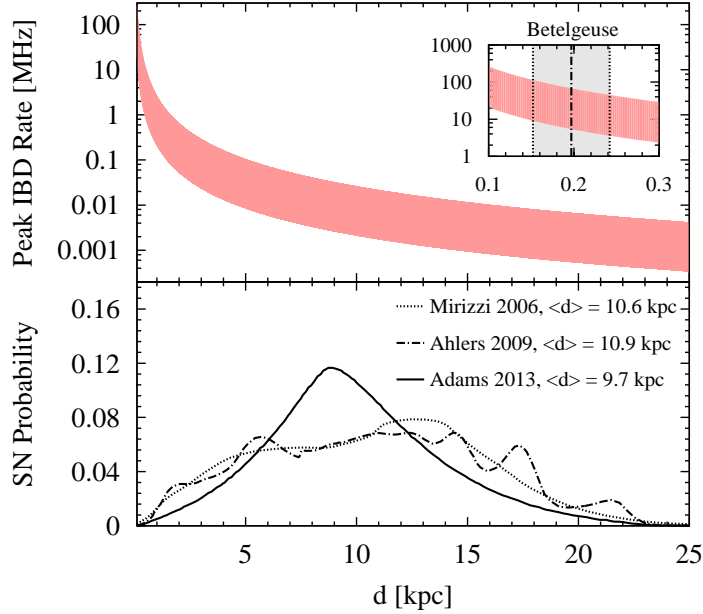}
    \vspace{-0.8cm}
    \caption{Maximum IBD rate at JUNO as a function of the distance to a galactic SN. In the upper panel, the shaded range has been obtained by considering a class of SN models from the Basel~\cite{Fischer:2009af}, Garching~\cite{Buras:2005tb,Serpico:2011ir} and Nakazato~\cite{Nakazato:2012qf} groups. The insert refers to the SN candidate Betelgeuse with a distance of $0.197 \pm 0.045~{\rm kpc}$~\cite{Harper:2008}. In the lower panel, the SN probability in our galaxy has been given according to three different evaluations~\cite{Mirizzi:2006xx,Ahlers:2009ae,Adams:2013ana}.}
\label{fig:daq}
\end{figure}
%%%%%%%%%%%%%%%%%%%%%%%%%%%%%%%%%%%%%%%%%%%%%%%%%%%%%%%%%%%%%%%%%%%%%%%%%%%%%%

The large scintillator detector of JUNO has great potential of detecting galactic SN neutrinos, which however could have diverse energies and intensities. In the previous discussions, a typical distance of $10~{\rm kpc}$ has been assumed for illustration. But it is important to notice that
a galactic SN may occur at a much shorter distance. For instance, the red supergiant Betelgeuse is a very promising SN candidate, which is located just about $0.2~{\rm kpc}$ away~\cite{Harper:2008}. Hence, the electronic and data acquisition systems of JUNO should be designed to properly handle a huge number of neutrino events even in this extreme case.

To clarify the requirement for the detector not to be blinded by SN neutrinos, we calculate the maximum IBD rate in JUNO for a large number of SN models from the Basel~\cite{Fischer:2009af}, Garching~\cite{Buras:2005tb,Serpico:2011ir} and Nakazato~\cite{Nakazato:2012qf} groups. The largest IBD rates for those models have been depicted as a function of the distance in Fig.~\ref{fig:daq}, where the wide band can be viewed as the uncertainties of SN models. The insert in the upper panel refers to the case when Betelgeuse finally turns into a core-collapse SN. Taking account of the distance uncertainty, we can see that the maximal IBD rate reaches $100~{\rm MHz}$, and even the minimum exceeds $1~{\rm MHz}$, which is several orders of magnitude larger than the baseline value $1~{\rm kHz}$ chosen for the determination of neutrino mass ordering. In the lower panel of Fig.~\ref{fig:daq}, the SN probability in our galaxy has been shown for three different theoretical evaluations~\cite{Mirizzi:2006xx,Ahlers:2009ae,Adams:2013ana}, where one can observe that the probability distribution is broad and the average distance is around $10~{\rm kpc}$.

In the best situation, the data acquisition system of JUNO should be able to work both for a ``typical" SN at the most probable distance $10~{\rm kpc}$ and the closest conceivable distance $0.2~{\rm kpc}$.

\subsection{Implications for astrophysics}

A high-statistics observation of galactic SN neutrinos is of crucial importance for astrophysics, in particular for the evolution of massive stars, the core-collapse SN explosion and the production of heavy chemical elements~\cite{Burrows:2012ew,Janka:2012sb,Fuller:1992,Qian:1993dg}. Moreover, the core-collapse SNe themselves are expected to be associated with the birth of neutron stars and black holes, and the emission of gravitational waves. Therefore, the neutrino signals at the JUNO detector could help us answer many fundamental questions in astrophysics:
\begin{itemize}
\item What are the conditions inside massive stars during their evolution, collapse, and explosion?

\item How does the SN explosion take place? Is the delayed neutrino-driven mechanism of SN explosion correct?

\item Is the compact remnant after the SN explosion a neutron star or a black hole?

\item Do SNe provide adequate conditions for producing various elements, especially those heavier than iron?
\end{itemize}
In the following, we elaborate on the implications of SN neutrino detection for several important astrophysical issues that are related to the above important questions. Although many discussions in this subsection are based on the JUNO detector, it should be remembered that there will be several SN neutrino detectors of scintillator, Cherenkov and liquid-argon~\cite{Adams:2013qkq} types in operation at the same time. Altogether, they will offer us valuable and complementary information.

\subsubsection{Pre-supernova neutrinos}

Before a massive star collapses and forms a SN, it experiences various ``stationary'' burning phases. In the standard stellar evolution model, one can obtain the temperature and density for a given radius at any time. However, it is difficult to test the stellar models solely by using optical observations: photons come out of the core by diffusion, and any information about the core will be lost. During the advanced burning stages, the flux of neutrinos becomes much larger than that of photons. As neutrinos interact only weakly with stellar matter, they record faithfully the inner structure of the stars. Therefore, the detection of pre-SN neutrinos is vital for the verification of stellar models.

Inside a massive star neutrinos are produced via thermal processes, out of which the dominant one is $e^+ + e^- \rightarrow \nu + \bar{\nu}$~\cite{Itoh:1996}. At any given time, a massive star can be divided into many thin layers -- each with approximately uniform temperature, density and chemical composition. One calculates neutrino production rate for each layer, and sum over all layers to derive the total neutrino flux and spectrum. In Table~\ref{table:massivestars}, neutrino fluxes and average energies at relevant burning stages are listed. The neutrino spectrum at the silicon-burning stage, in contrast with the solar neutrinos (e.g., $pp$ and $^{8}{\rm B}$ neutrinos), can be found in Ref.~\cite{Odrzywolek:2003vn,Odrzywolek:2010zz}. Generally speaking, the neutrino production rate increases significantly with temperature, and so do the average energies of neutrinos and antineutrinos. Therefore, the detection of pre-SN neutrinos becomes relatively easier for the nuclear burning at later stages.
%%%%%%%%%%%%%%%%%%%%%%%%%%%%%%%%%%%%%%%%%%%%%%%%%%%%%%%%%%%%%%%%%%%%%%%%%%%%%%
\begin{table}[!htb]
	\centering
	\begin{tabular}{cccc}
        \hline
		{\rm Burning phase} & Average energy [MeV] & Total energy [erg] & Duration
		[days]\\
		\hline
		{\rm C} & 0.71 & $7.0\times 10^{49}$ & $10^5$\\
		{\rm Ne} & 0.99 & $1.4\times 10^{50}$ & 140\\
		{\rm O} & 1.13 & $1.2\times 10^{51}$ & 180\\
		{\rm Si} & 1.85 & $5.4\times 10^{50}$ & 2\\
        \hline
	\end{tabular}
	\caption{Neutrino emission from a massive star of 20 $M_{\odot}$~\cite{Odrzywolek:2003vn,Odrzywolek:2010zz}.}
	\label{table:massivestars}
\end{table}
%%%%%%%%%%%%%%%%%%%%%%%%%%%%%%%%%%%%%%%%%%%%%%%%%%%%%%%%%%%%%%%%%%%%%%%%%%%%%%

The IBD reaction is the dominant channel to probe pre-SN neutrinos from massive stars. Fig.~\ref{fig:preSN} shows the expected event rates in JUNO for the nearest possible SN progenitor, i.e., the red supergiant Betelgeuse, whose mass is taken to be $20~M_\odot$ and distance is $0.2~{\rm kpc}$. The salient feature in the neutrino ``light curve" -- a quick rise starting at a few hours prior to core collapse -- makes the JUNO detector an ultimate pre-warning system of SN explosion. As shown in Fig.~\ref{fig:preSN}, the rate drops rapidly down to 60 events per day around 0.6 day before SN explosion. The reason is that the silicon has been burnt out in the center and the core temperature decreases, leading to a reduced neutrino production. This dip in the neutrino ``light curve" could serve as a discriminator for different progenitor star masses. The signal rate is time dependent, and there are about 400 events in the whole detector in the last day before core collapse.
%%%%%%%%%%%%%%%%%%%%%%%%%%%%%%%%%%%%%%%%%%%%%%%%%%%%%%%%%%%%%%%%%%%%%%%%%%%%%%
\begin{figure}[!t]
    \centering
    \includegraphics[width=0.7\textwidth]{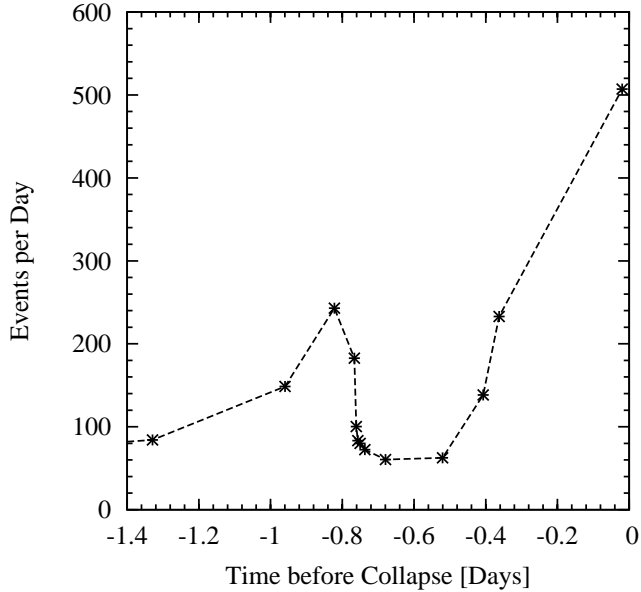}
    \caption{The neutrino event rate in JUNO for a massive star of $20~M_\odot$ at the silicon-burning stage, where the distance is assumed to be $0.2~{\rm kpc}$, the same as that of the nearest possible SN progenitor Betelgeuse.}
\label{fig:preSN}
\end{figure}
%%%%%%%%%%%%%%%%%%%%%%%%%%%%%%%%%%%%%%%%%%%%%%%%%%%%%%%%%%%%%%%%%%%%%%%%%%%%%%

Within the same energy window as for reactor neutrinos, any backgrounds of reactor neutrino experiments also act as backgrounds for detecting thermal neutrinos from massive stars. Those standard backgrounds include ${}^9$Li-${}^8$He events, fast-$n$ events, accidental coincidences, ($\alpha$, $n$) events, geo-neutrinos, and so on. Among all possible $\bar\nu_e$ sources, the reactor neutrino itself could be the main background for pre-SN neutrino detection. The event rates from reactor neutrinos and geoneutrinos are $0.25$ and $0.05$ per kiloton per day, respectively. Here the expected event rate is estimated in the energy region $1.8~{\rm MeV} < E < 3.3~{\rm MeV}$ and a 100\% detection efficiency is assumed.

However, it is worthwhile to point out that the detection of pre-SN neutrinos is only possible for a very close progenitor star. For a SN at $10~{\rm kpc}$, the event rate will be reduced by about four orders of magnitude, rendering the detection of pre-SN neutrinos in JUNO or the detector of similar size extremely difficult.

\subsubsection{Locating the supernova}

As mentioned before, the SN neutrinos arrive at the Earth several hours earlier than the optical signals, providing an early-time alert to the astronomical community. Moreover, if the optical display is hidden behind the dense gas and dust clouds of a star-forming region, the neutrino burst will be a unique tool to locate the SN~\cite{Beacom:1998fj,Tomas:2003xn}.

In general, there are two distinct methods to determine the location. First, the angular correlation between the final-state particle and the incident neutrino can be directly fixed in large Cherenkov detectors, such as Super-Kamiokande and its possible upgrade Hyper-Kamiokande. Second, two detectors separated at a long distance can locate the SN by neutrino triangulation.

According to the kinematics, the outgoing neutrons in the IBD reaction have a forward angular distribution. For a JUNO-like scintillator detector, it is possible to separately reconstruct the positions of positron and neutron. Therefore, the direction along the line connecting these two points is, at least statistically, related to the neutrino direction~\cite{Vogel:1999zy}. This method has been used by the CHOOZ experiment to determine the incoming direction of reactor antineutrinos in Ref.~\cite{Apollonio:1999jg}. It has been found that the direction of the neutrino source can be fixed within a cone of half-aperture angle $18^\circ$ at the $68\%$ confidence level, based on 2700 collected $\overline{\nu}_e$ events. For a galactic SN, if the total number of IBD events is about 5000, it is possible to measure the sky coordinates of the SN with an uncertainty of about $9^\circ$. It is expected that the JUNO detector will do better in this aspect. In the scintillator detector, it is also possible to observe the Cherenkov light from electrons, which are boosted to a high velocity in the elastic neutrino-electron scattering, so the direction of recoiled electron can help locate the SN.

On the other hand, in neutrino triangulation, the key point is to estimate the neutrino arrival time and figure out the time delay $\Delta t$ between two detectors at the distance $d$. In this case, the angle $\theta$ between the SN direction and the line connecting these two detectors is given by $\cos \theta = \Delta t/d$, and the uncertainty is dominated by the error in the measurement of $\Delta t$. Taking a detector similar to JUNO with $5000$ events and a distance of $30~{\rm ms}$, we obtain an uncertainty of about $\delta(\cos \theta) \sim 0.21$, following the strategy in Ref.~\cite{Beacom:1998fj}. Evidently, the precision for neutrino triangulation is not comparable to the statistical method.

\subsubsection{Coincidence with gravitational waves}

Besides neutrinos, the gravitational waves are good messengers to probe the interior of a SN. A gravitational-wave signal provides useful information on non-radial deformation and non-spherical hydrodynamic motions, whereas a high-statistics neutrino signal allows us to follow directly the different stages of core collapse without additional assumptions. As predicted by all theoretical SN models, the prompt $\nu_e$ burst is a robust and uniform landmark structure. Since the energy threshold of JUNO can be as low as 0.2 MeV, it offers a unique possibility of identifying this feature via elastic electron and proton scattering. One could use the prompt $\nu_e$ burst in JUNO for coincidence measurements with the gravitational wave burst that may arise at core bounce. Using the
prompt $\nu_e$ burst could provide an even sharper coincidence than
can be achieved with the onset of the $\bar\nu_e$ signal in
Super-Kamiokande and IceCube.

After core collapse, the luminosities of $\nu_e$ and $\bar\nu_e$ in the accretion phase depends on the mass infall rate and thus on the progenitor-dependent structure of the stellar core, with more massive cores producing higher luminosities. During this phase, the luminosity variations are accompanied by sizable gravitational-wave emission at several hundred~Hz, the observation of neutrino and gravitational-wave signals would confirm the presence of violent hydrodynamic instabilities stirring the accretion flow around the assembling neutron star~\cite{Mueller:2003fs}. Such activity and a several hundred millisecond delay of the onset of the explosion are expected within the framework of the delayed neutrino-driven
mechanism. A pronounced drop of the $\nu_e$ and $\bar\nu_e$
luminosities, followed by a close similarity to those of
heavy-lepton neutrinos, would finally signal the end of the
accretion phase and the launch of the outgoing SN blast wave. The
cooling signature of a nascent neutron star is characterized by a
monotonic and gradual decline of the neutrino emission. It would
be prolonged if additional energy was released by phase transitions
in the nuclear matter. Exotic scenarios might predict a secondary $\nu_e$ burst, such as a QCD phase transition, or an abrupt end of neutrino emission if the collapse to a black hole occurred.

\subsubsection{SN nucleosynthesis}

Another advantage of the JUNO detector is its superior energy
resolution, which could help to disentangle source-imposed spectral
features from those caused by neutrino-flavor conversions. Moreover,
detecting significant numbers not only of $\bar\nu_e$ but also of
$\nu_e$ and heavy-lepton neutrinos would yield at least time-averaged spectral information for different emission channels. Conceivably one could extract information on the neutron-to-proton ratio in the neutrino-processed SN outflows, presently also a sensitive result of numerical modeling of a
multitude of complex processes. The relative abundance of neutrons
and protons determines the conditions for nucleosynthesis and are
set by competing $\nu_e$ and $\bar\nu_e$ captures, which in turn
depend delicately on the relative fluxes and spectral distributions
of these neutrinos~\cite{Qian:1993dg}. A JUNO measurement of a SN burst may offer the only direct empirical test of the possibility for $r$-processing in the SN core, except for an extremely challenging in-situ measurement
of $r$-process nuclei in fresh SN ejecta~\cite{Qian:2003wd}.

In addition to the $r$-process for the production of heavy chemical elements, the neutrinos emitted from the core collapse and the subsequent cooling of the proto-neutron star will interact with outer layers of the SN, producing some rare nuclei such as $^{11}{\rm B}$ and $^7{\rm Li}$. This neutrino process is crucially important for the $^{11}{\rm B}$ and $^7{\rm Li}$ production~\cite{Woosley:1989bd,Heger:2003mm,Yoshida:2005uy}. In particular, the SN neutrinos of heavy-lepton flavors, which may have higher average energies, will dominate the generation of galactic light chemical elements. Therefore, the measurement of $\nu_x$ energy spectra at JUNO is necessary to directly pin down whether the neutrino process is really the true mechanism for producing the present galactic inventory of light nuclei.

One should notice that the flavor oscillations of SN neutrinos, which could lead to modification of neutrino energy spectra and fluxes, will affect both $r$- and neutrino processes~\cite{Wu:2014kaa}.

\subsection{Implications for particle physics}

Neutrino signals of a core-collapse SN have also profound implications for elementary particle physics. If the standard scenario of delayed neutrino-driven explosion is confirmed by future observations of SN  neutrinos and the cooling phase of neutrino emission is firmly established, the energy-loss argument will result in robust and restrictive constraints on a large number of particle physics models where new weakly-interacting particles are introduced, such as axions, majorons, and sterile neutrinos~\cite{Raffelt:1996bk}. Numerous results derived from the sparse SN~1987A data can be refined.

If neutrinos are Majorana particles, spin-flavor conversions caused by the combined action of magnetic fields and matter effects can transform some of the prompt $\nu_e$ burst to $\bar\nu_e$, leading to a huge inverse-beta signal. Such an observation would provide smoking-gun evidence for neutrino transition magnetic moments. Non-radiative decays would also produce a $\nu_e\to \bar\nu_e$ conversion during the prompt burst. On the other hand, null results of observing those conversions will set restrictive bounds on the neutrino transition magnetic moments~\cite{Raffelt:1999gv}. In addition, the existence of eV-mass sterile neutrinos may have great impact on the supernova explosion and nucleosynthesis~\cite{Tamborra:2011is,Wu:2013gxa}

The core-collapse SNe serve as laboratories to probe intrinsic properties of neutrinos themselves. The SN neutrino observations will shed light on important fundamental problems in particle physics:
\begin{itemize}
\item What is the absolute scale of neutrino masses?

\item What is the neutrino mass ordering?

\item Are there collective neutrino oscillations?

\item Are there exotic neutrino interactions?
\end{itemize}
In the following, we elaborate on a few important topics that are associated with these fundamental questions, and study the role that JUNO will play in this connection.

\subsubsection{Bound on neutrino masses}
Since neutrinos are massive, their flight time from a SN core to the detector at the Earth will be delayed, compared to massless particles~\cite{Zatsepin:1968kt}. For SN neutrinos, the time delay can be written as
\begin{equation}
\Delta t(m^{}_\nu, E^{}_\nu) = 5.14~{\rm ms}~\left( \frac{m^{}_\nu}{\rm eV} \right)^2 \left( \frac{10~{\rm MeV}}{E^{}_\nu} \right)^2 \frac{D}{10~{\rm kpc}} \; ,
\label{eq: Delay}
\end{equation}
where $E^{}_\nu$ is the neutrino energy, and $D$ is the distance between SN and detector. Thus a time-delay at the millisecond level is expected for neutrinos from a typical galactic SN. However, the average energy and luminosity of neutrinos evolve in time, which complicates the extraction of neutrino mass information. In order to derive a mass bound, one has to know the time evolution of neutrino energies and fluxes, and take into account neutrino flavor conversions.

The sharply rising and falling luminosity of prompt $\nu_e$ burst can be implemented to probe the time-delay effects of massive neutrinos, since the time structure is very unique and the characteristic timescale is around tens of milliseconds. However, the total number of neutrino-electron scattering events in JUNO is limited and it is impossible to reconstruct neutrino energy, which may reduce the sensitivity to neutrino masses. An abrupt termination of neutrino signals after the black-hole formation should be a perfect scenario to probe absolute neutrino masses~\cite{Beacom:2000ng}.

As the IBD is the dominant channel for SN neutrino detection, we concentrate on the $\overline{\nu}^{}_e$ flux $\Phi^0_{\overline{\nu}^{}_e}$ in the accretion and cooling phases. In Ref.~\cite{Pagliaroli:2008ur}, a simple parametrization of $\Phi^0_{\overline{\nu}^{}_e}$ has been proposed to capture essential physics of neutrino production and the main features of numerical simulations. In the cooling phase, the $\overline{\nu}^{}_e$ flux $\Phi^0_{\rm c}$ is parametrized by three model parameters: the initial temperature $T^{}_{\rm c}$, the radius of neutrino sphere $R^{}_{\rm c}$, and the cooling time scale $\tau^{}_{\rm c}$. More explicitly, we have
\begin{equation}
\Phi^0_{\rm c} (t, E^{}_\nu) = \frac{1}{4\pi D^2} \frac{1}{8\pi^2} \left\{ 4\pi R^2_{\rm c} \frac{E^2_\nu}{1 + \exp\left[E^{}_\nu/T^{}_{\rm c}(t)\right]}\right\} \; ,
\label{eq: PhiC}
\end{equation}
where the temperature evolves as $T^{}_{\rm c}(t) = T^{}_{\rm c} \exp[-t/(4\tau^{}_{\rm c})]$. In the accretion phase, one has to model the time evolution of neutron number and positron temperature in order to figure out the $\overline{\nu}^{}_e$ flux $\Phi^0_{\rm a}$. This can be done by introducing the accretion time scale $\tau^{}_{\rm a}$, and requiring the resultant neutrino energy and luminosity to follow numerical simulations. In addition, the initial number of neutrons depends on an initial accreting mass $M^{}_{\rm a}$, and a thermal energy spectrum of positrons with an initial temperature $T^{}_{\rm a}$ is reasonable. The flux of $\overline{\nu}_e$ is determined by the interaction between neutrons and positrons, so it can be written as
\begin{equation}
\Phi^0_{\rm a} (t, E^{}_\nu) = \frac{1}{4\pi D^2} \frac{1}{\pi^2} \left\{ N^{}_n(t) \sigma^{}_{n e^+}(E^{}_\nu) \frac{\overline{E}^2_{e^+}}{1 + \exp\left[\overline{E}^{}_{e^+}/T^{}_{\rm a}(t)\right]}\right\} \; ,
\label{eq: PhiA}
\end{equation}
where the average positron energy is evaluated as $\overline{E}^{}_{e^+} = (E^{}_\nu - 1.293~{\rm MeV})/(1 - E^{}_\nu/m^{}_n)$. For the neutrino energies of our interest (i.e., from 5 to 40 MeV), the cross section is approximately given by $\sigma^{}_{n e^+}(E^{}_\nu) \approx 4.8\times 10^{-44}~{\rm cm}^2/[1 + E^{}_\nu/(260~{\rm MeV})]$. The time evolution of neutron number $N^{}_n(t) = 0.6 M^{}_{\rm a} j^{}_k(t) T^6_{\rm a}/\{[1+t/(0.5~{\rm s})]T^6_{\rm a}(t)\}$ and the temperature function $T^{}_{\rm a}(t) = T^{}_{\rm a} + (0.6~T^{}_{\rm c} - T^{}_{\rm a}) (t/\tau^{}_{\rm a})^2$ are properly chosen to ensure a continuous neutrino average energy and to capture the numerical features~\cite{Pagliaroli:2008ur}. Putting all together, the total flux is~\cite{Pagliaroli:2008ur,Pagliaroli:2009qy}
\begin{equation}
\Phi^0_{\overline{\nu}^{}_e}(t, E^{}_\nu) = f^{}_{\rm r}(t) \Phi^0_{\rm a}(t, E^{}_\nu) + \left[1 - j^{}_k(t)\right] \Phi^0_{\rm c}(t, E^{}_\nu) \; ,
\label{eq: PhiT}
\end{equation}
where $f^{}_{\rm r}(t) = 1 - \exp(-t/\tau^{}_{\rm r})$ with the rising time scale $\tau^{}_{\rm r}$ further introduces fine early time structure, and $j^{}_k(t) = \exp[- (t/\tau^{}_{\rm a})^k]$ with a default value of $k = 2$ is the time function interpolating the accretion and cooling phases of neutrino emission.

However, neutrino flavor conversions take place when they propagate from the SN core to the detector. In particular, the Mikheyev-Smirnov-Wolfenstein (MSW) matter effects in the SN mantle will significantly change neutrino flavor content. Given a relatively large $\theta_{13}$, the electron antineutrino flux at the detector is $\Phi_{\bar{\nu}_e} = \cos^2 \theta_{12} \Phi_{\bar{\nu}_e}^0 + \sin^2 \theta_{12} \Phi_{\bar{\nu}_x}^0$ for normal neutrino mass ordering, while $\Phi_{\bar{\nu}_e} = \Phi_{\bar{\nu}_x}^0$ for inverted neutrino mass ordering.
%%%%%%%%%%%%%%%%%%%%%%%%%%%%%%%%% Fig. 4.9 %%%%%%%%%%%%%%%%%%%%%%%%%%%%%%%%%
\begin{figure}[!t]
    \centering
    \includegraphics[width=0.7\textwidth]{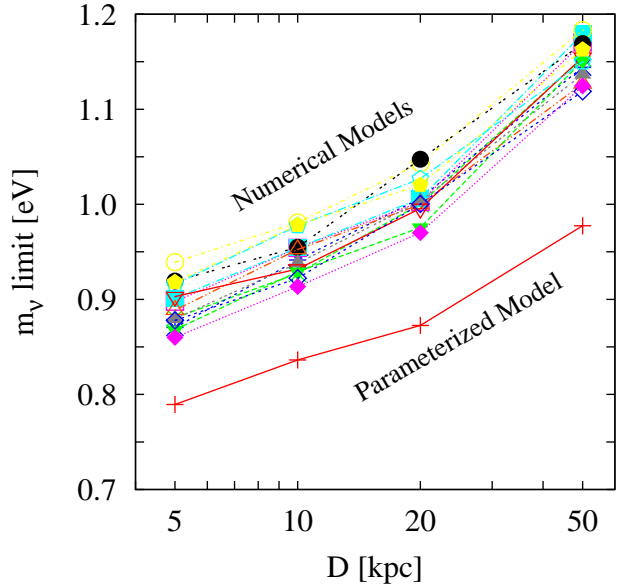}
    \caption{The upper bounds on the absolute scale of neutrino masses at the $95\%$ CL for a SN at a distance of $D = 5~{\rm kpc}$, $10~{\rm kpc}$, $20~{\rm kpc}$ and $50~{\rm kpc}$ in the parameterized model from Ref.~\cite{Pagliaroli:2008ur}, and a series of numerical models from Ref.~\cite{Nakazato:2012qf}, in which simulations have been performed for a progenitor-star mass $M = 13$, $20$, $30$ or $50$ solar masses, a metallicity $Z = 0.02$ or $0.004$, and a shock revival time $t^{}_{\rm revive} = 100~{\rm ms}$ or $300~{\rm ms}$. In our calculations, we have chosen fourteen numerical models, since a black hole is formed $842~{\rm ms}$ after bounce in two models with $M = 30$ solar masses and $Z = 0.02$. See Ref.~\cite{Nakazato:2012qf} for more details about the numerical models, and the SN neutrino data are publicly available at the website http://asphwww.ph.noda.tus.ac.jp/snn/. This figure is taken from Ref.~\cite{Lu:2014zma}}
\label{fig:models}
\end{figure}
%%%%%%%%%%%%%%%%%%%%%%%%%%%%%%%%%%%%%%%%%%%%%%%%%%%%%%%%%%%%%%%%%%%%%%%%%%%%%%

It is now straightforward to figure out the neutrino event rate $R(t, E)$ as a function of emission time and neutrino energy, which are related to the detection time and positron energy, given a definite neutrino mass. The strategy is to generate neutrino events, denoted by a detection time $t_i$ and a positron energy $E_i$, according to the event rate $R(t, E)$. The simulated data are fitted by the event rate with neutrino mass scale $m_\nu$ as an additional parameter. The maximum likelihood approach is implemented, and the likelihood function has been constructed by taking account of every single event~\cite{Pagliaroli:2010ik}. Assuming a nearly-degenerate neutrino mass spectrum and the normal mass ordering, we obtain a neutrino mass upper bound $m_\nu < (0.83 \pm 0.24)~{\rm eV}$ at $95\%$ confidence level, where $1\sigma$ Gaussian error has been attached to the best-fit value~\cite{Lu:2014zma}. In order to illustrate the impact of the SN distance on the mass limit, we have calculated the mass bound for different distances $D = 5$ kpc, 10 kpc, 20 kpc and 50 kpc, where the last one corresponds to the case of SN 1987A. It is generally expected that a closer SN offers a larger number of neutrino events, implying a better upper limit on neutrino masses. Furthermore, to take account of SN model uncertainties, we show in Fig.~\ref{fig:models} the results for the previous parametrized model of SN neutrino fluxes and a dozen of numerical SN models. In the latter case, the upper bound could be worsened by $0.15~{\rm eV}$ or so, which should be regarded as a kind of systematic uncertainty.

\subsubsection{Impact of mass ordering}

Flavor conversions of SN neutrinos are very interesting. Neutrinos
propagating through the SN mantle and envelope encounter a large
range of matter densities, allowing for MSW conversions driven first by the neutrino mass-squared difference $\Delta m^2_{31}$ and the mixing angle $\theta_{13}$, and then by $\Delta m^2_{21}$ and $\theta_{12}$. A SN neutrino signal is sensitive to the unknown neutrino mixing parameter: the ordering of neutrino masses that could be in the normal (NH) or inverted hierarchy~(IH). In the former case, the $\overline{\nu}_e$ flux outside the SN envelope will be a superposition of both initial $\overline{\nu}_e$ and $\nu_x$ fluxes, and the ratio between these two components is determined by the mixing angle $\theta^{}_{12}$. In the latter case, the final $\overline{\nu}_e$ flux at the Earth will be entirely the initial $\nu_x$ flux. Note in both cases, the final fluxes needed to be scaled by the square of SN distance.

Although the difference caused by flavor conversions can be seen from the third row of Fig.~\ref{fig:sn:SNburst}, where the IBD event rate is calculated and the complete flavor conversion has been assumed, it remains to find a model-independent way to extract the information of mass ordering directly from the experimental observations. As another example, in the fourth row of Fig.~\ref{fig:sn:SNburst}, we can observe that the neutrino mass ordering significantly affects the event rate of neutrino-electron scattering. One possible way to determine the neutrino mass ordering may be to compare the event rates in the IBD channel with that of elastic neutrino-proton scattering, since the latter is not changed by flavor conversions.

\subsubsection{Collective neutrino oscillations}
%%%%%%%%%%%%%%%%%%%%%%%%%%%%%%%%% Fig. 4.9 %%%%%%%%%%%%%%%%%%%%%%%%%%%%%%%%%
\begin{figure}[!t]
    \centering
    \includegraphics[width=0.9\textwidth]{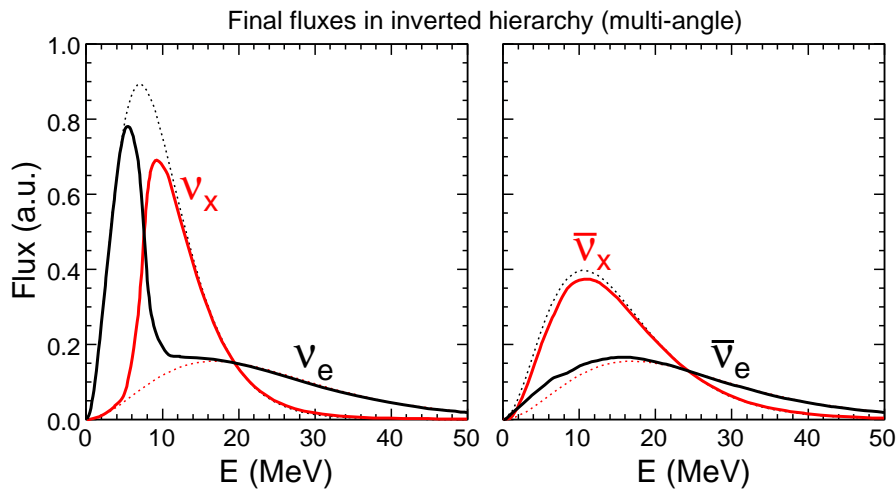}
    \vspace{-0.2cm}
    \caption{Numerical illustration for the splits in neutrino energy spectra after collective oscillations in the case of inverted mass hierarchy~\cite{Fogli:2007bk}. The initial spectra are indicated by dotted curves in the same color as the final ones, where one can observe a complete swap of the antineutrino spectra and a sharp split in the neutrino spectra.}
\label{fig:collective}
\end{figure}
%%%%%%%%%%%%%%%%%%%%%%%%%%%%%%%%%%%%%%%%%%%%%%%%%%%%%%%%%%%%%%%%%%%%%%%%%%%%%%

The picture of SN neutrino oscillations has recently been changed drastically by the insight that the neutrino-neutrino
refractive effect is crucial. These self-induced collective
flavor conversions occur within a few hundred km above the neutrino
sphere; see Ref.~\cite{Duan:2010bg} for a review of the recent
torrent of literature on this topic. The most important
observational consequence is a swap of the $\nu_e$ and $\bar\nu_e$
spectrum with that of $\nu_x$ and $\bar\nu_x$ in certain energy
intervals. The sharp spectral features at the edges of these swap intervals are known as ``spectral splits''. Their development depends on the neutrino mass hierarchy as well as on the ordering of the flavor fluxes at the source. Therefore, the split features can depend on time in interesting
ways. For instance, in Ref.~\cite{Fogli:2007bk}, it has been shown in the case of inverted neutrino mass hierarchy that SN neutrinos in the accretion phase will undergo collective oscillations, leading to a complete swap between $\overline{\nu}_e$ and $\overline{\nu}_x$ spectra and a sharp split around $7~{\rm MeV}$ in the neutrino spectra, as in Fig.~\ref{fig:collective}. The average neutrino energies $\langle E^{}_{\nu_e} \rangle = 10~{\rm MeV}$, $\langle E^{}_{\overline{\nu}_e} \rangle = 15~{\rm MeV}$ and $\langle E^{}_{\nu_x} \rangle = 24~{\rm MeV}$, together with a total neutrino energy of $3\times 10^{53}~{\rm erg}$ equally distributed in six neutrino species, are assumed. In the case of normal hierarchy, no significant flavor conversions are observed~\cite{Duan:2010bg,Fogli:2007bk}.

The main problem to detect oscillation features is that one can not
rely on detailed theoretical predictions of the flavor-dependent
fluxes and spectra. Therefore, model-independent signatures are
crucial. One case in point is the energy-dependent modulation of the
neutrino survival probability caused by Earth matter effects that
occur if SN neutrinos arrive at the detector ``from below''. The appearance  of Earth effects depends on the flux and mixing scenario. Therefore, its detection could give hints about the primary SN neutrino fluxes, as well as on the neutrino mass hierarchy~\cite{Dighe:2003be,Dighe:2003jg}.

The excellent energy resolution of JUNO leads to a particular advantage for
discovering small energy-dependent flux modulations caused by Earth
effects, but of course depends on seeing the SN shadowed by the Earth.
Additional signatures of flavor conversions can be imprinted by matter effects of the shock fronts in the SN envelope. The number of events, average energy, or the width of the spectrum may display dips or peaks for
short time intervals. Such signatures yield valuable information about shock-wave propagation and the neutrino mass hierarchy~\cite{Tomas:2004gr}. However, realistic chances to detect shock features remain unclear. The flavor-dependent spectral differences in the antineutrino channel are probably small during the cooling phase. Moreover, strong turbulence in the post-shock regions could affect these signatures~\cite{Borriello:2013tha}.

\subsubsection{Constraining new physics}

It is well known that the stars can be ideal places to constrain new physics scenarios beyond the standard model of elementary particles~\cite{Raffelt:1996bk}. If the standard picture of SN explosion is verified by a high-statistics observation, the duration of neutrino signal in the long cooling phase will be precisely measured. According to the standard energy-loss argument, any weakly-interacting exotic particles that can be copiously produced in the SN core will carry away a large amount of energies, significantly reducing the neutrino signal. To avoid excessive energy loss, the new particles should be interacting with matter and neutrinos so strongly that they are captured in the SN core, or so weakly that the production of new particles is inefficient. Therefore, by requiring the agreement between observations and theoretical prediction, we can draw restrictive constraints on the interaction strengths of new elementary particles. This approach has been applied to numerous new physics scenarios, such as axions, sterile neutrinos, majorons, and hidden photons.

As the neutrino energy spectra and flavors can be determined in future observations of SN neutrinos, new possibilities other than the energy-loss argument will be opened to set bounds on new physics.

\subsection{Summary}

A high-statistics detection of neutrinos from a galactic SN will provide us with precious information about the explosion mechanism and intrinsic properties of neutrinos themselves. For a galactic SN at a distance of 10 kpc, there are around 5000 events in the IBD channel, 2000 events for elastic neutrino-proton scattering, and 300 events for elastic neutrino-electron scattering in the JUNO detector. The time evolution, energy spectra and flavor contents of SN neutrinos can in principle be established and used to verify or disprove the neutrino-driven explosion mechanism. With experimental observations, many interesting questions in astronomy, astrophysics and particle physics, such as the early warning of SNe, the SN location, SN nucleosynthesis, absolute neutrino masses and neutrino mass ordering, can hopefully be revised or addressed.

\clearpage

\section{Diffuse Supernova Neutrino Background (DSNB)}
\label{sec:snd}

\blfootnote{Editors: Georg Raffelt (raffelt@mpp.mpg.de) and Michael Wurm (michael.wurm@uni-mainz.de)}
\blfootnote{Major contributor: Randolph M\"ollenberg, Zhimin Wang}

The integrated neutrino flux from all past core-collapse events in the visible universe forms the diffuse supernova neutrino background (DSNB), holding information on the cosmic star-formation rate, the average core-collapse neutrino spectrum, and the rate of failed SNe. The Super-Kamiokande water Cherenkov detector has provided first limits and eventually may achieve a measurement at the rate of a few events per year, depending on the implementation of its gadolinium upgrade. A JUNO-class detector has the potential to achieve a comparable measurement, benefitting from the excellent intrinsic capabilities of liquid scintillator detectors for antineutrino tagging and background rejection. The most critical background is created by neutral-current interactions of atmospheric neutrinos. Depending on the performance of pulse-shape discrimination techniques, this background may prove too high for a clear detection. However, a positive signal at the $3\sigma$ level is conceivable for a 10-year measurement and typical DSNB parameters. A non-detection would strongly improve current limits and exclude a significant range of DSNB parameter space.

\subsection{Motivation and Opportunities}
\label{subsec:snd:BasicPicture}

While core-collapse SNe are rare, and in a galaxy like our own Milky Way occur only once every few decades, the energy release is so large that the integrated neutrino flux from all past SNe, the diffuse SN neutrino background (DSNB), adds up to a large cosmic radiation density~\cite{Krauss:1983zn, Bisnovatyi-Kogan:1984, Woosley:1986, Ando:2004hc, Lunardini:2005jf, Beacom:2010kk, Lunardini:2012ne}. It is comparable to the extra-galactic background light, the photons emitted from all stars, and corresponds to around 10\% of the energy density of the cosmic microwave background. Note that the big-bang relic neutrinos are today non-relativistic, contributing a small hot dark matter component, so the DSNB is actually the largest cosmic neutrino {\em radiation\/} background, unless the lightest mass eigenstate of the big-bang neutrinos remains relativistic after all. It has been recognized for a long time that for $E_\nu\gtrsim 10$~MeV, above reactor neutrino energies, the DSNB is the dominant local anti-neutrino flux. The primary detection channel is inverse-beta decay (IBD), $\bar\nu_e+p\to n+e^+$, where the solar $\nu_e$ flux does not contribute even though it reaches up to about 15~MeV and dominates in this low-energy range.\footnote{In principle, $\nu_e\to\bar\nu_e$ transitions in the Sun caused by spin-flavor transitions of Majorana neutrinos could produce a solar background signal in this low-energy range \cite{Raffelt:2009mm}.} The DSNB flux component in $\bar\nu_e$ is around $20~{\rm cm}^{-2}~{\rm s}^{-1}$, details depending on the cosmic SN rate as a function of redshift and the average core-collapse $\bar\nu_e$ flux spectrum, including the effect of flavor conversion.

The DSNB can be detected if backgrounds, notably caused by atmospheric neutrinos, can be controlled. In this regard, a scintillator detector has an inherent advantage over the water-Cherenkov technique because neutron tagging of IBD events is a generic feature. The most restrictive current limits on the DSNB derive from Super-Kamiokande I--III without neutron tagging~\cite{Malek:2002ns, Bays:2011si}, already excluding the most extreme models for the DSNB flux spectrum. With a completely revised data-acquisition system, Super-Kamiokande~IV now has limited sensitivity to the 2.2\,MeV $\gamma$-ray caused by neutron capture on protons. A new DSNB limit based on this technique has recently become available~\cite{Zhang:2013tua}, but is not yet competitive with the earlier Super-Kamiokande I--III results. Limits by the much smaller SNO~\cite{Aharmim:2006wq} and KamLAND~\cite{Gando:2011jza} detectors are also not competitive. In future, gadolinium as an efficient neutron absorber may be dissolved in Super-Kamiokande \cite{Beacom:2003nk}, a technique currently studied at the EGADS facility~\cite{Watanabe:2008ru, Vagins:2012vta}, probably allowing them to detect the DSNB at a rate of perhaps a few events per year~\cite{Horiuchi:2008jz}. Super-Kamiokande has around $1.5\times10^{33}$ protons in its 22.5\,kt of fiducial volume for DSNB detection, whereas JUNO has around $1.2\times 10^{33}$ protons in the
fiducial volume of 17\,kt to be used for the DSNB analysis (see below). Therefore, JUNO will be comparable and complementary to the Super-Kamiokande effort, and it will be unique if the gadolinium upgrade would not be implemented after all.

Following Ref.~\cite{Beacom:2010kk}, the expected DSNB flux depends on two main ingredients, the core-collapse rate as a function of cosmic redshift and the number and spectrum of neutrinos and anti-neutrinos per flavor emitted by an average core-collapse event, including flavor conversion effects. These quantities bear significant uncertainties. The core-collapse SN rate rises from its local value by about an order of magnitude up to redshift $z\sim 1$, then levels off and eventually decreases \cite{Ando:2004hc, Lunardini:2005jf, Porciani:2000ag, Hopkins:2006bw, Yuksel:2008cu, Horiuchi:2011zz}. If the $\bar\nu_e$ detection threshold is above some 10~MeV, the overall detection rate is dominated by SNe with $z\lesssim 1$. However, the SN rate predicted from the observed star-formation rate appears to be systematically a factor of 2 smaller than the directly observed SN rate~\cite{Horiuchi:2011zz}. Additional contributions likely arise from sub-luminous SNe and from failed SNe, i.e. those core-collapse events leading to a black hole rather than an actual explosion \cite{Heger:2002by, Fischer:2008rh, Nakazato:2008vj, Kochanek:2008mp, Lunardini:2009ya, O'Connor:2010tk, Ugliano:2012kq, Keehn:2010pn, Kochanek:2013yca}. Both outcomes lead to a neutrino flux comparable to that of a standard SN.

The total liberated energy in a core collapse depends on the final neutron-star mass and the nuclear equation of state, introducing a significant uncertainty even if we ignore the unknown contribution by failed SNe. The predicted flux spectra of the different flavors depend on details of the neutrino interaction rates and the numerical implementation of neutrino transport. Flavor conversion during neutrino propagation in the SN environment partly swaps the spectra so that the detectable $\bar\nu_e$ flux is some combination of the original $\bar\nu_e$, $\bar\nu_\mu$ and $\bar\nu_\tau$ fluxes emitted by the SN core. One expects large variability for most of these effects between different core-collapse events because of the large range of progenitor masses and other properties.

In principle, the DSNB is a treasure trove of valuable information on the astrophysics of core collapse, failed SNe and black-hole formation, and flavor-dependent neutrino propagation and flavor conversion. However, a detailed spectral DSNB measurement is clearly out of reach with near-term detectors of the JUNO and Super-Kamiokande class. Neutrino astronomy is a field in its infancy where only solar neutrinos have been observed with robust statistical detail. At this stage, the very detection of the DSNB is the first milestone and will be a fundamental discovery in its own right.

\subsection{Parametric DSNB flux spectrum}
\label{subsec:snd:fluxspectrum}

In this situation, a detailed parameter study of the various DSNB ingredients is not warranted for a JUNO sensitivity forecast. Rather, we follow the practice adopted in the LENA physics study \cite{Wurm:2011zn} and adopt a simple parametric representation of the DSNB. Its isotropic flux spectrum is given by the line-of-sight integral over cosmic redshift~$z$
\begin{equation}
\frac{dF_{\bar\nu_e}}{dE_{\bar\nu_e}}=\frac{c}{H_0}\int_0^{z_{\rm max}}dz\,
\frac{R_{\rm SN}(z)}{\sqrt{\Omega_{\rm m}\left(1+z\right)^{3}+\Omega_{\Lambda}}}\,
\frac{dN_{\bar\nu_e}(E'_{\bar\nu_e})}{dE'_{\bar\nu_e}}\,.
\label{eq:dsnbflux}
\end{equation}
The maximum redshift of star-formation $z_{\rm max}$ is taken to be 5, but the exact value is irrelevant because $z\lesssim1$ dominates for the detectable part of the spectrum. $H_0=70~{\rm km}~{\rm s}^{-1}~{\rm Mpc}^{-1}$ is the present-day Hubble parameter and $c$ the speed of light. $R_{\rm SN}(z)$ is the redshift-dependent co-moving SN rate, i.e., the number of SNe per year and per cubic-Mpc. The cosmological model is taken to be flat with a matter-density parameter $\Omega_{\rm m}=0.30$ and cosmological-constant density parameter $\Omega_{\Lambda}=0.70$. The cosmological parameters are rather well established, but our results do not directly depend upon them because they also enter the determination of $R_{\rm SN}(z)$. Finally, $dN_{\bar\nu_e}(E_{\bar\nu_e})/dE_{\bar\nu_e}$ is the total number spectrum of $\bar\nu_e$ from an average core collapse, with flavor-conversion effects already included. Note that $E'_{\bar\nu_e}=(1+z)E_{\bar\nu_e}$ is the neutrino energy in the rest frame of a SN at redshift~$z$.

For the cosmic SN rate as a function of redshift we specifically adopt the functional form proposed by Porciani and Madau~\cite{Porciani:2000ag} in the form of equations~(4) and (5) of reference~\cite{Ando:2004hc} with the choice $f^*=1.5$. The SN rate in the local universe is thus taken to be $R_{\rm SN}(0)=1.25\times10^{-4}~{\rm yr}^{-1}~{\rm Mpc}^{-3}$ in agreement with reference~\cite{Horiuchi:2008jz}. A different parametrization provided in reference~\cite{Yuksel:2008cu} leads to similar DSNB results because the main contribution arises from $z\lesssim1$ so that differences of $R_{\rm SN}(z)$ at larger redshifts have little impact on the overall result. For each individual core collapse we assume that, after flavor conversion effects have been included, a total energy $E_{\rm tot}=0.5\times10^{53}$~erg is emitted in $\bar\nu_e$ with average energy $\langle E_{\bar\nu_e}\rangle$. While the instantaneous emission spectra tend to be ``pinched'', i.e., less broad than a thermal distribution, the time-averaged spectra may be close to thermal. So we assume a Maxwell-Boltzmann distribution, normalized to t$E_{\rm tot}$, of the form~\cite{Keil:2002in,Tamborra:2012ac}
\begin{equation}
\frac{dN_{\bar\nu_e}}{dE_{\bar\nu_e}}=\frac{0.5\times10^{53}~{\rm erg}}{\langle E_{\bar\nu_e}\rangle}
\,\frac{27}{2}\,\frac{E_{\bar\nu_e}^2}{\langle E_{\bar\nu_e}\rangle^3}
\,e^{-3E_{\bar\nu_e}/\langle E_{\bar\nu_e}\rangle}\,.
\end{equation}
Overall, this corresponds effectively to a two-parameter representation of the DSNB flux spectrum: one parameter is $\langle E_{\bar\nu_e}\rangle$, the other a global flux normalization $\Phi$ which depends on the product of the SN rate at $z=0$ and the average energy per SN emerging in the form of~$\bar\nu_e$. As a reference for the flux normalization, we define a default value $\Phi_0=E_{\rm tot}\cdot R_{\rm SN}(0)$ that corresponds to the most likely SN rate and total energy as described above.

In the following, we will explicitly consider $\langle E_{\bar\nu_e}\rangle=12$, 15, 18 and 21~MeV, where the upper end is certainly beyond what is nowadays expected for core-collapse SNe, but is meant to represent an extreme case. Figure~\ref{fig:snd:spectra} displays the corresponding DSNB antineutrino energy spectra for JUNO. Depending on $\langle E_{\bar\nu_e}\rangle$, the position of the spectral peak varies around 10\,MeV. The corresponding event rate is 0.2--0.5 per (kt$\cdot$yr) .

\begin{figure}[ht]
\centering
\includegraphics[width=0.49\textwidth]{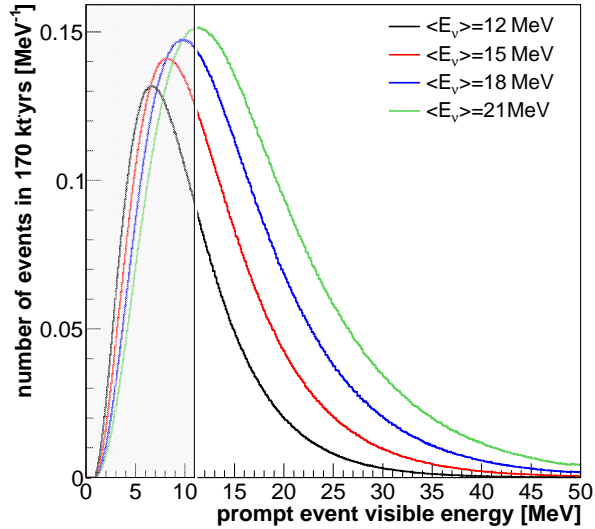}
\caption{Visible energy spectra of prompt DSNB events for $\langle E_{\bar\nu_e}\rangle=12$, 15, 18 and 21~MeV for a fiducial volume of 17\,kt (cf.~section~\ref{subsec:snd:sigbg}). Observation below 11\,MeV is obstructed by the overwhelming background from reactor antineutrinos and cosmogenic background.}
\label{fig:snd:spectra}
\end{figure}

\subsection{Signal and background sources in JUNO}
\label{subsec:snd:sigbg}

The following section lines out the expected signal and background sources for the DSNB search in JUNO and proposes a viable analysis strategy based on event selection cuts and event identification to enable a positive detection of the DSNB in spite of the significant backgrounds. The projected sensitivity is laid out in section~\ref{subsec:snd:sensitivity}.

\bigskip

\noindent {\bf DSNB antineutrino signal.} While the DSNB consists of approximately equal fluxes of neutrinos and antineutrinos of all flavors, the primary signal in liquid-scintillator detectors is the IBD reaction $\bar\nu_e+p\to e^+ + n$. The energy of the interacting antineutrino can be inferred by the signal of prompt positron that is shifted by $\sim0.8$\,MeV to lower energies. The expected event rates and prompt event spectra have been derived by the convolution of the DSNB spectrum described in section \ref{subsec:snd:fluxspectrum} and the parametrized cross-section proposed in \cite{Strumia:2003zx}. From this, the expected signal can be derived as a function of the mean energy of the SN spectrum $\langle E_{\bar\nu_e}\rangle$, the flux normalization $\Phi$, the number of protons contained in the fiducial mass $m_{\rm fid}$ and the detection efficiency $\varepsilon_\nu$. For $m_{\rm fid}=17\,$kt, $\Phi=\Phi_0$ and
$\langle E_{\bar\nu_e}\rangle \in [12;21]$\,MeV, we expect an event rate of 1.5 to 2.9 events per year. The DSNB signal spectra of figure~\ref{fig:snd:spectra} have been obtained by a toy MC assuming a 3\,\% energy resolution at 1\,MeV.

The DSNB provides a clear coincidence signature of prompt positron and delayed neutron capture on hydrogen ($E_\gamma = 2.2$\,MeV, $\tau_{np} \sim 220\,\mu$s) which allows for an efficient suppression of single-event background. However, there are several sources of correlated background events that must be taken into account when defining the energy window, fiducial volume and pulse shape selection criteria for DSNB detection (see below). The corresponding visible energy spectra are mostly derived from a modified version of the LENA detector simulation~\cite{Mollenberg:2014pwa} but provide a good approximation of the expected signal in JUNO. On the other hand, the background created by fast neutrons is much more dependent on the detector geometry and has therefore been taken from a dedicated simulation of the JUNO setup \cite{jileixu15}.

\bigskip

\noindent {\bf Reactor antineutrino background.} $\bar\nu_e$'s from neutron-rich fission products provide a large IBD signal in the energy region below 10\,MeV that effectively impedes a detection of the DSNB in this energy range. Due to the relative proximity to the nuclear power stations at Taishan and Yanjiang, the $\bar\nu_e$ background flux will be large compared to other proposed sites~\cite{Wurm:2007cy}.The expected rate is $\sim 800$ events per~(kt$\cdot$yr). In particular, this will lead to a substantial number of events in the high-energy tail of the reactor $\bar\nu_e$ spectrum that potentially extends to energies as high as 13\,MeV \cite{Wurm:2007cy}. To avoid this irreducible background, the energy threshold for DSNB detection has been set to a visible energy of 11\,MeV, indicated by the shaded region in figure~\ref{fig:snd:spectra}.

\bigskip

\noindent {\bf Cosmogenic isotopes.} The decay events of the $\beta n$-emitters {$^9$Li} and {$^8$He} that are induced by cosmic muons pose with $\sim 70\,{\rm d}^{-1}$ a large background for low-energy $\bar\nu_e$ detection. However, the spectral endpoints of both radioisotopes effectively lie below the lower analysis of 11\,MeV defined due to the reactor antineutrino signal. Therefore, this background can be safely neglected in the further discussion.

\bigskip

\noindent {\bf Atmospheric neutrino CC interactions.}  The events created by the IBDs of atmospheric $\bar\nu_e$'s on the target protons constitute a further source of indiscriminable background and start to dominate the DSNB signal for energies of 30\,MeV and higher. Compared to other sites, Jiangmen is at a low geographical latitude ($22.6^\circ$N), which leads to a relatively low atmospheric $\nu$ flux. The corresponding background spectrum of IBD events (scaled from reference~\cite{Wurm:2007cy}) is shown in figure~\ref{fig:snd:spectra}.

Unlike in water Cherenkov detectors, the background rate originating from the CC interactions of the large flux of atmospheric $\nu_\mu$'s and  $\bar\nu_\mu$'s is far less problematic in liquid scintillator. In this events, the presence of final state muons can be very efficiently tagged based on both the coincidence tag provided by the Michel electrons and the characteristic pulse shape of muon events. This background is therefore considered to be negligible in this analysis. However, a determination of a potential residual rate will be addressed in future, more elaborate studies.

\bigskip

\noindent {\bf Atmospheric neutrino NC interactions.} The search for antineutrinos at energies above 8\,MeV that has been performed by KamLAND has demonstrated that neutral current reactions of high-energy atmospheric neutrinos in liquid scintillator can result in an IBD-like signature, either by neutron knock-out or by more complicated processes~\cite{Gando:2011jza}. The corresponding event rate is more than an one order of magnitude above the DSNB signal in the region of interest. In many cases, a $^{11}$C-nucleus will remain in the end state of such reactions. Its delayed decay has the potential to reject about 50\,\% of this background. Moreover, a significant reduction can be achieved if pulse-shape discrimination (PSD) techniques are applied to the prompt signal \cite{Mollenberg:2014pwa}. In the following, we assume that the PSD efficiencies derived for the LENA setup in reference~\cite{Mollenberg:2014pwa} can be transferred without loss to JUNO. This approach is supported by the use of identical scintillators and neglects the positive impact of the considerably higher light yield in JUNO. In this way, the residual background rate within the observational window can be reduced to 0.6\,yr$^{-1}$ in the fiducial volume, corresponding to $\varepsilon_{\rm NC}=1.1\%$ of the original rate. In spite of this rather stringent cut, the signal efficiency is still at $\varepsilon_\nu=50\%$ \cite{Mollenberg:2014pwa}.

\bigskip

\noindent {\bf Fast neutron (FN) background.} Due to the relatively low depth of the JUNO detector cavern ($\sim 700$\,m of rock shielding), neutrons created by cosmic muons passing through the rock close-by the detector are relatively frequent and constitute a relevant source of background. While spallation neutrons can be easily discarded if the parent muon crosses the outer veto or the water pool, neutrons from further-out muons have a non-zero probability to pass undetected through the Cherenkov veto region and to mimic the IBD signature by a prompt proton recoil and a delayed capture of the thermalized neutron. Based on a muon flux of $\sim 20~{\rm m}^{-2}~{\rm h}^{-1}$ and a mean muon energy of $\langle E_\mu\rangle = 215\,$GeV, the expected FN background rate inside the JUNO target volume is of the order of 20 events per year.

Two factors allow a considerable reduction of this background: Because of the finite mean free path of the neutrons most of the events are concentrated towards the verge of the scintillator volume, clustering at the equator and the north pole of the spherical vessel. A soft fiducial volume cut rejecting all events at radii greater than 16.8\,m reduces the residual background rate to $\sim 1\,{\rm yr}^{-1}$ in the remaining target mass of 17\,kt. Moreover, the PSD analysis (see above) is sensitive to the prompt proton recoils, decreasing the residual rate to $\varepsilon_{\rm FN}=1.3$\,\% of the initial value, corresponding to a negligible FN rate of $\sim0.01\,{\rm yr}^{-1}$.

\bigskip

\noindent Reactor and atmospheric neutrino IBD signals define an observational window reaching from 11 to $\sim$30\,MeV. Corresponding numbers for signal and background events with and without PSD have been compiled in table~\ref{tab:snd:rates} for a 10-year measuring period. The beneficial impact of PSD is illustrated in figure~\ref{fig:snd:spectra:psd}: In the left panel before PSD, atmospheric NC and fast neutron backgrounds dominate the DSNB signal, while they are greatly reduced in the right panel that reflects the situation after application of the PSD. The signal-to-background ratio is expected to exceed 1:1, creating very favorable conditions for a positive detection of the DSNB.

\begin{figure}[ht]
\centering
\includegraphics[width=0.49\textwidth]{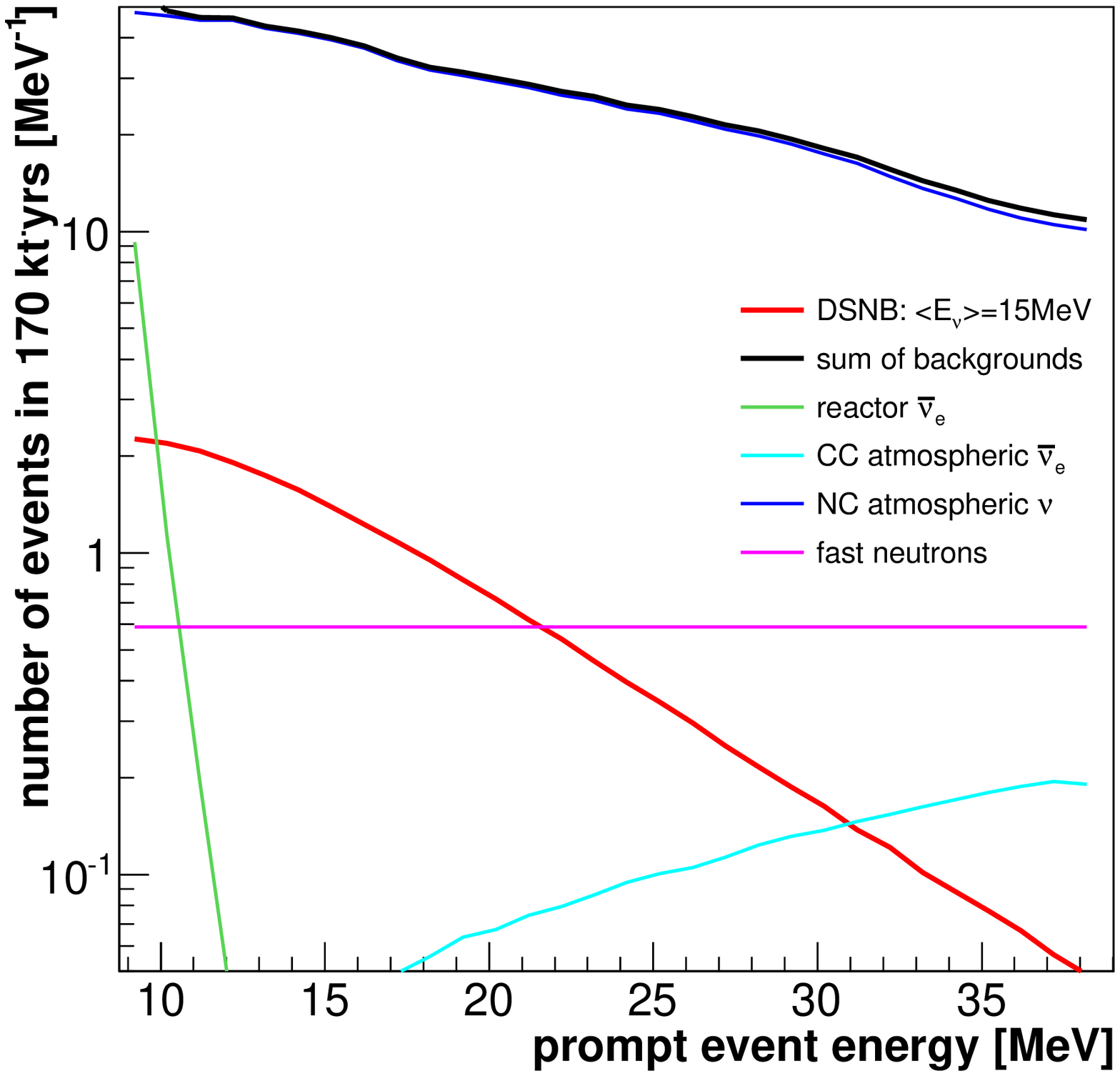}\hfill
\includegraphics[width=0.49\textwidth]{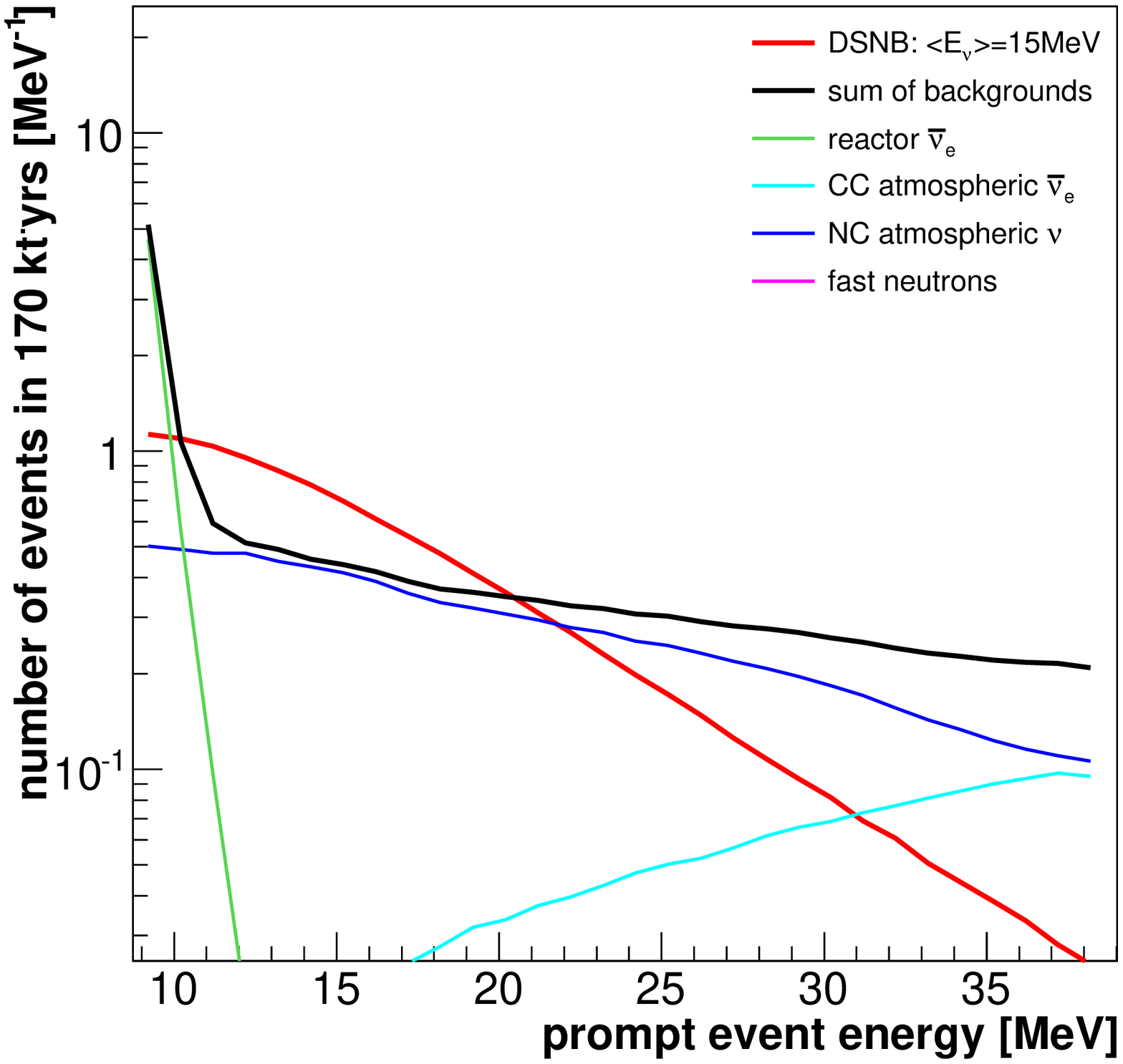}
\caption{Prompt DSNB signal ($\langle E_{\bar\nu_e}\rangle=15$\,MeV, $\Phi=\Phi_0$) and background spectra before ({\it left}) and after ({\it right}) the application of pulse-shape discrimination. The DSNB signal dominates all backgrounds for a large fraction of the observation window from 11 to 30~MeV.}
\label{fig:snd:spectra:psd}
\end{figure}

\begin{table}[!htb]
\begin{center}
\begin{tabular}[c]{ll|ccc} \hline
 Item		&          & Rate (no PSD) & PSD efficiency & Rate (PSD)  \\
 \hline\hline
 Signal	& $\langle E_{\bar\nu_e} \rangle = 12$\,MeV & 13 & $\varepsilon_\nu=50\,\%$ & 7 \\
 		& $\langle E_{\bar\nu_e} \rangle = 15$\,MeV & 23 & & 12 \\
 		& $\langle E_{\bar\nu_e} \rangle = 18$\,MeV & 33 & & 16 \\
 		& $\langle E_{\bar\nu_e} \rangle = 21$\,MeV & 39 & & 19 \\				
 \hline
 Background	& reactor $\bar\nu_e$ & 0.3 & $\varepsilon_\nu=50\,\%$ & 0.13 \\
 			& atm. CC & 1.3 & $\varepsilon_\nu=50\,\%$ & 0.7 \\
			& atm. NC & $6\cdot10^2$ & $\varepsilon_{\rm NC}=1.1\,\%$ & 6.2 \\
			& fast neutrons & 11 & $\varepsilon_{\rm FN}=1.3\,\%$ & 0.14\\
			& $\Sigma$ & & & 7.1 \\
 \hline
\end{tabular}
\caption{Signal and background event rates before and after PSD in 10 years of JUNO data taking. An energy window $11\,{\rm MeV} < E_\nu < 30\,{\rm MeV}$ and a fiducial volume cut corresponding to 17\,kt have been chosen for background suppression.}
\label{tab:snd:rates}
\end{center}
\end{table}

\subsection{Expected sensitivity}
\label{subsec:snd:sensitivity}

We have investigated two possible approaches for determining the potential of a positive DSNB detection by JUNO: Optimal sensitivity can be achieved in case the spectral shapes and rates of all backgrounds are well known, allowing for an energy-dependent fit of signal and background spectra to the data. Alternatively, we investigate a more conservative ansatz where detection significance is evaluated based on a rate-only analysis inside the observation window. Finally, the dependence of the sensitivity on the systematic uncertainty associated with the background normalizations is studied.

\bigskip

\noindent {\bf Spectral fit.} The sensitivity of the DSNB search will depend on the knowledge on spectral shape and normalization of the various background sources. Reactor and atmospheric antineutrino IBD spectra and fluxes can be extrapolated from the regions outside the observation window and will play only a minor role for most of the region of interest. For the more important FN and atmospheric NC backgrounds, rates and spectra can probably be determined with good accuracy if the energy-dependent PSD efficiencies are well understood.

Figure~\ref{fig:snd:sensitivity} depicts the significance of a positive DSNB measurement as a function of DSNB parametrization. Both the flux normalization $\Phi$ and the mean spectral energy have been varied around the most probable values $\Phi_0$ and $\langle E_{\bar\nu_e}\rangle = 15\,$MeV (cf.~section~\ref{subsec:snd:fluxspectrum}). Assuming 170\,kt$\cdot$yrs of exposure, event spectra based on the predictions for signal and all background sources have been generated for the energy range of 10-40\,MeV. Based on these data samples, likelihood fits have been performed to the spectra above a prompt-event visible energy of 11\,MeV, with and without including a contribution from the DSNB\footnote{In particular, the median sensitivities have been obtained by using the Asimov sample (without statistical fluctuations).}. The likelihood function employed is
\begin{equation}
\label{eq:snd:likelihood}
{\cal L}(\langle E_{\bar\nu_e}\rangle, \Phi, f_j) = \sum_i -2\log\bigg[P\Big(n_i, \Phi s_i+\sum_j f_j b_{j,i}\Big)\bigg] + \sum_j \frac{(f_j-1)^2}{\sigma_j^2},
\end{equation}
where $P$ is the Poissonian probability to obtain $n_i$ events in the $i^{th}$ bin based on the prediction for signal $s_i$ and backgrounds $b_{j,i}$, $\Phi$ and $f_j$ are the spectral normalizations of signal and backgrounds, respectively. The $\sigma_j$ are the systematic uncertainties on the background normalization. The profile of figure~\ref{fig:snd:sensitivity} assumes $\sigma_j=5\%$, while table~\ref{tab:snd:det_sig} also lists the expected sensitivities for $\sigma_j=20\%$. For the favored DSNB parameters, a $3\sigma$ evidence for the DSNB signal seems well within reach (cf.~table~\ref{tab:snd:det_sig}).

\bigskip

\noindent {\bf Rate-only analysis.} Assuming no accurate knowledge on spectral shapes of signal and background, the detection potential can be calculated based merely on the detected event rate (table \ref{tab:snd:rates}). This more conservative analysis is based on the integrated event rates within the nominal observational window of 11--30\,MeV. The detection significance is evaluated according to the same likelihood function (\ref{eq:snd:likelihood}), but assuming only a single energy bin $i$. Resulting sensitivities are listed in table \ref{tab:snd:det_sig}. Again, 5\,\% and 20\,\% are considered as systematic uncertainties for the individual background contributions. While the sensitivity is in all cases somewhat reduced, a $3\sigma$ evidence for $\langle E_{\bar\nu_e} \rangle=15$\,MeV can be still expected. Compared to the rate-only result, the inclusion of spectral information provides only a mild benefit due to the overall low statistics and the similarity in the spectral shapes of the DSNB and the background dominated by the atmospheric $\nu$ NC interactions.

\begin{table}[!htb]
\begin{center}
\begin{tabular}{|c|cc|cc|}
\hline
Syst. uncertainty BG & \multicolumn{2}{c|}{5\,\%}& \multicolumn{2}{c|}{20\,\%}\\
\hline
$\mathrm{\langle E_{\bar\nu_e}\rangle}$ & rate only & spectral fit & rate only & spectral fit \\
\hline
12\,MeV & $2.3\,\sigma$ & $2.5\,\sigma$ & $2.0\,\sigma$ & $2.3\,\sigma$\\
15\,MeV & $3.5\,\sigma$ & $3.7\,\sigma$ & $3.2\,\sigma$ & $3.3\,\sigma$\\
18\,MeV & $4.6\,\sigma$ & $4.8\,\sigma$ & $4.1\,\sigma$ & $4.3\,\sigma$\\
21\,MeV & $5.5\,\sigma$ & $5.8\,\sigma$ & $4.9\,\sigma$ & $5.1\,\sigma$\\
\hline
\end{tabular}
\caption{The expected detection significance after 10 years of data taking for different DSNB models
with $\langle E_{\bar\nu_e} \rangle$ ranging from 12\,MeV to 21\,MeV ($\Phi=\Phi_0$). Results are given based on either a rate-only or spectral fit analysis and assuming 5\% or 20\% for background uncertainty.}
\label{tab:snd:det_sig}
\end{center}
\end{table}

\noindent {\bf Upper limit on DSNB flux.} If there is no positive detection of the DSNB, the current limit can be significantly improved. Assuming that the detected event spectrum equals the background expectation in overall normalization and shape, the upper limit on the DSNB flux above 17.3\,MeV would be $\sim 0.2\,{\rm cm}^{-2}\,{\rm s}^{-1}$ (90\%~C.L.) after 10 years for $\langle E_{\bar\nu_e}\rangle=18$\,MeV. This limit is almost an order of magnitude better than the current value from Super-Kamiokande~\cite{Bays:2011si}. In figure~\ref{fig:snd:exclusion} we show the corresponding exclusion contour as function of $\langle E_{\bar\nu_e} \rangle$ (90\%~C.L.).

\subsection{Outlook on further studies}
\label{subsec:snd:outlook}

A precise estimate of the DSNB detection potential in JUNO requires further studies. Most important is the determination of the PSD efficiency. The current estimates for the detection potential (section~\ref{subsec:snd:sensitivity}) are based on the assumption that the pulse-shape cut has the same efficiency as in the LENA study. But due to the higher light yield in JUNO, the pulse shape cut should actually be more efficient. If the signal efficiency can be increased, the information extracted on the DSNB will be more detailed and might allow us to obtain some information on the spectral mean energy $\langle E_{\bar\nu_e}\rangle$. On the other hand, PSD efficiency will also greatly depend on the time resolution of the light sensors, for which we here assumed a rather optimistic value of 1\,ns ($1\sigma$). While first studies indicate that a moderate reduction in time resolution will have only a small impact on PSD efficiencies, more detailed investigations are mandatory.

\begin{figure}[htp]
\begin{minipage}{0.47\textwidth}
\centering
\includegraphics[width=\textwidth]{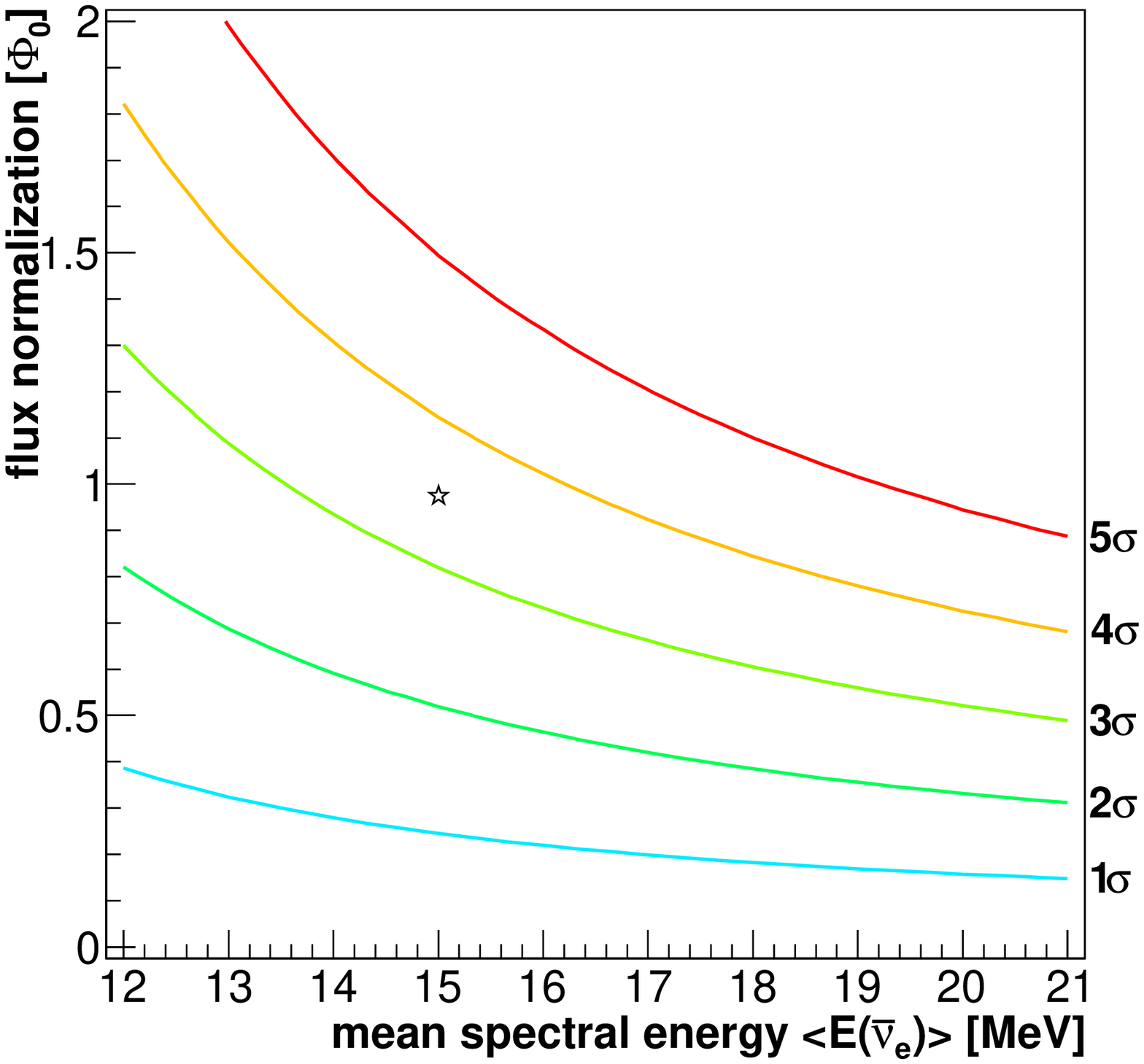}
\caption{JUNO's discovery potential for the DSNB as a function of the mean energy of the SN spectrum $\langle E_{\bar\nu_e}\rangle$ and the DSNB flux normalization $\Phi$ (cf.~section~\ref{subsec:snd:fluxspectrum}). We assume 10\,yrs measuring time, 5\% background uncertainty and a detected event spectrum corresponding to the sum of signal and background predictions. The significance is derived from a likelihood fit to the data. The star marks a theoretically well-motivated combination of DSNB parameters (cf.~section~\ref{subsec:snd:fluxspectrum}).}
\label{fig:snd:sensitivity}
%\end{figure}
\end{minipage}
\hfill
\begin{minipage}{0.47\textwidth}
%\begin{figure}
\centering
\includegraphics[width=\textwidth]{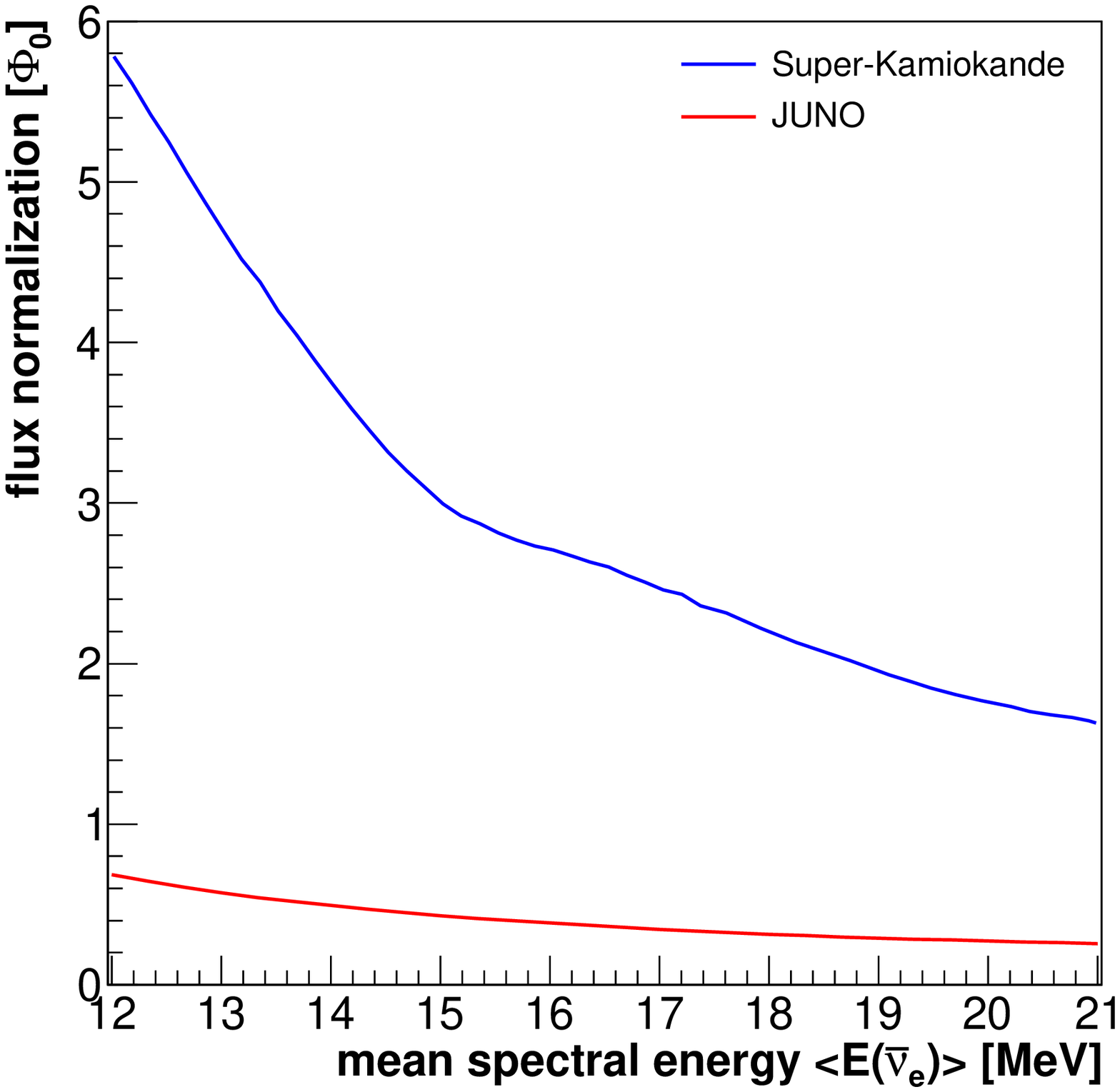}
\caption{Predicted exclusion contour (90\%~C.L.) if JUNO finds no signal of the DSNB above background. The upper limit is shown as a function of the mean energy of the SN spectrum $\langle E_{\bar\nu_e}\rangle$ and the DSNB flux normalization $\Phi$ (cf.~section~\ref{subsec:snd:fluxspectrum}). It has been derived from a spectral likelihood fit assuming 5\% background uncertainty, 10\,yrs of measurement time and $N_ {\rm det}=\langle N_{\rm bg} \rangle$. The upper limit derived from the Super-Kamiokande results presented in~\cite{Bays:2011si} is shown for comparison.}
\label{fig:snd:exclusion}
\end{minipage}
\end{figure}

Another important issue that needs to be studied is how precise the expected value for the number of background events can be determined.
While it is relatively easy to estimate the amount of radioactive, fast-neutron
and CC~atmospheric background events, the critical task will be to determine the expected value of the atmospheric NC~events
after the pulse-shape cut~\cite{Mollenberg:2014pwa}.

Finally, it should be investigated whether shape and rate uncertainties of the fast neutron background might be constrained by
a dedicated muon sampler above the detector. Such a device will allow us to track a sample of rock muons and to study the induced
signals of particle showers and especially neutrons in the water and scintillator volumes.

\subsection{Conclusions}
\label{subsec:snd:conclusions}

Large liquid-scintillator detectors are probably the most powerful
approach to measure the long-sought DSNB. Despite its shallow depth,
the relatively small size and the large atmospheric NC background,
the superior detector properties of JUNO may
be able to provide a DSNB detection at the $3\sigma$ level. If no
signal is detected, significant limits in the plausible parameter
space can be achieved, improving on the existing limits from Super-Kamiokande
which has approximately a 50\% larger fiducial mass for DSNB
detection. In case the gadolinium upgrade of Super-Kamiokande is
implemented, it will have a realistic chance of measuring the DSNB.
Adding the measurements of the two detectors will roughly double the overall
statistics while reducing systematic uncertainties due to the differences in background and
detection uncertainties. A combined analysis will either provide a more significant detection or more
restrictive limits for the DSNB.

A detailed spectral study of the DSNB will not be possible with
detectors of the JUNO or Super-Kamiokande size. In the more distant
future, larger detectors such as the original LENA concept may be
realized. For example, if the megaton Hyper-Kamiokande water Cherenkov
detector is built, Super-Kamiokande might be converted to a 50~kt
scintillator observatory.

The neutrino observation of a galactic SN is quite conceivable over
the time span until the JUNO measurements have been completed. Such an
observation would go a long way to give us confidence in theoretical
expectations of SN neutrino emission parameters, strongly improving on
the sparse SN~1987A data. However, the large variability of
core-collapse events, the signal even depending on observer direction
\cite{Tamborra:2014aua, Tamborra:2014hga}, will prevent one from an unambiguous prediction
of the average core-collapse neutrino emission properties.

In the more distant future, one may therefore construct a
multi-megaton detector that can measure SN neutrinos out to distances
of several Mpc, although only a few events per SN~\cite{Kistler:2008us}. Yet by stacking
such measurements, over time one can build up an average $\bar\nu_e$
flux spectrum that could be used to interpret DSNB measurements in
terms of SN rates, notably failed SNe and black-hole formation. One
possible implementation of such a program could be a dense infill of
the IceCube detector. At present, the PINGU infill is being considered
for studying atmospheric neutrino oscillations, and a further step
could bring the energy threshold down to the level required for
low-energy neutrino astronomy, the MICA concept~\cite{MICA:2013}.

Neutrino astronomy is a field in its infancy and high-statistics observations exist only for solar neutrinos.
JUNO will play a leading role in pushing the low-energy frontier of this exciting field and may be the first
instrument ever to measure low-energy neutrinos from the edge of the visible universe.

\clearpage

\section{Solar Neutrinos}
\label{sec:solar}

\blfootnote{Editors: Emanuela Meroni (emanuela.meroni@mi.infn.it), Lothar Oberauer (lothar.oberauer@tum.de), and Chao Zhang (chao@bnl.gov)}
\blfootnote{Major contributors: Vito Antonelli, Marco Grassi, Yufeng Li and Lino Miramonti}

The Sun is a powerful source of electron neutrinos with the energy of ${\mathcal O}(1)$ MeV, produced in the thermonuclear fusion
reactions in the solar core. The two fusion reactions involved in combining four protons into a $^4$He nucleus,
$4p \to {}^{4}_{2}{\rm He} + 2e^+ + 2\nu_e$, are known as the $pp$ chain and the CNO cycle.
Because the total mass of the four protons is larger than the total mass of the final state particles, extra energy is released
in the form of photons or of neutrino kinetic energy. The $pp$ chain constitutes around $99\%$ of the neutrino flux out of
the Sun, which includes the $pp$ neutrinos, $pep$ neutrinos, $hep$ neutrinos, $^7$Be neutrinos, and $^8$B neutrinos.
The names of these neutrinos are defined according to their respective fusion reactions.
On the other hand, neutrinos from nuclei decays (i.e., $^{13}$N, $^{15}$O and $^{17}$F) of the CNO cycle are often referred to as the CNO neutrinos.

The study of solar neutrinos has already contributed significantly to the development of elementary particle physics and of astrophysics, reinforcing the synergy between these two disciplines. Over the last decades, together with the atmospheric~\cite{Fukuda:1998mi}, accelerator~\cite{Ahn:2002up} and reactor~\cite{Eguchi:2002dm} neutrino experiments,
the experimental studies using solar neutrinos~\cite{Robertson:2012ib,Antonelli:2012qu} have obtained compelling evidence of neutrino oscillations.
Nowadays, the issue is not any more to prove that neutrinos are massive and oscillating particles. However, important questions are still open for relevant solutions.
Besides testing of the Mikheyev-Smirnov-Wolfenstein (MSW)~\cite{Wolfenstein:1977ue,Mikheev:1986gs} matter effect in particle physics,
we can also improve significantly our knowledge of fundamental solar physics, such as the mechanism ruling the dynamics of the Sun,
the solar metallicity problem, and the agreement between solar models and the data from helioseismology.

Using a liquid scintillator detector as KamLAND~\cite{Eguchi:2002dm} and Borexino~\cite{Alimonti:2008gc} but with a much larger mass comparable to the Super-Kamiokande
water Cherenkov detector~\cite{Fukuda:1998fd}, and with a very good energy resolution and hopefully an high radiopurity level,
the JUNO experiment can have the potentiality to contribute significantly to
the solar neutrino measurements, both from the astrophysical side and the elementary particle side.
In this section we discuss the possibilities of doing solar neutrino physics at JUNO, with particular attention to the $^7$Be and $^8$B neutrino measurements.

\subsection{History of solar neutrino experiments}
\label{subsec:solar:post2002}

The long-standing solar neutrino problem (SNP) was caused by the observed flux deficit in Homestake~\cite{Davis:1968cp,Cleveland:1998nv},
KamiokaNDE~\cite{Hirata:1989zj}, Super-Kamiokande~\cite{Fukuda:1998fd},  and Gallium experiments
(Gallex~\cite{Hampel:1998xg}, GNO~\cite{Altmann:2005ix}, and
SAGE~\cite{Abdurashitov:1999bv}) and it was solved finally by the SNO experiment~\cite{Jelley:2009zz} in 2002.
The SNO results confirmed the validity of the Standard Solar Models (SSM)~\cite{Bahcall:2004fg} using the neutral current signal that measured all active neutrino flavors.
By comparing these results with the $\nu_e$ flux recovered by the charged current signal, they offered a clear proof of the electron neutrino conversion into other active flavors~\cite{Ahmad:2001an}.
The combination of the SNO result with the previous solar neutrino experiments defined a complete oscillation solution to the SNP, and together with the data from the reactor experiment KamLAND~\cite{Eguchi:2002dm}, the so-called large mixing angle (LMA) solution was found as the correct neutrino mixing parameter set.
These results had great impact both on nuclear astrophysics and on elementary particle physics. In combination with atmospheric neutrino~\cite{Fukuda:1998mi} and long baseline accelerator experiments~\cite{Ahn:2002up}, they proved in an undeniable way that neutrinos oscillate and have non-zero masses. Therefore, they gave the first clear hint of the need to go beyond the Standard Model of elementary particle physics.

In the last years, the liquid scintillator experiments Borexino and KamLAND reached a low energy threshold in the sub-MeV region.
The solar $^7$Be-neutrinos have been measured with high precision~\cite{Arpesella:2008mt,Gando:2014wjd}, and for the first time solar $pep$ and $pp$ neutrinos have been observed~\cite{Collaboration:2011nga,Bellini:2014uqa}.
These measurements allow the new insight into the mechanism of thermal nuclear fusion processes in the center of the Sun.
In addition, the predicted MSW matter effect on neutrino oscillations was found to be in general agreement with the experimental observations.

Moreover, a global three neutrino analysis~\cite{Bellini:2013lnn}, including all the solar neutrino experiments (assuming the fluxes predicted by the
high-$Z$ version of SSM) and KamLAND data (assuming the $CPT$ invariance), gave the following values for the mixing angles and the mass-squared difference:
$
{\rm tan}^2 \, \theta_{12} = 0.457^{+0.038}_{-0.025} \, ;
{\rm sin}^2 \, \theta_{13} = 0.023^{+0.014}_{-0.018} \, ;
\Delta m_{21}^2 = 7.50^{+0.18}_{-0.21} \, \times \, 10^{-5} \, {\rm eV}^2\, ,
$
which are very similar to other global analyses of all neutrino data performed in the three neutrino framework~\cite{Capozzi:2013csa,Forero:2014bxa,Gonzalez-Garcia:2014bfa}
and other similar works published in 2012~\cite{Fogli:2012ua,Tortola:2012te,GonzalezGarcia:2012sz}.

\subsection{Relevant open questions in solar neutrino physics}

Despite the great achievements of the last decades, there are still important aspects of solar neutrino physics to clarify and some questions of great
relevance for astrophysics and elementary particle physics are waiting for definite solutions. The issues can be summarized as the need for a better
determination of the oscillation parameters, the solution of the solar metallicity problem, and the detailed analysis of the energy dependence for
the oscillation probability in the region corresponding to the low-energy solar $^8$B neutrinos.

For the measurements of oscillation parameters, there could be a significant improvement with respect to the present situation, thanks to the precision
measurements of reactor antineutrino oscillations at JUNO. As shown in the previous chapters, three of the oscillation parameters, including the solar
oscillation parameters $ {\rm sin}^2 \,\theta_{12}$, $\Delta m_{21}^2$ and the larger mass-squared difference $\Delta m_{ee}^2$ (i.e., a linear
combination of $\Delta m_{31}^2$ and $\Delta m_{32}^2$), can be measured with the precision level of $0.5\%$$-$$0.7\%$. In this respect,
one could enter the precision oscillation era in combination with other future long baseline experiments~\cite{Adams:2013qkq,Abe:2011ts,Abe:2015zbg}.
Moreover, solar neutrino oscillations itself can make the independent measurements of oscillation parameters without the assumption of $CPT$ invariance,
which is very important to test the consistency of the standard three neutrino framework and probe new physics beyond the Standard Model~\cite{Ohlsson:2012kf}.
As also discussed previously, the combination of the data from medium baseline reactor antineutrino experiments (like JUNO) and from short baseline reactor
(like Daya Bay) and solar experiments can offer a direct unitarity test of the MNSP mixing matrix, i.e., $|U_{e1}|^2 + |U_{e2}|^2 + |U_{e3}|^2 \stackrel{?}{=} 1$.
On the other hand, further improvement in the oscillation parameters from the solar experiment side is difficult in the post-SNO era.

Regarding the solar metallicity problem~\cite{Villante:2013mba,Bergemann:2014vaa}, so far we have observed neutrinos from the $pp$ chain and still missing is a measurement of the
sub-dominant solar CNO cycle. The best limits for its contribution to the solar energy generation are coming so far from the Borexino data,
however a CNO neutrino measurement with the accuracy of about 10\% would be necessary to shed light on the solar metallicity problem, which became apparent within the last years.
The former excellent agreement between the SSM and the solar data has been compromised by the revision of the solar surface heavy-element content from ($Z/X$) = 0.0229~\cite{Grevesse:1998bj} to ($Z/X$) = 0.0165~\cite{Asplund:2004eu}, leading to a discrepancy between the SSM and helioseismology results\footnote{$X$ and $Z$ are the mass fractions of hydrogen and metals respectively}. Solution to this puzzle would imply either to revise the physical inputs of the SSM or to modify the core abundances, in particular those of C, N, O, Ne, and Ar.
In 2009, a complete revision of the solar photospheric abundances for nearly all elements has been done~\cite{Asplund:2009fu},  including a new three dimensional hydrodynamical solar atmosphere model with the improved radiative transfer and opacity. The obtained results give a solar abundance ($Z/X$) = 0.0178.
The three different sets of solar abundances, GS98~\cite{Grevesse:1998bj}, AGS05~\cite{Asplund:2004eu}, and AGSS09~\cite{Asplund:2009fu},
have been used to construct different versions of the SSM~\cite{Serenelli:2009yc}.
The predictions of these SSM versions differ also for the $^8$B and $^7$Be neutrino fluxes. Therefore, a possible improvement at JUNO of the accuracy in the determination of these fluxes, together with data (coming from other future experiments) about the CNO fluxes could help solving this central problem of nuclear astrophysics, which will be explained further in the next subsection.

Coming, finally, to the study of energy dependence of the electron neutrino survival probability,
the vacuum-oscillation and matter-oscillation dominated regions are separated at around 1--3 MeV.
The continuous solar $^8$B neutrino spectrum is in principle a perfect tool to study the MSW-modulated energy dependence.
According to the standard LMA-MSW solution one would expect a continuous transition between the vacuum and matter related $\nu_e$-survival probabilities.
However, the data of existing detectors did not observe a clear evidence of this up-turn going towards the low energy part of the solar $^8$B spectrum, which gave rise to a series of theoretical discussions of sub-leading non-standard effects of the neutrino survival probability.
The Super-Kamiokande data published in 2014~\cite{Renshaw:2014awa}, including a combined analysis of all the four phases of this experiment, slightly indicated the presence of the up-turn.
However, an independent and high-significance test of the up-turn effect would be extremely important to confirm the consistency of the standard LMA-MSW solution,
or to indicate any possible deviations from this standard paradigm.

\subsection{Motivation of solar neutrino measurements at JUNO}
\label{subsec:solar:conclusions}

Solar neutrino measurement at JUNO is performed via the elastic neutrino electron scattering.
A low energy threshold and a very good energy resolution should be accomplished at JUNO.
Due to its much larger volume, significantly larger statistics can be reached as compared with Borexino.
The larger detector size will also help suppress external gamma background by defining a fiducial volume and
the self-shielding of the detector can be very good, provided that the intrinsic radiopurity of the scintillator is comparable to that reached in Borexino.
However, the overburden is significantly lower with respect to the Gran Sasso underground laboratory of Borexino and therefore one has to see how cosmogenic background events can be rejected at JUNO. In the following, we discuss motivations and prospects of the low (i.e., $E\sim$ 1\,MeV) and high energy solar neutrinos measurements.

A new, accurate and independent measurement of the $^7$Be flux, that presently is essentially determined by the Borexino data within $5 \%$ of accuracy,
would be interesting for a precise study of the vacuum dominated MSW-region. It could also shed some light on the solar metallicity problem,
which is one of the present central astrophysical puzzles~\cite{Villante:2013mba,Bergemann:2014vaa}.
In addition to the already cited high-$Z$~\cite{Grevesse:1998bj} and low-$Z$~\cite{Asplund:2004eu,Asplund:2009fu} versions of the SSM, in literature one can find also models in which the effect of reduced metallicity is partially compensated by an increase in the radiative opacity. In this way it is possible to reproduce the opacity profile and restore a good agreement with helioseismology, but this increased opacity can be only partially justified by theoretical arguments.

 The values of the different solar neutrino fluxes can be used as observables to solve the present ambiguity between the different versions of the SSM (high-$Z$, low-$Z$ and low-$Z$ with increased opacity), as shown in Fig.~\ref{fig:solar:Serenelli1}.
The relatively large uncertainty of theoretical models and the fact that present experimental results fall in the middle between different SSM predictions make it impossible to draw any final conclusion at present. It is also clear that, due to the ambiguity between the high-$Z$ SSM and a low-$Z$ solution with increased opacity, there is the need for complementary information, like the values of CNO neutrino fluxes, that would represent the real breakthrough in this field.
Nevertheless, a $^7$Be flux measurement with the reduced uncertainty, possibly complemented by a parallel reduction of the theoretical model uncertainties, could give an important contribution to the solution of this puzzle, especially in the case of central value moving towards one of the two solutions (high or low-$Z$).

\begin{figure}[!tb]
\begin{center}
\includegraphics[width=0.7\textwidth]{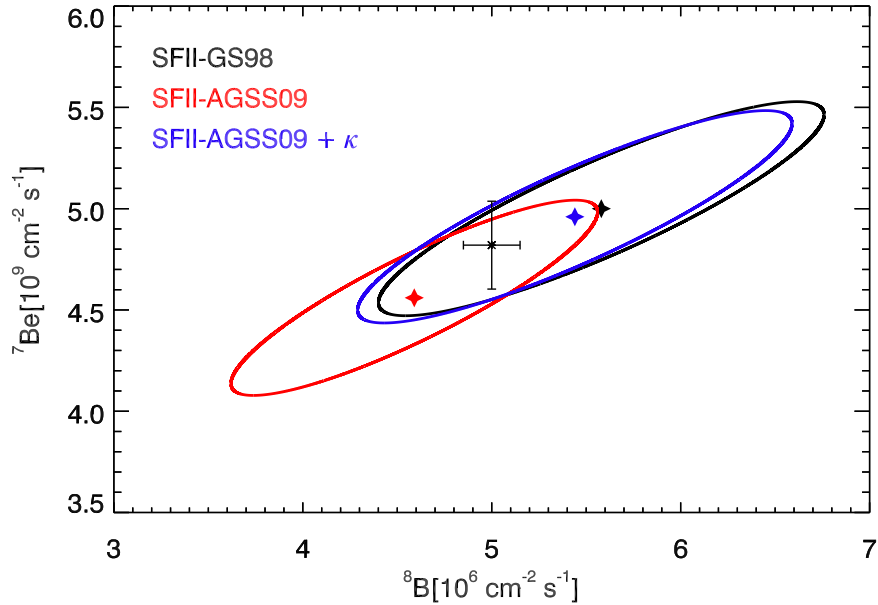}
\caption{Comparison between theoretical predictions and experimental results for the fluxes of $^8$B ($x$-axis) and $^7$Be ($y$-axis) solar neutrinos. The black, red, and blue ellipses represent the $1 \sigma$ allowed regions, respectively in the high-$Z$, low-$Z$ and low-$Z$ with increased opacity versions of the SSM. The $1 \sigma$ experimental results for the two fluxes correspond to the horizontal and vertical black bars.
Updated version of the figure from~\cite{Serenelli-talk}. See also~\cite{Zuber:2013uya} and~\cite{Serenelli:2009ww}.}
\label{fig:solar:Serenelli1}
\end{center}
\end{figure}

A further improved measurement of $^7$Be solar neutrinos could be also relevant to the search for the anomalous magnetic moment of neutrinos. In the Standard Model of electroweak interactions the magnetic moments of neutrinos are predicted to be very small, $\mu_{\nu} \leq 10^{-20} \mu_B$ (where $\mu_B = e/{2 m_e}$ is the Bohr-magneton), therefore, any significantly larger value of the
neutrino magnetic moment would be a clear signal of new physics~\cite{Voloshin:2010vm,Giunti:2014ixa,Broggini:2012df}.
The current experimental upper limit from particle physics experiments on the neutrino anomalous magnetic moment
has been obtained by experiments studying reactor antineutrinos~\cite{Beda:2013mta,Wong:2006nx,Daraktchieva:2005kn} and it is of the order of $\sim 3 \times 10^{-11} \mu_B$. Even more stringent (of almost one order of magnitude lower) indirect limits can be derived from astrophysical experiments and studies~\cite{Raffelt:1999gv,Viaux:2013lha}.
The Borexino experiment also studied the anomalous neutrino magnetic moment, using the
data of the elastic scattering of $^7$Be neutrinos on electrons. Due to the oscillations,
the incident neutrinos are a mixture of three flavors, and this means that what is measured
is an effective magnetic moment~\cite{Beacom:1999wx}.
A further improvement of the neutrino effective magnetic moment at JUNO could offer important information~\cite{Giunti:2015gga}
complementary to the reactor antineutrino studies and to the astrophysical studies.

A precise measurement of $^8$B solar neutrino flux from JUNO will also shed light on the metallicity problem with similar consideration as for the $^7$Be neutrinos.
It is worthwhile to notice that, as shown in Fig.~\ref{fig:solar:Serenelli2},
even in the case of CNO neutrino fluxes being accurately extracted by the future measurements, e.g. in Borexino or SNO+~\cite{Kraus:2010zzb,MOTTRAM:2014nna},
there could still be (in case of low values for these fluxes) an ambiguity in the solution between the low-$Z$ version of the SSM and the low-$Z$ with increased opacity.
Therefore, an additional measurement with the increased accuracy of at least one of the two fluxes of $^8$B and $^7$Be solar neutrinos would be useful to
break the degeneracy between the metallicity and opacity.

\begin{figure}[!tb]
\begin{center}
\includegraphics[width=0.7\textwidth]{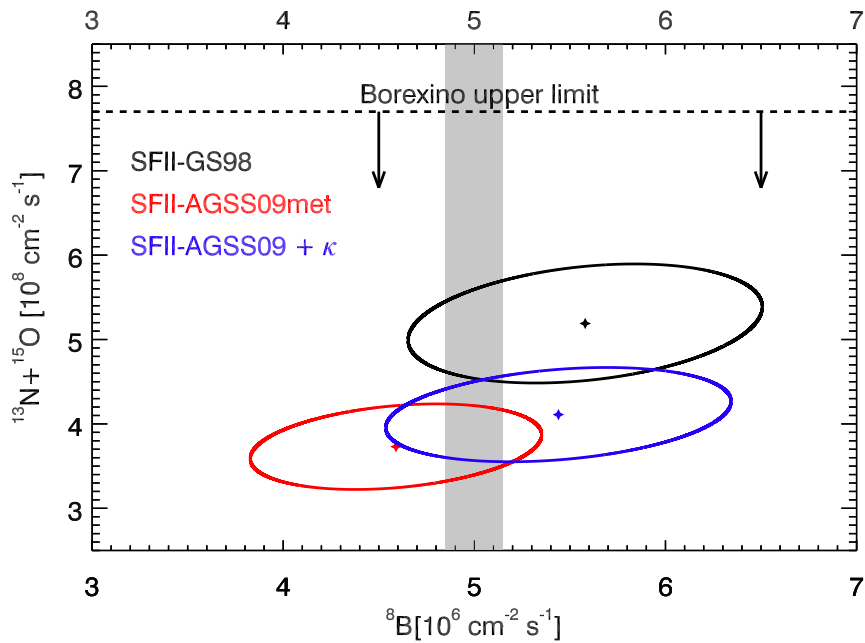}
\caption{Comparison between the theoretical predictions for the values of the $^8$B ($x$-axis) and $^{13}$N + $^{15}$O ($y$-axis) neutrino fluxes derived from different versions of the SSM (high-$Z$ in black, low-$Z$ in red, and low-$Z$ with increased opacity in blue).
The shaded grey vertical region represents the $1 \sigma$ region compatible with present data for $^8$B neutrinos. The horizontal line indicates the Borexino measured CNO flux limit.
Updated version of the figure from~\cite{Serenelli-talk}. See also~\cite{Zuber:2013uya,Serenelli:2009ww}.}
\label{fig:solar:Serenelli2}
\end{center}
\end{figure}

The measurement of the $^8$B solar neutrinos would be relevant also to test the consistency of the standard LMA-MSW paradigm. A first example is given by the study of the day-night asymmetry (i.e., defined as $A_{\rm DN}$). The outcome of the recent Super-Kamiokande analysis~\cite{Renshaw:2013dzu} found a value of $A_{\rm DN}$ different from zero at
$2.7 \, \sigma$ and an even more robust hint when combining with the SNO data. %
However, it still needs a confirmation with higher statistical significance. Even more compelling information comes from the detailed analysis of the lower energy part of the $^8$B neutrino spectrum (around 3 MeV),
which corresponds to the transition part between the matter enhanced and vacuum dominated regions.
Considering the current ambiguity~\cite{Renshaw:2014awa} in the up-turn behavior of the solar neutrino survival probability, an improved accuracy of this measurement in the region around 3 MeV would be essential to test the consistency of the LMA-MSW solution and definitely exclude (or confirm) more exotic sub-leading effects. Using the liquid scintillator and taking advantage of the low energy threshold, good energy resolution, low radioactive background, and large statistics, JUNO could be promising to arrive at a definite solution to this problem.

\subsection{Measurement of low energy solar neutrinos at JUNO}
\label{subsec:solar:be7}

In this section, we describe in details the requirements for low energy solar neutrino measurement at JUNO, in particular for $^7$Be neutrinos. The measurement of the higher energy $^{8}$B solar neutrinos will be discussed in the next section. In a liquid scintillator detector such as JUNO, the solar neutrinos of all flavors (via neutrino oscillations) are detected by means of their elastic scattering off electrons:
\begin{equation}
\nu_{e,\mu,\tau} + e^- \rightarrow  \nu_{e,\mu,\tau} + e^- \, .
\end{equation}
In contrast to the reactor $\bar\nu_e$ IBD reaction where a coincidence signature exists to largely suppress background, the detection of solar neutrinos appears as a single flash of light. Only a fraction of the neutrino energy is transferred to the electron, therefore the electron recoil spectrum is continuous even in the case of mono-energetic neutrinos. The expected solar neutrino rates at JUNO are summarized in Tab.~\ref{tab:solar:singles}. The rates are calculated using the BP05(OP)~\cite{Bahcall:2004pz} flux model, convolved with the neutrino-electron elastic scattering cross sections for all flavors. The standard three neutrino oscillation is applied with the solar LMA-MSW effect included. All the rates are estimated without any energy threshold cuts.

The emission of scintillation light is isotropic and any information about the initial direction of solar neutrinos is lost.
Neutrino elastic scattering events in a liquid scintillator are thus intrinsically indistinguishable on an event-by-event basis from the background due to $\beta$ or $\gamma$ decays. Therefore, high radiopurity is required in order for JUNO to have the capability of measuring low energy solar neutrinos. Two internal purity levels are considered in Table~\ref{tab:solar:singles} to calculate the intrinsic radioactive background. The ``baseline'' column is the minimum requirement of the purity level. The signal-to-background ratio at this level is approximately 1 : 3.  The ``ideal'' column is the purity level at which the signal-to-background ratio is approximately 2 : 1. As a comparison, the ``baseline'' requirement is at approximately the KamLAND solar phase purity level (in the cleanest region)~\cite{Gando:2014wjd}, and the ``ideal'' requirement is at about the Borexino phase-I (before 2010) purity level~\cite{Bellini:2013lnn}. The exceptions are $^{238}$U and $^{232}$Th, for which both KamLAND and Borexino have reached better than ``ideal'' requirement since the beginning.
\begin{table}[htb]
  \centering
  \caption{  \label{tab:solar:singles}
    The requirements of singles background rates for doing low energy solar neutrino measurements  and the estimated solar neutrino signal rates at JUNO.}
  \begin{tabular}{| r | c | c |}
    \hline \hline
    \multicolumn{3}{|c|}{Internal radiopurity requirements} \\
    \hline
    & baseline & ideal \\
    \hline
    $^{210}$Pb & $5\times10^{-24}$ [g/g] & $1\times10^{-24}$ [g/g] \\
    $^{85}$Kr  & $500$ [counts/day/kton] & $100$ [counts/day/kton] \\
    $^{238}$U & $1\times10^{-16}$ [g/g] & $1\times10^{-17}$ [g/g] \\
    $^{232}$Th & $1\times10^{-16}$ [g/g] & $1\times10^{-17}$ [g/g] \\
    $^{40}$K & $1\times10^{-17}$ [g/g] & $1\times10^{-18}$ [g/g] \\
    $^{14}$C & $1\times10^{-17}$ [g/g] & $1\times10^{-18}$ [g/g] \\
    \hline
    \multicolumn{3}{|c|}{Cosmogenic background rates [counts/day/kton]} \\
    \hline
    $^{11}$C & \multicolumn{2}{|c|}{$1860$}  \\
    $^{10}$C & \multicolumn{2}{|c|}{$35$}  \\
    \hline
    \multicolumn{3}{|c|}{Solar neutrino signal rates [counts/day/kton]} \\
    \hline
    pp $\nu$ &\multicolumn{2}{|c|}{$1378$ } \\
    $^{7}$Be $\nu$ &\multicolumn{2}{|c|}{$517$} \\
    pep $\nu$ &\multicolumn{2}{|c|}{$28$} \\
    $^{8}$B  $\nu$ &\multicolumn{2}{|c|}{$4.5$} \\
    $^{13}$N/$^{15}$O/$^{17}$F  $\nu$ &\multicolumn{2}{|c|}{$7.5 / 5.4 / 0.1$} \\
    \hline  \hline
  \end{tabular}
\end{table}

The expected cosmogenic $^{11}$C and $^{10}$C rates given in Table~\ref{tab:solar:singles} are scaled from KamLAND spallation measurements (Table IV of Ref.~\cite{Abe:2009aa}.) As an example, for $^{11}$C, the KamLAND measurement is $866\times10^{-7} \mu^{-1} {\rm g}^{-1} {\rm cm}^2$. At the JUNO site, the mean muon energy is smaller than KamLAND and the spallation production rate is about 0.9 times lower. The muon rate in the whole detector (20\,kton) is about 3\,Hz. The mean muon track length is about 23\,m. The density is about 0.8 g/cc. Therefore, the scaled $^{11}$C rate at JUNO is $\sim$1000 counts per day per kton. All other cosmogenic backgrounds are assumed to be minor at the low energy~\cite{Abe:2009aa,Bellini:2013pxa} and are ignored in the calculation.

Fig.~\ref{fig:solar:simul2} shows the expected singles spectra at JUNO with the ``baseline'' and the ``ideal'' radiopurity assumptions listed in Table~\ref{tab:solar:singles}. The energy resolution is assumed to be $\sigma(E) = 3\% \times \sqrt{E \textrm{(MeV)}}$. For simplicity, no energy non-linearity is applied to the spectrum.

\begin{figure}[!htb]
  \centering
  \subfigure[]
  {
      \label{fig:solar:simul2:a}
      \includegraphics[width=0.7\textwidth]{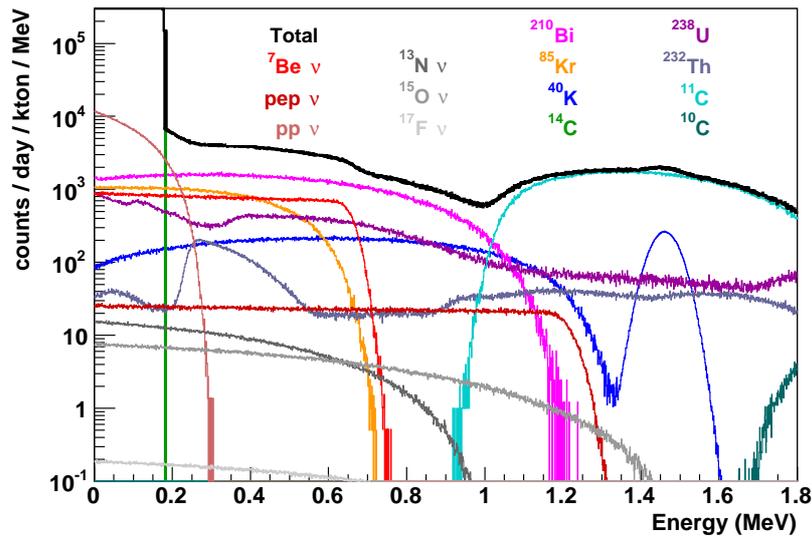}
  }
  \hfill
  \subfigure[]
  {
      \label{fig:solar:simul2:b}
      \includegraphics[width=0.7\textwidth]{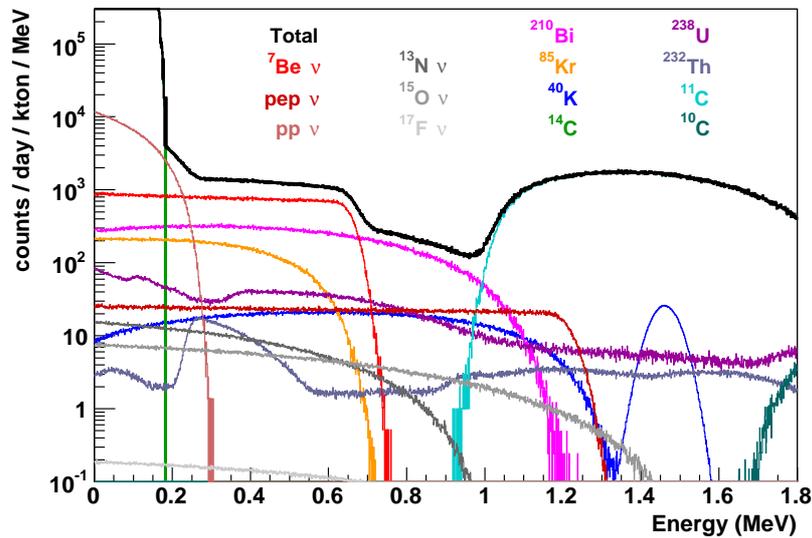}
  }
  \caption{The expected singles spectra at JUNO with (a) the ``baseline'' and (b) the ``ideal'' radiopurity assumptions listed in Table ~\ref{tab:solar:singles}. See text for details.}
  \label{fig:solar:simul2}
\end{figure}

In calculating $^{238}$U and $^{232}$Th decay spectra, secular equilibrium is assumed along the decay chains. The only non-equilibrium isotope considered is $^{210}$Pb (and the subsequent $^{210}$Bi decay) which has a 22-year half life and could break out of secular equilibrium by Rn contamination.
When calculating all beta decay spectra, the Fermi correction, screening correction, finite size correction, and weak magnetism correction are applied to the shape of allowed decay spectrum. A shape correction to forbidden decays is also applied based on the spin and parity difference between the initial and final states.

A few backgrounds are neglected when calculating the rate and spectrum for Table~\ref{tab:solar:singles} and Fig.~\ref{fig:solar:simul2}: $^{39}$Ar rate is assumed to be two orders of magnitude lower than $^{85}$Kr (as it is in the air), thus is neglected in the calculation. We assume that the external gamma background can be removed with fiducial volume cuts. Since the signal rate is high, we assume we can always afford to cut deep into the cleanest region of the detector, therefore only internal radioactivity from the liquid scintillator itself is considered. We assume that with FADC waveform analysis, pile-up events can be largely removed and thus are neglected in the calculation. We further assume that the alpha-decay events (e.g.~$^{210}$Po and alpha-decays in the $^{238}$U and $^{232}$Th decay chains) can be statistically subtracted from the singles spectrum with high precision, using the pulse-shape discrimination as demonstrated by Borexino~\cite{Bellini:2013lnn}. Therefore, only beta and gamma decays are considered when calculating the singles spectra. The reactor $\bar\nu_e$ elastic scattering rate is estimated to be only about 0.5 per day per kton and is neglected in the calculation.

As shown in Fig.~\ref{fig:solar:simul2} (a), the only solar neutrino branch that can be observed at the ``baseline'' purity level is the $^{7}$Be solar neutrino, which manifests itself as an edge around $T_{\rm max} = 665$ keV above the background. The extraction of this signal, however, requires a precise determination of the $^{210}$Bi background from the spectrum fit. On the other hand, the cosmogenic $^{11}$C doesn't cause background to the solar $^{7}$Be neutrinos, because it undergoes $\beta+$ decay, thus has a minimum energy of 1.022 MeV. JUNO's high energy resolution (3\%) makes sure that there is no leakage into the $^{7}$Be spectrum. At the ``baseline'' purity level, besides the dominating $^{210}$Bi background coming from the decay of $^{210}$Pb, $^{85}$Kr, $^{238}$U, and $^{40}$K all contribute non-negligibly in the $^{7}$Be signal range, thus need to be determined by other means (e.g., spectrum fitting, independent sample measurement, coincidence tagging, etc.) Detection of other low enegy solar neutrino branches is difficult. The solar $pep$ (and CNO) neutrino signal is overwhelmed by $^{210}$Bi at the low energy and by $^{11}$C at the high energy. The $pp$ solar neutrino signal is overwhelmed by $^{14}$C at the low energy and by $^{210}$Bi at the high energy.

Assuming that the ``ideal'' purity level in Table ~\ref{tab:solar:singles} can be achieved (via online distillation and other means), the signal-to-background ratio will be largely improved, as shown in Fig.~\ref{fig:solar:simul2} (b). The $^{7}$Be solar neutrino signal rate will be about three times higher than the total of all other backgrounds. The ES scattering edge is clearly visible in Fig.~\ref{fig:solar:simul2} (b).

JUNO's high energy resolution even makes it possible to observe the solar $pp$ neutrinos. This is because the intrinsic $^{14}$C background ends at 156 keV, therefore there exists a window from approximately 160 keV to 230 keV where the $pp$ neutrino flux is the dominating component of the singles spectrum. It manifests itself as a rising edge above the $^{7}$Be solar neutrino spectrum, as shown in Fig.~\ref{fig:solar:simul2} (b). The observation of $pp$ solar neutrinos, however, requires a good pulse-shape discrimination to remove the low energy quenched alpha events, a clean removal of pile-up events with waveform analysis, as well as good understanding of low energy noise events.

\subsection{Measurement of \texorpdfstring{$^8$B}{B-8} solar neutrinos at JUNO}
\label{subsec:solar:MSW}

The measurement of $^8$B solar neutrinos with a low energy threshold is possible at JUNO. Due to the much larger target mass, the counting statistics will be enlarged significantly with respect to the previous liquid scintillator experiments such as Borexino and KamLAND. The higher photoelectron yield makes it possible for JUNO to lower the energy threshold further. Similar to other solar neutrino components, the detection of $^8$B solar neutrinos is through the neutrino-electron elastic scattering (ES) channel. The signal is a single event in contrast to the prompt-delayed signal pair in the case of reactor antineutrino IBD reaction. Therefore, background need to be controlled to a low level. The intrinsic background, external background, reactor background, and cosmogenic background that are relevant for solar $^8$B neutrino detection will be discussed in the following.

The intrinsic background at high energy is dominated by the decay of $^{208}$Tl (Q = 5.0\,MeV, $\tau_{1/2}$ = 3 min), which comes from the contamination of $^{232}$Th in the liquid scintillator. As an internal radioactive impurity, the total energy from the cascading $\beta$- and $\gamma$-decays adds up to the Q-value of 5 MeV. This background cannot be removed by the fiducial volume cut. Assuming that secular equilibrium is reached, the $^{208}$Tl internal background can be measured via the $\beta$--$\alpha$ delayed coincidence from the $^{212}$Bi$-^{212}$Po decay chain and then statistically subtracted. Nonetheless, the internal $^{232}$Th contamination need to be controlled to $10^{-17}$ g/g level in order to lower the analysis threshold to much below 5 MeV.

The external background, at the low energy end, is dominated by the $2.6$ MeV $\gamma$-rays from $^{208}$Tl from the PMTs. Higher energy external gamma rays mainly come from the $(n,\gamma)$ reaction in the surrounding materials. For example, neutron captures on stainless steel yield 6-MeV and 8.5-MeV gamma rays that can penetrate into the central detector. Energetic neutrons can also excite heavy nuclei in the surrounding material and produce high energy gamma rays. The external background can be efficiently reduced by applying a fiducial volume cut. A preliminary Monte Carlo study indicates that at least 5\,m shielding is necessary~\cite{Mollenberg:2013qka}, which will significantly reduce the fiducial mass by more than 50\%.

The reactor background arises via elastic scattering (ES) of reactor antineutrinos off electrons in the liquid scintillator. The reactor ES event rate is estimated to be about 0.5 per day per kton, spanning over an energy range from 1 to 8 MeV. In total, the reactor ES background contribution is at about 5\% level of the $^8$B solar neutrino signal. Furthermore, its contribution can be measured accurately from the IBD reaction and statistically subtracted with high precision.

\begin{table}[!tb]
\begin{center}
\begin{tabular}{|c|c|c|c|c|c|}
\hline
Isotope& Decay Type & Q-Value  & Life time & Yield \cite{Abe:2009aa,Bellini:2013pxa}&Rate \\
 &   &  [MeV] &  & $10^{-7}$ $(\mu \, {\rm g}/{\rm cm}^2)^{-1}$ & [cpd/\,kton]\\
\hline
$^{11}$C& $\beta^+$ & 2.0 & 29.4\,min & 866 & 1860\\
$^{10}$C& $\beta^+$ & 3.7 & 27.8\,s & 16.5 & 35\\
$^{11}$Be& $\beta^-$ & 11.5 & 19.9\,s & 1.1& 2\\

\hline
\end{tabular}
\end{center}
\caption[Cosmogenic radioisotopes]
{List of the cosmogenic radioisotopes whose lifetimes are above 2\,s, which are the main backgrounds for $^8$B solar neutrino detection. Shorter-lifetime spallation products can be efficiently suppressed with proper muon veto cuts (see text for details). }
\label{tab:solar:cosmogenic_isotopes}
\end{table}

Above 5 MeV, the dominant background is caused by the cosmogenic isotopes, which are produced in-situ by spallation reactions of cosmic muons on the carbon nuclei in the liquid scintillator. The muon rate in the 20-kton JUNO central detector is 3\,Hz. The spallation products are typically unstable and undergo $\beta^-/\beta^+$ decays. From the Monte Carlo study, the short-lived spallation isotopes ($\tau \le \sim 1$\,s) can be efficiently suppressed by vetoing a cylindrical volume with 1\,m radius around each traversing muon track for 6.5 seconds. Since some of the muon events are muon bundles (multiple muons in one event), and some of the muons produce electromagnetic and hadronic showers (showering muons), good track reconstructions for muon bundles and showering muons are necessary (see the reactor neutrino chapter for details of muon reconstructions) for an efficient rejection of short-lived spallation backgrounds.

The remaining long-lived spallation radioisotopes are $^{11}$C ($\beta^+$, $\tau=29.4\,$min), $^{10}$C ($\beta^+$, $\tau=27.8\,$s), and $^{11}$Be ($\beta^-$, $\tau=19.9\,$s). Due to the long lifetime, they are difficult to remove without losing a large fraction of detector livetime. Therefore, their rates and spectra need to be measured accurately and subtracted. Their decay information, spallation yields, and estimated rates at JUNO are summarized in Table~\ref{tab:solar:cosmogenic_isotopes}. The expected rates are scaled from
the KamLAND and Borexino spallation measurements~\cite{Abe:2009aa,Bellini:2013pxa} similarly as described in the previous low energy solar neutrino section. Their expected energy spectra at JUNO are shown in Fig.~\ref{fig:solar:boro}, together with the expected $^8$B solar neutrino signal spectrum. The energy resolution is assumed to be $\sigma(E) = 3\% \times \sqrt{E \textrm{(MeV)}}$. For simplicity, no energy non-linearity is applied to the spectrum. One can see that due to the relatively shallow depth of JUNO, the long-lived spallation radioisotopes are the major background sources for $^8$B solar neutrino detection.
\begin{figure}[!tb]
\vspace{0 cm}
\begin{center}
\includegraphics[width=0.7\textwidth]{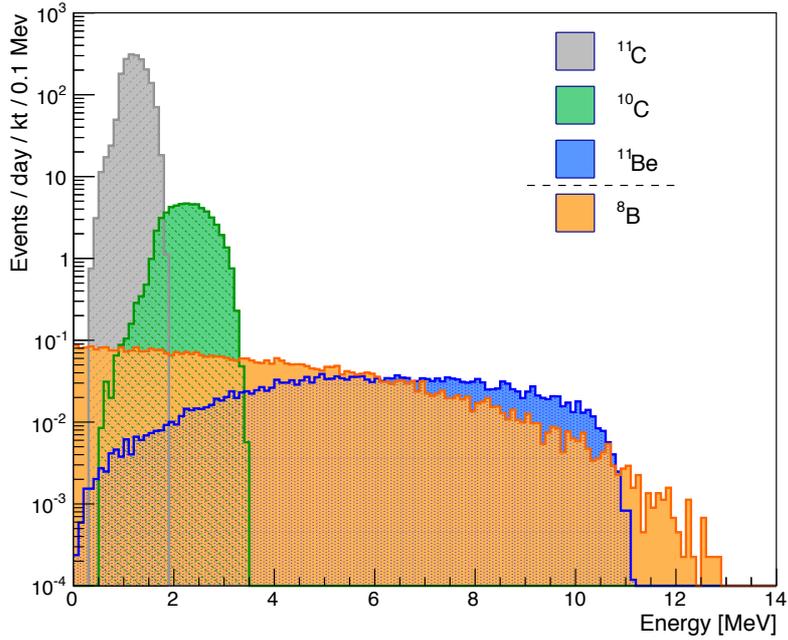}
\vspace{0 cm}
\caption{The simulated background spectra for the cosmogenics isotopes $^{11}$C, $^{10}$C, and $^{11}$Be at JUNO. Furthermore, the expected $^8$B ($\nu-e$) spectrum is shown for comparison. A reduction of $^{10}$C and $^{11}$C should be possible by Three-Fold Coincidence but is not applied in the figure (see text for details).}
\label{fig:solar:boro}
\end{center}
\end{figure}

Typically, neutrons are produced along with the spallation products. For example, the primary processes to produce the $^{11}$C and $^{10}$C isotopes are $(\gamma,n)$ and $(\pi^+, np)$, respectively. Therefore, a Three-Fold Coincidence (cosmic muon, neutron capture, and isotope decay) can be used to further suppress the $^{11}$C and $^{10}$C backgrounds. Since neutrons are captured with a lifetime of about 200\,$\mu$s, the electronics needs to recover within a few microseconds after high-multiplicity events in order to minimize the dead time and to have a high neutron tagging efficiency. The effect of neutrons leaking out of the scintillation region can be mitigated by a fiducial volume cut. Further rejection is possible by studying the underlining mechanism of spallation production and possible signatures from them~\cite{Li:2014sea,Li:2015kpa}.
We note that the rates and the spectra of the spallation products can also be measured in-situ with enhanced samples by selecting the candidates inside the veto volume close to the muon tracks.

\subsection{Conclusions}

The JUNO detector has many advantages in performing solar neutrinos measurements compared with previous detectors. Being a liquid scintillator detector similar to Borexino and KamLAND, it has the benefit of high light yield and, therefore, very high energy resolution and low energy threshold. Being a massive 20 kton detector it will have large statistics comparable to the Super-Kamiokande water Cherenkov detector. This makes JUNO an attractive detector to further improve the measurement precision of various components of the solar neutrino flux, shed light on the solar metallicity problem, and probe the transition region between the vacuum-dominated and MSW-dominated neutrino oscillations.  The solar neutrino measurements, however, demand challengingly low levels of radio-impurities and accurate determination of cosmogenic backgrounds. Since JUNO is optimized for reactor antineutrino measurements with relatively lenient background requirement, dedicated efforts to realize the low background phase for solar neutrino measurements are necessary.

\clearpage

\section{Atmospheric Neutrinos}
\label{sec:atm}

\blfootnote{Editors: Wanlei Guo (guowl@ihep.ac.cn) and Christopher Wiebusch (wiebusch@physik.rwth-aachen.de)}
\blfootnote{Major contributors: Michael Soiron and Zhe Wang}

\subsection{Introduction} \label{subsec:atm:Introduction}

A compelling three-flavor neutrino oscillation framework has been
established by various atmospheric, solar, reactor, and accelerator
neutrino experiments so far. Reviews of the progress can be found in
many articles, for example, in Ref.~\cite{Capozzi:2013csa}.
Precision of the oscillation parameters has been improved
significantly with recent experiments. However, there are still
several major unknowns in neutrino oscillation physics. Many new
experiments with various neutrino sources and detector technologies
are designed to address these questions~\cite{deGouvea:2013onf}.
Here, we explore the capabilities of JUNO to determine the neutrino
mass hierarchy (MH), the octant of atmospheric mixing angle
$\theta_{23}$ and the Dirac CP violation phase $\delta$ using
atmospheric neutrinos.

Atmospheric neutrinos are a very important neutrino source to study
the neutrino oscillation physics. In 1998, the Super-Kamiokande
experiment reported the first evidence of neutrino oscillations
based on a zenith angle dependent deficit of atmospheric muon
neutrinos~\cite{Fukuda:1998mi}. Atmospheric neutrinos have a broad
range in baseline (15 km $\sim$ 13000 km) and energy (0.1 GeV $\sim$
10 TeV), and contain neutrinos and antineutrinos of all flavors.
When they pass through the Earth, the Mikheyev-Smirnov-Wolfenstein
(MSW) matter effect~\cite{Wolfenstein:1977ue,Mikheev:1986gs} will play a key role
in answering the above three open questions. Super-Kamiokande has
reported the preliminary results on these issues based on $4538$
days of data~\cite{Wendell:2014dka}. Future atmospheric neutrino
experiments, such as PINGU~\cite{Aartsen:2014oha}, ORCA
\cite{Katz:2014tta} Hyper-Kamiokande~\cite{Abe:2011ts,Abe:2015zbg} and INO
\cite{Ahmed:2015jtv}, anticipate improved sensitivities.

The JUNO central detector as a liquid scintillator (LS) calorimeter
has a very low energy threshold and can measure atmospheric
neutrinos with excellent energy resolution. Characteristic signals
from Michel electrons, neutron captures and unstable daughter nuclei
are helpful for the particle recognition. Note that the JUNO LS
detector also has some capabilities to reconstruct the directions of
charged leptons in terms of the timing pattern of the first-hit on
the PMTs (See appendix and Refs.~\cite{Learned:2009rv,Wurm:2011zn}).
Based on the above capabilities, JUNO is a promising
detector for atmospheric neutrino oscillation measurements.

\subsection{Atmospheric Neutrino Oscillations}
\label{subsec:atm:Oscillation}

Atmospheric neutrinos are produced in the Earth's atmosphere as a
result of cosmic ray interactions and the weak decays of secondary
mesons, in particular pions and kaons. The competition between the
particle's decay and re-interaction probabilities leads to
characteristic shapes for neutrino's energy spectrum and angular
distributions. At energies relevant for JUNO the production of
$\nu_\mu$ and $\bar\nu_\mu$ is dominated by the decay chains
\begin{eqnarray}
 \pi^+  \to  \mu^+ + \nu_\mu, & &  \mu^+ \to  e^+ + \bar{\nu}_\mu + \nu_e; \\
 \pi^-  \to  \mu^- + \bar{\nu}_\mu, &  & \mu^-  \to  e^- +
\nu_\mu + \bar{\nu}_e.
\end{eqnarray}
At low energies below typically one GeV, all parent particles in the
decay chain decay at equal probability and the expected flux ratios
reflect the production ratios of parent mesons
$\frac{\phi_{\nu_\mu}}{\phi_{\bar{\nu}_\mu}} \simeq 1 $,
$\frac{\phi_{{\nu}_e}}{\phi_{\bar{\nu}_e}} \simeq 1 $ and $
\frac{\phi_{\nu_\mu} + \, \phi_{{\bar \nu}_\mu}}{\phi_{\nu_e} + \,
\phi_{\bar{\nu}_e}} \ge 2 $. These ratios increase at higher energies
as muons are less likely to decay before hitting ground. The
neutrino energy spectrum initially follows the primary cosmic ray
spectrum $\propto E^{-2.7} $ and becomes steeper at higher energy
reflecting the decreasing decay probability of parent particles. The
angular distribution exhibits a characteristic shape with an
increased flux towards the horizon due to the effects of the longer
path length together with lower interaction probability at the high
altitude. At even higher energies of the order of $\sim 100$\,GeV
the decay of kaons becomes more important, because of their shorter
lifetime with respect to pions. The flux of atmospheric neutrinos
has been simulated with increasing precision over the last decades.
The flux of atmospheric muon neutrinos ($\nu_\mu + \bar{\nu}_\mu $)
resulting from the recent calculation~\cite{Honda:2011nf} is shown
in Fig. \ref{fig:atm:Flux}. An up-down asymmetry appears at sub-GeV
energies due to the Earth magnetic field and depends on the
geographical location of the experiment. The suppression of the
electron neutrino flux becomes apparent above about $1$\,GeV. Note
that the above calculations show the fluxes at production level and
do not include flavor oscillations during propagation through the
Earth which are discussed below.

\begin{figure}[!htb]
\begin{center}
\includegraphics[scale=0.43]{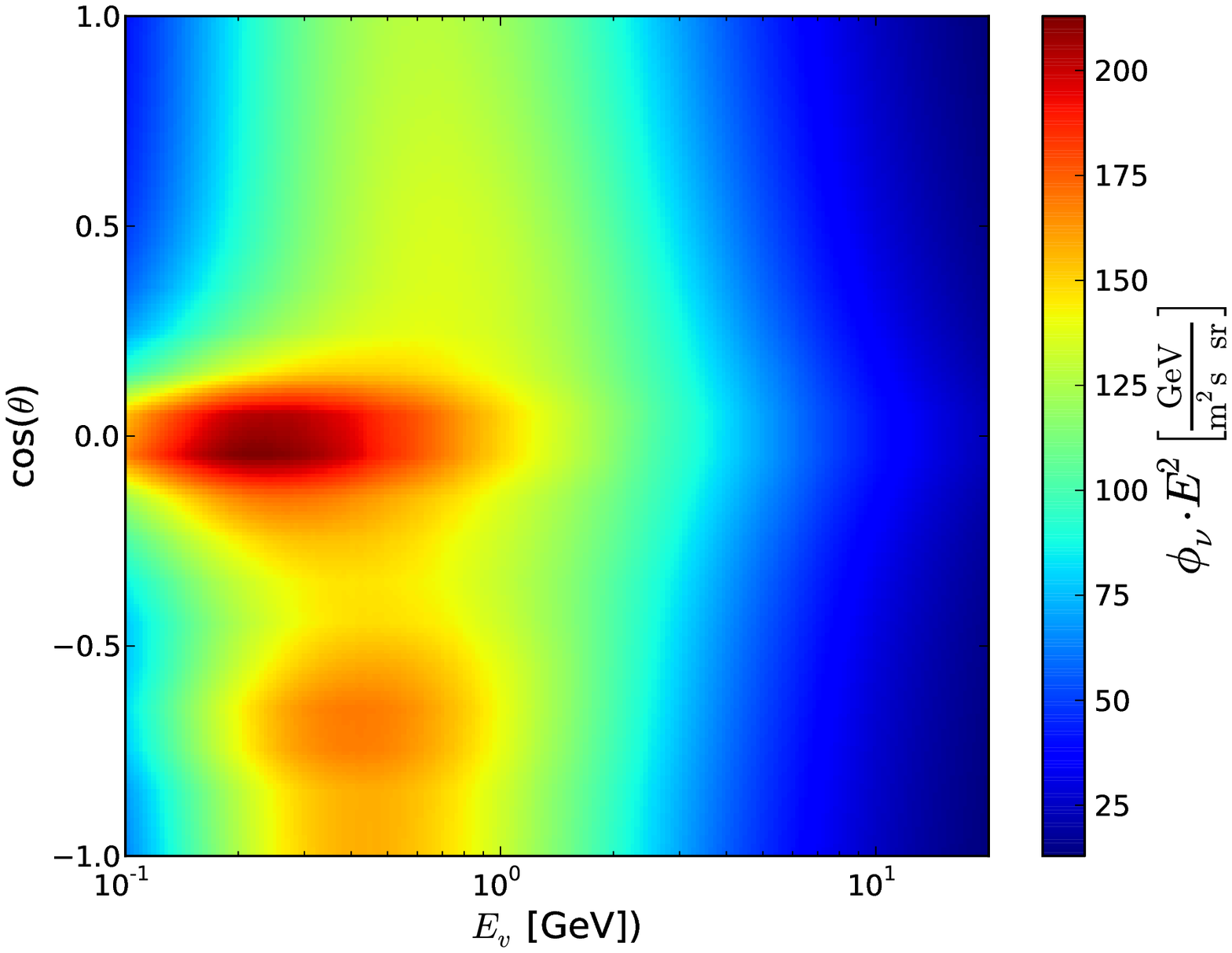}
\includegraphics[width=7.3cm,height=6.57cm]{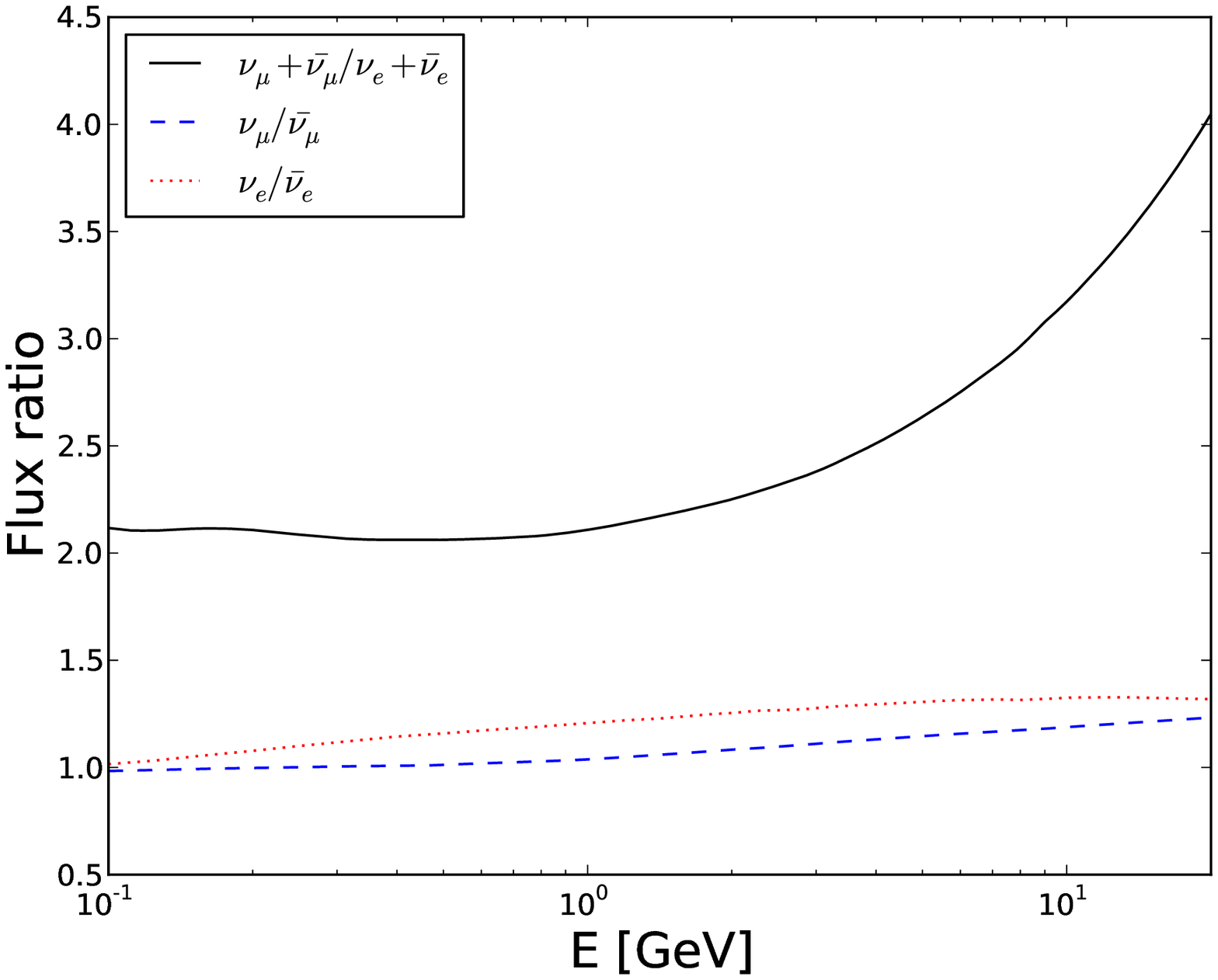}
\end{center}
\caption{The differential ($\nu_\mu + \bar{\nu}_\mu $) flux density multiplied with
$E^2$  versus energy and zenith angle (left) and flux ratios of
different flavors versus energy averaged over all zenith angles
(right). } \label{fig:atm:Flux}
\end{figure}

\begin{figure}[!htb]
\begin{center}
\includegraphics[scale=0.45]{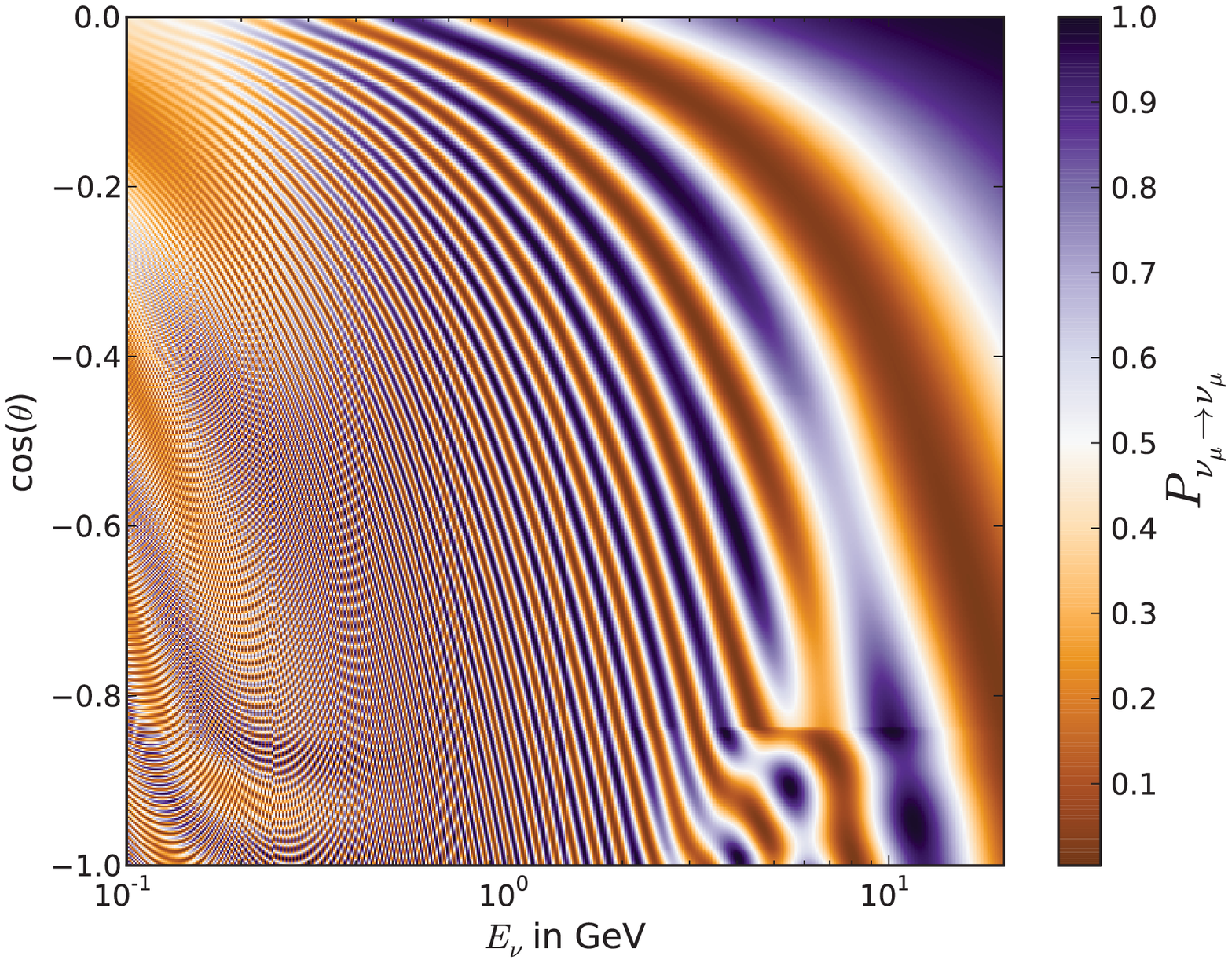}
\hspace{0.1cm}
\includegraphics[scale=0.45]{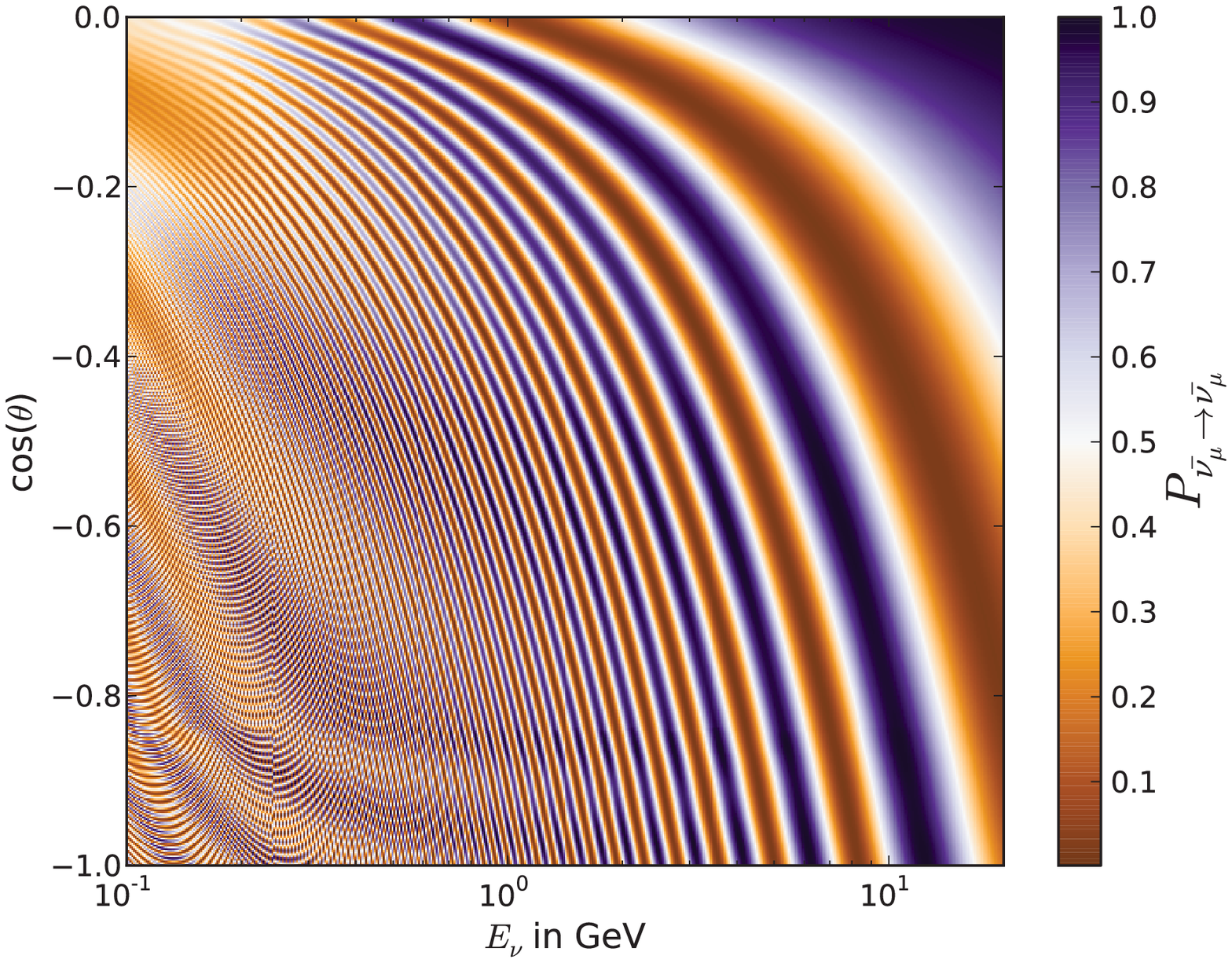}
\vspace{0.6cm}

\includegraphics[scale=0.45]{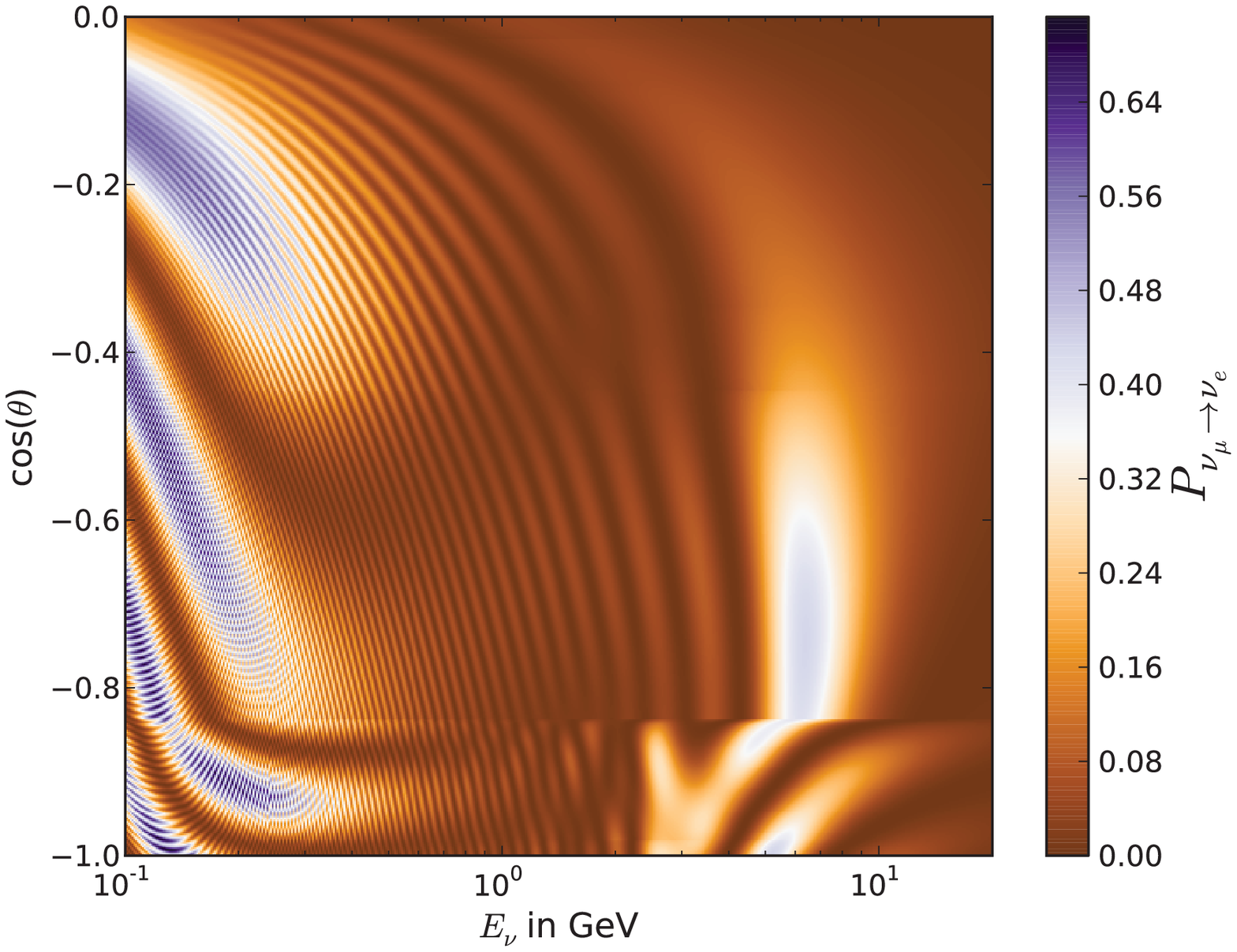}
\hspace{0.1cm}
\includegraphics[scale=0.45]{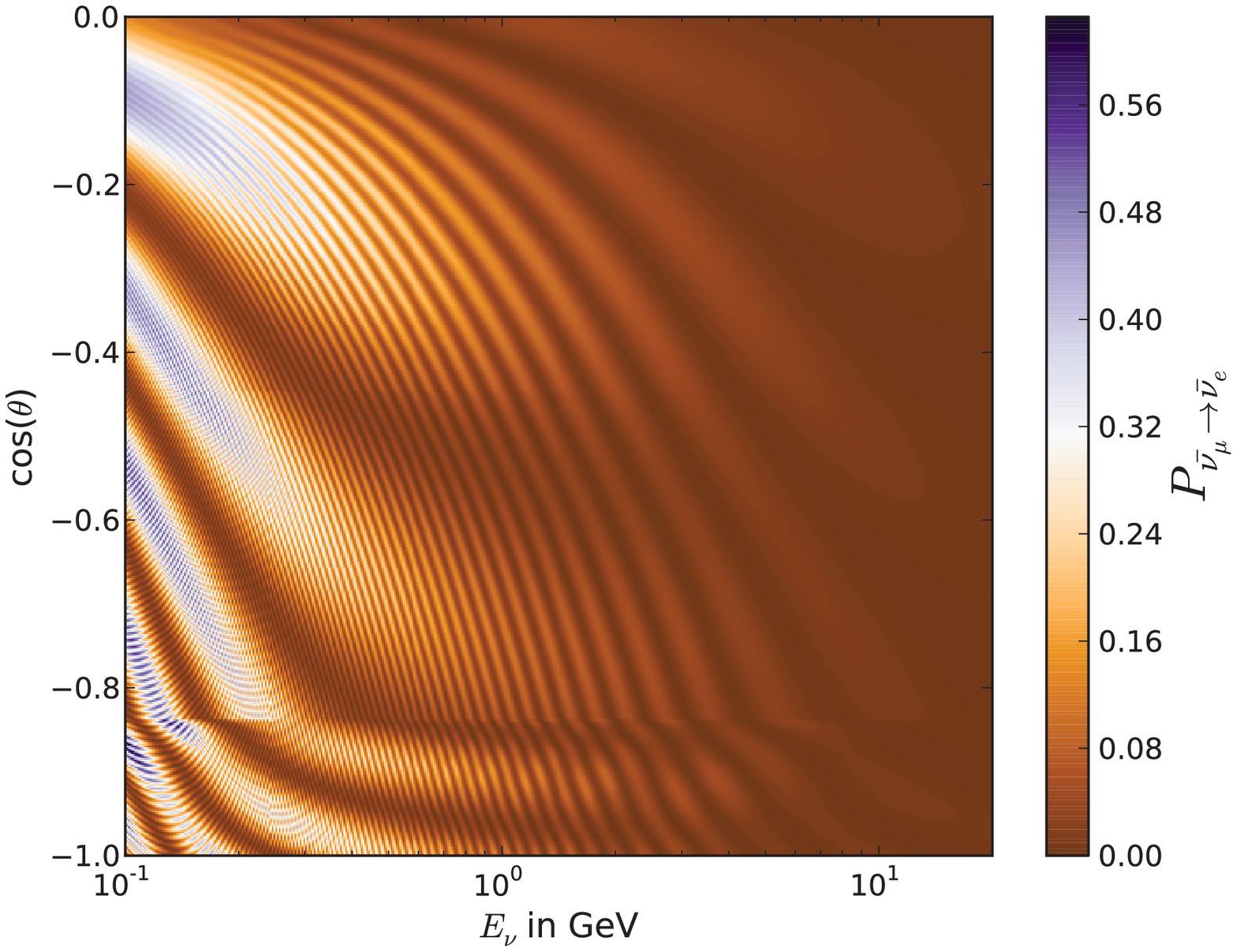}
\vspace{0.5cm}

\includegraphics[scale=0.45]{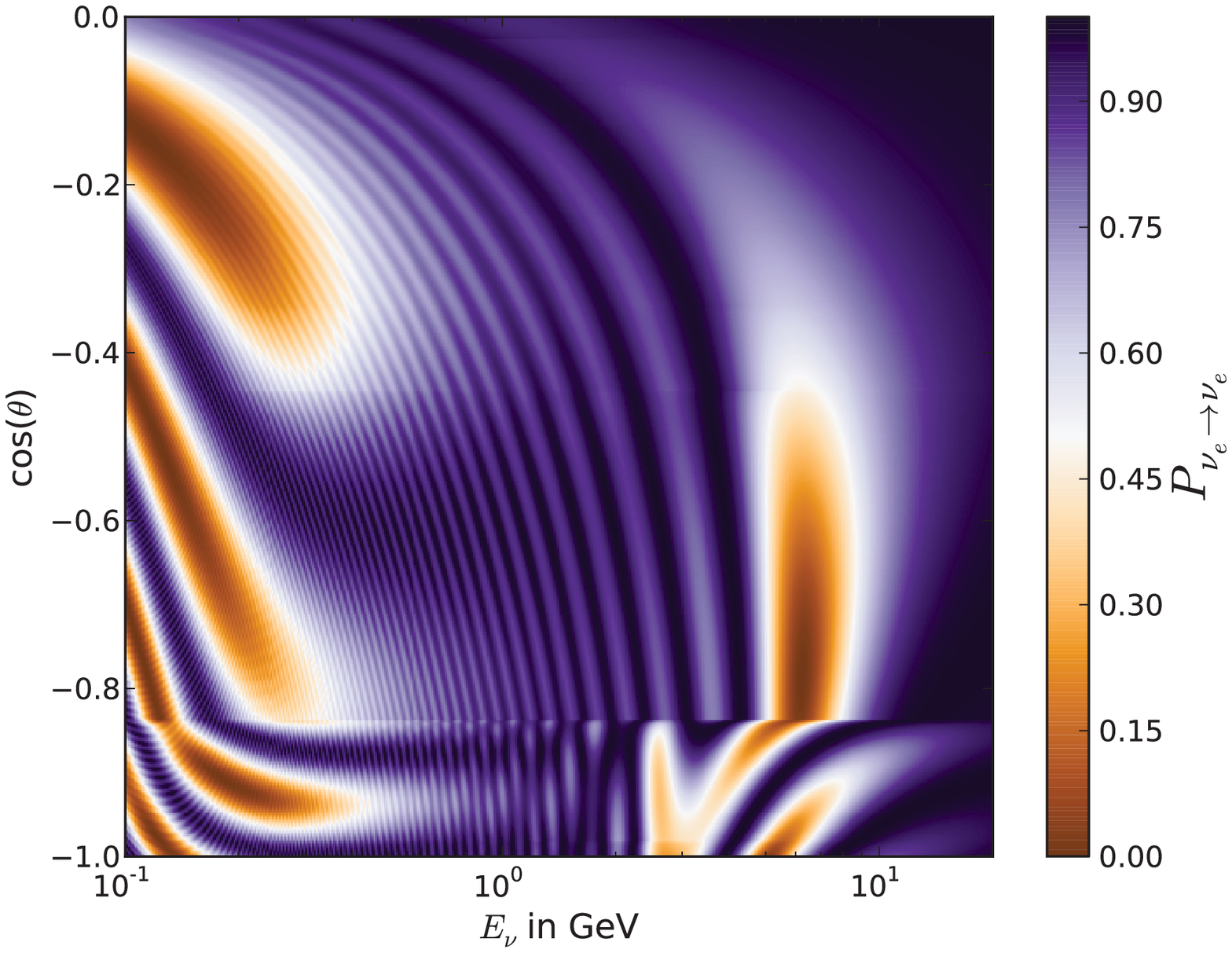}
\hspace{0.1cm}
\includegraphics[scale=0.45]{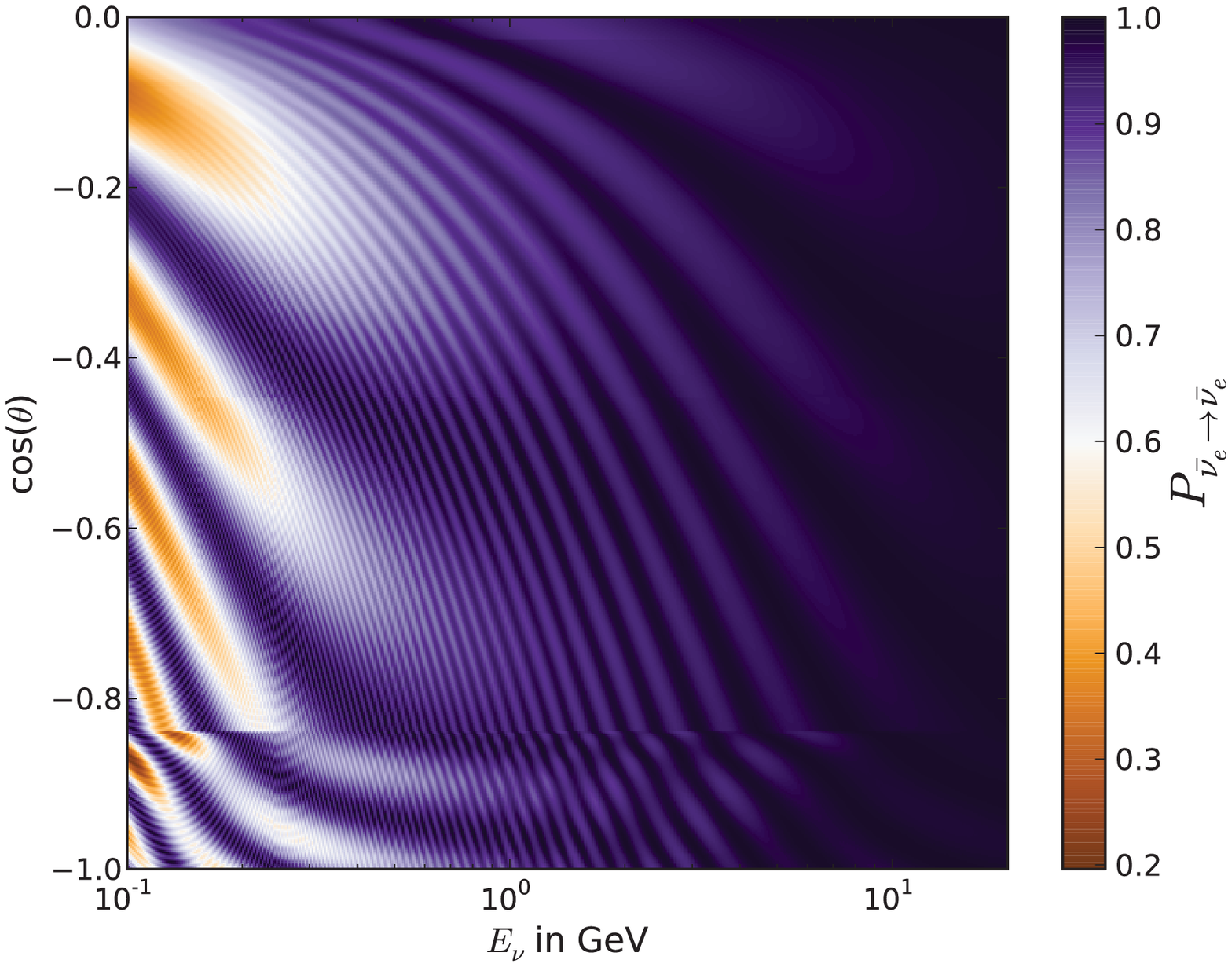}
\end{center}
\caption{Six relevant oscillograms of oscillation probabilities for
atmospheric neutrinos and antineutrinos in the normal hierarchy
hypothesis.} \label{fig:atm:probability}
\end{figure}

When atmospheric neutrinos propagate in the Earth, the evolution of
the flavor eigenstates is given by
\begin{eqnarray}
i \frac{d \nu_f}{d t} = \left ( U \frac{M^2}{2 E_{\nu_f}} U^\dag + V
\right ) \nu_f, \label{eqn:atm:evolution}
\end{eqnarray}
with $M^2 = {\rm diag} (m_1^2, m_2^2, m_3^2)$ and $V = {\rm diag} (
\sqrt{2} G_F N_e , 0, 0)$. For the antineutrinos,
Eq.~(\ref{eqn:atm:evolution}) is still valid when replacing the
leptonic mixing matrix $U \rightarrow U^*$ and the effective
potential $V \rightarrow -V$. $G_F$ is the Fermi coupling constant
and $N_e = Y_e \rho$ is the electron number density with $Y_e =
0.466$ for the core and $Y_e = 0.494$ for the mantle
\cite{Lisi:1997yc}. In terms of the PREM Earth density profile
\cite{Dziewonski:1981xy}, we numerically solve Eq.~(\ref{eqn:atm:evolution}) and
calculate six typical oscillation probabilities by use of nuCraft
\cite{Wallraff:2014qka} for the normal hierarchy (NH) hypothesis as
shown in Fig. \ref{fig:atm:probability}. Here we take the Dirac CP
phase $\delta = 0$  and use the current best known values of the
following oscillation parameters~\cite{Capozzi:2013csa}:
\begin{eqnarray}
\sin^2 \theta_{12} & = & 0.308^{+0.051}_{-0.049}, \nonumber\\
\sin^2 \theta_{13} & = & 0.0234^{+0.0061}_{-0.0058}, \nonumber\\
\sin^2 \theta_{23} & = & 0.437^{+0.187}_{-0.063}, \nonumber\\
\Delta m_{21}^2 &  = & 7.54^{+0.64}_{-0.55} \times 10^{-5} {\rm eV^2} ,\nonumber\\
\Delta m_{atm}^2 &  = & 2.43^{+0.18}_{-0.20} \times 10^{-3} {\rm
eV^2}, \label{eqn:atm:parameter}
\end{eqnarray}
where $\Delta m_{atm}^2 = m_3^2 - (m_1^2 +m_2^2)/2$. The matter
effects for neutrinos traveling through the dense Earth core ($ \cos
(\theta) \le -0.82$) is obvious. In particular, in the GeV region
the structures strongly differ for neutrinos and antineutrinos, and
these structures swap between neutrinos and antineutrinos for the
inverted hierarchy (IH) hypothesis.

A set of analytical expressions of the oscillation probabilities in
Fig. \ref{fig:atm:probability} are very convenient for us to
understand the oscillation features. For illustrative purposes we
adopt the ``one dominant mass scale" approximation ($\Delta m_{21}^2
\ll |\Delta m_{31}^2|$) and have the following formulas
\cite{Cervera:2000kp,Freund:2001pn,Akhmedov:2004ny, Choubey:2005zy}:
\begin{eqnarray}
 P(\nu_e\rightarrow \nu_e) & \approx & 1 - \sin^2 2\theta_{13}^m \sin^2 \left[
1.27
(\Delta m_{31}^2)^m \frac{L}{E_\nu} \right] ; \label{eqn:atm:Pee}\\
 P(\nu_e \rightarrow \nu_\mu)  & \approx & P(\nu_\mu \rightarrow \nu_e) \approx
\sin^2 \theta_{23} \sin^2 2\theta_{13}^m \sin^2 \left[ 1.27 (\Delta
m_{31}^2)^m \frac{L}{E_\nu} \right] ; \label{eqn:atm:Peu} \\
 P(\nu_\mu \rightarrow \nu_\mu) & \approx & 1 - \cos^2 \theta_{13}^m \sin^2
2\theta_{23} \sin^2 \left[ 1.27 \frac{\Delta m_{31}^2 + A +(\Delta
m_{31}^2)^m}{2}  \frac{L}{E_\nu} \right] \nonumber\\
 & & - \sin^2 \theta_{13}^m \sin^2 2\theta_{23} \sin^2 \left[ 1.27
\frac{\Delta m_{31}^2 + A -(\Delta m_{31}^2)^m}{2}  \frac{L}{E_\nu}
\right] \nonumber\\
& & - \sin^4 \theta_{23} \sin^2 2\theta_{13}^m \sin^2 \left[ 1.27
(\Delta m_{31}^2)^m \frac{L}{E_\nu} \right], \label{eqn:atm:Puu}
\end{eqnarray}
where $A = 2 \sqrt{2} G_F N_e E_\nu$ and the superscript $m$ denotes
the effective quantities in matter. The units of $L$, $E_\nu$ and
$\Delta m_{31}^2$ are in units of km, GeV and eV$^2$, respectively.
The effective mass squared difference $(\Delta m_{31}^2)^m$ and
mixing angle $\sin^2 2\theta_{13}^m$ are given by
\begin{eqnarray}
(\Delta m_{31}^2)^m & = & \Delta m_{31}^2 \sqrt{(\cos 2\theta_{13}
-A/\Delta m_{31}^2)^2 +  \sin^2 2\theta_{13}} \,, \label{eqn:atm:Delta_m31} \\
\sin^2 2\theta_{13}^m & = & \frac{\sin^2 2\theta_{13}}{( \cos
2\theta_{13} -A/\Delta m_{31}^2)^2 +  \sin^2 2\theta_{13}}\,.
 \label{eqn:atm:Theta13}
\end{eqnarray}
It should be worthwhile to stress that
Eqs.~(\ref{eqn:atm:Pee}-\ref{eqn:atm:Puu}) are valid when $\Delta
m_{21}^2 L/E_\nu \ll 1$, namely $L/E_\nu \ll 13263$ km/GeV. The
expansion in the small parameters ($\sin \theta_{13}$ and $\Delta
m_{21}^2$) can be found in Ref.~\cite{Akhmedov:2004ny}. For
antineutrinos, the corresponding oscillation probabilities
$P(\bar{\nu}_e \rightarrow \bar{\nu}_e), P(\bar{\nu}_e \rightarrow
\bar{\nu}_\mu)$ and $ P(\bar{\nu}_\mu \rightarrow \bar{\nu}_\mu)$
can be derived from Eqs.~(\ref{eqn:atm:Pee}-\ref{eqn:atm:Theta13})
with $A \rightarrow -A$. It is clear that neutrinos and
antineutrinos have the same oscillation probabilities for the
opposite mass hierarchies:
\begin{eqnarray}
P_{\rm NH}(\nu_\alpha \rightarrow \nu_\beta) = P_{\rm
IH}(\bar{\nu}_\alpha \rightarrow \bar{\nu}_\beta)\;, P_{\rm
IH}(\nu_\alpha \rightarrow \nu_\beta) = P_{\rm NH}(\bar{\nu}_\alpha
\rightarrow \bar{\nu}_\beta)\;. \label{eqn:atm:NHIHrelationship}
\end{eqnarray}
Therefore, one can image the oscillation probabilities in the IH
case from Fig. \ref{fig:atm:probability} if the condition $\Delta
m_{21}^2 L/E_\nu \ll 1$ is satisfied.

When the matter potential term $A \ll |\Delta m_{31}^2| \cos
2\theta_{13}$ for the downward atmospheric neutrinos ($\cos(\theta)
> 0$), we may neglect $A$ and obtain the vacuum oscillation
probabilities from the above equations. Then, these vacuum
oscillation probabilities are the same for the NH and IH cases. If
$A = \Delta m_{31}^2 \cos 2\theta_{13}$, the MSW resonance will
significantly enhance the effective mixing angle $\sin^2
2\theta_{13}^m \rightarrow 1$. Note that the MSW resonance
enhancement occurs for neutrinos in the normal mass hierarchy and
for antineutrinos in the inverted mass hierarchy. The resonance
energy can be written as
\begin{eqnarray}
E_\nu = \frac{\Delta m_{31}^2 \cos 2\theta_{13}}{2 \sqrt{2} G_F N_e}
= 32.1 {\rm GeV} \frac{g/{\rm cm^3} }{\rho} \frac{0.5}{Y_e}
\frac{\Delta m_{31}^2}{ 2.43 \times 10^{-3} {\rm eV^2}} \cos
2\theta_{13} \;.
 \label{eqn:atm:Ev_resonance}
\end{eqnarray}
Considering the Earth density profile, we can obtain the resonance
energy range $E_\nu \approx 3-10$ GeV for the atmospheric neutrinos
passing through the Earth. In the resonance case, the oscillation
probabilities in Eqs.~(\ref{eqn:atm:Pee}-\ref{eqn:atm:Puu}) have
significant differences for the NH and IH hypotheses. With the help
of Eq.~(\ref{eqn:atm:NHIHrelationship}), we can easily find these
differences from Fig.\ref{fig:atm:probability}. Therefore the upward
atmospheric neutrinos can be used to probe the neutrino mass
hierarchy.

The oscillation probabilities in
Eqs.~(\ref{eqn:atm:Pee}-\ref{eqn:atm:Puu}) do not include the CP
phase $\delta$ because we have ignored the subleading terms. Note
that these approximated expressions are only valid in the case of
$\Delta m_{21}^2 L/E_\nu \ll 1$. In order to estimate the impact of
the CP phase $\delta$, we numerically calculate the oscillation
probabilities in the neutrino energy and zenith angle plane while
scanning $\delta$ from 0 to 2$\pi$. In Fig.
\ref{fig:atm:CPVariation}, we plot the maximum variations due to
$\delta$ for $P(\nu_e \rightarrow \nu_\mu)$ and $P(\nu_\mu
\rightarrow \nu_\mu)$. Note that $P(\nu_\mu \rightarrow \nu_e)$ and
$P(\nu_e \rightarrow \nu_\mu)$ have the same results since
\begin{eqnarray}
P(\nu_\alpha \rightarrow \nu_\beta) = P(\nu_\beta \rightarrow
\nu_\alpha) (\delta \rightarrow - \delta)
 \label{eqn:atm:cp_relation}
\end{eqnarray}
for constant or symmetric matter density profiles
\cite{Akhmedov:2004ny}. In addition, the $P(\nu_e \rightarrow
\nu_e)$ case is not sensitive to $\delta$. Since only upward going neutrinos are
influenced by matter effects we only see differences for these directions as
shown in Fig. \ref{fig:atm:CPVariation}.
Because the expected rates are higher towards lower energies, we find that the
best region to search for the CP phase is the sub-GeV region.

\begin{figure}[!htb]
\begin{center}
\vspace{1.1cm}
\includegraphics[width=5.5cm,height=7.5cm,angle=270]{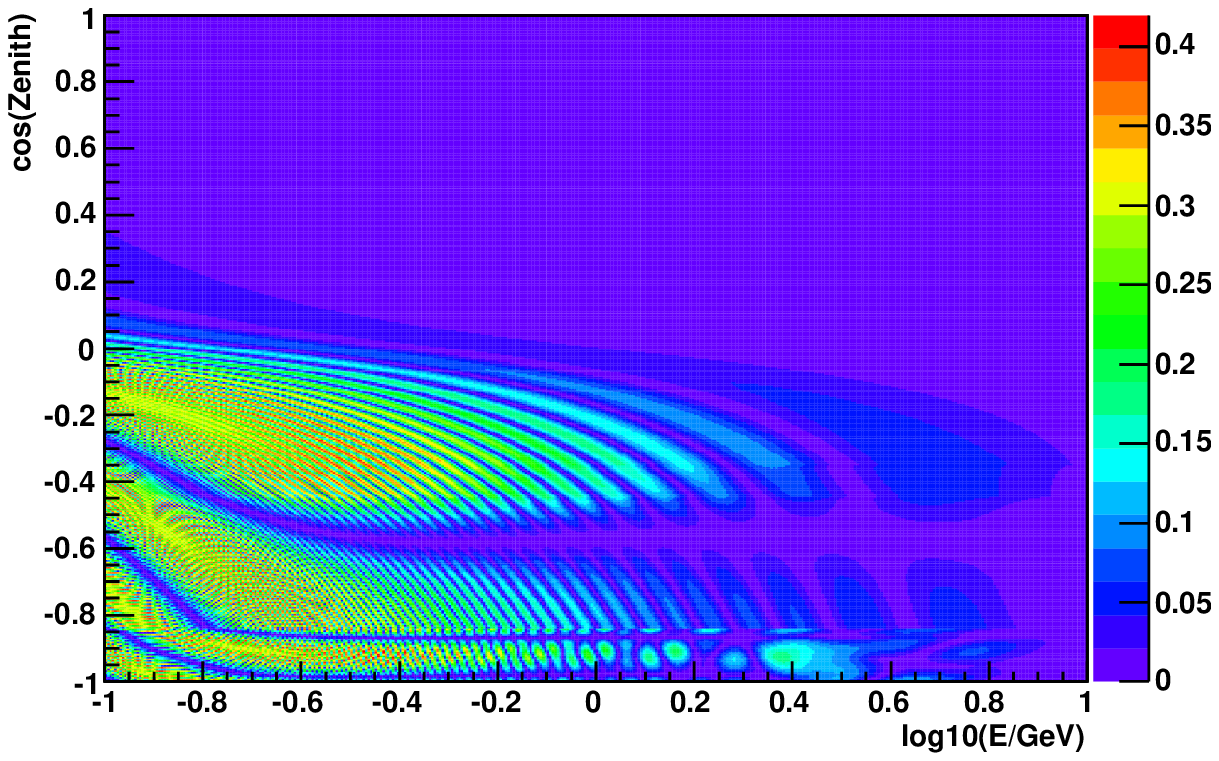}
\hspace{0.2cm}
\includegraphics[width=5.5cm,height=7.5cm,angle=270]{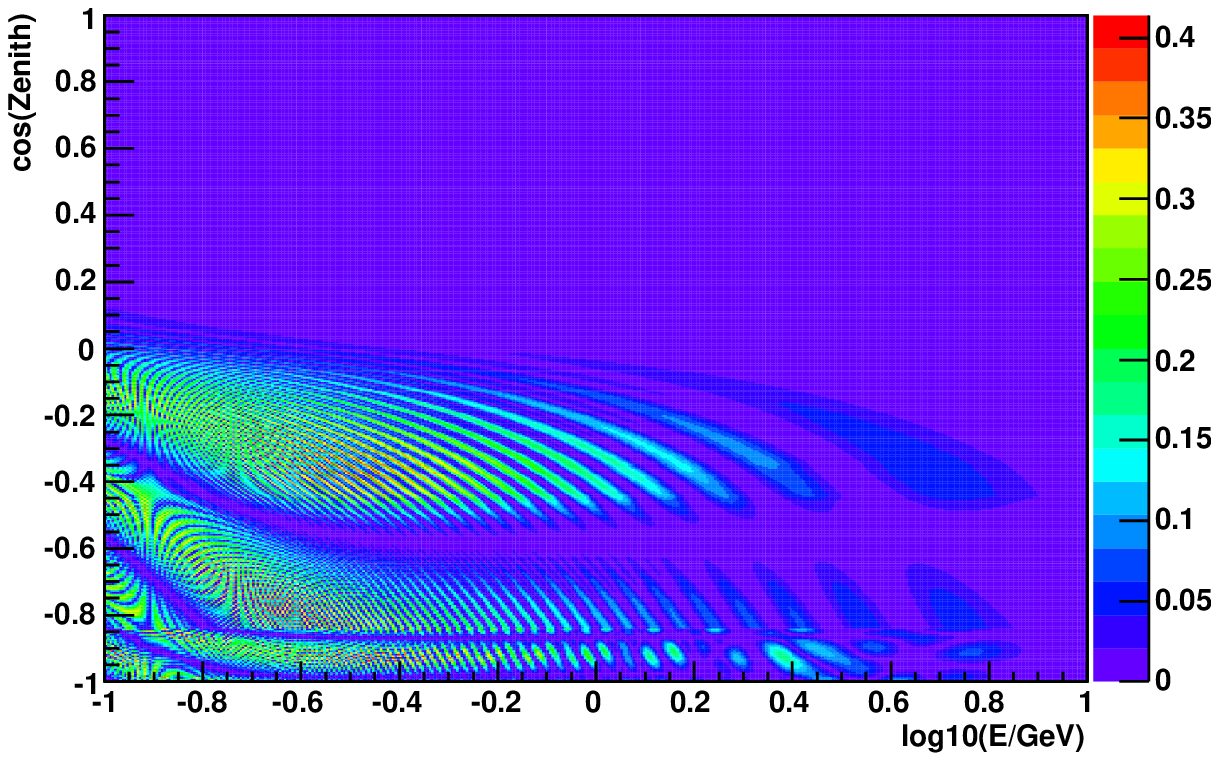}
\end{center}
\caption{Maximum variations of oscillation probabilities in the
energy and zenith angle plane for $P(\nu_e \rightarrow \nu_\mu)$ (
left) and $P(\nu_\mu \rightarrow \nu_\mu)$ (right) cases where we
scan the CP phase $\delta$ from 0 to 2$\pi$.}
\label{fig:atm:CPVariation}
\end{figure}

\subsection{Detector Performance} \label{subsec:atm:Detector}

\subsubsection{Monte Carlo Simulation}

According to the detector properties described in appendix, we
simulate atmospheric neutrino events using GENIE as generator
(Version 2.8.0)~\cite{Andreopoulos:2009rq} and Geant4 for
detector simulation. We use the atmospheric neutrino event
generation application in GENIE to generate 5 million events in JUNO
detector including oscillation effects and additionally 25 million
neutrino events which are later reweighted using the oscillation
probability with nuCraft~\cite{Wallraff:2014qka}. Unless otherwise
specified, we take the best fit values in
Eq.~(\ref{eqn:atm:parameter}), $\delta = 0$ and the normal hierarchy
to calculate the neutrino oscillation probabilities in this
subsection. For the first 5 million simulated events we have
performed a full detector simulation where the resulting particles
are propagated using the Geant4 simulation
\cite{Agostinelli:2002hh}. Geant4 then provide various quantities
for each neutrino event, such as the event vertex radius $R_\nu$,
visible energy $E_{vis}$, charged lepton's zenith angles $\theta_e$
and $\theta_\mu$, muon track length in the LS $L_\mu$, captured
neutron numbers $N_n$ and the Michel electron numbers $N_e$, etc.
The JUNO central detector can measure the neutrino energy very well.
For illustration, we plot the energy $E_\nu$ versus the visible
energy in the target region $E_{vis}$ for the
$\nu_\mu/\bar{\nu}_\mu$ CC interactions in the upper-left panel of
Fig. \ref{fig:atm:MC}. Since some particles can escape the LS region
with partially deposited energy, one finds $E_{vis} \ll E_\nu$ for
many events. If the event vertex $R_\nu < 16.7$ m and $\mu^\pm$
stops in the LS, the $E_{vis}$ smearing is small as shown in the
upper-right panel of Fig. \ref{fig:atm:MC}. In the lower panels of
Fig. \ref{fig:atm:MC}, we plot $\theta_\nu$ versus $\theta_\mu$ for
all (left) and $L_\mu \geq 5$ m (right) $\nu_\mu/\bar{\nu}_\mu$ CC
events. It is found that $\theta_\mu$ has smaller smearing for
$L_\mu \geq 5$ m case.

\begin{figure}[!htb]
\begin{center}
\vspace{1.1cm}
\includegraphics[scale=0.42]{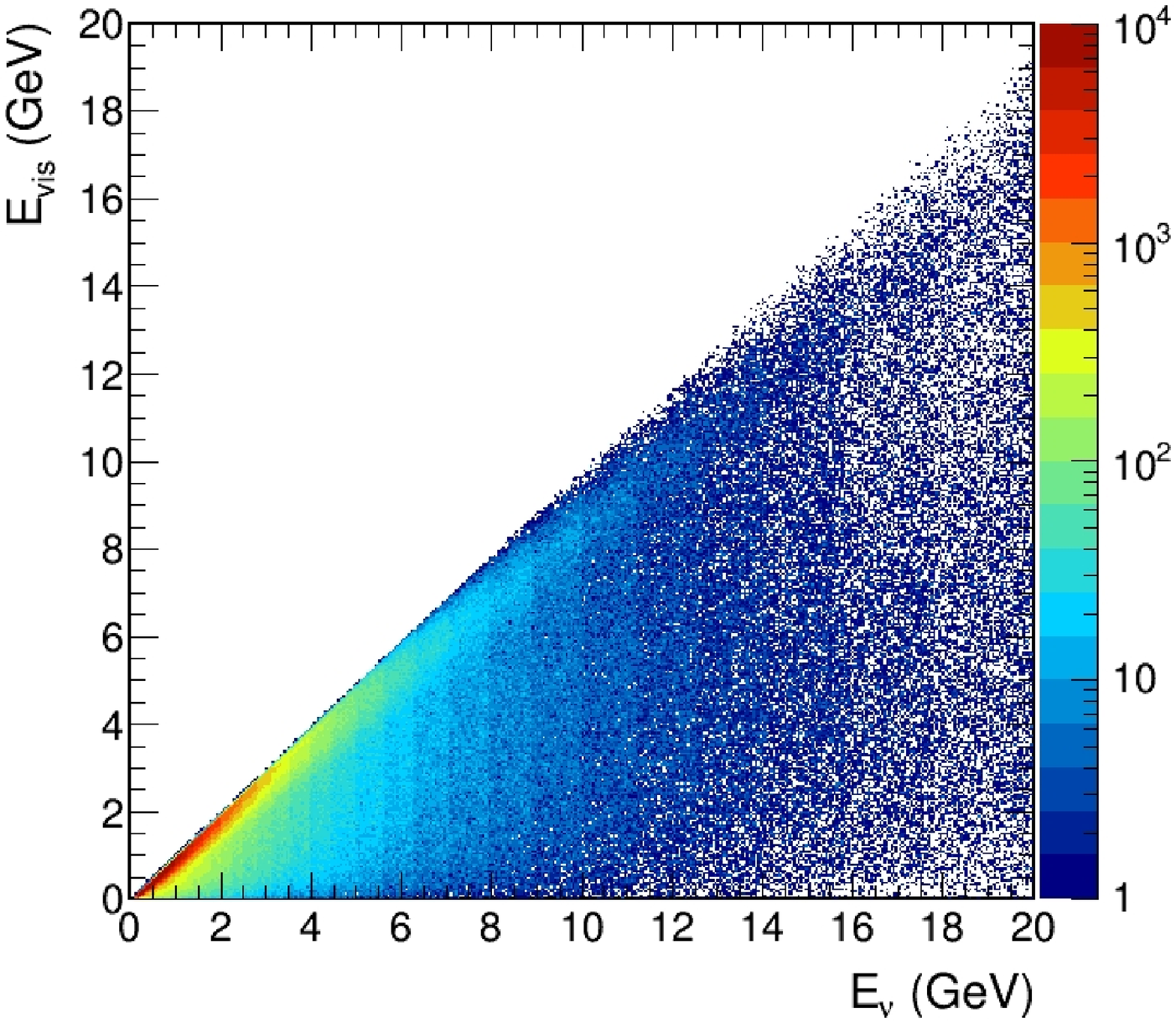}
\hspace{0.2cm}
\includegraphics[scale=0.42]{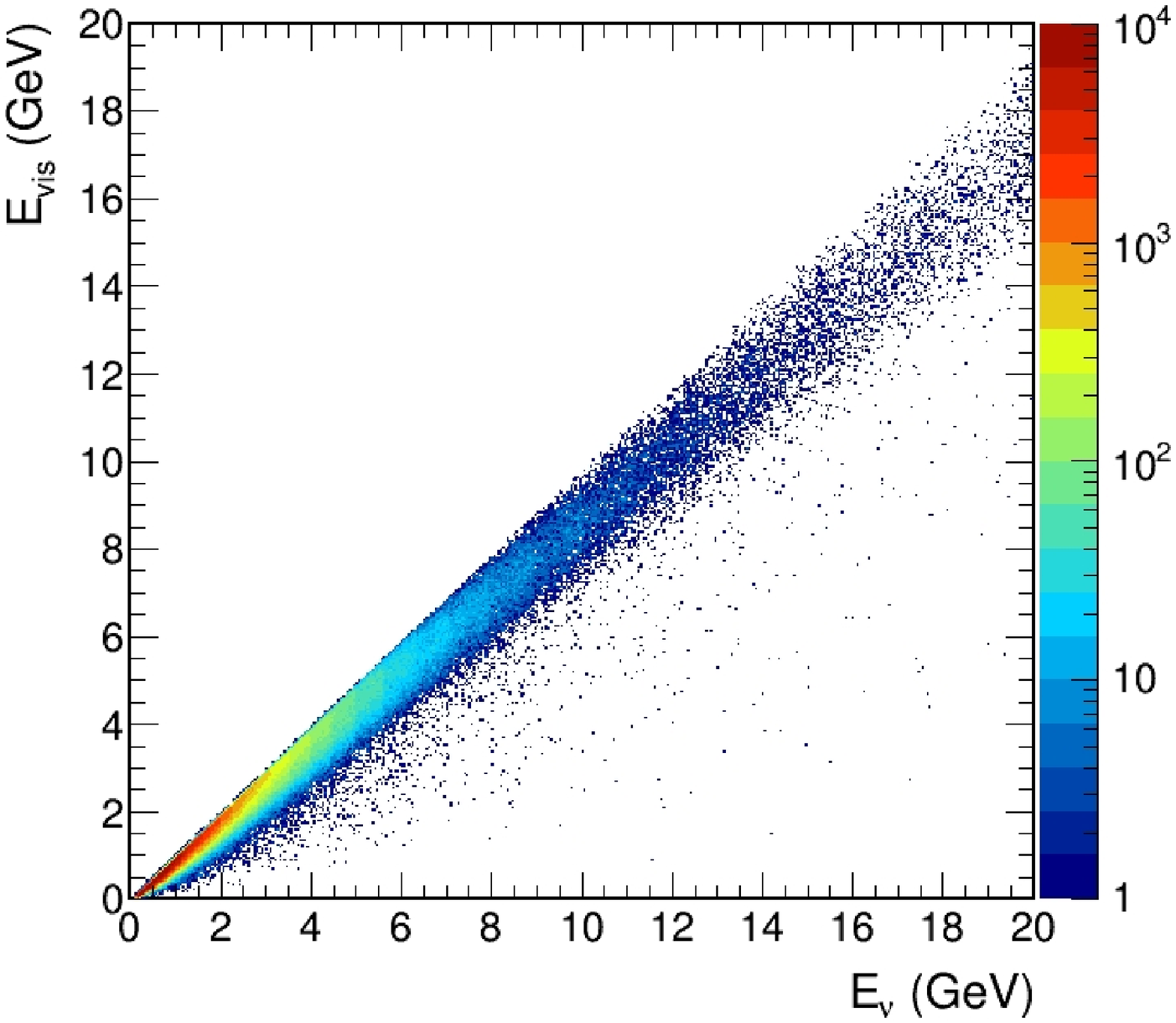}

\vspace{1.2cm}
\includegraphics[scale=0.42]{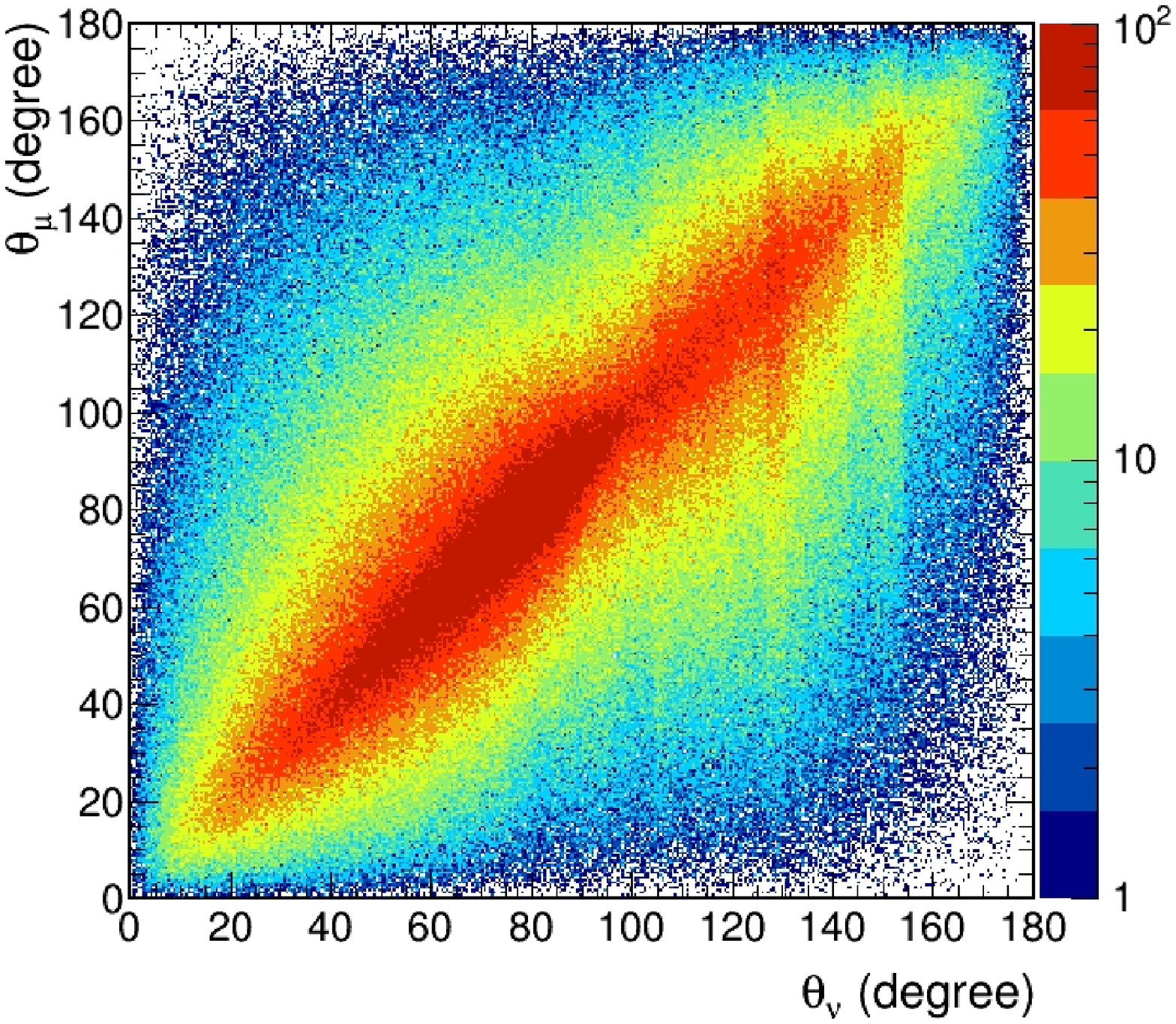}
\hspace{0.2cm}
\includegraphics[scale=0.42]{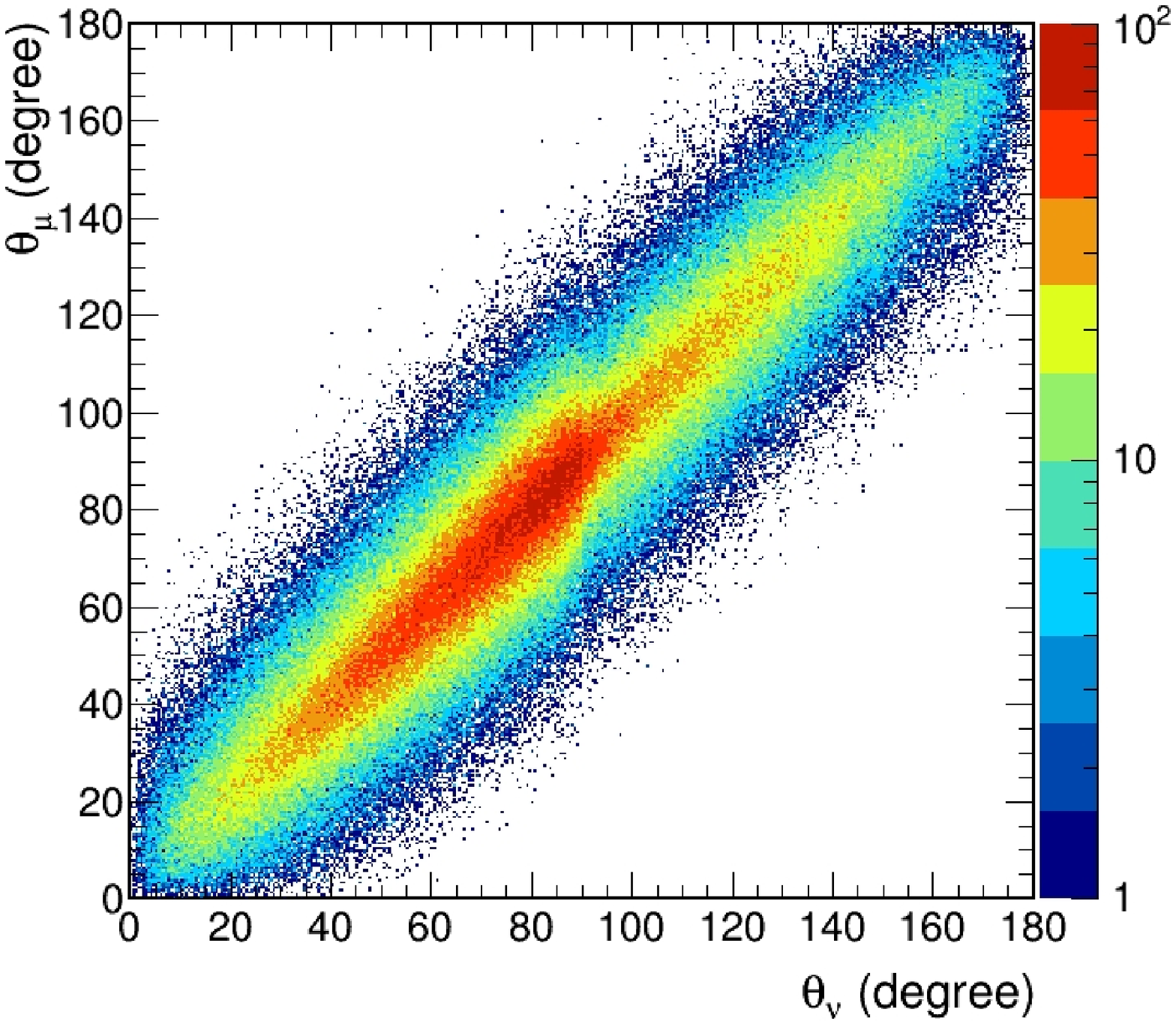}
\end{center}
\caption{ The initial neutrino energy $E_\nu$ versus the visible
energy $E_{vis}$ (upper panels) and the initial neutrino direction
$\theta_\nu$ versus the $\mu^\pm$ direction $\theta_\mu$ (lower
panels) for all (left) and selected (right) $\nu_\mu/\bar{\nu}_\mu$
CC events } \label{fig:atm:MC}
\end{figure}

The expected number of the atmospheric neutrino CC events in every
bin of the neutrino energy and zenith angle can be expressed as:
\begin{eqnarray}
T_{ij,\nu_\alpha} = 2 \pi M T \int
\limits_{\cos\theta_z^{i,min}}^{\cos\theta_z^{i,max}} d \cos
\theta_z \int \limits_{E_\nu^{j,min}}^{E_\nu^{j,max}} d E_{\nu} \,
F_\alpha(\cos\theta_z,E_{\nu}) \, \sigma_{\nu_\alpha} \;,
 \label{eqn:atm:Nij}
\end{eqnarray}
with
\begin{eqnarray}
F_{\alpha}(\cos\theta_z,E_{\nu})=\sum \limits_{l=(e,\mu,\tau)}
c_{\nu_l} \phi_{\nu_l} \cdot P(\nu_l \rightarrow \nu_\alpha),
\end{eqnarray}
where $\nu_\alpha = (\nu_e, \nu_\mu, \nu_\tau )$, $M = 20$ kton is
the LS target mass, $T$ is the exposure time,
$\cos\theta^{i,min/max}_z$ and $E_\nu^{j,min/max}$ are the borders
of the bin $i,j$ in zenith angle and energy. Unless otherwise
specified, the 200 kton-years exposure will be used in this section.
$\phi_{\nu_l}$ are the initial atmospheric neutrino fluxes from Ref.
\cite{Honda:2011nf}. For the flux normalization terms $c_{\nu_l}$,
we set $c_{\nu_e} = c_{\nu_\mu} = 1$ and $c_{\nu_{\tau}}=0$.
$\sigma_{\nu_\alpha}$ is the $\nu_\alpha$ per nucleon CC cross
section~\cite{Andreopoulos:2009rq} as shown in left panel of
Fig. \ref{fig:atm:spectrum}. Here we have considered that the LS
target includes 12\% $^1$H and 88\% $^{12}$C. Considering the
different cross sections, fluxes and oscillation probabilities the
above expression applies also to antineutrinos. Applying it to JUNO
we find that 8662 $\nu_\mu$ CC events, 3136 $\bar{\nu}_\mu$ CC
events, 6637 $\nu_e$ CC events, 2221 $\bar{\nu}_e$ CC events,  90
$\nu_\tau$ CC events, 44 $\bar{\nu}_\tau$ CC events and a total of
12255 NC events. Comparing the event numbers per bin of the MC
simulation and theoretical prediction, we determine the weight value
for every MC event. Then the expected event sample can be obtained
for the given neutrino interaction and oscillation parameters. In
the right panel of Fig. \ref{fig:atm:spectrum}, we plot the expected
spectra as a function of the visible energy $E_{vis}$. Note that
$88.5\%$ of the NC events have visible energies smaller than 1.0 GeV
since the final state neutrino does not deposit energy and the final
state hadrons have large quenching effect.

\begin{figure}[htb]
\begin{center}
\includegraphics[scale=0.47]{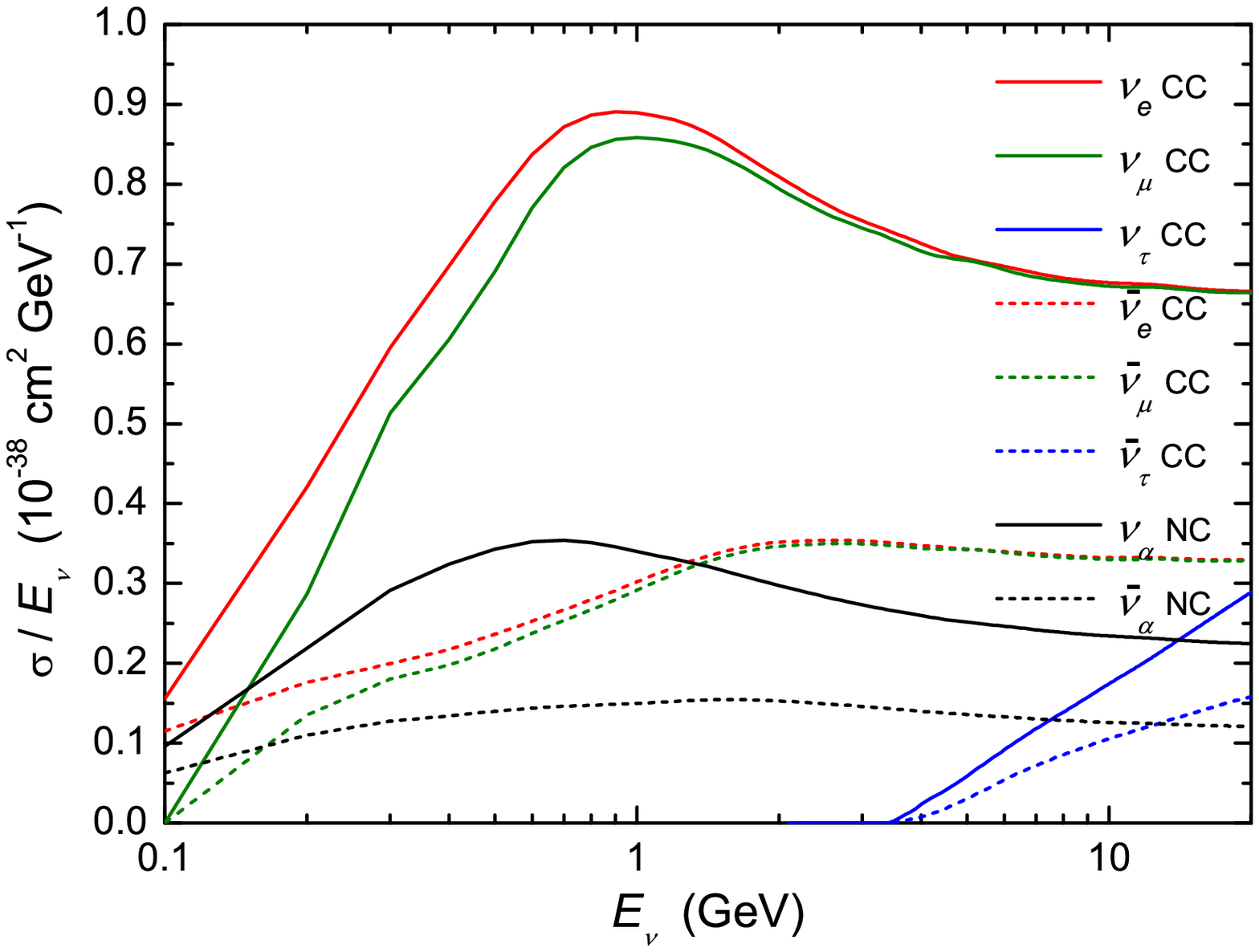}
\hspace{0.2cm}
\includegraphics[scale=0.47]{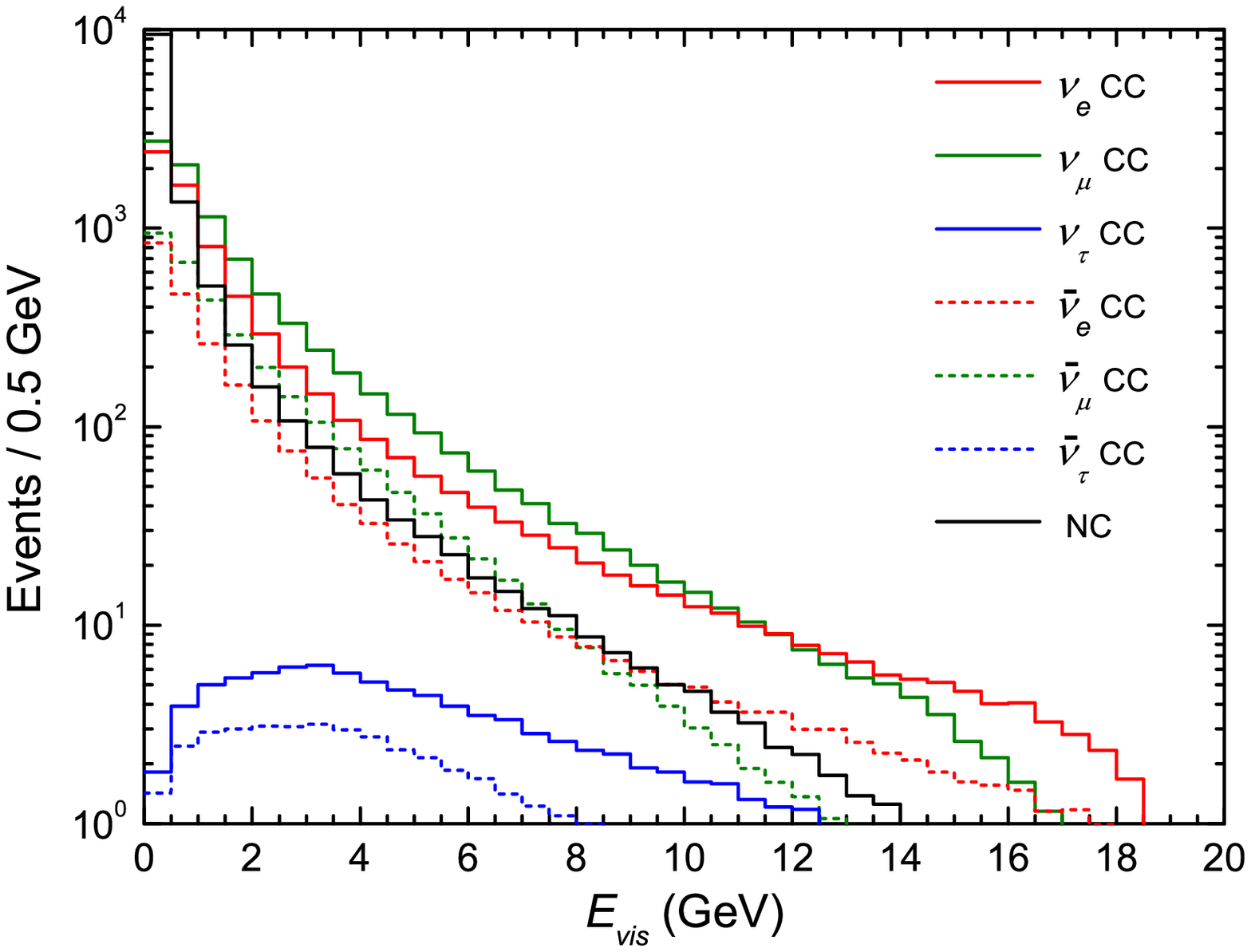}
\end{center}
\caption{Left panel: the neutrino and antineutrino per nucleon cross
sections for the LS target~\cite{Andreopoulos:2009rq}. Right
panel: the expected spectra as a function of $E_{vis}$ for the 200
kton-years exposure. } \label{fig:atm:spectrum}
\end{figure}

\subsubsection{Reconstruction Potential} \label{subsec:atm:reconstruction}

The JUNO central detector can measure the visible energy $E_{vis}$
of atmospheric neutrinos very well. In the Geant4 detector
simulation, we have considered the quenching effect for different
final state particles. The visible energy resolution depends on the
statistical fluctuation in the scintillation photon emission and the
quenching fluctuation. Here we assume $\sigma_{E_{vis}} = 0.01
\sqrt{E_{vis}/{\rm GeV}}$ for the following analysis. The JUNO
central detector can also reconstruct the track direction of the
energetic charged particle by use of the time profile per PMT. The
scintillation lights are emitted isotropically for every energy
deposition point of the track. However, the first photon arriving at
any PMT from this track are not isotropic when the energetic
particle travels faster than light in the LS~\cite{Learned:2009rv}.
This first photon surface coinciding with the Cherenkov surface is
directly correlated with the particle position and its time
evolution hence includes all information about the track. With the
help of the above principle, one may reconstruct the cosmic muon
direction and suppress the corresponding backgrounds. More
importantly, we can measure the charged lepton direction for the
atmospheric neutrino CC events.

We have made a toy MC simulation for single muon tracks in the
JUNO's central detector. For the track reconstruction, algorithms
using the first photons registered in each PMT show promising
results for JUNO (See the appendix for details). It is found that
the direction and track length $L_\mu$ resolutions mainly depend on
the intrinsic PMT timing resolution. If it is better than 4 ns, the
muon track length resolution is better than $0.5\%$ and the angular
resolution is better than $1^\circ$ when $L_\mu \geq 5$ m. For $1
\;{\rm m} < L_\mu < 5$ m, $L_\mu$ and angular resolution are better
than $1\%$ and $10^\circ$, respectively.

In the atmospheric neutrino CC interactions, the final states
consist of a charged lepton and several hadrons. The charged lepton
takes on average about $60\%$ of the initial neutrino energy and
several hadronic particles will share the residual energy. The
existence of hadronic final state particles will have a negative
effect on the expected resolutions for the charged lepton direction.
On the other hand, these hadronic particles can also affect the
charged lepton flavor recognition even if the signals from the
single $\mu^\pm$ track and $e^\pm$ shower have clear differences. In
addition, a substantial fraction of $\pi^\pm$ ($\gamma$ from $\pi^0
\rightarrow \gamma \gamma$) from the NC interactions will be
misidentified as $\mu^\pm$ ($e^\pm$)~\cite{Wurm:2011zn}. In this
case, the $\pi^\pm$ or $\gamma$ direction will be reconstructed as
the charged lepton direction and the atmospheric neutrino NC events
become the main backgrounds. Note that the direction reconstruction,
lepton flavor recognition and misidentification efficiencies have
not been fully explored so far~\cite{Wurm:2011zn,Peltoniemi:2009xx,Peltoniemi:2009hv}.
The related studies are under way.

To significantly increase the muon recognition capability and
suppress the misidentification efficiencies, we select tracks with a
length of at least a few meters. The final state hadrons ($p, n$ and
$\pi^\pm$) usually have the shorter track length than $\mu^\pm$
since they would undergo hadronic interactions. For the JUNO LS, the
nuclear and pion interaction lengths are about 0.9m and 1.3m,
respectively. The electromagnetic shower from $e^\pm$ and $\pi^0$
will not give a very long track because of the 0.5m radiation
length. It is clear that muons with longer track lengths are easier
to identify and reconstruct.

\subsubsection{Event Selection and Classification}
\label{subsec:atm:Classification}

To identify charged leptons, we conservatively require a track
length of $\geq 5$ m and only consider the $\nu_\mu/\bar{\nu}_\mu$
CC events. The selection efficiency of $L_\mu \geq 5$ m is $29.2\%$
for a MC sample of 11798 $\nu_\mu/\bar{\nu}_\mu$ CC events. For
these selected events, we assume that the final state $\mu^\pm$ can
be fully reconstructed and identified. The corresponding $\mu^\pm$
angular resolution is assumed to be $1^\circ$. Note that the Michel
electron from $\mu^\pm$ decay can also help us to optimize the
algorithm for distinguishing muon track from electron shower. For
the selected CC events, the NC backgrounds are negligible based on
the following three reasons. Firstly, event rates for the NC
interaction is far lower than those from the $\nu_\mu/\bar{\nu}_\mu$
CC interactions for $E_{vis} > 1$ GeV as shown in the right panel of
Fig. \ref{fig:atm:spectrum}. This is because that the NC events are
suppressed due to the $E_\nu^{-2.7}$ energy dependence of the
neutrino flux and hadron quenching effect for a given visible
energy. Secondly, the energetic $\pi^\pm$ production rates are
largely suppressed since several hadrons share this visible energy.
Finally, it is very rare for these $\pi^\pm$ to produce a long
straight track for the 1.3m pion interaction length. As shown in
right panel of Fig. \ref{fig:atm:spectrum}, the
$\nu_\tau/\bar{\nu}_\tau$ CC backgrounds can also be ignored when we
consider the 17.36\% branching ratio of $\tau^- \rightarrow \mu^-
\bar{\nu}_\mu \nu_\tau$.

According to the characteristics of the reconstructed muon track,
the selected $\nu_\mu$ and $\bar{\nu}_\mu$ CC events will be
classified as fully contained (FC) or partially contained (PC)
events, where FC and PC refers to muon track being fully or
partially contained in the LS region. Note that the Michel electron
can also help us to distinguish the FC from PC events. For the
specified sample we find 1932 FC and 1510 PC events with $L_\mu \geq
5$ m. The FC sample (green line) has a better correlation between
$E_{vis}$ and $E_\nu$ than those from the PC (violet line) sample
and all 11798 events (black line) as shown in left panel of Fig.
\ref{fig:atm:Smearing}. Here we have also plotted the spectrum
(orange line) of the FC sample as a function of the $\mu^\pm$
visible energy $E_\mu$ over $E_\nu$. It is found that the PC events
in the JUNO LS detector and the FC events in the Cherenkov detector
have the similar energy smearing. As shown in the right panel of
Fig. \ref{fig:atm:Smearing}, the PC sample has the smaller angle
smearing $\theta_\mu - \theta_\nu$ than the FC sample and all
$\nu_\mu/\bar{\nu}_\mu$ CC events. We will focus on the 1932 FC and
1510 PC events with $L_\mu \geq 5$ m and plot their expected spectra
as a function of $E_{vis}$ in Fig. \ref{fig:atm:FCPC}. The FC and PC
samples include $67.7\%$ and $63.6\%$ $\nu_\mu$, respectively.

\begin{figure}[htb]
\begin{center}
\vspace{1.1cm}
\includegraphics[scale=0.47]{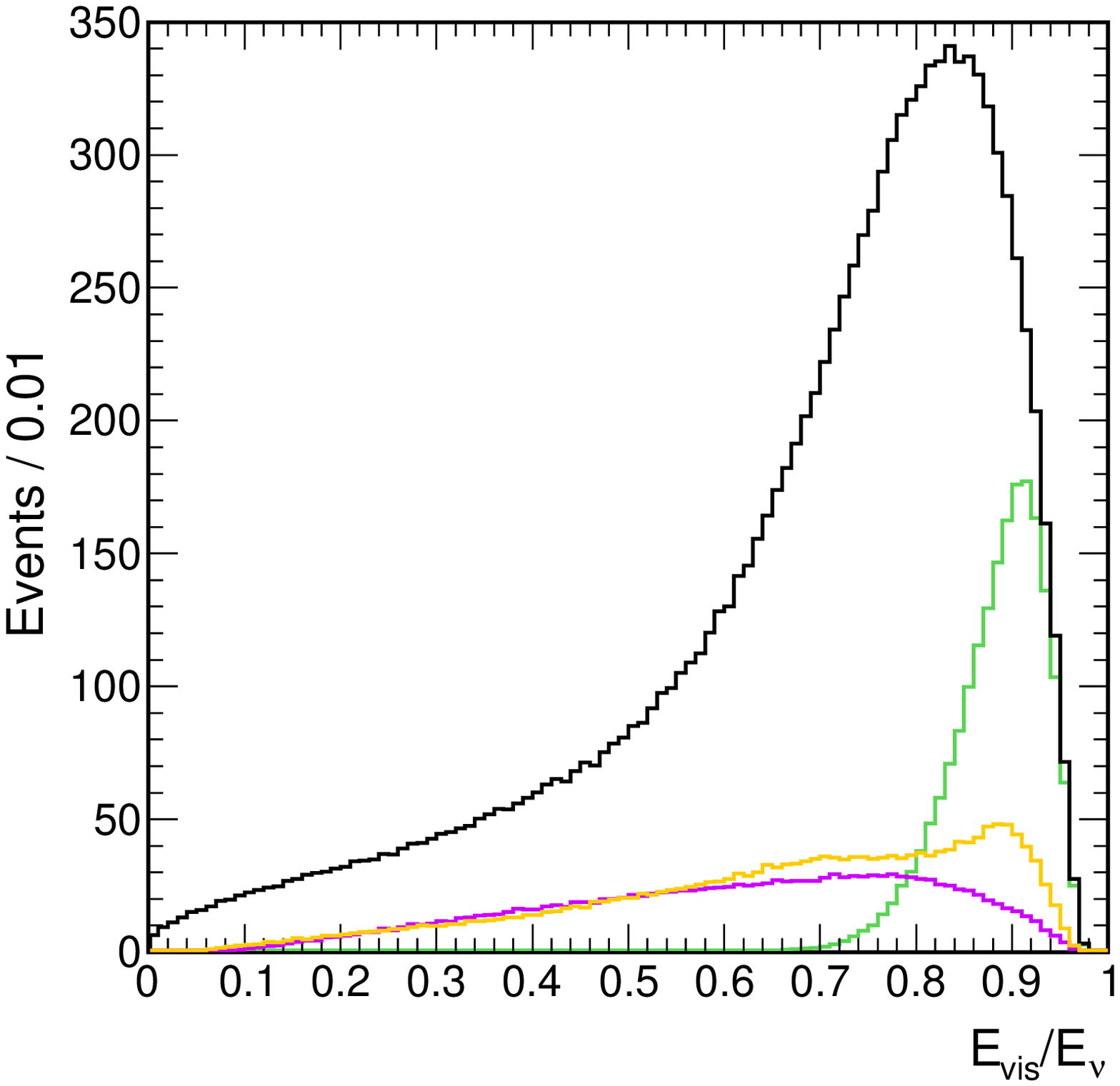}
\hspace{0.2cm}
\includegraphics[scale=0.47]{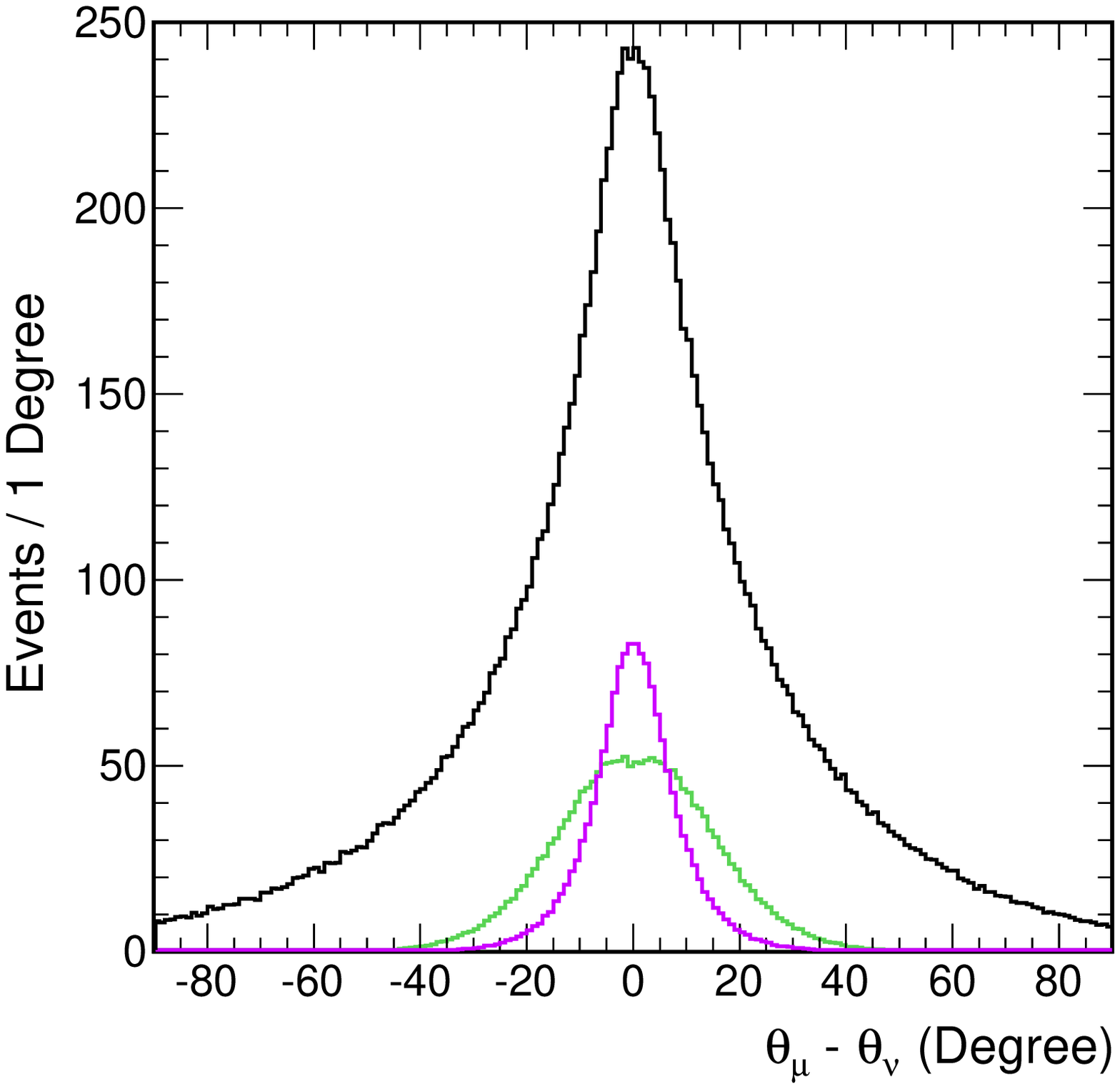}
\end{center}
\caption{ The calculated spectra as a function of $E_{vis}/E_\nu$
(left) and $\theta_\mu - \theta_\nu$ (right) for the selected FC
(green), PC (violet) and all (black) $\nu_\mu/\bar{\nu}_\mu$ CC
events. The orange line describes the $E_\mu/E_\nu$ distribution for
the FC events.} \label{fig:atm:Smearing}
\end{figure}

\begin{figure}[htb]
\begin{center}
\includegraphics[scale=0.47]{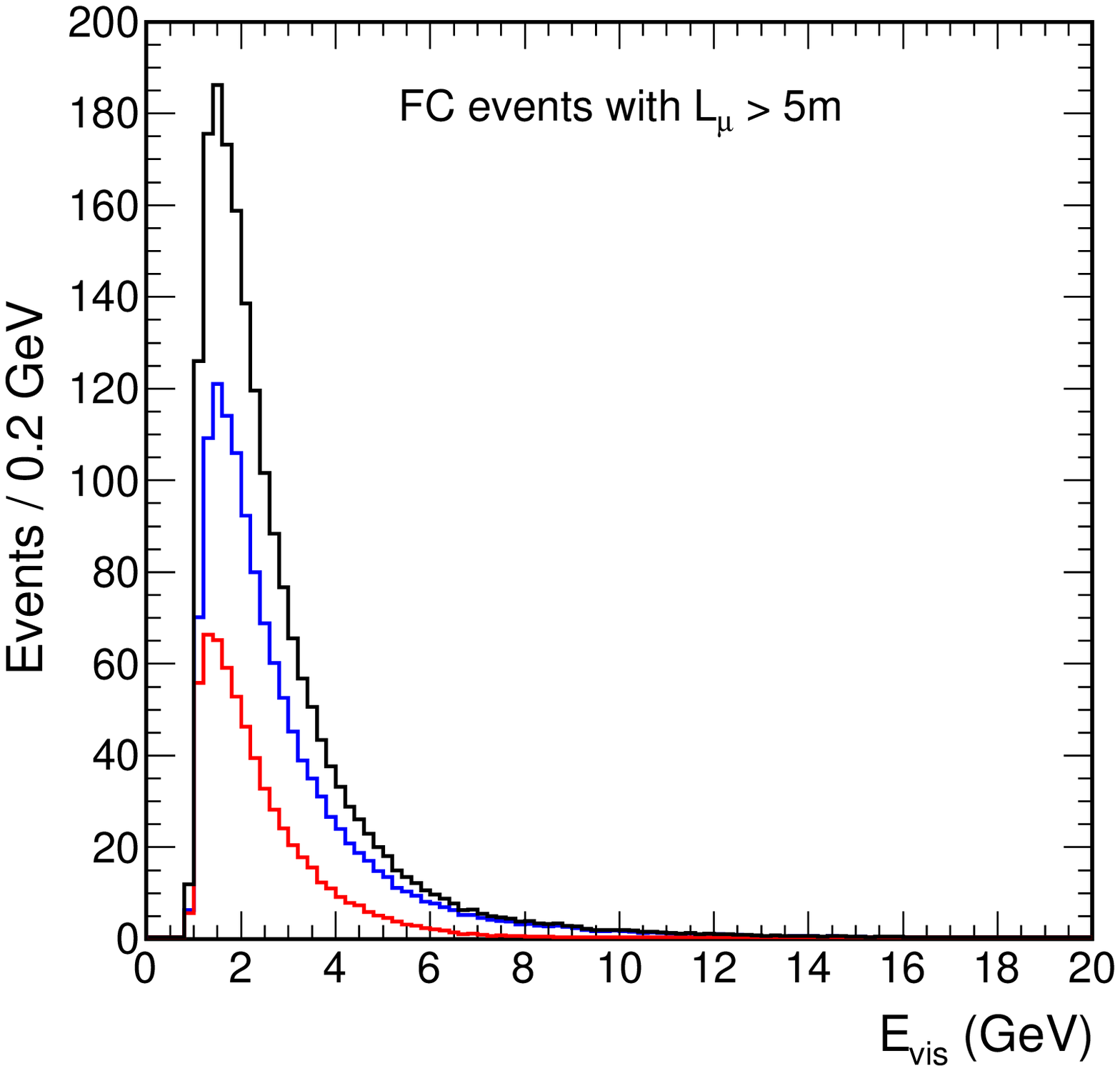}
\hspace{0.2cm}
\includegraphics[scale=0.47]{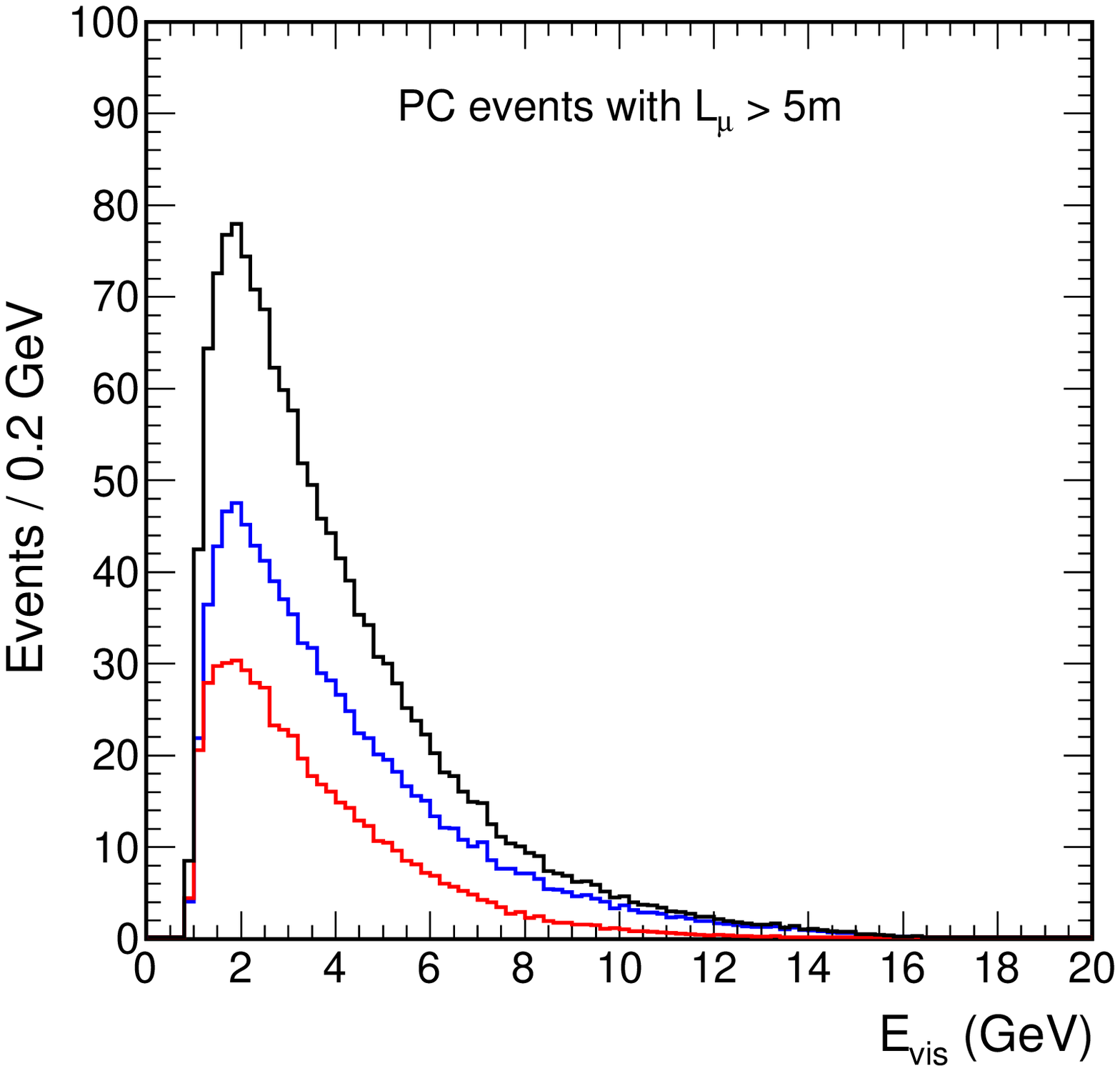}
\end{center}
\caption{ The expected FC (left) and PC (right) spectra as a
function of the visible energy $E_{vis}$ for the $L_\mu \geq 5$ m
$\nu_\mu/\bar{\nu}_\mu$ CC events. The blue and red lines denote
$\nu_\mu$ and $\bar{\nu}_\mu$, respectively.} \label{fig:atm:FCPC}
\end{figure}

Because of the oscillation effects of different neutrino flavors, a
neutrino and antineutrino mixed sample would have a smaller
sensitivity than each of its constituents. As a non-magnetic
detector, JUNO can not directly measure the charge of the muon
tracks to distinguish neutrino from antineutrino events. However, we
can choose other methods to statistically distinguish neutrinos from
antineutrinos~\cite{Huber:2008yx} which can improve the JUNO
sensitivities to the neutrino mass hierarchy, the octant of
$\theta_{23}$ and the CP phase $\delta$. In JUNO we could use the
following effects to discriminate between neutrinos and
antineutrinos. Firstly, the primary $\mu^-$ can be captured by a
nucleus to form a muonic atom after the $\mu^-$ losses its kinematic
energy. In this case, $\mu^-$ has a $8 \%$ capture probability on
the nucleus of $^{12}$C and no Michel electron is produced. This
feature allows a clear identification of $\nu_{\mu}$ if the primary
$\mu^-$ does not escape the LS region. Secondly, the neutrino
statistically shall transfer, on average, more momentum to the
hadronic final states than the antineutrino in CC interactions.
These primary hadrons may produce secondary particles through the
final state interaction (FSI) in the nuclear environment and the
inelastic scattering process in the LS. Therefore, the neutrino
events on average have more Michel electrons (from $\pi^\pm$ and
$\mu^\pm$ decay) than the antineutrino events~\cite{Abe:2011ph}.
Finally, we can deduce the visible energy of the hadron showers
$E_h$ from $ E_{vis}$ and the muon track length since the muons can
be assumed to be minimum ionizing which leads to a constant energy
deposition. Then the measured relative energy transfer $Y_{vis} =
E_h / E_{vis}$ may be used to statistically distinguish $\nu_\mu$
from $\bar{\nu}_\mu$ based on the same principle as in the second
method.

We now group the selected FC and PC events into $\nu_\mu$-like and
$\bar{\nu}_\mu$-like samples based on the $\mu^-$ capture of
nucleus, Michel electron numbers $N_e$ and the relative energy
transfer $Y_{vis}$. Firstly the FC events with $N_e \geq 2$ or the
$\mu^-$ capture on nucleus of $^{12}$C will be classified into the
FC $\nu_\mu$-like sample. Then, the other FC events with $Y_{vis}
\geq 0.5$ also get classified as FC $\nu_\mu$-like events. Finally,
we classify the residual FC events into the FC $\bar{\nu}_\mu$-like
sample. For the PC events, the PC $\nu_\mu$-like and PC
$\bar{\nu}_\mu$-like samples can be obtained when we only replace
$N_e \geq 2$ with $N_e \geq 1$. In Table \ref{tab:atmo:Samples}, we
list the expected $\nu_\mu$, $\bar{\nu}_\mu$ and total event numbers
of four samples for 200 kton-years exposure. It is worthwhile to
stress that the Michel electron numbers $N_e$ plays a key role in
the above classification. For example, the selection $N_e \geq 2$
will result in 524 $\nu_\mu$ and 62 $\bar{\nu}_\mu$ events in the FC
$\nu_\mu$-like sample. In addition to the above event
classification, we have also performed a simplified classification
in track-like and point-like events. Track-like events are all CC
$\nu_\mu$ events with an inelasticity $E_h/E_\nu < 0.65 $. All other
interactions are called point-like, because the cascade at the
interaction vertex dominates the visible energy. This classification
based on 25 million simulation events will be used for the
pessimistic analysis of the MH sensitivity.

\begin{table}[htb]
\centering
\begin{tabular}{|c|c|c|c|c|}\hline\hline
            & $\nu_\mu$ events & $\bar{\nu}_\mu$ events & Total events &
$\nu_\mu$ purity \\ \hline
 FC $\nu_\mu$-like       & 656 & 83  & 739  & 88.8\%  \\ \hline
 FC $\bar{\nu}_\mu$-like & 652 & 541 & 1193 & 54.6\%  \\ \hline
 PC $\nu_\mu$-like       & 577 & 166 & 743  & 77.7\%  \\ \hline
 PC $\bar{\nu}_\mu$-like & 383 & 384 & 767  & 50.0\%  \\ \hline
 \hline
\end{tabular}
\caption{The expected event numbers of four samples for 200
kton-years exposure. \label{tab:atmo:Samples}}
\end{table}

\subsubsection{Identification of \texorpdfstring{$\nu_\mu$}{v-mu},
\texorpdfstring{$\bar\nu_\mu$}{v-mubar}, \texorpdfstring{$\nu_e$}{v-e}, and
\texorpdfstring{$\bar\nu_e$}{v-ebar} for the CP-Violation analysis}
\label{subsec:atm:PID}

As shown in Fig. \ref{fig:atm:CPVariation}, the sub-GeV region is
most sensitive to the CP phase $\delta$. Below 100 MeV, the current
atmospheric neutrino flux prediction has a large uncertainty, and
the discrepancy between models can be as large as 50\%. Therefore,
it is a difficult region to study. Above 100 MeV, the CC quasi
elastic processes are open for $\nu_\mu$ and $\nu_e$. Around 300
MeV, the CC resonance production process is open. In this case, the
final states include the charged lepton, pions and the daughter
nuclei. For a liquid scintillator detector, which is weak in
detecting multiple tracks for the sub-GeV neutrino events, it is
also difficult to distinguish $\mu^\pm$ ($e^\pm$) from $\pi^\pm$
($\pi^0$).

For the identification of the sub-GeV $\nu_\mu$, $\bar\nu_\mu$,
$\nu_e$, and $\bar\nu_e$, the [100, 300] MeV region is very
interesting since it may be a region with relatively low background.
For example, the final state $\mu^-$ will on average take $74\%$ of
the initial 200 MeV energy in the $\nu_\mu$ CC interaction. It is
possible for us to reconstruct the track length and direction of the
charged lepton as discussed in appendix. Besides the quenching
effect, the energy of $\mu^\pm$ and $e^\pm$ can be measured. Another
advantage of the [100, 300] MeV region is that we may distinguish
neutrinos from antineutrinos by use of the different daughter
nuclei. In the [100, 300] MeV region, the $\bar\nu_e$ and $\nu_e$ CC
processes can be written as
\begin{eqnarray}
^{12}{\rm C}(\bar\nu_e, e^+) ^{12}{\rm B^*}, & & ^{12}{\rm C}(\bar\nu_e,e^+)^{12}{\rm B(g.s.)}, \\
^{12}{\rm C}(\nu_e, e^-) ^{12}{\rm N^*}, & & ^{12}{\rm C}(\nu_e,
e^-)^{12}{\rm N(g.s.)},
 \label{eqn:atm:nunubarid}
\end{eqnarray}
where the ground states $^{12}{\rm N(g.s.)}$ and $^{12}{\rm
B(g.s.)}$ can be identified by the JUNO detector since they have
different lifetimes and Q-values. The excited states $^{12}{\rm
B^*}$ and $^{12}{\rm N^*}$ have characteristic decay modes. For
example, $^{12}{\rm N^*}$ may decay into a proton and a $^{11}{\rm
C}$. The $\beta^+$ decay of $^{11}{\rm C}$ can be identified since
JUNO has a very low energy threshold. JUNO can also detect
$\bar\nu_e$ through IBD process $p(\bar\nu_e, e^+)n$. The same
principle can be applied to $\bar\nu_\mu$ and $\nu_\mu$. The final
state muon with a 2.2 $\mu$s lifetime can decay into a Michel
electron plus neutrinos. This delayed coincidence feature can be
used to distinguish muon neutrinos. In this case, a triple
coincidence from prompt muon production, the Michel electron from
muon decay, and the decay of the daughter nuclei are required.

It is promising to use double (triple) coincidence detection for
$\bar\nu_e$ and $\nu_e$ ($\bar\nu_\mu$ and $\nu_\mu$), and to rely
on different daughter nuclei to distinguish neutrinos from
antineutrinos. However it is very hard to precisely predict the
cross sections for different daughter nuclei in present nuclear
structure models. Some experiments only measure the ground state
cross section. Studies on the particle identification and
reconstruction are in progress for the [100, 300] MeV atmospheric
neutrinos. Therefore the NC and CC backgrounds are not yet certain.
In the [100, 300] MeV region, we shall take the idealized
assumptions and only present a very optimistic upper limit of the
JUNO detector for the CP research.

\subsection{Atmospheric Neutrino Analysis} \label{subsec:atm:Physics}

\subsubsection{The \texorpdfstring{$\chi^2$}{chi2} Function}
\label{subsec:atm:chi2}

In the following we will investigate different data samples using
the visible neutrino energy $E_{vis}$ and the $\mu^\pm$ zenith angle
$\theta_\mu$. For most of the analysis we will use a $\chi^2$
function, where we adopt the Poisson form:
\begin{eqnarray}
\chi^2 = 2 \sum_{ij} \left[  \tilde{T}_{ij}  - N_{ij} + N_{ij} \ln
\frac{N_{ij}}{ \tilde{T}_{ij}} \right] + \sum_{k=1}^4 \xi_k^2 \;,
\label{eqn:atm:chi2}
\end{eqnarray}
with
\begin{eqnarray}
\tilde{T}_{ij}  = T_{ij} \left(1 + \sum_{k=1}^4 \pi^k_{ij} \xi_k
\right) \;, \label{eqn:atm:Tij}
\end{eqnarray}
where the subscripts $i$ and $j$ run over the energy and angle bins,
respectively. For the range $-1 \leq \cos \theta_{\mu} \leq 1 $, the
number of bins is chosen to be 10, namely a bin width of 0.2. Then
we divide the visible energy into several bins and require the
experimental event numbers $ N_{ij} \geq 5$ for every bin. Adopting
a set of oscillation parameters as the true values, we can calculate
$N_{ij}$ based on the MC efficiency per bin. The expected event
numbers $T_{ij}$ can be obtained from Eq.~(\ref{eqn:atm:Nij}) with
any input values of oscillation parameters. Using the method of
pulls, we take into account the neutrino flux and cross section
systematic uncertainties~\cite{GonzalezGarcia:2004wg}. The pull
variable $\xi_1$ describes the $10\%$ normalization error of
neutrino cross section. $\xi_2$ is the pull for the overall $20\%$
neutrino flux error.  $\xi_3$ and $\xi_4$ parameterize the $5\%$
uncertainties on the energy and zenith angle dependence of
atmospheric neutrino fluxes, respectively. The ``coupling"
$\pi^k_{ij}$ describes the fractional change of $T_{ij}$ when the
corresponding uncertainty has $1\sigma$ deviation. Finally, we
minimize $\chi^2$ over the above four pull variables for a set of
input oscillation parameters and derive the best-fit minimal
$\chi^2({\rm test})$. For the test NH and IH hypotheses, the
difference $\Delta \chi^2 = |\chi^2 ({\rm IH}) - \chi^2 ({\rm NH})
|$ will be used to calculate the JUNO MH sensitivity with $N \sigma =
\sqrt{\Delta \chi^2}$.

\subsubsection{Neutrino Mass Hierarchy} \label{subsec:atm:MH}

As discussed in Sec.~\ref{subsec:atm:Oscillation}, $\nu_\mu$ and
$\bar{\nu}_\mu$ have opposite contributions to the MH sensitivity in
the absence of the charge identification. That is because $P(\nu_\mu
\rightarrow \nu_\mu)$ and $P(\nu_e \rightarrow \nu_\mu)$ in the NH
case is approximately equal to the corresponding antineutrino
oscillation probabilities in the IH case. Due to different fluxes
and cross sections one expects a higher neutrino rate compared to
antineutrinos which allows a measurement of the MH without
distinguishing between the two. The statistical separation of
$\nu_\mu$ and $\bar{\nu}_\mu$ events in
Sec.~\ref{subsec:atm:Classification} will obviously improve the
sensitivity further.

For the upward atmospheric neutrinos, the oscillation probabilities
$P(\nu_\mu \rightarrow \nu_\mu)$ and $P(\nu_e \rightarrow \nu_\mu)$
in the NH and IH cases have obvious differences due to the MSW
resonance effect. These differences can be easily found from
Eqs.~(\ref{eqn:atm:Peu}-\ref{eqn:atm:Puu}) and Fig.
\ref{fig:atm:probability} with the help of
Eq.~(\ref{eqn:atm:NHIHrelationship}). It is clear that the
approximated oscillation probabilities $P(\nu_\mu \rightarrow
\nu_\mu)$ and $P(\nu_e \rightarrow \nu_\mu)$ in
Eqs.~(\ref{eqn:atm:Peu}-\ref{eqn:atm:Puu}) do not depend on
$\theta_{12}, \delta $ and $\Delta m_{21}^2$. In addition, the MH
sensitivity is also insensitive to $\theta_{13}$ since the effective
mixing angle $\sin^2 2\theta_{13}^m $ will approach 1 in the MSW
resonance case. Unless otherwise specified, we shall take
$\theta_{12}, \theta_{13} $ and $\Delta m_{21}^2$ at their best
values of Eq.~(\ref{eqn:atm:parameter}), and the CP violation phase
$\delta = 0^\circ$ for the following analysis. Here we will analyze
the dependence on $\theta_{23}$ and $\Delta m_{atm}^2$. For three
typical $\sin^2 \theta_{23} = 0.4,0.5,0.6$, we have calculated the
theoretical and experimental event numbers per bin through varying
$\Delta m_{atm}^2$. Then the JUNO MH sensitivity can be determined
from Eqs.~(\ref{eqn:atm:chi2}) and (\ref{eqn:atm:Tij}). The minimal
$\Delta \chi^2$ has been found at $\Delta m_{atm}^2 = - 2.40 \times
10^{-3} {\rm eV}^2$ and $\Delta m_{atm}^2 = 2.47  \times 10^{-3}
{\rm eV}^2$ for the true $\Delta m_{atm}^2 = 2.43 \times 10^{-3}
{\rm eV}^2$ and $\Delta m_{atm}^2 = -2.43 \times 10^{-3} {\rm
eV}^2$, respectively. Using the two values as the test $\Delta
m_{atm}^2$, we calculate the MH sensitivity as a function of
livetime as shown in Fig. \ref{fig:atm:MHyears}. It is found that
JUNO has a  $0.9 \sigma$ MH sensitivity with 10 year data and
$\sin^2 \theta_{23} = 0.5$. The larger $\sin^2 \theta_{23}$ gives
better MH sensitivity. In the absence of the statistical separation
of neutrinos and antineutrinos, about $20\%$ reduction of the JUNO
MH sensitivity is expected.

\begin{figure}[htb]
\begin{center}
\includegraphics[scale=0.47]{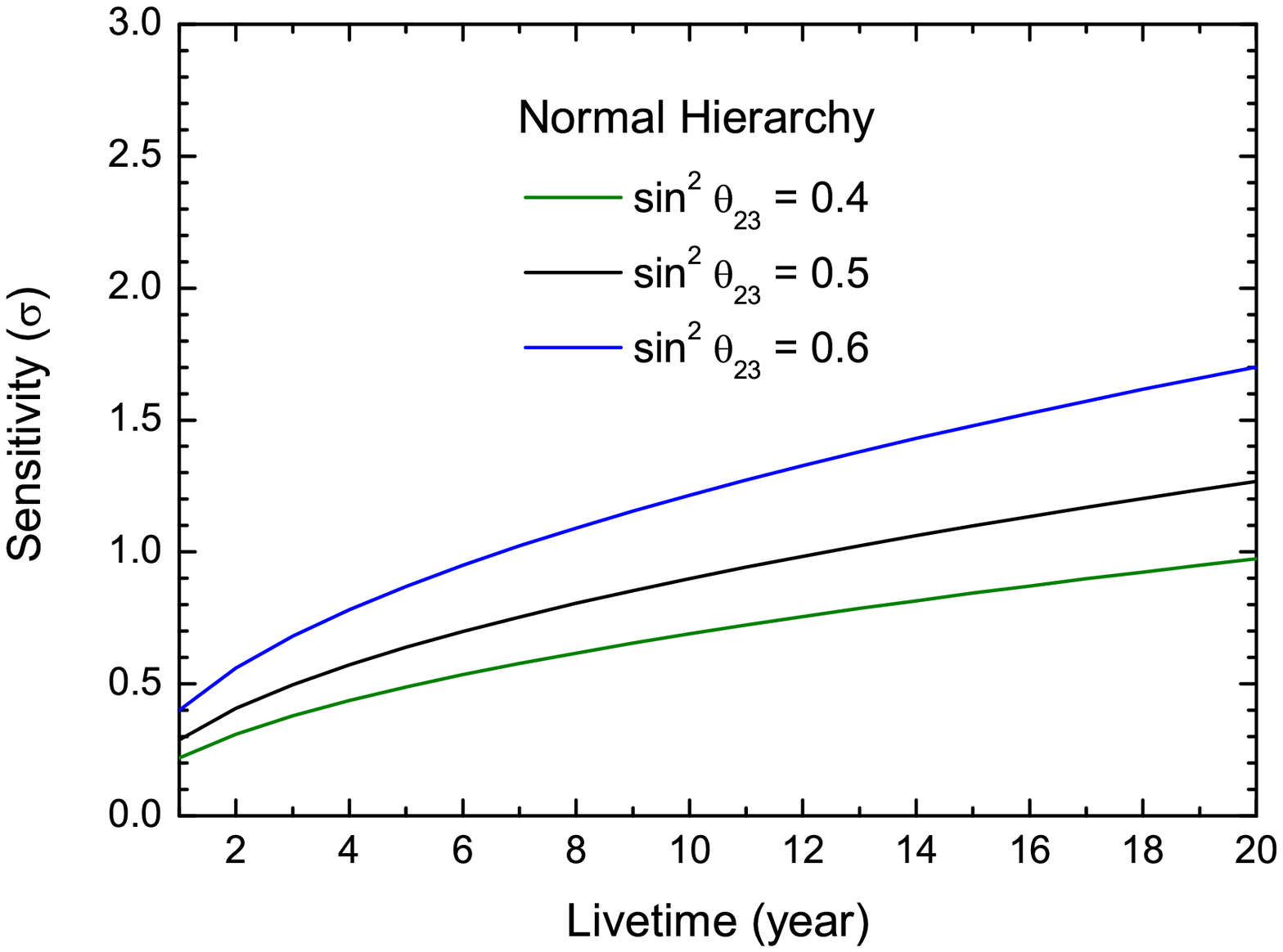}
\includegraphics[scale=0.47]{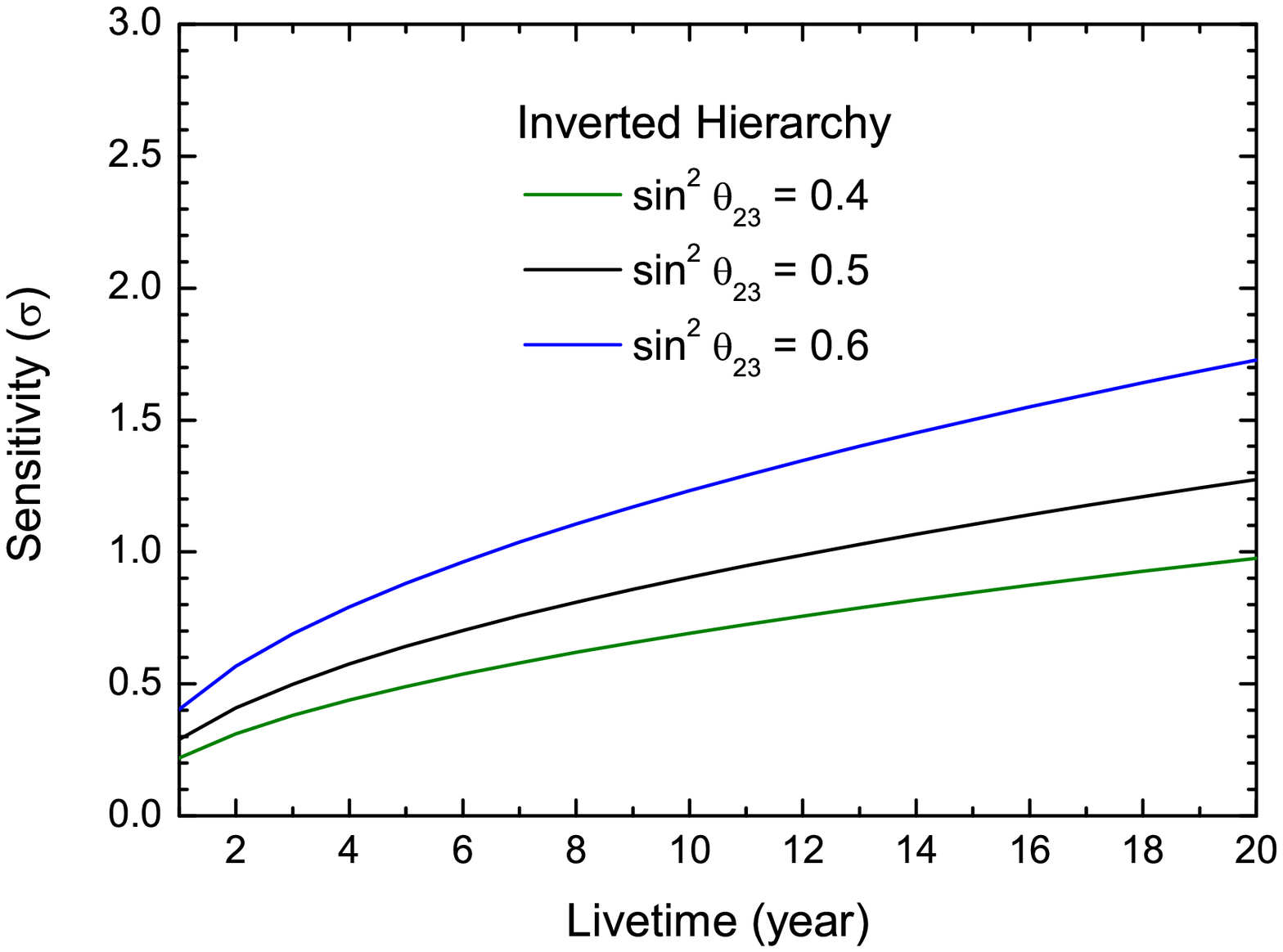}
\end{center}
\caption{The JUNO MH sensitivities from high energy muon neutrino
events as a function of livetime for the true NH (left) and IH
(right) hypotheses.} \label{fig:atm:MHyears}
\end{figure}

In the following, we will discuss an optimistic estimation for the
neutrino mass hierarchy based on few reasonable assumptions.
Firstly, the $\nu_e/\bar{\nu}_e$ CC events can be identified and
reconstructed very well in the $e^\pm$ visible energy $E^e_{vis} >
1$ GeV and $Y_{vis} < 0.5$ case. Secondly, we extend the selection
condition $L_{\mu} > 5 $m to $L_{\mu} > 3 $m for the
$\nu_\mu/\bar{\nu}_\mu$ CC events. Finally, we replace the charged
lepton direction with the neutrino direction for this physical
analysis and assume $10^\circ$ angular resolution. As shown in Fig.
\ref{fig:atm:potential}, we numerically calculate the optimistic MH
sensitivities from $\nu_\mu/\bar{\nu}_\mu$ (dashed lines) and
$\nu_e/\bar{\nu}_e$ (dotted lines) with a fixed $\sin^2 \theta_{23}
= 0.5$. It is found that the $\nu_\mu/\bar{\nu}_\mu$ contributions
have a small enhancement with respect to the results in Fig.
\ref{fig:atm:MHyears}. The $\nu_e/\bar{\nu}_e$ events have the
larger sensitivities than $\nu_\mu/\bar{\nu}_\mu$ events since most
of $\nu_e/\bar{\nu}_e$ events can deposit their energies in the LS
region. The selected $\nu_e/\bar{\nu}_e$ events only are divided
into two samples in terms of the Michel electron number $N_e = 0$
and $N_e \geq 1$. The combined sensitivities can reach 1.8 $\sigma$
and 2.6 $\sigma$ for 10 and 20 year data, respectively. Note that
the predicted MH sensitivity will have a $13\%$ reduction when we
take $20^\circ$ angular resolution.

\begin{figure}[htb]
\begin{center}
\includegraphics[scale=0.47]{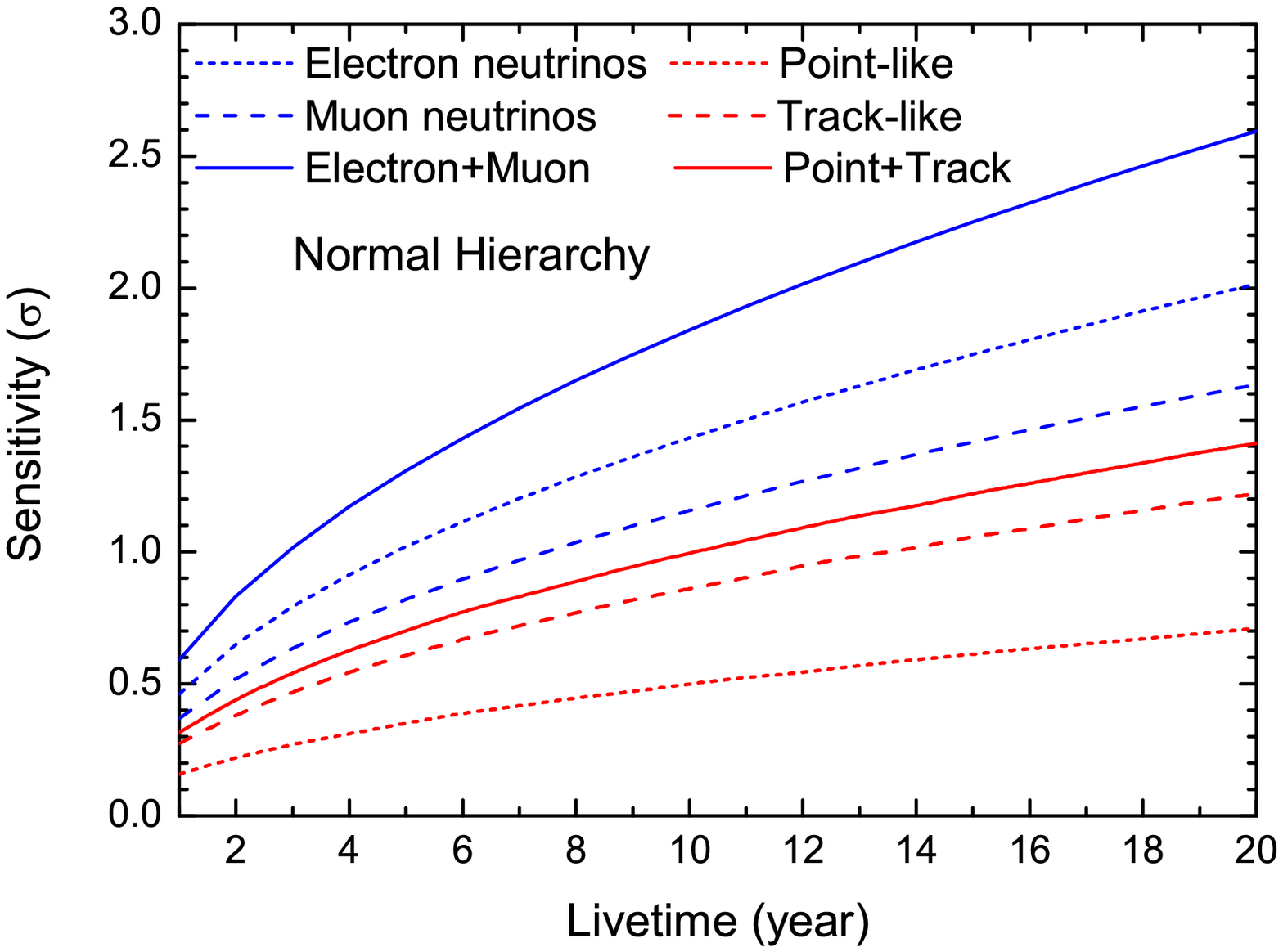}
\includegraphics[scale=0.47]{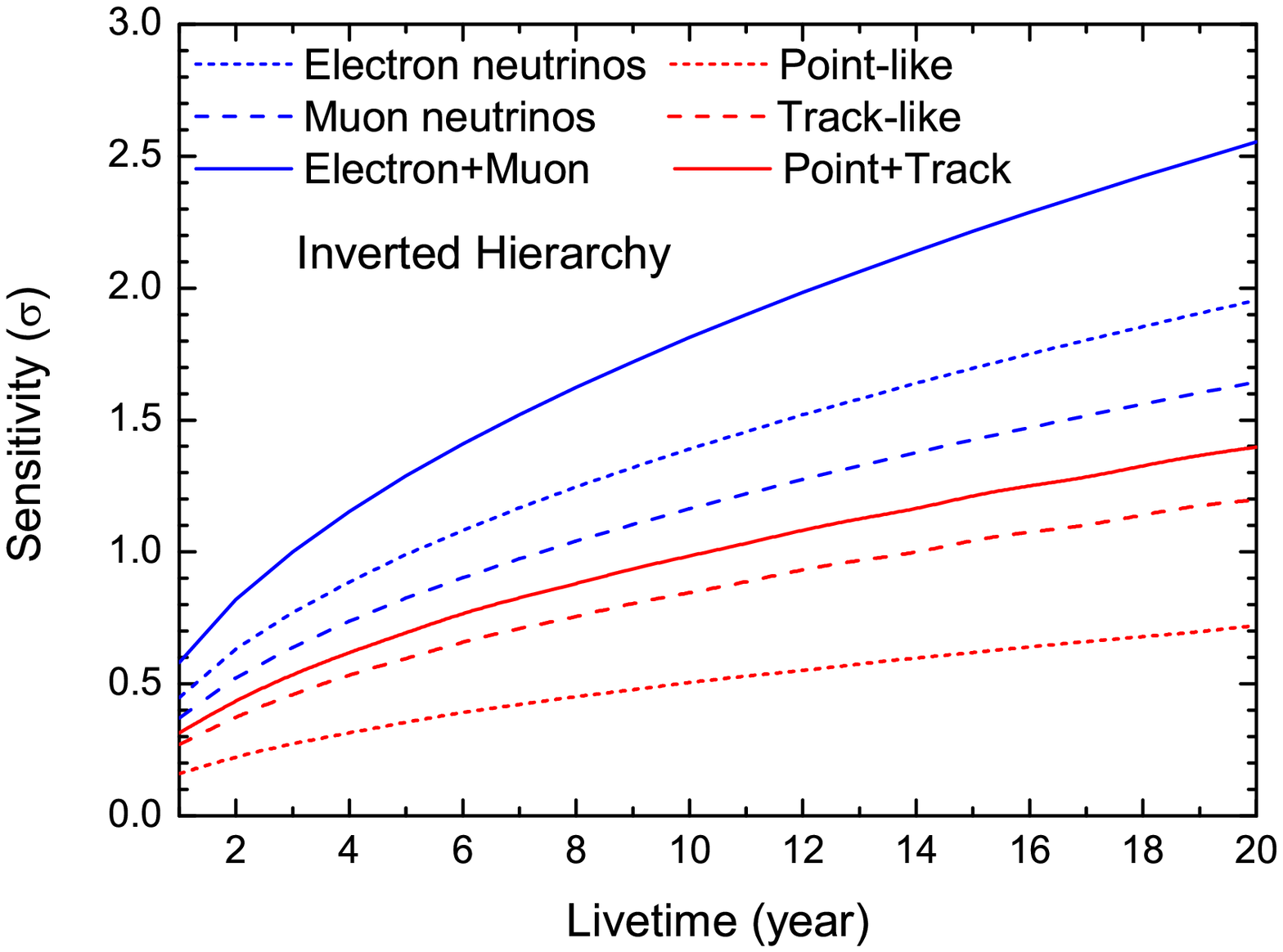}
\end{center}
\caption{The future optimistic (blue) and pessimistic (red) MH
sensitivities as a function of livetime for the true NH (left) and
IH (right) hypotheses. } \label{fig:atm:potential}
\end{figure}

Before the JUNO particle reconstruction and identification
capabilities are fully understood, we should consider a more
pessimistic case. For this analysis we adopt a different analytical
method and assume that it will be only possible to distinguish muon
tracks from other cascading events. We therefore divide the data
into the point-like and track-like samples. The track-like sample
contains only $\nu_\mu/\bar{\nu}_\mu$ CC events with a $E_h / E_\nu
<0.65$ inelasticity and the point-like sample all other CC and NC
events. Here we do not consider the statistical separation of
neutrinos and antineutrinos, and do not discriminate the FC and PC
events. In contrast to the optimistic case, we take the 5$\%
\sqrt{E_{vis}}$ and $37.2^\circ / \sqrt{E_{\nu}}$ for the visible
energy and the neutrino direction resolutions, respectively.
$37.2^\circ / \sqrt{E_{\nu}}$ corresponds to the mean angle between
the lepton and neutrino directions. In order to calculate the MH
sensitivity we weight a simulated dataset of 25 million events
according to the best fit parameters~\cite{Capozzi:2013csa} of
both NH and IH hierarchies. For the experimental event numbers $
N_{ij}$ in Eq.~(\ref{eqn:atm:chi2}), we dice pseudo experiments for
each hierarchy using a poisson distribution. This yields the
Gaussian distributed $\chi^2$ values. The MH sensitivity under the
assumption that one hierarchy is true is the distance between the
expectation values $\mu_{\mathrm{true}}$ and $\mu_{\mathrm{false}}$
expressed in units of the false hierarchy standard deviation
$\sigma_{\mathrm{false}}$. The estimated sensitivity $N
{\sigma}=|\mu_{\mathrm{true}}-\mu_{\mathrm{false}}|/
\sigma_{\mathrm{false}}$ can be seen in Fig.
\ref{fig:atm:potential}. After a 10 year measurement one would
expect a $1.0 \sigma$ combined sensitivity from the point and
track-like samples. The results are pessimistic compared to the
optimistic case which is mostly due to the assumed angular
uncertainties. Additionally, the sensitivity of the point-like
sample is decreased by a high contamination of NC events and deep
inelastic muon neutrino interactions while the track-like sample has
a higher uncertainty on energy resolution due to a high number of PC
events.

\subsubsection{Atmospheric Mixing Angle \texorpdfstring{$\theta_{23}$}{theta23}
} \label{subsec:atm:Theta23}

For the atmospheric mixing angle $\theta_{23}$, the MINOS
disappearance data indicates a non-maximal $\theta_{23}$
\cite{Adamson:2013whj}. However, the T2K disappearance data prefer a
nearly maximal mixing $\theta_{23} = 45^\circ$~\cite{Abe:2014ugx}.
It is an open question whether or not $\theta_{23}$ is maximal. If
$\theta_{23}$ deviates from $45^\circ$, one can get both the lower
octant (LO) $\theta_{23} < 45^\circ$ and higher octant (HO)
$\theta_{23}
> 45^\circ$ solutions, because the
$\nu_\mu/\bar{\nu}_\mu$ survival probability is mainly sensitive to
the $\sin^2 2\theta_{23}$ terms of Eq.~(\ref{eqn:atm:Puu}) for the
MINOS and T2K experiments. When the MSW resonance happens, the
$\sin^4 \theta_{23}$ term in Eq.~(\ref{eqn:atm:Puu}) will be
enlarged due to $\sin^2 2\theta_{13}^m \rightarrow 1$. Then the
$\sin^4 \theta_{23}$ term can help us to distinguish the
$\theta_{23}$ octant since $\sin^4 \theta_{23}$ is different for the
$\theta_{23}$ and $\pi/2- \theta_{23}$ solutions. In addition, we
should consider the oscillation probability $P(\nu_e \rightarrow
\nu_\mu)$ which is proportional to $\sin^2 \theta_{23}$ as shown in
Eq.~(\ref{eqn:atm:Peu}). It is worthwhile to stress that the octant
sensitivity from antineutrinos (neutrinos) is largely suppressed by
$\sin^2 2\theta_{13}^m \rightarrow \sin^2 2\theta_{13}$ when we take
the NH (IH) hypothesis as the true mass hierarchy. Therefore the
statistical separation of neutrinos and antineutrinos can suppress
the statistical errors and improve the octant sensitivity.

For the octant analysis, we assume a prior knowledge of mass
hierarchy and hence consider the normal (inverted) hierarchy as the
true hierarchy. For a given true $\theta_{23}$, we take the correct
octant solution $\theta_{23}$ and the wrong octant solution $\pi/2 -
\theta_{23}$ as the test values to calculate the best fit minimal
difference $\Delta \chi^2 =  \chi^2 (\pi/2 - \theta_{23}) - \chi^2
(\theta_{23})$. Then we can derive the JUNO octant sensitivity $
\sqrt{\Delta \chi^2}$. In Fig. \ref{fig:atm:T23octant}, we have
plotted the JUNO sensitivities to the octant for the true NH and IH
cases. The wrong $\theta_{23}$ octant could be ruled out at $1.8 \,
\sigma$ (NH) and $0.9 \,\sigma$ (IH) for the true $\theta_{23} =
35^\circ$. Note that the inverted hierarchy has the smaller octant
sensitivity than the normal hierarchy, because neutrinos have more
events than antineutrinos as shown in Fig. \ref{fig:atm:FCPC} and
the MSW resonance occurs in the NH (IH) case for neutrinos
(antineutrinos). We have also considered the $\Delta m^2_{atm}$
impact on the octant sensitivity. As shown in Fig.
\ref{fig:atm:T23octant}, the JUNO octant sensitivity has a weak
dependence on $\Delta m^2_{atm}$.

\begin{figure}[htb]
\begin{center}
\includegraphics[scale=0.47]{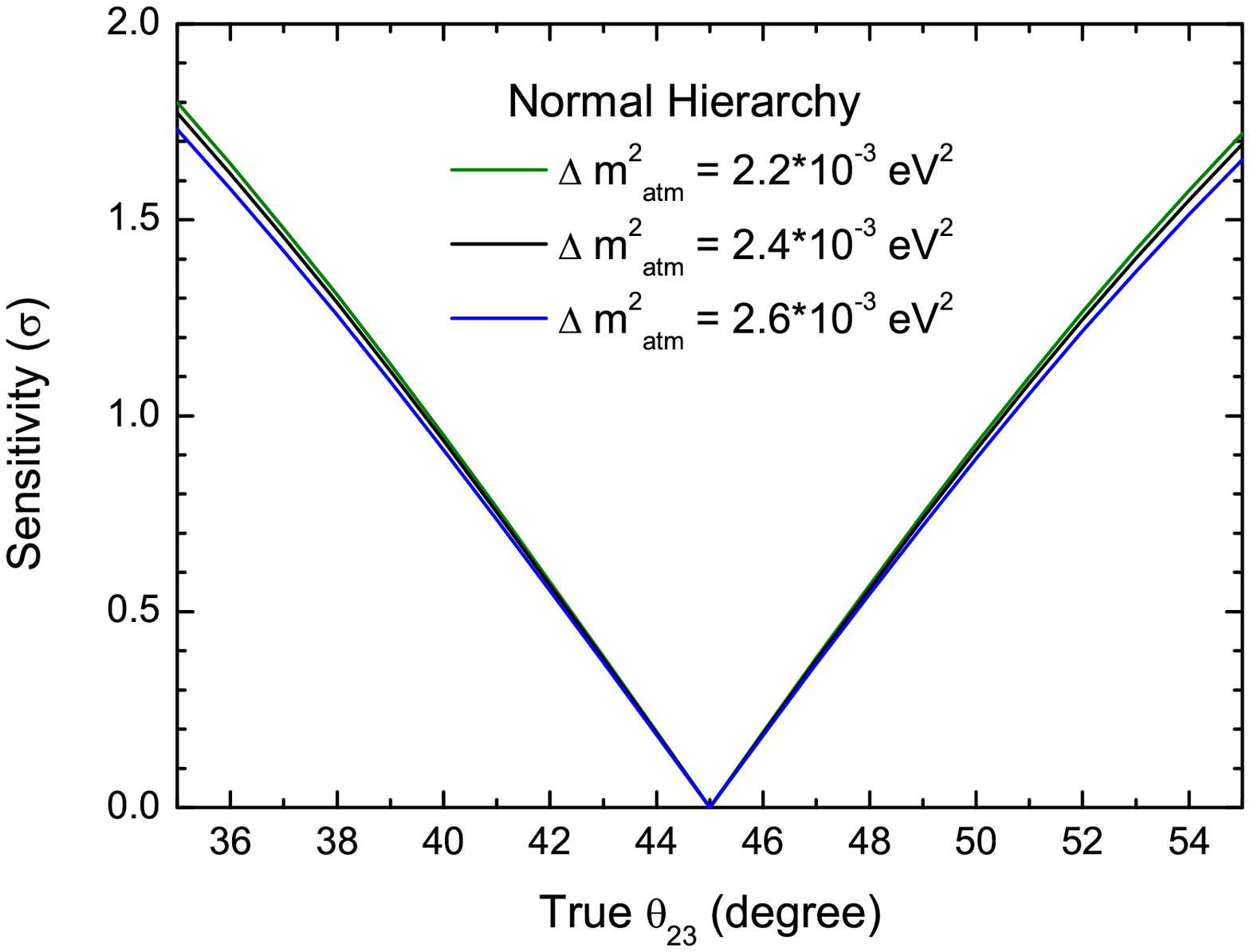}
\includegraphics[scale=0.47]{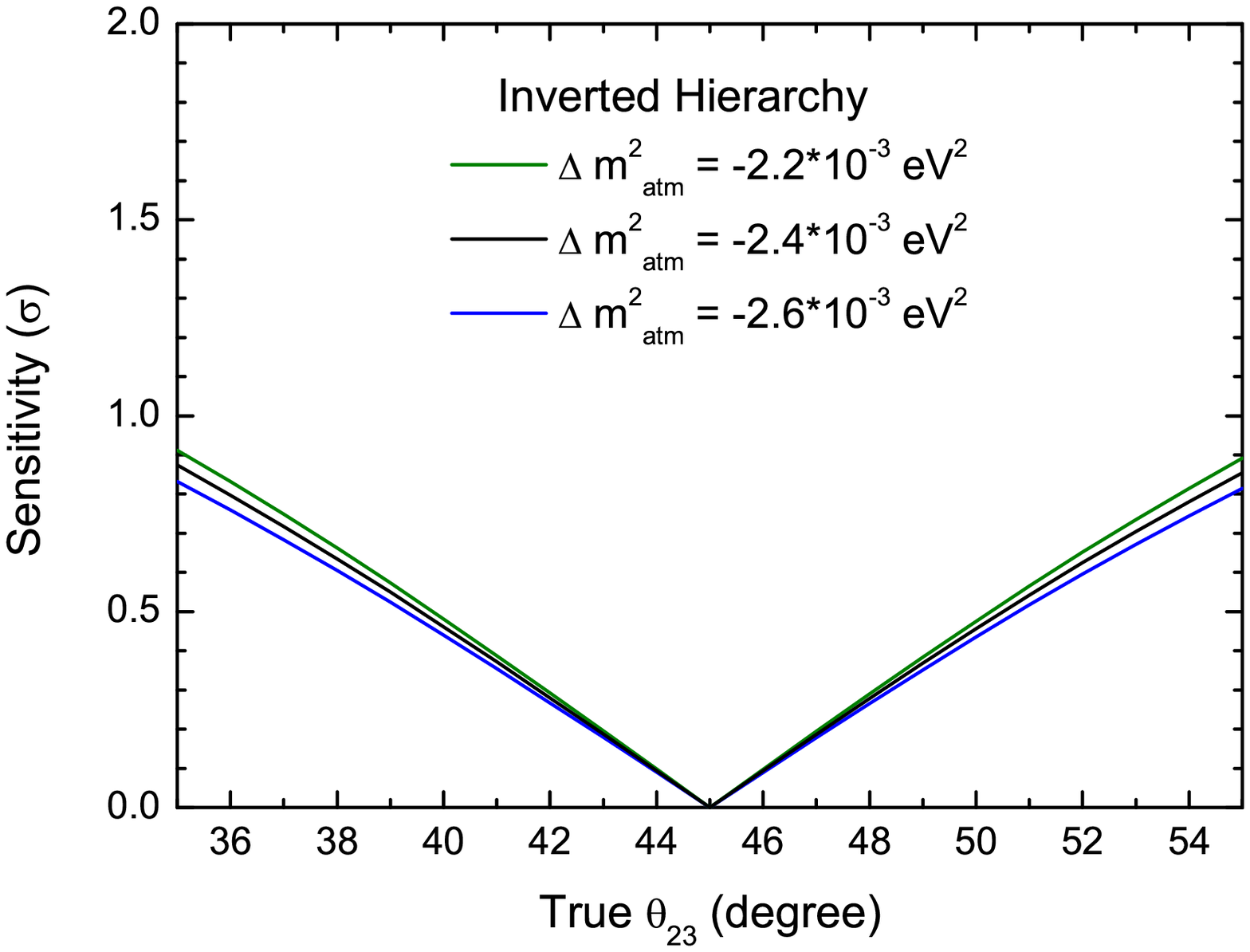}
\end{center}
\caption{The JUNO sensitivities to the octant for high energy muon
neutrino events as a function of true $\theta_{23}$ in the true NH
(left) and IH (right) cases. } \label{fig:atm:T23octant}
\end{figure}

\subsubsection{CP Phase \texorpdfstring{$\delta$}{delta-CP} }
\label{subsec:atm:CP}

As shown in Fig. \ref{fig:atm:CPVariation}, both $P(\nu_e
\rightarrow \nu_\mu)$ and $P(\nu_\mu \rightarrow \nu_\mu)$ can be
used to measure the CP phase $\delta$. However only the appearance
oscillation probabilities are possible for us to discover the CP
violation phenomenon, because $P(\nu_\alpha \rightarrow \nu_\alpha)
= P(\bar{\nu}_\alpha \rightarrow \bar{\nu}_\alpha)$ from the CPT
invariance. The disappearance $P(\nu_\mu \rightarrow \nu_\mu)$ only
contain the $\cos \delta$ terms which are invariant in the CP
transformation ($\delta \rightarrow -\delta$). The appearance
$P(\nu_e \rightarrow \nu_\mu)$ includes both $\sin \delta$ and $\cos
\delta$ terms as shown in the approximated expressions of Ref.
\cite{Akhmedov:2004ny}. The $\sin \delta$ terms can help us to
discover the CP violation phenomenon, namely $P(\nu_\alpha
\rightarrow \nu_\beta) \neq P(\bar{\nu}_\alpha \rightarrow
\bar{\nu}_\beta)$. Note that $\delta = 0$ and $\delta = \pi$
correspond to the CP conservation. For a given true $\delta$, the CP
violation sensitivity comes from whether or not it can be
distinguished from the CP conservation. We take $\delta = 0$ and
$\delta = \pi$ as the test values to calculate the best fit minimal
differences $\Delta \chi_0^2 = \chi^2 (0) - \chi^2 (\delta)$ and
$\Delta \chi_{\pi}^2 = \chi^2 (\pi) - \chi^2 (\delta)$. Then we
define the minimal value of $\sqrt{\Delta \chi_0^2}$ and
$\sqrt{\Delta \chi_{\pi}^2}$ as the JUNO CP violation sensitivity.
Since the $\sin \delta$ terms have the opposite signs for neutrinos
and antineutrinos, the statistical separation of $\nu_\mu$ and
$\bar{\nu}_\mu$ signals will improve the sensitivity.

In Sec.~\ref{subsec:atm:Classification}, we have required $L_\mu
\geq 5$ m which means the $\nu_\mu$ and $\bar{\nu}_\mu$ energy is
larger than 0.9 GeV. For these high energy  $\nu_\mu$ and
$\bar{\nu}_\mu$,  the oscillation probabilities $P(\nu_\mu
\rightarrow \nu_\mu)$ and $P(\nu_e \rightarrow \nu_\mu)$ are
sensitive to the CP phase $\delta$ as shown in Fig.
\ref{fig:atm:CPVariation}. However the kinematical smearing will
obviously suppress the CP violation sensitivity. As shown in Fig.
\ref{fig:atm:CP}, the JUNO sensitivity to CP violation is very small
for these high energy $\nu_\mu$ and $\bar{\nu}_\mu$ events where we
have taken the best fit values of $\Delta m_{atm}^2$. It is found
that different $\sin^2 \theta_{23}$ can change the predicted
sensitivity.

\begin{figure}[htb]
\begin{center}
\includegraphics[scale=0.46]{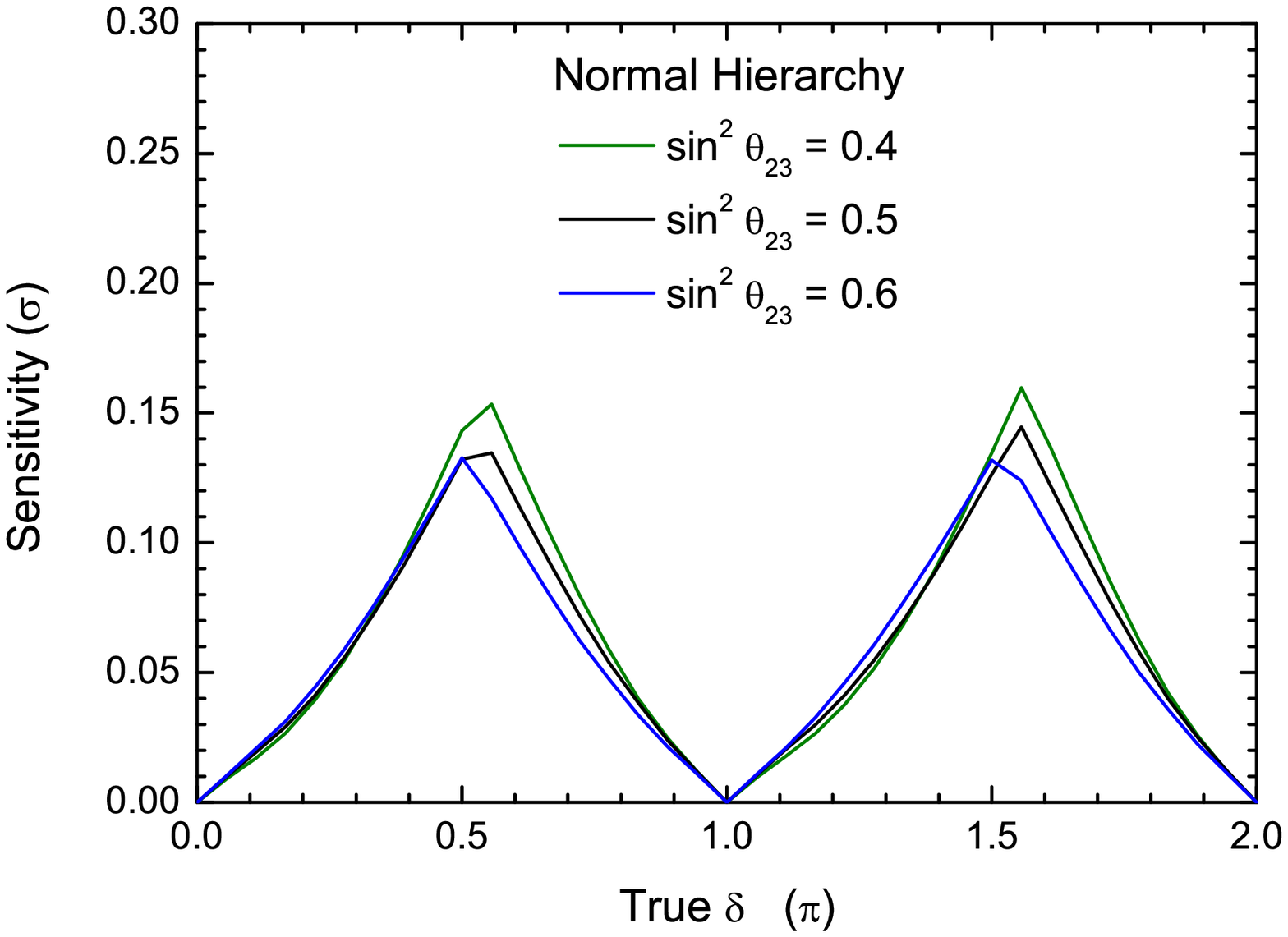}
\includegraphics[scale=0.46]{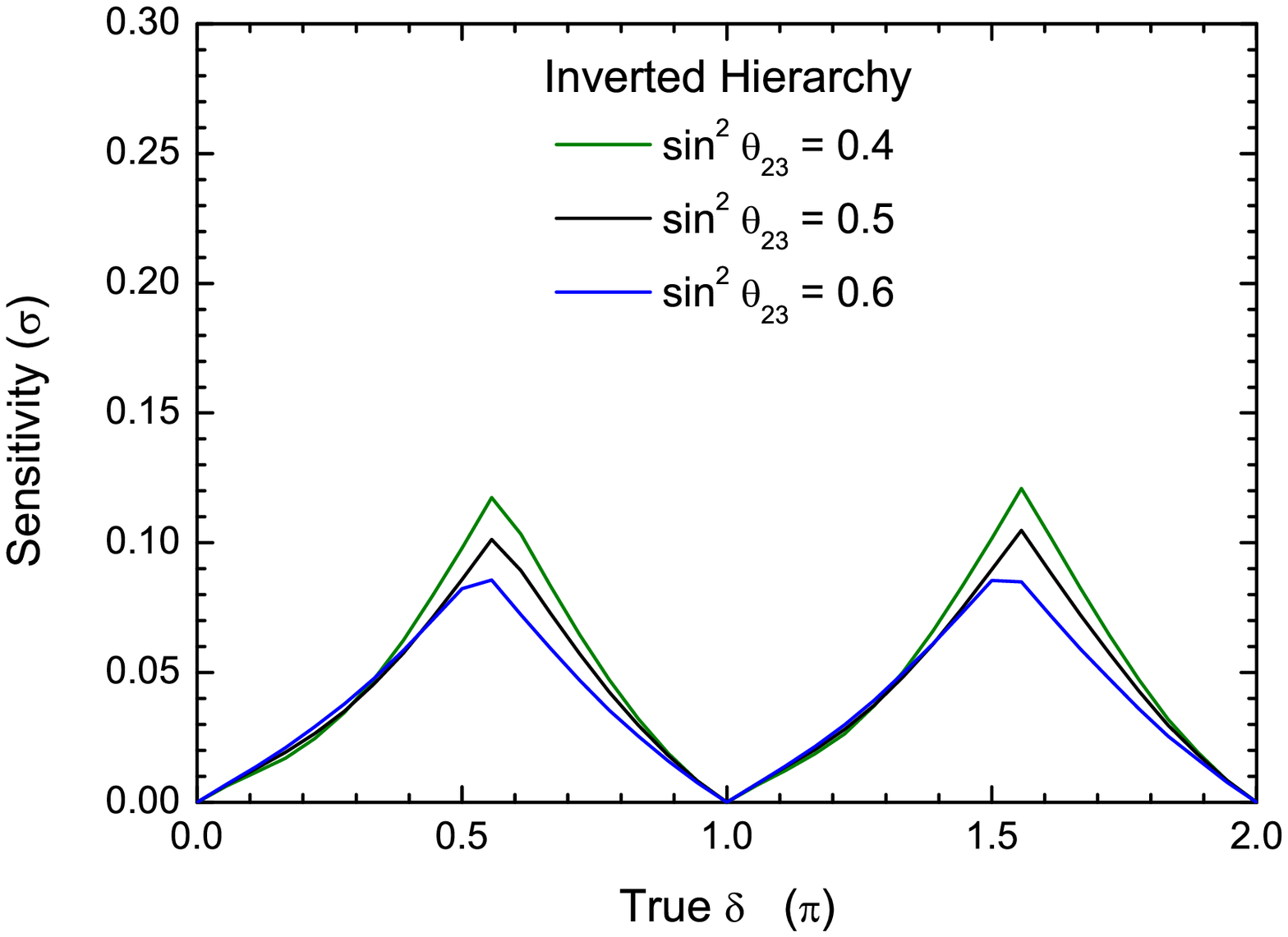}
\end{center}
\caption{The JUNO's sensitivities to CP violation for high energy
muon neutrino events as a function of true $\delta$ in the NH (left)
and IH (right) cases. } \label{fig:atm:CP}
\end{figure}

The lower energy [100, 300] MeV neutrinos  have more sensitivity to
measure CP phase $\delta$ as shown in Fig.
\ref{fig:atm:CPVariation}. With the techniques discussed in
Sec.~\ref{subsec:atm:PID}, it is possible to pursue such a
measurement. The major uncertainty comes from the lack of knowledge
of the daughter nuclei, which might spoil the idea. There might be
also some backgrounds from the NC and CC processes of high energy
neutrinos. An estimate of the maximal JUNO sensitivity is made by
ignoring the above unknowns. The pure statistical sensitivity of JUNO detector
in ten years is shown in Fig. \ref{fig:atm:CPSensitivity} where all
$\nu_\mu$, $\bar\nu_\mu$, $\nu_e$, and $\bar\nu_e$ are considered.
Direction reconstruction of electron and muon is not crucial,
because it is only weakly correlated to the initial neutrino
direction at these low energies.

\begin{figure}[!ht]
\begin{center}
\vspace{1.1cm}
\includegraphics[angle=270, scale=0.4]{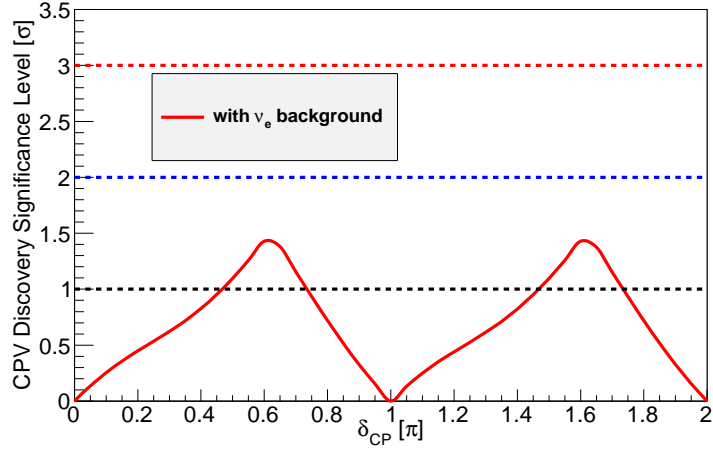}
\end{center}
\caption{The upper limit of the CP discovery sensitivity of JUNO in
ten years. All statistics of $\nu_\mu$, $\bar\nu_\mu$, $\nu_e$, and
$\bar\nu_e$ in low energy are considered.}
\label{fig:atm:CPSensitivity}
\end{figure}

\subsubsection{Summary} \label{subsec:atm:summary}

We have investigated atmospheric neutrinos in JUNO and discussed
their contributions to the MH, octant and CP violation. In terms of
the reconstruction potential of the JUNO detector, we conservatively
use the atmospheric $\nu_\mu$ and $\bar{\nu}_\mu$ with the track
length larger than 5 m for our physical analysis. These events have
been classified into the FC $\nu_\mu$-like, FC $\bar{\nu}_\mu$-like,
PC $\nu_\mu$-like and PC $\bar{\nu}_\mu$-like samples based on the
$\mu^\pm$ track and the statistical charge separation. Our numerical
results have shown that the JUNO's MH sensitivity can reach 0.9
$\sigma$ for a 200 kton-years exposure and $\sin^2 \theta_{23} =
0.5$, which is complementary to the JUNO reactor neutrino results.
The wrong $\theta_{23}$ octant could be ruled out at $1.8 \, \sigma$
($0.9 \,\sigma$) for the true normal (inverted) hierarchy  and
$\theta_{23} = 35^\circ$.  It is found that the JUNO sensitivity to
CP violation is very small when we only consider the high energy
$\nu_\mu$ and $\bar{\nu}_\mu$. In contrast to the high energy
neutrinos, the low energy neutrinos may give the dominant
contributions to the CP violation sensitivity even if we do not use
the direction information. In our analysis of the high-energy
neutrino events, all $\nu_e/\bar{\nu}_e$ and the sub-GeV
$\nu_\mu/\bar{\nu}_\mu$ data have been discarded from the
conservative point of view. According to the optimistic and
pessimistic estimations, atmospheric neutrinos in JUNO may give a
better sensitivity to the mass hierarchy. In the future, we shall
investigate the particle reconstruction and identification of the
JUNO detector in detail. In addition, we shall analyze the upward
through-going and stopping muon events those are produced by
atmospheric neutrinos in the rock and water pool. The future
atmospheric neutrino exploration in JUNO can also help us to probe
new physics beyond the standard model, such as the non-standard
neutrino interactions, sterile neutrinos and new long range forces,
etc.

\clearpage

\section{Geoneutrinos}
\label{sec:geo}

\blfootnote{Editors: Ran Han (hanran@ncepu.edu.cn), Livia Ludhova (ludhova@gmail.com), and Bill McDonough (mcdonoug@umd.edu)}
\blfootnote{Major contributors: Fabio Mantovani, Barbara Ricci, Liangjian Wen, Oleg Smirnov, Yufei Xi, and Sandra Zavatarelli}

\subsection{Introduction}
\label{subsec:geo:intro}

For half a century we have established with considerable precision the Earth's surface heat flow as $46 \pm 3$\,TW~\cite{Jaupart,Davies}.
However we are vigorously debating what fraction of this power comes from primordial versus radioactive sources.
This debate touches on the composition of the Earth, the question of chemical layering in the mantle, the nature of mantle convection, the
energy needed to drive plate tectonics, and the power source of the geodynamo, which powers the magnetosphere that shields the Earth from the harmful cosmic ray flux.

Over the last decade particle physicists have detected the Earth's geoneutrino flux~\cite{Bellini:2013nah,Gando:2013nba}, electron antineutrinos that
are derived from naturally occurring, radioactive beta-decay events inside the Earth~\cite{Krauss:1983zn,McDonough:2012zz}.
Matter, including the Earth, is mostly transparent to these elusive messengers that reveal the sources of heat inside the Earth as they
virtually escape detection, being that the Earth's geoneutrino flux is some $10^6$  cm$^{-2}$ s$^{-1}$. However, by detecting a few particles per years we are now measuring the geoneutrino flux from thorium and uranium inside the planet, which in turn allows us to estimate the amount of radiogenic power driving the Earth's engine.

Today we are experiencing a renaissance in neutrino studies, in part driven by the fact that 15 years ago physicists revealed the phenomenon
of neutrino oscillations and thus the fact that neutrinos have non-zero mass.  Neutrino science is now bringing us fundamental insights into
the nature of nuclear reactions, revealing for us the nuclear fusion in the core of the sun, nuclear fission in man-made nuclear reactors,
and identifying our planet's nuclear fuel cycle, as reflected in the heat produced during radioactive decay. Geology is the fortunate
recipient of the particle physicists' efforts to detect the Earth's emission of geoneutrinos.

How does geoneutrino detection serve the geological community and what transformational insights will it bring? It directly estimates the
radiogenic heating in the Earth from thorium and uranium.  These elements, along with potassium, account for more than $99\%$ of the
radiogenic heat production in the Earth. This component, along with the primordial energy of accretion and core segregation, define
the total planetary power budget of the planet.  By defining the absolute abundance of Th and U in the Earth, with accuracy and precision,
we can:
\begin{enumerate}
  \item define the building blocks, the chondritic meteorites, that formed the Earth,
  \item resolve ever vexing paradoxes (e.g. $\textsuperscript{4}$He-heat flow, Ar budget and degassing) that fuel the debates of compositionally layered mantle structures or not,
  \item discriminate models of parameterized mantle convection that define the thermal evolution (dT/dt) of the Earth,
  \item potentially identify and characterize deep, hidden reservoirs (or not) in the mantle, and
  \item fix the radiogenic contribution to the terrestrial heat flow.

\end{enumerate}
Moreover, such studies can place stringent limits on the power of any natural nuclear reactor in or near the Earth's core.

\subsection{Expected geoneutrino signal}
\label{subsec:geo:signal}

The amount of U and Th and their distribution in a reservoir (e.g. crust and mantle) affects the integrated geoneutrino flux at a location on the Earth's surface. The unoscillated geoneutrino flux produced by a source volume $\Delta V $ at the detector position\footnote{In the case of detector near the Earth surface $|\overrightarrow R|$ is the radius of the Earth.} can be calculated with:
\begin{equation}
\label{eq:geo:phi}
\Phi(\Delta V)=\frac{1}{4\pi}\int_{\Delta V}d^{3}r\frac{A(\overrightarrow{r})}{|\overrightarrow{R}-\overrightarrow{r}|^2} \\,
\end{equation}
where A is the rate of geoneutrinos produced in the volume $\Delta V $, which depends on density, antineutrino production rates per unit
mass and U and Th abundances. The intensity of the flux depends on the inverse-square of the distance to the sources, and thus the crust
surrounding the detector, which contains a relative small amount of the Earth's U and Th budget, gives a large contribution to the signal.
It is predicted that the first $\sim500$ km from the KamLAND detector contributes about $50\%$ of the measured geoneutrino signal, with this
volume containing $\sim0.1\%$ of the total crustal U and Th mass~\cite{Strati:2014kaa}. For the JUNO site it is predicted that the first 550 km
from the detector contribute some $50\%$ of the measured geoneutrino signal (Fig.~\ref{fig:geo:1}).

\begin{figure}[htb]
\begin{center}
\includegraphics[width=0.8\textwidth]{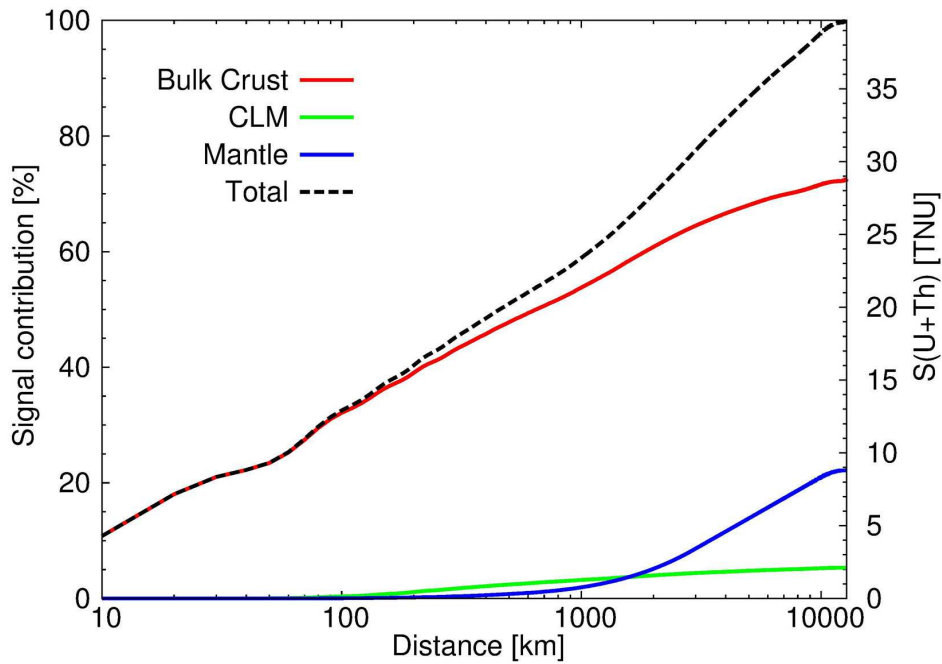}
\caption{Geoneutrino signal contribution at JUNO. The cumulative geoneutrino signal and the percentage contributions
of the Bulk Crust, Continental Lithospheric Mantle (CLM) and Mantle are represented as function of the distance from JUNO~\cite{Strati:2014kaa}.
\label{fig:geo:1}}
\end{center}
\end{figure}

The geophysical structure of the global crustal model described in~\cite{Huang13} takes into account data from different studies: body wave studies using reflection and refraction methods, surface wave dispersion studies and gravity surveys. On the basis of this model, the expected geoneutrino
signal at JUNO, which originates from the U and Th in the global Bulk Crust (BC) and in the Continental Lithospheric Mantle (CLM), are
S$_{\text{BC}} = 28.2 ^{+5.2}_{-4.5}$ TNU~\footnote{TNU: Terrestrial Neutrino Units. 1 TNU = 1 event/yr/$10^{32}$ target protons, which is approximately the number of free protons in 1 kiloton liquid scintillation.} and S$_{\text{CLM}} = 2.1 ^{+2.9}_{-1.3}$ TNU respectively (Tab.~\ref{tab:geo:GeoJuno})~\cite{Strati:2014kaa}. The 1$\sigma$ asymmetric uncertainties reflect the propagation of non-Gaussian distributions of U
and Th abundances in the deep BC and CLM, together with the Gaussian distributions of the errors associated with the geophysical inputs.

\begin{table}[ht]
\begin{center}
\caption{Geoneutrino signals from U and Th expected in JUNO. The signals from different reservoirs (CLM = Continental Lithospheric Mantle,
LS = Lithosphere, DM = Depleted Mantle, EM = Enriched Mantle) indicated in the first column are in TNU~\cite{Strati:2014kaa}. Total equals
Total LS + DM + EM. The mantle signal can span between 1-19 TNU according to the composition of the Earth: here the geoneutrino signal
produced by a primitive mantle having m(U)$ = 8.1 \cdot 10^{16}$ kg and m(Th)$ = 33\cdot 10^{16}$ kg is reported.\label{tab:geo:GeoJuno}}
\begin{tabular}{l c c c c  c }
\hline
\hline

               & S(U)                             &&  S(Th) && S(U+Th) \\
               \hline
&&&&&\\
\iffalse
Sed CC & $0.5^{+0.1}_{-0.1} $  &&   $0.2^{+0.0}_{-0.0} $  &&  $ 0.6^{+0.1}_{-0.1} $ \\
&&&&&\\
UC & $14.6^{+3.5}_{-3.4} $  &&   $3.9^{+0.5}_{-0.5} $  &&  $ 18.5^{+3.6}_{-3.4} $ \\
&&&&&\\
MC & $4.7^{+3.0}_{-1.8} $  &&   $1.7^{+1.6}_{-0.8} $  &&  $ 6.8^{+3.6}_{-2.3} $ \\
&&&&&\\
LC & $0.9^{+0.7}_{-0.4} $  &&   $0.4^{+0.7}_{-0.2} $  &&  $ 1.5^{+1.0}_{-0.6} $ \\
&&&&&\\
Sed OC& $0.08^{+0.02}_{-0.02} $  &&   $0.03^{+0.01}_{-0.01} $  &&  $ 0.11^{+0.02}_{-0.02} $ \\
&&&&&\\
OC& $0.05^{+0.02}_{-0.02} $  &&   $0.01^{+0.01}_{-0.00} $  &&  $ 0.06^{+0.02}_{-0.02} $ \\
&&&&&\\
\fi
Bulk crust & $21.3^{+4.8}_{-4.2} $  &&   $6.6^{+1.9}_{-1.2} $  &&  $ 28.2^{+5.2}_{-4.5} $ \\
&&&&&\\
CLM & $1.3^{+2.4}_{-0.9} $  &&   $0.4^{+1.0}_{-0.3} $  &&  $ 2.1^{+2.9}_{-1.3} $ \\
&&&&&\\
Total LS & $23.2^{+5.9}_{-4.8} $  &&   $7.3^{+2.4}_{-1.5} $  &&  $ 30.9^{+6.5}_{-5.2} $ \\
&&&&&\\
DM & $4.1$  &&   $0.8 $  &&  $ 4.9$ \\
&&&&&\\
EM & $2.9 $  &&   $0.9 $  &&  $ 3.8$ \\
&&&&&\\
\hline
&&&&&\\
Total & $30.3^{+5.9}_{-4.8} $  &&   $9.0^{+2.4}_{-1.5} $  &&  $ 39.7^{+6.5}_{-5.2} $ \\
&&&&&\\

\hline\hline
\end{tabular}
\end{center}
\end{table}

The Total predicted geoneutrino signal at JUNO S$_{\text{TOT}}$ $= 39.7^{+6.5}_{-5.2}$ TNU~\cite{Strati:2014kaa}, which is obtained assuming a
predicted Mantle contribution of S$_{\text{M}}$ $\sim$9 TNU according to a mass balance argument. The adopted mantle model, divided into two
spherically symmetric domains (Depleted Mantle and Enriched Mantle), refers to a Bulk Silicate Earth based on a primitive mantle having U
and Th mass of m(U)$ = 8.1 \cdot 10^{16}$ kg and m(Th)$ = 33\cdot 10^{16}$ kg, respectively ~\cite{McDonough:2012zz}.
 Alternatively, another class of Earth models~\cite{Javoy2010}, with a
global composition similar to that observed in enstatite chondrites, is characterized by smaller amounts of U and Th (i.e. m(U) $= 4.9\cdot
10^{16}$ kg and m(Th) $= 20 \cdot10^{16}$ kg).  Consequently, assuming the same lithospheric contribution and mass balance arguments, this model has a minimum signal from the mantle of S$_{\text{M}}$ $\sim$1 TNU~\cite{Fiorentini:2012yk}.
On the other hand, models based on the energetics of mantle convection and on the observed surface heat loss
\cite{Turcotte2002} require a high concentration of U and Th in the primitive mantle (e.g. m(U) = $12.5 \cdot 10^{16}$ kg and m(Th) = $52
\cdot 10^{16}$ kg). In this scenario the present mantle signal could reach S$_{\text{M}}$ $\sim$19 TNU.
Ranges of geoneutrino signals produced by symmetric and asymmetric distributions of U and Th in the mantle are extensively discussed in
\cite{Sramek:2012nk}.
The present crustal models are affected by uncertainties that are comparable to the mantle's contribution: this comes mainly from
understanding the abundance and distribution of Th and U in the regional crystalline rocks. Upper crustal estimates for Th and U abundances vary by 10\% and 22\%, respectively, whereas larger uncertainties exist for their estimates in middle and lower crustal
lithologies~\cite{Huang13}.

Thus, to understand the relative contributions from the crust and mantle to the total geoneutrino signal at JUNO an accurate estimation of
the geoneutrino flux from the crustal region surrounding JUNO is a priority. Detailed  geological, geochemical, and geophysical studies were
performed in the areas surrounding the KamLAND detector at Kamioka, Japan~\cite{Enomoto,Fiorentini:2005cu}, the Borexino detector at Gran Sasso,
Italy~\cite{Coltorti:2011gr}, and the SNO+ detector at Sudbury, Canada~\cite{Huang:2014dpa}. The present geoneutrino flux prediction for JUNO
\cite{Strati:2014kaa} is recognized as being a first step and is not based on data specifically from the local region.  This estimate was
generated from a set of global databases, including a compositional estimate of the average upper crust, physical structure of crustal
rocks, and models of the seismic and gravity properties of the crust (see~\cite{Huang13}).  To improve upon the accuracy and precision of
this estimate, future studies of JUNO will necessarily require input from geological, geochemical and geophysical studies of the area
surrounding JUNO, up to some 550 km away from the detector.  The survey distance from the JUNO detector can be fixed by resolution
studies, with trade offs between distance, limits on uncertainty, and costs of survey.

Thus, a refined model for predicting the geoneutrino flux at JUNO is needed. To meet this need, an integrated effort is required and will involve an exciting and collaborative research effort between particle physicists and Earth scientists.

\subsection{The local geology study around JUNO}
\label{subsec:geo:local}

The JUNO detector is to be built near the southern continental margin of China close to the South China Sea (Fig.~\ref{fig:geo:2}). This
passive margin area has an extensive continental shelf and represents the transition between continental and oceanic plates.

The JUNO detector will accumulate a signal from the regional distribution of Th and U, as well as from the rest of the planet.  The local
contribution (the immediate several hundred km of crust surrounding the detector) typically contributes about half of the signal at the
detector.  So it will be important for the geological community to map out carefully and identify the abundance and distribution of Th and U
in this region.  Models for calculating the total geoneutrino signal from an area surrounding a detector and the rest of the planet requires
extensive integration of spatially resolved local geological, geochemical, and geophysical data that is then integrated into a global model
for the distribution of geoneutrino sources. Detailed characterization of the local region also brings additional fundamental benefits to
the geoscience community.

\begin{figure}%[htb]
\begin{center}
\includegraphics[width=0.8\textwidth]{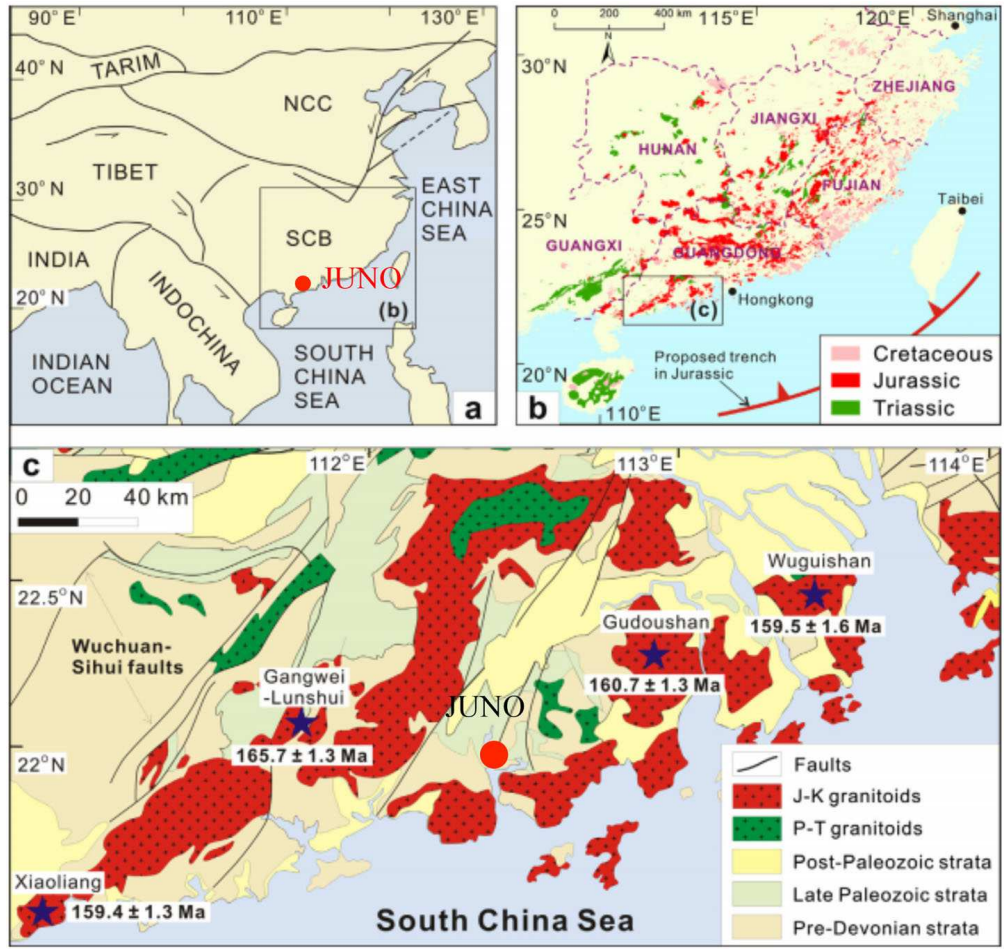}
\caption{(a) Sketch map showing major tectonic units around JUNO. NCC - North China Craton; SCB - South China Block. (b) Regional map showing the distribution of Mesozoic granites in the southeastern South China Block. Dashed purple lines show the provincial boundaries. (c) Geological map showing the age distribution of Mesozoic granites along the southern coastal region of the Guangdong Province, southeastern China). Stars highlight the plutons. New geochronological results are shown in bold. J-K-Jurassic and Cretaceous; P-T-Permian and Triassic~\cite{huang2013intraplate}.
\label{fig:geo:2}}
\end{center}
\end{figure}

Eastern China is documented, relative to a global perspective, as being a region of elevated heat flow and the Guangdong province
specifically, has hot springs and anomalously high heat flow zones associated with deep crustal fractures and fluid flow. Regionally,
southern and southeastern China have large surface areas covered with Mesozoic (circa 250 to 70 million years ago) granites containing high
concentrations of K, Th and U (the heat producing elements) (Fig.~\ref{fig:geo:2}b).

\begin{figure}%[htb]
\begin{center}
\includegraphics[width=0.8\textwidth]{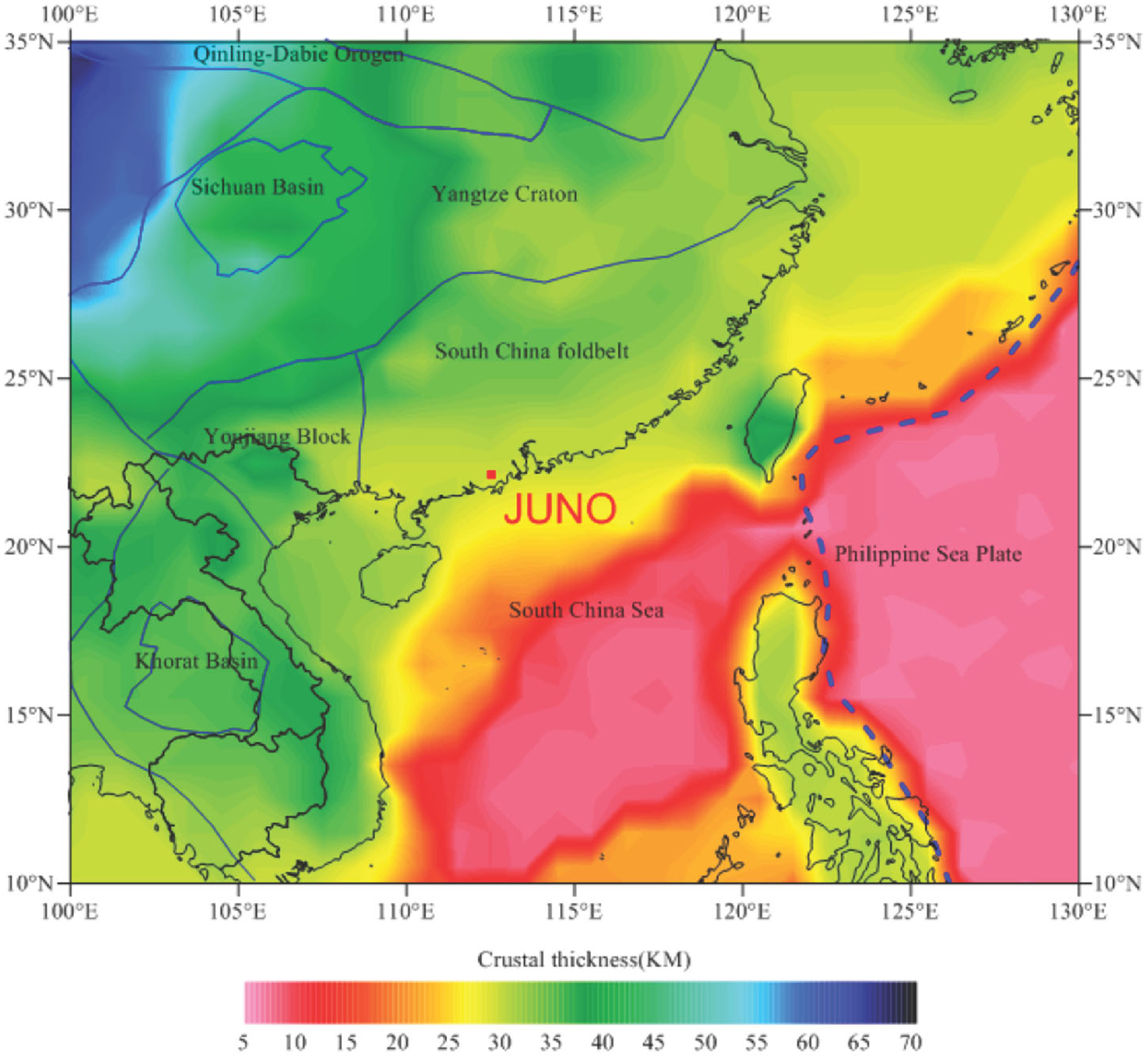}
\caption{Crustal thickness (exclusive water) in the surrounding area of JUNO from CRUST 1.0 model, 1$^\circ$ degree~\cite{Laske2013}.  Dashed thick blue line is plate boundary~\cite{DeMets1990}. Thin blue lines denote the main tectonic units. \label{fig:geo:3}}
\end{center}
\end{figure}

The Jurassic ($\sim$ 160 million years old) high-K granites are ubiquitous, extending from the continental interior to the coastal area
\cite{zhou2000origin} and
possess the highest amount of heat  producing elements.
The petrogenesis of granitic rocks and the tectonic evolution in the southeastern China continue to attract attention
(e.g.,\cite{gilder1996isotopic, zhou2000origin, niu2005generation, zhou2006petrogenesis, li2007formation, yang2013}) as they are a relevant
source of economic minerals and geothermal energy. Lithologically the granites are heterogeneous, rich in heat producing elements, and
important source rocks producing geoneutrinos~\cite{Strati:2014kaa}.

The thickness of the continental crust immediately surrounding JUNO has been modeled using CRUST 2.0 and 1.0~\cite{Strati:2014kaa,Huang13,Laske2013};
it varies between 26 and 32 km (Fig.~\ref{fig:geo:3}).
At some 350 km distance to the south of JUNO, the oceanic crust outcrops (approximately 8 km thick and containing $\sim$ 1/10 of the
concentration of Th and U as compared to the continental crust) and will contribute approximately to 0.2\% of the total expected
geoneutrino signal. On the basis of an existing global reference model for the Earth~\cite{Huang13}, an initial estimate of the local
geoneutrino signal expected at JUNO is S$_{\text{LOC}} = 17.4 ^{+3.3}_{-2.8}$ TNU~\cite{Strati:2014kaa}.
This flux contribution corresponds to $\sim$44\% of the expected geoneutrino signal at JUNO and exceeds the signal from the whole mantle.
The 2$^{\circ} \times2^{\circ}$ region including and to the north of JUNO is the area with the greatest thickness of sediments and
crystalline crust and it alone has been identified as contributing $\sim$27\% of the geoneutrino signal~\cite{Strati:2014kaa}.  Thus, these early studies
have identified significant regional targets that will require further intensive research to predict better the JUNO geoneutrino signal.

The main tasks for predicting the JUNO geoneutrino signal include surveys and descriptions of the geology, seismology, heat flow, and
geochemistry of the regional lithosphere (the Earth's outer conductive layer that includes the crust and the mechanically coupled layer of
mantle beneath it). Geological surveys define the surface lithologies, major structures, and geological provenances. Seismological
surveys include refraction and reflection mapping, receiver function and ambient noise tomography of the crust and lithospheric mantle;
these
studies provide a 3-dimensional image of the structures of the deep crust and lithospheric mantle, specifically in terms of major boundaries
and the vertical and horizontal variations in density, acoustic velocity and Poisson's ratio. Heat flow surveys will map out the surface
variation in heat flow, as well as the heat production and thermal conductivity of the immediate lithologies, and from this develop models
of near surface productivity versus reduced heat flow. Gravity surveys record the variations in density structure of the lithosphere and
mantle;
these surveys can be coupled with heat flow and seismic surveys to cross check and refine models of the lithosphere. Finally a wide range of
geochemical studies, including mapping the spatial variations in the abundance of K, Th and U, local helium gas flux measurements, and
gamma
ray spectroscopy, will define the nature and composition of the near surface and set up the potential for projecting the 3-dimensional
variation in the abundances of these elements at depth.
	
In addition to these survey studies it is vitally important to involve computational geological studies, where all of the
geological, geophysical and geochemical data are geo-located into an integrated, 3-dimensional model that is the essential physical and
chemical database of the Earth system.
The global and regional contributions of the geoneutrino flux need to be predicted for the area surrounding the JUNO detector, as have been
previously done at the existing detectors KamLAND (Japan)~\cite{Enomoto,Fiorentini:2005cu}, Borexino (Italy)~\cite{Coltorti:2011gr} and SNO+ (Canada)
\cite{Huang:2014dpa}.

The area of Kaiping, China, the site of the JUNO experiment, is close to the continental margin of south China. Beyond the shores of this
area is a significant region of continental shelf. This regional study therefore represents a golden opportunity to investigate in detail a passive
continental margin, a tectonic boundary that we know little of its nature. These broad continental margins typically host considerable
petroleum reserves and thus understanding their nature and structure might provide critical insights into identifying further energy
resources. Moreover, integration of the matrix of geophysical, geochemical and geological data into a single, self-consistent solution is
required in order to calculate the geoneutrino flux at JUNO.

\subsection{Detecting geoneutrino signal}
\label{subsec:geo:detection}

The full strength of geoneutrino studies comes from the synergistic activities of geology and physics, neither  acting independently of
the other and both accepting the challenge of knowing better the Earth and its secrets.  These studies represent the ultimate opportunity
for geologists and particle physicists to independently measure and test our predictions of the structure and composition of the planet's
interior~\cite{Bellini:2013nah}.  Thus, by 2020, when the JUNO detector will be built and begin data taking, we will have our best
prediction of the expected signal.  At that time we will wait and see what the measurement reveals.  It is a rare moment in geology where we
predict and then receive independent assessment of the strength of our prediction.

Detectors like JUNO are sited deep underground and are enormous (20 kiloton) structures; their size enhances their potential for detection and their overburden shields them from cosmic ray fluxes. These factors, combined with their ultra-clean construction, will allow these detectors to extract the geoneutrino signal.

Geoneutrinos from $^{238}$U and $^{232}$Th (not those from $^{235}$U and $^{40}$K) are above threshold for the classical antineutrino ($\bar
{\nu}_e$) detection reaction, the inverse beta decay (IBD) on free protons ($p$):
\begin{equation}
\label{eq:geo:1}
\bar{\nu_e} + p \rightarrow e^+ + n -1.806 ~{\mathrm {MeV}}.
\end{equation}
The characteristic temporal and spatial coincidence of prompt positron ($e^+$) and delayed neutron ($n$) flash events offers a clean signature that resolves them from background signals.

The first geoneutrino measurements were reported in 2005 by the KamLAND team~\cite{Araki:2005qa}, where they recorded 25 events over two years of
exposure. The KamLAND detector is sited deep in the Japanese Alps, opposite Tokyo, on the island of Honshu and is 1 kiloton in scale.
Another active detector is the Borexino experiment, located beneath the Apennine Mountains near the town of L'Aquila in Italy. This
detector, which is smaller (300 tons), reported in 2010 its measurement of the Earth's geoneutrino flux at the 99.997\% C.L (5$\sigma$)~\cite{Bellini:2010hy}. Both experiments
provided new updated measurements~\cite{Bellini:2013nah,Gando:2013nba}, on the occasion of the inter-disciplinary Neutrino Geoscience
conference series, in 2013 held in Takayama, Japan.

A third, new detector, SNO+ is a re-deployed Sudbury Neutrino Observatory. It is a second kiloton detector and it is situated in Sudbury, Ontario, Canada, on the ancient stable Superior Craton~\cite{Chen:2005zza}. JUNO will then join these detectors in 2020, being 20 times larger than all the existing devices.  In its first year of operation, JUNO will record more geoneutrino events than all other detectors combined will have accumulated to that time. Before discussing the detection of geoneutrino events we will discuss the various experimental backgrounds coming from nuclear reactors and other sources.

%********************************************************************************************

\subsection{Reactor antineutrino background}
\label{subsec:geo:reactor}

Determination of the expected signal from reactor $\bar{\nu}_e$'s  requires  a wide set of information, spanning from characteristics of nuclear cores placed around the world to neutrino properties. In this section, the number of expected reactor events and the corresponding spectral shape have been calculated as follows:
\begin{equation}
N_{ev}= \epsilon \,  N_p \,  \tau \,
 \sum_{r=1}^{N_{react}}
\frac { P_{r}}{4 \pi L_{r}^{2}} < LF_{r}>  \times
 \int dE_{\bar{\nu}_e} \sum_{i=1}^4 \frac {p_{i}}{Q_{i}} \phi_{i}(E_{\bar{\nu}_e})
 \sigma(E_{\bar{\nu}_e})
P_{ee}(E_{\bar{\nu}_e},\hat\theta, L_r),
\label{eqn:geo:reactor}
\end{equation}
where $\epsilon$ is the detection efficiency,  $N_p$ is the number of target protons, and $\tau$ is the data-taking time.
The index $r$ cycles over the number of reactors considered: $P_{r}$ is the nominal thermal power, $L_{r}$ is the reactor-detector distance,  $<LF_{r}>$ indicates the average Load Factor (LF)~\footnote{Load Factor reflects the ratio between the delivered and the nominal reactor power. The shut-down periods are taken into account.} in the period $\tau$,
$E_{\bar{\nu}_e}$ is the antineutrino energy.
The index $i$ stands for the different nuclear-fuel components ($^{235}$U, $^{238}$U, $^{239}$Pu, $^{241}$Pu),
$p_{i}$ is the power fraction,  $Q_i$ is the energy released per fission of the component $i$, and
$\phi_i(E_{\bar{\nu}})$ is the antineutrino energy spectrum originated by the fission of component $i$.
$\sigma(E_{\bar{\nu_e}})$ is the inverse-beta-decay cross section. $P_{ee}$ is the energy-dependent survival probability of electron antineutrinos traveling the baseline $L_r$, for  mixing parameters
$\hat\theta$ = ($\delta m^2$, $\Delta m^2$, $\sin^2\theta_{12}$, $\sin^2\theta_{13}$).

We express the expected reactor antineutrino signal in unit of TNU, same as the geoneutrinos. This means that in Eq.~(\ref{eqn:geo:reactor}) we assume a 100\% detection efficiency for a detector containing  $N_p=10^{32}$ target protons and operating continuously for $\tau=$ 1 year. We consider all the nuclear cores in the world operating in the year 2013, plus all the cores of the Taishan and Yangjiang nuclear power plants that will enter operation in 2020.
Information on the nominal thermal power and Load Factor for each existing nuclear core comes from the International Agency of Atomic Energy (IAEA)~\cite{IAEA}.
For each core the distance $L_r$ has been calculated taking into account the position of the JUNO detector (22.12 N,  112.52 E)~\cite{Li:2013zyd} and the positions of all the cores in the world according to the database adopted in~\cite{Baldoncini:2014vda}.
For the future Taishan and Yangjiang nuclear plants, we considered the thermal power of 18.4 GW and 17.4 GW, the distance of 52.7 and 53.0 km, respectively, and we assumed 80\% Load Factor.
Typical power fractions for the PWR and BWR type reactors are adopted to be equal to the averaged value of the Daya Bay nuclear cores:
$\mbox{$^{235}$U} : \mbox{$^{238}$U} : \mbox{$^{239}$Pu}  : \mbox{$^{241}$Pu}   =
0.577 :  0.076 : 0.295 : 0.052 $.
As in~\cite{Bellini:2013nah}, we have considered for the European reactors using Mixed OXide technology that 30\% of their thermal power originates from the fuel with power fractions $0.000 :  0.080  : 0.708 :  0.212$, respectively and for the cores in the world using heavy-water moderator $p_{i}$  = $\, 0.543: 0.024 : 0.411 : 0.022$, respectively.

The $\phi_i(E_{\bar{\nu}_e})$ energy spectra are taken from~\cite{Mueller:2011nm}, the $Q_i$  energy released per fission of component $i$ are from~\cite{Ma:2012bm}, and the interaction cross section $\sigma(E_{\bar{\nu}_e})$ for inverse beta decay reaction is from~\cite{Strumia:2003zx}.
For the vacuum survival probability $P_{ee}$ and the corresponding  mixing parameters we adopt the most recent determinations by Capozzi et al.~\cite{Capozzi:2013csa} for the Normal Hierarchy.

Tab~\ref{tab:geo:ReactorError} gives an overview of different contributions to the total error on the expected reactor antineutrino signal in the total energy window (1.8 MeV $< E_{\bar{\nu_{e}}} <$ 10 MeV).

\begin{table}[h]
\begin{center}
\caption{Systematic uncertainties on the expected reactor antineutrino signal in the total energy window.  See Eq.~(\ref{eqn:geo:reactor}) and accompanying text for details.}
\begin{tabular}{ll|ll}
\hline\hline
Source								&Error	&Source					&Error \\
								    &[\%]	&					&[\%] \\
\hline
Fuel composition						&1.0	&$\sin^2\theta_{12}$				& 3.6\\
$\phi_i(E_{\bar{\nu_e}})$				       &3.4    &$\sin^2\theta_{13}$				& 0.21\\
$P_{r}$				                                   &2.0    & $\Delta m^2$			        & $<10^{-2}$\\
$\sigma(E_{\bar{\nu}_e})$			              &0.4    & $\delta m^2$			        & 1.2\\
$Q_i$				                                &0.09	&				        & \\
\hline
Total								&		&				&5.6\\
\hline\hline
\end{tabular}
\label{tab:geo:ReactorError}
\end{center}
\end{table}

The error due to the fuel composition reflects the possible differences among different cores and the unknown stage of burn-up in each of them.  This error is determined as the variance of a uniform distribution, assumed for different signals calculated with the sets of power fractions available in the literature (see Tab.~1 of~\cite{Baldoncini:2014vda}).

The uncertainty due to $\phi_i(E_{\bar{\nu}_e})$ energy spectra is very conservatively quoted as 3.4\%, which is determined as the difference in the expected signal computed with the reference spectra of~\cite{Mueller:2011nm} and with the spectra published by~\cite{Huber:2004xh}.

The uncertainty due to the thermal power $P_r$ is conservatively assumed to be 2\%. Although thermal powers can be measured at sub-percent level, the adopted uncertainty reflects the regulatory specifications for Japan and United States (see e.g., Refs.~\cite{Cao:2011gb} and~\cite{Djurcic:2008ny}), furthermore allowing to account for the fact that the Load Factors reported in the IAEA database refer to the electrical power and not to the thermal one.

The effect of the uncertainties of oscillation parameters is calculated by varying each parameter one at the time in the $1\sigma$ range and assuming a uniform distribution of the obtained signals; note that the dominant contribution arises from the $\theta_{12}$ mixing angle.
A more extensive treatment of the uncertainties can be found in~\cite{Baldoncini:2014vda},
where a Monte Carlo based approach has been adopted in the determination of the uncertainty budget as well as of the signal central values.

In conclusion, the expected reactor antineutrino signal in the total energy window is $(1569 \pm 88)$\,TNU \footnote{The signal value dose not include the matter effect in the neutrino-oscillation mechanism, which represents an additional increase of +1.1\% with respect to the vacuum oscillation. Furthermore, the contribution of antineutrinos emitted from spent nuclear fuel (SNF) can be considered as +1.9\% signal increase~\cite{Zhou:2012zzc}}. The signal in the geoneutrino energy window (1.8 MeV $< E_{\bar{\nu_{e}}} <$ 3.27 MeV) is $(351 \pm 27)$\,TNU, where the Taishan and Yangjiang nuclear power stations contribute more than 90\% of the total.
%As the sensitivity in geoneutrino detection is mainly affected by the reactor signal in the geoneutrino energy window, it is interesting to note that according to the 2013 reactor operating status the expected reactor signal in the geoneutrino energy window is equal to $(26.3 \pm 1.5)$\,TNU, smaller by more than one order of magnitude with respect to the 2020 scenario.

We want to stress that JUNO will measure neutrino oscillations parameters with unprecedented precision. Thus, as a consequence, the total error on the expected antineutrino flux will be strongly reduced.
In the geoneutrino sensitivity studies presented here, we have adopted the 5.6\% uncertainty, since it reflects the current knowledge of the estimation of the reactor antineutrino background for JUNO geoneutrino measurement. As we will discuss below, the reactor antineutrino background is not constrained and is kept as a free fit parameter in our simulations of possible JUNO geoneutrino measurements. Thus, the error on the predicted reactor-antineutrino rate enters the simulations only as a parameter defining the signal-to-background ratio. We have checked, that considering an error of 2.4\% (instead of 5.6\%) does not significantly change our results on geoneutrino sensitivity.

%********************************************************************************************

\subsection{Non antineutrino background}
\label{subsec:geo:bgr}

The coincidence tag used in the $\bar{\nu}_e$-detection is a very powerful tool in background suppression.
Nevertheless, the event rate expected from geoneutrino interactions is quite small (few hundreds of events/year) and a number of non antineutrino background events faking geoneutrino interactions has to be properly accounted for in the sensitivity study.

\begin{table}[t]
\begin{center}
\caption{Main non antineutrino background components assumed in the geoneutrino sensitivity study: the rates are intended as number of events per day after all cuts (refer to Section~\ref{subsec:mh:sigbkg:bkgest}). The options  of acrylic vessel and balloon have been compared in the ($\alpha$, n) background evaluation.\label{tab:geo:bck}}
\begin{tabular}{l l l l l} \hline \hline
Background type           & Rate after IBD+  & Uncertainty & Uncertainty \\
                                          &  muon cuts  & ~in Rate & ~in Shape    \\
                                          & [events/day]   & [\%] & [\%] \\
\hline
$^{9}$Li - $^{8}$He  & 1.6 &20 & 10\\
%\hline
Fast neutrons  & 0.1  &100 & 20\\
%\hline
Accidental events & 0.9 &1 & negl.\\
%\hline
$^{13}$C($\alpha$, n)$^{16}$O (acrylic vessel) & 0.05  & 50 &50 \\
$^{13}$C($\alpha$, n)$^{16}$O (ballon)              &0.01 & 50    &50\\
\hline \hline
\end{tabular}
\end{center}
\end{table}

A detailed description of expected background in JUNO is given elsewhere. Here we briefly summarise the most relevant backgrounds for the geoneutrino analysis.
They can be divided into three main categories:
 \begin{itemize}
\item{Cosmic-muons spallation products, namely:}
\begin{itemize}
\item{ $^{9}$Li and $^{8}$He isotopes decaying in ($\beta$ + neutron) branches perfectly mimicking antineutrino interactions;}
\item{fast neutrons able to penetrate through construction materials and giving a delayed coincidence if scattered off by one or many protons before being captured;}
\end{itemize}
\item{Accidental coincidences of non-correlated events.}
\item{Backgrounds induced by radioactive contaminants of scintillator and detector materials. In particular, alpha particles emitted in the $^{238}$U and $^{232}$Th decay chains or by off-equilibrium $^{210}$Po can induce $^{13}$C ($\alpha$, n)$^{16}$O reactions on the scintillator $^{13}$C nuclides.  In these interactions, the prompt signal can be induced by three different processes: $i)$ de-excitation of $^{16}$O nuclides, if produced in an excited state; $ii)$ 4 MeV $\gamma$ ray from the de-excitation of $^{12}$C excited by neutron, and $iii)$ protons scattered off by the thermalizing neutron before its capture.}
\end{itemize}

In the second column of Tab.~\ref{tab:geo:bck} the rates of events surviving the IBD selection cuts and muon cuts are reported.
The muon cut properly vetoes in time and space the detector after each muon crossing the water pool or the central detector and it is effective in reducing the  $^{9}$Li and $^{8}$He backgrounds by a factor 44 (refer to Tab.~\ref{tab:mh:sigbkg}).

A fiducial volume of 18.35 ktons corresponding to a radial cut of $R$ = 17.2\,m was chosen since it is particularly effective in reducing the accidental background mostly arising from $^{238}$U/$^{232}$Th/$^{40}$K contamination of the acrylic vessel, PMT's glass, steel supports, and copper fasteners.

In the ($\alpha$, n) background evaluation, the options of acrylic vessel ($^{238}$U: 10\,ppt, $^{232}$Th: 10\,ppt) and ballon ($^{238}$U: 2\,ppt, $^{232}$Th: 4\,ppt) have been compared: after the fiducial volume cut,  the corresponding ($\alpha$, n) event rates  of 0.05 counts/day and negligible were found, respectively.
A 10$^{-15}$ g/g $^{238}$U/$^{232}$Th contamination of the liquid scintillator would be responsible for less than 0.01 ($\alpha$, n) events/day.
Considering the $^{210}$Pb contamination of the scintillator of the order of $1.4 \cdot 10^{-22}$ g/g, the $\alpha$-decay rate of $^{210}$Po  in equilibrium with $^{210}$Pb  would give a negligible contribution to ($\alpha$, n) background.

\begin{figure} %[htb]
\begin{center}
\includegraphics[width= 0.49\textwidth]{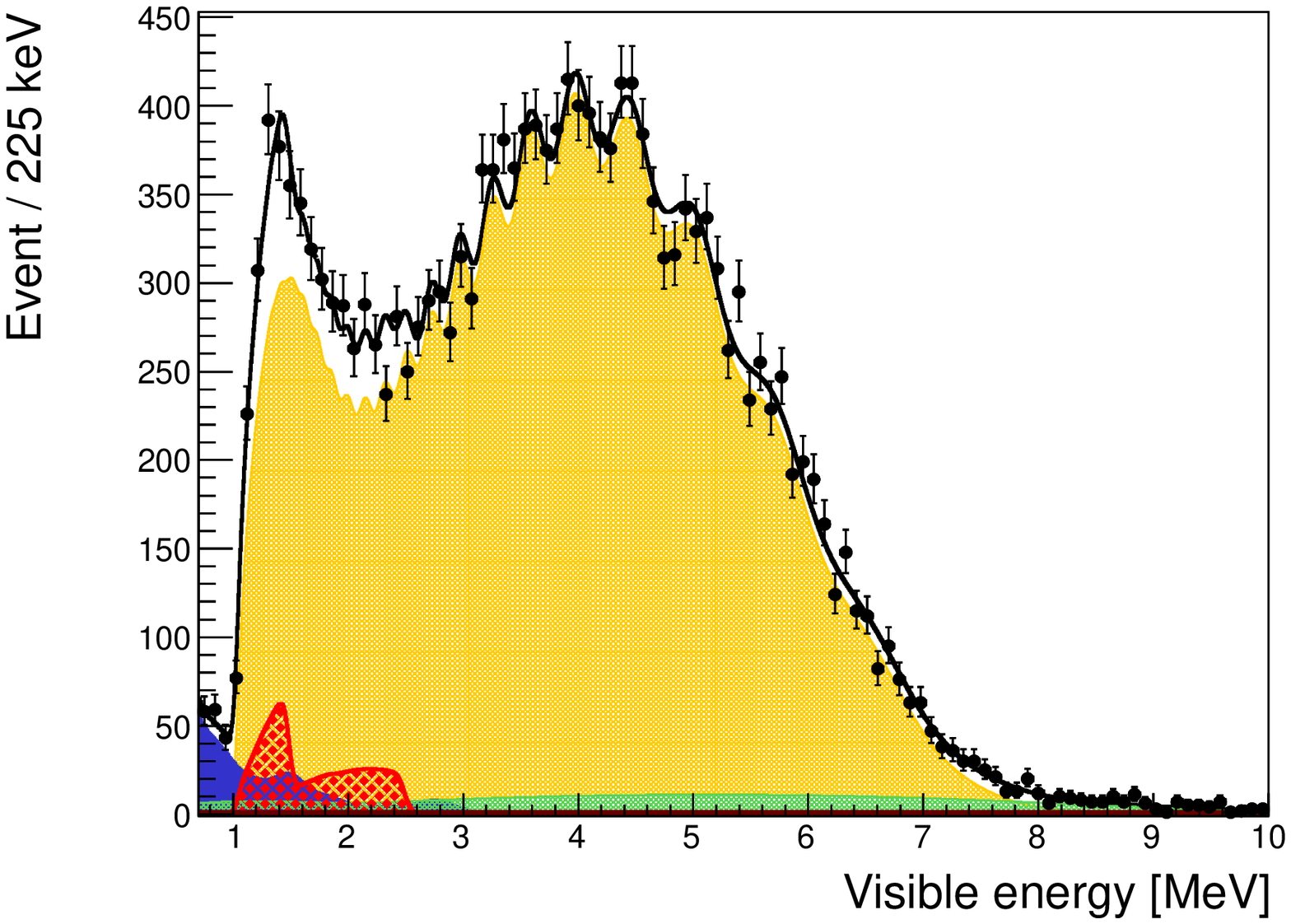}
\includegraphics[width= 0.49\textwidth]{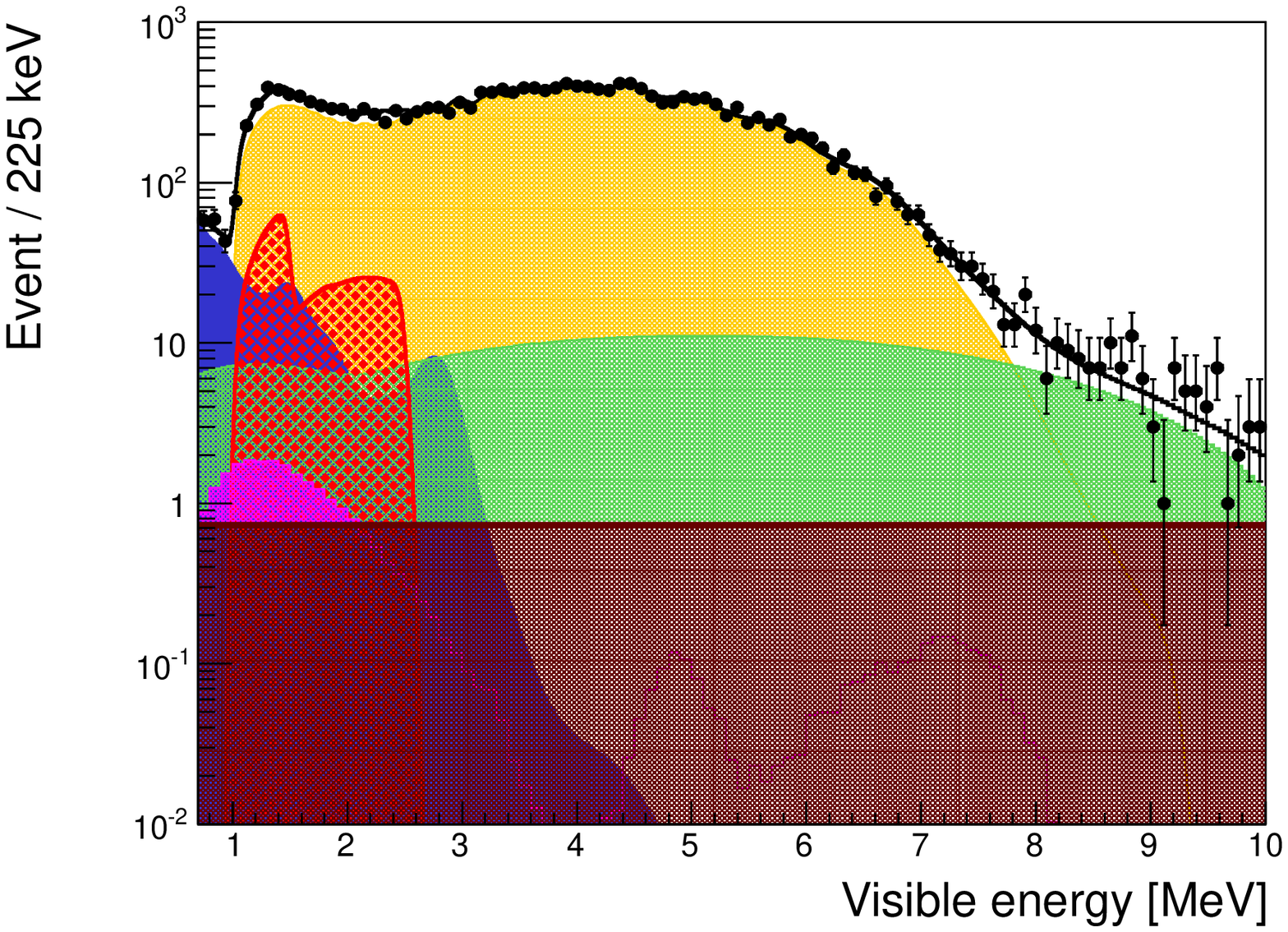}
\caption{Result of a single toy Monte Carlo for 1-year measurement with fixed chondritic Th/U mass ratio; the bottom plot is in logarithmic scale to show background shapes. The data points show the energy spectrum of prompt candidates of events passing IBD selection cuts. The different spectral components are shown as they result from the fit; black line shows the total sum for the best fit. The geoneutrino signal with Th/U fixed to chondritic ratio is shown in red. The following colour code applies to the backgrounds: orange (reactor antineutrinos), green ($^{9}$Li - $^{8}$He), blue (accidental), small magenta ($\alpha$, n). The flat contribution visible in the lower plot is due to fast neutron background.
\label{fig:geo:FitUThfixed1year}}
\end{center}
\end{figure}

%********************************************************************************************
\subsection{JUNO potential in measuring geoneutrinos}
\label{subsec:geo:potential}

JUNO will be able to detect geoneutrinos in spite of the noticeable background coming from various sources - and mainly from nearby nuclear power plants.  In order to assess the detectability and the measurement accuracy, the predicted total antineutrino spectrum (signal and backgrounds) has been studied under different possible conditions. The expected geoneutrino signal, as well as reactor antineutrino and non-antineutrino backgrounds have been described above in this section. For convenience, Tab.~\ref{tab:geo:Nev} summarizes the number of expected events for all components contributing to the IBD spectrum in the 0.7 - 12 MeV energy region of the prompt signal. We have assumed 80\% antineutrino detection efficiency and 17.2\,m radial cut (18.35\,kton of liquid scintillator). In the simulations described below, we have considered Gaussian distributions for all spectral components according to this table. For simplicity, this is valid also for the geoneutrino signal, taking the mean of the left and right variances described above as the variance of symmetric Gaussian.

\begin{table}[ht]
\begin{center}
\caption{Signal and backgrounds considered in the geoneutrino sensitivity study:  the number of expected events for all components contributing to the IBD spectrum in the 0.7 - 12 MeV energy region of the prompt signal. We have assumed 80\% antineutrino detection efficiency and 17.2\,m radial cut (18.35\,kton of liquid scintillator, $12.85 \times 10^{32}$ target protons).
\label{tab:geo:Nev}}
\vspace{0.4cm}
\begin{tabular}{ll}
\hline \hline
Source &
Events/year
\\ \hline
Geoneutrinos 	& $408\pm 60$   	\\
U chain      	& $311\pm 55$    	\\
Th chain 	& $92\pm 37$   		\\
Reactors 	& $16100\pm 900$ 	\\
Fast neutrons	& $36.5\pm 36.5$ 	\\
$^{9}$Li - $^{8}$He 		& $657\pm 130$ 		\\
$^{13}$C$(\alpha,n)^{16}$O & $18.2\pm 9.1$  \\
Accidental coincidences & $401\pm 4$ 	\\
\hline \hline
\end{tabular}
\end{center}
\end{table}

The main assumptions concerning the signal searched for (U and Th chains in the Earth, assumed to be in secular equilibrium) are the ones about their relative value (the Th/U ratio). One possibility, described in Sec.~\ref{subsubsec:geo:UThfixed}, is to consider in the analysis the fixed chondritic mass Th/U ratio of 3.9. Because the cross-section of the IBD detection interaction increases with energy, the ratio of the signals expected in the detector
is Th/U = 0.27. The other obvious choice in the analysis, described in Sec.~\ref{subsubsec:geo:UThfree}, is to leave both Th and U contributions free and independent (keeping both chains in secular equilibrium) and to study if the relative Th/U signal ratio is compatible with the expected chondritic mass ratio.

\subsubsection{Geoneutrino signal assuming chondritic mass Th/U ratio}
\label{subsubsec:geo:UThfixed}

We have simulated several thousand possible JUNO geoneutrino measurements using a toy Monte Carlo. In each simulation, we have attributed to each spectral component a rate randomly extracted from the Gaussian distributions according to Tab.~\ref{tab:geo:Nev}. We have used theoretical spectral shapes for geoneutrinos; Th and U components were summed in a proportion according to chondritic mass Th/U ratio of 3.9. The method of calculation of reactor antineutrino spectrum was described in Sec.~\ref{subsec:geo:reactor}. For non-antineutrino backgrounds (accidental coincidences, ($\alpha$, n), $^{9}$Li - $^{8}$He) we have used spectral shapes as they have been measured in Daya Bay. For fast neutrons, we have used a simple flat spectral shape.

In the simulations we have not included the shape uncertainties of the reactor neutrino spectrum. The shape uncertainty for the Huber-Muller model is about 2\% in average. Recently the reactor neutrino experiments, Daya Bay, Double Chooz, and RENO, have found discrepancies between the observed shape and the models, especially in the 4-6 MeV region. With these high precision measurements, it is possible to constrain the shape of the reactor neutrino spectrum to 1\%. Comparing to the statistic uncertainty of the reactor spectrum (several percent for each bin), the shape uncertainty is not critical for this sensitivity study.

The possible precision of geoneutrino measurement under these conditions was evaluated for 1, 3, 5, and 10 years of full live time after cuts, including muon-veto cuts that require 17\% loss of real DAQ time.

\begin{figure} %[htb]
\begin{center}
\includegraphics[width=0.45\textwidth]{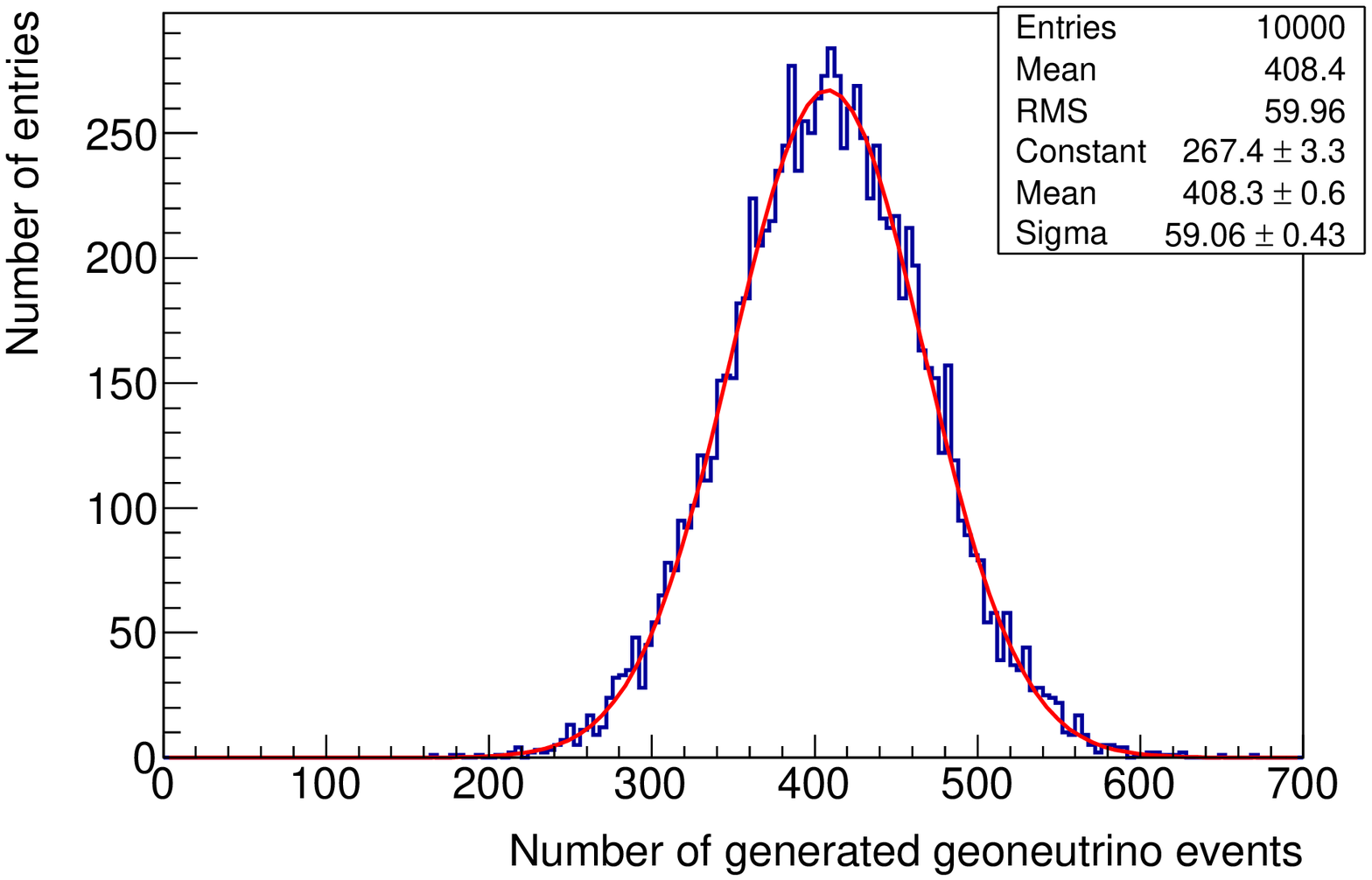}
\includegraphics[width=0.45\textwidth]{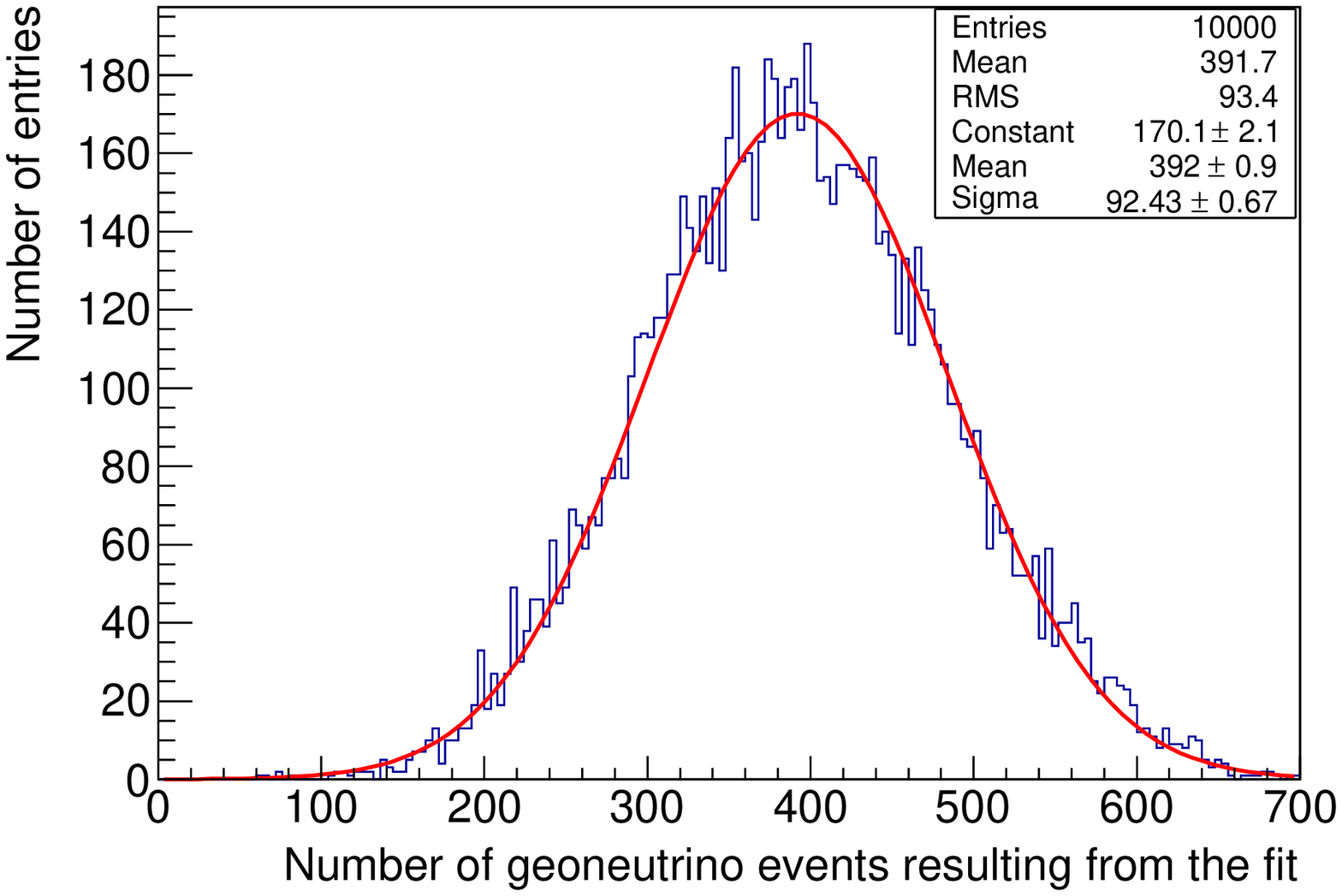}
\includegraphics[width=0.45\textwidth]{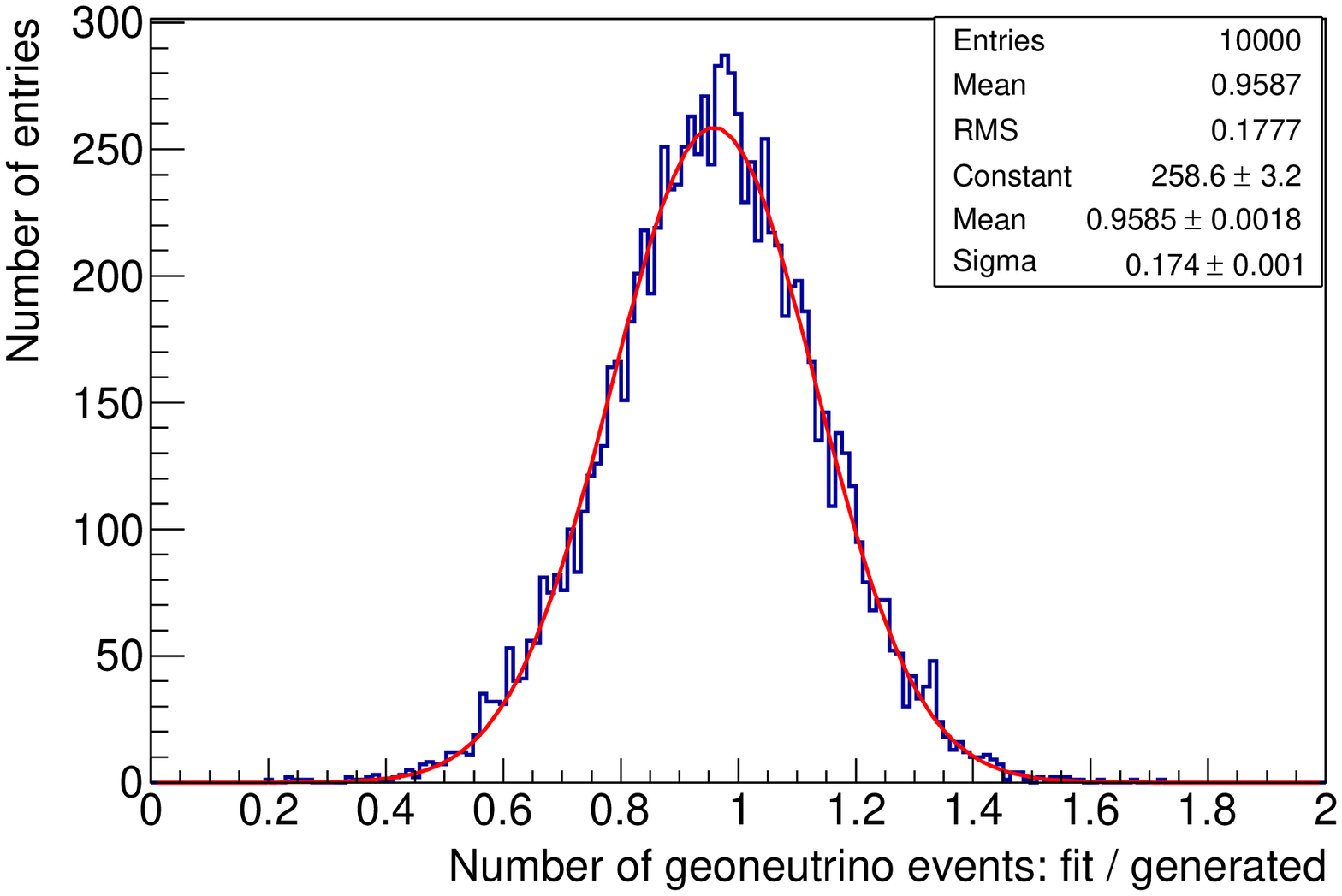}
\caption{The three plots demonstrate the procedure of the sensitivity study to measure geoneutrinos in JUNO. In particular, the capability of the fit to reproduce the correct number of geoneutrino events (fixed ratio U/Th ratio) after 1 year live time after all cuts. Ten thousand simulations and fits have been performed: the upper left plot shows the distribution of the number of generated geoneutrino events, while the upper right plot the distribution of fit results. Finally, the lower plot  gives the distribution of  the ratio between reconstructed and generated number events.
\label{fig:geo:Geo}}
\end{center}
\end{figure}

Figure~\ref{fig:geo:FitUThfixed1year} shows one of the possible total spectra (geoneutrinos and backgrounds) as simulated assuming a data taking period of one year (full livetime after cuts), with Th/U fixed at the chondritic value. After the simulation, the data are fit with the same spectral components used in the data generation. The amplitudes of geoneutrino and reactor antineutrino components are left as free fit parameters, while the non-antineutrino background components are constrained to their expectation value within $\pm 1 \sigma$ range.

This procedure was then repeated 10,000 times for 1, 3, 5, and 10 years of lifetime after cuts, as it is demonstrated in Fig.~\ref{fig:geo:Geo} for a specific case of 1 year. The upper left plot shows the number of geoneutrino events randomly generated in each simulation; the Gaussian fit is compatible with the values from the first input of Tab.~\ref{tab:geo:Nev}. The upper right plot is the distribution of the absolute number of reconstructed geoneutrino events as resulting from the fit. The goodness of the procedure can be appreciated by the third plot, showing the distribution of ratios, calculated for each simulation, between the number of geoneutrinos reconstructed by the fit and the true number of generated events. This distribution is centered at about 0.96, thus systematically shifted to lower values by about 4\% with respect to an ideal case centered at 1. Considering the width of this distribution, an error of about 17\% is expected for the JUNO geoneutrino measurement, assuming a fixed U/Th ratio and one year statistics (after cuts).
The correlation between geoneutrinos and reactor antineutrinos is demonstrated in Fig.~\ref{fig:geo:GeoVsRea} showing the distribution of the ratios of the reconstructed/generated number of events for geoneutrinos versus reactor antineutrinos for 1 year lifetime after cuts.

\begin{figure} %[htb]
\begin{center}
\includegraphics[width=0.7\textwidth]{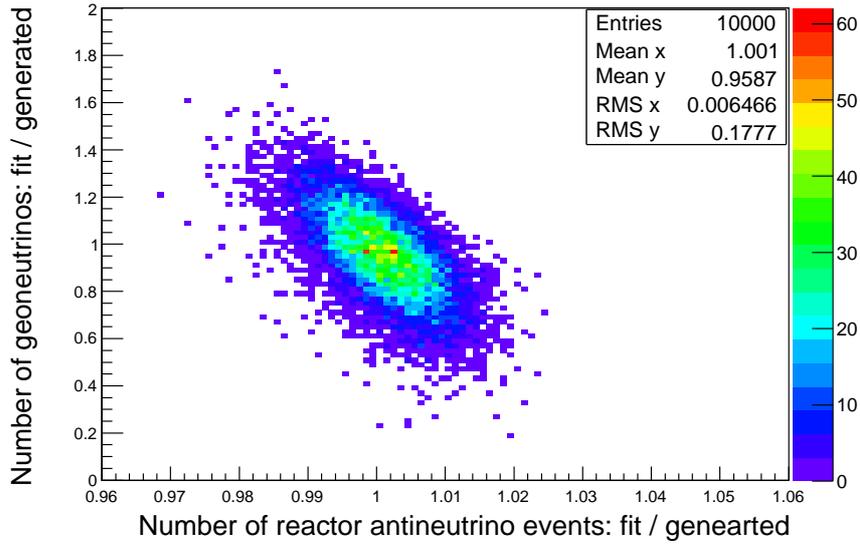}
\caption{Distribution of the ratios of the reconstructed/generated number of events for geoneutrinos versus reactor antineutrinos.
\label{fig:geo:GeoVsRea}}
\end{center}
\end{figure}

\begin{figure} %[htb]
\begin{center}
\includegraphics[width=0.49\textwidth]{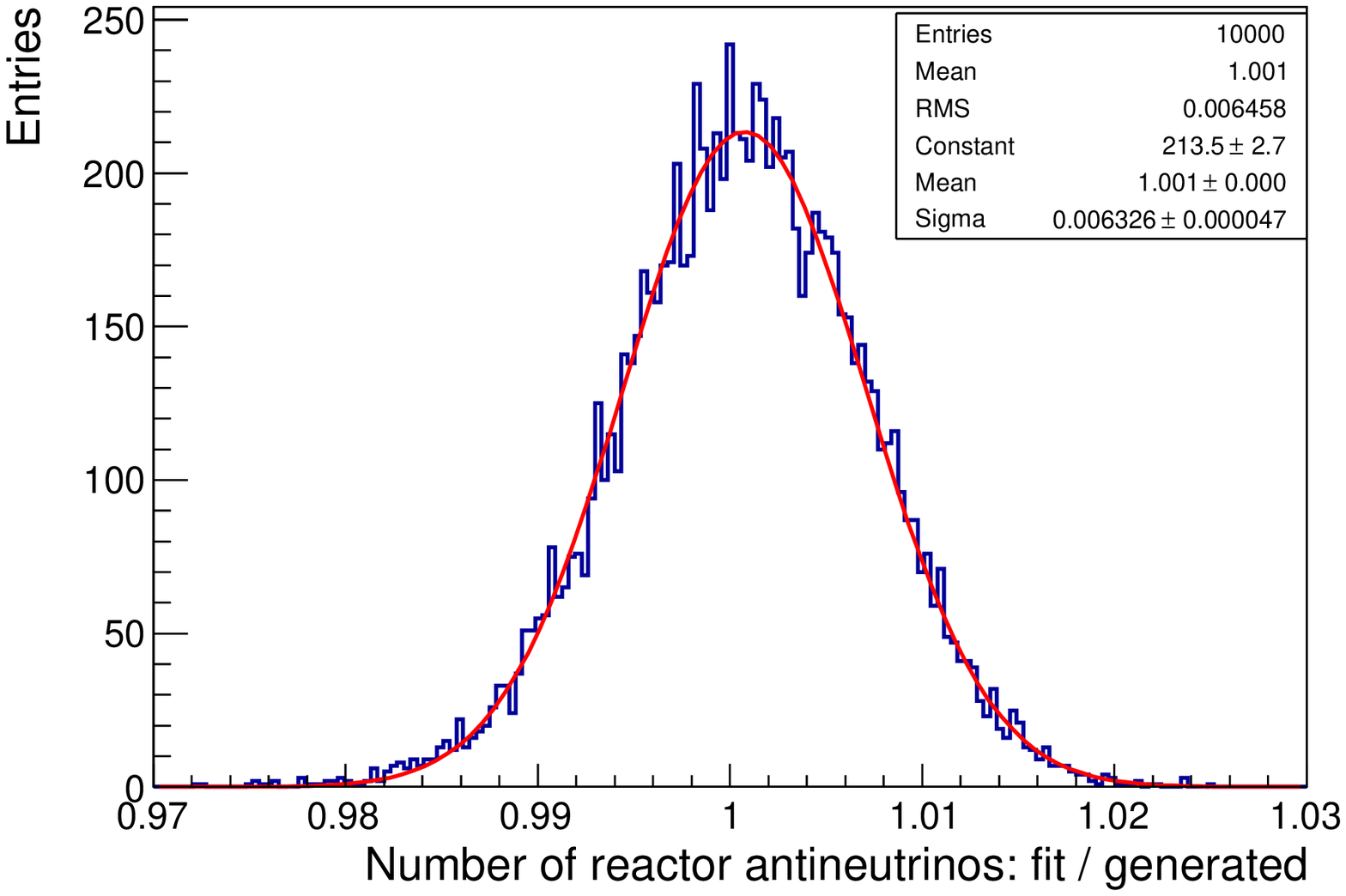}
\includegraphics[width=0.49\textwidth]{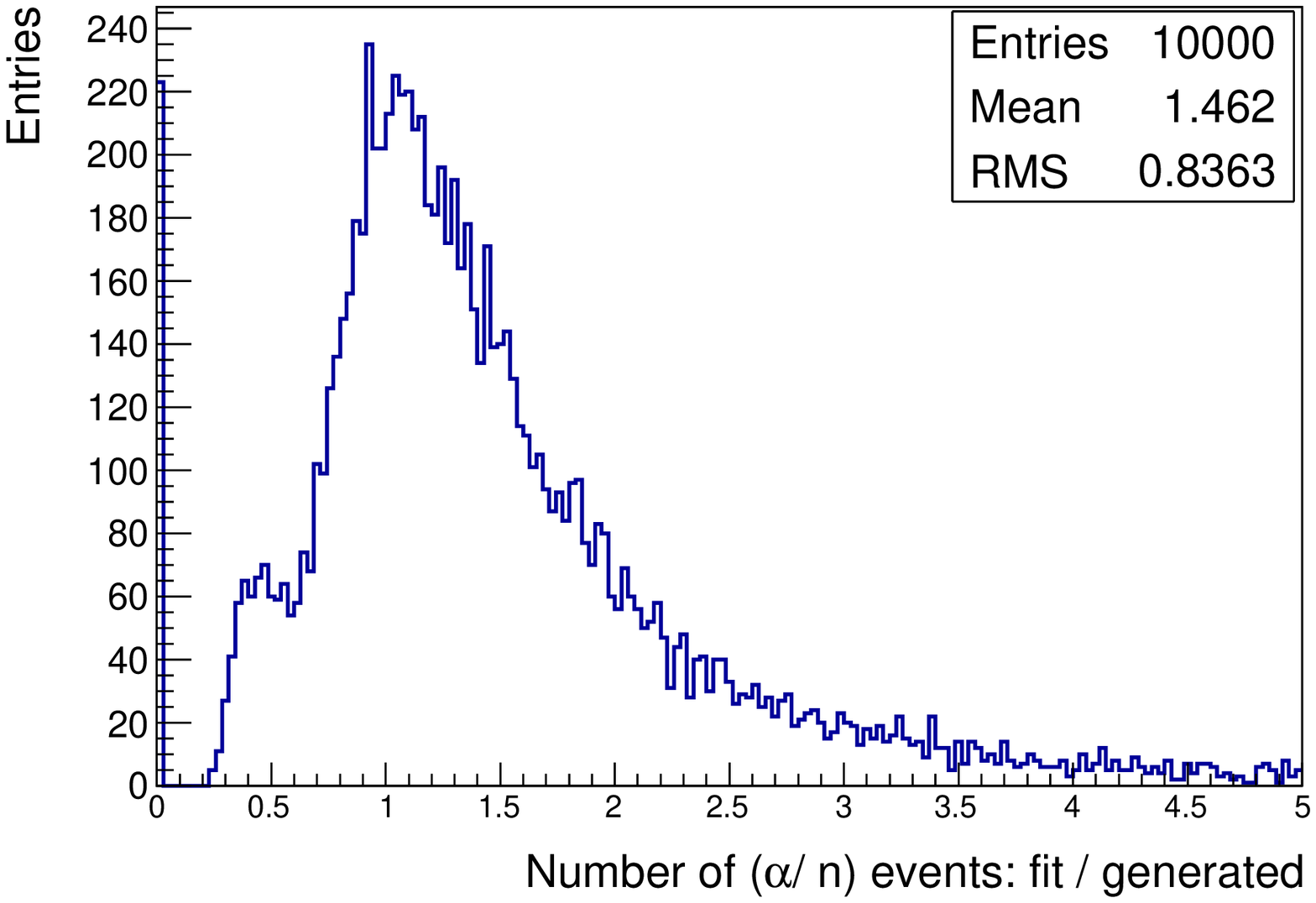}
\includegraphics[width=0.49\textwidth]{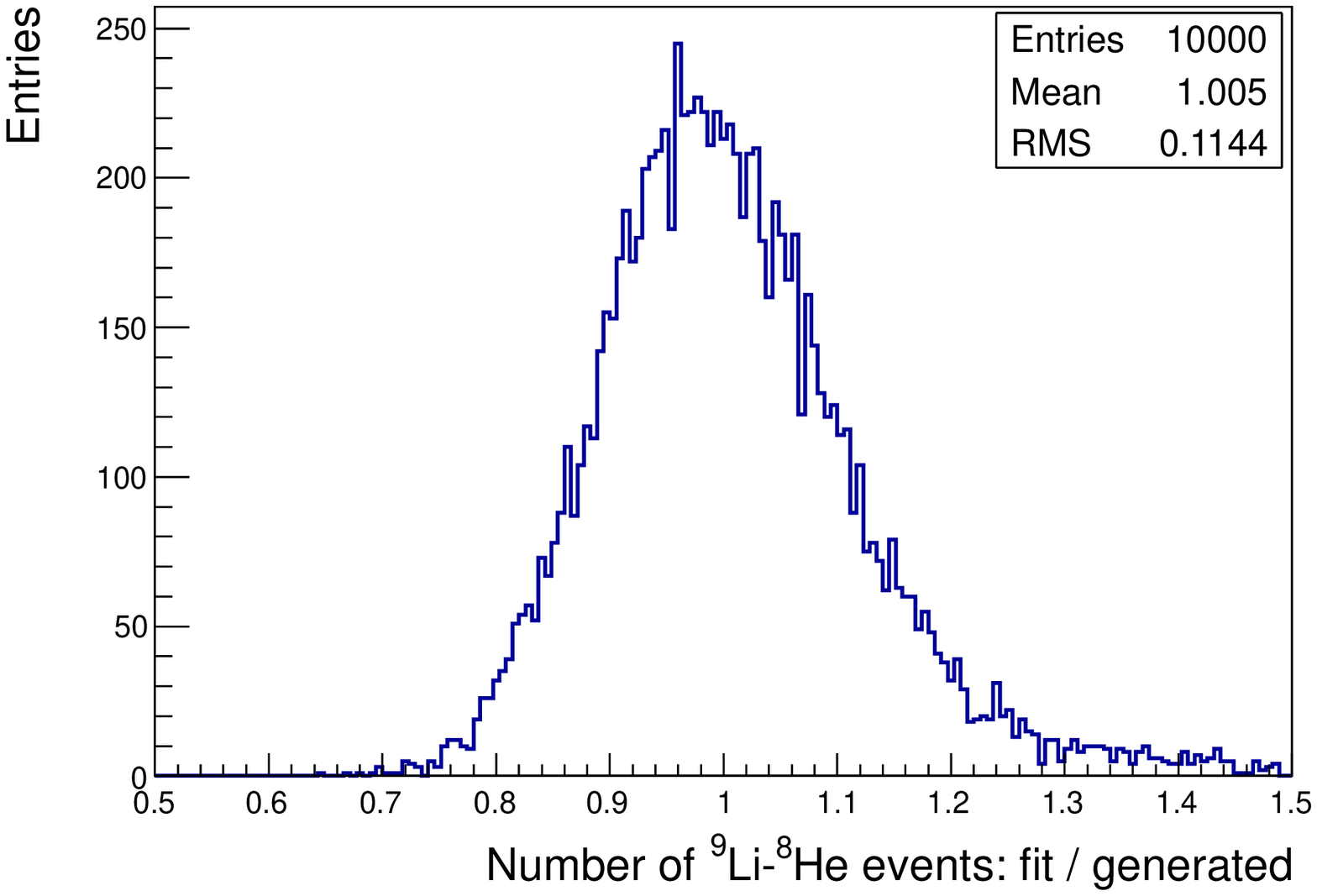}
\includegraphics[width=0.49\textwidth]{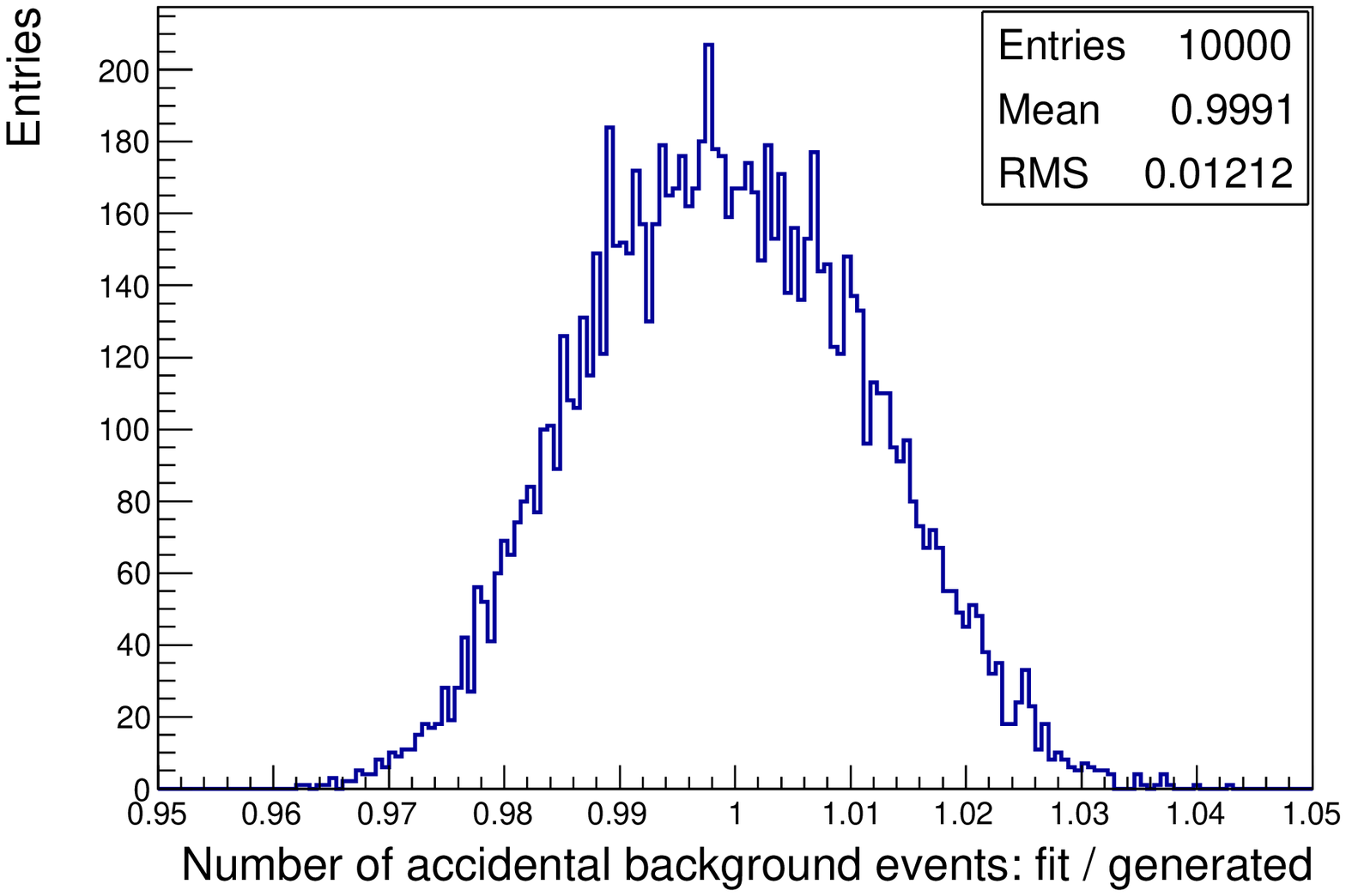}
\caption{Distributions of the ratios between reconstructed and generated number events for 1~year lifetime simulations, considering fixed Th/U ratio. Top left: reactor antineutrinos. Top right: ($\alpha$, n) background. Bottom left: $^9$Li - $^8$He events, Bottom right: accidental coincidences.
\label{fig:geo:Bgr}}
\end{center}
\end{figure}

The distributions of ratios between reconstructed and generated number events for reactor antineutrinos and for ($\alpha$, n), $^9$Li - $^8$He, and accidental backgrounds are shown in Fig.~\ref{fig:geo:Bgr}. The -4\% systematic shift in the reconstruction of geoneutrino signal (see lower part of Fig.~\ref{fig:geo:Geo}) is mostly due to the correlation with reactor antineutrino background: 0.1\% overestimate of the reactor background corresponds to 4\% decrease in the reconstructed geoneutrino signal.  The ($\alpha$, n) background tends to be overestimated and is correlated with the $^9$Li - $^8$He background. Background due to the accidental coincidences is well reconstructed.
\begin{table}[h]
\begin{center}
\caption{Precision of the reconstruction of geoneutrino signal, as it can be obtained in 1, 3, 5, and 10 years of lifetime, after cuts. Different columns refer to the measurement  of geoneutrino signal with fixed Th/U ratio, and U and Th signals fit as free and independent components. The given numbers are the position and RMS of the Gaussian fit to the distribution of the ratios between the number of reconstructed and generated events. It can be seen that while the RMS is decreasing with longer data acquisition time, there are some systematic effects which  do not depend on the acquired statistics and are described in text.\label{tab:geo:Fit}}
\vspace{0.4cm}
\begin{tabular}{cccc}
\hline \hline
Number of years & U + Th (fixed chondritic Th/U ratio) & U (free) & Th (free) \\
\hline
1	& $0.96\pm 0.17$   & $1.02\pm 0.32$	& $0.83\pm 0.60$	\\
3       & $0.96\pm 0.10$   & $1.03\pm 0.20$       & $0.80\pm 0.38$	\\
5 	 & $0.96\pm 0.08$   & $1.03\pm 0.16$       & $0.80\pm 0.28$  \\
10 	& $0.96\pm 0.06$   & $1.03\pm 0.11$	& $0.80\pm 0.19$	\\
\hline \hline
\end{tabular}
\end{center}
\end{table}

This analysis has been repeated also for 3, 5, and 10 years of lifetime (after cuts). The precision of the geoneutrino measurement with fixed Th/U ratio is summarised in the 2$^{\rm nd}$ column of Tab~\ref{tab:geo:Fit}. As it can be seen, the -4\% systematic shift in the reconstruction of the geoneutrino signal remains also for long data taking periods. The width of the distributions of the reconstructed/generated number of geoneutrino events decreases, and thus the statistical error on the measurement decreases with higher statistics, as expected. With 1, 3, 5, and 10 years of data, this error amounts to 17, 10, 8, and 6\%, respectively.

\subsubsection{Potential to measure Th/U ratio}
\label{subsubsec:geo:UThfree}

The large size of the JUNO detector and the significant number of geoneutrino events recorded each year offers the potential to measure individually the U and Th contributions.  The same study as described in Sec.~\ref{subsubsec:geo:UThfixed} has been repeated, but this time the constraint on the Th/U chondritic ratio has been removed and we allowed independent contributions from the two main natural radioactive chains. The Th and U signal has been generated from Gaussian distributions according to Tab.~\ref{tab:geo:Nev}.
As an example, Fig.~\ref{fig:geo:FitUThfree} shows the spectrum that could be obtained in 10 years of the JUNO data (after cuts).

The precision of the free U and Th signal measurements for 1, 3, 5, and 10 years of lifetime are summarised in the 3$^{\rm rd}$ and 4$^{\rm th}$ columns of Tab~\ref{tab:geo:Fit}, respectively.  One year of data collection would result in a a significant statistical error on the estimates of both U and Th, while the situation improves greatly with time. On the other hand, a systematic bias, at the level of 3\% overestimate of the U signal and 20\% underestimate of the Th signal is not eliminated with increased statistics. This is due to the correlations among different spectral components, as it is demonstrated on Fig.~\ref{fig:geo:ThSys} for 5 years of lifetime after cuts. The correlation between U and reactor antineutrinos is stronger than between Th and reactor antineutrinos (top plots). The correlation between U and Th is quite strong (bottom left). The ($\alpha$, n) background gets overestimated but is not strongly correlated with Th (and neither U), as it is shown on the bottom right plot, but it is correlated with $^9$Li - $^8$He background.

Accordingly, we have studied how well we can reconstruct the U/Th ratio. Figure~\ref{fig:geo:UThratio} shows the asymmetric distribution of the ratio reconstructed-to-generated U/Th ratio for 1 and 10 years of lifetime after cuts, excluding simulations  in which the Th contribution converges to 0. Tab.~\ref{tab:geo:UThratio} parametrizes such distributions for 1, 3, 5, and 10 years of lifetime after cuts and gives position of the peak, left and right RMS. The last column shows the fraction of fits in which the Th component converges to zero. We can see, that at least few years of lifetime are required in order to measure the Th/U ratio.
\begin{figure}%[htb]
\begin{center}
\includegraphics[width=0.8\textwidth]{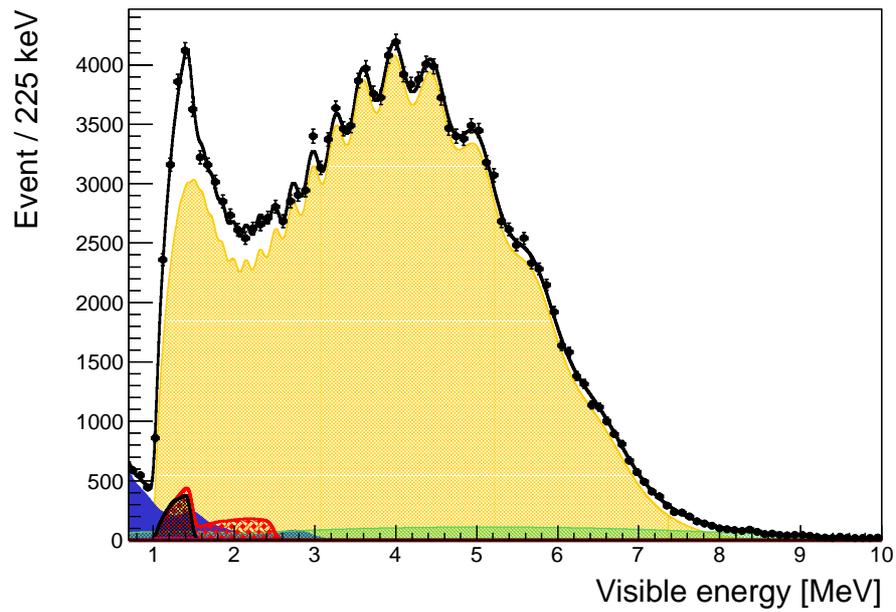}
\caption{Result of a single toy Monte Carlo for 10-year measurement with Th and U components left free and independent. The data points show the energy spectrum of prompt candidates of events passing IBD selection cuts. The different spectral components are shown as they result from the fit; black line shows the total sum for the best fit. The U and Th signal are shown in red and black areas, respectively. The following colour code applies to the backgrounds: orange (reactor antineutrinos), green ($^{9}$Li - $^{8}$He), blue (accidental), small magenta ($\alpha$, n).
\label{fig:geo:FitUThfree}}
\end{center}
\end{figure}
\begin{figure}%[htb]
\begin{center}
\includegraphics[width=0.49\textwidth]{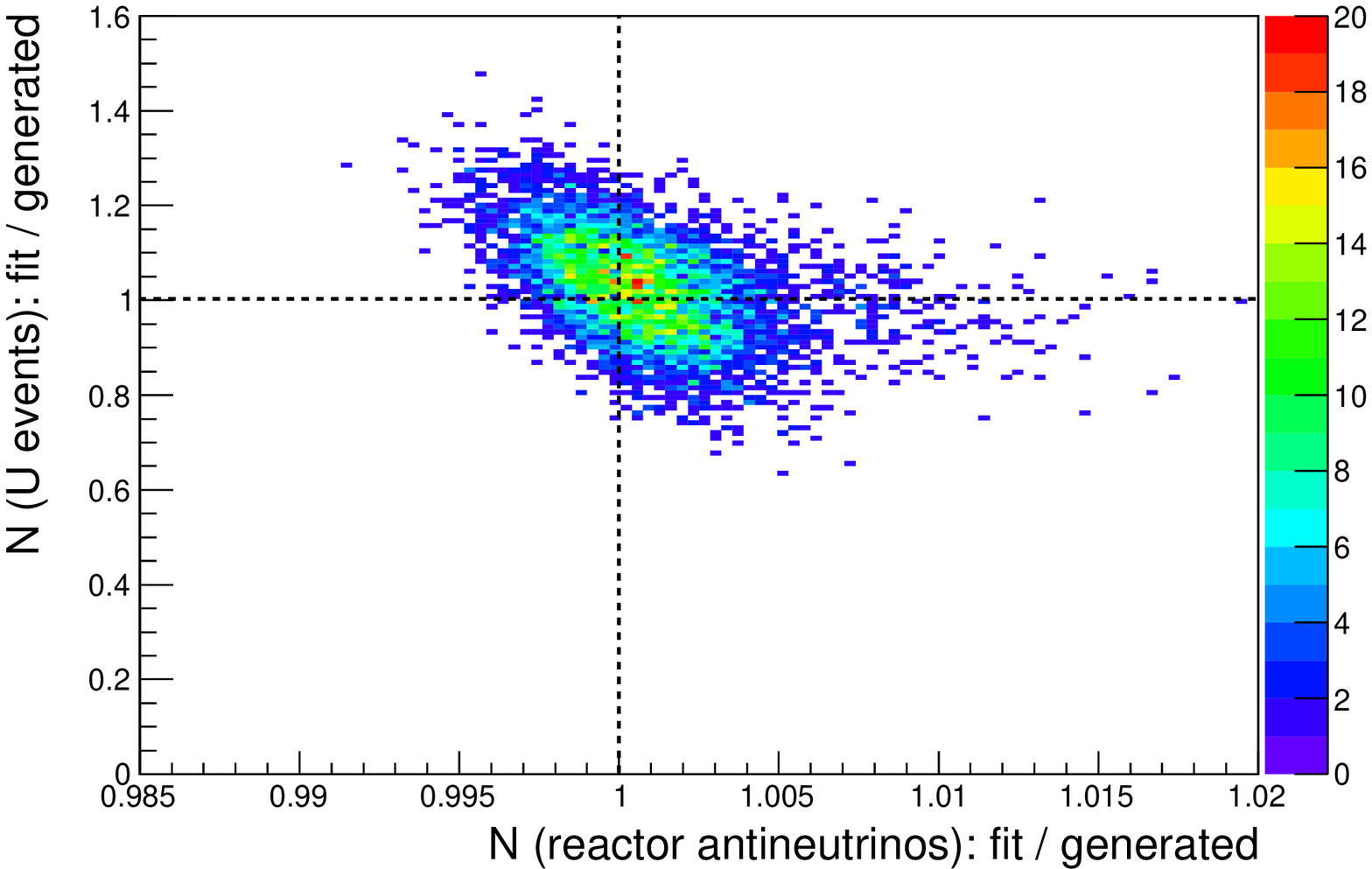}
\includegraphics[width=0.49\textwidth]{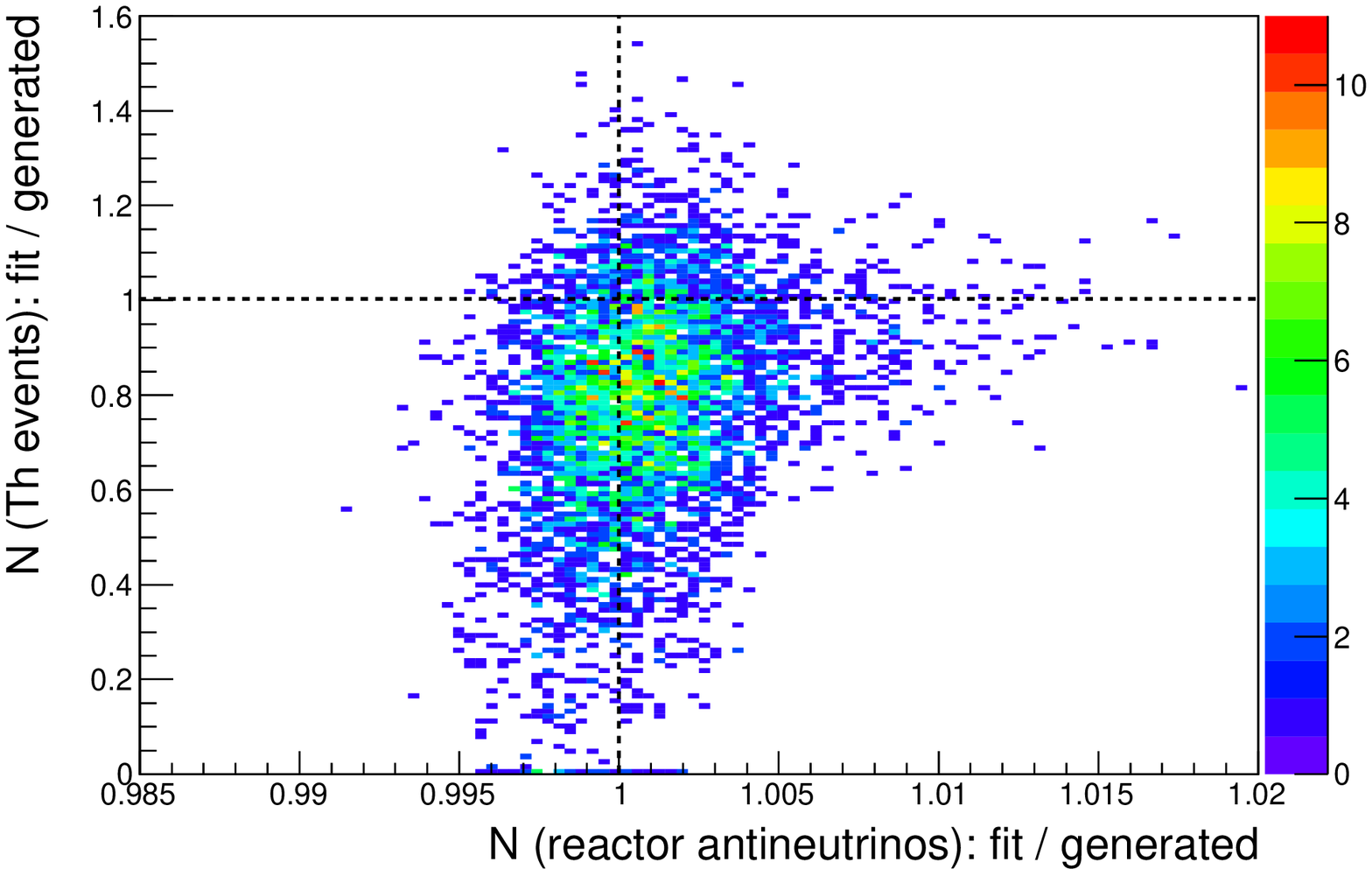}
\includegraphics[width=0.49\textwidth]{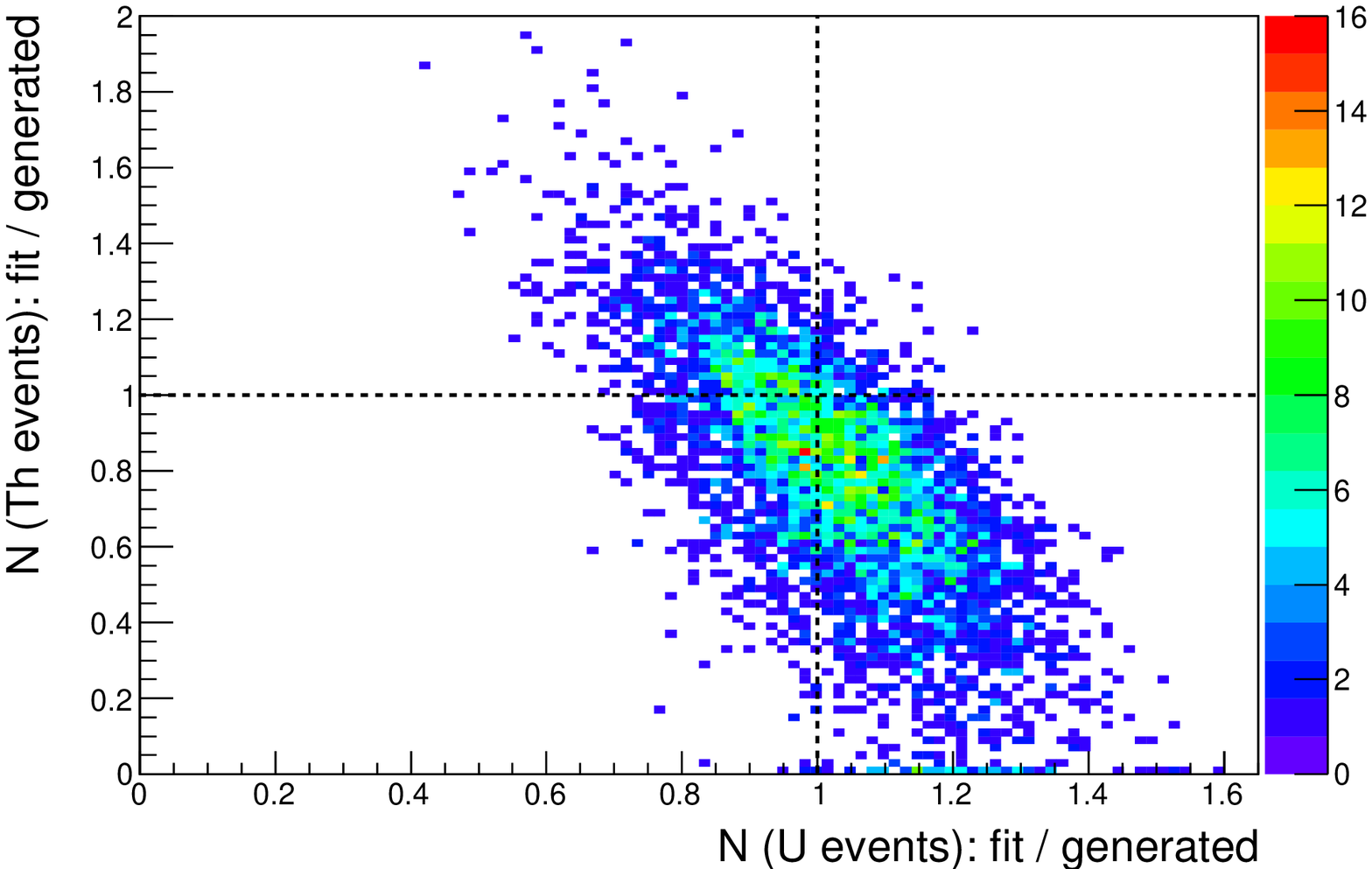}
\includegraphics[width=0.49\textwidth]{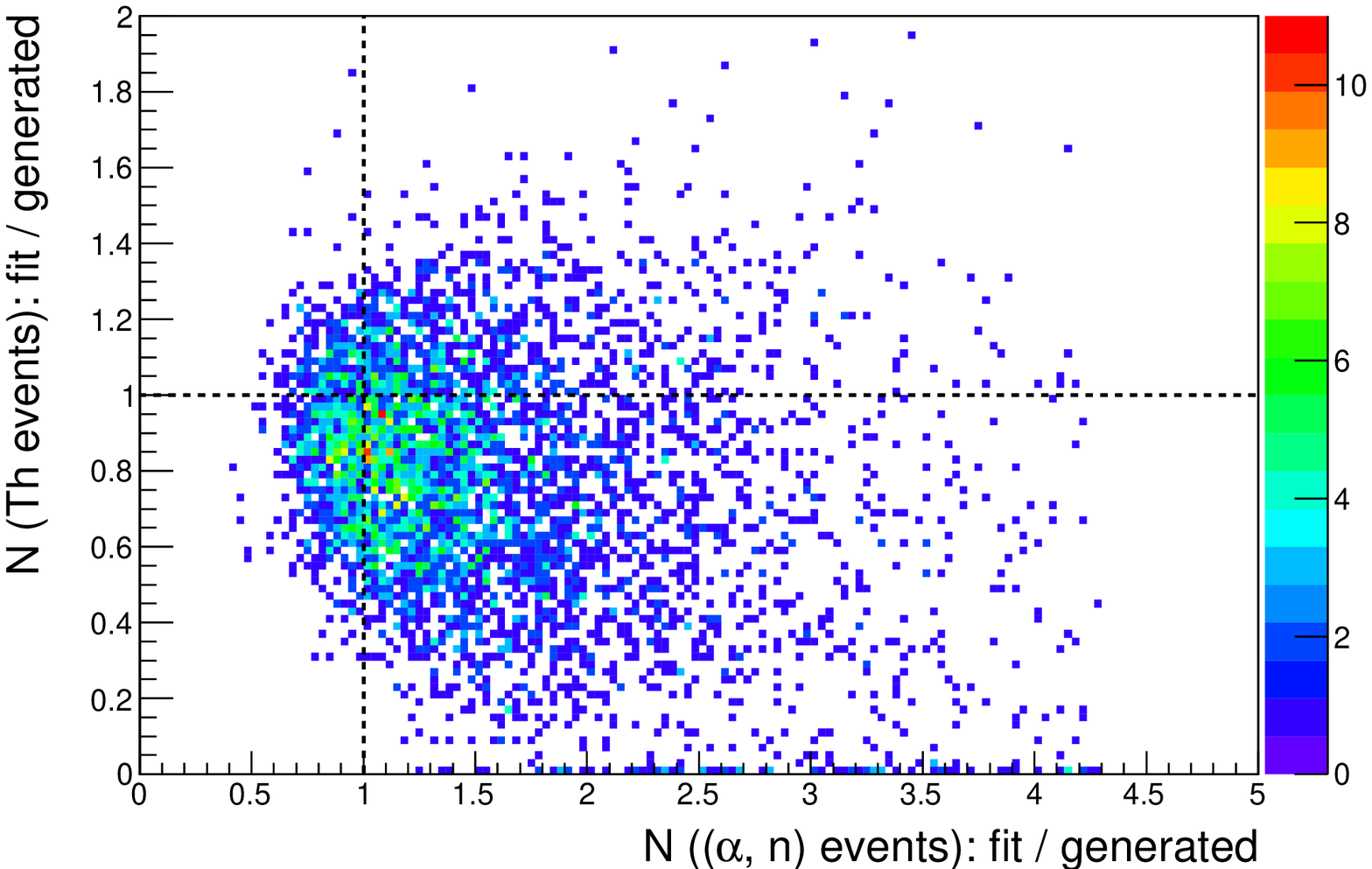}
\caption{Correlations between the ratios of reconstructed-to-generated number of events in 5\,years of lifetime (after cuts): Top left: U versus reactors antineutrinos. Top right: Th vs reactor antineutrinos. Bottom left: Th versus U. Bottom right: Th versus ($\alpha$, n) background.
\label{fig:geo:ThSys}}
\end{center}
\end{figure}
\begin{table}%[ht]
\begin{center}
\caption{Parametrization of asymmetric distributions of the reconstructed-to-generated U/Th ratio for 1, 3, 5, and 10 years of lifetime after cuts (see examples for 1 and 10 years on Fig.~\ref{fig:geo:UThratio}). Different columns give the position of the peak,  RMS, and the fraction of fits in which Th component converges to zero. \label{tab:geo:UThratio}}
\vspace{0.4cm}
\begin{tabular}{cccc}
\hline \hline
Number of years & Peak position  & RMS & Fits with Th(fit) = 0 \\
               & &  & [\%] \\
\hline
1	&  0.7   &  2.43	     & 8.2 \\
3	&  0.9   & 2.23        & 3.6 \\
5	&  1.0   & 2.22        & 2.7 \\
10 	&  1.1   &  1.84       & 1.8 	\\
\hline \hline
\end{tabular}
\end{center}
\end{table}
\begin{figure}%[htb]
\begin{center}
\includegraphics[width=0.7\textwidth]{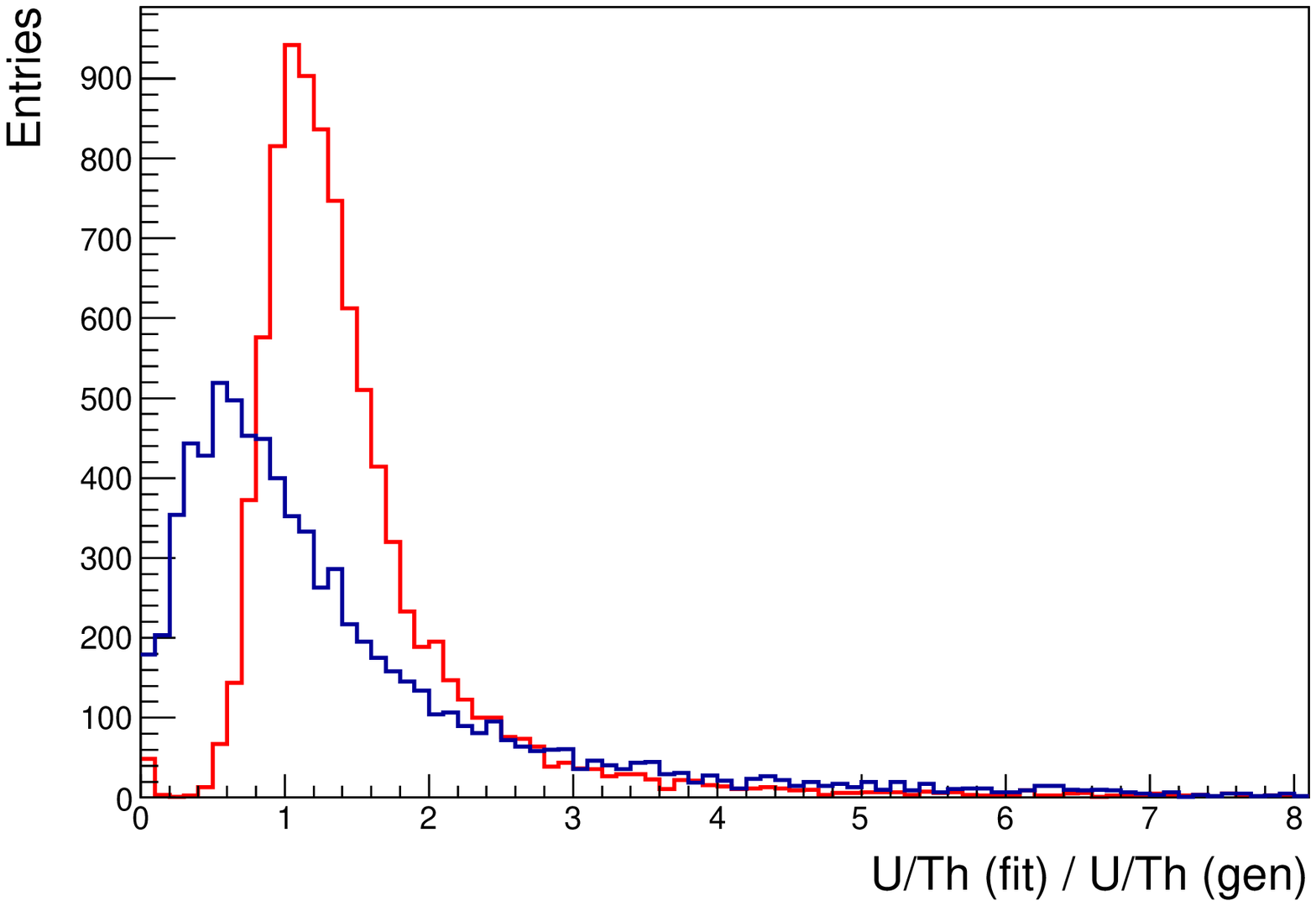}
\caption{Distribution of the ratio reconstructed-to-generated U/Th ratio for 1 (blue line) and 10 (red line) years of lifetime after cuts.  The simulations resulting in zero Th contribution are not plot here.
\label{fig:geo:UThratio}}
\end{center}
\end{figure}

\subsection{Directionality measurement}
\label{subsec:geo:directionality}

The average forward shift of neutrons in the direction of incoming antineutrinos have been observed by reactor experiments (i.e. by CHOOZ collaboration~\cite{Apollonio:1999jg}).  More recent theoretical considerations with respect to the geoneutrino detection can be found in~\cite{Batygov}. The basic idea is to search for the statistical displacement of the capture vertex of the neutron with respect to the vertex of the prompt positron event ($\overrightarrow{\Delta R}=\overrightarrow{R_{prompt}}-\overrightarrow{R_{delayed}}$).

The neutron from the inverse beta decay of geoneutrino carries energy up to tens of keV and is emitted in a relatively narrow range (below $\sim\,$55 degrees~\cite{Batygov}) of angles around the incoming antineutrino. Emitted neutron is thermalized and then captured on hydrogen in liquid scintillator. The average forward displacement of the neutron capture vertex is about 1.7 cm, as observed by CHOOZ for reactor neutrinos, while the spread due to neutron drifting is about 10 cm. The gravity of energy deposit in liquid scintillator of the gamma released by the neutron capture will further smear the vertex by about 20 cm, together with the vertex reconstruction resolution, a few to 10 cm for JUNO. The prompt event has relatively smaller vertex smearing, a few cm due to the energy deposit and a few to 10 cm due to vertex reconstruction. Given the small displacement ($\sim$1.7 cm) and the large intrinsic smearing ($\sim$25 cm), the direction of the reconstructed antineutrino is only meaningful statistically and needs large statistics.

The positron is almost isotropic in direction with respect to the incoming antineutrino. It can not provide directional information of the antineutrino directly. Optimistically, if the PMT time response is good enough, JUNO could reconstruct the direction of the positron, which could constrain the reconstruction of the neutron displacement since kinematically they are in the same plane. A simulation shows that the reconstructed direction of a supernova could be improved from 11.2$^\circ$ to 8.6$^\circ$ with 5000 events.

Because the direction to the reactors in JUNO is known, it looks promising exploiting the fit of displacement ($\overrightarrow{\Delta R})$ distribution with predicted separate distributions from geo and reactor neutrinos in conjunction with the spectral fit. An attempt to separate the crust and mantle geoneutrino components could be made.
Both tasks need extensive MC studies.

\subsection{Conclusions}
\label{subsec:geo:concl}

JUNO represents a fantastic opportunity to measure geoneutrinos.
Its unprecedented size and sensitivity allows for the recording of 300 to 500 geoneutrino interactions per year.
In approximately six months JUNO would match the present world sample of recorded geoneutrino interactions, which is less than 150 events.

This contribution investigates the prospects for extracting the geoneutrino signal at JUNO from the considerable background of reactor antineutrinos and non-antineutrino sources. Using a well constrained estimate of the reactor signal and reasonable estimates of the non-antineutrino sources, the conclusion of the presented analysis is that
geoneutrinos are indeed observable at JUNO. This encouraging result motivates continued studies of the potential to perform neutrino geoscience with JUNO.

With this illumination of the potential importance of JUNO to geological sciences it becomes important to explore fully the sensitivity of JUNO to geoneutrino observations. For example, radiogenic heating in the mantle, which is closely related to the mantle geoneutrino signal, is of critical geological significance. Maximizing the precision of the mantle geoneutrino measurement at JUNO requires detailed knowledge of the uranium and thorium content in the crust within several hundreds of kilometers of JUNO. Moreover, the statistical power of the geoneutrino signal at JUNO enables a measurement of the thorium to uranium ratio, which provides valuable insight to the Earth's origin and evolution.

Fully realizing JUNO's potential contributions to neutrino geosciences recommends the following studies: detailed investigations of the sensitivity of JUNO to geoneutrinos, including refining details of the geoneutrino signal and background Monte Carlo; and a geological examination of the local crust.

In summary, this contribution reveals the unprecedented opportunity to explore the origin and thermal evolution of the Earth by recording geoneutrino interactions with JUNO. It motivates continued studies to fully realize the indicated potential. There is an experienced and dedicated community of neutrino geoscientists that is eager to take advantage of this unique opportunity.

\clearpage

%\newcommand{\nua}[1]{\ensuremath{\rlap
%           {\kern-2.5pt\ensuremath
%           {\overset{\scriptscriptstyle(-)}{\phantom{\nu}}}}
%           {\ensuremath{{\nu}_{#1}}}}}
%% this is needed when compile this chapter only.

\section{Sterile Neutrinos}
\label{sec:sterile}

\blfootnote{Editors: Yufeng Li (liyufeng@ihep.ac.cn), Weili Zhong (zhongwl@ihep.ac.cn)}
\blfootnote{Major contributors: Janet Conrad, Chao Li, Jiajie Ling, Gioacchino Ranucci, and Mike Shaevitz}

\subsection{Introduction}
\label{subsec:sterile:introduction}

Sterile neutrinos are hypothesized gauge singlets of the Standard Model. They do not participate in standard weak
interactions but couple to the active neutrinos through non-zero mixing between active and sterile flavors. Heavy sterile
neutrinos featuring masses near the Grand Unified Theory (GUT) scale can explain the smallness of the three active neutrino masses via
the traditional type-I seesaw mechanism~\cite{Minkowski:1977sc,Yanagida:1979as,GellMann:1980vs,Glashow:1980unknown1,Mohapatra:1979ia,Schechter:1980gr},
and play a pivotal role in the leptogenesis explanation of
the cosmological matter-antimatter asymmetry ~\cite{Fukugita:1986hr,Davidson:2008bu}. On the other hand, light sterile neutrinos are also a hot topic in
particle physics and cosmology. A sterile neutrino at the
keV mass scale is a promising candidate for Warm Dark Matter~\cite{Kusenko:2009up}. Moreover,
sterile neutrinos at the eV or sub-eV scale~\cite{Abazajian:2013oma,Gariazzo:2015rra} are
well-motivated by the short-baseline neutrino oscillation anomalies, which include the LSND~\cite{Aguilar:2001ty} and
MiniBooNE~\cite{AguilarArevalo:2010wv} $\bar\nu_{\mu}\to\bar\nu_{e}$ event excess,
the Reactor Antineutrino Anomaly~\cite{Mention:2011rk,Huber:2011wv} and the Gallium anomaly~\cite{Giunti:2010zu}.
Finally, the standard LMA-MSW solution~\cite{Wolfenstein:1977ue, Mikheev:1986gs} of solar neutrino oscillations predicts an energy-dependent variation (upturn) in the mid-energy regime of the solar neutrino spectrum that is not observed in current solar neutrino data~\cite{Aharmim:2011vm,Renshaw:2014awa}.
A hypothetical super-light sterile neutrino at the $\Delta m^2$ scale of ${\cal O}(10^{-5})$ eV$^2$ has been proposed to explain the suppression of the upturn in the 5-MeV region \cite{deHolanda:2003tx, deHolanda:2010am}.

In the following, we investigate JUNO's potential to search for sterile neutrino oscillation with $\Delta m^2$ values on the scale of eV$^2$ and ${\cal O}(10^{-5})$ eV$^2$. For the former case, we consider both a radioactive (anti-)neutrino source inside or near the JUNO detector and a cyclotron-produced
neutrino beam to directly test the short baseline anomalies.
For the latter one, an additional $\Delta m^2$ near the solar mass-squared difference could be discovered in JUNO by studying reactor antineutrino oscillations as the experiment is very sensitive to both the solar and atmospheric mass-squared differences.
We will restrict ourselves to the simplest $(3+1)$ scheme, which includes three active and
one sterile neutrino mass eigenstate. Depending on the neutrino mass hierarchy, there are four different scenarios of
neutrino mass spectra, which are shown in Fig.~\ref{fig:sterile:mh_sterile}.
\begin{figure}[htb]
\begin{center}
\begin{tabular}{cc}
\includegraphics*[width=0.45\textwidth]{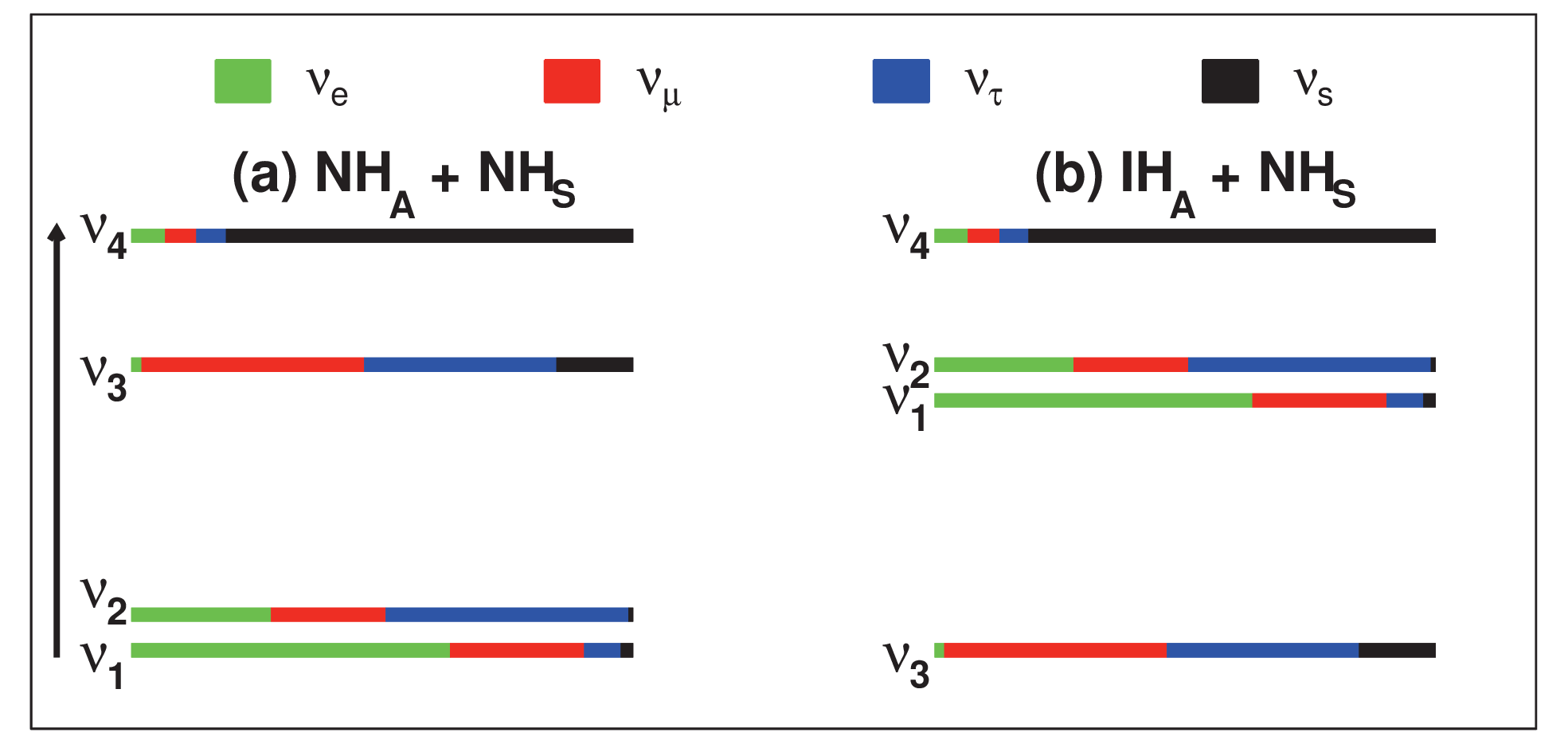}%pdf}
\includegraphics*[width=0.45\textwidth]{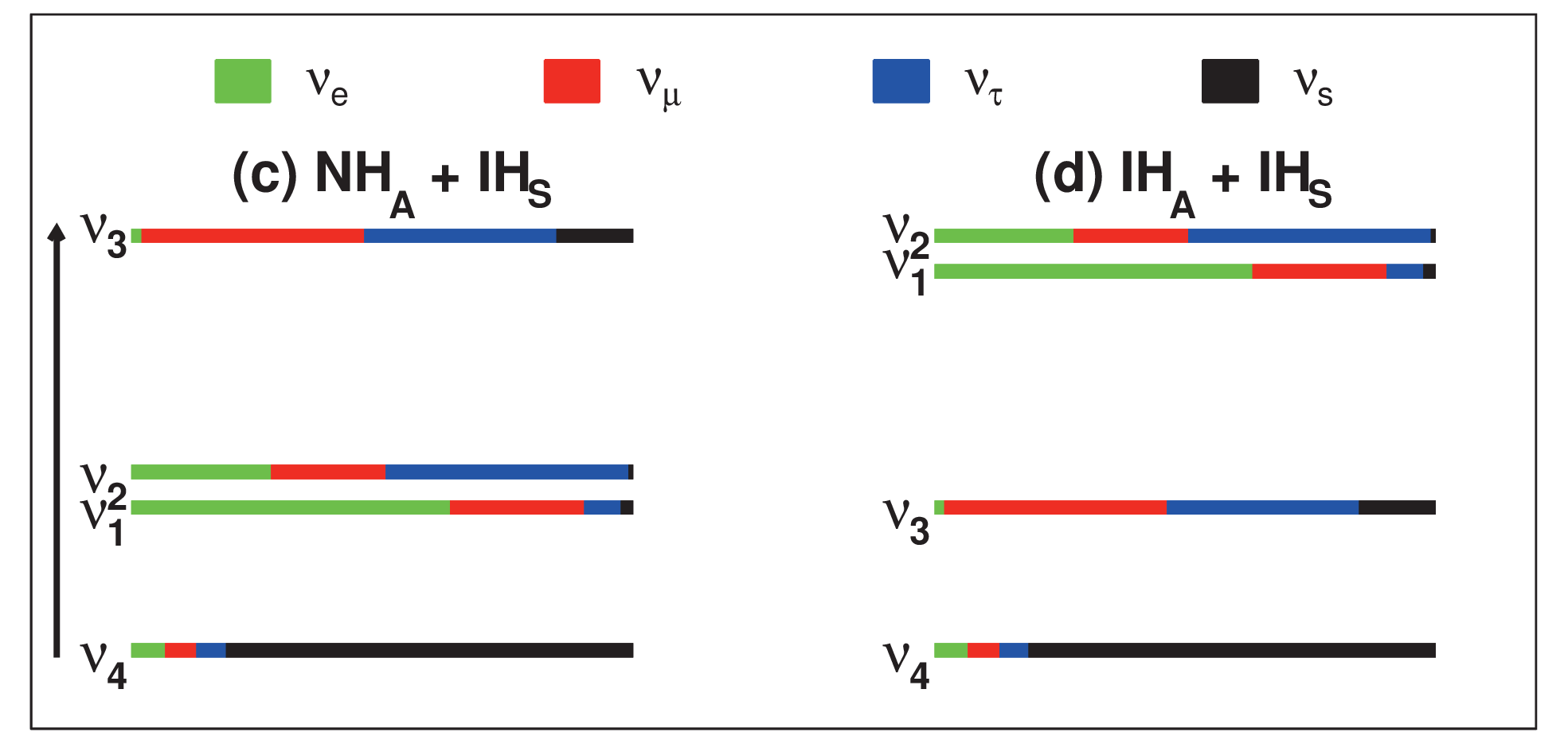}%pdf}
\end{tabular}
\end{center}
\vspace{-0.5cm}\caption{Four different scenarios of neutrino mass spectra in the (3+1) scheme, depending on the neutrino mass hierarchies
 of three active and one sterile neutrino mass eigenstates. The (2+2) scheme with two separate groups of mass eigenstates are already ruled out
 with the global analysis of solar and atmospheric neutrino data~\cite{Maltoni:2002xd}.}
\label{fig:sterile:mh_sterile}
\end{figure}

According to the general expression in Eq.~\ref{eq:intro:peegeneral}, the electron
(anti)neutrino survival probability in the (3+1) scheme is given by
\begin{eqnarray}
P(\nua{e} \rightarrow \nua{e}\;) = 1 - 4
\sum^{3}_{i=1}\sum^{4}_{j > i}|U_{ei}|^{2}|U_{ej}|^{2}\sin^{2}\frac{\Delta m^{2}_{ji} L}{4 E}\;,
\label{eq:ste:pee4t4}
\end{eqnarray}
where $U_{}$ is the neutrino mixing matrix defined in Eq.~\ref{eq:intro:UPMNS}, and $\Delta m^{2}_{ji} \equiv m^{2}_{j}-m^{2}_{i}$
for ($1\leq i < j \leq 4$).
For the elements of $U_{}$, we define $U_{e1} \equiv \cos \theta_{14} \cos \theta_{13} \cos \theta_{12}\,$,
$U_{e2} \equiv \cos \theta_{14} \cos \theta_{13} \sin \theta_{12}\,$,
$U_{e3} \equiv \cos \theta_{14} \sin \theta_{13}\,$, and
$U_{e4} \equiv \sin \theta_{14}$. Using the mixing angles defined above,
the survival probability in Eq.~\ref{eq:ste:pee4t4} can be written as
\begin{eqnarray}
\label{eq:ste:pee4t4new}
P(\;\nua{e} \rightarrow \nua{e}\;) =1
&-&\cos^{4}\theta_{14}\cos^{4}\theta_{13}\sin^{2}2\theta_{12}\sin^{2}\Delta_{21}\nonumber\\
&-&\cos^{4}\theta_{14}\sin^{2}2\theta_{13}(\cos^{2}\theta_{12}\sin^{2}\Delta_{31}+\sin^{2}\theta_{12}\sin^{2}{\Delta_{32}})\nonumber\\
&-&\cos^{4}\theta_{13}\sin^{2}2\theta_{14}(\cos^{2}\theta_{12}\sin^{2}\Delta_{41}+\sin^{2}\theta_{12}\sin^{2}{\Delta_{42}})\nonumber\\
&-&\sin^{2}\theta_{13}\sin^{2}2\theta_{14}\sin^{2}\Delta_{43}\;,
\end{eqnarray}
with $\Delta_{ij}\equiv\Delta m_{ij}^{2}L/4E$. The oscillation parameters from the three neutrino framework~\cite{Agashe:2014kda}
are assumed to remain unchanged by the inclusion of the light sterile neutrino.

In the case of an eV-scale sterile neutrino, we can simplify Eq.~\ref{eq:ste:pee4t4}
to an effective two-neutrino survival probability,
\begin{equation}
P(\;\nua{e} \rightarrow \nua{e}\;) = 1 -\sin^{2}2\theta_{14}\sin^{2}\frac{\Delta m^{2}_{41} L}{4 E}\;,
\label{eq:ste:pee4sim2}
\end{equation}
and the neutrino transition probability to
\begin{equation}
P(\;\nua{\mu} \rightarrow \nua{e}\;) = 4\sin^{2}\theta_{14}\sin^{2}\theta_{24}\sin^{2}\frac{\Delta m^{2}_{41} L}{4 E}\;,
\label{eq:ste:pmue4sim2}
\end{equation}
with $\theta_{24}$ being a further active-sterile mixing angle defined by $|U_{\mu 4}|^2\equiv\sin^{2}\theta_{24}$. Comparing the
above oscillation probabilities, the amplitudes of the $\nua{e} \rightarrow \nua{e}$ and
$\nua{\mu} \rightarrow \nua{e}$ channels are related with each other. A test of one of these channels can be used to restrict the other and vice versa.

\subsection{Indications of eV-scale sterile neutrinos}
\label{subsec:sterile:indications}

In the past decades, there has been a number of experimental results that appear as anomalies in the context of the standard three-neutrino framework. This includes,
\begin{itemize}
\item  the event excess of electron antineutrinos in a muon antineutrino beam~\cite{Aguilar:2001ty} in the LSND experiment;

\item  the event excess of electron antineutrinos in a muon antineutrino beam~\cite{AguilarArevalo:2010wv} in the MiniBooNE experiment;

\item  the antineutrino rate deficit in reactor short-baseline experiments based on a recent evaluation of the reactor antineutrino flux model~\cite{Mention:2011rk,Huber:2011wv};

\item  the neutrino rate deficit in the calibration runs performed with radioactive sources in the Gallium solar neutrino experiments~\cite{Giunti:2010zu}.
\end{itemize}
It is quite interesting that all the above anomalies can be consistently described by short-baseline oscillations via a
$\Delta m^2$ of around 1 eV$^2$. Given that solar neutrino oscillations indicate a value of $\Delta m^{2}_{21} \simeq 7.5 \times 10^{-5} \;{\rm eV}^{2} $ and atmospheric neutrino oscillations correspond to $|\Delta m^{2}_{31}| \simeq 2.4 \times 10^{-3}$ eV$^{2}$, the new mass-squared difference requires a fourth neutrino mass and thus also flavor eigenstate. However, results from the Large Electron-Positron collider (LEP) at CERN on the invisible decay width of $Z$ bosons show
that there are only three species of light neutrinos coupling to the $Z$ featuring masses lower than half of the $Z$ mass~\cite{ALEPH:2005ab}.
Therefore, the fourth neutrino flavor, if it exists, must be a sterile neutrino without direct coupling to standard model gauge bosons.

The LSND experiment used a pion decay-at-rest beam in the 20-60 MeV energy range to study neutrino oscillations at a baseline of 30 meters. Within the muon antineutrino beam, LSND observed an excess of electron-antineutrino events. The statistical significance of the observed excess is about 3.8 $\sigma$~\cite{Aguilar:2001ty}.
In the follow-up, the MiniBooNE experiment was designed  to test the LSND signal at a similar $L/E$ ratio but both baseline and energy larger by about one order of magnitude. The result observed by MiniBooNE was inconclusive with regards to the LSND signal as there was no $\bar\nu_e$ excess observed in the same $L/E$ region. However,
an anomalous excess was identified in the low-energy bins~\cite{AguilarArevalo:2010wv}.

In the electron (anti-)neutrino disappearance channel, the recent re-evaluations of the reactor antineutrino spectrum resulted in a 3$\%$ increase in the interaction rates expected for neutrino detectors with respect to earlier calculations. Meanwhile, the experimental value for neutron lifetime decreased, which in turn implied
a larger inverse $\beta$-decay cross section. Including previously-neglected effects on the rate expectation caused by long-lived isotopes not reaching
equilibrium in a nuclear reactor core during the detector running time, the overall expectation value for the antineutrino rate from nuclear
reactors increased by about $6\%$. As a result, more than 30 years of data from short-baseline reactor antineutrino experiments,
which formerly agreed well with the flux prediction, featured now an apparent $6\%$ deficit in their electron antineutrino event rates.

Another hint consistent with short-baseline oscillations arose from the source calibrations performed for the radio-chemical
Gallium experiments for solar neutrino detection. High-intensity sources based on the radioisotopes $^{51}$Cr and $^{37}$Ar, which both decay via electron capture
and emit mono-energetic electron neutrinos, were placed in close proximity to the detector and the resulting event rates were measured.
Compared to the rates predicted based on the source strengths and reaction cross-sections, a 5-20$\%$ deficit of the measured count rate was observed.

At the same time, a number of measurements disfavor the interpretation of a sterile neutrino state with a mass of around 1\,eV.  Neither muon neutrino disappearance in MINOS~\cite{Sousa:2015bxa}, nor electron neutrino appearance was observed in
KARMEN~\cite{Armbruster:2002mp}, NOMAD~\cite{Astier:2003gs}, ICARUS~\cite{Antonello:2013gut} and OPERA~\cite{Agafonova:2013xsk}.
Several global analyses~\cite{Kopp:2013vaa,Giunti:2013aea,Conrad:2012qt} that include
all the available short-baseline data show strong tension between the appearance and disappearance data. A careful analysis of the existing data on the cosmic microwave background (CMB) anisotropies, galaxy clustering and supernovae Ia seems
to favor one additional sterile neutrino species at the sub-eV mass scale~\cite{Hamann:2010bk,Hamann:2011ge,Giusarma:2011ex}. However,
the full thermalization of the additional degree of freedom before the CMB decoupling is disfavored.

In summary, a number of short baseline neutrino experiments support the sterile neutrino hypothesis, but some others do not. Considering the ambiguity of the available data, an unambiguous confirmation or refusal of the existence of light sterile neutrinos by a dedicated short-baseline oscillation experiments is a pressing requirement.

\subsection{Requirements for future measurements}
\label{subsec:sterile:future}

The neutrino community is engaging itself in an experimental effort to confirm or defy the existence of light sterile neutrinos
(or an equivalent mechanism which would be consistent with the observed phenomena).
In this context, the question arises how to define satisfactory criteria for a positive measurement.
The current situation seems to suggest that a mere rate excess or deficit, although statistically significant, may not be sufficient.
Therefore, future experiments are required to test the anomalies with as little model-dependence as possible
and by complementary methods to obtain as complete a picture as possible.

Future short-baseline accelerator neutrino experiments, such as the three detector short-baseline neutrino oscillation program laid out for
the Fermilab Booster neutrino beam~\cite{Antonello:2015lea}, will use charged-current reactions to observe both appearance and disappearance signals.
Meanwhile, reactor antineutrino experiments may provide a charged-current measurement of the baseline-dependence of $\bar\nu_e$ disappearance oscillations, as has been recently demonstrated by the Daya Bay experiment~\cite{Dwyer:2011xs}.
Currently, a new strategy emerges based on experiments with high-intensity radioactive neutrino sources at large liquid-scintillator detectors. The intense sources currently proposed are either based on electron-capture nuclei emitting mono-energetic electron neutrinos detected via elastic scattering of electrons or on high-endpoint $\beta^-$-emitters providing a sizable flux of electron antineutrinos above the detection threshold of the inverse beta decay. Corresponding experiments like SOX~\cite{Borexino:2013xxa} or CeLAND~\cite{Gando:2013zoa} provide a test of both the distance and energy dependence expected for short-baseline oscillations.
If very low energy thresholds were reached, a detection of all active neutrino flavors via neutral-current coherent scattering might become available, providing the most direct test of active-neutrino disappearance.
Only multiple experiments using different channels will allow for a robust interpretation of the data as sterile neutrino oscillations. In addition, cosmological constraints~\cite{Abazajian:2013oma} as well as a direct neutrino mass measurement at KATRIN~\cite{Riis:2010zm}  may provide the final orthogonal check for the existence of
light sterile neutrinos.

Under the assumption of {\it CPT} invariance, the $\bar \nu_e$ disappearance parameter space must contain the $\bar\nu_\mu \rightarrow \bar \nu_e$
appearance parameter space. This is because {\it CPT} invariance requires that $P(\bar \nu_\mu \rightarrow
\bar \nu_e) =P(\nu_e \rightarrow \nu_\mu$) and $P(\nu_e \rightarrow \nu_e) = P(\bar\nu_e \rightarrow \bar\nu_e)$.
However, it must also hold $P(\nu_e \rightarrow \nu_e)> P(\nu_e \rightarrow \nu_\mu)$, since the latter is only a single
appearance channel. Thus, if a $\bar\nu_e$ disappearance experiment such as IsoDAR$@$JUNO will provide sufficient sensitivity to
 cover the entire parameter space for the observation of $\bar \nu_\mu \rightarrow \bar \nu_e$ appearance,
this experiment offers a decisive test.

\subsection{JUNO potential for light sterile neutrino searches}
\label{subsec:sterile:juno}

The JUNO experiment is designed to deploy a single (far) detector at baselines of about 53 km from both the Yangjiang and Taishan NPPs.
Without an additional near detector, reactor antineutrino oscillations cannot be used to search for eV-scale sterile neutrinos.
However, the diameter of the JUNO central detector will be around 35 meters,
which is perfectly suitable for a short-baseline oscillation experiment with a radioactive neutrino source sensitive to eV-scale sterile neutrinos.

Source-based initiatives add a complementary approach to the reactor and accelerator short baseline programs because of both the purity of their source and the possibility to probe the baseline-dependence of the oscillation signal.
In addition, source-based experiments are less susceptible to uncertainties arising from the normalization of both flux and cross-section. The possibility to perform orthogonal measurements by using multiple detection channels (i.e. elastic, neutral current, and charged current scattering)  may provide the necessary orthogonal measurement to positively claim the existence of sterile neutrinos.

However, the production of MCi neutrino sources (e.g., $^{51}$Cr, $^{37}$Ar) depends on reactor facilities for irradiation, while kCi antineutrino sources (e.g., $^{144}$Ce-$^{144}$Pr) are based on fission isotopes won from spent nuclear fuel.
Therefore, they can only be produced in countries that feature the corresponding facilities for the reprocessing of nuclear fuel.
Further, short half-lives and regulatory issues complicate the transport of such sources, especially across international borders.

There are two basic options for the radioactive sources that can be deployed in a liquid scintillator detector like JUNO:
A monochromatic $\nu_{e}$ emitter, such as $^{51}$Cr or $^{37}$Ar, or a $\bar \nu_{e}$ emitter with a continuous $\beta$-spectrum.
Regarding the former case, the signature is provided by $\nu_{e}$ elastic scattering off electrons in the LS target.
This signature can be mimicked by beta-decays or the Compton scattering of gamma-rays induced by radioactive and cosmogenic background sources and as well the elastic scattering of solar neutrinos.
Especially if deployed outside of the detector target, an oscillation experiment with a $\nu_{e}$ source requires a very high source activity of $5-10$\,MCi to overcome these backgrounds.
In the second approach, $\bar\nu_{e}$ are detected via the inverse beta decay,
in which the delayed coincidence signature of positron and neutron provides efficient rejection power for the above-mentioned backgrounds.
Therefore, we focus our studies on this second option.

A suitable $\bar\nu_{e}$ source must feature a $Q_{\beta}$ value larger than 1.8 MeV (i.e., threshold of the inverse beta decay)
and a sufficiently long lifetime $\tau_{1/2}$ ($\geq$ 1 month) to allow for the production and transportation to the detector.
All candidate sources that have been proposed feature a long-lived low-$Q$ nucleus that decays to a short-lived high-$Q$ nucleus (e.g.~\cite{Kornoukhov:1994zq}).
Several suitable pairs have been identified:

\begin{itemize}
\item $^{144}$Ce-$^{144}$Pr, with $Q_{\beta}$(Pr)=2.996~MeV and $\tau_{1/2}$(Ce)=285~d,
\item $^{106}$Ru-$^{106}$Rh, with $Q_{\beta}$(Rh)=3.54~MeV and $\tau_{1/2}$(Ru)=372~d,
\item $^{90}$Sr-$^{90}$Y, with $Q_{\beta}$(Y)=2.28~MeV and $\tau_{1/2}$(Sr)=28.9~y,
\item $^{42}$Ar-$^{42}$K, with $Q_{\beta}$(K)=3.52~MeV and $\tau_{1/2}$(Ar)=32.9~y,
\end{itemize}
The first three candidates are common fission products from nuclear reactors, which can be extracted from spent fuel rods.
Since the $Q_{\beta}$ of $^{144}$Ce is larger than that of $^{90}$Sr, and the chemical production of $^{144}$Ce is easier than for $^{106}$Ru, we will focus in the following on a $^{144}$Ce-$^{144}$Pr $\bar\nu_{e}$ source to study the sensitivity of a sterile neutrino search in JUNO. This source can be either positioned at the center of the detector or  outside the central detector, right next to the wall of the detector vessel.

%%  source at detector center %%
\subsubsection{Sensitivity using an antineutrino source at the detector center}
\label{subsec:sterile:anti_juno_center}

In this section, we study the deployment of a $^{144}$Ce-$^{144}$Pr $\bar\nu_{e}$ antineutrino source featuring an energy spectrum $S$($E_\nu$), a mean lifetime $\tau$, and an initial activity $A_{0}=50$\,kCi. The source is encapsulated by a tungsten and copper shielding sphere and positioned at the detector center.
We consider a running time $t_{e}$ and a detection efficiency of 76$\%$ .
The expected number of inverse beta decay interactions at radius $R$ from the source and energy $E_{\nu}$ can be written as
\begin{equation}\label{eqn:sterile:dnde}
\frac{d^{2}N(R, E_{\nu})}{dRdE_{\nu}} = A_{0}\cdot n \cdot \sigma(E_{\nu}) \cdot S(E_{\nu}) \cdot P(R, E_{\nu}) \int^{t_{e}}_{0} e^{-t/\tau}dt;
\end{equation}
where $n$ is the number density of free protons in the detector target, 6.4$\times$10$^{28}$ protons/m$^{3}$ for JUNO, $\sigma(E_{\nu})$ is the IBD cross-section, and $P(R, E_{\nu})$ is the oscillation probability given in Eq.~\ref{eq:ste:pee4t4}.

The radius range considered for the oscillation search ranges from 1~m to 16~m with respect to the source.
The total event number from inverse beta decays will reach $2\times 10^5$ for 900 days of data taking.
The main background will be caused by coincidences among the 2.185 MeV gamma rays emitted in the decay of $^{144}$Pr with a branching ratio of $0.7\%$ branch.
This background will be reduced by a factor of $5\times 10^{9}$ when shielded by 33~cm of tungsten and 2~cm of copper.
The remaining background is mainly given by the reactor antineutrinos emitted from the Taishan and Yangjiang reactor cores. Corresponding event numbers can be derived based on the respective thermal powers (total power of 35.8~GW) and baselines given in Tab.~\ref{tab:intro:NPP}.

The expected sensitivity has been calculated based on a standard least-squares function,
\begin{eqnarray}\label{eqn:sterile:chi2eq1}
\chi^2 = \sum_i\sum_j \frac{\left(N^{\rm obs}_{i,j} - N^{\rm exp}_{i,j}\right)^2}{N^{\rm exp}_{i,j}\left(1+\sigma^2_b N^{\rm exp}_{i,j}\right)}
+ \left(\frac{\alpha_s}{\sigma_s}\right)^2 +  \left(\frac{\alpha_r}{\sigma_r}\right)^2 + \left(\frac{\alpha_d}{\sigma_d}\right)^2\,,
\end{eqnarray}
where the indices $i$ and $j$ refer to bins in the detected energy and position respectively. $N^{\rm obs}_{i,j}$ is the antineutrino events detected in each bin, including the possible sterile neutrino oscillation effect.
The expected number of events assuming no sterile neutrino oscillation $N^{\rm exp}_{i,j}$ is the sum of events from the source $S^{\rm exp}_{i,j}$
and the background from reactor antineutrinos $R^{\rm exp}_{i,j}$:
\begin{eqnarray}
\label{eqn:sterile:chi2eq2}
N^{\rm exp}_{i,j} = (1+\alpha_d)[(1+\alpha_s)S^{\rm exp}_{i,j} + (1+ \alpha_r)R^{\rm exp}_{i,j}] \,.
\end{eqnarray}
For the observed and expected number of events, the survival probabilities for reactor and source antineutrinos are both calculated with the full 3+1 scheme oscillation probability
assuming normal hierarchies for three active and one sterile mass eigenstates (as shown in panel (a)  of Fig.~\ref{fig:sterile:mh_sterile}). The expected number of events is allowed to vary within the systematic uncertainties via nuisance parameters. $\alpha_d$ accounts for detector efficiency variations ($\sigma_d$, $2\%$), while $\alpha_s$ and $\alpha_r$ account for the source ($\sigma_s$, $2\%$) and reactor ($\sigma_d$, $3\%$) normalization uncertainties. In addition, $\sigma_b$ is the bin-to-bin uncertainty of $0.3\%$.

Fig.~\ref{fig:sterile:juno_spectra} shows the distance and energy distributions expected for the antineutrino source signal (first row) as well as the reactor antineutrino background (third row). The second row illustrates the spectral features introduced by the presence of short-baseline disappearance oscillations, assuming a mass-squared difference of $\Delta m^2_{41}$ = 1 eV$^2$, and an active-sterile mixing ampltiude of $\sin^22\theta_{14} =0.1$.
\begin{figure}%[htb]
\begin{center}
\includegraphics[width=0.9\textwidth]{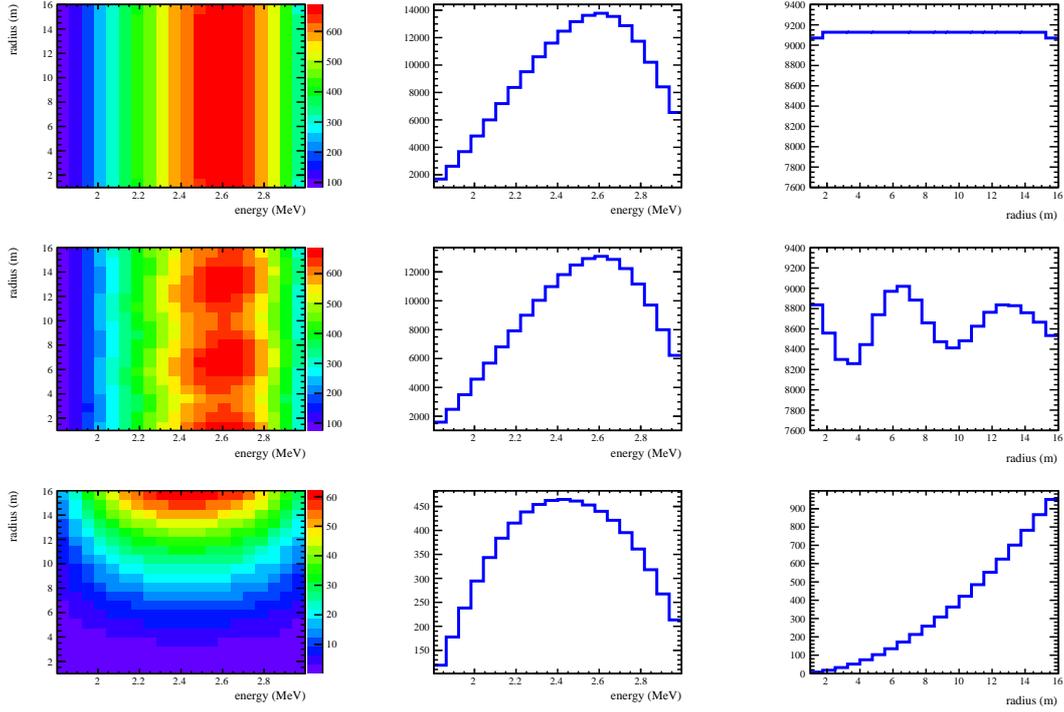}
\end{center}
\vspace{-0.25in}
\caption{Energy and position dependence of the event numbers expected for a $^{144}$Ce $\bar\nu_e$ source (50~kCi, 900~days) at the center of the detector. The top row of panels shows the signal of the antineutrino source without oscillations. The middle row illustrates the effect of $\bar{\nu}_e \rightarrow \bar\nu_s$ oscillations. The bottom row illustrates the event distribution expected for the reactor antineutrino background.
In each row, the left panel shows the 2-dimension distribution of the
event rates as a function of signal energy and the distance from the source. The middle panels are
 are one-dimensional projections of the energy spectrum, while right panels show the distance distributions. }
\label{fig:sterile:juno_spectra}
\end{figure}
Based on the above $\chi^2$ function, we have performed a scan of the sensitivity to $\sin^{2}2\theta_{14}$ as a function of running time, energy resolution and spatial resolution,
assuming a fixed mass-squared difference $\Delta m^2_{41}$ is 1 eV$^2$. The results are shown in Fig.~\ref{fig:sterile:juno_scan}.
%%%%%%%%%%%%%%%%%%%%%%%%%%%%%%%%%%%%%%%%%%%%%%%%%%%%%%%%%%%
\begin{figure}%[htb]
\begin{center}
\begin{tabular}{ccc}
\includegraphics*[width=0.3\textwidth]{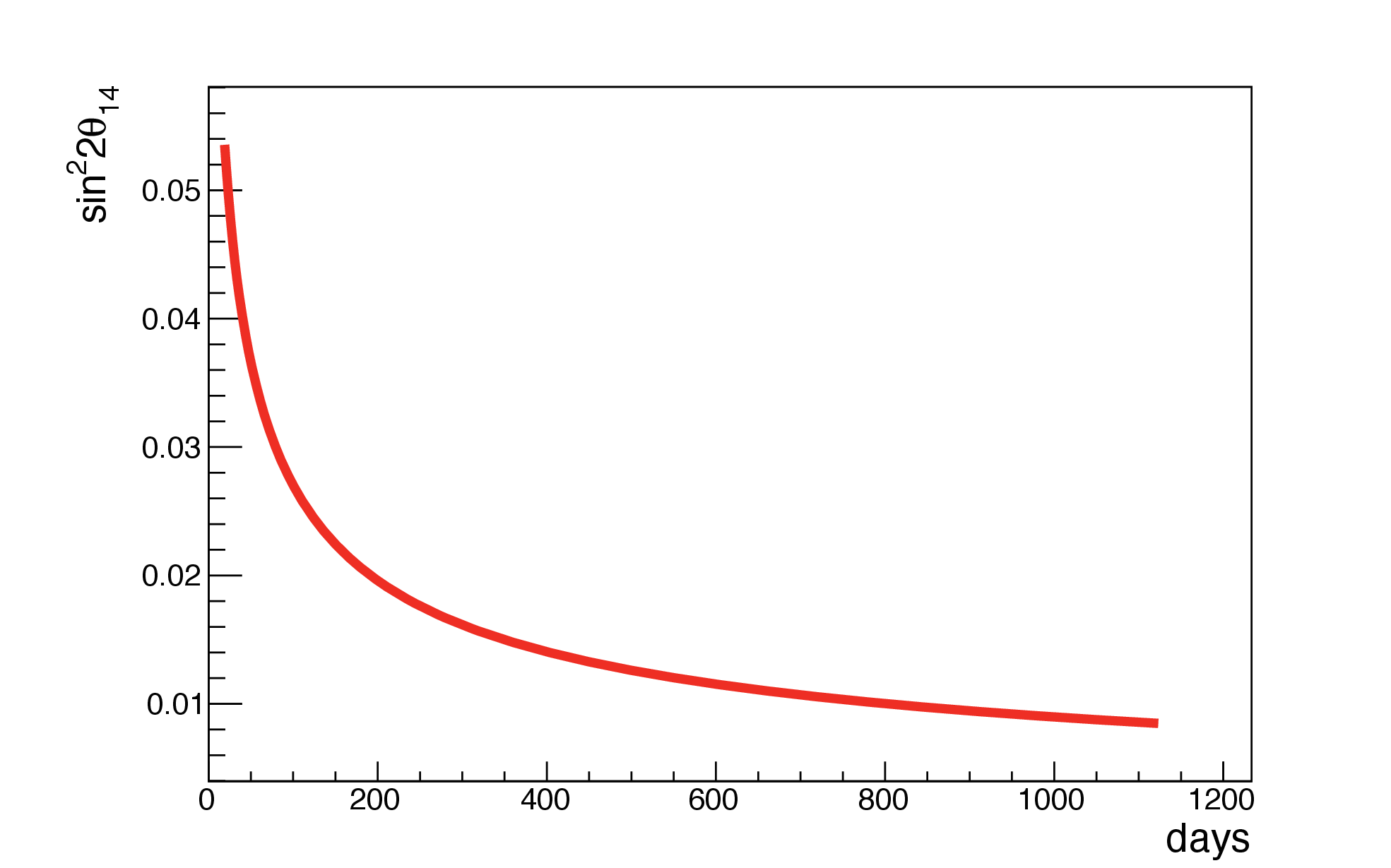}
&
\includegraphics*[width=0.3\textwidth]{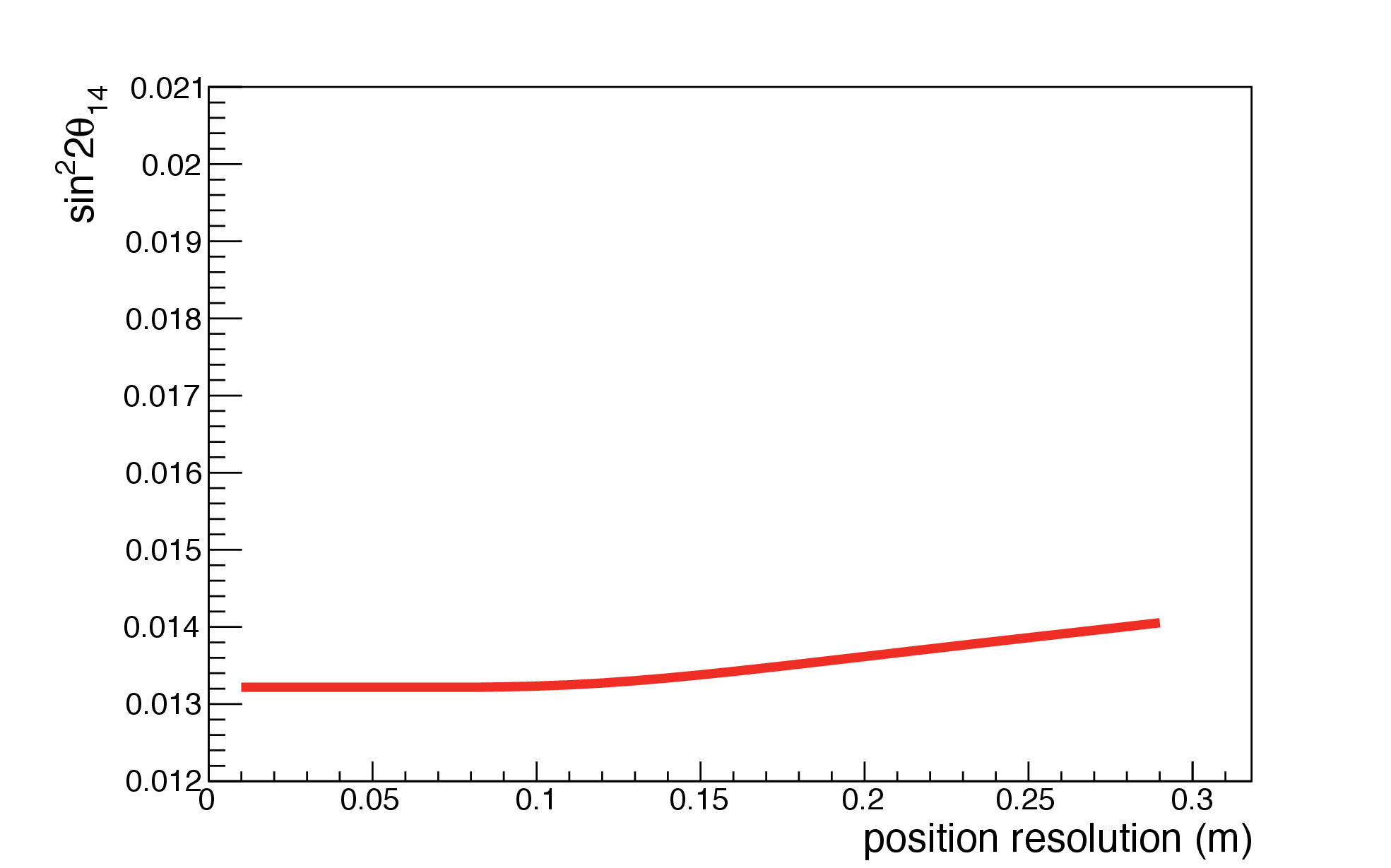}
&
\includegraphics*[width=0.3\textwidth]{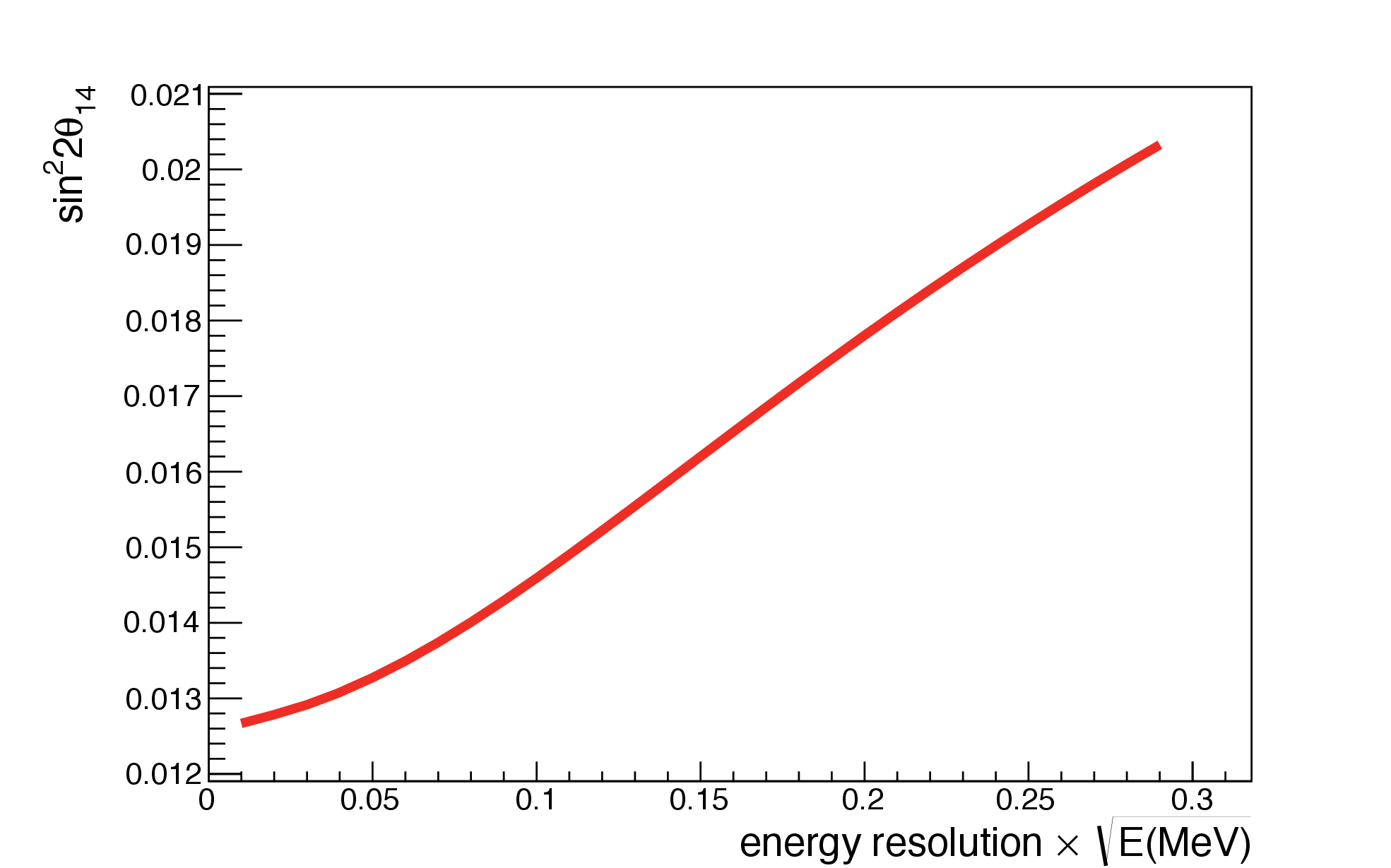}
\end{tabular}
\end{center}
%\vspace{-0.25in}
\caption{Sensitivity of $\sin^{2}2\theta_{14}$ at the 90\% C.L. (for $\Delta m^2_{41}$ fixed at 1 eV$^2$): Default values are a running time of 450 days, an energy resolution of $3\%/\sqrt{E  (\rm MeV)}$, and a
spatial resolution of 12~cm. While the other values are kept fixed, the left panel shows the dependence of the sensitivity on the running time, the middle on the energy resolution, and the right on the spatial resolution.}
\label{fig:sterile:juno_scan}
\end{figure}
As the default value for the calculation of the final sensitivity, we choose 450 days as a practical data-taking time, a spatial resolution of 12~cm according to the current detector design, and an energy resolution of $3\%/\sqrt{E  (\rm MeV)}$.
When varying these parameters, the resulting sensitivity shows only weak dependence on the spatial resolution, while the dependence on the energy resolution is significant.

Assuming these default values, the final sensitivity as a function of both  $\sin^{2}2\theta_{14}$ and $\Delta m^2_{41}$ is shown in Fig.~\ref{fig:sterile:juno_center}. Despite the lower activity of the source and a shorter running time, JUNO will surpass the sensitivity of CeLAND~\cite{Gando:2013zoa} in the $\Delta m^2_{41}$ region from 0.1 to 10~eV$^2$.
\begin{figure}[htb]
\begin{center}
\includegraphics*[width=0.5\textwidth]{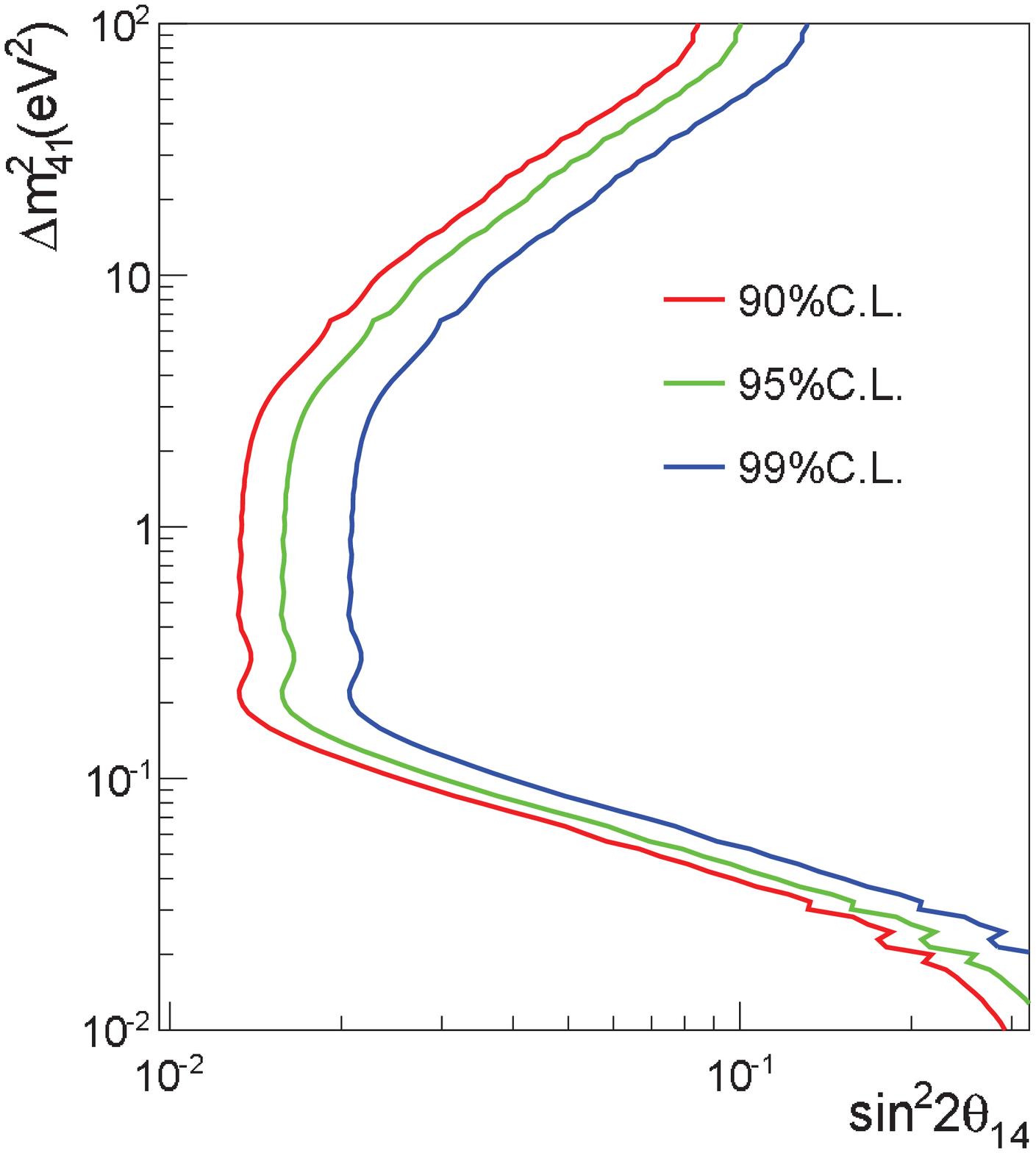}%pdf}
\end{center}
\vspace{-0.25in}
\caption{Sensitivity of a $\bar{\nu}_e$ disappearance search at JUNO to the oscillation parameters $\Delta m^2_{41}$ and $\sin^22\theta_{14}$ assuming a 50 kCi $^{144}$Ce source at the detector center, with 450 days of data-taking. We show the 90, 95 and 99\% confidence levels, with the reactor antineutrino background taken into account.}
\label{fig:sterile:juno_center}
\end{figure}

%% source is outside of the detector %%
\subsubsection{Sensitivity with the antineutrino sources outside the detector}
\label{subsec:sterile:anti_juno_out}
In case the kCi $^{144}$Ce source is deployed outside the central detector, we determine the sensitivity of a sterile neutrino search following the approach described in~\cite{Cribier:2011fv} . Specifically, the sensitivity is quantified using a $\chi^2$ analysis along the lines of Eq.~\ref{eqn:sterile:chi2eq1}.
The corresponding exclusion contours have been derived using the null-oscillation hypothesis, in which the observed event number $N^{\rm obs}_{i,j}$ is calculated assuming the absence of oscillations.
On the other hand, the predicted $N^{\rm exp}_{i,j}$ profile includes the possibility of short-baseline oscillations with the oscillation parameters $\Delta m^2_{41}$ and $\sin^22\theta_{14}$. The background from reactor antineutrinos is included in the analysis.
The uncorrelated systematic bin-to-bin error is assumed to be 2\%, and an additional pull term has been included to quantify the uncertainty in the knowledge of the source intensity, which has been set to 1.5\%.
Other inputs used in the calculations are the distance from the center of the detector, 20~m, the strength of the source, 100~kCi, the overall data taking period, 1.5 years, and the spatial resolution, 12~cm. For illustration, the energy resolution is set to be $5\%/\sqrt{E  (\rm MeV)}$ here as that of the Borexino detector.
The exclusion contours, stemming from Eq.~\ref{eqn:sterile:chi2eq1} and reported in Fig.~\ref{fig:sterile:juno_outside}, are obtained
with the 90, 95 and 99\% C.L. of the $\chi^2$ distribution for two degrees of freedom.

Comparing to the experiment placing the antineutrino source at the detector center as described before, the sensitivity in $\sin^22\theta_{14}$ seems comparable. Note, however, that the experiment with the external source assumes a twice higher source activity as well as slightly different systematic uncertainties.
\begin{figure}%[htb]
\begin{center}
\includegraphics[width=0.65\textwidth]{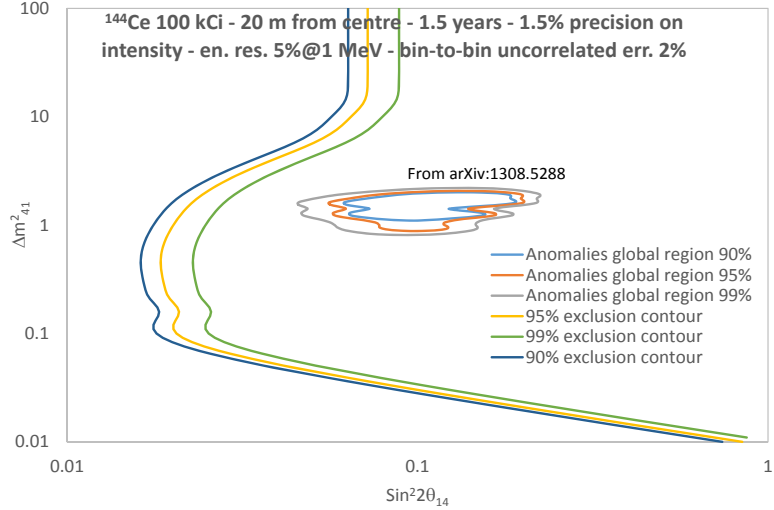}
\end{center}
\vspace{-0.25in}
\caption{Sensitivity of a $\bar{\nu}_e$ disappearance search at JUNO to the oscillation parameters $\Delta m^2_{41}$ and $\sin^22\theta_{14}$, placing a 100 kCi $^{144}$Ce source outside the central detector. We assume a data taking time of 450 days and take the reactor antineutrino background into account. The exclusion limits shown represent 90, 95 and 99\% confidence levels. The allowed regions are taken from the global analysis in Ref.~\cite{Giunti:2013aea}.}
\label{fig:sterile:juno_outside}
\end{figure}

%% IsoDAR %%
\subsection{Sensitivity of a cyclotron-driven \texorpdfstring{$^8$Li}{8Li} source (IsoDAR)}
\label{subsec:sterile:isodar}

IsoDAR$@$JUNO uses the decay of cyclotron-produced $^8$Li as a pure $\bar\nu_e$ source located at short distance from the JUNO detector. The combination of high antineutrino flux and the large target volume of the JUNO detector
allows the experiment to make an extremely sensitive search for antineutrino disappearance
in the parameter region favored by the sterile-neutrino anomalies.
Based on the small spatial diameter of the $^8$Li production region and the excellent position resolution of JUNO, such an experiment will be able to detect oscillations by observing the rate deficit that is induced by the oscillation wave as a function of the ratio of distance over energy ($L/E$).
An analysis of survival probabilities in $L/E$ is well suited to detect oscillations but also highly effective in reducing the systematic uncertainties introduced by background and flux normalizations.
In the following, we investigate the sensitivity of IsoDAR$@$JUNO for the oscillation parameter range indicated by the current experimental anomalies.
\paragraph{Neutrino source.} A 60 MeV/amu cyclotron accelerator produces deuterons that impinge on a beryllium target, such producing copious amounts of neutrons. The target is surrounded on all sides by a high-purity (99.99\%)  $^7$Li sleeve, where neutron
capture results in $^8$Li production. The subsequent $\beta^-$-decay of the $^8$Li nuclei created produces an isotropic
$\bar \nu_e$ flux with an average energy of $\sim$6.5~MeV and an endpoint of $\sim$13~MeV.
The spatial extension of the source leads to an uncertainty of 40~cm (1$\sigma$) regarding the origin of the antineutrinos emitted.
The source requires substantial iron and concrete shielding to contain the neutrons that escape the $^7$Li sleeve.
For this analysis, we assume that the center of the source is located 5.0 m from the edge of the active region of JUNO.
\begin{figure}%[t]
\begin{center}
{\includegraphics[width=0.5\textwidth]{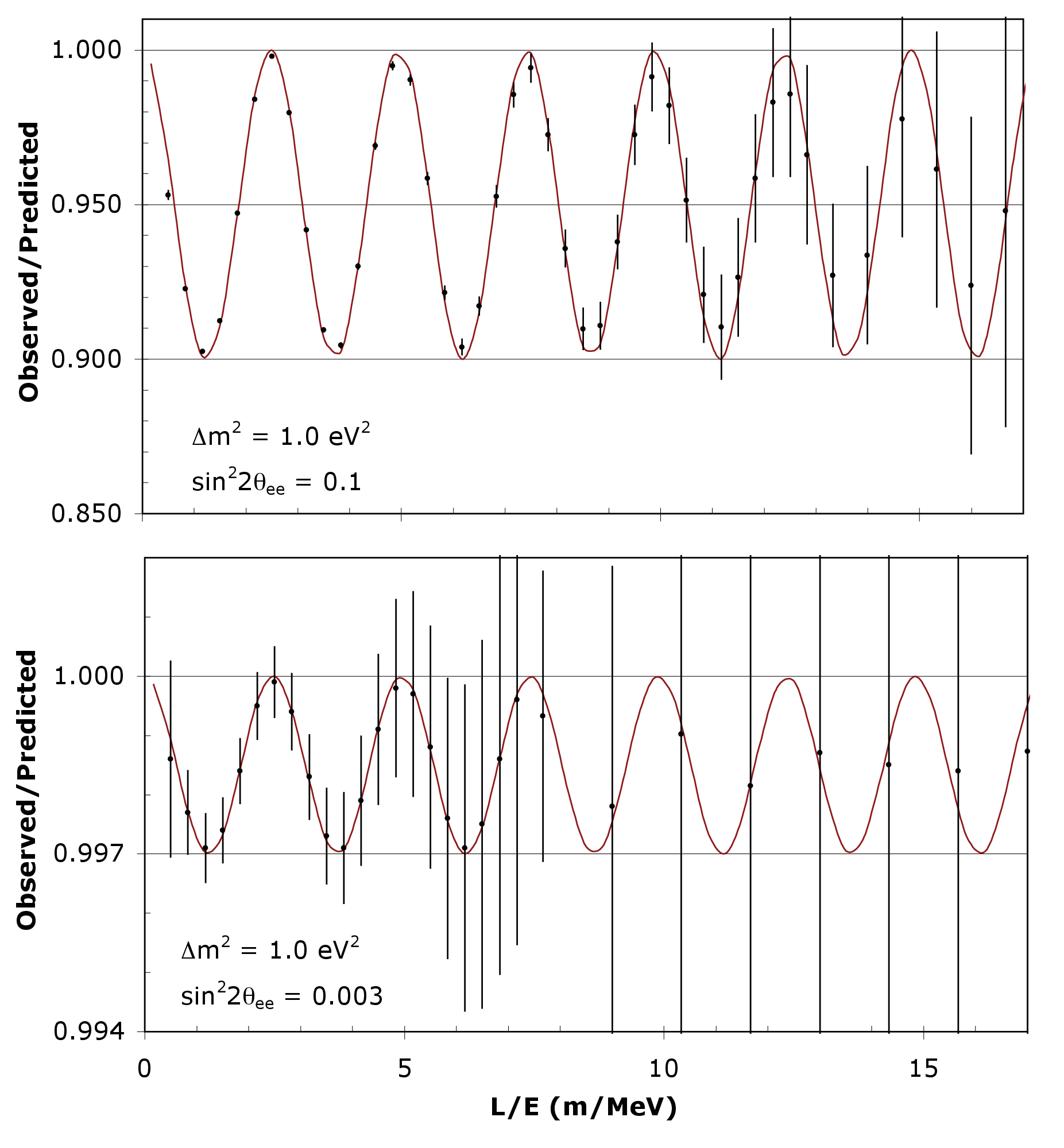}%pdf}
}
\end{center}
\vspace{-0.25in}
\caption{Oscillation signature in the $L/E$ space, with five years
running, assuming (top) $\Delta m^2 (\equiv\Delta m^2_{41})= 1.0 \ {\rm eV}^2$ and $\sin^2
2\theta_{ee}(\equiv\sin^22\theta_{14}) = 0.1$ and (bottom) $\sin^2
2\theta_{ee} = 0.003$.
Black points corresponds to the simulation data, while the
solid curve represents the oscillation probability without energy and position
smearing. Plots from Ref.~\cite{Conrad:2013ova}
\label{fig:sterile:JUNO_Wave} }
\end{figure}
Electron antineutrinos are observed via the inverse beta decay (IBD),
$\bar \nu_e +p \rightarrow e^+ + n$.
The neutrino energy is reconstructed from the visible energy of the positron:
$E_\nu = E_{e^+} + 0.8 \ {\rm MeV } $.
The vertex position is reconstructed using the arrival time of the scintillation light.
For the preliminary analysis published in Ref.~\cite{Conrad:2013ova}, the
following parameters have been used for JUNO:
\begin{itemize}
\item an active target of 34.5~m in diameter, with a fiducial volume of 20~ktons.
\item PMTs are located at a diameter of 37.5~m.
\item Veto region extends 1.5~m beyond the PMTs
\item rock overburden corresponding to $\sim 2000$~m.w.e.
\item an energy resolution of $3\%/\sqrt{E  (\rm MeV)}$.
\item a vertex resolution of $12~{\rm cm}/\sqrt{E  (\rm MeV)}$.
\end{itemize}

The analysis for the sterile-neutrino sensitivity follows the method presented in Ref.~\cite{Agarwalla:2011qf},
that assumes anormalization uncertainty of  5\% and a detection efficiency of 90\%.
Since $L$ and $E$ can be measured rather precisely, this analysis exploits the $L/E$ dependence of the possible
oscillation probability %$P = 1 - \sin^2 2\theta \sin^2[1.27\Delta m^2(L/E)]$,
to estimate the $\Delta m^2_{41}$-$\sin^22\theta_{14}$ regions where the null-oscillation
hypothesis can be excluded at the 5$\sigma$ confidence level.

We assume 5 years of running time. Fig.~\ref{fig:sterile:JUNO_Wave} shows example oscillation waves for parameter combinations motivated by the global oscillation analyses for sterile neutrinos.
The  parameters for the top example are $\Delta m^2_{41} = 1.0 \ {\rm eV}^2$ and $\sin^22\theta_{14} = 0.1$,
which corresponds to the allowed parameter region for $\bar\nu_e$ disappearance derived in Ref.~\cite{Giunti:2013aea}.
The bottom example assumes parameter values of $\Delta m^2_{41} = 1.0 \ {\rm eV}^2$ and $\sin^22\theta_{14} = 0.003$, which is a solution in agreement with the allowed region for the $\bar \nu_\mu\to\bar \nu_e$ appearance signal~\cite{Conrad:2012qt}.
The bottom plot illustrates that even for the lowest values $\sin^22\theta_{e\mu}$($\equiv4\sin^2\theta_{14}\sin^2\theta_{24}$) allowed by the LSND anomaly,
IsoDAR$@$JUNO will be able to definitively map out the oscillation wave.

Fig.~\ref{fig:sterile:JUNO_sense} shows the sensitivity curve for IsoDAR$@$JUNO (blue line).
Such an experiment could not only perform a decisive test of the parameter regions favored by the reactor anomaly (light gray) and the corresponding global analysis region
for electron-flavor disappearance in a (3+1) model (dark gray). It also covers completely the ``global $\bar\nu_e$ appearance'' region (purple) at a significance level greater than 5$\sigma$. If no oscillation signal was observed, sterile neutrinos mixing could be excluded at $5\sigma$ level as the cause of all of the present anomalies, including
LSND and MiniBooNE.
In case of a positive detection of short-baseline oscillations, the precise measurement of the $L/E$ dependence
will allow to quantify the emerging oscillation pattern in order to trace the complicated wave behavior and assess the consistency with
oscillation models featuring one (3+1) or two (3+2) sterile in addition to the three active neutrino flavors. Currently, there are no other experiments proposed that would match the sensitivity of IsoDAR$@$JUNO.

\begin{figure}%[t]
\begin{center}
{\includegraphics[width=0.45\textwidth]{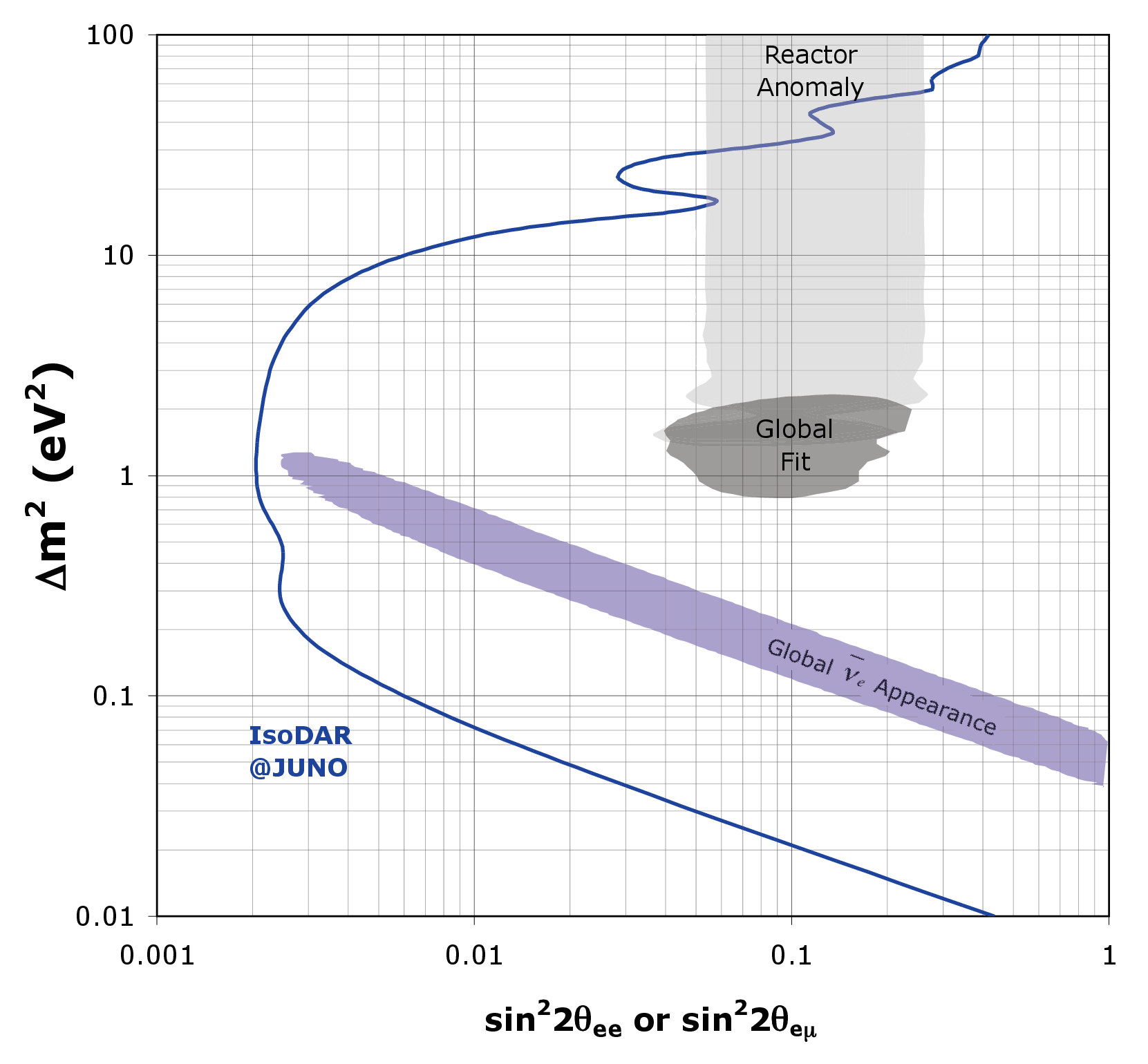}%pdf}
}
\end{center}
\vspace{-0.25in}
\caption{The sensitivity of an oscillation search of JUNO@IsoDAR experiment in comparison to the parameter regions favored by the current anomalies.
The blue curve indicates the $\Delta m^2$($\equiv\Delta m^2_{41}$) vs. $\sin^2 2\theta_{ee}$($\equiv\sin^2 2\theta_{14}$) contours for which the
null-oscillation hypothesis can be excluded at more than 5$\sigma$ with IsoDAR@JUNO
after five years of data-taking.
The light (dark)
gray area is the 99\% allowed region for the Reactor Anomaly~\cite{Mention:2011rk}
(Global Oscillation Fit~\cite{Giunti:2013aea}) plotted as $\Delta
m^2$ vs. $\sin^2 2\theta_{ee}$.  The purple area is the 99\% CL for a combined
fit to all  $\bar\nu_e$ appearance data\cite{Conrad:2012qt}, plotted as
$\Delta m^2$ vs. $\sin^2 2\theta_{e\mu}$($\equiv4\sin^2\theta_{14}\sin^2\theta_{24}$).
Plot is adapted from Ref.~\cite{Conrad:2013ova},
\label{fig:sterile:JUNO_sense} }
\end{figure}

\subsection{Sensitivity with reactor antineutrino oscillations}
\label{subsec:sterile:reactor_juno}

Super-light sterile neutrinos on a $\Delta m^2$ scale of ${\cal O}(10^{-5})$ eV$^2$ \cite{Bakhti:2013ora, Girardi:2014wea} could be discovered not only by a precision measurement of the solar MSW-LMA oscillation transition region but also by a
medium-baseline reactor neutrino experiment such as JUNO. In the following, we will study the sensitivity of a corresponding search in JUNO.

The JUNO detector, a 20~kton liquid scintillator detector, is located at a baseline of 53~km from two reactor complexes, with a total thermal power of 36~GW. The targeted energy resolution is $3\%/\sqrt{E({\rm MeV})}$.
We assume the uncertainty of the absolute energy scale for positron signals is similar to the Daya Bay experiment~\cite{An:2013zwz},
which corresponds to about 1\% between 1 and 10~MeV. With 6 years of running at full reactor power, a total of $\sim\,$100,000 IBD events will be collected. The reactor neutrino data will be used to determine the mass hierarchy as well as precision measurement of three oscillation parameters. However, if super-light sterile neutrinos exist, additional distortion could be observed in the reactor neutrino spectrum.

For the three active neutrinos, two neutrino mass hierarchies are still allowed by experiment.
They are usually referred to as the normal mass hierarchy if $m_{3} > m_{1}$, and inverted mass hierarchy if $m_{3} < m_{1}$.
Similarly, there are also two possible mass orderings for the new sterile neutrino mass eigenstate, $m_{4}$, with respect to $m_{1}$.
Following the convention, we define $m_{4} > m_{1}$ as the sterile neutrino normal mass hierarchy,
and $m_{4} < m_{1}$ as the inverted mass hierarchy. In order to distinguish those two arrangements,
we denote the conventional active neutrino normal and inverted mass hierarchies as NH$_{\rm A}$ and IH$_{\rm A}$,
and NH$_{\rm S}$ and IH$_{\rm S}$ for the sterile neutrino cases. As illustrated in Fig.~\ref{fig:sterile:mh_sterile},
there are in the extended case four possible combinations for the neutrino mass hierarchy.
Note that since the JUNO experiment has a 3-4$\sigma$ medium sensitivity to determine the mass hierarchy of the active neutrinos
we may end up with only two scenarios.

In order to calculate the sensitivity for oscillations into super-light sterile neutrinos,
we follow the method used in the Daya Bay reactor sterile neutrino search~\cite{An:2014bik}.
A binned log-likelihood method is adopted with nuisance parameters reflecting the uncertainty on the detector response
and a covariance matrix encapsulating the reactor flux uncertainties as given in the Huber~\cite{Huber:2011wv} and Mueller~\cite{Mueller:2011nm} flux models.
Since there is no near detector to constrain the flux~\cite{Hayes:2013wra}, the rate uncertainty of the reactor neutrino flux was enlarged to 6\%.
The fit uses $|\Delta m^{2}_{32}| = (2.41 \pm 0.10) \times 10^{-3} {\rm eV}^{2}$~\cite{Adamson:2013whj}.
The values of $\sin^{2}2\theta_{14}$, $\sin^{2}2\theta_{13}$, $\sin^{2}2\theta_{12}$, $\Delta m^{2}_{41}$ and $\Delta m^{2}_{32}$ are unconstrained.
\begin{figure}%[htb]
\begin{center}
\includegraphics[width=0.5\textwidth]{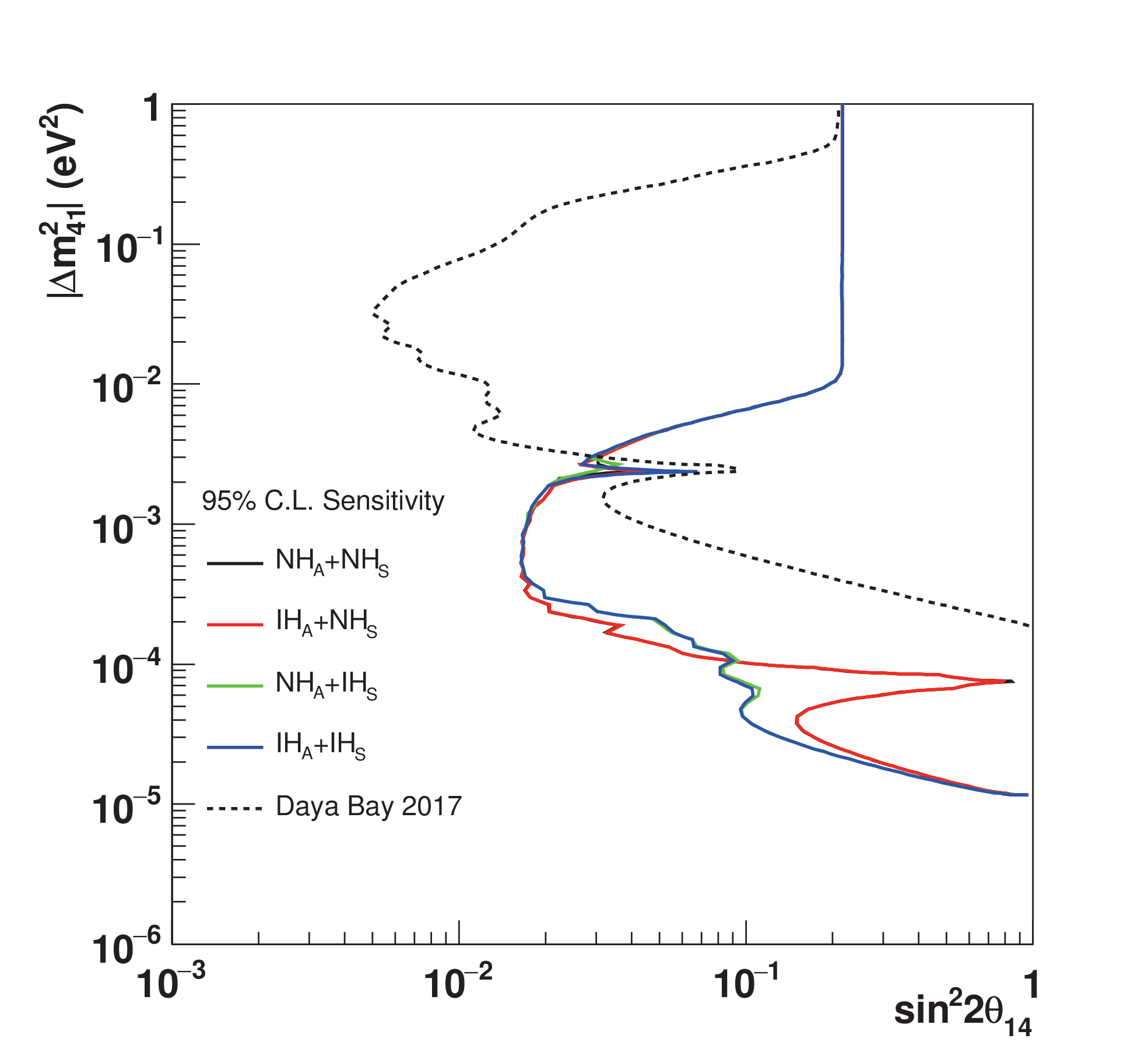}%pdf}
\end{center}
\vspace{-0.5cm}\caption{Comparison of the 95\% C.L. sensitivities for
various combinations of neutrino mass hierarchies.}
\label{fig:sterile:reactor}
\end{figure}
Fig.~\ref{fig:sterile:reactor} shows the 95\% C.L. sensitivity contours in the oscillation parameters
$\sin^{2}2\theta_{14}$ and $|\Delta m^{2}_{41}|$ for different combinations
of the neutrino mass hierarchy. %Due to the designed optimal baseline for the neutrino mass hierarchy measurement,
The search is most sensitive rin the region $10^{-5} < |\Delta m^{2}_{41}| < 10^{-2}$ eV$^{2}$.
For the region of $|\Delta m^{2}_{41}| > 10^{-2}$ eV$^{2}$, the fast oscillation pattern is smeared out in energy as the detector resolution is no longer sufficient. Therefore, the sensitivity in this region depends only on the uncertainty of the flux normalization, and no further information on $|\Delta m^{2}_{41}|$ can be obtained.

For different mass hierarchies, there are some clearly distinctive features. For the sterile neutrino normal hierarchy NH$_{\rm S}$, there appears a large dip in sensitivity at
$|\Delta m^{2}_{41}|=7 \times \;10^{-5}$eV$^{2}$. This is caused by $|\Delta m^{2}_{41}| \approx \Delta m^{2}_{21}$ as the masses of
$m_{4}$ and $m_{2}$ become degenerate. However, this degeneracy is broken for other mass configurations that result in a different value of $|\Delta m^{2}_{42}|$. As shown in Fig.~\ref{fig:sterile:mh_sterile},
for the NH$_{\rm S}$ case, $|\Delta m^{2}_{42}| = |\Delta m^{2}_{41}| - \Delta m^{2}_{21}$;
and for the IH$_{\rm S}$ case, $|\Delta m^{2}_{42}| = |\Delta m^{2}_{41}| + \Delta m^{2}_{21}$.
According to the oscillation probability in Eq.~\ref{eq:ste:pee4t4}, the $|U_{e2}|^{2}|U_{e4}|^{2}\sin^{2}(\Delta m^{2}_{42}L/4E)$ term
becomes quite different for those two cases.
%Note that the degeneracy for $m_{2}$ and $m_{4}$ can be broken if we constrain the mixing angle $\sin^{2}\theta_{12}$ with the solar measurements.
Interestingly, there is a smaller dip structure at $|\Delta m^{2}_{41}|= 2.4 \times 10^{-3}$eV$^{2}$
for all neutrino mass hierarchies. Similarly when masses $m_{4}$ and $m_{3}$ become degenerate at
$|\Delta m^{2}_{41}| \approx |\Delta m^{2}_{31}|$, $|\Delta m^{2}_{43}|$ itself is different for the four cases.
However, the oscillation term associated with $|\Delta m^{2}_{43}|$, $|U_{e3}|^{2}|U_{e4}|^{2}\sin^{2}(\Delta m^{2}_{43}L/4E)$,
is suppressed by the smallness of $|U_{e3}|^{2} \approx 0.02$,
compared with $|U_{e2}|^{2} \approx 0.3$ for the case of degeneracy of $m_{2}$ and $m_{4}$.
Since JUNO has sensitivity to the active neutrino mass hierarchy determination,
the dip structure shows up at slightly different locations for the NH$_{\rm A}$ and IH$_{\rm A}$ cases.

In summary, the precision measurement of reactor antineutrino oscillations in JUNO provides sensitivity to search for super-light sterile neutrinos at a $\Delta m^2$ scale of ${\cal O}(10^{-5})$ eV$^2$.
The most sensitive region is located at $10^{-5} < |\Delta m^{2}_{41}| < 10^{-2}$ eV$^{2}$.
If it is combined with the results of the sterile neutrino search conducted by the Daya Bay experiment~\cite{An:2014bik},
which is sensitive to $10^{-3} < |\Delta m^{2}_{41}| < 0.3$ eV$^{2}$, the sensitive region will cover about four orders of magnitude in
$|\Delta m^{2}_{41}|$, from $10^{-5}$ eV$^{2}$ to 0.3 eV$^{2}$. Combined with one of the experiment approaches searching for eV-scale sterile neutrinos that have been described above, the whole region of interest for light sterile neutrinos can be covered.

\subsection{Conclusion}
\label{subsec:sterile:conclusion}
The JUNO detector has multiple advantages in searching for light sterile neutrinos, including its large dimensions, unprecedent energy resolution and
excellent position accuracy.
Placing a 50 to 100~kCi source of antineutrinos extracted from spent reactor fuel, inside or outside the detector for a 1.5 year run,
JUNO is sensitive to the entire global analysis region for electron-flavor disappearance in the (3+1) scheme, at a more than 3$\sigma$ confidence level.
It has the greatest sensitivity for 0.1 - 10~eV$^2$-scale sterile neutrinos among all current and planned experiments with various proposed sources.
Using a $^8$Li antineutrino source produced from a 60~MeV/amu cyclotron accelerator (IsoDAR@JUNO) located 5~m away
from the detector would provide sensitivity to 1~eV$^2$-scale sterile neutrinos.
Assuming 5 years of data-taking, the sensitivity curve of IsoDAR@JUNO covers the allowed reactor anomaly region,
the 3+1 scheme global analysis region for electron-flavor disappearance,
and the global $\bar\nu_e$ appearance region (i.e., all present anomaly regions) at a greater than 5$\sigma$ confidence level.
In addition to the excellent sensitivity to 1~eV$^2$-scale light sterile neutrinos,
the JUNO experiment can search for super-light sterile neutrinos at the $\Delta m^2$ scale of ${\cal O}(10^{-5})$~eV$^2$ using reactor antineutrinos.
With a total of 100,000 IBD events from reactor antineutrinos collected over 6 years of full-power running,
the most sensitive region is $10^{-5} < |\Delta m^{2}_{41}| < 10^{-2}$~eV$^{2}$.
Combined with the sterile neutrino exclusion region of the Daya Bay experiment, i.e., $10^{-3} < |\Delta m^{2}_{41}| < 0.3$~eV$^{2}$,
the JUNO experiment will have good sensitivity across the entire range of light sterile neutrino searches,
about seven orders of magnitude in $|\Delta m^{2}_{41}|$, when using antineutrinos both from sources and reactors.

\clearpage

\section{Nucleon Decays}
\label{sec:nd}

\blfootnote{Editor: Chao Zhang (chao@bnl.gov)}

Being a large liquid scintillator detector deep underground, JUNO is
in an excellent position to search for nucleon decays. In
particular, in the SUSY-favored decay channel $p \to K^+ +
\overline\nu$, JUNO will be competitive with or complementary to
those experiments using either water Cherenkov or liquid argon
detectors.

\subsection{A brief overview of nucleon decays}
\label{subsec:nd:review}
% \begin{itemize}
%     \item Theoretical interest
%     \item Past experimental limits (from Super-Kamiokande)
%     \item JUNO's sensitive channels
% \end{itemize}

To explain the observed matter-antimatter asymmetry of the universe,
baryon number violation is one of the prerequisites \cite{Sakharov:1967rr}.
However, there has been no experimental evidence for baryon number
violation. Unlike the electric charge, the baryon number is only a
global symmetry in the Standard Model (SM) and its natural
extensions. In many Grand Unified Theories (GUTs) that unify strong
and weak interactions, baryon number conservation is only an approximate symmetry (i.e., it is slightly broken). A recent review of the GUT models can
be found in Ref.~\cite{Nath:2006ut}. The gauge coupling unification
scale of such GUT models is typically of the order of $10^{16}$ GeV,
which can never be touched by any man-made particle accelerators in the foreseeable future.
Fortunately, an indirect experimental test of the GUTs is possible
through an observation of the proton decays --- one of the unique
predictions of the GUTs.

The first experiment in search of the proton decay dates back to
1954, when Reines {\it et al} ~\cite{Reines:1954pg} set a lower limit on
the lifetime of protons with the help of a 300-liter liquid
scintillator detector: $\tau>10^{22}$ yrs. Since then many
larger-scale experiments, such as Kolar Gold Field~\cite{Krishnaswamy:1981uc},
NUSEX~\cite{Battistoni:1985na}, FREJUS~\cite{Berger:1987ke}, SOUDAN~\cite{Thron:1989cd},
IMB~\cite{BeckerSzendy:1992hr} and Kamiokande~\cite{Hirata:1989kn}, have been done for
this purpose. Today the largest running experiment of this kind is
the Super-Kamiokande, a 50 kton (22.5 kton fiducial mass) water
Cherenkov detector located in Kamioka in Japan. Thanks to these
experiments, the lower bound on the lifetime of the proton has been
improved by many orders of magnitude in the past 60 years.

The two decay modes which have often been searched for are $p \to
\pi^{0}e^{+}$ and $p \to K^{+} \overline\nu$. The first one is
expected to be the leading mode in many GUTs, in particular in those
non-SUSY GUTs which typically predict the lifetime of the proton to
be about $10^{35}$ yrs. This decay mode has also been constrained to
the best degree of accuracy because of the high efficiency and
background rejection of the water Cherenkov detectors. The current
limit is $\tau(p \to \pi^{0}e^{+}) > 1.4\times10^{34}$ yrs at 90\%
C.L. from the Super-Kamiokande experiment~\cite{kearns:isoup}.

In comparison, the decay mode $p \to K^{+} \overline\nu$ is favored
by a number of SUSY GUTs which typically predict the lifetime of the
proton to be less than a few $\times ~10^{34}$ yrs. The search for
this mode in a large water Cherenkov detector is hindered by the
decay kinematics. The momentum of the $K^+$ in this two-body decay
is 339 MeV/c (kinetic energy of 105 MeV), which is below the
Cherenkov threshold in water. Therefore, both of the final-state
particles are invisible. Nonetheless, the daughter particles
originating from the two main decay channels of $K^{+}$ (i.e.,
$K^{+} \to \mu^+ \nu_{\mu}$ and $K^{+} \to \pi^+  \pi^0$) can be
reconstructed in a water Cherenkov detector. A photon-tagging
technique from the de-excitation of $^{15}$N, which is emitted from
the decay of a proton bound in $^{16}$O, can be further used to
increase the detection efficiency. But the overall efficiency
remains rather low in a water Cherenkov detector. Today's best limit
is $\tau(p \to K^{+} \bar\nu) > 5.9\times10^{33}$ yrs at 90\% C.L.
reported by the Super-Kamiokande collaboration~\cite{kearns:isoup}.

There are some other decay modes of the proton, which are also
interesting from a theoretical point of view. In some GUTs, for
example, the branching ratios of $p \to \mu^{+} \pi^0$ and $p \to
e^+ \eta$ are of the order of 10---20\%. Note that the lifetime of
bound neutrons is expected to be comparable with that of protons.
The 90\% C.L. limits on the nucleon lifetimes in various two-body
decay modes of the nucleons are graphically presented in
Fig.~\ref{fig:nd:pdk_limits}. Most of the most stringent limits are
achieved from the Super-Kamiokande
experiment~\cite{Nishino:2012ipa,Regis:2012sn,Abe:2013lua}.
%%%%%%%%%%%%%%%%%%%%%%%%%%%%% Figure 10-1 %%%%%%%%%%%%%%%%%%%%%%%%%%%%%
\begin{figure}[htb!]
\centering
\includegraphics[width=0.7\textwidth]{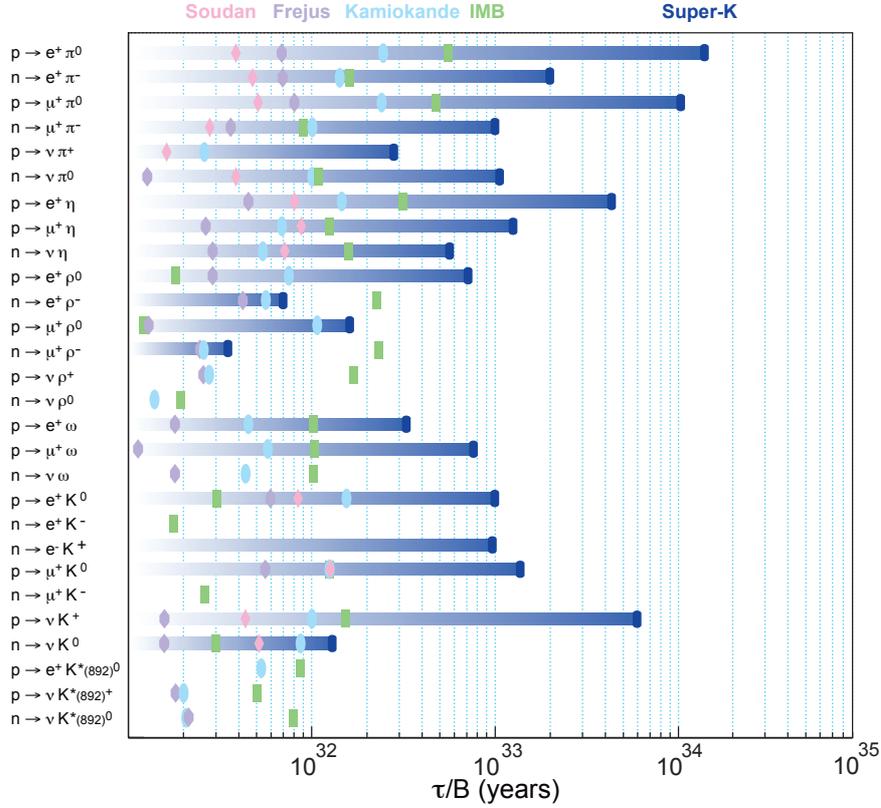}
\caption{The 90\% C.L. limits on the nucleon lifetimes in a variety
of two-body decay modes of the nucleons~\cite{kearns:isoup}. Today's
best limits are mostly from the Super-Kamiokande experiment using
the water Cherenkov detector.} \label{fig:nd:pdk_limits}
\end{figure}
%%%%%%%%%%%%%%%%%%%%%%%%%%%%%%%%%%%%%%%%%%%%%%%%%%%%%%%%%%%%%%%%%%%%%%%%

The Super-Kamiokande detector (22.5 kt) and the future
Hyper-Kamiokande detector (560 kt) are certainly larger than the
JUNO detector (20 kt) in mass. Compared with the water Cherenkov
detectors, however, the liquid scintillator detectors have the
unique advantages in detecting some proton decay modes, in
particular $p \to K^{+} \overline\nu$~\cite{Undagoitia:2005uu}. The tagging
efficiency for the proton decay can be largely improved due to the
large scintillation signal created by the $K^{+}$ itself, which is
invisible in a water Cherenkov detector. For most other decay modes,
the liquid scintillator does not provide any immediate advantages
over water in the aspects of the signal efficiency and background. Thus we
shall mainly focus on the $p \to K^{+} \overline\nu$ channel in the
following.

\subsection{Detection principle in liquid scintillator}
\label{subsec:nd:det}
% \begin{itemize}
%     \item Energy resolution, vertex resolution, pulse readout and PMT timing resolution
%     \item Signature of K, pi, mu, etc.
%     \item Signature of $p \to K^+ + \bar\nu$ and other channels
%     \item Impact of nuclear effects???
%     \item A comparison with water Cherenkov and LAr
% \end{itemize}

Let us focus on the $p \to K^{+} \overline\nu$ channel. The protons
in the JUNO detector are from both hydrogen and carbon nuclei. Using
the Daya Bay liquid scintillator as a reference, the H-to-C molar
ratio is 1.639. For a 20 kt fiducial volume detector, the number of
protons from hydrogen (free protons) is $1.45\times10^{33}$, and
that from carbon (bound protons) is $5.30\times10^{33}$.

$p \to K^{+} \overline\nu$ is a two-body decay. If the decaying
proton is from hydrogen, then it decays at rest. In this case the
kinetic energy of $K^+$ is fixed to be 105 MeV, which yields a
prompt signal in the liquid scintillator. The $K^{+}$ meson has a
lifetime of 12.4 nanoseconds and can quickly decay via the following
major channels:
\begin{itemize}
\item $K^{+} \to \mu^+ \nu_{\mu} \quad (63.43\%)$,
\item $K^{+} \to \pi^+ \pi^0  \quad (21.13\%)$,
\item $K^{+} \to \pi^+ \pi^+ \pi^-  \quad (5.58\%)$,
\item $K^{+} \to \pi^0 e^+ \nu_e  \quad (4.87\%)$,
\item $K^{+} \to \pi^+ \pi^0 \pi^0  \quad (1.73\%)$.
\end{itemize}
We mainly consider the two most important decay modes: $K^{+} \to
\mu^+ \nu_{\mu}$ and $K^{+} \to \pi^+  \pi^0$. In either case there
is a shortly delayed ($\sim$12~ns) signal from the daughter
particle(s). If the $K^+$ meson decays into $\mu^+ \nu_{\mu}$, the
delayed signal comes from $\mu^+$, which has a fixed kinetic energy
of 152 MeV as required by kinematics. Then the decay $\mu^+ \to e^+
\nu_e \overline\nu_{\mu}$ happens about 2.2~$\mu s$ later, leading
to the third long-delayed signal with a well-known (Michel electron)
energy spectrum. If the $K^+$ meson decays into $\pi^+ \pi^0$, the
$\pi^+$ deposits its kinetic energy (108 MeV) and the $\pi^0$
instantaneously decays into two gamma rays with the sum of the
energies equal to the total energy of $\pi^0$ (246 MeV).
The delayed signal includes all of the aforementioned deposited
energies. Then the $\pi^+$ meson decays primarily into $\mu^+
\nu_{\mu}$. The $\mu^+$ itself has very low kinetic energy (4.1
MeV), but it decays into $e^+ \nu_e \overline\nu_{\mu}$ about
2.2~$\mu s$ later, yielding the third long-delayed decay positron
signal. The simulated hit time distribution of a $K^{+} \to \mu^+
\nu_{\mu}$ event is shown in Figure.~\ref{fig:nd:pdk_event}, which
displays a clear three-fold coincidence.
%%%%%%%%%%%%%%%%%%%%%%%%%%%%%%%%%%%%%%%%%%%
\begin{figure}[htb]
\centering
\includegraphics[width=0.7\textwidth]{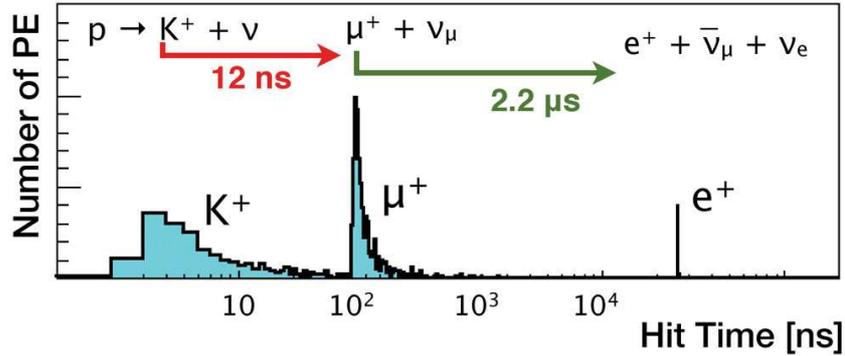}
\caption{The simulated hit time distribution of photoelectrons (PEs)
from a $K^{+} \to \mu^+ \nu_{\mu}$ event at JUNO.}
\label{fig:nd:pdk_event}
\end{figure}
%%%%%%%%%%%%%%%%%%%%%%%%%%%%%%%%%%%%%%%%%%%%

If a proton decays in a carbon nucleus, the nuclear effects have
to be taken into account. In particular, the binding energy and Fermi
motion modify the decaying proton's effective mass and momentum,
leading to a change of the kinematics of the decay process. In
Ref.~\cite{Undagoitia:2005uu}, the limiting values for the ranges of the
kinetic energy of $K^+$ are calculated to be 25.1---198.8 MeV for
protons in the s-state and 30.0---207.2 MeV for protons in the p-state. The
$K^+$ meson may also rescatter inside the nucleus, producing the intranuclear
cascades. This possibility has been discussed in Ref.~\cite{Stefan:2008zi}.

In summary, the signatures of $p \to K^{+} \overline\nu$ in the
JUNO experiment are:
\begin{itemize}
\item A prompt signal from $K^+$ and a delayed signal from
its decay daughters with a time coincidence of 12 ns.

\item Both the prompt and delayed signals have well-defined
energies.

\item There is one and only one decay positron with a time coincidence
of 2.2 $\mu$s from the prompt signals.
\end{itemize}
The time coincidence and well-defined energies provide a powerful
tool to reject the background, which is crucial in the search for
proton decays.

\subsection{Signal selection and efficiency}
\label{subsec:nd:sig}
% \begin{itemize}
%     \item Time cut efficiency
%     \item Energy cut efficiency
%     \item $p \to K^+ + \bar\nu$ efficiency
%     \item Other channels
% \end{itemize}

Because the decay time of $K^+$ is very short (i.e., about 12 ns), the
signal pulses from $K^+$ and from its daughter particles (e.g. $\mu^+$) will
typically be in fast sequence or even partially overlapped. A fast
response and high resolution waveform digitizer (e.g. a Giga-Hz flash
ADC) is required to separate the prompt and delayed signals. Even
so, if the $K^+$ decays early, the prompt and delayed pulses may mix
together and will resemble a single muon event from the
quasi-elastic interactions of atmospheric muon neutrinos.

The combined photon hit time distribution recorded by all the PMTs can
be used to distinguish a two-pulse event (the signal of proton decays)
from a one-pulse event (the atmospheric muon neutrino background).
The hit time distribution depends on the scintillation light time profile,
light yield, light attenuation, PMT time jitter and
electronic signal response. Here we use Ref.~\cite{Undagoitia:2005uu} as a
baseline for the efficiency estimation. In Ref.~\cite{Undagoitia:2005uu} a pulse
shape discrimination criterion ($\Delta T_{15\%-85\%}$) is
constructed using the time difference between the moments for which
the pulse reaches 15\% and 85\% of the maximum pulse height. $\Delta
T_{15\%-85\%}$ is typically longer for a two-pulse signal-like event
than a one-pulse background-like event. A cut at $\Delta
T_{15\%-85\%} > 7$ ns retains 65\% of the signals while rejecting
99.995\% of the muon background induced by the quasi-elastic
interactions of atmospheric neutrinos.

Scaled from the Super-Kamiokande detector~\cite{Hayato:1999az},
the upper limit on the charged-current interaction rate of atmospheric
muon neutrinos in the related energy range is estimated to be $4.8 \times 10^{-2}~
{\rm MeV}^{-1}{\rm kt}^{-1}{\rm y}^{-1}$. Further background rejection can be
achieved by an energy cut on the two shortly correlated events, i.e.
the sum of the deposited energies ($E_{\rm sum}$) from $K^{+}$ and the
daughter particles of its decay. For $K^{+} \to \mu^+ \nu_{\mu}$,
$E_{\rm sum}$ includes $E_{K^+}$ (105 MeV) and $E_{\mu^+}$ (152 MeV),
a total of 257 MeV. For $K^{+} \to \pi^+ \pi^0$, $E_{\rm sum}$ includes
$E_{K^+}$ (105 MeV), $E_{\pi^+}$ (108 MeV) and E$_{\pi^0 \to \gamma\gamma}$
(246 MeV), a total of 459 MeV. When defining the visible energy
window, one has to take into account that the light output is
quenched, especially for heavy ionizing particles. The quenching
mechanism can be reasonably well described by the Birk's formula:
\begin{eqnarray}
\left. \frac{{\rm d} E}{{\rm d} x} \right|^{\rm dep} =
\frac{{\rm d} E}{{\rm d}x} \left(1 + k^{}_B \cdot
\frac{{\rm d} E}{{\rm d}x}\right)^{-1} \; ,
%     (1)
\end{eqnarray}
where $k_B$ is around 0.01~cm/MeV for typical liquid scintillators.
A cut of $150 ~\mathrm{MeV}<E_{\rm sum}<650 ~\mathrm{MeV}$ retains 99\%
of the signals~\cite{Undagoitia:2005uu}. In a more realistic simulation, the hadronic
interactions will reduce the energy resolution and slightly decrease
the energy cut efficiency. The background rate from the quasi-elastic
interactions of atmospheric muon neutrinos in this 500 MeV energy window
is estimated to be at most 0.024 yr$^{-1}$ by
taking into account the previously described pulse shape
discrimination.

In the proton decay $p \to K^{+} \overline\nu$, there will be only a single
decay positron from the decay of $\mu^+$ --- the descendant of $K^{+}$. This
is in contrast with most of the hadronic processes from atmospheric
neutrinos that produce kaons, which are the potential backgrounds of
the proton decay. Apart from kaons, there are some other hadrons produced
from these processes, such as $\Lambda^{0}$. By
requiring precisely one single decay positron, we can efficiently
reject the backgrounds producing more than one decay positron. We
require $\mu^+$ to decay later than 100 ns to prevent its light leaking
into $E_{\rm sum}$, also to separate it from the daughter positron. We
further require the energy of the decay positron to be larger than 1
MeV to reduce the contribution from random coincidences with the
radioactive background events. The efficiency of the decay positron
cut is estimated to be 99\%. The total signal efficiency by combining
the pulse shape cut, energy cut and decay positron cut, is estimated to be 64\%.

\subsection{Background estimation}
\label{subsec:nd:bkgd}
% \begin{itemize}
%     \item muon produced by atmospheric neutrino
%     \item kaon produced by atmospheric neutrino
%     \item cosmic-ray muon caused background
% \end{itemize}

In the search of a rare event, such as the proton decay, the understanding of
the background is crucial. JUNO is located deep underground with an
overburden of $\sim$1900 m.w.e, thus the cosmic-muon-related
background is largely suppressed. The muon rate in the detector is
about 3 Hz with the average muon energy of about 215~GeV. The outer muon
system can efficiently veto muons which are in the energy range that
can mimic the events of $p \to K^{+} \overline\nu$, as well as any
short-lived particles and isotopes produced by the muon spallation. For the
long-lived isotopes, their beta decays are in the low energy region
($<20$ MeV) and do not cause any background to the proton decay. On the other
hand, neutral particles (e.g., neutrons, neutral kaons and $\Lambda^0$
particles produced by muons outside the veto system) can penetrate
the scintillator before being tagged. These neutral particles
have energy spectra extending from a few MeV to a few GeV~\cite{Bueno:2007um}.
Fast neutrons will be thermalized inside the detector and captured on
hydrogen within 200 $\mu$s. The capture process emits a 2.2 MeV gamma ray and can
be used to veto fast neutrons. The lack of the decay $e^+$ signal
further reduces the fast neutron background to a negligible level.
The production of $\Lambda^0$ is about $10^{-4}$ relative to the
production of neutrons~\cite{Bueno:2007um} therefore is negligible. Finally, $K^0_L$ could mimic a signal
event either through the charge exchange and converting to a $K^+$ in the
detector, or via its decay (e.g., $K^0_L \to \pi^+ e^- \overline\nu_e$ and
$K^0_L \to \pi^- e^+ \nu_e$) with
the right amount of deposited energy. The production rate of $K^0_L$
is estimated to be about 1:500 relative to that of the neutron in
Ref.~\cite{Bueno:2007um}. However, due to the passive shielding from the water
pool, we expect a stronger suppression of the $K^0_L$ (and $\Lambda^0$) background as
compared with the estimate given in Ref.~\cite{Bueno:2007um}.

This leaves atmospheric neutrinos as the main background for the
proton decay search. The background from $\nu_{\mu} n \to \mu^- p$
and $\overline\nu_{\mu} p \to \mu^+ n$ has been discussed in the previous
subsection. Such a background can be largely rejected by combining the
prompt energy cut and the pulse shape cut. The background
rate from the quasi-elastic interactions of atmospheric muon neutrinos is estimated to be
0.024 yr$^{-1}$. The charged-current single pion production has
smaller cross sections and can be further suppressed by the decay
positron cut, and hence it is negligible.

Now let us consider the neutrino-triggered production of strange
particles, in particular with the $K^+$ mesons in the final state. The first
category is relevant for the processes in which the strangeness is
conserved (i.e., $\Delta S = 0$). The typical process of this kind
is $\nu_{\mu} n \to \mu^- K^+ \Lambda^0$. In such
processes a $\Lambda^0$ particle is usually produced. The $\Lambda^0$ decays shortly ($\tau_{\Lambda^0} = 0.26$ ns) primarily via
$\Lambda^0 \to p \pi^-$ and $\Lambda^0 \to n \pi^0$. The decay daughters of $\Lambda^0$ combined with $\mu^-, K^+$ result in a larger prompt signal than that
expected from a true proton decay. These events can be further rejected because there are
three or more decay electrons or positrons from the decay chains. The second
category is associated with the processes in which the strangeness
is not conserved. The representative process of this kind
is $\nu_{\mu} p \to \mu^- K^+ p$. Although the cross sections of
such processes are Cabbibo-suppressed as compared with those $\Delta S = 0$
reactions, they are the potential background events in case the prompt
energy is similar to the expectation from the proton decay and one of
the decay electrons is missed due to the low pulse height or the overlap
with the prompt signal. Such a background can be largely removed if
sophisticated particle ID algorithms can be developed to distinguish
a muon-like event from a kaon-like event in the scintillator (e.g.,
using the Cherenkov and range information). Here
we simply scale the result obtained in Ref.~\cite{Undagoitia:2005uu}.
The estimated background in
this channel that produces strange particles in the final state is  0.026 yr$^{-1}$. The total background, combining the
quasi-elastic channel, is estimated to be 0.05 yr$^{-1}$.

\subsection{Sensitivity}
\label{subsec:nd:sens}
% \begin{itemize}
%     \item Nominal sensitivity versus time
%     \item Sensitivity versus time resolution (maybe light yield)
%     \item Comparison with LBNE and SuperK (HyperK)
% \end{itemize}

The sensitivity to the proton lifetime can be expressed as
\begin{eqnarray}
\tau (p \to K^+ + \overline\nu) = N_p T R \frac{\epsilon}{S} \; ,
%     (2)
\end{eqnarray}
where $N_p = 6.75\times10^{33}$ is the total number of protons,
$T$ is the measurement time which is assumed to be 10 years,
$R = 84.5\%$ is the branching ratio of the $K^+$ decays
being included in the analysis, $\epsilon = 65\%$ is the total
signal efficiency, and $S$ is the upper limit on the
number of signal events at a certain confidence interval which
depends on the number of observed events and the expected
number of background events. We follow the Feldman-Cousins
approach~\cite{Feldman:1997qc} in our analysis. The expected background
is 0.5 events in 10 years. If no event is observed, the 90\% C.L.
upper limit is $S = 1.94$. The corresponding sensitivity to
the proton lifetime is $\tau >
1.9\times10^{34}$ yrs. This represents ``a factor of three"
improvement over today's best limit from the
Super-Kamiokande experiment~\cite{kearns:isoup}, and starts to approach
the region of interest predicted by some GUT models \cite{Nath:2006ut}.

In a realistic experiment, the sensitivity may decrease if there is an
upward fluctuation of the background. In the case that one event is
observed (30\% probability), the 90\% C.L. upper limit is $S = 3.86$.
The corresponding sensitivity to the proton lifetime is $\tau >
9.6\times10^{33}$ yrs. If two events are observed (7.6\%
probability), the sensitivity is further reduced to $\tau >
6.8\times10^{33}$ yrs.
%%%%%%%%%%%%%%%%%%%%%%%%%%%%%%%%%%%%%%%%%%%%%%%%%%%%%%%%%%%%%%%
\begin{figure}[htb]
\centering
\includegraphics[width=0.7\textwidth]{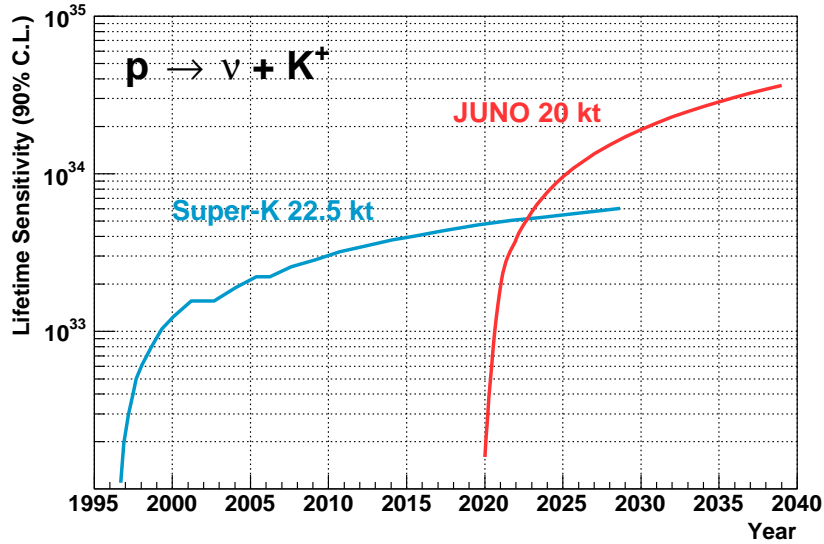}
\caption{The 90\% C.L. sensitivity to the proton lifetime in the decay
mode $p \to K^+ + \overline\nu$ at JUNO as a function of time. In
comparison, Super-Kamiokande's sensitivity is also projected to year
2030.} \label{fig:nd:pdk_sens}
\end{figure}
%%%%%%%%%%%%%%%%%%%%%%%%%%%%%%%%%%%%%%%%%%%%%%%%%%%%%%%%%%%%%%%%

In Figure.~\ref{fig:nd:pdk_sens} we plot the 90\% C.L. sensitivity
to the proton lifetime in the decay mode $p \to K^+ + \overline\nu$ at JUNO
as a function of time. Due to the high efficiency in measuring this
mode, JUNO's sensitivity will surpass Super-Kamiokande's in only 3
years since its data taking.

\clearpage

\section{Neutrinos from Dark Matter}
\label{sec:idm}

\blfootnote{Editor: Guey-Lin Lin (glin@mail.nctu.edu.tw)}
\blfootnote{Major contributors: Wanlei Guo, Yen-Hsun Lin, and Yufeng Zhou}

\subsection{Introduction}
% Evidence for DM
The existence of non-baryonic dark matter (DM) in the Universe has been
well established by astronomical observations.
% Galaxy: Rotation curve
For most spiral galaxies,
the rotation curve of stars or gas far from the galactic center
does not decline with increasing distance
but rather stays as a  constant.
This strongly indicates the existence of a massive dark halo
which envelops the galactic disk and extends well beyond the size of
the visible part of the galaxy
\cite{Rubin:1970zza}.
% Cluster: X-rays and lensing
At cluster scales,
the mass of a cluster determined by
the radial velocity distribution
\cite{Zwicky:1933gu,Zwicky:1937zza},
or the gravitational lensing of background galaxies in the cluster
\cite{Hoekstra:2002nf},
or
the measurements of the X-ray temperature of hot gas in the cluster,
is typically an order of magnitude larger than that inferred from
the visible  part of the cluster.
The observation of the bullet cluster (1E0657-558)
\cite{Clowe:2006eq}
and the large-scale structure filament
in superclusters (Abell~222/223)
\cite{Dietrich:2012mp}
also strongly support the existence of DM.

% CMB
The most accurate determination of the DM energy density comes from
the measurement of the anisotropy of the cosmic microwave
background.
The recent measurement gives \cite{Ade:2013zuv}
\begin{equation}
\Omega_{\textrm{DM}} h^{2}=0.1198\pm 0.0026 \; ,
%     (11.1)
\end{equation}
where $h = 0.673 \pm 0.012$ is the Hubble constant.
The contribution from baryonic matter is $\Omega_{\rm b}
h^{2}=0.02207\pm 0.00027$ \cite{Ade:2013zuv}.
Thus DM contributes to nearly $85\%$ of the total mass in the Universe.
%

%% DM and SM %%
The DM candidate must be stable on he cosmological time scale,
and must only have very weak interactions with ordinary matter and
electromagnetic radiation.
The N-body simulations suggest that DM particles should be cold (or
non-relativistic) at the time of galaxy formation.
The Standard Model (SM) of particle physics, although extremely
successful in explaining the data, can not provide a viable cold DM
candidate.
In the particle content of the SM, only the neutrinos are
electromagnetically neutral and interact with matter very weakly.
However, the relic density of the neutrinos is too low today:
$\Omega_{\nu} h^{2}\leq 0.0062$ at $95\%$ CL \cite{Ade:2013zuv}, and
they could only serve as hot (or relativistic) DM particles in the
early Universe.
Therefore, the existence of cold DM is also a striking indication of
new physics beyond the SM.

%% WIMPS %%
A widely studied class of DM candidates are the Weakly Interacting
Massive Particles (WIMPs).
The masses of WIMPs are in the range of ${\cal O}(1)$ GeV to ${\cal
O}(1)$ TeV, and their interaction strengths with the SM particles
are around the weak interaction  scale.
Such WIMPs can naturally fit the observed DM density through
decoupling from the thermal equilibrium with electromagnetic
radiation fields, which is due to the expansion and cooling of the early
Universe.
Well-motivated WIMP candidates exist in the supersymmetric (SUSY)
extensions of the SM.
If the $R$-parity is conserved, the lightest neutral supersymmetric
particle (LSP) is stable and can be a cold DM candidate.
The LSP can be the lightest neutralino or sneutrino, and the latter
is stringently constrained by the direct detection experiments.
% UED
In models with universal extra dimensions, the Kaluza-Klein (KK)
excitations of the SM particles have odd KK-parity while the SM
particles are all KK-even.
Thus the lightest neutral KK excitation particle (LKP) is stable and
can be a WIMP.
A possible DM candidate of this kind is the first KK excitation of
the hypercharge field $B^{(1)}$.
% Little Higgs
In a variety of the little Higgs models, a new discrete symmetry $T$
is introduced to avoid the constraints from the electroweak
precision measurements.
The lightest $T$-odd particle is also a viable WIMP candidate.
Besides the well-motivated WIMP candidates, one can construct
``minimal'' DM models, such as the SM singlet scalar models, singlet
fermion models, etc.
The motivation of such DM models does not seem to be compelling,
but they can have rich phenomenology.
%
%% Others %%
Non-WIMP candidates, such as the primordial black holes, axions and keV
sterile neutrinos, also exist and attract some attentions~\cite{Feng:2010gw}.
Gravitinos and axionos, the respective supersymmetric partners of
gravitons and axions, can be the cold DM candidates with the relic
density obtained through the nonthermal processes.

DM can be detected either directly or indirectly. The former is to
observe the nucleus recoil as DM interacts with the target nuclei in
the detector, and the latter is to detect the final-state particles
resulting from DM annihilation or decays. A direct detection of DM
is possible because the DM particles constantly bombard the Earth as
the Earth sweeps through the local halos.
Some of the direct detection experiments have observed the
preliminary excesses of candidate events over known backgrounds,
such as DAMA~\cite{Bernabei:2010mq},
CoGeNT~\cite{Aalseth:2010vx,Aalseth:2011wp,Aalseth:2014eft},
CRESS-II~\cite{Angloher:2011uu}, and CDMS-Si~\cite{Agnese:2013rvf}.
In contrast, other experiments have only obtained stringent upper
limits on DM, such as those from
XENON100~\cite{Aprile:2012nq,Aprile:2013doa},
LUX~\cite{Akerib:2013tjd}, superCDMS~\cite{Agnese:2014aze},
SIMPLE~\cite{Felizardo:2011uw}, and CDEX~\cite{Yue:2014qdu}.

DM can be indirectly detected  by measuring the cosmic-ray particle
fluxes which may receive extra contributions from DM annihilation or
decays in the galactic halo.
The recent measurements by PAMELA~\cite{Adriani:2008zr},
ATIC-2~\cite{Chang:2008aa}, Fermi-LAT~\cite{Abdo:2009zk}, and
AMS-02~\cite{Aguilar:2013qda} have indicated an excess in the
fraction of cosmic-ray positrons, but whether it points to DM or not
remains an open question.

DM can also be detected indirectly by looking for the neutrino
signature from DM annihilation or decays in the galactic halo, the
Sun or the Earth. In particular, the search for the DM-induced
neutrino signature from the Sun has given quite tight constraints on
the spin-dependent (SD) DM-proton scattering cross section $\sigma^{\rm
SD}_{\chi p}$~\cite{Aartsen:2012kia}.
In the following we shall study the sensitivities of the JUNO detector
to $\sigma^{\rm SD}_{\chi p}$ and spin-independent (SI) DM-proton scattering cross section $\sigma^{\rm SI}_{\chi p}$ by
focusing on the neutrino signature from the Sun as well.

\subsection{The neutrino flux from DM annihilation in the Sun}

To facilitate our discussions, let us define ${\rm d}
N^f_{\nu^{}_\alpha}/{\rm d} E_\nu $ as the energy spectrum of
$\nu^{}_\alpha$ (for $\alpha = e, \mu, \tau$) produced  per DM
annihilation $\chi\chi\to f\bar{f}$ in the Sun. The differential
neutrino flux of flavor $\beta$ (for $\beta = e, \mu, \tau$)
arriving at the terrestrial neutrino detector is
\begin{equation}
\frac{{\rm d} \Phi^{\textrm{DM}}_{\nu^{}_\beta}}{{\rm d} E_\nu} =
P_{\nu^{}_\alpha\to \nu^{}_\beta}(E_\nu, D) \frac{\Gamma_A}{4\pi
D^2} \sum_{f} B^{f}_{\chi}\frac{{\rm d} N^f_{\nu^{}_\alpha}}{{\rm d}
E_\nu} \label{eq:dNdE} \; ,
%     (11.2)
\end{equation}
where $\Gamma_A$ is the DM annihilation rate, $B^{f}_{\chi}$ is the
branching ratio for the DM annihilation channel $\chi\chi\to
f\bar{f}$, $D$ is the distance between the source and detector, and
$P_{\nu^{}_\alpha\to \nu^{}_\beta}(E_\nu, D)$ is the neutrino
oscillation probability from the source to the detector.
$P_{\nu^{}_\alpha\to \nu^{}_\beta}(E_\nu, D)$ can be calculated with
the best fit neutrino oscillation parameters  given
by Ref.~\cite{Capozzi:2013csa}. The DM
annihilation rate $\Gamma_A$ can be determined by the following
argument. When the Sun sweeps through the DM halo, a WIMP could
collide with matter inside the Sun and lose its speed. If the WIMP
speed becomes less than its escape velocity, the WIMP can be
captured by the Sun's gravitational force and then sinks into the
core of the Sun. After a long period of accumulation, WIMPs inside
the Sun can begin to annihilate into the SM particles at an
appreciable rate. Among the final states of DM annihilation,
neutrinos can be detected by a neutrino telescope, such as the JUNO
detector. Defining $N(t)$ as the number of WIMPs in the Sun at time
$t$, we have
\begin{equation}
\frac{{\rm d} N}{{\rm d} t} = C_c - C_a N^2  -C_e N
\label{eq:x2} \; ,
%     (11.3)
\end{equation}
where $C_c$ is the capture rate, $C_a N^2$ is the annihilation rate,
and $C_e N$ is the evaporation rate. It has been shown that  WIMPs
with masses below $3$ to $4$ GeV may evaporate from the
Sun~\cite{Krauss:1985aaa,Griest:1986yu,Nauenberg:1986em,Gould:1987ju}.
The critical mass scale for WIMP evaporation may be raised by about
$1$ GeV if the WIMP self-interaction is taken into
account~\cite{Chen:2014oaa}. In this case Eq.~(\ref{eq:x2}) should
be modified by including the self-interaction effect. Here we shall
not address the issue of DM self-interaction.

Taking the nominal values of $C_c$, $C_a$ and $C_e$, one can show that the total DM number in the Sun has already reached to the equilibrium, i.e.,
the right hand side of Eq.~(\ref{eq:x2}) vanishes..
For the mass range in which DM evaporation is negligible (i.e.,
$C_e\to 0$), it is easy to see that $\Gamma_A(t)=C_c/2$, i.e., the DM
annihilation rate in this case only depends on the capture rate
$C_c$. Since $C_c$ depends on the DM-nucleon scattering cross
section and the chemical composition of the Sun, the constraint on
the SD cross section $\sigma^{\rm SD}_{\chi p}$ and that
on the SI cross section $\sigma^{\rm SI}_{\chi p}$ can
be extracted by searching for the DM-induced neutrinos from the Sun.

For spin-dependent (SD) interactions, the
capture rate is given by~\cite{Jungman:1995df,Bertone:2004pz}
\begin{eqnarray}
C_{c}^{\rm SD}&\simeq& 3.35\times10^{24}\textrm{
s}^{-1}\left(\frac{\rho^{}_{0}}{0.3\textrm{
GeV/cm}^{3}}\right)\left(\frac{270\textrm{
km/s}}{\bar{v}}\right)^{3}\left(\frac{\textrm{GeV}}
{m_{\chi}}\right)^{2} \nonumber \\
&\times &\left(\frac{\sigma_{\rm H}^{\rm SD}}{10^{-6}
\textrm{ pb}}\right) \; , \label{eq:capture_SD}
%     (11.8)
\end{eqnarray}
where $\rho^{}_{0}$ is the local DM density, $\bar{v}$ is the
velocity dispersion, $\sigma_{\rm H}^{\rm SD}\equiv \sigma_{\chi p}^{\rm SD}$ is the SD
DM-hydrogen scattering cross section, and $m_{\chi}$ is the DM mass.
The capture rate for spin-independent (SI)  scattering behaves like
~\cite{Jungman:1995df,Bertone:2004pz}
\begin{eqnarray}
C_c\propto \frac{\rho_{0}}{\rm{GeV}\,cm^{-3}} \times
\frac{\rm km\,s^{-1}}{\bar{v}}\times
\frac{\rm{GeV}}{m_{\chi}}\times
\frac{\sigma^{\rm SI}_{\chi p}}{\rm{pb}}\times \sum_{A_i} F_{A_i}^{*}(m_\chi),
\label{eq:capture_SI}
\end{eqnarray}
with $A_i$ the atomic number of chemical element $i$ in the Sun,
$F_{A_i}^{*}(m_\chi)$ is the product of various factors
including the mass fraction of element $i$,
the gravitational potential for element $i$,
kinematic suppression factor, form factor, and a factor
of reduced mass.

The DM evaporation rate in the Sun, $C_e$, has been well
investigated in Refs.~\cite{Griest:1986yu,Gould:1987ju}. Here we do not discuss it further since
we shall focus on the DM mass range in which the neutrino flux from the Sun is not suppressed.

\subsection{The sensitivities of the JUNO detector}

In general, DM inside the Sun can annihilate into leptons, quarks
and gauge bosons. The neutrino flux results from the decays of such
final-state particles. Here we consider two annihilation channels
$\chi\chi\to \tau^+\tau^-$ and $\chi\chi\to \nu\bar{\nu}$  as a
benchmark. We shall focus on the detection of muon-neutrino flux from the above two annihilation channels.
Hence we consider track events induced by charged current
interactions of $\nu_{\mu}$ and $\bar{\nu}_{\mu}$. The differential muon-neutrino flux arriving at the
terrestrial neutrino detector can be calculated with the help of
Eq.~(\ref{eq:dNdE}). For simplicity, we set the branching fraction
$B^{\tau,\nu}_{\chi}=1$. We employ \texttt{WimpSim}
\cite{Blennow:2007tw} with a total of 50,000 Monte Carlo generated
events. The muon-neutrino event rate in the detector reads
\begin{equation}
N_{\rm DM}=\int_{E_{\rm th}}^{m_{\chi}}\frac{{\rm d} \Phi_{\nu}^{\rm DM}}{ {\rm
d} E_{\nu}}A_{\nu}(E_{\nu}) \ {\rm d} E_{\nu} \ {\rm d} \Omega \; ,
\label{eq:nu_event}
%     (11.13)
\end{equation}
where $E_{\rm{th}}$ is the detector's threshold energy, ${\rm d}
\Phi_{\nu}^{\rm DM}/{\rm d} E_{\nu}$ is the muon-neutrino flux from DM
annihilation, $A_{\nu}$ is the detector's effective area, and
$\Omega$ is the solid angle.
The angular resolution of muon-neutrino track event has been presented in Chapter~7.  It is shown there that, to reconstruct the muon direction in a good precision,
one requires the muon track length to be greater than $5$ m within the
JUNO detector. In such a condition, the angular resolution of the muon track is better than $1^{\circ}$.
This sets the lower limit of $\nu_{\mu}$ ($\bar{\nu}_{\mu}$) energy to be about $1$ GeV. Hence we set $E_{\rm th}=1$ GeV. The effective area of the JUNO
detector is expressed as
\begin{equation}
A_{\nu}(E_{\nu})=M_{\rm{LS}}\left(\frac{N_{A}}{m}\right)
\left[n_{p}\sigma_{\nu p}(E_{\nu})+n_{n}\sigma_{\nu
n}(E_{\nu})\right]\epsilon (E_{\nu}) \; ,
%     (11.14)
\end{equation}
where $M_{\rm{LS}}$ is the target mass of the JUNO detector, $N_{A}$
is the Avogadro constant, $m$ is the mass of liquid scintillator per
mole, $n_{p}$ and $n_n$ are the number of protons and neutrons, respectively, in a liquid
scintillator molecule, $\sigma_{\nu p}$ and $\sigma_{\nu n}$ are the neutrino-proton
and neutrino-neutron scattering cross sections, respectively, and $\epsilon(E_{\nu})$ is the energy-dependent efficiency for selecting
muon events with track length greater than $5$ m in the detector. This efficiency has been studied in Chap. 7 where $\epsilon$ is found to be less than $1\%$ for $E_{\nu}=1$ GeV and
it is greater than $70\%$ for $E_{\nu}> 5$ GeV.  We plot the efficiency $\epsilon(E_{\nu})$ in Fig.~\ref{fig:efficiency}.
The selected track events are classified as fully contained (FC) and partially contained (PC) events. PC events begin to dominate FC events for $E_{\nu}>5$ GeV.
At $E_{\nu}=20$ GeV, $N_{\rm PC}/N_{\rm FC}> 10$.
%%%%%%%%%%%%%%%%%%%%%%%%%   F   I   G   U   R   E   %%%%%%%%%%%%%%%%%%%%%%%%%%%%
\begin{figure}[htb]
\begin{center}
\includegraphics[width=0.7\textwidth]{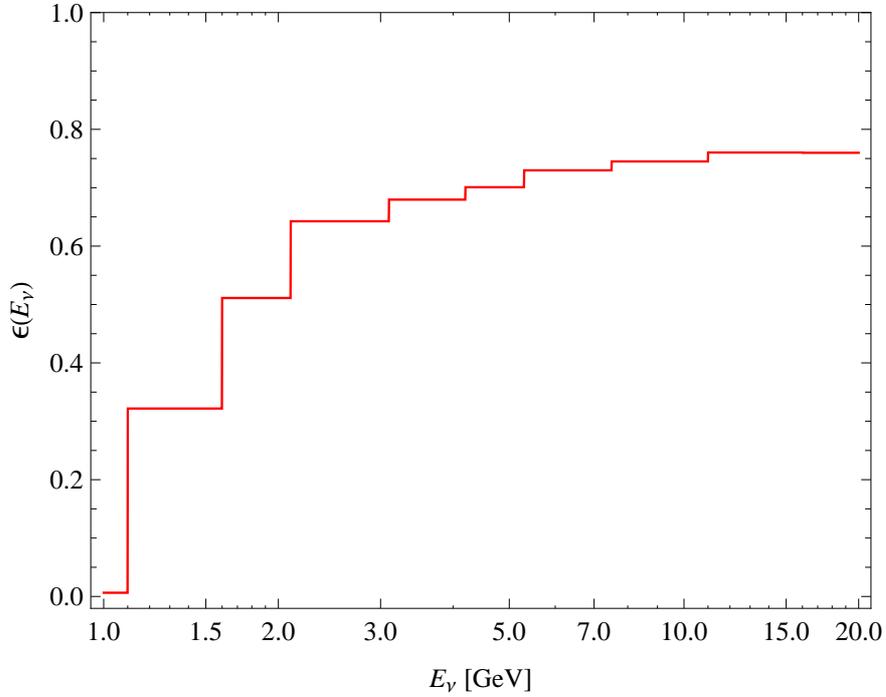}
\end{center}
\caption{The energy-dependent efficiency $\epsilon(E_{\nu})$ for selecting
muon events with track length greater than $5$ m in the detector. \label{fig:efficiency}}
\end{figure}
%%%%%%%%%%%%%%%%%%%%%%%%%%%%%%%%%%%%%%%%%%%%%%%%%%%%%%%%%%%%%%%%%%%%%%%%%%%%%%%%

The effective area for anti-neutrino interactions can be calculated by the same equation
with the replacements $\sigma_{\nu p}\to \sigma_{\bar{\nu} p}$ and $\sigma_{\nu n}\to \sigma_{\bar{\nu} n}$.
 We use the results in~\cite{Andreopoulos:2009rq}
for the above cross sections in our interested energy range $1\textrm{ GeV}\leq E_{\nu}\leq 20\textrm{ GeV}$. Here we take
$\rho=0.86$~g/cm$^3$ for the density of liquid scintillator which
has been used in the Daya Bay experiment. The mass fractions of
carbon and hydrogen in the liquid scintillator are taken to be
$88\%$ and $12\%$, respectively. These fractions are also quoted
from those used in the Daya Bay experiment. As the neutrinos
propagate from the source to the detector, they encounter the
high-density medium in the Sun, the vacuum in space, and the Earth
medium. The matter effects on neutrino oscillations have been
considered in $P_{\nu^{}_\alpha \to \nu^{}_\beta}$ in
Eq.~(\ref{eq:dNdE}).

The atmospheric background event rate can also be calculated by
using Eq.~(\ref{eq:nu_event}) with ${\rm d} \Phi_{\nu}^{\rm DM}/{\rm d}
E_{\nu}$ replaced by the atmospheric neutrino flux. Hence
\begin{equation}
N_{\rm{atm}}=\int_{E_{\rm{th}}}^{E_{\rm{max}}}\frac{{\rm d}
\Phi_{\nu}^{\rm{atm}}}{{\rm d} E_{\nu}}A_{\nu}(E_{\nu}) \ {\rm d}
E_{\nu} \ {\rm d} \Omega \; . \label{eq:atm_event}
%     (11.16)
\end{equation}
In our calculations the atmospheric neutrino flux ${\rm d}
\Phi_{\nu}^{\rm{atm}}/{\rm d} E_{\nu}$ is quoted from
Refs.~\cite{Aartsen:2012uu,Honda:2015fha}. We set
$E_{\rm{max}}=m_{\chi}$ so as to compare with the DM signal. The
threshold energy $E_{\rm{th}}$ is taken to be $1$ GeV.

To estimate the JUNO sensitivity to the DM-induced neutrino flux
from the Sun, we focus on events coming from a chosen solid angle range surrounding the direction of the Sun.
The solid angle range $\Delta \Omega$ is related to the half angle of the observation cone by
$\Delta{\Omega}=2\pi(1-\cos\psi)$. Due to the spreading of muon direction relative to the direction of original $\nu_{\mu}$, which has been extensively studied in Chap.7, there is an efficiency factor corresponding to each choice of $\psi$ for collecting the neutrino events from the Sun. This collection efficiency is plotted in Fig.~\ref{fig:collection}.
%%%%%%%%%%%%%%%%%%%%%%%%%   F   I   G   U   R   E   %%%%%%%%%%%%%%%%%%%%%%%%%%%%
\begin{figure}[htb]
\centering
\includegraphics[width=0.7\textwidth]{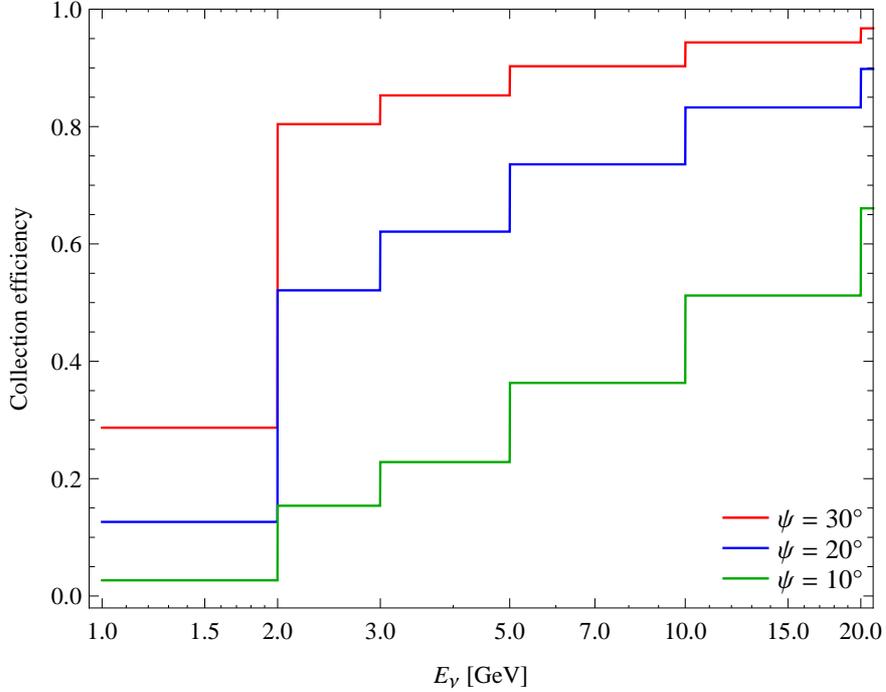}
\caption{ The efficiency for collecting DM-induced neutrino events from the Sun with the cone half angle $\psi=10^{\circ}$, $20^{\circ}$ and $30^{\circ}$. \label{fig:collection}}
\end{figure}
%%%%%%%%%%%%%%%%%%%%%%%%%%%%%%%%%%%%%%%%%%%%%%%%%%%%%%%%%%%%%%%%%%%%%%%%%%%%%%%
We choose $\psi=30^{\circ}$ for our sensitivity calculation since the collection efficiency is greater than $80\%$
for the relevant DM mass range, $m_{\chi}> 3$ GeV. The DM-induced neutrino event rate given by Eq.~(\ref{eq:nu_event}) is then corrected by the collection
efficiency. The sensitivities to $\sigma^{\rm SD}_{\chi p}$ and $\sigma^{\rm
SI}_{\chi p}$ are taken to be the $2\sigma$ significance for 5 years
of data taking. We extract the sensitivities using
\begin{equation}
\frac{s}{\sqrt{s+b}}=2.0 \; ,
\label{eq:convention_eq}
%     (11.17)
\end{equation}
%%%%%%%%%%%%%%%%%%%%%%%%%   F   I   G   U   R   E   %%%%%%%%%%%%%%%%%%%%%%%%%%%%
\begin{figure}[htb]
\centering
\includegraphics[width=0.7\textwidth]{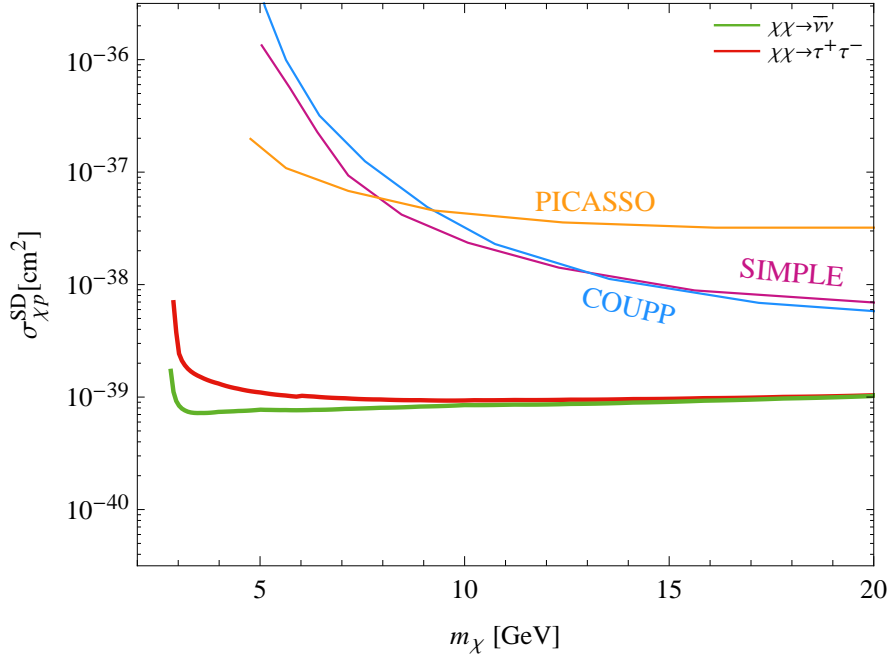}
\caption{ The JUNO $2\sigma$ sensitivity in 5 years to the spin-dependent
cross section $\sigma^{\rm SD}_{\chi p}$ in 5 years. The
constraints from the direct detection experiments are also
shown for comparison.\label{fig:SD}}
\end{figure}
%%%%%%%%%%%%%%%%%%%%%%%%%%%%%%%%%%%%%%%%%%%%%%%%%%%%%%%%%%%%%%%%%%%%%%%%%%%%%%%%
%%%%%%%%%%%%%%%%%%%%%%%%%   F   I   G   U   R   E   %%%%%%%%%%%%%%%%%%%%%%%%%%%%
\begin{figure}[htb]
\centering
\includegraphics[width=0.7\textwidth]{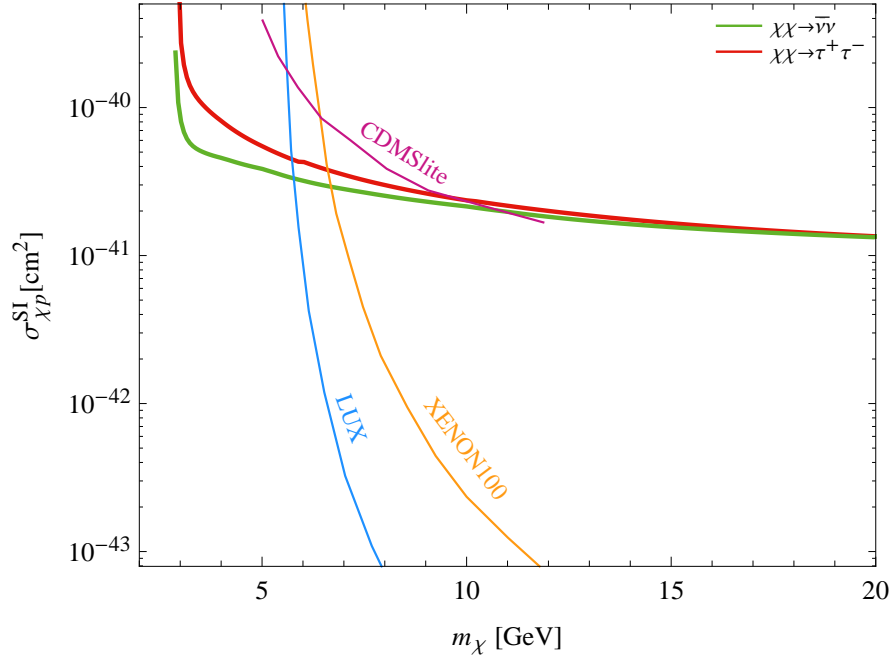}
\caption{ The JUNO $2\sigma$ sensitivity in 5 years to the spin-independent
cross section $\sigma^{\rm SI}_{\chi p}$.  The
recent constraints from the direct detection experiments are also
shown for comparison. \label{fig:SI}}
\end{figure}
%%%%%%%%%%%%%%%%%%%%%%%%%%%%%%%%%%%%%%%%%%%%%%%%%%%%%%%%%%%%%%%%%%%%%%%%%%%%%%%%
where $s$ is the DM signal, $b$ is the atmospheric background, and
$2.0$ refers to the $2\sigma$ detection significance. The
results of our sensitivity calculations are given in Fig.~\ref{fig:SD} and Fig.~\ref{fig:SI}.
It is clearly seen that the JUNO sensitivities to $\sigma^{\rm SD, SI}_{\chi p}$ through $\chi\chi\to \nu\bar{\nu}$ channel
are essentially identical to those through $\chi\chi\to \tau^+\tau^-$ channel. This follows from our definition of sensitivity
as the excess of total neutrino event rate integrated from $E_{\rm th}$ to $m_{\chi}$. If we consider the spectral feature of $\chi\chi\to \nu\bar{\nu}$ annihilation channel,
the JUNO sensitivities to $\sigma^{\rm SD, SI}_{\chi p}$ can be significantly improved.
In Fig.~\ref{fig:SD} one can see that the JUNO sensitivity to
$\sigma^{\rm SD}_{\chi p}$ is much better than the current direct
detection constraints set by COUPP~\cite{Behnke:2012ys},
SIMPLE~\cite{Felizardo:2011uw} and PICASSO~\cite{Archambault:2012pm}
experiments. For $m_{\chi}< 3$ GeV, the sensitivity
becomes poor due to the DM evaporation from the Sun.
Fig.~\ref{fig:SI} shows that the direct detection experiments, such
as XENON100~\cite{Aprile:2012nq}, LUX~\cite{Akerib:2013tjd} and
CDMSlite~\cite{Agnese:2013jaa}, can set very stringent constraints
on $\sigma^{\rm SI}_{\chi p}$ for $m_{\chi}> 7$ GeV. On the other
hand, JUNO is more sensitive than the direct detection experiments
for $m_{\chi} < 7$  GeV. This advantage continues until the
evaporation mass scale.

\subsection{Summary}

In this chapter we first reviewed the developments of dark
matter physics and the status of dark matter detection search. We then discussed the
sensitivities of the JUNO detector to spin-dependent DM-nucleon
scattering cross section $\sigma^{\rm SD}_{\chi p}$ and the
spin-independent one $\sigma^{\rm SI}_{\chi p}$ by searching for
DM-induced neutrino events from the core of the Sun. The sensitivities
are calculated based upon the excess of neutrino events beyond the
atmospheric background. We focus on muon neutrino events resulting from DM
annihilation channels $\chi\chi\to \tau^+\tau^-$ and $\chi\chi\to \nu\bar{\nu}$. We select those events with the muon track length greater than 5 m within the detector.  In such a case, the direction of the muon track can be reconstructed with an accuracy better than $1^{\circ}$. To search for DM-induced neutrino events from the Sun, we choose an observation cone
with a cone half angle $\psi=30^{\circ}$.  We have found that JUNO sensitivity to
$\sigma^{\rm SD}_{\chi p}$ is much better than the current direct
detection constraints. In the case of $\sigma^{\rm SI}_{\chi p}$, JUNO is competitive with direct detection experiments
for $m_{\chi}< 7$ GeV.

\clearpage

\renewcommand{\baselinestretch}{1.15}
\def\thefootnote{\fnsymbol{footnote}}

\newcommand{\bq}{\begin{eqnarray}}
\newcommand{\nq}{\end{eqnarray}}
\newcommand{\nub}{\overline{\nu}}
\newcommand{\epsilont}{\tilde{\epsilon}}
\newcommand{\depsilon}{\delta\epsilon}
\newcommand{\dvarepsilon}{\delta\varepsilon}
\newcommand{\Ut}{\tilde{U}}
\newcommand{\dU}{\delta{U}}
\newcommand{\Nt}{\tilde{N}}
\newcommand{\dN}{\delta{N}}
\newcommand{\cost}{\tilde{c}}
\newcommand{\sint}{\tilde{s}}
\newcommand{\thetat}{\tilde{\theta}}
\newcommand{\deltat}{\tilde{\delta}}
\newcommand{\alphat}{\tilde{\alpha}}
\newcommand{\phit}{\tilde{\phi}}
\newcommand{\rhot}{\tilde{\rho}}
\newcommand{\sigmat}{\tilde{\sigma}}
\newcommand{\Pt}{\tilde{P}}
\newcommand{\eb}{\overline{e}}
\newcommand{\Deltat}{\tilde{\Delta}}
\newcommand{\Phit}{\tilde{\Phi}}
\newcommand{\At}{\tilde{A}}
\newcommand{\St}{\tilde{S}}

\newcommand{\alphab}{\overline{\alpha}}
\newcommand{\betab}{\overline{\beta}}
\newcommand{\gammab}{\overline{\gamma}}

\addtolength{\arraycolsep}{-3pt}

%\section{New Physics Searches}
\section{Exotic Searches with Neutrinos}
\label{sec:exotic}

\blfootnote{Editor: Yufeng Li (liyufeng@ihep.ac.cn)}

The standard three-neutrino mixing paradigm, characterized by two mass-squared differences ($\Delta m^2_{21}$ and $\Delta m^2_{31}$),
three mixing angles ($\theta_{12}$, $\theta_{13}$ and $\theta_{23}$) and one Dirac CP-violating phase $\delta$,
can describe most of the phenomena of solar, atmospheric, reactor, and accelerator neutrino oscillations \cite{Agashe:2014kda}. However, other new
physics mechanisms can operate at a sub-leading level, and may appear at the stage of precision measurements of neutrino oscillations.

Among different possibilities, the hypotheses of light sterile neutrinos, unitarity violation, non-standard interactions (NSIs), and Lorentz invariance violation (LIV),
are of particular importance at JUNO.
\begin{itemize}
\item The light sterile neutrinos in light of short baseline oscillation anomalies have attracted active research attention in both the experimental
searches and phenomenological implications. At JUNO, the hypothesis of light sterile neutrinos could be tested either using radioactive sources or with
reactor antineutrinos. This part is thoroughly explored in Sec.~\ref{sec:sterile}.
\item Unitarity violation in the lepton mixing matrix is the generic consequence of a large class of seesaw models, where the light active neutrinos are
mixed with heavy fermion degrees of freedom. Tests of unitarity violation have been discussed in Sec.~3.3 in the MUV framework.% of minimal unitarity violation (MUV). %\ref{sec:prec:uv}.
\item High-dimensional operators from the new physics contributions can affect the neutrino oscillation in the form of NSIs~\cite{Antusch:2008tz,Ohlsson:2012kf,Severijns:2006dr,Ohlsson:2013nna,Li:2014mlo},
which emerge as effective four fermion interactions after integrating out the heavy particles beyond the SM.
NSIs can modify both the neutrino production and detection processes, and induce shifts for both the mixing angles and mass-squared differences at reactor
antineutrino oscillations.
\item LIV is one of the most important evidence for Quantum Gravity \cite{Kostelecky:1988zi}. The low energy phenomena of LIV can be systematically
studied in the framework of the Standard Model Extension (SME) \cite{Colladay:1996iz,Colladay:1998fq}. In reactor antineutrino experiments, LIV can be tested in terms of both the
spectral distortion and sidereal variation effects\cite{Li:2014rya}.
\end{itemize}
In this section, we shall concentrate on the tests of NSIs and LIV at JUNO with reactor antineutrino oscillations. We shall also
give brief discussions on other aspects of exotic searches beyond the SM.

\subsection{Non-standard Interactions}
\label{subsec:exotic:NSI}

NSIs can occur in the neutrino production, propagation and detection processes in the experiments of neutrino oscillations.
The NSI effects in the propagation process are negligible because of the suppression of the neutrino energy and terrestrial matter density.
The neutrino and antineutrino states produced in the source and observed in the detector are superpositions of neutrino and antineutrino flavor states
\cite{Ohlsson:2013nna,Li:2014mlo}
%%%%%%%%%%%%%%%%%%%%%%%%%%%%%%%%%%%%%%%%
\bq &&|\nu^{\rm s}_\alpha\rangle = \frac{1}{N^{\rm s}_\alpha}
\Big(|\nu_\alpha\rangle+\sum_{\beta}
\epsilon^{\rm s}_{\alpha\beta}|\nu_\beta\rangle \Big),\qquad
|\nub^{\rm s}_\alpha\rangle = \frac{1}{N^{\rm s}_\alpha} \Big(|\nub_\alpha\rangle+\sum_{\beta}
\epsilon^{\rm s*}_{\alpha\beta}|\nub_\beta\rangle \Big), \nonumber\\
&&\langle\nu^{\rm d}_\beta| = \frac{1}{N^{\rm d}_\beta}
\Big(\langle\nu_\beta|+\sum_{\alpha}
\epsilon^{\rm d}_{\alpha\beta}\langle\nu_\alpha| \Big), \qquad
\langle\nub^{\rm d}_\beta| = \frac{1}{N^{\rm d}_\beta}
\Big(\langle\nub_\beta|+\sum_{\alpha}
\epsilon^{\rm d*}_{\alpha\beta}\langle\nub_\alpha| \Big),\nq
%%%%%%%%%%%%%%%%%%%%%%%%%%%%%%%%%%%%%%%%%%%%%%%%%%%%%%%%%%%%%%%%%%%%%
where $\epsilon^{\rm s}_{\alpha\beta}$ and $\epsilon^{\rm d}_{\alpha\beta}$ denote the NSI parameters at the source and
detector, respectively, and \bq &N^{\rm s}_\alpha=\sqrt{\sum_\beta
|\delta_{\alpha\beta}+\epsilon^{\rm s}_{\alpha\beta}|^2}, \qquad
N^{\rm d}_\beta=\sqrt{\sum_\alpha
|\delta_{\alpha\beta}+\epsilon^{\rm d}_{\alpha\beta}|^2}
\label{eq:normalize_factor_sd} \nq
are normalization factors.
In order to measure the average and difference between neutrino
production and detection processes, we introduce two sets of new NSI parameters as
%%%%%%%%%%%%%%%%%%%%%%%%%%%%%%%%%%%%%%%%%%%%%%%%%%%%%%%%%%%%%%%%%%%%%
\bq
&&\epsilont_{\alpha\beta}=(\epsilon^{\rm s}_{\alpha\beta}+\epsilon^{\rm d*}_{\beta\alpha})/2\,,\qquad
\depsilon_{\alpha\beta}=(\epsilon^{\rm s}_{\alpha\beta}-\epsilon^{\rm d*}_{\beta\alpha})/2\,,
\label{eq:definition} \nq
%%%%%%%%%%%%%%%%%%%%%%%%%%%%%%%%%%%%%%%%%%%%%%%%%%%%%%%%%%%%%%%%%%%%%
to rewrite the NSI effects.

Including the NSI effects at the source and detector, the amplitude of the $\nub_\alpha\to\nub_\alpha$ transition can be written as
%%%%%%%%%%%%%%%%%%%%%%%%%%%%%%%%%%%%%%%%%%%%%%%%%%%%%%%%%%%%%%%%%%%%%
\begin{eqnarray}
\tilde{\mathcal{A}}_{\alphab\alphab}&=&\langle \nub^{\rm d}_\alpha| e^{-iHL}
|\nub^{\rm s}_\alpha \rangle= \sum_{\beta\gamma}\langle
\nub^{\rm d}_\alpha|\nub_\gamma \rangle \langle \nub_\gamma| e^{-iHL}
|\nub_\beta \rangle \langle \nub_\beta|\nub^{\rm s}_\alpha \rangle
=(\delta_{\gamma\alpha}+\epsilon^{\rm d*}_{\gamma\alpha})
\mathcal{A}_{\betab\gammab}
(\delta_{\alpha\beta}+\epsilon^{\rm s*}_{\alpha\beta}),
\end{eqnarray}
%%%%%%%%%%%%%%%%%%%%%%%%%%%%%%%%%%%%%%%%%%%%%%%%%%%%%%%%%%%%%%%%%%%%%
where $L$ is the baseline, and
%%%%%%%%%%%%%%%%%%%%%%%%%%%%%%%%%%%%%%%%%%%%%%%%%%%%%%%%%%%%%%%%%%%%%
\bq \mathcal{A}_{\betab\gammab} =\langle \nub_\gamma| e^{-iHL}
|\nub_\beta \rangle = \sum_i U^*_{\gamma i}U_{\beta i}
\exp\left(-i\frac{m^2_iL}{2E}\right) \nq
%%%%%%%%%%%%%%%%%%%%%%%%%%%%%%%%%%%%%%%%%%%%%%%%%%%%%%%%%%%%%%%%%%%%%
is the amplitude of $\nub_\beta\to\nub_\gamma$ without NSIs. It is
useful to define
%%%%%%%%%%%%%%%%%%%%%%%%%%%%%%%%%%%%%%%%%%%%%%%%%%%%%%%%%%%%%%%%%%%%%
\bq \Ut_{\alpha i}=\frac{1}{\Nt_\alpha}
\sum_\beta(\delta_{\alpha\beta}+\epsilont_{\alpha\beta}^*)U_{\beta
i}, \qquad \dU_{\alpha i}=\frac{1}{\Nt_\alpha}
\sum_\beta\depsilon_{\alpha\beta}^*U_{\beta i},
\label{eq:effectivePMNS} \nq
%%%%%%%%%%%%%%%%%%%%%%%%%%%%%%%%%%%%%%%%%%%%%%%%%%%%%%%%%%%%%%%%%%%%%
where
%%%%%%%%%%%%%%%%%%%%%%%%%%%%%%%%%%%%%%%%%%%%%%%%%%%%%%%%%%%%%%%%%%%%%
\bq &\Nt_\alpha =\sqrt{\sum_\beta
|\delta_{\alpha\beta}+\epsilont_{\alpha\beta}|^2}=\sqrt{N^{\rm s}_\alpha
N^{\rm d}_\alpha} +\mathcal{O}(\depsilon^2)\,, \label{eq:normalize_factor}
\nq
%%%%%%%%%%%%%%%%%%%%%%%%%%%%%%%%%%%%%%%%%%%%%%%%%%%%%%%%%%%%%%%%%%%%%
and $\sum_{i}|\Ut_{\alpha i}|^2=1$ is required. Thus we can obtain
$\tilde{\mathcal{A}}_{\alphab\alphab}$ as
%%%%%%%%%%%%%%%%%%%%%%%%%%%%%%%%%%%%%%%%%%%%%%%%%%%%%%%%%%%%%%%%%%%%%
\bq \tilde{\mathcal{A}}_{\alphab\alphab}&=& \sum_i(\Ut-\dU)^*_{\alpha
i}(\Ut+\dU)_{\alpha i} \exp\left(-i\frac{m^2_iL}{2E}\right)
+\mathcal{O}(\depsilon^2)\,, \nq
%%%%%%%%%%%%%%%%%%%%%%%%%%%%%%%%%%%%%%%%%%%%%%%%%%%%%%%%%%%%%%%%%%%%%
and the survival probability for $\nub_\alpha\to\nub_\alpha$ is
expressed as
%%%%%%%%%%%%%%%%%%%%%%%%%%%%%%%%%%%%%%%%%%%%%%%%%%%%%%%%%%%%%%%%%%%%%
\bq \Pt_{\alphab\alphab} =|\tilde{\mathcal{A}}_{\alphab\alphab}|^2=
1-4\sum_{i<j}|\Ut_{\alpha i}|^2|\Ut_{\alpha j}|^2\ \left[
\sin^2\Delta_{ji} +{\rm Im}\left(\frac{\dU_{\alpha i}}{\Ut_{\alpha
i}}-\frac{\dU_{\alpha j}}{\Ut_{\alpha j}}\right) \sin2\Delta_{ji}
\right] +\mathcal{O}(\depsilon^2)\,.\nq
%%%%%%%%%%%%%%%%%%%%%%%%%%%%%%%%%%%%%%%%%%%%%%%%%%%%%%%%%%%%%%%%%%%%%
Due to the smallness of $\dU_{\alpha i}/\Ut_{\alpha i}$, we can
rewrite the above equation as
%%%%%%%%%%%%%%%%%%%%%%%%%%%%%%%%%%%%%%%%%%%%%%%%%%%%%%%%%%%%%%%%%%%%%
\bq \Pt_{\alphab\alphab} = 1-4\sum_{i<j}|\Ut_{\alpha
i}|^2|\Ut_{\alpha j}|^2\ \sin^2\Deltat_{ji}^\alpha
+\mathcal{O}(\depsilon^2) \label{eq:palphavaccumlike} \nq
%%%%%%%%%%%%%%%%%%%%%%%%%%%%%%%%%%%%%%%%%%%%%%%%%%%%%%%%%%%%%%%%%%%%%
with
%%%%%%%%%%%%%%%%%%%%%%%%%%%%%%%%%%%%%%%%%%%%%%%%%%%%%%%%%%%%%%%%%%%%%
\bq \Deltat_{ji}^\alpha=\Delta_{ji}+{\rm Im}\left(\frac{\dU_{\alpha
i}}{\Ut_{\alpha i}}-\frac{\dU_{\alpha j}}{\Ut_{\alpha j}}\right) .
\nq
%%%%%%%%%%%%%%%%%%%%%%%%%%%%%%%%%%%%%%%%%%%%%%%%%%%%%%%%%%%%%%%%%%%%%
For the effective mass-squared differences we have
%%%%%%%%%%%%%%%%%%%%%%%%%%%%%%%%%%%%%%%%%%%%%%%%%%%%%%%%%%%%%%%%%%%%%
\bq \Delta \tilde{m}^{2\alpha}_{ji}({E}/{L})=\Delta m^2_{ji}+{\rm
Im}\left(\frac{\dU_{\alpha i}}{\Ut_{\alpha i}}-\frac{\dU_{\alpha
j}}{\Ut_{\alpha j}}\right)\frac{4E}{L}\,, \nq
%%%%%%%%%%%%%%%%%%%%%%%%%%%%%%%%%%%%%%%%%%%%%%%%%%%%%%%%%%%%%%%%%%%%%
which is an energy/baseline- and flavor-dependent effective quantity.

With Eq.~(\ref{eq:palphavaccumlike}) we have obtained a
standard-like expression for the antineutrino survival probability in
vacuum in the presence of NSIs. The corresponding NSI effects are
encoded in the effective mass and mixing parameters. The average
parts induce constant shifts for the neutrino mixing elements, and
the difference parts generate energy and baseline dependent
corrections to the mass-squared differences.

In reactor antineutrino oscillations, only the electron antineutrino
survival probability is relevant because of the high threshold of
the $\mu$/$\tau$ production. We can rewrite $\Pt_{\eb\eb}$ with
these effective mixing parameters as
%%%%%%%%%%%%%%%%%%%%%%%%%%%%%%%%%%%%%%%%%%%%%%%%%%%%%%%%%%%%%%%%%%%%%
\bq
\Pt_{\eb\eb}=1&-&\cost^4_{13}\sin^22\thetat_{12}[\sin^2\Delta_{21}+(\dvarepsilon_1-\dvarepsilon_2)\sin2\Delta_{21}] \nonumber\\
&-&\cost^2_{12}\sin^22\thetat_{13}[\sin^2\Delta_{31}+(\dvarepsilon_1-\dvarepsilon_3)\sin2\Delta_{31}]\nonumber\\
&-&\sint^2_{12}\sin^22\thetat_{13}[\sin^2\Delta_{32}+(\dvarepsilon_2-\dvarepsilon_3)\sin2\Delta_{32}]
+\mathcal{O}(\depsilon^2) \nonumber\\
=1&-&\cost^4_{13}\sin^22\thetat_{12}\sin^2\Deltat_{21}
-\cost^2_{12}\sin^22\thetat_{13}\sin^2\Deltat_{31}\nonumber\\
&-&\sint^2_{12}\sin^22\thetat_{13}\sin^2\Deltat_{32}
+\mathcal{O}(\dvarepsilon^2)\,, \label{eq:reactorprobability} \nq
%%%%%%%%%%%%%%%%%%%%%%%%%%%%%%%%%%%%%%%%%%%%%%%%%%%%%%%%%%%%%%%%%%%%%
with
%%%%%%%%%%%%%%%%%%%%%%%%%%%%%%%%%%%%%%%%%%%%%%%%%%%%%%%%%%%%%%%%%%%%%
\bq
&&\Deltat_{ji}=\Delta_{ji}+\dvarepsilon_i-\dvarepsilon_j, \nonumber\\
&&\dvarepsilon_1=\frac{{\rm
Im}(\dU_{e1})}{\cost_{12}\cost_{13}},\quad \dvarepsilon_2=\frac{{\rm
Im}(\dU_{e2})}{\sint_{12}\cost_{13}},\quad \dvarepsilon_3=\frac{{\rm
Im}(\dU_{e3})}{\sint_{13}}\,,
\label{eq:epsilon_definition}
\nq
%%%%%%%%%%%%%%%%%%%%%%%%%%%%%%%%%%%%%%%%%%%%%%%%%%%%%%%%%%%%%%%%%%%%%
where the superscript $\alpha=e$ in $\Deltat_{ji}^e$ has been
ignored, and the effective mixing angles $\thetat_{13}$ and $\thetat_{12}$ are defined through
$\sint_{13}=|\Ut_{e3}|$ and $\sint_{12}=|\Ut_{e2}|/\cost_{13}$. Correspondingly we have
$|\Ut_{e1}|=\cost_{12}\cost_{13}$ by using the normalization relation
$|\Ut_{e1}|^2+|\Ut_{e2}|^2+|\Ut_{e3}|^2=1$.

The average part $\epsilont$ can be treated as constant shifts to
mixing angles $\theta_{12}$ and $\theta_{13}$, and the difference
part $\depsilon$ leads to energy- and baseline-dependent shifts to
the mass-squared differences $\Delta m^2_{ji}$\, as
%%%%%%%%%%%%%%%%%%%%%%%%%%%%%%%%%%%%%%%%%%%%%%%%%%%%%%%%%%%%%%%%%%%%%
\bq \Delta \tilde{m}^2_{ji}(E/L)=\Delta
m^2_{ji}+(\dvarepsilon_i-\dvarepsilon_j)4E/L\,.
\label{eq:mass_shift} \nq
%%%%%%%%%%%%%%%%%%%%%%%%%%%%%%%%%%%%%%%%%%%%%%%%%%%%%%%%%%%%%%%%%%%%%
However, only two combinations of the three parameters
$\dvarepsilon_i$ contribute to the oscillation probability thanks to
the relation
$(\dvarepsilon_2-\dvarepsilon_3)=(\dvarepsilon_1-\dvarepsilon_3)-(\dvarepsilon_1-\dvarepsilon_2)$.
It is notable that one cannot distinguish the effects of mixing angle
shifts from the scenario of three neutrino mixing using reactor
antineutrino oscillation experiments. On the other hand, the shifts of mass-squared
differences are clearly observable due to the baseline- and
energy-dependent corrections in the reactor antineutrino spectrum.

%%%%%%%%%%%%%%%%%
\begin{figure}%[h!]
\begin{center}
\includegraphics[width=0.45\textwidth]{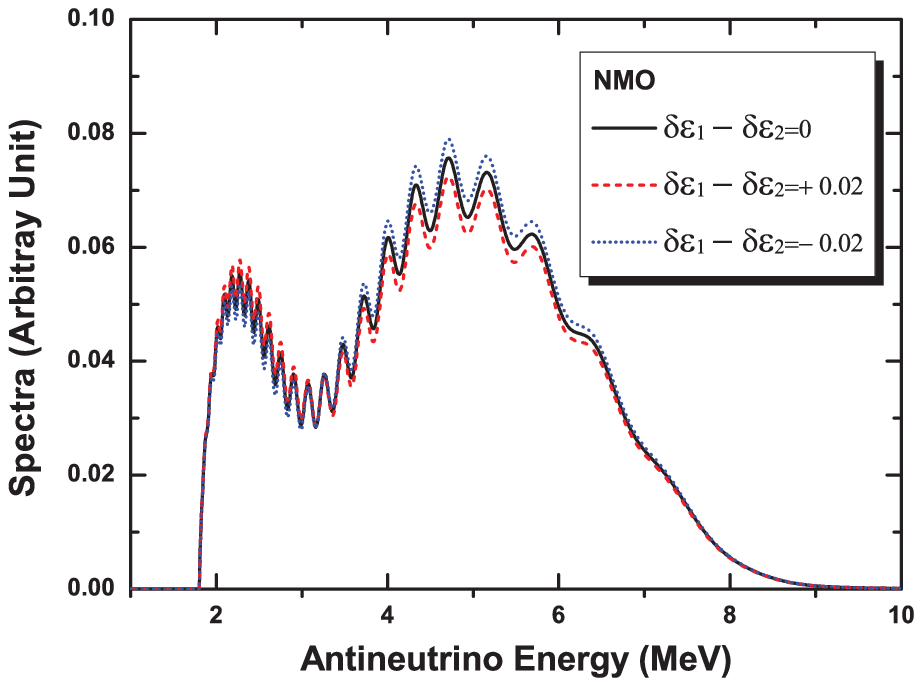}
\includegraphics[width=0.45\textwidth]{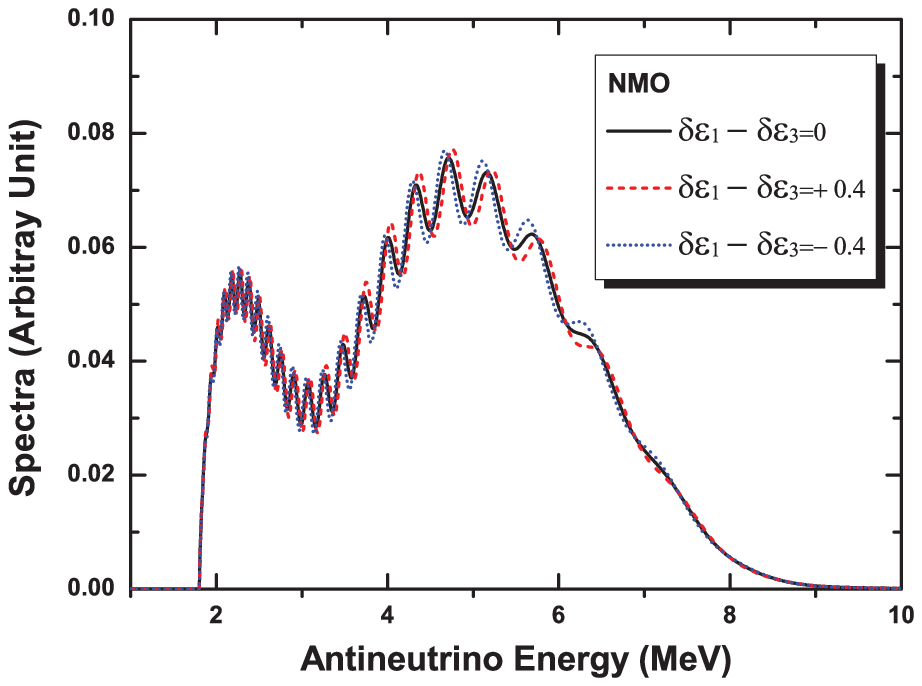}
\caption{The effects of NSIs in reactor $\nub_e$
spectra at a baseline of 52.5 km \cite{Li:2014mlo}. For visualization,  we set
$\dvarepsilon_1-\dvarepsilon_2=0,\pm0.02$ in the upper panel and
$\dvarepsilon_1-\dvarepsilon_3=0,\pm0.4$ in the lower panel.
$\dvarepsilon_1-\dvarepsilon_3$ is fixed at zero in the upper panel,
and $\dvarepsilon_1-\dvarepsilon_2$ is fixed at zero in the lower
panel. The NH is assumed for illustration.}
\label{fig:spectrum}
\end{center}
\end{figure}
%%%%%%%%%%%%%%%%%
Following the nominal setup defined in the MH sensitivity study, we numerically show the NSI effects at JUNO. We
illustrate how the NSI effects shift the mixing angles and mass-squared differences,
how it influences the mass ordering measurement and to what extent
we can constrain the NSI parameters. The relevant oscillation
parameters are $\thetat_{12}$, $\thetat_{13}$, $\Delta m^2_{21}$ and
$\Delta m^2_{ee}$ and the NSI parameters are $\dvarepsilon_1-\dvarepsilon_2$, $\dvarepsilon_1-\dvarepsilon_3$.
We directly employ the mixing angles measured in recent reactor antineutrino experiments as our effective mixing angles, which can
be shown as
%%%%%%%%%%%%%%%%%%%%%%%%%%%%%%%%%%%%%%%%%%%%%%%%%%%%%%%%%%%%%%%%%%%%%
\bq
\sin^22\thetat_{13}=\sin^22\theta_{13}^{\rm{D}}=0.084, \quad
\tan^2\thetat_{12}=\tan^2\theta_{12}^{\rm{K}}=0.481\,,
\label{eq:datamixing}
\nq
%%%%%%%%%%%%%%%%%%%%%%%%%%%%%%%%%%%%%%%%%%%%%%%%%%%%%%%%%%%%%%%%%%%%%
where the measured mixing angles $\theta_{13}^{\rm{D}}$ and
$\theta_{12}^{\rm{K}}$ are from Daya Bay~\cite{An:2015rpe} and
KamLAND~\cite{Gando:2013nba}, respectively.
On the other hand, because the current uncertainties of mass-squared differences from Daya Bay and KamLAND are much larger than the NSI corrections,
we simply take the measured mass-squared differences as the true parameters,
%%%%%%%%%%%%%%%%%%%%%%%%%%%%%%%%%%%%%%%%%%%%%%%%%%%%%%%%%%%%%%%%%%%%%
\bq
\Delta m^2_{ee}&=&\Delta m^{2\,\rm{D}}_{ee}=2.44 \times 10^{-3}\,\rm{eV}^2\,,\\
\Delta m^2_{21}&=&\Delta m^{2\,\rm{K}}_{21}=7.54 \times 10^{-5}\,\rm{eV}^2. 
\label{eq:datamass} 
\nq
%%%%%%%%%%%%%%%%%%%%%%%%%%%%%%%%%%%%%%%%%%%%%%%%%%%%%%%%%%%%%%%%%%%%%
We show the effects of NSIs in the reactor $\nub_e$ spectra at a baseline of 52.5 km in Fig.~\ref{fig:spectrum}, where the influences
of $\dvarepsilon_1-\dvarepsilon_2$ and $\dvarepsilon_1-\dvarepsilon_3$ are presented in the upper panel
and lower panel, respectively.
%The scenario of standard three-neutrino oscillations with $\dU=0$ is also shown for comparison.
In the upper panel, we fix $\dvarepsilon_1-\dvarepsilon_3=0$ and find that non-zero
$\dvarepsilon_1-\dvarepsilon_2$ introduces the spectral distortion to the slow oscillation term $\Delta_{21}$. For
$\dvarepsilon_1-\dvarepsilon_2=0.02$, the spectrum is suppressed in the high energy region with $E>3$ MeV and enhanced for the low
energy range 2 MeV $<E<3$ MeV. In comparison, negative $\dvarepsilon_1-\dvarepsilon_2$ gives the opposite effect on the
spectrum distortion. In the lower panel, we set $\dvarepsilon_1-\dvarepsilon_2=0$ and observe that
$\dvarepsilon_1-\dvarepsilon_3$ can affect the spectral distribution for the fast oscillation term $\Delta_{31}$. The non-trivial NSI effects
will contribute a small phase advancement or retardance to the fast oscillation depending on the sign of $\dvarepsilon_1-\dvarepsilon_3$.

\begin{figure}%[!h]
\includegraphics[width=0.45\textwidth]{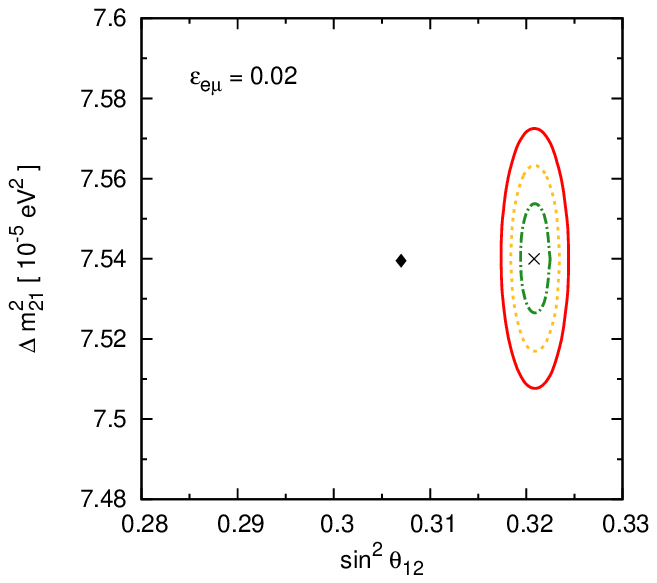}
\includegraphics[width=0.45\textwidth]{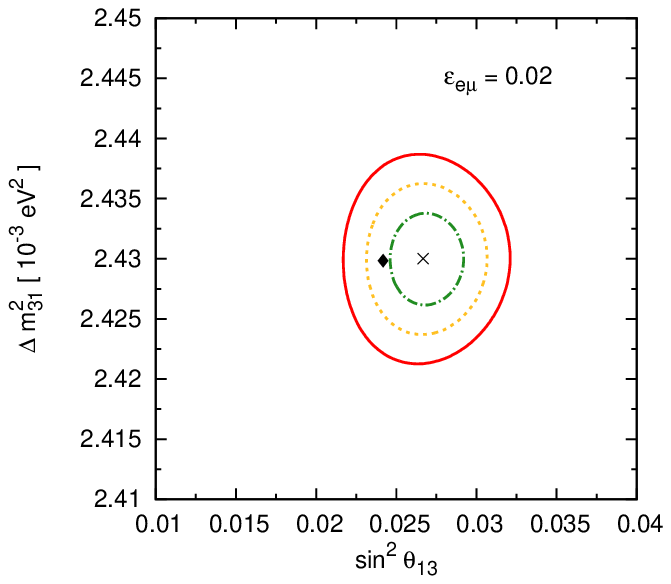}
\caption{\label{fig:thetashift} The shifts of mixing angles $\sin^2\theta_{12}$ (left) and $\sin^2\theta_{13}$ (right)
induced by NSIs in fitting ($\theta_{12}$, $\Delta m^2_{21}$) and
($\theta_{13}$, $\Delta m^2_{31}$) to the simulated data \cite{Ohlsson:2013nna}. The NSI effects are neglected in the fitting process. The black diamonds indicate the true values,
whereas the crosses correspond to the extracted parameters. The dotted-dashed (green), dotted (yellow), and solid (red) curves stand for the 1$\sigma$, 2$\sigma$, and 3$\sigma$ C.L., respectively.
}
\end{figure}
%%%%%%%%%%%%%%%%%
\begin{figure}%[ht!]
\includegraphics[width=0.45\textwidth]{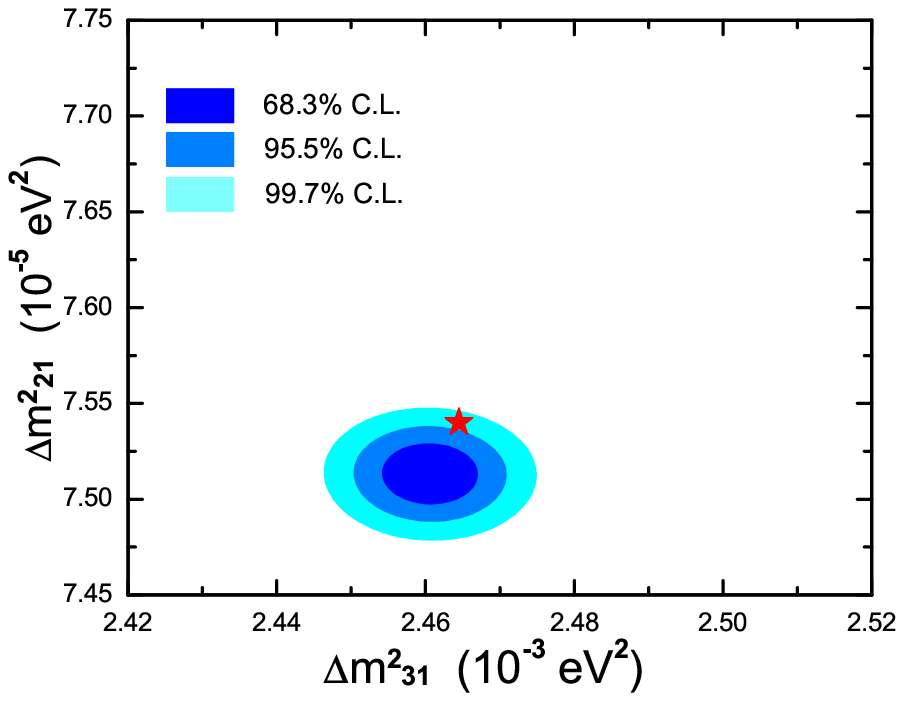}
\includegraphics[width=0.45\textwidth]{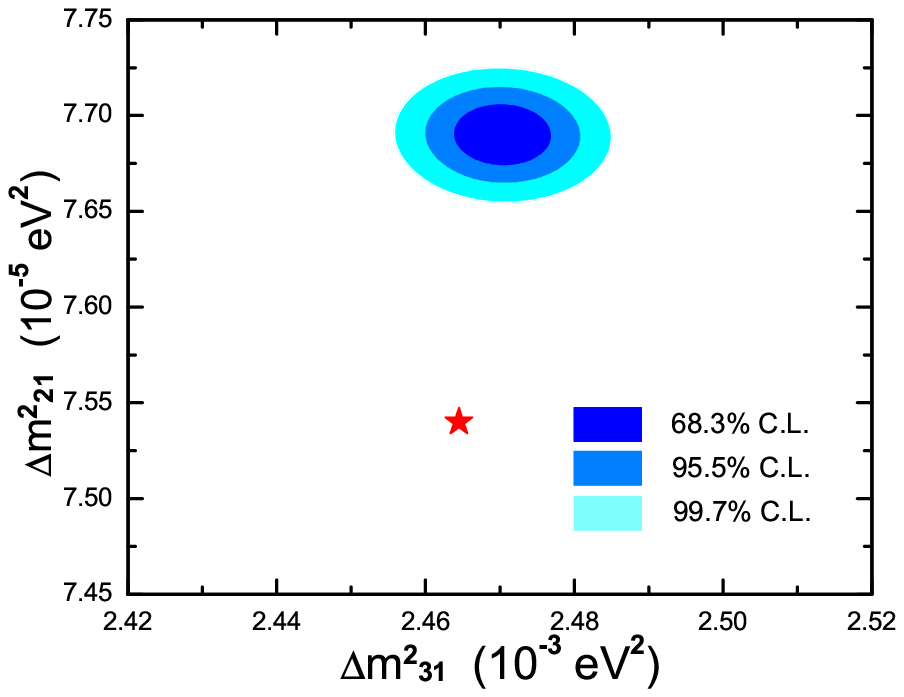}
\caption{The shifts of mass-squared differences induced by NSIs in fitting ($\Delta m^2_{21}$, $\Delta m^2_{31}$) to the simulated data.
The left and right panels are illustrated for $\dvarepsilon_1-\dvarepsilon_2=-0.54\,\dU,\,\dvarepsilon_1-\dvarepsilon_3=-5.60\,\dU$ and
$\dvarepsilon_1-\dvarepsilon_2=+3.00\,\dU,\,\dvarepsilon_1-\dvarepsilon_3=+8.06\,\dU$ respectively, with $\dU=0.01$ \cite{Li:2014mlo}.
The NSI effects are neglected in the fitting process.}\label{fig:massshift}
\end{figure}
%%%%%%%%%%%%%%%%%
The NSI-induced shifts of mixing angles and mass squared differences are presented in Figs.~\ref{fig:thetashift} and~\ref{fig:massshift} respectively,
where the true scenario is the generic scenario with the NSI effects, and the best-fit values of $\theta_{12}$, $\theta_{13}$, $\Delta m^2_{21}$ and $\Delta m^2_{31}$ are
obtained from the minimization process without the NSI effects. One can observe from Fig.~\ref{fig:thetashift} that the average part (i.e., $\varepsilon_{e\mu}=0.02$)
of NSIs introduces constant shifts of $\theta_{12}$ and $\theta_{13}$ \cite{Ohlsson:2013nna}.
On the other hand, from Fig.~\ref{fig:massshift}, the difference part of NSIs induce
the shifts to $\Delta m^2_{21}$ and $\Delta m^2_{31}$, where the left and right panels are illustrated for
$\dvarepsilon_1-\dvarepsilon_2=-0.54\,\dU,\,\dvarepsilon_1-\dvarepsilon_3=-5.60\,\dU$ and
$\dvarepsilon_1-\dvarepsilon_2=+3.00\,\dU,\,\dvarepsilon_1-\dvarepsilon_3=+8.06\,\dU$ respectively, with $\dU=0.01$ \cite{Li:2014mlo}.

Next we shall discuss the NSI effects on the MH measurement. The iso-$\Delta \chi^2$ contours are illustrated in Fig.~\ref{fig:mh_iso}
for the MH sensitivity as a function of two effective NSI parameters
$\dvarepsilon_1-\dvarepsilon_2$ and $\dvarepsilon_1-\dvarepsilon_3$.
The NH is assumed for illustration. We can learn from the figure that the smaller $\dvarepsilon_1-\dvarepsilon_2$ and larger
$\dvarepsilon_1-\dvarepsilon_3$ will reduce the possibility of the
MH measurement. If $\dvarepsilon_1-\dvarepsilon_2$ decreases by 0.03 or $\dvarepsilon_1-\dvarepsilon_3$ increases by 0.05, $\Delta
\chi^2$ will be suppressed by 2 units.
%%%%%%%%%%%%%%%%%
\begin{figure}%[h!]
\begin{center}
\includegraphics[width=0.5\textwidth]{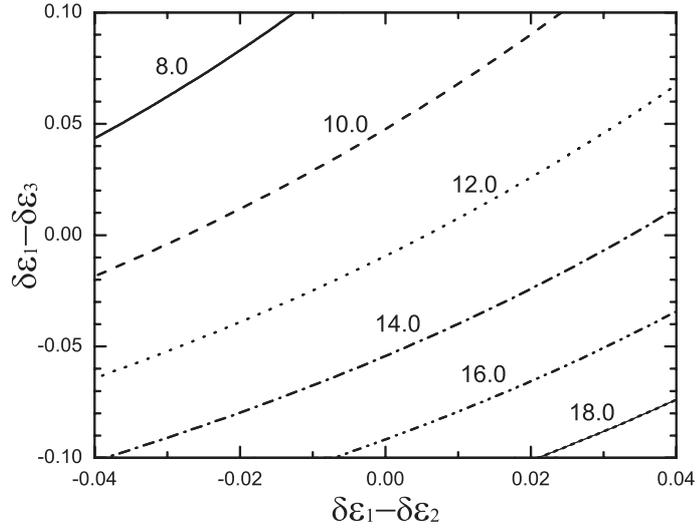}
\caption{The iso-$\Delta \chi^2$ contours for the MH
sensitivity as a function of two effective NSI parameters $\dvarepsilon_1-\dvarepsilon_2$ and
$\dvarepsilon_1-\dvarepsilon_3$. \cite{Li:2014mlo} The NH is assumed for
illustration.} \label{fig:mh_iso}
\end{center}
\end{figure}
%%%%%%%%%%%%%%%%%
%%%%%%%%%%%%%%%%%
\begin{figure}%[h]
\begin{center}
\includegraphics[width=0.55\textwidth]{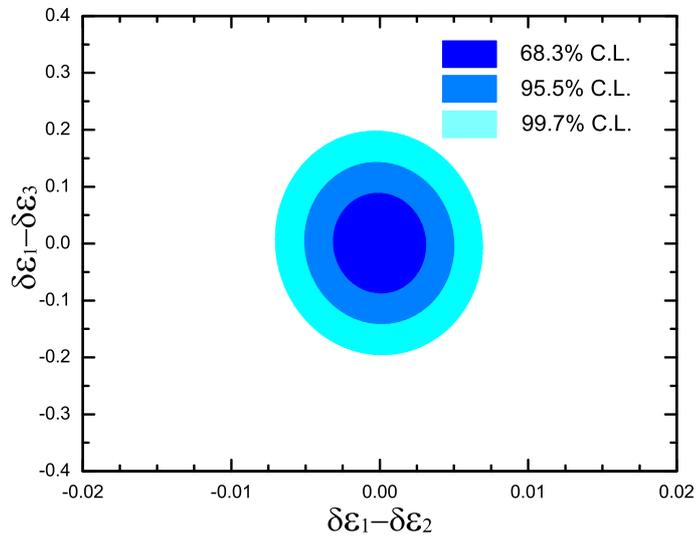}
\caption{The experimental constraints on the generic
NSI parameters $\dvarepsilon_1-\dvarepsilon_2$ and
$\dvarepsilon_1-\dvarepsilon_3$, where the true values are fixed at
$\dvarepsilon_1-\dvarepsilon_2=\dvarepsilon_1-\dvarepsilon_3=0$. \cite{Li:2014mlo}
The NH is assumed for illustration.}
\label{fig:paralimit}
\end{center}
\end{figure}
%%%%%%%%%%%%%%%%%

Finally we want to present the constraints on NSI parameters with JUNO.
In our calculation, the true NSI parameters are taken as $\dvarepsilon_1-\dvarepsilon_2=\dvarepsilon_1-\dvarepsilon_3=0$,
but the corresponding fitted ones are free. The constraints on the NSI parameters are shown in Fig.~\ref{fig:paralimit},
where the limits on these two parameters are at the 1$\sigma$, 2$\sigma$, 3$\sigma$ confidence levels respectively. For
$\dvarepsilon_1-\dvarepsilon_2$, the precision is much better than 1\%.
However, the precision for $\dvarepsilon_1-\dvarepsilon_3$ is around
the $10\%$ level. JUNO is designed for a precision spectral measurement at the
oscillation maximum of $\Delta m^2_{21}$. From
Eq.~(\ref{eq:reactorprobability}), the precision for
$\dvarepsilon_1-\dvarepsilon_2$ can be compatible with that of
$\sin^2 2\theta_{12}$, where a sub-percent level can be achieved. On the other hand, the
precision for $\sin^2 2\theta_{13}$ is also at the $10\%$ level,
also consistent with that of $\dvarepsilon_1-\dvarepsilon_3$ in our
numerical simulation. Because $\dvarepsilon_1-\dvarepsilon_3$ is
suppressed by $\sin\thetat_{13}$, the above two constraints are
actually compatible if we consider the physical NSI parameters
$\depsilon_{\alpha\beta}$ defined in Eq.~(\ref{eq:definition}).

\subsection{Lorentz Invariance Violation}
\label{subsec:exotic:LV}

Special Relativity is a fundamental theory describing the Lorentz space-time symmetry, which is a consequence of the homogeneous and
isotropic space-time and the relativity principle among different inertial frames. Although Lorentz invariance has been widely accepted,
well-motivated models with LIV are anticipated from the principle of Quantum Gravity \cite{Kostelecky:1988zi}.
As one of the most important low energy phenomena, neutrino oscillations provide an opportunity to test the Lorentz invariance violation.

The low energy phenomena of LIV can be systematically studied in the framework of the Standard Model Extension (SME) \cite{Colladay:1996iz,Colladay:1998fq},
which includes all possible LIV terms formed by the SM fields in the Lagrangian,
\begin{equation}
{\mathcal L}_{LV} \hspace{0.2cm}\simeq - (a_{L})_\mu \overline{\psi_{L}} \gamma^\mu \psi_{L}
-(c_{L})_{\mu \nu} \overline{\psi_{L}} i \gamma^\mu \partial^\nu \psi_{L}.\label{eq:lagrangian}
\end{equation}
Notice that $a_{L}$ violates both the Lorentz and CPT symmetries, but $c_{L}$ is CPT-even and only violates the Lorentz invariance.
Although the LIV coefficients are very small due to the suppression factor of the order of the electro-weak scale divided by the Planck scale (i.e., $10^{-17}$),
they provide an accessible test to the Planck scale physics.

The SME framework predicts distinct behaviors of neutrino flavor conversions,
which are different from the standard picture of three neutrino oscillations. The transition probability depends on the ratio
of neutrino propagation baseline $L$ to energy $E$ (i.e., $L/E$) in the standard oscillation framework,
but in the SME it depends on either $L$ or $L\times E$ for the contribution induced by $a_{L}$ or $c_{L}$.
On the other hand, LIV also predicts the breakdown of space-time's isotropy, which manifests as a sidereal modulation
of the neutrino events for experiments with both the neutrino source and detector fixed on the earth. The sidereal time is
defined on the basis of the orientation of the earth with respect to the sun-centered reference inertial frame.
At JUNO both the sidereal variation and spectral distortion effects will be employed to test the LV.
We shall analyze the LIV effects in the reference frame of the sun~\cite{Kostelecky:2001mb,Kostelecky:2002hh} and present the sensitivity of JUNO to LIV coefficients.

According to the SME, the effective Hamiltonian for Lorentz violating neutrino oscillations is written as \cite{Kostelecky:2003cr}
\begin{equation}
H_{\alpha \beta}=\displaystyle \frac{1}{E}\hspace{0.05cm}\left[\hspace{0.05cm}\displaystyle \frac{m^2}{2}+(a_{L})_\mu
p^{\mu}+(c_{L})_{\mu\nu} p^{\mu}p^{\nu}\hspace{0.05cm}\right]_{\alpha \beta}, \label{eq:hamitonian}
\end{equation}
where $E$ and $p^{\mu}$ are the neutrino energy and 4-momentum,  $\alpha$ and $\beta$ are flavor indices, and $(m^2)_{\alpha \beta}=U^{}{\rm diag}(m^{2}_{1},m^{2}_{2},m^{2}_{3})U^{\rm \dagger}$ is the mass-squared matrix in the flavor basis.
To derive the oscillation probabilities from the Hamiltonian, we employ the perturbation treatment~\cite{Diaz:2009qk} to factorize the LIV part from the
conventional neutrino oscillation part.

By defining $\delta H = [(a_{L})_\mu p^{\mu}+(c_{L})_{\mu\nu} p^{\mu}p^{\nu}]/E$ and taking the perturbative expansion,
we derive the time evolution operator as
\begin{equation}\label{sfull}
\begin{array}{ll}
\vspace{0.2cm}
S&=\hspace{0.2cm} e^{-iH t}\hspace{0.2cm}= \hspace{0.2cm}\left(e^{-i H_0 t-i\delta H t}\hspace{0.05cm}e^{i H_0 t}\right)\hspace{0.05cm}e^{-iH_0t}
=\hspace{0.2cm}\left(1-i\int_0^t dt_1 \hspace{0.05cm} e^{-i H_0 t_1}\hspace{0.05cm}\delta H \hspace{0.05cm} e^{i H_0 t_1}+\cdots\right)S^{(0)}\,. % &=\hspace{0.2cm}S^{(0)}+S^{(1)}+\cdots,
\end{array}
\end{equation}
Therefore, the oscillation probability is expanded as
\begin{equation}\label{pab}
P_{\alpha\rightarrow\beta}\hspace{0.1cm}=\hspace{0.1cm}\left|S_{\alpha \beta}\right|^2\hspace{0.1cm}=\hspace{0.1cm}\left|\hspace{0.05cm}S^{(0)}_{\alpha \beta}\hspace{0.05cm}\right|^2+2\hspace{0.05cm}{\rm Re}\hspace{0.05cm}\left[\hspace{0.05cm}(S^{(0)}_{\alpha \beta})^*\hspace{0.05cm}S^{(1)}_{\alpha \beta}\hspace{0.05cm}\right]\hspace{0.1cm}=\hspace{0.1cm} P^{(0)}_{\alpha\rightarrow\beta}+P^{(1)}_{\alpha\rightarrow\beta}.
\end{equation}
In particular, the expression of $P^{(1)}_{\alpha\rightarrow\beta}$ can be written as \cite{Diaz:2009qk}
\begin{equation}\label{p1}
P^{(1)}_{\alpha\rightarrow\beta}=\displaystyle{\sum_{i,j}} \hspace{0.05cm} {\displaystyle\sum_{\rho,\sigma}}\hspace{0.05cm}2\hspace{0.05cm} L\hspace{0.05cm} {\rm Im}\hspace{0.05cm}\left[\hspace{0.05cm}(S^{(0)}_{\alpha \beta})^*\hspace{0.05cm}U_{\alpha i}\hspace{0.05cm} U_{\rho i}^* \hspace{0.05cm}\tau_{ij} \hspace{0.05cm}\delta H_{\rho \sigma}\hspace{0.05cm} U_{\sigma j} \hspace{0.05cm}U_{\beta j}^*\hspace{0.05cm}\right],
\end{equation}
with
\begin{equation}\label{tauij}
\tau_{ij}=\left \{
\begin{array}{cc}
\vspace{0.2cm}
\exp{\{-iE_i L\}}&\hspace{0.3cm} {\rm when}\hspace{0.3cm}i=j\\
\displaystyle \frac{\exp{\{-iE_i L\}}-\exp{\{-iE_j L\}}}{-i(E_i-E_j)L}&\hspace{0.3cm} {\rm when}\hspace{0.3cm}i\neq j
\end{array}
\right..
\end{equation}
$\delta H_{\rho \sigma}$ is required to be real when $\rho =\sigma$ and $\delta H_{\rho \sigma}=\delta H_{ \sigma \rho}^*$ when $\rho \neq \sigma$.
For the former case the LIV terms can be extracted directly as
\begin{equation}\label{p1eq}
P^{(1)}_{\alpha\rightarrow\beta}=\displaystyle{\sum_{i,j}}\hspace{0.05cm}{\displaystyle\sum_{\rho=\sigma}}\hspace{0.05cm} 2\hspace{0.05cm}L\hspace{0.05cm} {\rm Im}\hspace{0.05cm}\left[(S^{(0)}_{\alpha \beta})^*\hspace{0.05cm}U_{\alpha i}\hspace{0.05cm} U_{\rho i}^* \hspace{0.05cm}\tau_{ij} \hspace{0.05cm}\hspace{0.05cm} U_{\sigma j} \hspace{0.05cm}U_{\beta j}^*\right]\delta H_{\rho \sigma}.
\end{equation}
On the other hand, the LIV terms with $\rho \neq \sigma$ can be expressed as
\begin{equation}\label{p1neq}
\begin{array}{ll}
\vspace{0.2cm}
P^{(1)}_{\alpha\rightarrow\beta}&={\displaystyle\sum_{i,j}}\hspace{0.05cm}\displaystyle{\sum_{\rho \neq \sigma}}\hspace{0.05cm} L\hspace{0.05cm} {\rm Im}\hspace{0.05cm}\left[(S^{(0)}_{\alpha \beta})^*\hspace{0.05cm}U_{\alpha i}\hspace{0.05cm} U_{\rho i} \hspace{0.05cm}\tau_{ij} \hspace{0.05cm} U_{\sigma j} \hspace{0.05cm}U_{\beta j}(\delta H_{\rho \sigma}+\delta H_{\sigma \rho})\right]\\
&={\displaystyle\sum_{i,j}}\hspace{0.05cm}\displaystyle{\sum_{\rho \neq \sigma}}\hspace{0.05cm}2\hspace{0.05cm} L\hspace{0.05cm} {\rm Im}\hspace{0.05cm}\left[(S^{(0)}_{\alpha \beta})^*\hspace{0.05cm}U_{\alpha i}\hspace{0.05cm} U_{\rho i} \hspace{0.05cm}\tau_{ij} \hspace{0.05cm}\hspace{0.05cm} U_{\sigma j} \hspace{0.05cm}U_{\beta j}\right] \hspace{0.05cm}{\rm Re}\hspace{0.05cm} \delta H_{\rho \sigma}\,,
\end{array}
\end{equation}
where the CP-violating phase in the MNSP matrix has been neglected for simplicity.
We notice from Eqs.~(\ref{p1eq}) and (\ref{p1neq}) that contributions from the mass term and LIV terms can be factorized.
Therefore, we define the quantity $I_{\alpha \beta}^{\rho \sigma}$ as an indicator of the sensitivity to $\delta H_{\rho \sigma}$
in a certain oscillation channel $P_{\alpha\to\beta}$,
\begin{equation}\label{iterm}
I_{\alpha \beta}^{\rho \sigma}= {\displaystyle\sum_{i,j}}\hspace{0.05cm}2\hspace{0.05cm}L\hspace{0.05cm} {\rm Im}\hspace{0.05cm}\left[(S^{(0)}_{\alpha \beta})^*\hspace{0.05cm}U_{\alpha i}\hspace{0.05cm} U_{\rho i} \hspace{0.05cm}\tau_{ij} \hspace{0.05cm} U_{\sigma j} \hspace{0.05cm}U_{\beta j}\right].
\end{equation}
This indicator can be used to understand the distinct properties of different LIV components $\delta H_{\rho \sigma}$ in the following numerical analysis.

Since the LIV coefficients $(a_{L})_\mu$ and $(c_{L})_{\mu \nu}$ are direction-dependent, a reference frame should be specified when an experiment
is going to report the LIV results. The sun-centered system \cite{Kostelecky:2001mb,Kostelecky:2002hh} can take on this responsibility,
which is depicted in the left panel of Fig.~\ref{fig:sun} and defined in the following way:
\begin{itemize}
\item Point $S$ is the center of the sun;
\item The $Z$ axis has the same direction as the earth's rotational axis, so the $X$-$Y$ plane is parallel to the earth's equator;
\item The $X$ axis is parallel to the vector pointed from the sun to the autumnal equinox,
while the $Y$ axis completes the right-handed system.
\end{itemize}
%%%%%%%%%%%%%%%%%
\begin{figure}%[ht!]
\includegraphics[width=0.45\textwidth]{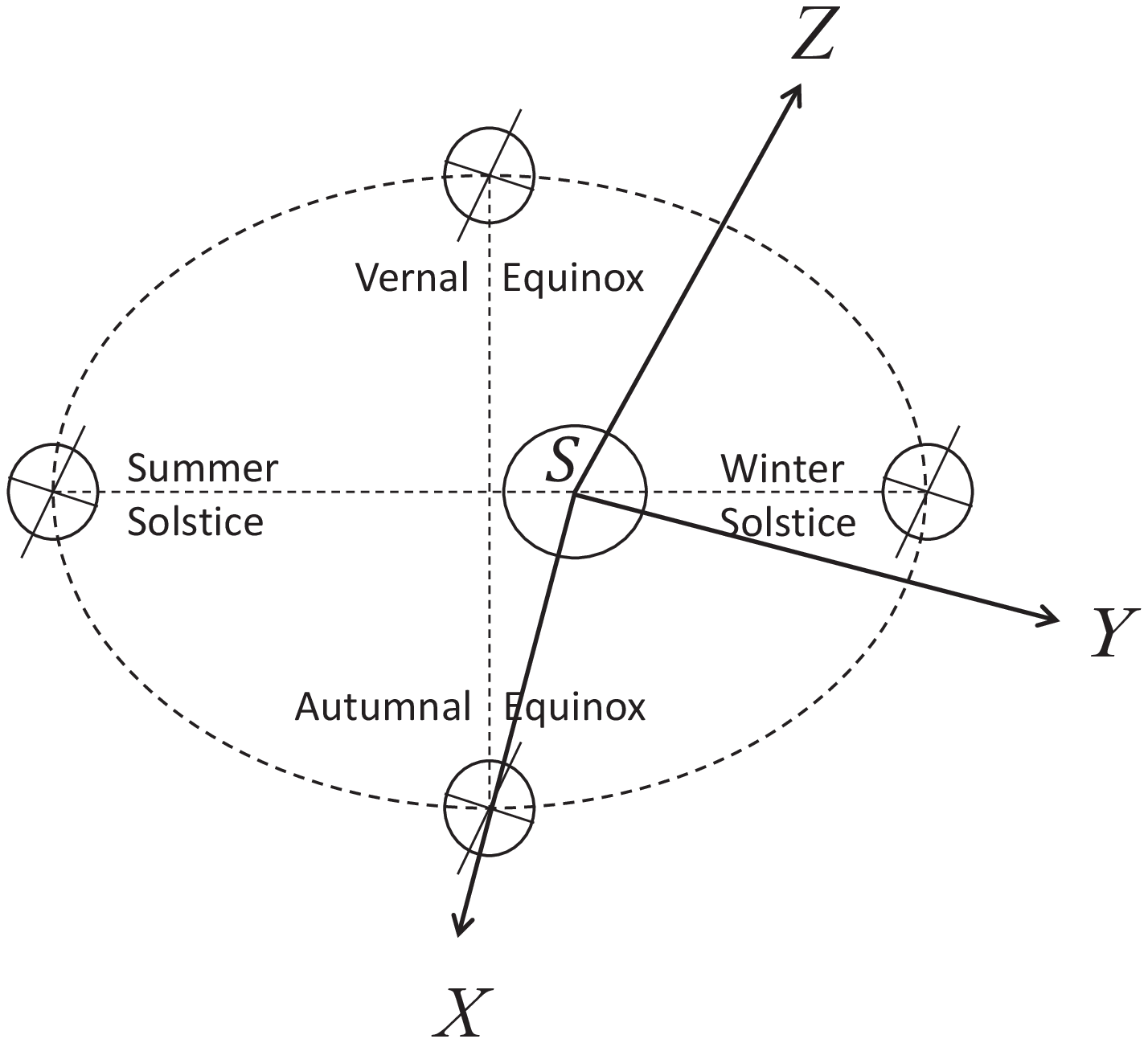}
\includegraphics[width=0.45\textwidth]{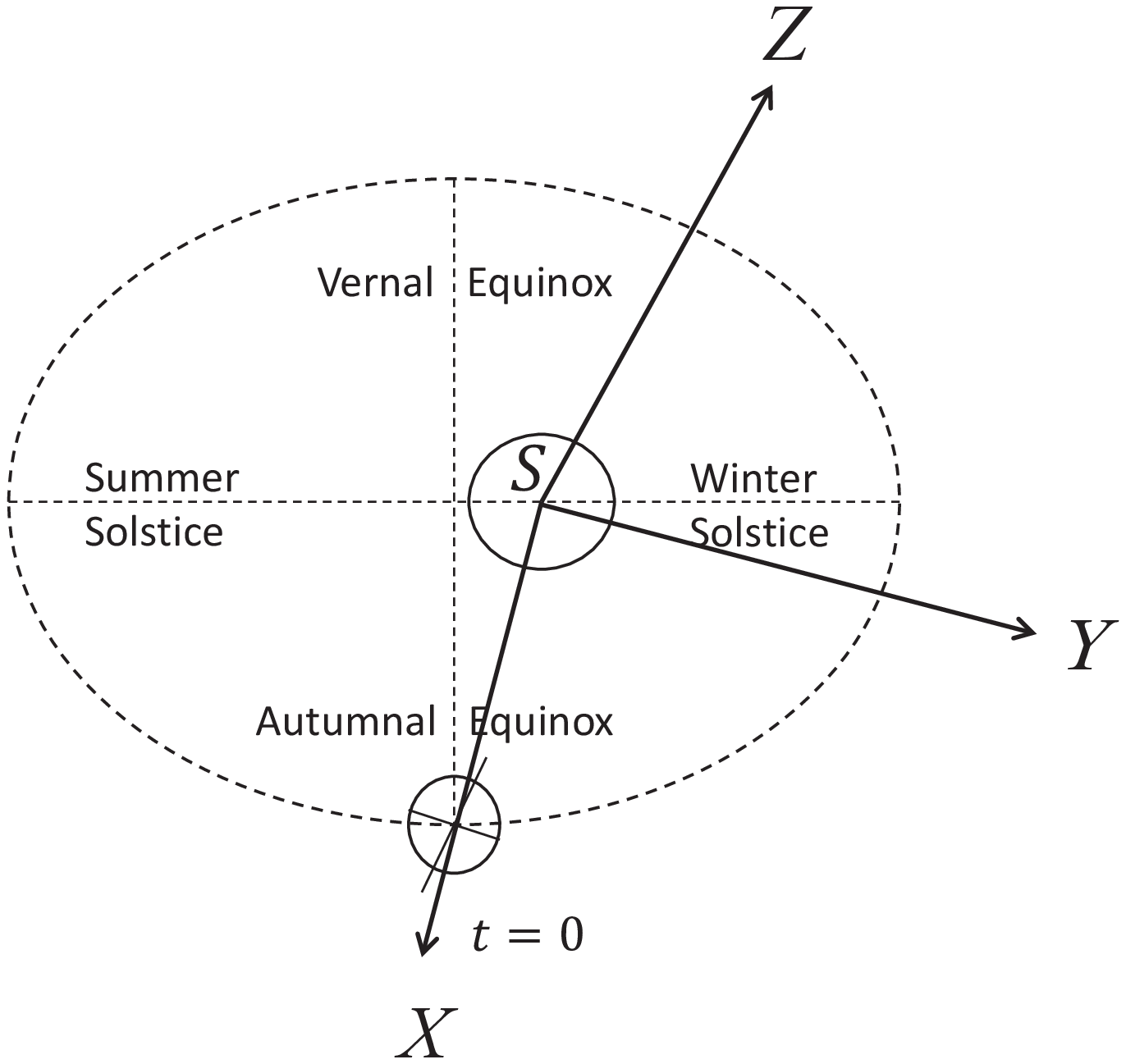}
\caption{The definition (left panel) of the sun-centered system \cite{Kostelecky:2001mb,Kostelecky:2002hh}, and definition (right panel) of $t=0$ in the sun-centered system. }\label{fig:sun}
\end{figure}
%%%%%%%%%%%%%%%%%
%%%%%%%%%%%%%%%%%
\begin{figure}%[ht!]
\includegraphics[width=0.45\textwidth]{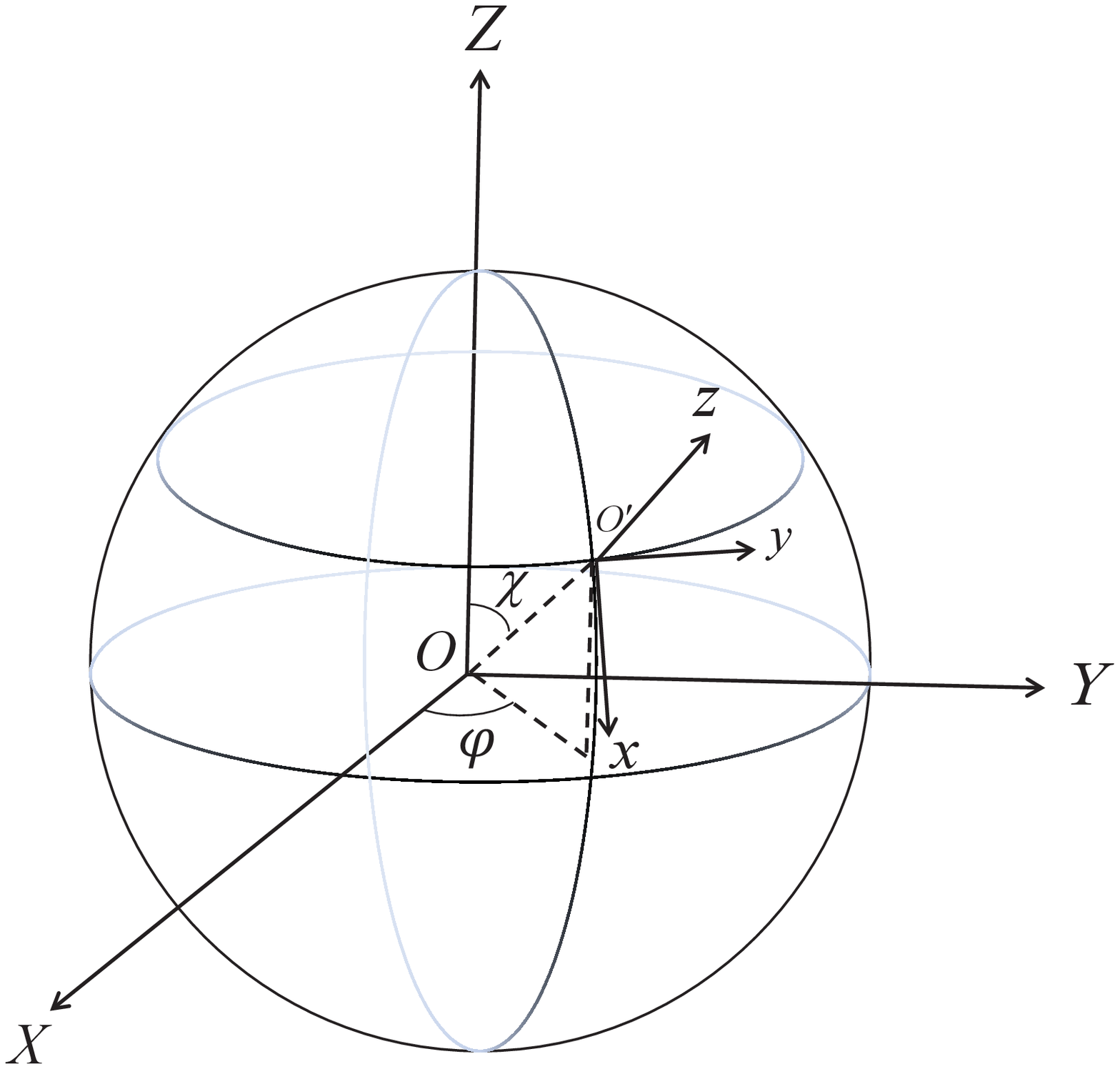}
\includegraphics[width=0.45\textwidth]{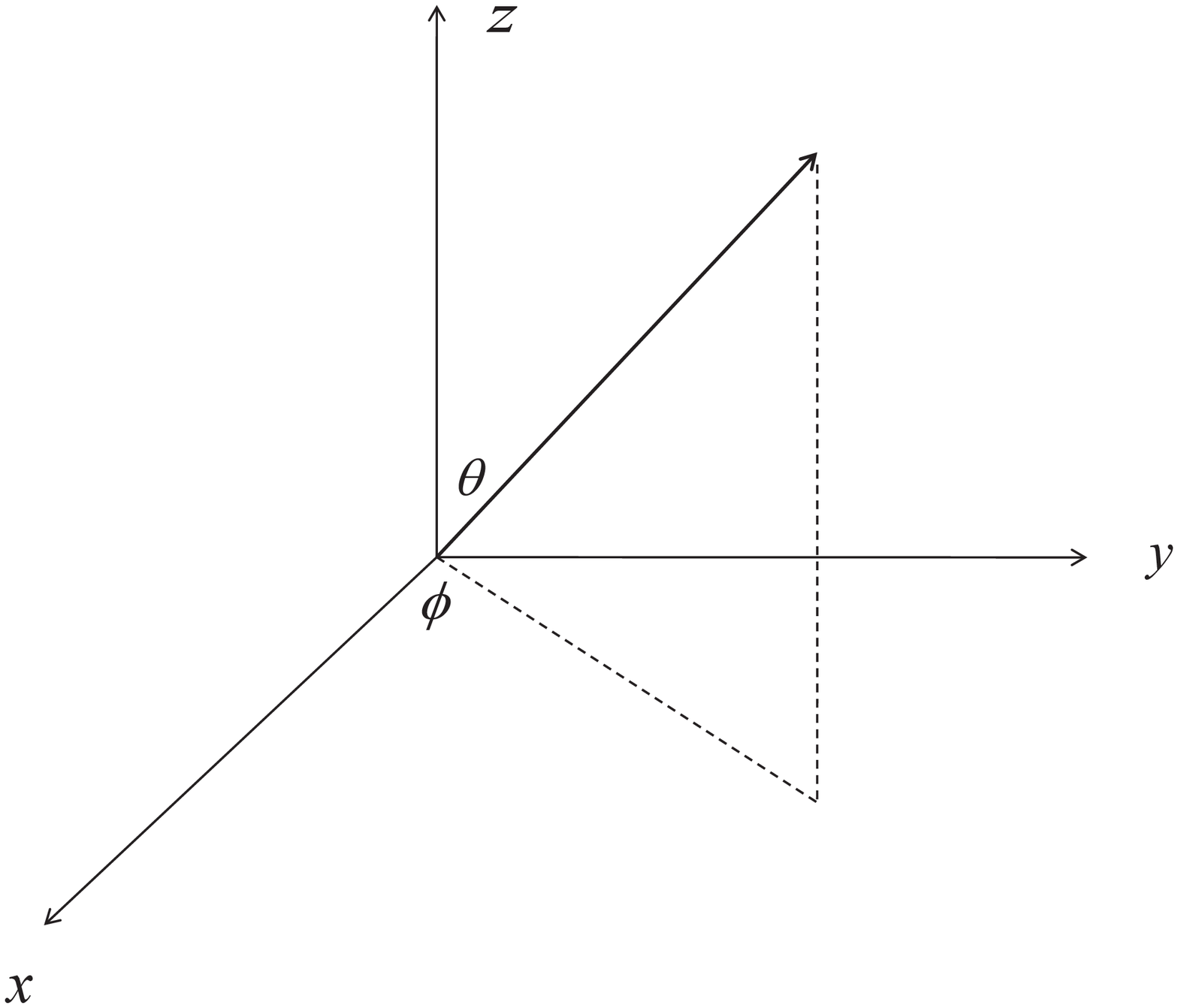}
\caption{The definition of the earth-centered system (left panel) \cite{Kostelecky:2001mb,Kostelecky:2002hh}, and definition of the local coordinate system (right panel). }\label{fig:earth}
\end{figure}
%%%%%%%%%%%%%%%%%
In a terrestrial experiment, the direction of neutrino propagation is described by
its components along the $X,Y,Z$ axes (i.e., $\hat N^{X}, \hat N^{Y}$ and $\hat N^{Z}$).
For convenience, the origin of the sun-centered system can be defined to sit in
the center of the earth $O$ due to the invariance of spatial translation, as shown in the left panel of Fig.~\ref{fig:earth}.
In order to express $\hat N^{X}, \hat N^{Y}$ and $\hat N^{Z}$ in terms of local geographical information,
a local coordinate system $(x,y,z)$ is also introduced:
\begin{itemize}
\item point $O^\prime$ is the site of the neutrino source, and $\chi$ (i.e., the angle between $OO^\prime$ and $Z$ axis) denotes its colatitude;
\item the $z$ axis is defined to be upward;
\item the $x$ and $y$ axes point to the south and east, respectively.
\end{itemize}
In the local coordinate system depicted in the right panel of Fig.~\ref{fig:earth}, the direction of neutrino propagation is parameterized by two angles, of which
$\theta$ is the angle between the beam direction and $z$ axis and $\phi$ is the angle between the beam direction and $x$ axis.

Because the neutrino source and detector are fixed on the earth, the rotation of the earth will
induce a periodic change of the neutrino propagation directions relative to the standard reference frame,
with an angular frequency $\omega=2\pi/T_{\rm s}$.
Here $T_{\rm s}\simeq23 \hspace{0.1cm} {\rm h} \hspace{0.1cm}56\hspace{0.1cm} {\rm min}$ is the period of a sidereal day.
For this reason, a reference time origin should be specified.
Without loss of generality, we can set the local midnight when the earth arrives at the autumnal equinox to be $t=0$ (see the right panel of Fig.~\ref{fig:sun}).
At this moment, the $x$-$z$ plane coincides with the $X$-$Z$ plane, resulting in the coordinate transformation
between the coordinate in the sun-centered system and that in the local system as
\begin{equation}
\begin{array}{lll}
\hat N^{X}&= & \cos{\chi} \sin{\theta} \cos{\phi}+ \sin{\chi} \cos{\theta},\\
\hat N^{Y}&= &\sin{\theta} \sin{\phi},\\
\hat N^{Z}&= &-\sin{\chi} \sin{\theta} \cos{\phi}+ \cos{\chi} \cos{\theta}.
\end{array}
\label{systrans}
\end{equation}
Therefore, the direction of neutrino propagation $\widehat{p^{\mu}}=(1,\ \widehat{p^{X}},\ \widehat{p^{Y}},\ \widehat{p^{Z}})$
is a periodic function of the time $t$
\begin{equation}
\begin{array}{lll}
\widehat{p^{X}}&=& \hat N^{X} \cos{\omega  t }- \hat N^{Y} \sin{\omega t },\\
\widehat{p^{Y}}&=& \hat N^{X} \cos{\omega  t }+ \hat N^{Y} \sin{\omega t },\\
\widehat{p^{Z}}&=& \hat N^{Z}.
\end{array}
\label{dirtrans}
\end{equation}
Accordingly, $\delta H_{\rho \sigma}$ can be decomposed as
\begin{equation}
\begin{array}{ll}
\delta H_{{\rho \sigma}}=&\mathcal C_{{\rho \sigma}}+(\mathcal A_s)_{\rho \sigma} \sin{\omega  t}+(\mathcal A_c)_{\rho \sigma} \cos{\omega  t}\\
&+(\mathcal B_s)_{\rho \sigma} \sin{2\omega  t}+ (\mathcal B_c)_{\rho \sigma} \cos{2\omega  t}.
\end{array}
\label{dh}
\end{equation}
Note that the above coefficients (i.e., $\mathcal C$, $\mathcal A_s$, $\mathcal A_c$,
$\mathcal B_s$ and $\mathcal B_c$) are linear combinations of the LIV coefficients
$(a_{L})_\mu$ and $(c_{L})_{\mu \nu}$.
$\mathcal A_s$, $\mathcal A_c$, $\mathcal B_s$ and $\mathcal B_c$ (written as $\mathcal {A/B}$ for short) associated terms
are time-dependent and can induce periodic variations for the oscillation probability.
On the other hand, the $\mathcal C$ term can modify the absolute value of the oscillation probability with
unconventional energy and baseline dependence,
while the contributions of $\mathcal {A/B}$ cancel out in a full sidereal period.

In our numerical calculation of the JUNO sensitivity to LIV coefficients, we take the same setup as that in the MH studies when using the spectral information.
However, we simplify the reactor complexes in Yangjiang and Taishan NPPs as two virtual reactors with equal baselines (i.e., 52.5 km). When
we discuss the sidereal variation of IBD event rates, a sidereal day is divided into 24 bins. Notice that a normalization
factor has a negligible effect in the sidereal variation analysis. Thus we only consider the statistical uncertainty and the time-dependent uncorrelated uncertainty
of the $1\%$ level for this study.

The upper limits at the $95\%$ confidence level for those LIV coefficients responsible for the spectral distortion
(i.e., $\mathcal C_{\rho \sigma}^{\rm eff}$) are listed in the first row of Tab.~\ref{tab:limit}.
The relative differences in the power of constraining these six coefficients can be understood by the quantity defined in
Eq.~({\ref{iterm}}), which indicates that the flavor components with larger $I_{\alpha \beta}$ will get more severe constraints.
The $95\%$ upper limits for those LIV coefficients from the sidereal variation (i.e., ${\mathcal A/B}_{\rho \sigma}^{\rm eff}$) are listed in the second row of Tab.~\ref{tab:limit},
where the $\tau\tau$ component gets the worst sensitivity, but the $e\tau$ coefficient turns out to be the most severely constrained parameter.
The order of magnitude for ${\mathcal A/B}_{\rho \sigma}^{\rm eff}$ is $10^{-24}$ GeV, which is much smaller than that for $\mathcal C_{\rho \sigma}^{\rm eff}$ (i.e., $10^{-22}$ GeV).
This is because the uncertainties of the spectrum and oscillation parameters do not enter the sidereal variation of IBD events.

It is straightforward to transfer the limits for $\mathcal C_{\rho \sigma}^{\rm eff}$ and $\mathcal {A/B}_{\rho \sigma}^{\rm eff}$
to that for each space component of the physical parameters $a_{L}^\mu$ and $c_{L}^{\mu \nu}$.
Using the real positions of the reactor cores and experiment site, we can present explicit relations between the effective parameters
and the physical parameters \cite{Li:2014rya}. The expansion coefficients for antineutrinos coming from Yangjiang are given as follows:
\begin{equation}
\begin{array}{lll}
\vspace{0.2cm}
(\mathcal C_{\rho \sigma}^{*})_1&=&-a^{T}_{\rho \sigma}+0.81\  a^{Z}_{\rho \sigma}
+[\ 0.42\  c^{TT}_{\rho \sigma}-0.48\ c^{ZZ}_{\rho \sigma}\ ]\ E,\\
\vspace{0.2cm}
((\mathcal A_s)_{\rho \sigma}^{*})_1&=&-0.49\ a^{X}_{\rho \sigma}-0.33\ a^{Y}_{\rho \sigma}
-[\ 0.98\ c^{TX}_{\rho \sigma} +0.65\ c^{TY}_{\rho \sigma}-0.79\ c^{XZ}_{\rho \sigma}-
0.52\ c^{YZ}_{\rho \sigma}\ ]\ E,\\
\vspace{0.2cm}
((\mathcal A_c)_{\rho \sigma}^{*})_1&=&-0.33\  a^{X}_{\rho \sigma}+0.49\ a^{Y}_{\rho \sigma}
-[\ 0.65\ c^{TX}_{\rho \sigma} -0.98\ c^{TY}_{\rho \sigma}-0.52\ c^{XZ}_{\rho \sigma}+
0.79\ c^{YZ}_{\rho \sigma}\ ]\ E,\\
\vspace{0.2cm}
((\mathcal B_s)_{\rho \sigma}^{*})_1&=&-0.16\ c_{\rho \sigma}^{XX}+0.16\ c_{\rho \sigma}^{YY}
+0.13\ c_{\rho \sigma}^{XY},\\
\vspace{0.2cm}
((\mathcal B_c)_{\rho \sigma}^*)_1&=&-0.067\ c_{\rho \sigma}^{XX}+0.067\ c_{\rho \sigma}^{YY}
+0.32\ c_{\rho \sigma}^{XY}.
\label{yj}
\end{array}
\end{equation}
Meanwhile, expansion coefficients for antineutrinos coming from Taishan are given as
\begin{equation}
\begin{array}{lll}
\vspace{0.2cm}
(\mathcal C_{\rho \sigma}^{*})_2&=&-a^{T}_{\rho \sigma}+0.41\  a^{Z}_{\rho \sigma}
-[\ 0.59\  c^{TT}_{\rho \sigma}-0.25\ c^{ZZ}_{\rho \sigma}\ ]\ E,\\
\vspace{0.2cm}
((\mathcal A_s)_{\rho \sigma}^{*})_2&=&0.90\ a^{X}_{\rho \sigma}-0.17\ a^{Y}_{\rho \sigma}
+[\ 1.8\ c^{TX}_{\rho \sigma} -0.34\ c^{TY}_{\rho \sigma}-0.73\ c^{XZ}_{\rho \sigma}+
0.14\ c^{YZ}_{\rho \sigma}\ ]\ E,\\
\vspace{0.2cm}
((\mathcal A_c)_{\rho \sigma}^{*})_2&=&-0.17\  a^{X}_{\rho \sigma}-0.90\ a^{Y}_{\rho \sigma}
-[\ 0.34\ c^{TX}_{\rho \sigma} +1.8\ c^{TY}_{\rho \sigma}-0.14\ c^{XZ}_{\rho \sigma}-
0.73\ c^{YZ}_{\rho \sigma}\ ]\ E,\\
\vspace{0.2cm}
((\mathcal B_s)_{\rho \sigma}^{*})_2&=&0.15\ c_{\rho \sigma}^{XX}-0.15\ c_{\rho \sigma}^{YY}
+0.77\ c_{\rho \sigma}^{XY},\\
\vspace{0.2cm}
((\mathcal B_c)_{\rho \sigma}^*)_2&=&-0.39\ c_{\rho \sigma}^{XX}+0.39\ c_{\rho \sigma}^{YY}
-0.30\ c_{\rho \sigma}^{XY}.
\label{ts}
\end{array}
\end{equation}
Here the subscript ``$_L$" for $a_L$ and $c_L$ is omitted for simplicity.

One should keep in mind that the degrees of freedom in $a_{L}^\mu$ and $c_{L}^{\mu \nu}$ are much larger than those in the effective LIV coefficients.
In order to obtain independent constraints, it is thus more reasonable to use the effective coefficients.
On the other hand, it is better to derive the limits of $a_{L}^\mu$ and $c_{L}^{\mu \nu}$
when we compare and combine the limits from different oscillation experiments.

\begin{table}
\centering
\begin{tabular}{|c|c|c|c|c|c|c|}
\hline\hline
&$\mathcal C_{ e e}^{{\rm eff}}$ &$ \hspace{0.1cm}\mathcal C_{ e \mu}^{{\rm eff}}$  &$\hspace{0.1cm}\mathcal C_{ e  \tau}^{{\rm eff}}$
&$\mathcal C_{\mu \mu}^{{\rm eff}}$ &$\hspace{0.1cm}\mathcal C_{ \mu  \tau}^{{\rm eff}}$  &$\mathcal C_{ \tau  \tau}^{{\rm eff}}$ \\
\hline
$10^{-22}\ {\rm GeV}$  & 1.5  &   0.5   & 0.4 & 4.4   &  3.4   & 1.1  \\
\hline
\hline
&$\mathcal {A/B}_{e e}^{{\rm eff}}$ &$\hspace{0.1cm}\mathcal {A/B}_{e \mu}^{{\rm eff}}$  &$\hspace{0.1cm}\mathcal {A/B}_{e \tau}^{{\rm eff}}$
&$\mathcal {A/B}_{\mu \mu}^{{\rm eff}}$ &$\hspace{0.1cm}\mathcal {A/B}_{\mu \tau}^{{\rm eff}}$  &$ \mathcal {A/B}_{\tau \tau}^{{\rm eff}}$ \\
\hline
$10^{-24}\ {\rm GeV}$  & 3.7 &   3.7   &  2.7  & 5.4   & 6.6   &  11.6    \\
\hline
\end{tabular}
\caption{The JUNO sensitivity at the $95\%$ confidence level for the effective LIV coefficients $\mathcal C$, and $\mathcal {A/B}$ from
the effects of spectral distortion and sidereal variation \cite{Li:2014rya}.}
\label{tab:limit}
\end{table}

\subsection{Discussions}
\label{subsec:exotic:dis}

Besides the exotic search with the reactor neutrino oscillation, there are many other possibilities in the aspect of
new physics studies.

Rare decays and similar processes related to the baryon number and lepton number violation,
constitute a large group of fundamental issues in the new physics searches beyond the SM. Future prospects of
nucleon decays at JUNO are discussed in Sec.~\ref{sec:nd}, with the special attention to the $p\rightarrow K^{+}+{\bar \nu}$ channel of proton decays.
Solar and supernova neutrino studies at JUNO also provide us useful astrophysical probes for testing new physics scenarios, among which tests of neutrino
electromagnetic properties are of fundamental importance~\cite{Giunti:2015gga}. In solar neutrinos, effects of neutrino magnetic dipole
moments could appear in the $\nu_{e}$-$e^{}$ scattering cross section and spin-flavor precession mechanism. Through the solar $^7$Be and $^8$B neutrino
observations at JUNO, the neutrino magnetic moment can be tested at a level comparable to that using reactor antineutrinos. On the other hand, with
the high-precision supernova neutrino detection at JUNO, it might be possible to reveal the collective spin-flavor oscillation effects
with the Majorana transition magnetic moments as small as $10^{-22}\;\mu_{\rm B}$~\cite{deGouvea:2012hg,deGouvea:2013zp}.

In summary, as a multi-purpose underground neutrino observatory, JUNO will provide us the unique opportunity to study new physics beyond the SM.

\clearpage

\section{Appendix}
\label{sec:app}

\blfootnote{Editors: Jun Cao (caoj@ihep.ac.cn), Miao He (hem@ihep.ac.cn), Liangjian Wen (wenlj@ihep.ac.cn), and Weili Zhong (zhongwl@ihep.ac.cn)}
\blfootnote{Major Contributors: Ziyan Deng and Xinying Li}

\subsection{Reactor Neutrinos}
\label{subsec:mh:flux}
\subsubsection{Introduction}

Reactor neutrinos are electron antineutrinos that emitted from subsequent $\beta$-decays of instable fission fragments. All reactors close to JUNO are pressurized water reactors (PWR). In these reactors, fission of four fuel isotopes, $^{235}\rm U$, $^{238}\rm U$, $^{239}\rm Pu$, and $^{241}\rm Pu$, makes up more than 99.7\% of the thermal power and reactor antineutrinos. Reactor neutrino fluxes per fission of each isotope are determined by inversion of the measured $\beta$ spectra of fissioning~\cite{VonFeilitzsch:1982jw,Schreckenbach:1985ep,Hahn:1989zr,Huber:2011wv,Mueller:2011nm}
or by calculation with nuclear database~\cite{Vogel:1980bk,Dwyer:2014eka}. Their fission rates in a reactor can be estimated with the core simulation and the thermal power measurements. Therefore, the reactor neutrino flux can be predicted as $\Phi(E_\nu)=\sum_i F_i S_i(E_\nu)$, where $F_i$ is the fission rate of isotope $i$ and $S_i(E_\nu)$ is the neutrino flux per fission, summing over the four isotopes~\cite{Cao:2011gb}.
Such a prediction is expected to carry an uncertainty of 2-3\%~\cite{An:2012eh,An:2012bu,An:2013zwz}.
Recently, reactor neutrino experiments
(i.e., Daya Bay~\cite{DYBbump}, RENO~\cite{Seon-HeeSeofortheRENO:2014jza} Double Chooz~\cite{Abe:2014bwa})
found a large discrepancy between the predicted and measured spectra in 4-6 MeV.
Model independent prediction based on the new precision measurements could avoid this bias, and might be able to improve the precision to 1\%.

\subsubsection{Neutrino flux per fission}

Neutrino flux per fission for each isotope could be calculated by superposing thousands of beta decays of the fission fragments. However, such a first-principle calculation is challenging due to missing or inaccurate data even with modern nuclear databases. In general the uncertainty is $\sim\,$10\%~\cite{Vogel:1980bk}.

Several direct measurements were done at ILL in 1980s to determine the neutrino fluxes and energy spectra of the thermal fissile isotopes $^{235}$U, $^{239}$Pu and $^{241}$Pu. In these measurements, sample foils were placed into a reactor and exposed to neutrons for one or two days. A high precision electron spectrometer recorded the emitted $\beta$ spectra, which were then inverted to the antineutrino spectra by fitting the observed $\beta$ spectra to a set of 30 virtual $\beta$-branches. With the Q-value and the branching ratios of the virtual $\beta$-branches, the corresponding neutrino spectra could be computed out~\cite{VonFeilitzsch:1982jw,Schreckenbach:1985ep,Hahn:1989zr}.

These measurements didn't include $^{238}$U which fissions only with fast neutrons. A theoretical calculation of the $^{238}$U neutrino spectrum has been computed by Vogel {\it et al.}~\cite{Vogel:1980bk}. Since $^{238}$U contributes only $\sim\,$10\% antineutrinos in a typical PWR, the error by using this calculated $^{238}$U neutrino spectrum should be less than 1\%. The neutrino fluxes per fission of the four isotopes determined in above literature, which is called ILL+Vogel model, are shown in Fig.~\ref{fig:rcs}. The inverted neutrino spectrum were compared with the Bugey-3 data~\cite{Achkar:1996rd} and showed reasonable agreement in both the total rate and energy spectrum.

The neutrino flux per fission was recently improved by Huber~\cite{Huber:2011wv}, in which the ILL $\beta$ spectra were re-analyzed with higher order corrections in the $\beta$ decay spectrum taken into account.
The spectrum of $^{238}$U was also updated with an {\it ab initio} calculation by Mueller {\it et al.}~\cite{Mueller:2011nm}. Comparing to the ILL+Vogel model, the Huber+Mueller model shows a 3.5\% increase in total flux and a small excess in the high energy part of the spectra. The flux uncertainty is reduced to 2\% from 2.7\% of the ILL+Vogel model. The upward shift in the total flux introduces tension with short baseline reactor neutrino experiments, which is called Reactor Neutrino Anomaly~\cite{Mention:2011rk}.

The reactor antineutrinos are generally detected via the inverse beta decay (IBD) reaction $\bar\nu_{e} + p \to n + e^{+}$. The reaction cross section $\sigma(E_\nu)$ is calculated to the order of $1/M$ in Ref.~\cite{Vogel:1999zy}. The observable reactor neutrino spectrum is the multiplication of the neutrino flux per fission and the cross section, which is shown in Fig.~\ref{fig:rcs} for the four isotopes.
\begin{figure}[thb]
\begin{centering}
\includegraphics[width=0.6\textwidth]{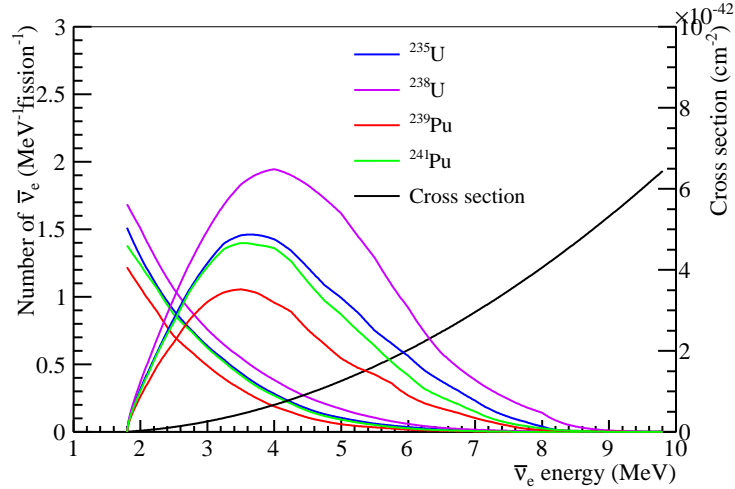}
\par\end{centering}
\caption{\label{fig:rcs} Neutrino yield per fission,
the interaction cross section of the inverse beta decay, and the
observable spectra of the listed isotopes.}
\end{figure}

\subsubsection{Reactor Power and Fuel Evolution}

Fission rates of isotopes (at nominal power) as a function of time, as well as the fuel composition, can be obtained via core simulation. Since the fission rates are correlated with the reactor power, normally we use fission fraction in the core simulation, which is the ratio of the fission rate of an isotope over the total fission rate. Fresh fuel contains only uranium. The plutonium isotopes are gradually generated via the neutron capture of $^{238}$U and the subsequent evolution. Generally a PWR core refuels every 12-18 months, and replaces 1/4 to 1/3 fuel assemblies each time. To describe the fuel evolution as a function of time, burnup of the fuel is defined as
\begin{eqnarray}
B(t)=\frac{W \cdot D}{M_{init\text{-}U}}\,,
\label{eq:mh:burnup}
\end{eqnarray}
where $W$ is the fission power of the fuel, $D$ is the fissioning days, and $M_{init\text{-}U}$ is the initial mass of the uranium. The unit of the burnup is ${\rm MW}\cdot {\rm d}/{\rm ton}$. Since a fuel assembly will stay in the core for 3-4 refueling cycles, and fuel assemblies have different burnup, a more convenient variable ``cycle burnup'' is defined to describe the aging of a reactor core within a refueling cycle. The cycle burnup has the same expression as Eq.~(\ref{eq:mh:burnup}), but with $W$ being the total nuclear power of the reactor core, $D$ being the fissioning days since the beginning of the refueling cycle, and $M_{init\text{-}U}$ being the total initial uranium mass in the reactor core. Cycle burnup can be calculated by using the daily thermal power which are obtained by the power monitoring system.

The most accurate thermal power measurement is the Secondary Heat Balance method. Detailed description of this measurement can be found, for example, in Ref.~\cite{kme}. It is an offline measurement, normally done weekly or monthly. Primary Heat Balance test are online thermal power measurement. Normally it is calibrated to the Secondary Heat Balance measurement weekly. Daya Bay power plants control the difference between these two measurements to less than 0.1\% of the full power. This data is good for neutrino flux analysis. To 0.1\% level, it can be taken as the Secondary Heat Balance measurement. The power plants also monitor the ex-core neutron flux, which gives the nuclear power. This monitoring is online, for safety and reactor operation control. It is normally calibrated to the Primary Heat Balance measurement daily. The ex-core neutron measurement is less accurate, controlled to be less than 1.5\% of the full power by the Daya Bay power plant. Using the Primary Heat Balance measurement, the thermal power uncertainty is estimated to be 0.5\% per core for the Daya Bay experiment~\cite{An:2012eh,An:2012bu,An:2013zwz}. The Yangjiang and Taishan NPPs use similar reactors as the Daya Bay NPP. The power uncertainty is also taken as 0.5\% for JUNO in the analyses.

The reactor power plants simulate the fuel evolution in every refueling cycle to reconfigure the location of the fuel assemblies of different burnup in the core. Simulations are done for possible configurations before the refueling to optimize the safety factor and operation efficiency, and are redone with actual power history and in-core neutron flux measurements to better estimate the burnup of the fuel assemblies when the cycle completes. Fission fraction and fuel composition can be extracted from these simulations. The simulation is performed by Daya Bay power plants with a validated and licensed commercial software SCIENCE developed by CEA, France. Its core component is APOLLO2~\cite{apollo2}.  Uncertainties of the simulation were estimated by comparing the simulated fuel composition at different burnup with isotopic analyses of spent fuel samples taking from the reactors. The uncertainties of the fission fractions reported by SCIENCE are estimated to be $\sim\,$5\%. Another simulation package based on DRAGON~\cite{dragon}, a public available software, was developed by the Daya Bay experiment to cross check the simulation done by the power plant and to evaluate the correlation of uncertainties.

The Yangjiang reactors are CPR1000+. The nuclear cores are almost identical to the Daya Bay cores (Framatone M310) and Ling Ao cores (CP1000, a derivative of M310). The fresh fuel enrichment is $\sim\,$4.5\%. A refueling cycle lasts for 12-18 months. The Taishan reactors use the EPR technology of the French AREVA company. The fresh fuel enrichment is about 7.44\%. A refueling cycle could be 18-22 months. When refueling for a new cycle, the fuel elements are configured in the reactor core around the center as symmetrical as possible, which makes the center of gravity being stable at the core center. The fission fractions versus burnup for the Tainshan core are simulated with DRAGON packages, and compared to the Daya Bay core, as shown in Fig.~\ref{fig:MH:fisfrac}. Due to the different enrichment, there are slight differences between the Daya Bay (Yangjiang) core and the Taishan core. The uncertainty of the fission fraction will be taken as 5\% for both core types, while further investigation will be done in the future.
\begin{figure}[htb]
\begin{centering}
\includegraphics[width=0.6\textwidth]{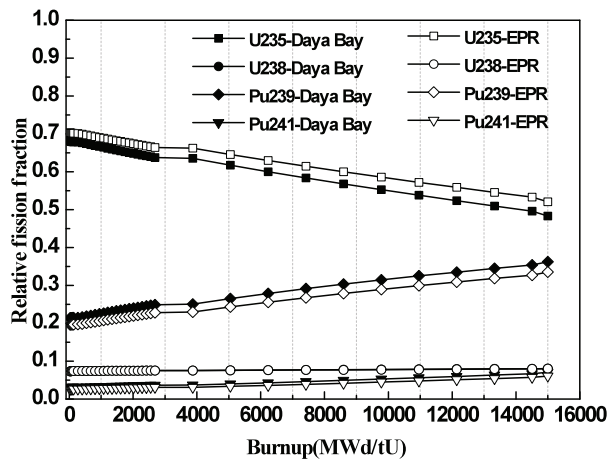}
\par\end{centering}
\caption{\label{fig:MH:fisfrac}
Simulation of the fission fractions in the Daya Bay and Taishan cores with DRAGON packages
for a normal burnup cycle. The simulation assumes a 18 months refueling cycle and average disassembly burnup at 45~${\rm MW}\cdot {\rm d}/{\rm ton}$. Differences between the Daya Bay and Taishan cores are due to fuel enrichment. }
\end{figure}

\subsubsection{Antineutrino Spectrum and Uncertainties}

With the thermal power $W_{th}$ and fission fractions of four main isotopes $f_i$, the expected neutrino flux emitted from a reactor is predicted as:
\begin{eqnarray}
\label{eq:nuflux}
\Phi(E_{\nu})=\frac{W_{th}}{\sum_i f_i e_i}\cdot\sum_i f_i \cdot S_{i}(E_{\nu}),
\end{eqnarray}
Where $e_i$ and $S_i(E_{\nu})$ are the thermal energy released in each fission and the neutrino flux per fission for the $i$-th isotope, respectively.
Eq.~(\ref{eq:nuflux}) is equivalent to the expression $\sum_i F_i S_i(E_\nu)$ in the introduction since ${W_{th}}/{\sum_i f_i e_i}$ is the total fission rate, and the fission rate of the $i$-th isotope is $F_i={W_{th}}/{\sum_i f_i e_i} \cdot f_i$.

Fission energies of the four isotopes could be obtained by reactor core simulation or analytical calculation.
The typical values and their uncertainties are listed in Tab.~\ref{tab:mh:fissionE} which were calculated by Kopeikin{\it et al.}~\cite{Kopeikin:2004cn} and updated by Ma {\it et al.}~\cite{Ma:2012bm} recently.
\begin{table}[!htb]
\begin{center}
\begin{tabular}[c]{l|l|l|l|l} \hline\hline
isotope &  $^{235}\rm U$  & $^{238}\rm U$ & $^{239}\rm Pu$ &
$^{241}\rm Pu$ \\ \hline
 Kopeikin (MeV) & $201.92\pm0.46$  & $205.52\pm0.96$ & $209.99\pm0.60$ & $213.60\pm0.65$  \\ \hline
 Ma (MeV ) & $202.36\pm0.26$  & $205.99\pm0.52$ & $211.12\pm0.34$ & $214.26\pm0.33$  \\
\hline\hline
\end{tabular}
\caption{\label{tab:mh:fissionE} Energy release per fission of the
main fissile isotopes.}
\end{center}
\end{table}

The fission fractions predicted by the core simulation carry relatively large uncertainty.
But they are not independent. On one hand, they are strongly constrained by the more accurate total thermal power. As a consequence, the uncertainty is greatly reduced. On the other hand, they are correlated in the fuel evolution. For example, if $^{239}$Pu is over-estimated, $^{241}$Pu will also be over-estimated since they come from the same neutron capture process on $^{238}$U. The correlations were studied with the DRAGON simulation by systematically varying the intermediate fuel composition. Tab.~\ref{tab:mh:fissfraction} shows the correlation of the fission fractions that were simulated at several burnup stages for Daya Bay cores~\cite{Ma:2014bpa}. Similar simulations will be done for JUNO.
\begin{table}[!htb]
\begin{center}
\begin{tabular}[c]{l|l|l|l|l} \hline\hline
 &  $^{235}\rm U$  & $^{238}\rm U$ & $^{239}\rm Pu$ & $^{241}\rm Pu$
\\ \hline
 $^{235}\rm U$ & 1.0  & -0.22 & -0.53 & -0.18  \\ \hline
 $^{238}\rm U$ & -0.22  & 1.0 & 0.18 & 0.26  \\ \hline
 $^{239}\rm Pu$ & -0.53  & 0.18 & 1.0 & 0.49  \\ \hline
 $^{241}\rm Pu$ & -0.18  & 0.26 & 0.49 & 1.0  \\ \hline\hline
\end{tabular}
\caption{\label{tab:mh:fissfraction} Correlation coefficients of
isotope fission fraction uncertainties.}
\end{center}
\end{table}

The expected neutrino spectrum in the detector without oscillation effect is
\begin{eqnarray}
S(E_{\nu})=\frac{1}{4\pi L^{2}}\cdot\Phi(E_{\nu})\cdot \epsilon \cdot N_{p}
\cdot \sigma(E_{\nu})\,,
\end{eqnarray}
where $L$ is the distance from the reactor to the detector, $\epsilon$ is the detection efficiency, $N_{p}$ is the target proton number,
and $\sigma(E_{\nu})$ is the inverse $\beta$ decay cross section. The calculated cross section relates to the neutron lifetime, whose uncertainty is $0.2\%$. The multiplication of cross section and total isotope antineutrino spectrum is defined as 'reaction cross section'.

In ILL measurements, fissile samples were exposed in neutrons for one to two days. Therefore, the beta decays from some of the long-lived fission fragments not reaching equilibrium were missed in the ILL measurements. The long lived fission fragments accumulate in core during operation and contribute an extra antineutrino flux at low energy. This effect can be evaluated with nuclear database by using the cumulative yields of the known long-lived ( e.g. lifetime longer than 10 hours) fission fragments which has Q-values above the inverse $\beta$ decay threshold ($1.8$~MeV )~\cite{Mueller:2011nm}. On the other hand, after refueling the reactor cores, the spent fuel taken out from the previous cycle are moved to a cooling pool adjacent to the core. The long-lived isotopes will continue decay and generate antineutrinos. This contributions to the total antineutrinos can also be evaluated by using the cumulative yields~\cite{Zhou:2012zzc,Kopeikin:2001qv,Afanasev:2003ci,Kopeikin:2004fd,An:2009zz}. The contribution to IBD events from off-equilibrium of long-lived isotopes in reactor cores and from spent fuel pools in the Daya Bay experiment were evaluated to be both less than 0.3\%. The uncertainties of them were taken as 100\%, or 0.3\% in terms of the total IBD events. For JUNO experiment, the similar calculation could be done with the reactor core and spent fuel pool information. The total neutrino spectrum should then be modified by including the contributions from the off-equilibrium long-lived isotopes in reactor cores and from the spent fuel pools,
\begin{equation}
S_{\nu}(E_{\nu})=S(E_{\nu})+S_{\rm offEq}+S_{\rm SNF}\,.
\end{equation}

The event rate of the IBD reactions in the $k$-th energy bin $[E^{\rm min}_{k},E^{\rm
max}_{k}]$ of a detector from several reactor cores is given by
\begin{equation}
R_{k}(E_{\nu})=\sum_r \frac{1}{4\pi
L^{2}_{r}}\cdot\frac{W^{r}_{th}}{\sum_i f^{r}_i e^{r}_i}\cdot
\sum_{i} f^{r}_i \int^{E^{\rm max}_{k}}_{E^{\rm min}_{k}}\cdot
S_{i}(E_{\nu}) \cdot \sigma(E_{\nu})dE_{\nu} \cdot \epsilon \cdot
N_{p}\,,\label{eq:mh:binrate}
\end{equation}
where $k$ is the bin index and $r$ is the index of the reactor
cores.

The flux uncertainties can be propagated using
Eq.~(\ref{eq:mh:binrate}). The reaction cross section uncertainties
are considered to be correlated between isotopes. The
reactor-related uncertainties are categorized into the
reactor-correlated ones and reactor-uncorrelated ones, respectively.
The correlated uncertainties includes those from energy per fission
($0.2\%$) and reaction cross section ($2.7\%$). The uncorrelated
uncertainties are the combination of the thermal power ($0.5\%$),
fission fraction ($0.6\%$ with the thermal power constraint and correlation among isotopes), off-Equilibrium ($0.3\%$) and spent fuel ($0.3\%$).

The shape uncertainties can also be propagated using Eq.~(\ref{eq:mh:binrate}).
There are two categories basically, energy dependent and energy independent uncertainties.
The energy independent uncertainties are same as flux uncertainties, such as power and fission energies. The energy dependent uncertainties includes the uncertainties of antineutrino flux per fission of each fuel isotope, the uncertainties of off-equilibrium and SNF corrections,
and the uncertainties introduced by fission fraction uncertainty.
These energy dependent uncertainties are categorized as bin-to-bin correlated and uncorrelated uncertainties.
The $^{235}{\rm U}$, $^{239}{\rm Pu}$ and $^{241}{\rm Pu}$ spectra have both bin-to-bin uncorrelated uncertainties (statistical uncertainties) and bin-to-bin correlated uncertainties (systematic uncertainties from ILL measurements, and beta-antineutrino conversion), while the $^{238}{\rm U}$ spectrum only has bin-to-bin correlated uncertainties since it is from theoretical calculation. The off-equilibrium uncertainties, spent fuel uncertainties and fission fraction uncertainties are treated as energy correlated, but reactor uncorrelated.

\subsubsection{Model Independent Prediction of the Antineutrino Spectrum}
\label{subsubsec:mh:generic}
Recently, Daya Bay~\cite{DYBbump}, RENO~\cite{Seon-HeeSeofortheRENO:2014jza} Double Chooz~\cite{Abe:2014bwa} have found significant local inconsistency (at $4\sim6$ MeV) between the measured reactor neutrino spectrum and the predicted spectrum, no matter using the ILL$+$Vogel or the Huber$+$Mueller flux model. The largest deviation reaches $\sim\,$10\%, significantly larger than the expected uncertainty 2-3\%. In 2014, there were new evaluations of the shape uncertainties of the antineutrino flux per fission from ILL measurement, which claimed that the uncertainties of the inversion from the $\beta$ spectra to the antineutrino spectrum is underestimated. The energy dependent uncertainty should be no less than 4\%~\cite{Hayes:2013wra}. Further more, the latest calculation of the antineutrino flux per fission with the $\beta$ branch information in database of the fission fragments indicates similar local structure of isotope antineutrino spectrum as the measurements in the reactor antineutrino experiments~\cite{Dwyer:2014eka}.

One possible way to do more precise prediction of the expected antineutrino spectrum for JUNO is to use the measured positron spectrum of the Daya Bay experiment directly. The principle is to treat Daya Bay as a virtual near detector of the JUNO experiment. The reactor cores of Yangjiang and Taishan will be constructed and operated by the same company as Daya Bay. Similar core technology and simulation enable a virtual near/far relative measurement, where the reactor related uncertainties are almost cancelled, and the relative detector uncertainties between near and far site is the main contribution of final uncertainty. Detector simulation would help to determine the relative uncertainty of energy scale between the antineutrino detectors of JUNO and Daya Bay. The current absolute energy scale uncertainty of Daya Bay is within 1\% and the energy non-linearity has been determined to $\sim\,$1\%.

We expect that the shape uncertainties of the measured positron spectrum for JUNO could be determined to 1\% with further studies. As an alternative method, the antineutrino spectrum from unfolding of the direct measurement of Daya Bay could be also used for spectrum prediction. The unfolded spectrum is independent of the Daya Bay detector effect, could be used more generally, such as comparing with different reactor models of antineutrino spectrum, and predicting the measured positron spectrum by applying their own detector effects of the experiments~\cite{DYBbump}.

\subsection{Monte Carlo Simulation and Reconstruction}

\subsubsection{Monte Carlo Simulation}
\label{subsec:intro:sim}
A Geant4~\cite{Agostinelli:2002hh,Allison:2006ve} based computer simulation (Monte Carlo, MC) of the detectors and readout electronics is used to study the detector response and optimize the detector design. It consists of five components: kinematic generator, detector simulation, electronics simulation, trigger simulation and readout simulation. The JUNO MC is developed based on the Daya Bay MC simulation, which has been carefully tuned to match observed detector distributions, such as the liquid scintillator light yield, charge response, and energy non-linearity.

\par
The antineutrino generator reads from a database that stored the reactor antineutrino spectra.  The cosmic muons in the underground laboratory are simulated using a digitized topographic map of the site and Muon Simulation Code~\cite{Antonioli:1997qw} (MUSIC), which calculates the energy loss and multiple scattering due to the rock overburden. The muon generator for Geant4 reads randomly from a library of muon events generated with MUSIC. The software generators for the calibration sources and the simulation of the decay sequences for natural radionuclides found in our detectors are customized based on data from the ENDF database~\cite{ENDF}.

\par
All physical processes in Geant4 relevant to the reactor neutrino experiment have been validated. The gamma spectra of neutron capture and muon capture on many nuclei are incorrectly modeled in Geant4. Since a systematic correction is complex, we implement corrections on a case by case basis. Furthermore, the simulation of thermal neutron scattering is improved by considering the molecular binding energy of the scattering nuclei.

\par
The details of the electronics simulation can be found in Ref.~\cite{Jiang:2012zze,Jetter:2012xp}. Using the timing and number of photoelectron (p.e.) generated in PMTs, an analog signal pulse for each PMT is generated and tracked through the digitization process, taking into account the non-linearity, dark rate, pre-pulsing, after-pulsing and ringing of the waveform.  The simulated analog pulse is then used as input to a trigger system simulation.

\par
Most of the properties of the detector materials are borrowed from the Daya Bay experiment.
The elemental concentrations of the liquid scintillator were measured and incorporated into the MC. All relevant optical properties of the detector components are derived from measurements, including refractive indices of the liquids as well as the acrylic or nylon components, time constants and photon emission spectra of the liquid scintillator, and the reflectivity of the detector materials. Photon absorption and re-emission processes in liquid scintillator are modeled based on measurements in order to properly simulate the propagation of scintillation photons and contributions from Cherenkov process.

\par
The major differences from the Daya Bay MC include the detector geometry, the attenuation length and the rayleigh scattering length of the liquid scintillator, and the PMT geometry and quantum efficiency. The detector is spherical, of a diameter of 35.4 m for the liquid scintillator container. Currently there are two options for the central detector to be determined with R\&D. One use an acrylic ball as the LS container. PMTs are installed on a truss and look inward in the water buffer. The water cherenkov detector for muon veto is optically separated from the water buffer for the center detector. Another option is a nylon balloon as the LS container. PMTs are installed on the inner wall of a stainless steel sphere and shielded with non-scintillation liquid. Both options are implemented in the MC and simulated to compare the detector performance. To achieve an energy resolution of $3\%/\sqrt{E}$, 17746 20-inch PMTs corresponding to 77\% photocathode coverage are used. The quantum efficiency is set to be 35\%. The attenuation length of the LS is set to be 20 m, larger than 15 m for the Daya Bay. Recently we have measured the rayleigh scattering length of the liquid scintillator to be $\sim\,$30 m. The light yield of the liquid scintillator is tuned to the data of the Daya Bay detector. We find that the $3\%/\sqrt{E}$ energy resolution is achieved with the above settings. Therefore, they served as the requirements for the JUNO detector design. The baseline parameters of the JUNO MC are shown in Tab.~\ref{tab:intro:parameters}, as well as in the following.
\begin{table}[htb]
\centering
\begin{tabular}{|c|c|}\hline
target mass & 20 kt \\ \hline
target radius & 17.7 m \\ \hline
target density & 0.856 g/cm$^3$ \\ \hline
mass fraction of C & 0.8792 \\ \hline
mass fraction of H & 0.1201 \\ \hline
light yield & 10400 /MeV \\ \hline
birks1 & $6.5 \times {10^{ - 3}}$ g/cm$^2$/MeV \\ \hline
birks2 & $1.5 \times {10^{ - 6}}$ (g/cm$^2$/MeV)$^2$ \\ \hline
emission time ${\tau _1}$ & 4.93 ns \\ \hline
emission time ${\tau _2}$ & 20.6 ns \\ \hline
emission time ${\tau _3}$ & 190 ns \\ \hline
${\tau _1}$ weight & 0.799 \\ \hline
${\tau _2}$ weight & 0.171 \\ \hline
${\tau _3}$ weight & 0.03 \\ \hline
attenuation length & 20 m \\ \hline
absorption length & 60 m \\ \hline
rayleigh scat. length & 30 m \\ \hline
optical coverage & 75\% \\ \hline
quantum efficiency & 35\% \\ \hline
%photonelectron yield & 1200 pe/MeV at (0,0,0) \\ \hline
\end{tabular}
\caption{Baseline parameters in the JUNO MC simulation.
\label{tab:intro:parameters}}
\end{table}

{\bf Detector dimensions.} The scintillator volume is 35.4 m in diameter, surrounded by a buffer medium with a thickness of 1.5 m, either water in the acrylic ball option or non-scintillation oil in the balloon option. PMTs are assumed to be 20-inch PMTs, with their bulk center located at 19.5 m in radius. The number of PMTs is 17746. The water pool, as the cherenkov detector for muons, is a cylinder of 42.5 m in diameter and 42.5 m in height.

{\bf Light emission.} The light yield of the liquid scintillator is about 10000 photons per MeV. The exact value may vary by the order of 10\% depending on the scintillator solvent and fluor concentrations. For different settings in the simulation, we normalize the light yield to the response of the Daya Bay detector. The light output also depends on the energy loss rate $dE/dx$ of the ionizing particle, resulting in a quenched visible energy. This effect is taken into account by the Birks' law,
\begin{equation}
{\rm Quenched\ Energy} = \frac{\rm Energy\ Deposit}{1 + C_1 \frac{dE}{dx} + C_2 \left(\frac{dE}{dx}\right)^2}\,.
\end{equation}
The light is emitted following a time profile described by a superposition of exponential decays. The simulation uses a description of three components with time constants ${\tau _1}$, ${\tau _2}$, and ${\tau _3}$, respectively.

{\bf Light propagation.} The scintillation light is produced in a relatively broad span of wavelengths. The emission spectrum of liquid scintillator is from the Daya Bay measurements. The photon absorption, re-emission, and Rayleigh scattering are simulated.

{\bf Light detection.} The baseline design for JUNO assumes an photocathode coverage of 77\% and 35\% peak quantum efficiency of the PMTs. The quantum efficiency spectrum is scaled from the Daya Bay PMTs.

Based on these configurations, the p.e.\ yield per deposited energy can be obtained as a function of the vertex position inside the target volume. The p.e.\ yield significantly depends on vertex position, the transparency of liquid scintillator container, and the refractive indices of liquid materials, as shown in Figure~\ref{fig:intro:gammaTotalPE}.
%%%%%%%%%%%%%%%%%%%%%%%%%%%%%%%%%%%%%%%%
\begin{figure}[!htb]
\centering
\includegraphics[width=.5\textwidth]{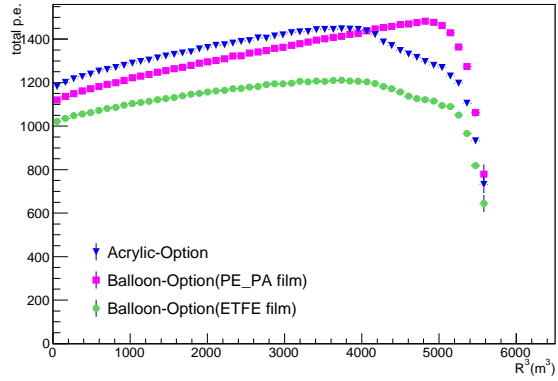}
\vspace{-0.15cm}
\caption{Total p.e. for 1 MeV gamma uniformly generated in the central detector.}
\label{fig:intro:gammaTotalPE}
\end{figure}
%%%%%%%%%%%%%%%%%%%%%%%%%%%%%%%%%%%%%%%%%

\subsubsection{Reconstruction and performance}
\label{subsubsec:reconstruction}
Since the energy of the reactor antineutrino is mainly below 10 MeV, it can be approximated as a point-like source in the JUNO detector. The vertex can be determined by the measurement of the time of flight to each PMT. In case of multiple photoelectrons that hit on the same PMT, only the first hit time is used in the reconstruction because the latter ones may be distorted due to the response of electronics. The vertex resolution depends on the PMT time resolution as well as the time spread of the light emission of the liquid scintillator.

The performance of the vertex reconstruction is studied using single $\gamma$ or positron uniformly distributed in the detector.
Fig.~\ref{fig:intro:VertexResolution} (left) shows the vertex resolution as a function of visible energy, assuming a PMT time resolution of 1~ns. Fig.~\ref{fig:intro:VertexResolution} (right) shows the impact of the PMT time resolution. It dominates the vertex performance when larger than 4~ns, while the decay time of the scintillation light becomes significant when it is better than 2~ns.

\begin{figure}[!htb]
\centering
\includegraphics[width=.42\textwidth]{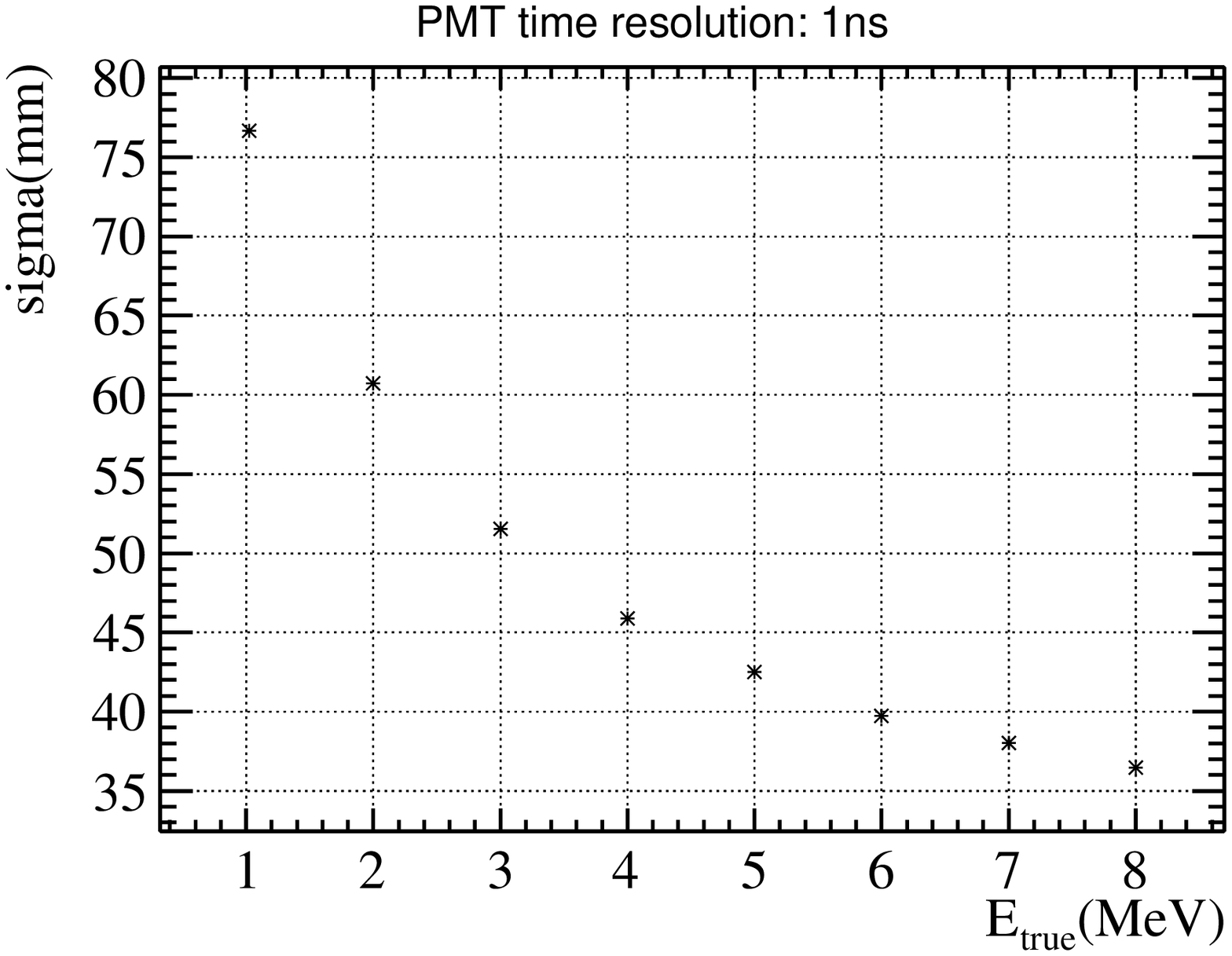}
\includegraphics[width=.45\textwidth]{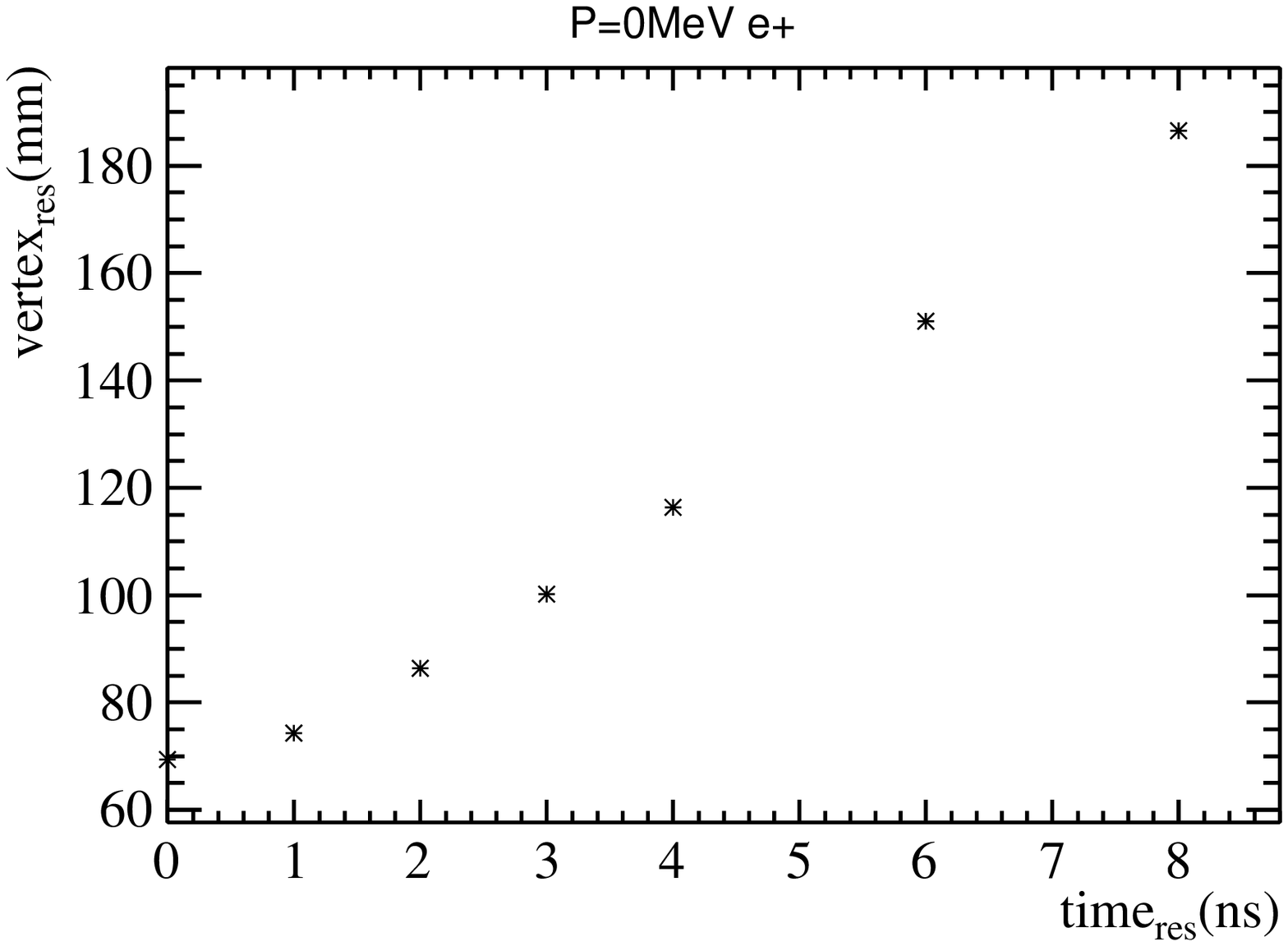}
\vspace{-0.15cm}
\caption{Vertex resolution as a function of energy (left) and PMT time resolution (right).}
\label{fig:intro:VertexResolution}
\end{figure}

The visible energy in the detector is reconstructed by comparing the measured charge of each PMT with the expectation, which relies on the event vertex, the light emission and propagation, and the photon detection with PMTs. The energy solution as a function of the visible energy can be described with a 3-parameter function
\begin{equation}
\frac{\sigma}{E}=\sqrt{\left(\frac{p_0}{\sqrt{E}}\right)^2+p_1^2+\left(\frac{p_2}{E}\right)^2}\,,
\end{equation}
where $E$ refers to the visible energy in MeV, $p_0$ is the leading term dominated by the photon statistics, $p_1$ and $p_2$ come from detector effects such as PMT dark noise, variation of the PMT QE, and the reconstructed vertex smearing. Toy MC samples are generated to study each term, which is shown in Tab.~\ref{tab:intro:energyres}.

\begin{table}[htb]
\centering
\begin{tabular}{|c|c|c|c|}\hline\hline
effects & p0 & p1 & p2 \\ \hline
statistical fluctuation & 2.68 & 0 & 0\\
PMT charge resolution(30\%) & 2.8 & 0 & 0\\
PMT dark noise(50kHz per PMT) & 2.68 & 0 & 0.9\\
PMT QE difference(20\%)& 2.68 & 0.26 & 0\\
vertex smearing(11cm @1MeV) & 2.68 & 0.17 & 0\\
\hline\hline
\end{tabular}
\caption{Factors that impact to the energy resolution.
\label{tab:intro:energyres}}
\end{table}

Muon tracking is important for the cosmogenic background rejection and the atmospheric neutrino study. Similar to the vertex reconstruction, a track is determined by the the first hit time of each PMT. For the muons that go through the detector, the PMTs around the injection point and the outgoing point see more lights and form two clusters, which can be used to estimate the initial tracking parameters. Toy MC samples are used to estimate the tracking performance. Assuming the intrinsic PMT time resolution is better than 4~ns,
for a muon track longer than 5~m (i.e. $>$ 1~GeV), the track length resolution is better than 0.5\% and the angular resolution is better than 1~degree. For a muon track between 1~m and 5~m (i.e. 0.2~GeV $<E<$ 1~GeV), the track length resolution is better than 1\% and the angular resolution is better than 10~degree. Tracking of electrons and identification of muon/electron are still under development.

\subsection{Antineutrino Detection in JUNO}
\label{subsec:intro:antiNeu}

Antineutrinos from reactors are detected by LS via the inverse beta decay reaction (shorten as IBD in the following contents):
\begin{equation*}
\bar\nu_e + p \rightarrow e^+ + n
\end{equation*}
The prompt positron signal and delayed neutron capture signal constitute an anti-neutrino event. In LS, neutron are captured by free protons or Carbon with a capture time of $\sim$216$\mu$s obtained by MC simulation. The delayed signal, 2.2 MeV $\gamma$-ray emitted after neutron captured on proton, can be contaminated by the natural radioactivity. Thus, a triple-coincidence criteria of energy, time and space is necessary to suppress the accidental background.

Full MC simulation and reconstruction is performed to obtain the antineutrino detection efficiency. The efficiency of (0.7 MeV, 12 MeV) energy cut on prompt signal is 100\%. The efficiency of (1.9 MeV, 2.5 MeV) cut on the delayed signal is 97.8\%, and the in-efficiency is mainly due to neutron-capture on Carbon. The efficiency for different time-, and space- correlation cuts are listed in Tab.~\ref{tab:intro:effT} and Tab.~\ref{tab:intro:effD}.

\begin{table}[!htb]
\centering
\begin{tabular}{|c|c|c|c|c|c|}\hline\hline
 Time Cut ($\mu$s) & 300 & 500 & 1000 & 1500 & 2000 \\ \hline
Efficiency (\%) & 74.55 & 89.57 & 98.68 & 99.49 & 99.60 \\ \hline\hline
\end{tabular}
\caption{IBD efficiency of different time cuts from MC simulation.
\label{tab:intro:effT}}
\end{table}

\begin{table}[!htb]
\centering
\begin{tabular}{|c|c|c|c|c|c|}\hline\hline
 Distance Cut (m) & 0.5 & 1 & 1.5 & 2.0 & 2.5 \\ \hline
Efficiency (\%) & 76.66 & 95.90 & 99.08 & 99.74 & 99.89 \\ \hline\hline
\end{tabular}
\caption{IBD efficiency of different distance cuts from MC simulation.
\label{tab:intro:effD}}
\end{table}

\subsection{Backgrounds in JUNO}

\subsubsection{Cosmic Muons at JUNO experimental site}
\label{subsec:intro:bkg:muon}

For underground neutrino observatories, sufficient amount of overburden above the detector is the most effective approach to suppress the cosmogenic backgrounds. The JUNO experimental site is chosen to be under a 286-m high mountain, and the detector will be at -480 m depth. The mountain profile is shown in Figure.~\ref{fig:intro:junomap}, converted from a high precision, digitized topographic map. The JUNO experimental site is at the coordinate of (0, 0, -480 m) in the map. The shortest distance from the mountain surface is 664 m. A modified Gaisser formula~\cite{Guo:2007ug} is used to describe the muon flux at sea level. With the mountain profile data, the cosmic muons are transported from the sea level to the underground JUNO detector site using the MUSIC~\cite{Antonioli:1997qw} package. A uniform rock density of 2.60g/cm$^3$ is assumed. The simulated muon rate and flux is shown in Tab.~\ref{tab:intro:muflux}.

\begin{figure}[htb!]
\centering
\includegraphics[width=0.8\textwidth]{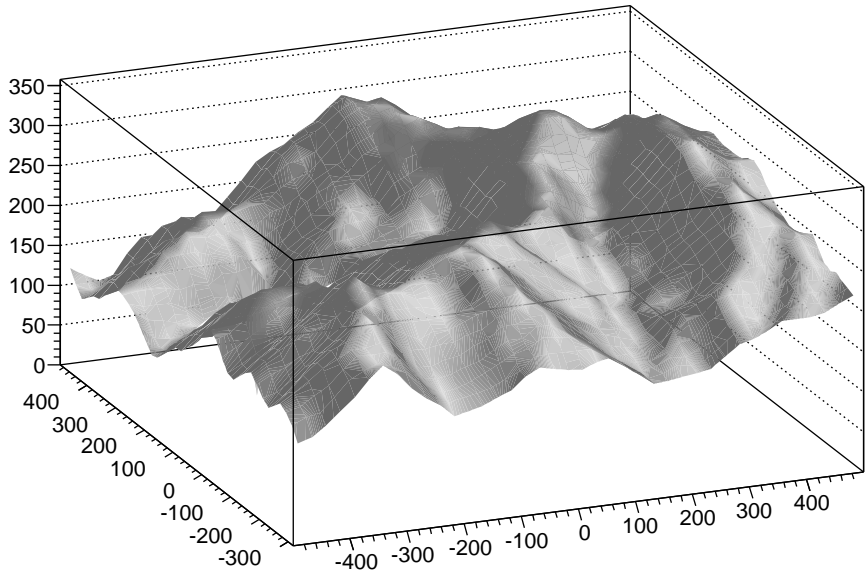}
\vspace{-0.15cm}
\caption{Mountain profile at the JUNO experimental site.
\label{fig:intro:junomap}}
\end{figure}

\begin{table}[htb]
\centering
\begin{tabular}{|c|c|c|c|c|}\hline\hline
Overburden & Muon flux & $<E_\mu>$ & $R_\mu$ in CD & $R_\mu$ in WP \\ \hline
748 m & 0.003 Hz/m$^{2}$ & 215 GeV & 3.0 Hz & 1.0 Hz  \\ \hline\hline
\end{tabular}
\caption{The simulated muon flux and mean energy at JUNO site.
\label{tab:intro:muflux}}
\end{table}

A parametrization method~\cite{Becherini:2005sr} is used to investigate the muon bundles, and it gives 20\% probability of muon bundle at the depth of JUNO detector. The probability of multiple muon bundle reduces to 10\% after taking into account the JUNO detector geometry. MC simulation indicates the muons in the bundle are almost in parallel within 0.2$^\circ$. The multiplicity of muons going through JUNO detector is shown in Table.~\ref{tab:intro:muMulti}.

\begin{table}[htb]
\centering
\begin{tabular}{|c|c|c|c|c|c|c|}\hline\hline
Multiplicity & 1 & 2 & 3 & 4 & 5 & 6 \\ \hline
Fraction & 89.6\% & 7.7\% & 1.8\% & 0.6\% & 0.3\% & 0.07\% \\ \hline\hline
\end{tabular}
\caption{The multiplicity of muons going through JUNO detector.
\label{tab:intro:muMulti}}
\end{table}

\subsubsection{Neutrons}
\label{subsec:intro:bkg:neutron}

The neutron production depends on the muon flux and the average energy at the JUNO detector. Full muon simulation using Geant4 gives $\sim$1.8 Hz spallation neutrons in JUNO LS.

Neutron produced by muons passing through the JUNO LS will be tagged with almost 100\% efficiency. Neutron produced in water buffer can be tagged with an efficiency of 99.8\%, since their parent muons pass through the muon systems. The 0.2\% inefficiency is mainly from the corner clipping muons. The tagged neutrons can be rejected by sufficient time veto after the tagged muons to suppress the possible correlated background.

Neutrons produced by the un-tagged corner clipping muons and in the surrounding rocks arise from the muons missing muon systems, have to traverse at least 3.2 m for the ``Sphere Acrylic" option (or 2.5 m for ``Balloon" option) to reach JUNO LS. The ``un-tagged" neutrons can enter LS and produce prompt proton-recoil signal then captured by H or C in LS. These events, namely the fast neutron events, can mimic the anti-neutrino signal as correlated backgrounds. A full MC simulation is performed to propagate the neutrons produced by cosmic muons in water pool and surrounding rocks, and the fast neutrons mimicking the anti-neutrino events are estimated to be $\sim$0.1/day.

\subsubsection{Cosmogenic Isotopes}
\label{subsec:intro:bkg:cosmogenic}

In the liquid scintillator, the energetic cosmic muons and subsequent showers can interact with $^{12}$C and produce radioactive isotopes with $Z\leq6$ by electromagnetic or hadronic processes. Among them, $^{9}$Li and $^{8}$He with half-lives of 0.178 s and 0.119 s, respectively, are the most serious correlated background source to reactor anti-neutrinos, because they can decay by emitting both a beta and a neutron which mimics as an anti-neutrino signal. The estimation of $^{9}$Li and $^{8}$He is discussed in detail in Section~\ref{subsec:mh:sigbkg:bkgest}. Other isotopes, such as $^{11}$Li, $^{12}$Be, $^{14}$B, $^{16}$C, $^{17}$N and $^{18}$N, are also beta-neutron emitters but have much less contribution to the background.

The other long-lived cosmogenic isotopes have beta decay without an accompanying neutron. They can not form correlated backgrounds by themselves but can contribute the neutron-like signal if they have beta decay energy in the 1.9-2.5 MeV range. Measuring the isotopes production yield in JUNO detector is useful for understanding the muon shower spallation processes. The rates of those cosmogenic radioactive isotopes from FLUKA MC simulation are listed in Table.~\ref{tab:intro:isotopes}. Recently there are measurements from KamLAND~\cite{Abe:2009aa} and Borexino~\cite{Bellini:2013pxa}. Physics-driven model is built to analyze the cosmogenic backgrounds in Super-K~\cite{Li:2014sea,Li:2015kpa}, which could significantly reduce the backgrounds, and also increase the live time by well defining the shower position. Cosmogenic backgrounds in JUNO detector will be elaborated with these studies.

\begin{table}[!htb]
\centering
\begin{tabular}{|c|c|c|c|}\hline\hline
Isotopes & $Q$ (MeV) & $T_{1/2}$ & Rate (per day) \\ \hline
$^{3}$H & 0.0186 ($\beta^-$) & 12.31 year & 1.14$\times10^4$ \\
$^{6}$He & 3.508 ($\beta^-$) & 0.807 s & 544 \\

$^{7}$Be & $Q_{EC}$=0.862 (10.4\% $\gamma$, $E_\gamma=0.478$) & 53.22 day & 5438 \\

$^{8}$He & 10.66 ($\beta^-\gamma:84\%$), 8.63 ($\beta^-n:16\%$) & 0.119 s & 11\\
$^{8}$Li & 16.0 ($\beta^-$) & 0.839 s & 938 \\
$^{8}$B & 16.6 ($\beta^+$) & 0.770 s & 225  \\
%{\it $^{8}$Be} & {\it 0.092 ($\alpha$ decay)} & {\it 5.57 eV} & - \\

$^{9}$Li & 13.6 ($\beta^-:49\%$), 11.94 ($\beta^-n:51\%$) & 0.178 s & 94 \\
$^{9}$C & 15.47 ($\beta^+p:61.6\%, \beta^+\alpha:38.4\%$) & 0.126 s & 31 \\
%{\it $^{9}$B} & {\it 0.186 ($p$+$^{8}$Be)} & {\it 0.54 keV} & - \\

$^{10}$Be & 0.556 ($\beta^-$) & 1.51e6 year & 1419  \\
$^{10}$C & 2.626 ($\beta^+\gamma$) & 19.29 s & 482  \\

$^{11}$Li & 20.55 ($\beta^-n:83\%$, $\beta^-2n:4.1\%$) & 0.00875 s & 0.06\\
$^{11}$Be & 11.51 ($\beta^-\gamma:96.9\%$), 2.85 ($\beta^-\alpha:3.1\%$) & 13.76 s &  24 \\
$^{11}$C & 0.960 ($\beta^+$) & 20.36 min & 1.62$\times10^4$ \\

$^{12}$Be & 11.708 ($\beta^-\gamma$, $\beta^-n:0.5\%$) & 0.0215 s & 0.45 \\
$^{12}$B & 13.37 ($\beta^-\gamma$) & 0.0202 s & 966 \\
$^{12}$N & 16.316 ($\beta^+\gamma$) & 0.0110 s & 17 \\

$^{13}$B & 13.437 ($\beta^-\gamma$) & 0.0174 s & 12 \\
$^{13}$N & 1.198 ($\beta^+$) & 9.965 min & 19 \\

$^{14}$B & 20.644 ($\beta^-\gamma$, $\beta^-n:6.1\%$) & 0.0126 s & 0.021 \\
$^{14}$C & 0.156 ($\beta^-$) & 5730 year & 132 \\

$^{15}$C & 9.772 ($\beta^-$) & 2.449 s & 0.6 \\

$^{16}$C & 8.010 ($\beta^-n:99\%$) & 0.747 s & 0.012 \\
$^{16}$N & 10.42 ($\beta^-\gamma$) & 7.130 s & 13 \\

$^{17}$N & 8.680 ($\beta^-\gamma:5\%$), 4.536 ($\beta^-n:95\%$) & 4.173 s & 0.42 \\
$^{18}$N & 13.896 ($\beta^-\gamma:93\%$), 5.851 ($\beta^-n:7\%$) & 0.620 s & 0.009 \\
\hline
neutron & & & 155 000 \\
\hline\hline
\end{tabular}
\caption{The estimated rates for cosmogenic isotopes in JUNO LS by FLUKA simulation, in which the oxygen isotopes are neglected. The decay modes and Q values are from TUNL Nuclear Data Group~\cite{TUNLndg}.
\label{tab:intro:isotopes}}
\end{table}

\subsubsection{Natural Radioactivity}
\label{subsec:intro:bkg:natural}

Natural radioactivity exists in the material of JUNO detector components and its surroundings. Particular care need be taken to select low radioactivity materials and design the passive shielding to control the radioactivity background. For JUNO experiment, the radioactivity comes from various sources. For simplicity, the radioactive isotopes $^{238}$U, $^{232}$Th, $^{222}$Rn, $^{85}$Kr, $^{60}$Co, $^{40}$K are shorten as U, Th, Rn, Kr, Co, K in the following text. The radioactive sources include:
\begin{itemize}
  \item U/Th/K in the rocks around the detector hall
  \item U/Th/K and Rn dissolved in the water buffer
  \item U/Th/K/Co in the stainless steel (vessel or strut)
  \item Rn and Kr in air
  \item U/Th/K in the PMT glass
  \item U/Th/K in the liquid scintillator container (acrylic or polymer film)
  \item U/Th/K/Kr/Ar in the liquid scintillator
  \item Dust and other impurities
\end{itemize}
In the following, the rate of singles means signals from radioactivity depositing $>$0.7 MeV of visible energy in LS.

Assuming the radioactivity of the rock at JUNO experimental site is similar as that measured at Daya Bay site: $\sim$10 ppm for U, $\sim$30 ppm for Th and $\sim$5 ppm for K. Since a full MC simulation would be extremely time consuming, thus a numerical calculation is performed to estimate the effect of the rock radioactivity: 1) divide the 50 cm-thick rock around the water pool into small voxels; 2) the activities of $\gamma$ rays with different energies are calculated for each voxel by using the radioactivity generator; 3) then for each gamma energy, the effective solid angle of each voxel to the LS detector is calculated by taking into account the attenuation of different water thickness; 4) in the end the singles rate is obtained by summing the contributions from all voxels. With the shielding of 3.2 meter buffer, there are $\sim$0.61 Hz, $\sim$6.74 Hz and $\sim$0.07 Hz singles rates in all LS volume for U/Th/K, respectively. After fiducial volume cut (R$<17.2$ m), the total singles rate reduces to 0.98 Hz.

The water buffer of JUNO will be circulated and purified to achieve a long absorption length for Cherenkov photons as well as low radioactivity. In addition, there will be nitrogen flow on the top of water pool and anti-Rn liner (e.g, 2 mm HDPE film) on the water pool walls to control Radon permeation into water buffer. In the ``Acrylic Sphere" option of central detector, water acts as the buffer liquid and is just outside LS, thus the Radon dissolved in water will contribute more singles than the ``Balloon" option. A MC simulation gives 16 Hz singles rate in all LS volume if the Radon concentration in water is 0.2 Bq/m$^3$, and the rate will reduce to 1.3 Hz inside the $R<17.2$m volume.

Based on the experience from the existing neutrino experiments, the projected radioactivity of detector materials such as PMT glass, acrylic, polymer film, steel and copper is listed in Table.~\ref{tab:intro:rad}.

\begin{table}[htb]
\centering
\begin{tabular}{|c|c|c|c|c|}\hline\hline
Detector material & $^{238}$U  & $^{238}$Th & $^{40}$K & $^{60}$Co \\ \hline
PMT glass & 22 ppb & 20 ppb & 3.54 ppb & -   \\
Acrylic & 10 ppt & 10 ppt & 10 ppt & -  \\
Polymer film & 2 ppt & 4 ppt & 1 ppt & - \\
Steel & 0.096 ppb & 1.975 ppb & 0.049 ppb & 0.002 Bq/kg \\
Copper & 1.23 mBq/kg & 0.405 mBq/kg & 0.0377 mBq/kg & - \\
\hline\hline
\end{tabular}
\caption{The estimated radioactivity of JUNO detector construction materials.
\label{tab:intro:rad}}
\end{table}

Above external radioactivity can be rejected by proper fiducial volume cut since their energy deposits are mainly at the LS edge. Thus, the internal LS radio-purity is very important to the JUNO experiment and should be well controlled. The fractional distillation process at the last step of raw LAB production and water extraction of the fluors are necessary to improve the radio-purity of raw LS materials. There will be nitrogen protection during LS production and handling to suppress Radon contamination. In addition, the residual Radon contamination will lead to non-equilibrium isotope $^{210}$Pb (and the subsequent $^{210}$Bi decay) which has 22-year half life. The $^{210}$Pb isotope is the dominant background in searching solar neutrinos, as discussed in Section.6. On-line purification, such as distillation, is required to remove $^{210}$Pb from LS to open the opportunity to observe $^{7}$Be solar neutrino. In the JUNO experiment, an initial purity level of $10^{-15}$ g/g U/Th, $10^{-16}$ g/g K and $1.4\cdot10^{-22}$ g/g $^{210}$Pb can be achieved without distillation. After setting up the on-line distillation, we believe two orders of magnitude better purity level can be achieved: $10^{-17}$ g/g U/Th, $10^{-18}$ g/g K and $10^{-24}$ g/g $^{210}$Pb. The purity level of other isotope such as $^{85}$Kr and $^{39}$Ar is also listed.

\begin{table}[htb]
\centering
\begin{tabular}{|c|c|c|c|c|c|c|}\hline\hline
 LS & $^{238}$U  & $^{238}$Th & $^{40}$K & $^{210}$Pb & $^{85}$Kr & $^{39}$Ar \\ \hline
No distillation & $10^{-15}$ g/g & $10^{-15}$ g/g & $10^{-16}$ g/g & $1.4\cdot10^{-22}$ g/g & 50 $\mu$Bq/m$^{3}$ & 50 $\mu$Bq/m$^{3}$  \\
After distillation & $10^{-17}$ g/g & $10^{-17}$ g/g & $10^{-18}$ g/g & $10^{-24}$ g/g & 1 $\mu$Bq/m$^{3}$ & - \\
\hline\hline
\end{tabular}
\caption{The estimated radioactivity of JUNO LS.
\label{tab:intro:radls}}
\end{table}

Full MC simulation is performed to obtain the singles rates from LS and other detector construction materials. Taking the ``Acrylic Sphere" option of central detector, the singles rates in different volumes are listed in Table.~\ref{tab:intro:rate1}.

\begin{table}[htb]
\centering
\begin{tabular}{|c|c|c|c|c|c|c|}\hline\hline
Fiducial Cut & LS (Hz) & PMT (Hz) & Acrylic (Hz) & Strut (Hz) & Fastener (Hz) & Sum (Hz) \\ \hline
R$<$17.7 m & 2.39 & 2.43 & 69.23 & 0.89 & 0.82 & 75.75  \\ \hline
R$<$17.6 m & 2.35 & 1.91 & 41.27 & 0.66 & 0.55 & 46.74  \\ \hline
R$<$17.5 m & 2.31 & 1.03 & 21.82 & 0.28 & 0.32 & 25.75  \\ \hline
R$<$17.4 m & 2.27 & 0.75 & 12.23 & 0.22 & 0.19 & 15.66  \\ \hline
R$<$17.3 m & 2.24 & 0.39 & 6.47 & 0.13 &  0.12 & 9.33   \\ \hline
R$<$17.2 m & 2.20 & 0.33 & 3.61 & 0.083 & 0.087 & 6.31  \\ \hline
R$<$17.1 m & 2.16 & 0.23 & 1.96 & 0.060 & 0.060 & 4.47  \\ \hline
R$<$17.0 m & 2.12 & 0.15 & 0.97 & 0.009 & 0.031 & 3.28  \\ \hline\hline
\end{tabular}
\caption{The simulated singles rates from different detector components.
\label{tab:intro:rate1}}
\end{table}

A fiducial volume cut is necessary to reject the external radioactivity thus reduce the accidental background and $(\alpha, n)$ background, as discussed in Section.2. The total singles rate will reduce to $\sim$7.6 Hz if applying an R$<17.2$ m fiducial volume cut, as shown in Table.~\ref{tab:intro:rate2}.

\begin{table}[htb]
\centering
\begin{tabular}{|c|c|c|c|c|}\hline\hline
Fiducial Cut & Detector Components & Radon in water & Rock & Total \\ \hline
R$<$17.2 m & 6.3 Hz & 1.3 Hz & 0.98 Hz & 7.63 Hz  \\ \hline\hline
\end{tabular}
\caption{The summary of singles rates inside the Fiducial volume from the detector components, Radon in water and rock.
\label{tab:intro:rate2}}
\end{table}

\clearpage

\section*{Acknowledgement}

The JUNO Experiment is supported by
the Chinese Academy of Sciences,
the Ministry of Science and Technology of China,
the National Natural Science Foundation of China,
Shanghai Jiao Tong University,
and Tsinghua University in China,
the Belgian Fonds de la Recherche Scientifique, FNRS in Belgium,
the Charles University in Prague, Czech Republic,
the CNRS National Institute of Nuclear and Particle Physics (IN2P3) in France,
the Deutsche Forschungsgemeinschaft (DFG),
Helmholtz Alliance for Astroparticle Physics (HAP),
and Helmholtz Gemeinschaft (HGF) in Germany,
the Istituto Nazionale di Fisica Nucleare (INFN) in Italy,
the Joint Institute for Nuclear Research (JINR) in Russia,
and the NSFC-RFBR joint research program.
We acknowledge the generous support of the Guangdong provincial government
and the Jiangmen and Kaiping Municipal Government.
We are grateful for the ongoing cooperation from the China General Nuclear Power Group (CGNPC).

\clearpage
%the Ministry of Education, Youth and Sports of the Czech Republic, Charles University in Prague, the Czech Science Foundation,

% chap: intro
%%%%%%%%%%%%%%%%%%%%%%%%%%%%%%%%%%%%%%%%%%%%%%%%%%%%%%%%%%%%%%%%%%%%%%%
%%%%%%%%%%%%%%%%%%%%%%%%%%%%%%%%%%%%%%%%%%%%%%%%%%%%%%%%%%%%%%%%%%%%%%%
% chap: intro


\begin{thebibliography}{999}

%\cite{Weinberg:1967tq}
\bibitem{Weinberg:1967tq}
  S.~Weinberg,
  %``A Model of Leptons,''
  Phys.\ Rev.\ Lett.\  {\bf 19}, 1264 (1967).  %%CITATION = PRLTA,19,1264;%%

%\cite{Davis:1968cp}
\bibitem{Davis:1968cp}
  R.~Davis, Jr., D.~S.~Harmer and K.~C.~Hoffman,
  %``Search for neutrinos from the sun,''
  Phys.\ Rev.\ Lett.\  {\bf 20}, 1205 (1968).  %%CITATION = PRLTA,20,1205;%%

%\cite{Bahcall:1968hc}
\bibitem{Bahcall:1968hc}
  J.~N.~Bahcall, N.~A.~Bahcall and G.~Shaviv,
  %``Present status of the theoretical predictions for the Cl-36 solar neutrino experiment,''
  Phys.\ Rev.\ Lett.\  {\bf 20}, 1209 (1968).  %%CITATION = PRLTA,20,1209;%%

%\cite{Bahcall:1981zh}
\bibitem{Bahcall:1981zh}
  J.~N.~Bahcall, W.~F.~Huebner, S.~H.~Lubow, P.~D.~Parker and R.~K.~Ulrich,
  %``Standard Solar Models and the Uncertainties in Predicted Capture Rates of Solar Neutrinos,''
  Rev.\ Mod.\ Phys.\  {\bf 54}, 767 (1982).  %%CITATION = RMPHA,54,767;%%

%\cite{Maki:1962mu}
\bibitem{Maki:1962mu}
  Z.~Maki, M.~Nakagawa and S.~Sakata,
  %``Remarks on the unified model of elementary particles,''
  Prog.\ Theor.\ Phys.\  {\bf 28}, 870 (1962).  %%CITATION = PTPKA,28,870;%%

%\cite{Pontecorvo:1967fh}
\bibitem{Pontecorvo:1967fh}
  B.~Pontecorvo,
  %``Neutrino Experiments and the Problem of Conservation of Leptonic Charge,''
  Sov.\ Phys.\ JETP {\bf 26}, 984 (1968)
  [Zh.\ Eksp.\ Teor.\ Fiz.\  {\bf 53}, 1717 (1967)].  %%CITATION = SPHJA,26,984;%%

%\cite{Cabibbo:1963yz}
\bibitem{Cabibbo:1963yz}
  N.~Cabibbo,
  %``Unitary Symmetry and Leptonic Decays,''
  Phys.\ Rev.\ Lett.\  {\bf 10}, 531 (1963).  %%CITATION = PRLTA,10,531;%%

%\cite{Kobayashi:1973fv}
\bibitem{Kobayashi:1973fv}
  M.~Kobayashi and T.~Maskawa,
  %``CP Violation in the Renormalizable Theory of Weak Interaction,''
  Prog.\ Theor.\ Phys.\  {\bf 49}, 652 (1973).  %%CITATION = PTPKA,49,652;%%

%\cite{Xing:2011zza}
\bibitem{Xing:2011zza}
  Z.~z.~Xing and S.~Zhou,
  ``Neutrinos in particle physics, astronomy and cosmology,''
  Springer-Verlag, Berlin Heidelberg (2011).

%\cite{Konetschny:1977bn}
\bibitem{Konetschny:1977bn}
  W.~Konetschny and W.~Kummer,
  %``Nonconservation of Total Lepton Number with Scalar Bosons,''
  Phys.\ Lett.\ B {\bf 70}, 433 (1977).  %%CITATION = PHLTA,B70,433;%%

%\cite{Magg:1980ut}
\bibitem{Magg:1980ut}
  M.~Magg and C.~Wetterich,
  %``Neutrino Mass Problem and Gauge Hierarchy,''
  Phys.\ Lett.\ B {\bf 94}, 61 (1980).  %%CITATION = PHLTA,B94,61;%%

%%\cite{Schechter:1980gr}
%\bibitem{Schechter:1980gr}
%  J.~Schechter and J.~W.~F.~Valle,
%  %``Neutrino Masses in SU(2) x U(1) Theories,''
%  Phys.\ Rev.\ D {\bf 22}, 2227 (1980).  %%CITATION = PHRVA,D22,2227;%%

%\cite{Cheng:1980qt}
\bibitem{Cheng:1980qt}
  T.~P.~Cheng and L.~F.~Li,
  %``Neutrino Masses, Mixings and Oscillations in SU(2) x U(1) Models of Electroweak Interactions,''
  Phys.\ Rev.\ D {\bf 22}, 2860 (1980).  %%CITATION = PHRVA,D22,2860;%%

%\cite{Lazarides:1980nt}
\bibitem{Lazarides:1980nt}
  G.~Lazarides, Q.~Shafi and C.~Wetterich,
  %``Proton Lifetime and Fermion Masses in an SO(10) Model,''
  Nucl.\ Phys.\ B {\bf 181}, 287 (1981).  %%CITATION = NUPHA,B181,287;%%

%\cite{Mohapatra:1980yp}
\bibitem{Mohapatra:1980yp}
  R.~N.~Mohapatra and G.~Senjanovic,
  %``Neutrino Masses and Mixings in Gauge Models with Spontaneous Parity Violation,''
  Phys.\ Rev.\ D {\bf 23}, 165 (1981).  %%CITATION = PHRVA,D23,165;%%

%\cite{Minkowski:1977sc}
\bibitem{Minkowski:1977sc}
  P.~Minkowski,
  %``$\mu \to e\gamma$ at a Rate of One Out of $10^{9}$ Muon Decays?,''
  Phys.\ Lett.\ B {\bf 67}, 421 (1977).  %%CITATION = PHLTA,B67,421;%%

%\cite{Yanagida:1979as}
\bibitem{Yanagida:1979as}
T.~Yanagida, in {\it Proceedings of the Workshop on Unified Theory and the Baryon Number of the Universe},
edited by O.~Sawada and A.~Sugamoto (KEK, Tsukuba, 1979), p. 95.

%\cite{GellMann:1980vs}
\bibitem{GellMann:1980vs}
M.~Gell-Mann, P.~Ramond, and R.~Slansky, in {\it Supergravity},
edited by P. van Nieuwenhuizen and D. Freedman (North Holland, Amsterdam, 1979), p. 315.

\bibitem{Glashow:1980unknown1}
S.~L.~Glashow, in {\it Quarks and Leptons},
edited by M. L$\acute{\rm  e}$vy {\it et al.} (Plenum, New York, 1980), p. 707.

%\cite{Mohapatra:1979ia}
\bibitem{Mohapatra:1979ia}
  R.~N.~Mohapatra and G.~Senjanovic,
  %``Neutrino Mass and Spontaneous Parity Violation,''
  Phys.\ Rev.\ Lett.\  {\bf 44}, 912 (1980).  %%CITATION = PRLTA,44,912;%%

%\cite{Schechter:1980gr}
\bibitem{Schechter:1980gr}
  J.~Schechter and J.~W.~F.~Valle,
  %``Neutrino Masses in SU(2) x U(1) Theories,''
  Phys.\ Rev.\ D {\bf 22}, 2227 (1980).  %%CITATION = PHRVA,D22,2227;%%

%\cite{Foot:1988aq}
\bibitem{Foot:1988aq}
  R.~Foot, H.~Lew, X.~G.~He and G.~C.~Joshi,
  %``Seesaw Neutrino Masses Induced by a Triplet of Leptons,''
  Z.\ Phys.\ C {\bf 44}, 441 (1989).  %%CITATION = ZEPYA,C44,441;%%

%\cite{Antusch:2006vwa}
\bibitem{Antusch:2006vwa}
  S.~Antusch, C.~Biggio, E.~Fernandez-Martinez, M.~B.~Gavela and J.~Lopez-Pavon,
  %``Unitarity of the Leptonic Mixing Matrix,''
  JHEP {\bf 0610}, 084 (2006)
  [hep-ph/0607020].  %%CITATION = HEP-PH/0607020;%%

%\cite{Antusch:2014woa}
\bibitem{Antusch:2014woa}
  S.~Antusch and O.~Fischer,
  %``Non-unitarity of the leptonic mixing matrix: Present bounds and future sensitivities,''
  JHEP {\bf 1410}, 94 (2014)
  [arXiv:1407.6607 [hep-ph]].  %%CITATION = ARXIV:1407.6607;%%

%\cite{Xing:2012kh}
\bibitem{Xing:2012kh}
  Z.~z.~Xing,
  %``Towards testing the unitarity of the 3X3 lepton flavor mixing matrix in a precision reactor antineutrino oscillation experiment,''
  Phys.\ Lett.\ B {\bf 718}, 1447 (2013)
  [arXiv:1210.1523 [hep-ph]].  %%CITATION = ARXIV:1210.1523;%%

%\cite{Li:2015oal}
\bibitem{Li:2015oal}
  Y.~F.~Li and S.~Luo,
  %``Neutrino Oscillation Probabilities in Matter with Direct and Indirect Unitarity Violation in the Lepton Mixing Matrix,''
  arXiv:1508.00052 [hep-ph].
  %%CITATION = ARXIV:1508.00052;%%

%\cite{Agashe:2014kda}
\bibitem{Agashe:2014kda}
  K.~A.~Olive {\it et al.}  [Particle Data Group Collaboration],
  %``Review of Particle Physics,''
  Chin.\ Phys.\ C {\bf 38}, 090001 (2014).  %%CITATION = CHPHD,C38,090001;%%


%\cite{Capozzi:2013csa}
\bibitem{Capozzi:2013csa}
  F.~Capozzi, G.~L.~Fogli, E.~Lisi, A.~Marrone, D.~Montanino and A.~Palazzo,
  %``Status of three-neutrino oscillation parameters, circa 2013,''
  Phys.\ Rev.\ D {\bf 89}, 093018 (2014)
  [arXiv:1312.2878 [hep-ph]].  %%CITATION = ARXIV:1312.2878;%%

%\cite{Forero:2014bxa}
\bibitem{Forero:2014bxa}
  D.~V.~Forero, M.~Tortola and J.~W.~F.~Valle,
  %``Neutrino oscillations refitted,''
  Phys.\ Rev.\ D {\bf 90}, no. 9, 093006 (2014)
  [arXiv:1405.7540 [hep-ph]].  %%CITATION = ARXIV:1405.7540;%%

%\cite{Gonzalez-Garcia:2014bfa}
\bibitem{Gonzalez-Garcia:2014bfa}
  M.~C.~Gonzalez-Garcia, M.~Maltoni and T.~Schwetz,
  %``Updated fit to three neutrino mixing: status of leptonic CP violation,''
  JHEP {\bf 1411}, 052 (2014)
  [arXiv:1409.5439 [hep-ph]].  %%CITATION = ARXIV:1409.5439;%%

%\cite{An:2012eh}
\bibitem{An:2012eh}
  F.~P.~An {\it et al.}  [Daya Bay Collaboration],
  %``Observation of electron-antineutrino disappearance at Daya Bay,''
  Phys.\ Rev.\ Lett.\  {\bf 108}, 171803 (2012)
  [arXiv:1203.1669 [hep-ex]].  %%CITATION = ARXIV:1203.1669;%%

%\cite{An:2012bu}
\bibitem{An:2012bu}
  F.~P.~An {\it et al.}  [Daya Bay Collaboration],
  %``Improved Measurement of Electron Antineutrino Disappearance at Daya Bay,''
  Chin.\ Phys.\ C {\bf 37}, 011001 (2013)
  [arXiv:1210.6327 [hep-ex]].  %%CITATION = ARXIV:1210.6327;%%

%\cite{An:2013zwz}
\bibitem{An:2013zwz}
  F.~P.~An {\it et al.}  [Daya Bay Collaboration],
  %``Spectral measurement of electron antineutrino oscillation amplitude and frequency at Daya Bay,''
  Phys.\ Rev.\ Lett.\  {\bf 112}, 061801 (2014)
  [arXiv:1310.6732 [hep-ex]].  %%CITATION = ARXIV:1310.6732;%%

%\cite{Majorana:1937vz}
\bibitem{Majorana:1937vz}
  E.~Majorana,
  %``Theory of the Symmetry of Electrons and Positrons,''
  Nuovo Cim.\  {\bf 14}, 171 (1937).  %%CITATION = NUCIA,14,171;%%

%\cite{Furry:1939qr}
\bibitem{Furry:1939qr}
  W.~H.~Furry,
  %``On transition probabilities in double beta-disintegration,''
  Phys.\ Rev.\  {\bf 56}, 1184 (1939).  %%CITATION = PHRVA,56,1184;%%

%\cite{Rodejohann:2011mu}
\bibitem{Rodejohann:2011mu}
  W.~Rodejohann,
  %``Neutrino-less Double Beta Decay and Particle Physics,''
  Int.\ J.\ Mod.\ Phys.\ E {\bf 20}, 1833 (2011)
  [arXiv:1106.1334 [hep-ph]].  %%CITATION = ARXIV:1106.1334;%%

%\cite{Dell'Oro:2014yca}
\bibitem{Dell'Oro:2014yca}
  S.~Dell'Oro, S.~Marcocci and F.~Vissani,
  %``New expectations and uncertainties on neutrinoless double beta decay,''
  Phys.\ Rev.\ D {\bf 90}, no. 3, 033005 (2014)
  [arXiv:1404.2616 [hep-ph]].  %%CITATION = ARXIV:1404.2616;%%

%\cite{Bilenky:2014uka}
\bibitem{Bilenky:2014uka}
  S.~M.~Bilenky and C.~Giunti,
  %``Neutrinoless Double-Beta Decay: a Probe of Physics Beyond the Standard Model,''
  Int.\ J.\ Mod.\ Phys.\ A {\bf 30}, no. 04n05, 1530001 (2015)
  [arXiv:1411.4791 [hep-ph]].  %%CITATION = ARXIV:1411.4791;%%

\bibitem{Li:2010vy}
  Y.~F.~Li and Z.~z.~Xing,
  %``Possible Capture of keV Sterile Neutrino Dark Matter on Radioactive $\beta$-decaying Nuclei,''
  Phys.\ Lett.\ B {\bf 695}, 205 (2011)
  [arXiv:1009.5870 [hep-ph]].  %%CITATION = ARXIV:1009.5870;%%

%\cite{Bernabeu:2003yp}
\bibitem{Bernabeu:2003yp}
  J.~Bernabeu, S.~Palomares Ruiz and S.~T.~Petcov,
  %``Atmospheric neutrino oscillations, theta(13) and neutrino mass hierarchy,''
  Nucl.\ Phys.\ B {\bf 669}, 255 (2003)
  [hep-ph/0305152].
  %%CITATION = HEP-PH/0305152;%%

%\cite{Dighe:1999bi}
\bibitem{Dighe:1999bi}
  A.~S.~Dighe and A.~Y.~Smirnov,
  %``Identifying the neutrino mass spectrum from the neutrino burst from a supernova,''
  Phys.\ Rev.\ D {\bf 62}, 033007 (2000)
  [hep-ph/9907423].
  %%CITATION = HEP-PH/9907423;%%

%\cite{Petcov:2001sy}
\bibitem{Petcov:2001sy}
  S.~T.~Petcov and M.~Piai,
  %``The LMA MSW solution of the solar neutrino problem, inverted neutrino mass hierarchy and reactor neutrino experiments,''
  Phys.\ Lett.\ B {\bf 533}, 94 (2002)
  [hep-ph/0112074].
  %%CITATION = HEP-PH/0112074;%%

%\cite{Qian:2015waa}
\bibitem{Qian:2015waa}
  X.~Qian and P.~Vogel,
  %``Neutrino Mass Hierarchy,''
  Prog.\ Part.\ Nucl.\ Phys.\  {\bf 83}, 1 (2015)
  [arXiv:1505.01891 [hep-ex]].
  %%CITATION = ARXIV:1505.01891;%%

%\cite{Patterson:2015xja}
\bibitem{Patterson:2015xja}
  R.~B.~Patterson,
  %``Prospects for Measurement of the Neutrino Mass Hierarchy,''
  arXiv:1506.07917 [hep-ex].

%\cite{Fukugita:1986hr}
\bibitem{Fukugita:1986hr}
  M.~Fukugita and T.~Yanagida,
  %``Baryogenesis Without Grand Unification,''
  Phys.\ Lett.\ B {\bf 174}, 45 (1986).  %%CITATION = PHLTA,B174,45;%%

%\cite{Buchmuller:2004nz}
\bibitem{Buchmuller:2004nz}
  W.~Buchmuller, P.~Di Bari and M.~Plumacher,
  %``Leptogenesis for pedestrians,''
  Annals Phys.\  {\bf 315}, 305 (2005)
  [hep-ph/0401240].  %%CITATION = HEP-PH/0401240;%%

%\cite{Davidson:2008bu}
\bibitem{Davidson:2008bu}
  S.~Davidson, E.~Nardi and Y.~Nir,
  %``Leptogenesis,''
  Phys.\ Rept.\  {\bf 466}, 105 (2008)
  [arXiv:0802.2962 [hep-ph]].  %%CITATION = ARXIV:0802.2962;%%

%\cite{Agostini:2013mzu}
\bibitem{Agostini:2013mzu}
  M.~Agostini {\it et al.}  [GERDA Collaboration],
  %``Results on Neutrinoless Double-$\beta$ Decay of $^{76}$Ge from Phase I of the GERDA Experiment,''
  Phys.\ Rev.\ Lett.\  {\bf 111}, no. 12, 122503 (2013)
  [arXiv:1307.4720 [nucl-ex]].  %%CITATION = ARXIV:1307.4720;%%

%\cite{Bornschein:2003xi}
\bibitem{Bornschein:2003xi}
  L.~Bornschein [KATRIN Collaboration],
  %``KATRIN: Direct measurement of neutrino masses in the sub-Ev region,''
  eConf C {\bf 030626}, FRAP14 (2003)
  [hep-ex/0309007].  %%CITATION = HEP-EX/0309007;%%

%\cite{Ade:2013zuv}
\bibitem{Ade:2013zuv}
  P.~A.~R.~Ade {\it et al.}  [Planck Collaboration],
  %``Planck 2013 results. XVI. Cosmological parameters,''
  Astron.\ Astrophys.\  {\bf 571}, A16 (2014)
  [arXiv:1303.5076 [astro-ph.CO]].  %%CITATION = ARXIV:1303.5076;%%

%%\cite{An:2012eh}
%\bibitem{An:2012eh}
%  F.~P.~An {\it et al.}  [Daya Bay Collaboration],
%  %``Observation of electron-antineutrino disappearance at Daya Bay,''
%  Phys.\ Rev.\ Lett.\  {\bf 108}, 171803 (2012)
%  [arXiv:1203.1669 [hep-ex]].  %%CITATION = ARXIV:1203.1669;%%
%
%%\cite{An:2012bu}
%\bibitem{An:2012bu}
%  F.~P.~An {\it et al.}  [Daya Bay Collaboration],
%  %``Improved Measurement of Electron Antineutrino Disappearance at Daya Bay,''
%  Chin.\ Phys.\ C {\bf 37}, 011001 (2013)
%  [arXiv:1210.6327 [hep-ex]].  %%CITATION = ARXIV:1210.6327;%%
%
%%\cite{An:2013zwz}
%\bibitem{An:2013zwz}
%  F.~P.~An {\it et al.}  [Daya Bay Collaboration],
%  %``Spectral measurement of electron antineutrino oscillation amplitude and frequency at Daya Bay,''
%  Phys.\ Rev.\ Lett.\  {\bf 112}, 061801 (2014)
%  [arXiv:1310.6732 [hep-ex]].  %%CITATION = ARXIV:1310.6732;%%

%\cite{Xing:2014zka}
\bibitem{Xing:2014zka}
  Z.~z.~Xing and S.~Zhou,
  %``A partial ¦Ì ¨C ¦Ó symmetry and its prediction for leptonic CP violation,''
  Phys.\ Lett.\ B {\bf 737}, 196 (2014)
  [arXiv:1404.7021 [hep-ph]].  %%CITATION = ARXIV:1404.7021;%%

%\cite{Luo:2014upa}
\bibitem{Luo:2014upa}
  S.~Luo and Z.~z.~Xing,
  %``Resolving the octant of $\theta_{23}$ via radiative $\mu-\tau$ symmetry breaking,''
  Phys.\ Rev.\ D {\bf 90}, no. 7, 073005 (2014)
  [arXiv:1408.5005 [hep-ph]].  %%CITATION = ARXIV:1408.5005;%%

%\cite{Zhou:2014sya}
\bibitem{Zhou:2014sya}
  Y.~L.~Zhou,
  %``$\mu$-$\tau$ reflection symmetry and radiative corrections,''
  arXiv:1409.8600 [hep-ph].  %%CITATION = ARXIV:1409.8600;%%

%\cite{Xing:2013woa}
\bibitem{Xing:2013woa}
  Z.~z.~Xing and Y.~L.~Zhou,
  %``Majorana CP-violating phases in neutrino-antineutrino oscillations and other lepton-number-violating processes,''
  Phys.\ Rev.\ D {\bf 88}, 033002 (2013)
  [arXiv:1305.5718 [hep-ph]].  %%CITATION = ARXIV:1305.5718;%%

%\cite{Xing:2015zha}
\bibitem{Xing:2015zha}
  Z.~z.~Xing, Z.~h.~Zhao and Y.~L.~Zhou,
  %``How to interpret a discovery or null result of the $0\nu 2\beta$ decay,''
  arXiv:1504.05820 [hep-ph].  %%CITATION = ARXIV:1504.05820;%%

%\cite{Abazajian:2012ys}
\bibitem{Abazajian:2012ys}
  K.~N.~Abazajian, M.~A.~Acero, S.~K.~Agarwalla, A.~A.~Aguilar-Arevalo, C.~H.~Albright, S.~Antusch, C.~A.~Arguelles and A.~B.~Balantekin {\it et al.},
  %``Light Sterile Neutrinos: A White Paper,''
  arXiv:1204.5379 [hep-ph].  %%CITATION = ARXIV:1204.5379;%%

%\cite{Aguilar:2001ty}
\bibitem{Aguilar:2001ty}
  A.~Aguilar-Arevalo {\it et al.}  [LSND Collaboration],
  %``Evidence for neutrino oscillations from the observation of anti-neutrino(electron) appearance in a anti-neutrino(muon) beam,''
  Phys.\ Rev.\ D {\bf 64}, 112007 (2001)
  [hep-ex/0104049].  %%CITATION = HEP-EX/0104049;%%

%\cite{AguilarArevalo:2010wv}
\bibitem{AguilarArevalo:2010wv}
  A.~A.~Aguilar-Arevalo {\it et al.}  [MiniBooNE Collaboration],
  %``Event Excess in the MiniBooNE Search for $\bar \nu_\mu \rightarrow \bar \nu_e$ Oscillations,''
  Phys.\ Rev.\ Lett.\  {\bf 105}, 181801 (2010)
  [arXiv:1007.1150 [hep-ex]].  %%CITATION = ARXIV:1007.1150;%%

%\cite{Mention:2011rk}
\bibitem{Mention:2011rk}
  G.~Mention, M.~Fechner, T.~Lasserre, T.~A.~Mueller, D.~Lhuillier, M.~Cribier and A.~Letourneau,
  %``The Reactor Antineutrino Anomaly,''
  Phys.\ Rev.\ D {\bf 83}, 073006 (2011)
  [arXiv:1101.2755 [hep-ex]].  %%CITATION = ARXIV:1101.2755;%%

%\cite{Kopp:2013vaa}
\bibitem{Kopp:2013vaa}
  J.~Kopp, P.~A.~N.~Machado, M.~Maltoni and T.~Schwetz,
  %``Sterile Neutrino Oscillations: The Global Picture,''
  JHEP {\bf 1305}, 050 (2013)
  [arXiv:1303.3011 [hep-ph]].  %%CITATION = ARXIV:1303.3011;%%

%\cite{Giunti:2013aea}
\bibitem{Giunti:2013aea}
  C.~Giunti, M.~Laveder, Y.~F.~Li and H.~W.~Long,
  %``Pragmatic View of Short-Baseline Neutrino Oscillations,''
  Phys.\ Rev.\ D {\bf 88}, 073008 (2013)
  [arXiv:1308.5288 [hep-ph]].  %%CITATION = ARXIV:1308.5288;%%

%\cite{Hamann:2010bk}
\bibitem{Hamann:2010bk}
  J.~Hamann, S.~Hannestad, G.~G.~Raffelt, I.~Tamborra and Y.~Y.~Y.~Wong,
  %``Cosmology seeking friendship with sterile neutrinos,''
  Phys.\ Rev.\ Lett.\  {\bf 105}, 181301 (2010)
  [arXiv:1006.5276 [hep-ph]].  %%CITATION = ARXIV:1006.5276;%%

%\cite{Hamann:2011ge}
\bibitem{Hamann:2011ge}
  J.~Hamann, S.~Hannestad, G.~G.~Raffelt and Y.~Y.~Y.~Wong,
  %``Sterile neutrinos with eV masses in cosmology: How disfavoured exactly?,''
  JCAP {\bf 1109}, 034 (2011)
  [arXiv:1108.4136 [astro-ph.CO]].  %%CITATION = ARXIV:1108.4136;%%

%\cite{Giusarma:2011ex}
\bibitem{Giusarma:2011ex}
  E.~Giusarma, M.~Corsi, M.~Archidiacono, R.~de Putter, A.~Melchiorri, O.~Mena and S.~Pandolfi,
  %``Constraints on massive sterile neutrino species from current and future cosmological data,''
  Phys.\ Rev.\ D {\bf 83}, 115023 (2011)
  [arXiv:1102.4774 [astro-ph.CO]].  %%CITATION = ARXIV:1102.4774;%%

%\cite{Bode:2000gq}
\bibitem{Bode:2000gq}
  P.~Bode, J.~P.~Ostriker and N.~Turok,
  %``Halo formation in warm dark matter models,''
  Astrophys.\ J.\  {\bf 556}, 93 (2001)
  [astro-ph/0010389].  %%CITATION = ASTRO-PH/0010389;%%

%%\cite{Agashe:2014kda}
%\bibitem{Agashe:2014kda}
%  K.~A.~Olive {\it et al.}  [Particle Data Group Collaboration],
%  %``Review of Particle Physics,''
%  Chin.\ Phys.\ C {\bf 38}, 090001 (2014).  %%CITATION = CHPHD,C38,090001;%%

%\cite{Zhan:2008id}
\bibitem{Zhan:2008id}
  L.~Zhan, Y.~Wang, J.~Cao and L.~Wen,
  %``Determination of the Neutrino Mass Hierarchy at an Intermediate Baseline,''
  Phys.\ Rev.\ D {\bf 78}, 111103 (2008)
  [arXiv:0807.3203 [hep-ex]].  %%CITATION = ARXIV:0807.3203;%%

%\cite{Zhan:2009rs}
\bibitem{Zhan:2009rs}
  L.~Zhan, Y.~Wang, J.~Cao and L.~Wen,
  %``Experimental Requirements to Determine the Neutrino Mass Hierarchy Using Reactor Neutrinos,''
  Phys.\ Rev.\ D {\bf 79}, 073007 (2009)
  [arXiv:0901.2976 [hep-ex]].  %%CITATION = ARXIV:0901.2976;%%

\bibitem{yfwang2008}
Yifang Wang, talk at ICFA seminar, (2008). \url{http://www-conf.slac.stanford.edu/icfa2008/Yifang\_Wang\_102808.pdf}

\bibitem{caoj2009}
Jun Cao, talk at Neutrino Telescope, (2009). \url{http://neutrino.pd.infn.it/NEUTEL09/Talks/Cao.pdf}

%\cite{Li:2013zyd}
\bibitem{Li:2013zyd}
  Y.~F.~Li, J.~Cao, Y.~Wang and L.~Zhan,
  %``Unambiguous Determination of the Neutrino Mass Hierarchy Using Reactor Neutrinos,''
  Phys.\ Rev.\ D {\bf 88}, 013008 (2013)
  [arXiv:1303.6733 [hep-ex]].
  %%CITATION = ARXIV:1303.6733;%%

%\cite{Liu:2015hwa}
\bibitem{Liu:2015hwa}
  Q.~Liu, X.~Zhou, W.~Huang, Y.~Zhang, W.~Wu, W.~Luo, M.~Yu and Y.~Zheng {\it et al.},
  %``Rayleigh scattering and depolarization ratio in linear alkylbenzene,''
  Nucl.\ Instrum.\ Meth.\ A {\bf 795}, 284 (2015)
  [arXiv:1504.01001 [physics.ins-det]].  %%CITATION = ARXIV:1504.01001;%%

%\cite{Adam:2007ex}
\bibitem{Adam:2007ex}
  T.~Adam, E.~Baussan, K.~Borer, J.~E.~Campagne, N.~Con-Sen, C.~de La Taille, N.~Dick and M.~Dracos {\it et al.},
  %``The OPERA experiment target tracker,''
  Nucl.\ Instrum.\ Meth.\ A {\bf 577}, 523 (2007)
  [physics/0701153].  %%CITATION = PHYSICS/0701153;%%

%\end{thebibliography}


% chap: MH
%%%%%%%%%%%%%%%%%%%%%%%%%%%%%%%%%%%%%%%%%%%%%%%%%%%%%%%%%%%%%%%%%%%%%%%%%%%%%%%%%%%
%%%%%%%%%%%%%%%%%%%%%%%%%%%%%%%%%%%%%%%%%%%%%%%%%%%%%%%%%%%%%%%%%%%%%%%%%%%%%%%%%%%
% chap: MH

%\begin{thebibliography}{99}

%\bibitem{mDYB}
%Daya Bay Collaboration, (F.P.~An {\it et al.}), Phys. Rev. Lett.
%{\bf 108}, 171803 (2012).

%%\cite{An:2012eh}
%\bibitem{An:2012eh}
%  F.~P.~An {\it et al.}  [Daya Bay Collaboration],
%  %``Observation of electron-antineutrino disappearance at Daya Bay,''
%  Phys.\ Rev.\ Lett.\  {\bf 108}, 171803 (2012)
%  [arXiv:1203.1669 [hep-ex]].
%  %%CITATION = ARXIV:1203.1669;%%


%\bibitem{mDYBcpc}
%Daya Bay Collaboration, (F.P. An {\it et al.}), Chin. Phys. C {\bf
%37}, 011001 (2013).

%%\cite{An:2012bu}
%\bibitem{An:2012bu}
%  F.~P.~An {\it et al.}  [Daya Bay Collaboration],
%  %``Improved Measurement of Electron Antineutrino Disappearance at Daya Bay,''
%  Chin.\ Phys.\ C {\bf 37}, 011001 (2013)
%  [arXiv:1210.6327 [hep-ex]].
%  %%CITATION = ARXIV:1210.6327;%%

%\bibitem{mDC}
%Double Chooz Collaboration, (Y.~Abe {\it et al.}), Phys. Rev. Lett.
%{\bf 108}, 131801 (2012).

%\cite{Abe:2011fz}
\bibitem{Abe:2011fz}
  Y.~Abe {\it et al.}  [Double Chooz Collaboration],
  %``Indication for the disappearance of reactor electron antineutrinos in the Double Chooz experiment,''
  Phys.\ Rev.\ Lett.\  {\bf 108}, 131801 (2012)
  [arXiv:1112.6353 [hep-ex]].
  %%CITATION = ARXIV:1112.6353;%%

%\bibitem{mRENO}
%RENO Collaboration, (J.K.~Ahn {\it et al.}), Phys. Rev. Lett. {\bf
%108}, 191802 (2012).

%\cite{Ahn:2012nd}
\bibitem{Ahn:2012nd}
  J.~K.~Ahn {\it et al.}  [RENO Collaboration],
  %``Observation of Reactor Electron Antineutrino Disappearance in the RENO Experiment,''
  Phys.\ Rev.\ Lett.\  {\bf 108}, 191802 (2012)
  [arXiv:1204.0626 [hep-ex]].
  %%CITATION = ARXIV:1204.0626;%%

%\bibitem{mT2K}
%T2K Collaboration, (K.~Abe {\it et al.}), Phys. Rev. Lett. {\bf
%107}, 041801 (2011).

%\cite{Abe:2011sj}
\bibitem{Abe:2011sj}
  K.~Abe {\it et al.}  [T2K Collaboration],
  %``Indication of Electron Neutrino Appearance from an Accelerator-produced Off-axis Muon Neutrino Beam,''
  Phys.\ Rev.\ Lett.\  {\bf 107}, 041801 (2011)
  [arXiv:1106.2822 [hep-ex]].
  %%CITATION = ARXIV:1106.2822;%%

%\bibitem{mMINOS}
%MINOS Collaboration, (P.~Adamson {\it et al.}), Phys. Rev. Lett.
%{\bf 107}, 181802 (2011).

%\cite{Adamson:2011qu}
\bibitem{Adamson:2011qu}
  P.~Adamson {\it et al.}  [MINOS Collaboration],
  %``Improved search for muon-neutrino to electron-neutrino oscillations in MINOS,''
  Phys.\ Rev.\ Lett.\  {\bf 107}, 181802 (2011)
  [arXiv:1108.0015 [hep-ex]].
  %%CITATION = ARXIV:1108.0015;%%

%\bibitem{mGF}
%D.~V.~Forero, M.~Tortola, J.W.F.~Valle, Phys.\ Rev.\ D {\bf 86},
%073012 (2012); F.~Capozzi {\it et al.}, arXiv:1312.2878 [hep-ph];
%M.~C.~Gonzalez-Garcia, M.~Maltoni, J.~Salvado and T.~Schwetz, JHEP
%{\bf 12}, 123 (2012).%, and update NUFIT 1.2 at
%%http://www.nu-fit.org/.

%%\cite{Capozzi:2013csa}
%\bibitem{Capozzi:2013csa}
%  F.~Capozzi, G.~L.~Fogli, E.~Lisi, A.~Marrone, D.~Montanino and A.~Palazzo,
%  %``Status of three-neutrino oscillation parameters, circa 2013,''
%  Phys.\ Rev.\ D {\bf 89}, 093018 (2014)
%  [arXiv:1312.2878 [hep-ph]].  %%CITATION = ARXIV:1312.2878;%%

%%\cite{Forero:2014bxa}
%\bibitem{Forero:2014bxa}
%  D.~V.~Forero, M.~Tortola and J.~W.~F.~Valle,
%  %``Neutrino oscillations refitted,''
%  Phys.\ Rev.\ D {\bf 90}, no. 9, 093006 (2014)
%  [arXiv:1405.7540 [hep-ph]].  %%CITATION = ARXIV:1405.7540;%%
%
%%\cite{Gonzalez-Garcia:2014bfa}
%\bibitem{Gonzalez-Garcia:2014bfa}
%  M.~C.~Gonzalez-Garcia, M.~Maltoni and T.~Schwetz,
%  %``Updated fit to three neutrino mixing: status of leptonic CP violation,''
%  JHEP {\bf 1411}, 052 (2014)
%  [arXiv:1409.5439 [hep-ph]].  %%CITATION = ARXIV:1409.5439;%%

%\cite{Fogli:2012ua}
\bibitem{Fogli:2012ua}
  G.~L.~Fogli, E.~Lisi, A.~Marrone, D.~Montanino, A.~Palazzo and A.~M.~Rotunno,
  %``Global analysis of neutrino masses, mixings and phases: entering the era of leptonic CP violation searches,''
  Phys.\ Rev.\ D {\bf 86}, 013012 (2012)
  [arXiv:1205.5254 [hep-ph]].  %%CITATION = ARXIV:1205.5254;%%

%\cite{Tortola:2012te}
\bibitem{Tortola:2012te}
  D.~V.~Forero, M.~Tortola and J.~W.~F.~Valle,
  %``Global status of neutrino oscillation parameters after Neutrino-2012,''
  Phys.\ Rev.\ D {\bf 86}, 073012 (2012)
  [arXiv:1205.4018 [hep-ph]].
  %%CITATION = ARXIV:1205.4018;%%

  %\cite{GonzalezGarcia:2012sz}
\bibitem{GonzalezGarcia:2012sz}
  M.~C.~Gonzalez-Garcia, M.~Maltoni, J.~Salvado and T.~Schwetz,
  %``Global fit to three neutrino mixing: critical look at present precision,''
  JHEP {\bf 1212}, 123 (2012)
  [arXiv:1209.3023 [hep-ph]].
  %%CITATION = ARXIV:1209.3023;%%

%\bibitem{mPDG}
%%Particle Data Group, (J.~Beringer {\it et al.}), Phys. Rev. D {\bf 86}, 010001 (2012).
%  K.~A.~Olive {\it et al.}  [Particle Data Group Collaboration],
%  %``Review of Particle Physics,''
%  Chin.\ Phys.\ C {\bf 38}, 090001 (2014).

%  %\cite{Agashe:2014kda}
%\bibitem{Agashe:2014kda}
%  K.~A.~Olive {\it et al.}  [Particle Data Group Collaboration],
%  %``Review of Particle Physics,''
%  Chin.\ Phys.\ C {\bf 38}, 090001 (2014).
%  %%CITATION = CHPHD,C38,090001;%%

%\bibitem{mMNSP}
%Z.~Maki, M.~Nakagawa, and S.~Sakata, Prog.\ Theor.\ Phys.\ {\bf 28},
%870 (1962); B.~Pontecorvo, Sov.\ Phys.\ JETP {\bf 26}, 984 (1968).

%%\cite{Maki:1962mu}
%\bibitem{Maki:1962mu}
%  Z.~Maki, M.~Nakagawa and S.~Sakata,
%  %``Remarks on the unified model of elementary particles,''
%  Prog.\ Theor.\ Phys.\  {\bf 28}, 870 (1962).  %%CITATION = PTPKA,28,870;%%

%%\cite{Pontecorvo:1967fh}
%\bibitem{Pontecorvo:1967fh}
%  B.~Pontecorvo,
%  %``Neutrino Experiments and the Problem of Conservation of Leptonic Charge,''
%  Sov.\ Phys.\ JETP {\bf 26}, 984 (1968)
%  [Zh.\ Eksp.\ Teor.\ Fiz.\  {\bf 53}, 1717 (1967)].  %%CITATION = SPHJA,26,984;%%

%\bibitem{mDYB2t}
%L. Zhan, Y. Wang, J. Cao and L. Wen, Phys. Rev. D {\bf 78}, 111103
%(2008).

%%\cite{Zhan:2008id}
%\bibitem{Zhan:2008id}
%  L.~Zhan, Y.~Wang, J.~Cao and L.~Wen,
%  %``Determination of the Neutrino Mass Hierarchy at an Intermediate Baseline,''
%  Phys.\ Rev.\ D {\bf 78}, 111103 (2008)
%  [arXiv:0807.3203 [hep-ex]].
%  %%CITATION = ARXIV:0807.3203;%%

%\bibitem{mDYB2e}
%L. Zhan, Y. Wang, J. Cao and L. Wen, Phys. Rev. D {\bf 79}, 073007
%(2009).

%%\cite{Zhan:2009rs}
%\bibitem{Zhan:2009rs}
%  L.~Zhan, Y.~Wang, J.~Cao and L.~Wen,
%  %``Experimental Requirements to Determine the Neutrino Mass Hierarchy Using Reactor Neutrinos,''
%  Phys.\ Rev.\ D {\bf 79}, 073007 (2009)
%  [arXiv:0901.2976 [hep-ex]].
%  %%CITATION = ARXIV:0901.2976;%%

%\bibitem{mJUNO}
%Y.~F.~Li, J.~Cao, Y.~F.~Wang and L.~Zhan, Phys.\ Rev.\ D {\bf 88},
%013008 (2013).

%%\cite{Li:2013zyd}
%\bibitem{Li:2013zyd}
%  Y.~F.~Li, J.~Cao, Y.~Wang and L.~Zhan,
%  %``Unambiguous Determination of the Neutrino Mass Hierarchy Using Reactor Neutrinos,''
%  Phys.\ Rev.\ D {\bf 88}, 013008 (2013)
%  [arXiv:1303.6733 [hep-ex]].
%  %%CITATION = ARXIV:1303.6733;%%

%\bibitem{mXQ}
%X.~Qian, D.~Dwyer, R.~McKeown, P.~Vogel, W.~Wang, {\it et al.},
%Phys.\ Rev.\ D {\bf 87}, 033005 (2013);
%A.~B.~Balantekin, {\it et al.}, arXiv: 1307.7419.

%\cite{Qian:2012xh}
\bibitem{Qian:2012xh}
  X.~Qian, D.~A.~Dwyer, R.~D.~McKeown, P.~Vogel, W.~Wang and C.~Zhang,
  %``Mass Hierarchy Resolution in Reactor Anti-neutrino Experiments: Parameter Degeneracies and Detector Energy Response,''
  Phys.\ Rev.\ D {\bf 87}, 033005 (2013)
  [arXiv:1208.1551 [physics.ins-det]].
  %%CITATION = ARXIV:1208.1551;%%

%\cite{Kettell:2013eos}
\bibitem{Kettell:2013eos}
  A.~B.~Balantekin, H.~Band, R.~Betts, J.~J.~Cherwinka, J.~A.~Detwiler, S.~Dye, K.~M.~Heeger and R.~Johnson {\it et al.},
  %``Neutrino mass hierarchy determination and other physics potential of medium-baseline reactor neutrino oscillation experiments,''
  arXiv:1307.7419 [hep-ex].
  %%CITATION = ARXIV:1307.7419;%%

%\bibitem{mRENO50}
%S.~B.~Kim, Proposal for RENO-50: detector design and goals,
%International Workshop on ¡±RENO-50¡± toward Neutrino Mass
%Hierarchy, Seoul, June 13-14, (2013).

%\cite{Kim:2014rfa}
\bibitem{Kim:2014rfa}
  S.~B.~Kim,
  %``New results from RENO and prospects with RENO-50,''
  arXiv:1412.2199 [hep-ex].
  %%CITATION = ARXIV:1412.2199;%%

%\bibitem{mNOvApro}
%D.~Ayres {\it et al.}, [NOvA Collaboration] arXiv: hep-ex/0503053.

%\cite{Ayres:2004js}
\bibitem{Ayres:2004js}
  D.~S.~Ayres {\it et al.}  [NO$\nu$A Collaboration],
  %``NOvA: Proposal to build a 30 kiloton off-axis detector to study nu(mu) ---> nu(e) oscillations in the NuMI beamline,''
  hep-ex/0503053.
  %%CITATION = HEP-EX/0503053;%%

%\bibitem{mLBNE}
%C.~Adams {\it et al.}, [LBNE Collaboration], arXiv: 1307.7335
%[hep-ex].

%\cite{Adams:2013qkq}
\bibitem{Adams:2013qkq}
  C.~Adams {\it et al.}  [LBNE Collaboration],
  %``The Long-Baseline Neutrino Experiment: Exploring Fundamental Symmetries of the Universe,''
  arXiv:1307.7335 [hep-ex].
  %%CITATION = ARXIV:1307.7335;%%

%\bibitem{mINO}
%A~Samanta, Phys.\ Lett.\ B {\bf 673}, 37 (2009); Phys.\ Rev.\ D {\bf
%81}, 037302 (2010).

%\cite{Ahmed:2015jtv}
\bibitem{Ahmed:2015jtv}
  S.~Ahmed {\it et al.}  [ICAL Collaboration],
  %``Physics Potential of the ICAL detector at the India-based Neutrino Observatory (INO),''
  arXiv:1505.07380 [physics.ins-det].
  %%CITATION = ARXIV:1505.07380;%%

%\bibitem{mPINGU}
%M.~G.~Aartsen {\it et al.}, [IceCube-PINGU Collaboration],
%arXiv:1401.2046 [physics.ins-det].

%\cite{Aartsen:2014oha}
\bibitem{Aartsen:2014oha}
  M.~G.~Aartsen {\it et al.}  [IceCube PINGU Collaboration],
  %``Letter of Intent: The Precision IceCube Next Generation Upgrade (PINGU),''
  arXiv:1401.2046 [physics.ins-det].
  %%CITATION = ARXIV:1401.2046;%%

%\cite{VanElewyck:2015una}
\bibitem{VanElewyck:2015una}
  V.~Van Elewyck [KM3NeT Collaboration],
  %``ORCA: measuring the neutrino mass hierarchy with atmospheric neutrinos in the Mediterranean,''
  J.\ Phys.\ Conf.\ Ser.\  {\bf 598}, 012033 (2015).
  %%CITATION = 00462,598,012033;%%

%\bibitem{mHK}
%K.~Abe {\it et al.}, [HyperK Collaboration], arXiv: 1109.3262
%[hep-ex].

%\cite{Abe:2011ts}
\bibitem{Abe:2011ts}
  K.~Abe, T.~Abe, H.~Aihara, Y.~Fukuda, Y.~Hayato, K.~Huang, A.~K.~Ichikawa and M.~Ikeda {\it et al.},
  %``Letter of Intent: The Hyper-Kamiokande Experiment --- Detector Design and Physics Potential ---,''
  arXiv:1109.3262 [hep-ex].
  %%CITATION = ARXIV:1109.3262;%%

%%\cite{Abe:2014oxa}
%\bibitem{Abe:2014oxa}
%  K.~Abe {\it et al.} [Hyper-Kamiokande Working Group Collaboration],
%  %``A Long Baseline Neutrino Oscillation Experiment Using J-PARC Neutrino Beam and Hyper-Kamiokande,''
%  arXiv:1412.4673 [physics.ins-det].
%  %%CITATION = ARXIV:1412.4673;%%

%\cite{Abe:2015zbg}
\bibitem{Abe:2015zbg}
  K.~Abe {\it et al.} [Hyper-Kamiokande Proto- Collaboration],
  %``Physics potential of a long-baseline neutrino oscillation experiment using a J-PARC neutrino beam and Hyper-Kamiokande,''
  PTEP {\bf 2015}, no. 5, 053C02 (2015)
  [arXiv:1502.05199 [hep-ex]].
  %%CITATION = ARXIV:1502.05199;%%



%\cite{Bilenky:2012qi}
\bibitem{Bilenky:2012qi}
  S.~M.~Bilenky and C.~Giunti,
  %``Neutrinoless double-beta decay: A brief review,''
  Mod.\ Phys.\ Lett.\ A {\bf 27}, 1230015 (2012)
  [arXiv:1203.5250 [hep-ph]].
  %%CITATION = ARXIV:1203.5250;%%

%\bibitem{mMOMENT}
%J.~Cao {\it et al.}, Phys.\ Rev.\ ST\ Accel.\ Beams {\bf 17}, 090101 (2014), arXiv:1401.8125 [physics.acc-ph].

%\cite{Cao:2014bea}
\bibitem{Cao:2014bea}
  J.~Cao, M.~He, Z.~L.~Hou, H.~T.~Jing, Y.~F.~Li, Z.~H.~Li, Y.~P.~Song and J.~Y.~Tang {\it et al.},
  %``Muon-decay medium-baseline neutrino beam facility,''
  Phys.\ Rev.\ ST Accel.\ Beams {\bf 17}, 090101 (2014)
  [arXiv:1401.8125 [physics.acc-ph]].
  %%CITATION = ARXIV:1401.8125;%%

%\bibitem{mSplit}
%G.~G.~Raffelt, A.~Yu.~Smirnov, Phys.\ Rev.\ D {\bf 76}, 081301
%(2007), Erratum-ibid.\ D {\bf 77}, 029903 (2008).

%\cite{Raffelt:2007cb}
\bibitem{Raffelt:2007cb}
  G.~G.~Raffelt and A.~Y.~Smirnov,
  %``Self-induced spectral splits in supernova neutrino fluxes,''
  Phys.\ Rev.\ D {\bf 76}, 081301 (2007)
  [Phys.\ Rev.\ D {\bf 77}, 029903 (2008)]
  [arXiv:0705.1830 [hep-ph]].
  %%CITATION = ARXIV:0705.1830;%%

%\bibitem{mKajino}
%T.~Kajino, G.~J~Mathews, T.~Hayakawa, J.\ Phys.\ G:\ Nucl.\ Part.\
%Phys.\ {\bf 41}, 044007 (2014).

%\cite{Kajino:2014bra}
\bibitem{Kajino:2014bra}
  T.~Kajino, G.~J.~Mathews and T.~Hayakawa,
  %``Neutrinos in core-collapse supernovae and nucleosynthesis,''
  J.\ Phys.\ G {\bf 41}, 044007 (2014).
  %%CITATION = JPAGA,G41,044007;%%

%\cite{Abazajian:2013oma}
\bibitem{Abazajian:2013oma}
  K.~N.~Abazajian {\it et al.} %[Topical Conveners: K.N. Abazajian, J.E. Carlstrom, A.T. Lee Collaboration],
  %``Neutrino Physics from the Cosmic Microwave Background and Large Scale Structure,''
  Astropart.\ Phys.\  {\bf 63}, 66 (2015)
  [arXiv:1309.5383 [astro-ph.CO]].
  %%CITATION = ARXIV:1309.5383;%%


%\cite{Minakata:2007tn}
\bibitem{Minakata:2007tn}
  H.~Minakata, H.~Nunokawa, S.~J.~Parke and R.~Zukanovich Funchal,
  %``Determination of the neutrino mass hierarchy via the phase of the disappearance oscillation probability with a monochromatic anti-electron-neutrino source,''
  Phys.\ Rev.\ D {\bf 76}, 053004 (2007)
  [Phys.\ Rev.\ D {\bf 76}, 079901 (2007)]
  [hep-ph/0701151].
  %%CITATION = HEP-PH/0701151;%%


%\cite{Nunokawa:2005nx}
\bibitem{Nunokawa:2005nx}
  H.~Nunokawa, S.~J.~Parke and R.~Zukanovich Funchal,
  %``Another possible way to determine the neutrino mass hierarchy,''
  Phys.\ Rev.\ D {\bf 72}, 013009 (2005)
  [hep-ph/0503283].
  %%CITATION = HEP-PH/0503283;%%


%\cite{deGouvea:2005hk}
\bibitem{deGouvea:2005hk}
  A.~de Gouvea, J.~Jenkins and B.~Kayser,
  %``Neutrino mass hierarchy, vacuum oscillations, and vanishing |U(e3)|,''
  Phys.\ Rev.\ D {\bf 71}, 113009 (2005)  [hep-ph/0503079].
  %%CITATION = HEP-PH/0503079;%%


%\cite{Learned:2006wy}
\bibitem{Learned:2006wy}
  J.~Learned, S.~T.~Dye, S.~Pakvasa and R.~C.~Svoboda,
  %``Determination of neutrino mass hierarchy and theta(13) with a remote detector of reactor antineutrinos,''
  Phys.\ Rev.\ D {\bf 78}, 071302 (2008)
  [hep-ex/0612022].  %%CITATION = HEP-EX/0612022;%%


%\cite{VonFeilitzsch:1982jw}
\bibitem{VonFeilitzsch:1982jw}
  F.~Von Feilitzsch, A.~A.~Hahn and K.~Schreckenbach,
  %``Experimental Beta Spectra From Pu-239 And U-235 Thermal Neutron Fission Products And Their Correlated Anti-neutrinos Spectra,''
  Phys.\ Lett.\ B {\bf 118}, 162 (1982).  %%CITATION = PHLTA,B118,162;%%

  %\cite{Schreckenbach:1985ep}
\bibitem{Schreckenbach:1985ep}
  K.~Schreckenbach, G.~Colvin, W.~Gelletly and F.~Von Feilitzsch,
  %``Determination Of The Anti-neutrino Spectrum From U-235 Thermal Neutron Fission Products Up To 9.5-mev,''
  Phys.\ Lett.\ B {\bf 160}, 325 (1985).  %%CITATION = PHLTA,B160,325;%%

%\cite{Hahn:1989zr}
\bibitem{Hahn:1989zr}
  A.~A.~Hahn, K.~Schreckenbach, G.~Colvin, B.~Krusche, W.~Gelletly and F.~Von Feilitzsch,
  %``Anti-neutrino Spectra From $^{241}$Pu and $^{239}$Pu Thermal Neutron Fission Products,''
  Phys.\ Lett.\ B {\bf 218}, 365 (1989).  %%CITATION = PHLTA,B218,365;%%


%\cite{Huber:2011wv}
\bibitem{Huber:2011wv}
  P.~Huber,
  %``On the determination of anti-neutrino spectra from nuclear reactors,''
  Phys.\ Rev.\ C {\bf 84}, 024617 (2011)
  [Phys.\ Rev.\ C {\bf 85}, 029901 (2012)]  [arXiv:1106.0687 [hep-ph]].  %%CITATION = ARXIV:1106.0687;%%


%\cite{Mueller:2011nm}
\bibitem{Mueller:2011nm}
  T.~A.~Mueller, D.~Lhuillier, M.~Fallot, A.~Letourneau, S.~Cormon, M.~Fechner, L.~Giot and T.~Lasserre {\it et al.},
  %``Improved Predictions of Reactor Antineutrino Spectra,''
  Phys.\ Rev.\ C {\bf 83}, 054615 (2011)  [arXiv:1101.2663 [hep-ex]].  %%CITATION = ARXIV:1101.2663;%%


%\cite{Vogel:1980bk}
\bibitem{Vogel:1980bk}
  P.~Vogel, G.~K.~Schenter, F.~M.~Mann and R.~E.~Schenter,
  %``Reactor Anti-neutrino Spectra and Their Application to Anti-neutrino Induced Reactions. 2.,''
  Phys.\ Rev.\ C {\bf 24}, 1543 (1981).  %%CITATION = PHRVA,C24,1543;%%


%\cite{Dwyer:2014eka}
\bibitem{Dwyer:2014eka}
  D.~A.~Dwyer and T.~J.~Langford,
  %``Spectral Structure of Electron Antineutrinos from Nuclear Reactors,''
  Phys.\ Rev.\ Lett.\  {\bf 114}, 012502 (2015)
  [arXiv:1407.1281 [nucl-ex]].  %%CITATION = ARXIV:1407.1281;%%


%\cite{DYBbump}
\bibitem{DYBbump}
W.~L.~Zhong, for Daya Bay collaboration, talk at ICHEP 2014;
F.~P.~An, for Daya Bay collaboration, talk at NuFact 2014; See also
%\cite{Zhan:2015aha}
%\bibitem{Zhan:2015aha}
  L.~Zhan [Daya Bay Collaboration],
  %``Recent Results from Daya Bay,''
  arXiv:1506.01149 [hep-ex].

%\cite{Seon-HeeSeofortheRENO:2014jza}
\bibitem{Seon-HeeSeofortheRENO:2014jza}
  S.~H.~Seo [RENO Collaboration],
  %``New Results from RENO and The 5 MeV Excess,''
  arXiv:1410.7987 [hep-ex].  %%CITATION = ARXIV:1410.7987;%%

%\cite{Abe:2014bwa}
\bibitem{Abe:2014bwa}
  Y.~Abe {\it et al.}  [Double Chooz Collaboration],
  %``Improved measurements of the neutrino mixing angle $\theta_{13}$ with the Double Chooz detector,''
  JHEP {\bf 1410}, 086 (2014)  [JHEP {\bf 1502}, 074 (2015)]
  [arXiv:1406.7763 [hep-ex]].  %%CITATION = ARXIV:1406.7763;%%


%\cite{Vogel:2015wua}
\bibitem{Vogel:2015wua}
  P.~Vogel, L.~Wen and C.~Zhang,
  %``Neutrino Oscillation Studies with Reactors,''
  %Vogel, P., Wen, L. J. & Zhang, C.,
  Nature Communications {\bf 6}, 6935 (2015)
  [arXiv:1503.01059 [hep-ex]].
  %%CITATION = ARXIV:1503.01059;%%


%\cite{Abe:2009aa}
\bibitem{Abe:2009aa}
  S.~Abe {\it et al.}  [KamLAND Collaboration],
  %``Production of Radioactive Isotopes through Cosmic Muon Spallation in KamLAND,''
  Phys.\ Rev.\ C {\bf 81}, 025807 (2010)
  [arXiv:0907.0066 [hep-ex]].  %%CITATION = ARXIV:0907.0066;%%


%\cite{Wen:2006hx}
\bibitem{Wen:2006hx}
  L.~Wen, J.~Cao, K.~Luk, Y.~Ma, Y.~Wang and C.~Yang,
  %``Measuring cosmogenic Li-9 background in a reactor neutrino experiment,''
  Nucl.\ Instrum.\ Meth.\ A {\bf 564}, 471 (2006)
  [hep-ex/0604034].  %%CITATION = HEP-EX/0604034;%%


%\cite{Schwetz:2006md}
\bibitem{Schwetz:2006md}
  T.~Schwetz,
  %``What is the probability that theta(13) and CP violation will be discovered in future neutrino oscillation experiments?,''
  Phys.\ Lett.\ B {\bf 648}, 54 (2007)
  [hep-ph/0612223].  %%CITATION = HEP-PH/0612223;%%


%\cite{Blennow:2013oma}
\bibitem{Blennow:2013oma}
  M.~Blennow, P.~Coloma, P.~Huber and T.~Schwetz,
  %``Quantifying the sensitivity of oscillation experiments to the neutrino mass ordering,''
  JHEP {\bf 1403}, 028 (2014)
  [arXiv:1311.1822 [hep-ph]].  %%CITATION = ARXIV:1311.1822;%%


%\cite{Qian:2012zn}
\bibitem{Qian:2012zn}
  X.~Qian, A.~Tan, W.~Wang, J.~J.~Ling, R.~D.~McKeown and C.~Zhang,
  %``Statistical Evaluation of Experimental Determinations of Neutrino Mass Hierarchy,''
  Phys.\ Rev.\ D {\bf 86}, 113011 (2012)
  [arXiv:1210.3651 [hep-ph]].  %%CITATION = ARXIV:1210.3651;%%

%\cite{Ge:2012wj}
\bibitem{Ge:2012wj}
  S.~F.~Ge, K.~Hagiwara, N.~Okamura and Y.~Takaesu,
  %``Determination of mass hierarchy with medium baseline reactor neutrino experiments,''
  JHEP {\bf 1305}, 131 (2013)
  [arXiv:1210.8141 [hep-ph]].
  %%CITATION = ARXIV:1210.8141;%%

%\cite{Ciuffoli:2013rza}
\bibitem{Ciuffoli:2013rza}
  E.~Ciuffoli, J.~Evslin and X.~Zhang,
  %``Confidence in a neutrino mass hierarchy determination,''
  JHEP {\bf 1401}, 095 (2014)
  [arXiv:1305.5150 [hep-ph]].  %%CITATION = ARXIV:1305.5150;%%

%\bibitem{mminos_plus}
%A.~Sousa, on behalf of MINOS/MINOS+ collaboration, Neutrino 2014, ``Results from MINOS/MINOS+''.
%
%%\cite{Sousa:2015bxa}
%\bibitem{Sousa:2015bxa}
%  A.~Sousa [MINOS and MINOS+ Collaborations],
%  %``First MINOS+ Data and New Results from MINOS,''
%  arXiv:1502.07715 [hep-ex].  %%CITATION = ARXIV:1502.07715;%%
%
%\bibitem{mt2k_latest}
%T2K Collaboration, (K.~Abe {\it et al.}), Phys. Rev. Lett. {\bf
%112}, 181801 (2014).
%
%%\cite{Abe:2014ugx}
%\bibitem{Abe:2014ugx}
%  K.~Abe {\it et al.}  [T2K Collaboration],
%  %``Precise Measurement of the Neutrino Mixing Parameter \theta_{23} from Muon Neutrino Disappearance in an Off-Axis Beam,''
%  Phys.\ Rev.\ Lett.\  {\bf 112}, 181801 (2014)
%  [arXiv:1403.1532 [hep-ex]].  %%CITATION = ARXIV:1403.1532;%%
%
%\bibitem{mLBNO}
%LAGUNA-LBNO Collaboration (S. K. Agarwalla {\it et al.}), arXiv:1312.6520 (2013).
%
%%\cite{::2013kaa}
%\bibitem{::2013kaa}
%  S.~K.~Agarwalla {\it et al.}  [LAGUNA-LBNO Collaboration],
%  %``The mass-hierarchy and CP-violation discovery reach of the LBNO long-baseline neutrino experiment,''
%  JHEP {\bf 1405}, 094 (2014)
%  [arXiv:1312.6520 [hep-ph]].  %%CITATION = ARXIV:1312.6520;%%
%
%\bibitem{mSK_neutrino2014}
%R.~Wendell, on behalf of Super-Kamiokande collaboration, Neutrino 2014, ``Atmospheric Results from Super-Kamiokande''.
%
%%\cite{Wendell:2014dka}
%\bibitem{Wendell:2014dka}
%  R.~Wendell [Super-Kamiokande Collaboration],
%  %``Atmospheric Results from Super-Kamiokande,''
%  arXiv:1412.5234 [hep-ex].  %%CITATION = ARXIV:1412.5234;%%


%\end{thebibliography}


% chap: prec
%%%%%%%%%%%%%%%%%%%%%%%%%%%%%%%%%%%%%%%%%%%%%%%%%%%%%%%%%%%%%%%%%%%%%%%%%%%%%%%%%%%%
%%%%%%%%%%%%%%%%%%%%%%%%%%%%%%%%%%%%%%%%%%%%%%%%%%%%%%%%%%%%%%%%%%%%%%%%%%%%%%%%%%%%
% chap: prec

%\begin{thebibliography}{99}

%\bibitem{DBp}
%F.~P.~An {\it et al.}, [Daya Bay Collaboration], Phys.\ Rev.\ Lett.\
%{\bf 108}, 171803 (2012).

%%\cite{An:2012eh}
%\bibitem{An:2012eh}
%  F.~P.~An {\it et al.}  [Daya Bay Collaboration],
%  %``Observation of electron-antineutrino disappearance at Daya Bay,''
%  Phys.\ Rev.\ Lett.\  {\bf 108}, 171803 (2012)
%  [arXiv:1203.1669 [hep-ex]].
%  %%CITATION = ARXIV:1203.1669;%%

%\bibitem{JUNO}
%Y.~F.~Li, J.~Cao, Y.~F.~Wang and L.~Zhan, Phys.\ Rev.\ D {\bf 88},
%013008 (2013).

%%\cite{Li:2013zyd}
%\bibitem{Li:2013zyd}
%  Y.~F.~Li, J.~Cao, Y.~Wang and L.~Zhan,
%  %``Unambiguous Determination of the Neutrino Mass Hierarchy Using Reactor Neutrinos,''
%  Phys.\ Rev.\ D {\bf 88}, 013008 (2013)
%  [arXiv:1303.6733 [hep-ex]].
%  %%CITATION = ARXIV:1303.6733;%%

%\bibitem{XQ}
%X.~Qian, D.~Dwyer, R.~McKeown, P.~Vogel, W.~Wang, {\it et al.},
%Phys.\ Rev.\ D {\bf 87}, 033005 (2013),
%A.~B.~Balantekin, {\it et al.}, arXiv: 1307.7419.

%%\cite{Qian:2012xh}
%\bibitem{Qian:2012xh}
%  X.~Qian, D.~A.~Dwyer, R.~D.~McKeown, P.~Vogel, W.~Wang and C.~Zhang,
%  %``Mass Hierarchy Resolution in Reactor Anti-neutrino Experiments: Parameter Degeneracies and Detector Energy Response,''
%  Phys.\ Rev.\ D {\bf 87}, no. 3, 033005 (2013)
%  [arXiv:1208.1551 [physics.ins-det]].
%  %%CITATION = ARXIV:1208.1551;%%
%
%%\cite{Kettell:2013eos}
%\bibitem{Kettell:2013eos}
%  A.~B.~Balantekin, H.~Band, R.~Betts, J.~J.~Cherwinka, J.~A.~Detwiler, S.~Dye, K.~M.~Heeger and R.~Johnson {\it et al.},
%  %``Neutrino mass hierarchy determination and other physics potential of medium-baseline reactor neutrino oscillation experiments,''
%  arXiv:1307.7419 [hep-ex].
%  %%CITATION = ARXIV:1307.7419;%%

%\bibitem{LBNE}
%C.~Adams {\it et al.}, [LBNE Collaboration], arXiv: 1307.7335
%[hep-ex].

%%\cite{Adams:2013qkq}
%\bibitem{Adams:2013qkq}
%  C.~Adams {\it et al.}  [LBNE Collaboration],
%  %``The Long-Baseline Neutrino Experiment: Exploring Fundamental Symmetries of the Universe,''
%  arXiv:1307.7335 [hep-ex].
%  %%CITATION = ARXIV:1307.7335;%%

%\bibitem{HK}
%K.~Abe {\it et al.}, [HyperK Collaboration], arXiv: 1109.3262
%[hep-ex].

%%\cite{Abe:2011ts}
%\bibitem{Abe:2011ts}
%  K.~Abe, T.~Abe, H.~Aihara, Y.~Fukuda, Y.~Hayato, K.~Huang, A.~K.~Ichikawa and M.~Ikeda {\it et al.},
%  %``Letter of Intent: The Hyper-Kamiokande Experiment --- Detector Design and Physics Potential ---,''
%  arXiv:1109.3262 [hep-ex].
%  %%CITATION = ARXIV:1109.3262;%%
%
%%\cite{Abe:2014oxa}
%\bibitem{Abe:2014oxa}
%  K.~Abe {\it et al.}  [Hyper-Kamiokande Working Group Collaboration],
%  %``A Long Baseline Neutrino Oscillation Experiment Using J-PARC Neutrino Beam and Hyper-Kamiokande,''
%  arXiv:1412.4673 [physics.ins-det].
%  %%CITATION = ARXIV:1412.4673;%%

%\bibitem{SNO}
%B.~Aharmim {\it et al.}, [SNO Collaboration], Phys.\ Rev.\ C {\bf
%88}, 025501 (2013).

%\cite{Aharmim:2011vm}
\bibitem{Aharmim:2011vm}
  B.~Aharmim {\it et al.}  [SNO Collaboration],
  %``Combined Analysis of all Three Phases of Solar Neutrino Data from the Sudbury Neutrino Observatory,''
  Phys.\ Rev.\ C {\bf 88}, 025501 (2013)
  [arXiv:1109.0763 [nucl-ex]].  %%CITATION = ARXIV:1109.0763;%%


%\bibitem{UT1}
%S.~Antusch, C.~Biggio, E.~Fernandez-Martinez, M.~B.~Gavela, {\it et
%al.}, JHEP {\bf 0610}, 084 (2006).

%%\cite{Antusch:2006vwa}
%\bibitem{Antusch:2006vwa}
%  S.~Antusch, C.~Biggio, E.~Fernandez-Martinez, M.~B.~Gavela and J.~Lopez-Pavon,
%  %``Unitarity of the Leptonic Mixing Matrix,''
%  JHEP {\bf 0610}, 084 (2006)  [hep-ph/0607020].  %%CITATION = HEP-PH/0607020;%%

%\bibitem{UT2}
%Z.~Z.~Xing, Phys.\ Lett.\ B {\bf 718}, 1447 (2013).

%%\cite{Xing:2012kh}
%\bibitem{Xing:2012kh}
%  Z.~z.~Xing,
%  %``Towards testing the unitarity of the 3X3 lepton flavor mixing matrix in a precision reactor antineutrino oscillation experiment,''
%  Phys.\ Lett.\ B {\bf 718}, 1447 (2013)
%  [arXiv:1210.1523 [hep-ph]].  %%CITATION = ARXIV:1210.1523;%%

%\bibitem{UT3}
%X.~Qian, C.~Zhang, M.~Diwan and P.~Vogel, arXiv:1308.5700 [hep-ex].

%\cite{Qian:2013ora}
\bibitem{Qian:2013ora}
  X.~Qian, C.~Zhang, M.~Diwan and P.~Vogel,
  %``Unitarity Tests of the Neutrino Mixing Matrix,''
  arXiv:1308.5700 [hep-ex].  %%CITATION = ARXIV:1308.5700;%%

%\bibitem{UT4}
%S.~Antusch and O.~Fischer, JHEP {\bf 1410}, 096 (2014).

%%\cite{Antusch:2014woa}
%\bibitem{Antusch:2014woa}
%  S.~Antusch and O.~Fischer,
%  %``Non-unitarity of the leptonic mixing matrix: Present bounds and future sensitivities,''
%  JHEP {\bf 1410}, 94 (2014)  [arXiv:1407.6607 [hep-ph]].  %%CITATION = ARXIV:1407.6607;%%

%\cite{Lindner:2005kr}
\bibitem{Lindner:2005kr}
  M.~Lindner, A.~Merle and W.~Rodejohann,
  %``Improved limit on theta(13) and implications for neutrino masses in neutrino-less double beta decay and cosmology,''
  Phys.\ Rev.\ D {\bf 73}, 053005 (2006) [hep-ph/0512143].
  %%CITATION = HEP-PH/0512143;%%


%\cite{Dueck:2011hu}
\bibitem{Dueck:2011hu}
  A.~Dueck, W.~Rodejohann and K.~Zuber,
  %``Neutrinoless Double Beta Decay, the Inverted Hierarchy and Precision Determination of theta(12),''
  Phys.\ Rev.\ D {\bf 83}, 113010 (2011)  [arXiv:1103.4152 [hep-ph]].  %%CITATION = ARXIV:1103.4152;%%

%\cite{Ge:2015bfa}
\bibitem{Ge:2015bfa}
  S.~F.~Ge and W.~Rodejohann,
  %``JUNO and Neutrinoless Double Beta Decay,''
  arXiv:1507.05514 [hep-ph].
  %%CITATION = ARXIV:1507.05514;%%

%\cite{Rodejohann:2012xd}
\bibitem{Rodejohann:2012xd}
  W.~Rodejohann,
  %``Neutrinoless double beta decay and neutrino physics,''
  J.\ Phys.\ G {\bf 39}, 124008 (2012)  [arXiv:1206.2560 [hep-ph]].  %%CITATION = ARXIV:1206.2560;%%

%\bibitem{TBM}
%P.~F.~Harrison, D.~H.~Perkins and W.~G.~Scott, Phys.\ Lett.\ B {\bf
%530}, 167 (2002); Z.~Z.~Xing, Phys.\ Lett.\ B {\bf 533}, 85 (2002);
%X.~G.~He and A.~Zee, Phys.\ Lett.\ B {\bf 560}, 87 (2003).

%\cite{Harrison:2002er}
\bibitem{Harrison:2002er}
  P.~F.~Harrison, D.~H.~Perkins and W.~G.~Scott,
  %``Tri-bimaximal mixing and the neutrino oscillation data,''
  Phys.\ Lett.\ B {\bf 530}, 167 (2002)  [hep-ph/0202074].  %%CITATION = HEP-PH/0202074;%%

%\cite{Xing:2002sw}
\bibitem{Xing:2002sw}
  Z.~z.~Xing,
  %``Nearly tri bimaximal neutrino mixing and CP violation,''
  Phys.\ Lett.\ B {\bf 533}, 85 (2002)  [hep-ph/0204049].  %%CITATION = HEP-PH/0204049;%%

%\cite{He:2003rm}
\bibitem{He:2003rm}
  X.~G.~He and A.~Zee,
  %``Some simple mixing and mass matrices for neutrinos,''
  Phys.\ Lett.\ B {\bf 560}, 87 (2003)  [hep-ph/0301092].  %%CITATION = HEP-PH/0301092;%%

%\bibitem{TM}
%C.~H.~Albright, W.~Rodejohann, Eur.\ Phys.\ J.\ C {\bf 62}, 599
%(2009).

%\cite{Albright:2008rp}
\bibitem{Albright:2008rp}
  C.~H.~Albright and W.~Rodejohann,
  %``Comparing Trimaximal Mixing and Its Variants with Deviations from Tri-bimaximal Mixing,''
  Eur.\ Phys.\ J.\ C {\bf 62}, 599 (2009)  [arXiv:0812.0436 [hep-ph]].  %%CITATION = ARXIV:0812.0436;%%

%\bibitem{Nunokawa}
%H.~Nunokawa, S.~J.~Parke, and R.~Zukanovich Funchal, Phys.\ Rev.\ D
%{\bf 72}, 013009, (2005).

%%\cite{Nunokawa:2005nx}
%\bibitem{Nunokawa:2005nx}
%  H.~Nunokawa, S.~J.~Parke and R.~Zukanovich Funchal,
%  %``Another possible way to determine the neutrino mass hierarchy,''
%  Phys.\ Rev.\ D {\bf 72}, 013009 (2005)
%  [hep-ph/0503283].
%  %%CITATION = HEP-PH/0503283;%%

%\bibitem{Kayser}
%A.~de~Gouvea, J.~Jenkins and B.~Kayser, Phys.Rev. D {\bf 71}, 113009
%(2005).

%%\cite{deGouvea:2005hk}
%\bibitem{deGouvea:2005hk}
%  A.~de Gouvea, J.~Jenkins and B.~Kayser,
%  %``Neutrino mass hierarchy, vacuum oscillations, and vanishing |U(e3)|,''
%  Phys.\ Rev.\ D {\bf 71}, 113009 (2005)  [hep-ph/0503079].
%  %%CITATION = HEP-PH/0503079;%%

%\bibitem{T2Kpro}
%Y.~Itow {\it et al.}, [T2K Collaboration], Nucl.\ Phys.\ Proc.\
%Suppl.\ {\bf 111}, 146 (2001).

%\cite{Itow:2001ee}
\bibitem{Itow:2001ee}
  Y.~Itow {\it et al.}  [T2K Collaboration],
  %``The JHF-Kamioka neutrino project,''
  hep-ex/0106019.  %%CITATION = HEP-EX/0106019;%%

%\bibitem{Novapro}
%I.~Ambats {\it et al.}, [NOvA Collaboration] (2004), hep-ex/0503053.

%%\cite{Ayres:2004js}
%\bibitem{Ayres:2004js}
%  D.~S.~Ayres {\it et al.}  [NOvA Collaboration],
%  %``NOvA: Proposal to build a 30 kiloton off-axis detector to study nu(mu) ---> nu(e) oscillations in the NuMI beamline,''
%  hep-ex/0503053.  %%CITATION = HEP-EX/0503053;%%

%\bibitem{TplusN}
%S.~K,~Agarwalla, S.~Prakash AND W.~Wang, arXiv:1312.1477 [hep-ph].

%\cite{Agarwalla:2013qfa}
\bibitem{Agarwalla:2013qfa}
  S.~K.~Agarwalla, S.~Prakash and W.~Wang,
  %``High-precision measurement of atmospheric mass-squared splitting with T2K and NOvA,''
  arXiv:1312.1477 [hep-ph].  %%CITATION = ARXIV:1312.1477;%%

%\bibitem{SKloe}
%Y.~Ashie {\it et al.}, [Super-Kamiokande Collaboration], Phys.\
%Rev.\ Lett.\ {\bf 93}, 101801 (2004).

%\cite{Ashie:2004mr}
\bibitem{Ashie:2004mr}
  Y.~Ashie {\it et al.}  [Super-Kamiokande Collaboration],
  %``Evidence for an oscillatory signature in atmospheric neutrino oscillation,''
  Phys.\ Rev.\ Lett.\  {\bf 93}, 101801 (2004)  [hep-ex/0404034].  %%CITATION = HEP-EX/0404034;%%

%\bibitem{KLloe}
%A.~Gando {\it et al.}, [KamLAND Collaboration], Phys.\ Rev.\ D {\bf
%83}, 052002 (2011).

%\cite{Gando:2010aa}
\bibitem{Gando:2010aa}
  A.~Gando {\it et al.}  [KamLAND Collaboration],
  %``Constraints on $\theta_{13}$ from A Three-Flavor Oscillation Analysis of Reactor Antineutrinos at KamLAND,''
  Phys.\ Rev.\ D {\bf 83}, 052002 (2011)  [arXiv:1009.4771 [hep-ex]].  %%CITATION = ARXIV:1009.4771;%%

%\bibitem{DBloe}
%F.~P.~An {\it et al.}, [Daya Bay Collaboration], Phys.\ Rev.\ Lett.\
%{\bf 112}, 061801 (2014);
%F.~P.~An {\it et al.},  [Daya Bay Collaboration],
%%``A new measurement of antineutrino oscillation with the full detector configuration at Daya Bay,''
%arXiv:1505.03456 [hep-ex].

%\cite{An:2015rpe}
\bibitem{An:2015rpe}
  F.~P.~An {\it et al.}  [Daya Bay Collaboration],
  %``A new measurement of antineutrino oscillation with the full detector configuration at Daya Bay,''
  arXiv:1505.03456 [hep-ex].  %%CITATION = ARXIV:1505.03456;%%

%\bibitem{MINOS}
%P.~Adamson {\it et al.}, [MINOS Collaboration], Phys.\ Rev.\ Lett.\
%{\bf 110}, 251801 (2013).

%\cite{Adamson:2013whj}
\bibitem{Adamson:2013whj}
  P.~Adamson {\it et al.}  [MINOS Collaboration],
  %``Measurement of Neutrino and Antineutrino Oscillations Using Beam and Atmospheric Data in MINOS,''
  Phys.\ Rev.\ Lett.\  {\bf 110}, 251801 (2013)  [arXiv:1304.6335 [hep-ex]].  %%CITATION = ARXIV:1304.6335;%%

%\bibitem{SKth23}
%R.~Wendell {\it et al.}, [Super-Kamiokande Collaboration], Phys.\
%Rev.\ D {\bf 81}, 092004 (2010).

%%\cite{Wendell:2010md}
%\bibitem{Wendell:2010md}
%  R.~Wendell {\it et al.}  [Super-Kamiokande Collaboration],
%  %``Atmospheric neutrino oscillation analysis with sub-leading effects in Super-Kamiokande I, II, and III,''
%  Phys.\ Rev.\ D {\bf 81}, 092004 (2010)  [arXiv:1002.3471 [hep-ex]].  %%CITATION = ARXIV:1002.3471;%%

%\cite{Wendell:2014dka}
\bibitem{Wendell:2014dka}
  R.~Wendell [Super-Kamiokande Collaboration],
  %``Atmospheric Results from Super-Kamiokande,''
  arXiv:1412.5234 [hep-ex].  %%CITATION = ARXIV:1412.5234;%%

%\bibitem{T2Kth23}
%K.~Abe {\it et al.}, [T2K Collaboration], Phys.\ Rev.\ Lett.\ {\bf
%111}, 211803 (2013); for the latest update see arXiv:1403.1532
%[hep-ex].

%\cite{Abe:2014ugx}
\bibitem{Abe:2014ugx}
  K.~Abe {\it et al.}  [T2K Collaboration],
  %``Precise Measurement of the Neutrino Mixing Parameter \theta_{23} from Muon Neutrino Disappearance in an Off-Axis Beam,''
  Phys.\ Rev.\ Lett.\  {\bf 112}, 181801 (2014)  [arXiv:1403.1532 [hep-ex]].  %%CITATION = ARXIV:1403.1532;%%

%\bibitem{GF1}
%D.~V.~Forero, M.~Tortola, J.W.F.~Valle, Phys.\ Rev.\ D {\bf 86},
%073012 (2012); F.~Capozzi {\it et al.}, arXiv:1312.2878 [hep-ph];
%M.~C.~Gonzalez-Garcia, M.~Maltoni, J.~Salvado and T.~Schwetz, JHEP
%{\bf 12}, 123 (2012).%, and update NUFIT 1.2 at
%%http://www.nu-fit.org/.

%%\cite{Tortola:2012te}
%\bibitem{Tortola:2012te}
%  D.~V.~Forero, M.~Tortola and J.~W.~F.~Valle,
%  %``Global status of neutrino oscillation parameters after Neutrino-2012,''
%  Phys.\ Rev.\ D {\bf 86}, 073012 (2012)
%  [arXiv:1205.4018 [hep-ph]].
%  %%CITATION = ARXIV:1205.4018;%%
%
%  %\cite{Capozzi:2013csa}
%\bibitem{Capozzi:2013csa}
%  F.~Capozzi, G.~L.~Fogli, E.~Lisi, A.~Marrone, D.~Montanino and A.~Palazzo,
%  %``Status of three-neutrino oscillation parameters, circa 2013,''
%  Phys.\ Rev.\ D {\bf 89}, 093018 (2014)
%  [arXiv:1312.2878 [hep-ph]].
%  %%CITATION = ARXIV:1312.2878;%%

%  %\cite{Gonzalez-Garcia:2014bfa}
%\bibitem{Gonzalez-Garcia:2014bfa}
%  M.~C.~Gonzalez-Garcia, M.~Maltoni and T.~Schwetz,
%  %``Updated fit to three neutrino mixing: status of leptonic CP violation,''
%  JHEP {\bf 1411}, 052 (2014)
%  [arXiv:1409.5439 [hep-ph]].
%  %%CITATION = ARXIV:1409.5439;%%
%
%  %\cite{GonzalezGarcia:2012sz}
%\bibitem{GonzalezGarcia:2012sz}
%  M.~C.~Gonzalez-Garcia, M.~Maltoni, J.~Salvado and T.~Schwetz,
%  %``Global fit to three neutrino mixing: critical look at present precision,''
%  JHEP {\bf 1212}, 123 (2012)
%  [arXiv:1209.3023 [hep-ph]].
%  %%CITATION = ARXIV:1209.3023;%%

%\end{thebibliography}


% chap: SN burst
%%%%%%%%%%%%%%%%%%%%%%%%%%%%%%%%%%%%%%%%%%%%%%%%%%%%%%%%%%%%%%%%%%%%%%%%%%%%%%%%%%%%
%%%%%%%%%%%%%%%%%%%%%%%%%%%%%%%%%%%%%%%%%%%%%%%%%%%%%%%%%%%%%%%%%%%%%%%%%%%%%%%%%%%%
% chap: SN burst


%\begin{thebibliography}{99}

%\cite{Langer:2012jz}
\bibitem{Langer:2012jz}
  N.~Langer,
  %``Pre-Supernova Evolution of Massive Single and Binary Stars,''
  Ann.\ Rev.\ Astron.\ Astrophys.\  {\bf 50}, 107 (2012)
  [arXiv:1206.5443].
  %%CITATION = ARXIV:1206.5443;%%
  %82 citations counted in INSPIRE as of 13 Jul 2015

%\cite{Burrows:2000mk}
\bibitem{Burrows:2000mk}
  A.~Burrows,
  %``Supernova explosions in the universe,''
  Nature {\bf 403}, 727 (2000).
  %%CITATION = NATUA,403,727;%%
  %65 citations counted in INSPIRE as of 19 Apr 2014

%\cite{Janka:2012wk}
\bibitem{Janka:2012wk}
  H.-Th.~Janka,
  %``Explosion Mechanisms of Core-Collapse Supernovae,''
  Ann.\ Rev.\ Nucl.\ Part.\ Sci.\  {\bf 62}, 407 (2012)
  [arXiv:1206.2503].
  %%CITATION = ARXIV:1206.2503;%%
  %70 citations counted in INSPIRE as of 19 Apr 2014

  %\cite{Cappellaro:2000ez}
\bibitem{Cappellaro:2000ez}
  E.~Cappellaro and M.~Turatto,
  %``Supernova types and rates,''
  Astrophys.\ Space Sci.\ Libr.\  {\bf 264}, 199 (2001)
  [astro-ph/0012455].  %%CITATION = ASTRO-PH/0012455;%%

%\cite{Mazzali:2007et}
\bibitem{Mazzali:2007et}
  P.~A.~Mazzali, F.~K.~R\"{o}pke, S.~Benetti and W.~Hillebrandt,
  %``A Common Explosion Mechanism for Type Ia Supernovae,''
  Science {\bf 315}, 825 (2007)
  [astro-ph/0702351].
  %%CITATION = ASTRO-PH/0702351;%%
  %129 citations counted in INSPIRE as of 19 Apr 2014

%\cite{Beacom:2010kk}
\bibitem{Beacom:2010kk}
  J.~F.~Beacom,
  %``The Diffuse Supernova Neutrino Background,''
  Ann.\ Rev.\ Nucl.\ Part.\ Sci.\  {\bf 60}, 439 (2010)
  [arXiv:1004.3311].
  %%CITATION = ARXIV:1004.3311;%%
  %49 citations counted in INSPIRE as of 03 Jul 2015

%\cite{Koshiba:1992yb}
\bibitem{Koshiba:1992yb}
  M.~Koshiba,
  %``Observational neutrino astrophysics,''
  Phys.\ Rept.\  {\bf 220}, 229 (1992).
  %%CITATION = PRPLC,220,229;%%
  %72 citations counted in INSPIRE as of 19 Apr 2014

%\cite{Mirizzi:2006xx}
\bibitem{Mirizzi:2006xx}
  A.~Mirizzi, G.~G.~Raffelt and P.~D.~Serpico,
  %``Earth matter effects in supernova neutrinos: Optimal detector locations,''
  JCAP {\bf 0605}, 012 (2006)
  [astro-ph/0604300].
  %%CITATION = ASTRO-PH/0604300;%%
  %44 citations counted in INSPIRE as of 20 Apr 2014

%\cite{Ahlers:2009ae}
\bibitem{Ahlers:2009ae}
  M.~Ahlers, P.~Mertsch and S.~Sarkar,
  %``On cosmic ray acceleration in supernova remnants and the FERMI/PAMELA data,''
  Phys.\ Rev.\ D {\bf 80}, 123017 (2009)
  [arXiv:0909.4060].
  %%CITATION = ARXIV:0909.4060;%%
  %63 citations counted in INSPIRE as of 20 Apr 2014

%\cite{Adams:2013ana}
\bibitem{Adams:2013ana}
  S.~M.~Adams, C.~S.~Kochanek, J.~F.~Beacom, M.~R.~Vagins and K.~Z.~Stanek,
  %``Observing the Next Galactic Supernova,''
  Astrophys.\ J.\  {\bf 778}, 164 (2013)
  [arXiv:1306.0559].
  %%CITATION = ARXIV:1306.0559;%%
  %4 citations counted in INSPIRE as of 18 Apr 2014

%\cite{Harper:2008}
\bibitem{Harper:2008}
  G.~M.~Harper, A.~Brown and E.~F.~Guinan,
  %``A NEW VLAIPPARCOS DISTANCE TO BETELGEUSE AND ITS IMPLICATIONS''
  Astron.\ J.\ {\bf 135}, 1430 (2008).

%\cite{vandenBergh:1994}
\bibitem{vandenBergh:1994}
  S.~van den Bergh and R.~D.~McClure,
  Astrophys.\ J.\ {\bf 425}, 205 (1994).

%\cite{Li:2010kd}
\bibitem{Li:2010kd}
  W.~Li, R.~Chornock, J.~Leaman, A.~V.~Filippenko, D.~Poznanski, X.~Wang, M.~Ganeshalingam
  and F.~Mannucci,
  %``Nearby Supernova Rates from the Lick Observatory Supernova Search. III. The Rate-Size
  %Relation, and the Rates as a Function of Galaxy Hubble Type and Colour,''
  Mon.\ Not.\ R.\ Astron.\ Soc.\ {\bf 412}, 1473 (2011)
  [arXiv:1006.4613].
  %%CITATION = ARXIV:1006.4613;%%
  %28 citations counted in INSPIRE as of 20 Apr 2014

%\cite{Reed:2005en}
\bibitem{Reed:2005en}
  B.~C.~Reed,
  %``New estimates of the solar-neighborhood massive-stars birthrate and the Galactic
  %supernova rate,''
  Astron.\ J.\  {\bf 130}, 1652 (2005)
  [astro-ph/0506708].
  %%CITATION = ASTRO-PH/0506708;%%
  %14 citations counted in INSPIRE as of 20 Apr 2014

%\cite{FaucherGiguere:2005ny}
\bibitem{FaucherGiguere:2005ny}
  C.-A.~Faucher-Gigu\`ere and V.~M.~Kaspi,
  %``Birth and evolution of isolated radio pulsars,''
  Astrophys.\ J.\  {\bf 643}, 332 (2006)
  [astro-ph/0512585].
  %%CITATION = ASTRO-PH/0512585;%%
  %209 citations counted in INSPIRE as of 20 Apr 2014

%\cite{Keane:2008jj}
\bibitem{Keane:2008jj}
  E.~F.~Keane and M.~Kramer,
  %``On the birthrates of Galactic neutron stars,''
  Mon.\ Not.\ R.\ Astron.\ Soc.\ {\bf 391}, 2009 (2008)
  [arXiv:0810.1512].
  %%CITATION = ARXIV:0810.1512;%%
  %5 citations counted in INSPIRE as of 20 Apr 2014

%\cite{Diehl:2006cf}
\bibitem{Diehl:2006cf}
  R.~Diehl {\it et al.},
  %``Radioactive Al-26 and massive stars in the galaxy,''
  Nature {\bf 439}, 45 (2006).
  %[astro-ph/0601015].
  %%CITATION = ASTRO-PH/0601015;%%

%\cite{Strom:1994}
\bibitem{Strom:1994}
  R.~G.~Strom,
  Astron.\ Astrophys.\ {\bf 288}, L1 (1994).

%\cite{Tammann:1994ev}
\bibitem{Tammann:1994ev}
  G.~A.~Tammann, W.~L\"offler and A.~Schr\"oder,
  %``The Galactic supernova rate,''
  Astrophys.\ J.\ Suppl.\  {\bf 92}, 487 (1994).
  %%CITATION = APJSA,92,487;%%

%\cite{Alekseev:1993dy}
\bibitem{Alekseev:1993dy}
  E.~N.~Alekseev {\it et al.},
  %``Upper bound on the collapse rate of massive stars in the Milky Way given by neutrino
  %observations with the Baksan underground telescope,''
  Zh.\ Eksp.\ Teor.\ Fiz.\  {\bf 104}, 2897 (1993)
  [J.\ Exp.\ Theor.\ Phys.\ {\bf 77}, 339 (1993)].
  %%CITATION = JTPHE,77,339;%%

%\cite{Green:2003ir}
\bibitem{Green:2003ir}
  D.~A.~Green and F.~R.~Stephenson,
  %``The Historical supernovae,''
  Lect.\ Notes Phys.\ {\bf 598}, 7 (2003)
  [astro-ph/0301603].
  %%CITATION = ASTRO-PH/0301603;%%
  %2 citations counted in INSPIRE as of 20 Apr 2014

%\cite{Antonioli:2004zb}
\bibitem{Antonioli:2004zb}
  P.~Antonioli {\it et al.},
  %``SNEWS: The Supernova Early Warning System,''
  New J.\ Phys.\  {\bf 6}, 114 (2004)
  [astro-ph/0406214].
  For the SNEWS project homepage see
  http://snews.bnl.gov/
  %%CITATION = ASTRO-PH/0406214;%%
  %74 citations counted in INSPIRE as of 20 Apr 2014

%\cite{Scholberg:2012id}
\bibitem{Scholberg:2012id}
  K.~Scholberg,
  %``Supernova Neutrino Detection,''
  Ann.\ Rev.\ Nucl.\ Part.\ Sci.\  {\bf 62}, 81 (2012).
  %[arXiv:1205.6003 [astro-ph.IM]].
  %%CITATION = ARXIV:1205.6003;%%

%\cite{Abe:2013gga}
\bibitem{Abe:2013gga}
  K.~Abe {\it et al.} (Super-Kamiokande Collaboration),
  %``Calibration of the Super-Kamiokande Detector,''
  Nucl.\ Instrum.\ Meth.\ A {\bf 737}, 253 (2014).
  %[arXiv:1307.0162 [physics.ins-det]].
  %%CITATION = ARXIV:1307.0162;%%

%\cite{Mori:2013wua}
\bibitem{Mori:2013wua}
  T.~Mori (for Super-Kamiokande Collaboration),
  %``Status of the Super-Kamiokande gadolinium project,''
  Nucl.\ Instrum.\ Meth.\ A {\bf 732}, 316 (2013).
  %%CITATION = NUIMA,A732,316;%%

%\cite{Abbasi:2011ss}
\bibitem{Abbasi:2011ss}
  R.~Abbasi {\it et al.} (IceCube Collaboration),
  %``IceCube Sensitivity for Low-Energy Neutrinos from Nearby Supernovae,''
  Astron.\ Astrophys.\  {\bf 535}, A109 (2011).
  %[arXiv:1108.0171 [astro-ph.HE]].

%\cite{Demiroers:2011am}
\bibitem{Demiroers:2011am}
  M.~Salathe, M.~Ribordy and L.~Demirors,
  %``Novel technique for supernova detection with IceCube,''
  Astropart.\ Phys.\ {\bf 35}, 485 (2012).
  %[arXiv:1106.1937 [astro-ph.IM]].

%\cite{Aartsen:2013nla}
\bibitem{Aartsen:2013nla}
  M.~G.~Aartsen {\it et al.} (IceCube Collaboration),
  %``The IceCube Neutrino Observatory Part V: Neutrino Oscillations and %Supernova Searches,''
  arXiv: 1309.7008.
  %%CITATION = ARXIV:1309.7008;%%

%\cite{Wurm:2011zn}
\bibitem{Wurm:2011zn}
  M.~Wurm {\it et al.} (LENA Collaboration),
  %``The next-generation liquid-scintillator neutrino observatory LENA,''
  Astropart.\ Phys.\  {\bf 35}, 685 (2012)
  [arXiv:1104.5620].
  %%CITATION = ARXIV:1104.5620;%%
  %73 citations counted in INSPIRE as of 18 Apr 2014

%\cite{Bethe:1984ux}
\bibitem{Bethe:1984ux}
  H.~A.~Bethe and J.~Wilson, R.,
  %``Revival of a stalled supernova shock by neutrino heating,''
  Astrophys.\ J.\  {\bf 295}, 14 (1985).
  %%CITATION = ASJOA,295,14;%%

%\cite{Bethe:1990mw}
\bibitem{Bethe:1990mw}
  H.~A.~Bethe,
  %``Supernova mechanisms,''
  Rev.\ Mod.\ Phys.\  {\bf 62}, 801 (1990).
 %%CITATION = RMPHA,62,801;%%

%%\cite{Dasgupta:2009yj}
%\bibitem{Dasgupta:2009yj}
%  B.~Dasgupta, T.~Fischer, S.~Horiuchi, M.~Liebendorfer, A.~Mirizzi, I.~Sagert and J.~Schaffner-Bielich,
%  %``Detecting the QCD phase transition in the next Galactic supernova neutrino burst,''
%  Phys.\ Rev.\ D {\bf 81}, 103005 (2010)  [arXiv:0912.2568].
%  %%CITATION = ARXIV:0912.2568;%%
%  %24 citations counted in INSPIRE as of 03 May 2014

%%\cite{Beacom:2000qy}
%\bibitem{Beacom:2000qy}
%  J.~F.~Beacom, R.~N.~Boyd and A.~Mezzacappa,
%  %``Black hole formation in core collapse supernovae and time-of-flight measurements of the neutrino masses,''
%  Phys.\ Rev.\ D {\bf 63}, 073011 (2001)  [astro-ph/0010398].
%  %%CITATION = ASTRO-PH/0010398;%%
%  %85 citations counted in INSPIRE as of 03 May 2014

%\cite{Blondin:2002sm}
\bibitem{Blondin:2002sm}
  J.~M.~Blondin, A.~Mezzacappa and C.~DeMarino,
  %``Stability of standing accretion shocks, with an eye toward core collapse supernovae,''
  Astrophys.\ J.\  {\bf 584}, 971 (2003)
  [astro-ph/0210634].
  %%CITATION = ASTRO-PH/0210634;%%
  %270 citations counted in INSPIRE as of 14 juil. 2015

%\cite{Tamborra:2014aua}
\bibitem{Tamborra:2014aua}
  I.~Tamborra, F.~Hanke, H.~T.~Janka, B.~M¨¹ller, G.~G.~Raffelt and A.~Marek,
  %``Self-sustained asymmetry of lepton-number emission: A new phenomenon during the supernova shock-accretion phase in three dimensions,''
  Astrophys.\ J.\  {\bf 792}, no. 2, 96 (2014)
  [arXiv:1402.5418].
  %%CITATION = ARXIV:1402.5418;%%
  %15 citations counted in INSPIRE as of 14 Jul 2015

%\cite{Janka:2012sb}
\bibitem{Janka:2012sb}
  H.-Th.~Janka, F.~Hanke, L.~H\"{u}depohl, A.~Marek, B.~M\"{u}ller and M.~Obergaulinger,
  %``Core-Collapse Supernovae: Reflections and Directions,''
  PTEP {\bf 2012}, 01A309 (2012)  [arXiv:1211.1378].
  %%CITATION = ARXIV:1211.1378;%%
  %7 citations counted in INSPIRE as of 08 Apr 2014

\bibitem{GarchingModel}
  L.~H\"{u}depohl, PhD Thesis, Technische Universit\"at M\"unchen (2013).
  Neutrino model data can be made available upon request at http://www.mpa-garching.mpg.de/ccsnarchive/

%\cite{Mirizzi:2015eza}
\bibitem{Mirizzi:2015eza}
  A.~Mirizzi, I.~Tamborra, H.~T.~Janka, N.~Saviano, K.~Scholberg, R.~Bollig, L.~Hudepohl and S.~Chakraborty,
  %``Supernova Neutrinos: Production, Oscillations and Detection,''
  arXiv:1508.00785 [astro-ph.HE].
  %%CITATION = ARXIV:1508.00785;%%

%\cite{Burrows:2005dv}
\bibitem{Burrows:2005dv}
  A.~Burrows, E.~Livne, L.~Dessart, C.~Ott and J.~Murphy,
  %``A new mechanism for core-collapse supernova explosions,''
  Astrophys.\ J.\  {\bf 640}, 878 (2006)
  [astro-ph/0510687].
  %%CITATION = ASTRO-PH/0510687;%%
  %232 citations counted in INSPIRE as of 14 juil. 2015

%\cite{Akiyama:2002xn}
\bibitem{Akiyama:2002xn}
  S.~Akiyama, J.~C.~Wheeler, D.~L.~Meier and I.~Lichtenstadt,
  %``The magnetorotational instability in core collapse supernova explosions,''
  Astrophys.\ J.\  {\bf 584}, 954 (2003)
  [astro-ph/0208128].
  %%CITATION = ASTRO-PH/0208128;%%
  %202 citations counted in INSPIRE as of 14 juil. 2015

%\cite{Gentile:1993ma}
\bibitem{Gentile:1993ma}
  N.~A.~Gentile, M.~B.~Aufderheide, G.~J.~Mathews, F.~D.~Swesty and G.~M.~Fuller,
  %``The QCD phase transition and supernova core collapse,''
  Astrophys.\ J.\  {\bf 414}, 701 (1993).
  %%CITATION = ASJOA,414,701;%%
  %51 citations counted in INSPIRE as of 14 juil. 2015

%\cite{Dai:1995uj}
\bibitem{Dai:1995uj}
  Z.~G.~Dai, Q.~H.~Peng and T.~Lu,
  %``The conversion from two - flavor to three - flavor quark matter in a supernova core,''
  Astrophys.\ J.\  {\bf 440}, 815 (1995).
  %%CITATION = ASJOA,440,815;%%
  %46 citations counted in INSPIRE as of 14 juil. 2015

%\cite{Sagert:2008ka}
\bibitem{Sagert:2008ka}
  I.~Sagert, T.~Fischer, M.~Hempel, G.~Pagliara, J.~Schaffner-Bielich, A.~Mezzacappa, F.-K.~Thielemann and M.~Liebendorfer,
  %``Signals of the QCD phase transition in core-collapse supernovae,''
  Phys.\ Rev.\ Lett.\  {\bf 102}, 081101 (2009)
  [arXiv:0809.4225].
  %%CITATION = ARXIV:0809.4225;%%
  %111 citations counted in INSPIRE as of 14 Jul 2015

%\cite{Keil:2002in}
\bibitem{Keil:2002in}
  M.~Th.~Keil, G.~G.~Raffelt and H.-Th.~Janka,
  %``Monte Carlo study of supernova neutrino spectra formation,''
  Astrophys.\ J.\  {\bf 590}, 971 (2003)  [astro-ph/0208035].
  %%CITATION = ASTRO-PH/0208035;%%
  %252 citations counted in INSPIRE as of 21 Apr 2014

%\cite{Strumia:2003zx}
\bibitem{Strumia:2003zx}
  A.~Strumia and F.~Vissani,
  %``Precise quasielastic neutrino/nucleon cross-section,''
  Phys.\ Lett.\ B {\bf 564}, 42 (2003)  [astro-ph/0302055].
  %%CITATION = ASTRO-PH/0302055;%%
  %116 citations counted in INSPIRE as of 09 Apr 2014

%\cite{Vogel:1999zy}
\bibitem{Vogel:1999zy}
  P.~Vogel and J.~F.~Beacom,
  %``Angular distribution of neutron inverse beta decay, anti-neutrino(e) + p ---> e+ + n,''
  Phys.\ Rev.\ D {\bf 60}, 053003 (1999)  [hep-ph/9903554].
  %%CITATION = HEP-PH/9903554;%%
  %317 citations counted in INSPIRE as of 10 Apr 2014

%\cite{Apollonio:1999jg}
\bibitem{Apollonio:1999jg}
  M.~Apollonio {\it et al.}  [CHOOZ Collaboration],
  %``Determination of neutrino incoming direction in the CHOOZ experiment and supernova explosion location by scintillator detectors,''
  Phys.\ Rev.\ D {\bf 61}, 012001 (2000)  [hep-ex/9906011].
  %%CITATION = HEP-EX/9906011;%%
  %44 citations counted in INSPIRE as of 17 Sep 2014

%\cite{Laha:2014yua}
\bibitem{Laha:2014yua}
  R.~Laha, J.~F.~Beacom and S.~K.~Agarwalla,
  %``New Power to Measure Supernova $\nu_e$ with Large Liquid Scintillator Detectors,''
  arXiv:1412.8425.
  %%CITATION = ARXIV:1412.8425;%%
  %1 citations counted in INSPIRE as of 14 juil. 2015

%\cite{Fukugita:1988hg}
\bibitem{Fukugita:1988hg}
  M.~Fukugita, Y.~Kohyama and K.~Kubodera,
  %``Neutrino Reaction Cross-sections On C-12 Target,''
  Phys.\ Lett.\ B {\bf 212}, 139 (1988).
  %%CITATION = PHLTA,B212,139;%%
  %60 citations counted in INSPIRE as of 11 Apr 2014

%\cite{Volpe:2000zn}
\bibitem{Volpe:2000zn}
  C.~Volpe, N.~Auerbach, G.~Colo, T.~Suzuki and N.~Van Giai,
  %``Microscopic theories of neutrino C-12 reactions,''
  Phys.\ Rev.\ C {\bf 62}, 015501 (2000)  [nucl-th/0001050].
  %%CITATION = NUCL-TH/0001050;%%
  %83 citations counted in INSPIRE as of 11 Apr 2014

%\cite{Auerbach:2001hz}
\bibitem{Auerbach:2001hz}
  L.~B.~Auerbach {\it et al.}  [LSND Collaboration],
  %``Measurements of charged current reactions of nu(e) on 12-C,''
  Phys.\ Rev.\ C {\bf 64}, 065501 (2001)  [hep-ex/0105068].
  %%CITATION = HEP-EX/0105068;%%
  %51 citations counted in INSPIRE as of 11 Apr 2014

%\cite{Marciano:2003eq}
\bibitem{Marciano:2003eq}
  W.~J.~Marciano and Z.~Parsa,
  %``Neutrino electron scattering theory,''
  J.\ Phys.\ G {\bf 29}, 2629 (2003)  [hep-ph/0403168].
  %%CITATION = HEP-PH/0403168;%%
  %33 citations counted in INSPIRE as of 14 Apr 2014

%\cite{Beacom:2002hs}
\bibitem{Beacom:2002hs}
  J.~F.~Beacom, W.~M.~Farr and P.~Vogel,
  %``Detection of supernova neutrinos by neutrino proton elastic scattering,''
  Phys.\ Rev.\ D {\bf 66}, 033001 (2002)  [hep-ph/0205220].
  %%CITATION = HEP-PH/0205220;%%
  %86 citations counted in INSPIRE as of 14 Apr 2014

%\cite{Dasgupta:2011wg}
\bibitem{Dasgupta:2011wg}
  B.~Dasgupta and J.~F.~Beacom,
  %``Reconstruction of supernova $\nu_\mu$, $\nu_\tau$, anti-$\nu_\mu$, and anti-$\nu_\tau$ neutrino spectra at scintillator detectors,''
  Phys.\ Rev.\ D {\bf 83}, 113006 (2011)  [arXiv:1103.2768].
  %%CITATION = ARXIV:1103.2768;%%
  %19 citations counted in INSPIRE as of 14 Apr 2014

%\cite{Weinberg:1972tu}
\bibitem{Weinberg:1972tu}
  S.~Weinberg,
  %``Effects of a neutral intermediate boson in semileptonic processes,''
  Phys.\ Rev.\ D {\bf 5}, 1412 (1972).
  %%CITATION = PHRVA,D5,1412;%%
  %671 citations counted in INSPIRE as of 14 Apr 2014

%\cite{vonKrosigk:2013sa}
\bibitem{vonKrosigk:2013sa}
  B.~von Krosigk, L.~Neumann, R.~Nolte, S.~Rottger and K.~Zuber,
  %``Measurement of the proton light response of various LAB based scintillators and its implication for supernova neutrino detection via neutrino-proton scattering,''
  Eur.\ Phys.\ J.\ C {\bf 73}, 2390 (2013)  [arXiv:1301.6403].
  %%CITATION = ARXIV:1301.6403;%%
  %1 citations counted in INSPIRE as of 14 Apr 2014

\bibitem{PSTAR}
PSTAR Database: http://physics.nist.gov/PhysRefData/Star/Text/PSTAR.html

%\bibitem{Basel}
%T. Fischer, S. C. Whitehouse, A. Mezzacappa, F.-K. Thielemann and M. Liebend\"{o}rfer, Astron. Astrophys. {\bf 517}, A80 (2010).

%\cite{Fischer:2009af}
\bibitem{Fischer:2009af}
  T.~Fischer, S.~C.~Whitehouse, A.~Mezzacappa, F.-K.~Thielemann and M.~Liebendorfer,
  %``Protoneutron star evolution and the neutrino driven wind in general relativistic neutrino radiation hydrodynamics simulations,''
  Astron.\ Astrophys.\  {\bf 517}, A80 (2010)
  [arXiv:0908.1871].  %%CITATION = ARXIV:0908.1871;%%

%\bibitem{Garching1}
%
%R. Buras, H.-Th. Janka, M. Rampp and K. Kifonidis,
%Astron. Astrophys. {\bf 457}, 281 (2006);

%\cite{Buras:2005tb}
\bibitem{Buras:2005tb}
  R.~Buras, H.~T.~Janka, M.~Rampp and K.~Kifonidis,
  %``Two-dimensional hydrodynamic core-collapse supernova simulations with spectral neutrino transport. 2. models for different progenitor stars,''
  Astron.\ Astrophys.\  {\bf 457}, 281 (2006)
  [astro-ph/0512189].  %%CITATION = ASTRO-PH/0512189;%%

%\bibitem{Garching2}
%P. D. Serpico, S. Chakraborty, T. Fischer, L. H\"{u}depohl, H.-Th. Janka and A. Mirizzi, Phys. Rev. D {\bf 85}, 085031 (2012).

%\cite{Serpico:2011ir}
\bibitem{Serpico:2011ir}
  P.~D.~Serpico, S.~Chakraborty, T.~Fischer, L.~Hudepohl, H.~T.~Janka and A.~Mirizzi,
  %``Probing the neutrino mass hierarchy with the rise time of a supernova burst,''
  Phys.\ Rev.\ D {\bf 85}, 085031 (2012)  [arXiv:1111.4483 [astro-ph.SR]].  %%CITATION = ARXIV:1111.4483;%%

%\bibitem{Nakazato}
%K. Nakazato, K. Sumiyoshi, H. Suzuki, T. Totani, H. Umeda and S. Yamada, Astrophys. J. Supp. {\bf 215}, 2 (2013).

%\cite{Nakazato:2012qf}
\bibitem{Nakazato:2012qf}
  K.~Nakazato, K.~Sumiyoshi, H.~Suzuki, T.~Totani, H.~Umeda and S.~Yamada,
  %``Supernova Neutrino Light Curves and Spectra for Various Progenitor Stars: From Core Collapse to Proto-neutron Star Cooling,''
  Astrophys.\ J.\ Suppl.\  {\bf 205}, 2 (2013)
  [arXiv:1210.6841].  %%CITATION = ARXIV:1210.6841;%%

%%\cite{Bellini:2009jr}
%\bibitem{Bellini:2009jr}
%  G.~Bellini {\it et al.}  [Borexino Collaboration],
%  %``New experimental limits on the Pauli forbidden transitions in C-12 nuclei obtained with 485 days Borexino data,''
%  Phys.\ Rev.\ C {\bf 81}, 034317 (2010)  [arXiv:0911.0548].
%  %%CITATION = ARXIV:0911.0548;%%
%  %19 citations counted in INSPIRE as of 19 Apr 2014

%\cite{Burrows:2012ew}
\bibitem{Burrows:2012ew}
  A.~Burrows,
  %``Colloquium: Perspectives on core-collapse supernova theory,''
  Rev.\ Mod.\ Phys.\  {\bf 85}, 245 (2013)
  [arXiv:1210.4921 [astro-ph.SR]].  %%CITATION = ARXIV:1210.4921;%%

%\cite{Qian:1993dg}
\bibitem{Fuller:1992}
  G.~M.~Fuller, R.~Mayle, B.~S.~Meyer, and J.~R.~Wilson,
  %``Can a closure mass neutrino help solve the supernova shock reheating problem?''
  Astrophys.\ J.\  {\bf 389}, 517 (1992).
  %%CITATION = PRLTA,71,1965;%%
  %202 citations counted in INSPIRE as of 16 juin 2015

%\cite{Qian:1993dg}
\bibitem{Qian:1993dg}
  Y.~Z.~Qian, G.~M.~Fuller, G.~J.~Mathews, R.~Mayle, J.~R.~Wilson and S.~E.~Woosley,
  %``A Connection between flavor mixing of cosmologically significant neutrinos and heavy element nucleosynthesis in supernovae,''
  Phys.\ Rev.\ Lett.\  {\bf 71}, 1965 (1993).
  %%CITATION = PRLTA,71,1965;%%
  %202 citations counted in INSPIRE as of 16 juin 2015

%\bibitem{Adams:2013qkq}
%  C.~Adams {\it et al.}  [LBNE Collaboration],
%  %``The Long-Baseline Neutrino Experiment: Exploring Fundamental Symmetries of the Universe,''
%  arXiv:1307.7335 [hep-ex].
%  %%CITATION = ARXIV:1307.7335;%%

%\cite{Itoh:1996}
\bibitem{Itoh:1996}
  N.~Itoh, H.~Hayashi, A.~Nishikawa and Y.~Kohyama,
  %``Neutrino Energy Loss in Stellar Interiors. VII. Pair, Photo-, Plasma, Bremsstrahlung, and Recombination Neutrino Processes,''
  Astrophys.\ J.\ Suppl.\  {\bf 102}, 411 (1996).

%\cite{Odrzywolek:2003vn}
\bibitem{Odrzywolek:2003vn}
  A.~Odrzywolek, M.~Misiaszek and M.~Kutschera,
  %``Detection possibility of the pair - annihilation neutrinos from the neutrino - cooled pre-supernova star,''
  Astropart.\ Phys.\  {\bf 21}, 303 (2004)  [astro-ph/0311012].
  %%CITATION = ASTRO-PH/0311012;%%
  %20 citations counted in INSPIRE as of 21 Apr 2014

%\cite{Odrzywolek:2010zz}
\bibitem{Odrzywolek:2010zz}
  A.~Odrzywolek and A.~Heger,
  %``Neutrino signatures of dying massive stars: From main sequence to the neutron star,''
  Acta Phys.\ Polon.\ B {\bf 41}, 1611 (2010).
  %%CITATION = APPOA,B41,1611;%%
  %2 citations counted in INSPIRE as of 15 Jul 2015

%\cite{Beacom:1998fj}
\bibitem{Beacom:1998fj}
  J.~F.~Beacom and P.~Vogel,
  %``Can a supernova be located by its neutrinos?,''
  Phys.\ Rev.\ D {\bf 60}, 033007 (1999)  [astro-ph/9811350].
  %%CITATION = ASTRO-PH/9811350;%%
  %81 citations counted in INSPIRE as of 02 Aug 2014

%\cite{Tomas:2003xn}
\bibitem{Tomas:2003xn}
  R.~Tom\`{a}s, D.~Semikoz, G.~G.~Raffelt, M.~Kachelrie{\ss} and A.~S.~Dighe,
  %``Supernova pointing with low-energy and high-energy neutrino detectors,''
  Phys.\ Rev.\ D {\bf 68}, 093013 (2003)  [hep-ph/0307050].
  %%CITATION = HEP-PH/0307050;%%
  %48 citations counted in INSPIRE as of 02 Aug 2014

%\cite{Mueller:2003fs}
\bibitem{Mueller:2003fs}
  E.~M\"{u}ller, M.~Rampp, R.~Buras, H.-Th.~Janka and D.~H.~Shoemaker,
  %``Towards gravitational wave signals from realistic core collapse supernova models,''
  Astrophys.\ J.\  {\bf 603}, 221 (2004)  [astro-ph/0309833].
  %%CITATION = ASTRO-PH/0309833;%%
  %93 citations counted in INSPIRE as of 02 Aug 2014

%\cite{Qian:2003wd}
\bibitem{Qian:2003wd}
  Y.~Z.~Qian,
  %``The origin of the heavy elements: recent progress in the understanding of the r-process,''
  Prog.\ Part.\ Nucl.\ Phys.\  {\bf 50}, 153 (2003)
  [astro-ph/0301422].
  %%CITATION = ASTRO-PH/0301422;%%

%\bibitem{Woosley:1990}
% S.~E.~Woosley, D.~H.~Hartmann, R.~D.~Hoffman and W.~C.~Haxton, Astrophys. J. {\bf 356}, 272 (1990).

 %\cite{Woosley:1989bd}
\bibitem{Woosley:1989bd}
  S.~E.~Woosley, D.~H.~Hartmann, R.~D.~Hoffman and W.~C.~Haxton,
  %``The Neutrino Process,''
  Astrophys.\ J.\  {\bf 356}, 272 (1990).  %%CITATION = ASJOA,356,272;%%

%\cite{Heger:2003mm}
\bibitem{Heger:2003mm}
  A.~Heger, E.~Kolbe, W.~C.~Haxton, K.~Langanke, G.~Mart\'{\i}nez-Pinedo and S.~E.~Woosley,
  %``Neutrino nucleosynthesis,''
  Phys.\ Lett.\ B {\bf 606}, 258 (2005)  [astro-ph/0307546].
  %%CITATION = ASTRO-PH/0307546;%%
  %85 citations counted in INSPIRE as of 17 Sep 2014

%\cite{Yoshida:2005uy}
\bibitem{Yoshida:2005uy}
  T.~Yoshida, T.~Kajino and D.~H.~Hartmann,
  %``Constraining the spectrum of supernova neutrinos from neutrino-process-induced light-element synthesis,''
  Phys.\ Rev.\ Lett.\  {\bf 94}, 231101 (2005)  [astro-ph/0505043].
  %%CITATION = ASTRO-PH/0505043;%%
  %38 citations counted in INSPIRE as of 17 Sep 2014

%\cite{Wu:2014kaa}
\bibitem{Wu:2014kaa}
  M.~R.~Wu, Y.~Z.~Qian, G.~Mart\'{\i}nez-Pinedo, T.~Fischer and L.~Huther,
  %``Effects of neutrino oscillations on nucleosynthesis and neutrino signals for an 18 $M$$_{\odot}$ supernova model,''
  Phys.\ Rev.\ D {\bf 91}, no. 6, 065016 (2015)
  [arXiv:1412.8587].
  %%CITATION = ARXIV:1412.8587;%%
  %5 citations counted in INSPIRE as of 16 Jun 2015

\bibitem{Raffelt:1996bk}
  G.~G.~Raffelt, {\it Stars as Laboratories for Fundamental Physics} (Chicago University Press, Chicago, 1996).

\bibitem{Raffelt:1999gv}
  G.~G.~Raffelt,
  %``Limits on neutrino electromagnetic properties: An update,''
  Phys.\ Rept.\  {\bf 320}, 319 (1999).
  %%CITATION = PRPLC,320,319;%%

%\cite{Tamborra:2011is}
\bibitem{Tamborra:2011is}
  I.~Tamborra, G.~G.~Raffelt, L.~H\"{u}depohl and H.-Th.~Janka,
  %``Impact of eV-mass sterile neutrinos on neutrino-driven supernova outflows,''
  JCAP {\bf 1201}, 013 (2012)
  [arXiv:1110.2104].
  %%CITATION = ARXIV:1110.2104;%%
  %13 citations counted in INSPIRE as of 17 Jun 2015

%\cite{Wu:2013gxa}
\bibitem{Wu:2013gxa}
  M.~R.~Wu, T.~Fischer, L.~Huther, G.~Mart\'{\i}nez-Pinedo and Y.~Z.~Qian,
  %``Impact of active-sterile neutrino mixing on supernova explosion and nucleosynthesis,''
  Phys.\ Rev.\ D {\bf 89}, 061303 (2014)
  [arXiv:1305.2382].
  %%CITATION = ARXIV:1305.2382;%%
  %12 citations counted in INSPIRE as of 16 juin 2015

%\cite{Zatsepin:1968kt}
\bibitem{Zatsepin:1968kt}
  G.~T.~Zatsepin,
  %``On the possibility of determining the upper limit of the neutrino mass by means of the flight time,''
  Pisma Zh.\ Eksp.\ Teor.\ Fiz.\  {\bf 8}, 333 (1968).
  %%CITATION = ZFPRA,8,333;%%  %15 citations counted in INSPIRE as of 11 Feb 2014

%\cite{Beacom:2000ng}
\bibitem{Beacom:2000ng}
  J.~F.~Beacom, R.~N.~Boyd and A.~Mezzacappa,
  %``Technique for direct eV scale measurements of the mu and tau neutrino masses using supernova neutrinos,''
  Phys.\ Rev.\ Lett.\  {\bf 85}, 3568 (2000)
  [hep-ph/0006015].
  %%CITATION = HEP-PH/0006015;%%
  %66 citations counted in INSPIRE as of 15 Jul 2015

%\cite{Pagliaroli:2008ur}
\bibitem{Pagliaroli:2008ur}
  G.~Pagliaroli, F.~Vissani, M.~L.~Costantini and A.~Ianni,
  %``Improved analysis of SN1987A antineutrino events,''
  Astropart.\ Phys.\  {\bf 31}, 163 (2009)  [arXiv:0810.0466].
  %%CITATION = ARXIV:0810.0466;%%
  %38 citations counted in INSPIRE as of 13 Feb 2014

%\cite{Pagliaroli:2009qy}
\bibitem{Pagliaroli:2009qy}
  G.~Pagliaroli, F.~Vissani, E.~Coccia and W.~Fulgione,
  %``Neutrinos from Supernovae as a Trigger for Gravitational Wave Search,''
  Phys.\ Rev.\ Lett.\  {\bf 103}, 031102 (2009)  [arXiv:0903.1191].
   %%CITATION = ARXIV:0903.1191;%%
   %33 citations counted in INSPIRE as of 13 Feb 2014

%\cite{Pagliaroli:2010ik}
\bibitem{Pagliaroli:2010ik}
  G.~Pagliaroli, F.~Rossi-Torres and F.~Vissani,
  %``Neutrino mass bound in the standard scenario for supernova electronic antineutrino emission,''
  Astropart.\ Phys.\  {\bf 33}, 287 (2010)  [arXiv:1002.3349].
  %%CITATION = ARXIV:1002.3349;%%
  %14 citations counted in INSPIRE as of 11 Feb 2014

%\cite{Lu:2014zma}
\bibitem{Lu:2014zma}
  J.~S.~Lu, J.~Cao, Y.~F.~Li and S.~Zhou,
  %``Constraining Absolute Neutrino Masses via Detection of Galactic Supernova Neutrinos at JUNO,''
  JCAP {\bf 1505}, 044 (2015)
  [arXiv:1412.7418].
  %%CITATION = ARXIV:1412.7418;%%
  %2 citations counted in INSPIRE as of 17 juin 2015

%\cite{Duan:2010bg}
\bibitem{Duan:2010bg}
  H.~Duan, G.~M.~Fuller and Y.~Z.~Qian,
  %``Collective Neutrino Oscillations,''
  Ann.\ Rev.\ Nucl.\ Part.\ Sci.\  {\bf 60}, 569 (2010)  [arXiv:1001.2799].
  %%CITATION = ARXIV:1001.2799;%%
  %121 citations counted in INSPIRE as of 04 May 2014

\bibitem{Fogli:2007bk}
  G.~L.~Fogli, E.~Lisi, A.~Marrone and A.~Mirizzi,
  %``Collective neutrino flavor transitions in supernovae and the role of %trajectory averaging,''
  JCAP {\bf 0712}, 010 (2007)
  [arXiv:0707.1998].
  %%CITATION = ARXIV:0707.1998;%%

%\cite{Dighe:2003be}
\bibitem{Dighe:2003be}
  A.~S.~Dighe, M.~T.~Keil and G.~G.~Raffelt,
  %``Detecting the neutrino mass hierarchy with a supernova at IceCube,''
  JCAP {\bf 0306}, 005 (2003)
  [hep-ph/0303210].
  %%CITATION = HEP-PH/0303210;%%
  %80 citations counted in INSPIRE as of 15 Jul 2015

%\cite{Dighe:2003jg}
\bibitem{Dighe:2003jg}
  A.~S.~Dighe, M.~T.~Keil and G.~G.~Raffelt,
  %``Identifying earth matter effects on supernova neutrinos at a single detector,''
  JCAP {\bf 0306}, 006 (2003)
  [hep-ph/0304150].
  %%CITATION = HEP-PH/0304150;%%
  %88 citations counted in INSPIRE as of 15 Jul 2015

%\cite{Tomas:2004gr}
\bibitem{Tomas:2004gr}
  R.~Tomas, M.~Kachelriess, G.~Raffelt, A.~Dighe, H.-T.~Janka and L.~Scheck,
  %``Neutrino signatures of supernova shock and reverse shock propagation,''
  JCAP {\bf 0409}, 015 (2004)
  [astro-ph/0407132].
  %%CITATION = ASTRO-PH/0407132;%%
  %104 citations counted in INSPIRE as of 15 juil. 2015

%\cite{Borriello:2013tha}
\bibitem{Borriello:2013tha}
  E.~Borriello, S.~Chakraborty, H.~T.~Janka, E.~Lisi and A.~Mirizzi,
  %``Turbulence patterns and neutrino flavor transitions in high-resolution supernova models,''
  JCAP {\bf 1411}, no. 11, 030 (2014)
  [arXiv:1310.7488].
  %%CITATION = ARXIV:1310.7488;%%
  %7 citations counted in INSPIRE as of 15 juil. 2015

% chap: SN diffuse
%%%%%%%%%%%%%%%%%%%%%%%%%%%%%%%%%%%%%%%%%%%%%%%%%%%%%%%%%%%%%%%%%%%%%%%%%%%%%%%%%%%%
%%%%%%%%%%%%%%%%%%%%%%%%%%%%%%%%%%%%%%%%%%%%%%%%%%%%%%%%%%%%%%%%%%%%%%%%%%%%%%%%%%%%
% chap: SN diffuse


%\begin{thebibliography}{99}

%\cite{Krauss:1983zn}
\bibitem{Krauss:1983zn}
  L.~M.~Krauss, S.~L.~Glashow and D.~N.~Schramm,
  %``Anti-neutrinos Astronomy and Geophysics,''
  Nature {\bf 310}, 191 (1984).
  %%CITATION = NATUA,310,191;%%
  %105 citations counted in INSPIRE as of 30 Apr 2014

%\cite{Bisnovatyi-Kogan:1984}
\bibitem{Bisnovatyi-Kogan:1984}
  G.~S.~Bisnovatyi-Kogan and S.~F.~Seidov,
  %``Supernovae, Neutrino Rest Mass, and the Middle-Energy Neutrino Background in the Universe''
  Annals N.~Y.\ Acad.\ Sci.\ {\bf 422}, 319 (1984).

%\cite{Woosley:1986}
\bibitem{Woosley:1986}
  S.~E.~Woosley, J.~R.~Wilson and R.~Mayle,
  %``Gravitational collapse and the cosmic antineutrino background''
  Astrophys.\ J.\ {\bf 302}, 19 (1986).


%\cite{Ando:2004hc}
\bibitem{Ando:2004hc}
  S.~Ando and K.~Sato,
  %``Relic neutrino background from cosmological supernovae,''
  New J.\ Phys.\  {\bf 6}, 170 (2004)
  [astro-ph/0410061].
  %%CITATION = ASTRO-PH/0410061;%%
  %70 citations counted in INSPIRE as of 30 Apr 2014

%\cite{Lunardini:2005jf}
\bibitem{Lunardini:2005jf}
  C.~Lunardini,
  %``The diffuse supernova neutrino flux, supernova rate and sn1987a,''
  Astropart.\ Phys.\  {\bf 26}, 190 (2006)
  [astro-ph/0509233].
  %%CITATION = ASTRO-PH/0509233;%%
  %35 citations counted in INSPIRE as of 01 May 2014

%\cite{Lunardini:2012ne}
\bibitem{Lunardini:2012ne}
  C.~Lunardini and I.~Tamborra,
  %``Diffuse supernova neutrinos: oscillation effects, stellar cooling and progenitor mass dependence,''
  JCAP {\bf 1207}, 012 (2012)
  [arXiv:1205.6292].
  %%CITATION = ARXIV:1205.6292;%%
  %9 citations counted in INSPIRE as of 03 Jul 2015


%\cite{Raffelt:2009mm}
\bibitem{Raffelt:2009mm}
  G.~Raffelt and T.~Rashba,
  %``Mimicking diffuse supernova antineutrinos with the Sun as a source,''
  Phys.\ Atom.\ Nucl.\  {\bf 73}, 609 (2010)
  [arXiv:0902.4832].
  %%CITATION = ARXIV:0902.4832;%%
  %6 citations counted in INSPIRE as of 01 May 2014

%\cite{Malek:2002ns}
\bibitem{Malek:2002ns}
  M.~Malek {\it et al.} (Super-Kamiokande Collaboration),
  %``Search for supernova relic neutrinos at SUPER-KAMIOKANDE,''
  Phys.\ Rev.\ Lett.\  {\bf 90}, 061101 (2003)
  [hep-ex/0209028].
  %%CITATION = HEP-EX/0209028;%%
  %164 citations counted in INSPIRE as of 28 Apr 2014

%\cite{Bays:2011si}
\bibitem{Bays:2011si}
  K.~Bays {\it et al.} (Super-Kamiokande Collaboration),
  %``Supernova Relic Neutrino Search at Super-Kamiokande,''
  Phys.\ Rev.\ D {\bf 85}, 052007 (2012)
  [arXiv:1111.5031].
  %%CITATION = ARXIV:1111.5031;%%
  %24 citations counted in INSPIRE as of 28 Apr 2014

%\cite{Zhang:2013tua}
\bibitem{Zhang:2013tua}
  H.~Zhang {\it et al.} (Super-Kamiokande Collaboration),
  %``Supernova Relic Neutrino Search with Neutron Tagging at Super-Kamiokande-IV,''
  Astropart.\ Phys.\ {\bf 60}, 41 (2014)
  [arXiv:1311.3738].
  %%CITATION = ARXIV:1311.3738;%%
  %8 citations counted in INSPIRE as of 03 Jul 2015

%\cite{Aharmim:2006wq}
\bibitem{Aharmim:2006wq}
  B.~Aharmim {\it et al.}  (SNO Collaboration),
  %``A Search for Neutrinos from the Solar hep Reaction and the Diffuse Supernova Neutrino %Background with the Sudbury Neutrino Observatory,''
  Astrophys.\ J.\  {\bf 653}, 1545 (2006)
  [hep-ex/0607010].
  %%CITATION = HEP-EX/0607010;%%
  %41 citations counted in INSPIRE as of 01 May 2014

%\cite{Gando:2011jza}
\bibitem{Gando:2011jza}
  A.~Gando {\it et al.} (KamLAND Collaboration),
  %``A study of extraterrestrial antineutrino sources with the KamLAND detector,''
  Astrophys.\ J.\  {\bf 745}, 193 (2012)
  [arXiv:1105.3516].
  %%CITATION = ARXIV:1105.3516;%%
  %5 citations counted in INSPIRE as of 01 May 2014

%\cite{Beacom:2003nk}
\bibitem{Beacom:2003nk}
  J.~F.~Beacom and M.~R.~Vagins,
  %``GADZOOKS! Anti-neutrino spectroscopy with large water Cherenkov detectors,''
  Phys.\ Rev.\ Lett.\  {\bf 93}, 171101 (2004)
  [hep-ph/0309300].
  %%CITATION = HEP-PH/0309300;%%
  %189 citations counted in INSPIRE as of 01 May 2014

%\cite{Watanabe:2008ru}
\bibitem{Watanabe:2008ru}
  H.~Watanabe {\it et al.} (Super-Kamiokande Collaboration),
  %``First Study of Neutron Tagging with a Water Cherenkov Detector,''
  Astropart.\ Phys.\  {\bf 31}, 320 (2009)
  [arXiv:0811.0735].
  %%CITATION = ARXIV:0811.0735;%%
  %33 citations counted in INSPIRE as of 01 May 2014

%\cite{Vagins:2012vta}
\bibitem{Vagins:2012vta}
  M.~R.~Vagins,
  %``Detection of Supernova Neutrinos,''
  Nucl.\ Phys.\ Proc.\ Suppl.\  {\bf 229--232}, 325 (2012).
  %%CITATION = NUPHZ,229-232,325;%%
  %2 citations counted in INSPIRE as of 01 May 2014

%\cite{Horiuchi:2008jz}
\bibitem{Horiuchi:2008jz}
  S.~Horiuchi, J.~F.~Beacom and E.~Dwek,
  %``The Diffuse Supernova Neutrino Background is detectable in Super-Kamiokande,''
  Phys.\ Rev.\ D {\bf 79}, 083013 (2009)
  [arXiv:0812.3157].
  %%CITATION = ARXIV:0812.3157;%%
  %64 citations counted in INSPIRE as of 01 May 2014

%\cite{Porciani:2000ag}
\bibitem{Porciani:2000ag}
  C.~Porciani and P.~Madau,
  %``On the Association of gamma-ray bursts with massive stars: implications for number counts and lensing %statistics,''
  Astrophys.\ J.\ {\bf 548}, 522 (2001)
  [astro-ph/0008294].
  %%CITATION = ASTRO-PH/0008294;%%
  %211 citations counted in INSPIRE as of 01 May 2014

%\cite{Hopkins:2006bw}
\bibitem{Hopkins:2006bw}
  A.~M.~Hopkins and J.~F.~Beacom,
  %``On the normalisation of the cosmic star formation history,''
  Astrophys.\ J.\  {\bf 651}, 142 (2006)
  [astro-ph/0601463].
  %%CITATION = ASTRO-PH/0601463;%%
  %778 citations counted in INSPIRE as of 01 May 2014

%\cite{Yuksel:2008cu}
\bibitem{Yuksel:2008cu}
  H.~Y\"uksel, M.~D.~Kistler, J.~F.~Beacom and A.~M.~Hopkins,
  %``Revealing the High-Redshift Star Formation Rate with Gamma-Ray Bursts,''
  Astrophys.\ J.\  {\bf 683}, L5 (2008)
  [arXiv:0804.4008].
  %%CITATION = ARXIV:0804.4008;%%
  %82 citations counted in INSPIRE as of 01 May 2014

%\cite{Horiuchi:2011zz}
\bibitem{Horiuchi:2011zz}
  S.~Horiuchi, J.~F.~Beacom, C.~S.~Kochanek, J.~L.~Prieto, K.~Z.~Stanek and T.~A.~Thompson,
  %``The Cosmic Core-collapse Supernova Rate does not match the Massive-Star Formation Rate,''
  Astrophys.\ J.\  {\bf 738}, 154 (2011)
  [arXiv:1102.1977].
  %%CITATION = ARXIV:1102.1977;%%
  %43 citations counted in INSPIRE as of 01 May 2014

%\cite{Heger:2002by}
\bibitem{Heger:2002by}
  A.~Heger, C.~L.~Fryer, S.~E.~Woosley, N.~Langer and D.~H.~Hartmann,
  %``How massive single stars end their life,''
  Astrophys.\ J.\  {\bf 591}, 288 (2003)
  [astro-ph/0212469].
  %%CITATION = ASTRO-PH/0212469;%%
  %679 citations counted in INSPIRE as of 01 May 2014

%\cite{Fischer:2008rh}
\bibitem{Fischer:2008rh}
  T.~Fischer, S.~C.~Whitehouse, A.~Mezzacappa, F.-K.~Thielemann and M.~Liebend\"orfer,
  %``The neutrino signal from protoneutron star accretion and black hole formation,''
  Astron.\ Astrophys.\ {\bf 499}, 1 (2009)
  [arXiv:0809.5129].
  %%CITATION = ARXIV:0809.5129;%%
  %15 citations counted in INSPIRE as of 01 May 2014

%\cite{Nakazato:2008vj}
\bibitem{Nakazato:2008vj}
  K.~Nakazato, K.~Sumiyoshi, H.~Suzuki and S.~Yamada,
  %``Oscillation and Future Detection of Failed Supernova Neutrinos from Black Hole Forming Collapse,''
  Phys.\ Rev.\ D {\bf 78}, 083014 (2008);
  Erratum {\em ibid.} {\bf 79}, 069901 (2009)
  [arXiv:0810.3734].
  %%CITATION = ARXIV:0810.3734;%%
  %22 citations counted in INSPIRE as of 01 May 2014

%\cite{Kochanek:2008mp}
\bibitem{Kochanek:2008mp}
  C.~S.~Kochanek, J.~F.~Beacom, M.~D.~Kistler, J.~L.~Prieto, K.~Z.~Stanek, T.~A.~Thompson and H.~Y\"uksel,
  %``A Survey About Nothing: Monitoring a Million Supergiants for Failed Supernovae,''
  Astrophys.\ J.\  {\bf 684}, 1336 (2008)
  [arXiv:0802.0456].
  %%CITATION = ARXIV:0802.0456;%%
  %67 citations counted in INSPIRE as of 01 May 2014

%\cite{Lunardini:2009ya}
\bibitem{Lunardini:2009ya}
  C.~Lunardini,
  %``Diffuse neutrino flux from failed supernovae,''
  Phys.\ Rev.\ Lett.\ {\bf 102}, 231101 (2009)
  [arXiv:0901.0568].
  %%CITATION = ARXIV:0901.0568;%%
  %20 citations counted in INSPIRE as of 01 May 2014

%\cite{O'Connor:2010tk}
\bibitem{O'Connor:2010tk}
  E.~O'Connor and C.~D.~Ott,
  %``Black Hole Formation in Failing Core-Collapse Supernovae,''
  Astrophys.\ J.\  {\bf 730}, 70 (2011)
  [arXiv:1010.5550].
  %%CITATION = ARXIV:1010.5550;%%
  %82 citations counted in INSPIRE as of 01 May 2014

%\cite{Ugliano:2012kq}
\bibitem{Ugliano:2012kq}
  M.~Ugliano, H.-T.~Janka, A.~Marek and A.~Arcones,
  %``Progenitor-Explosion Connection and Remnant Birth Masses for Neutrino-Driven Supernovae of Iron-Core %Progenitors,''
  Astrophys.\ J.\ {\bf 757}, 69 (2012)
  [arXiv:1205.3657].
  %%CITATION = ARXIV:1205.3657;%%
  %35 citations counted in INSPIRE as of 01 May 2014

%\cite{Keehn:2010pn}
\bibitem{Keehn:2010pn}
  J.~G.~Keehn and C.~Lunardini,
  %``Neutrinos from failed supernovae at future water and liquid argon detectors,''
  Phys.\ Rev.\ D {\bf 85}, 043011 (2012)
  [arXiv:1012.1274].
  %%CITATION = ARXIV:1012.1274;%%
  %6 citations counted in INSPIRE as of 01 May 2014

%\cite{Kochanek:2013yca}
\bibitem{Kochanek:2013yca}
  C.~S.~Kochanek,
  %``Failed Supernovae Explain the Compact Remnant Mass Function,''
  Astrophys.\ J.\  {\bf 785}, 28 (2014)
  [arXiv:1308.0013].
  %%CITATION = ARXIV:1308.0013;%%
  %3 citations counted in INSPIRE as of 01 May 2014

%%\cite{Wurm:2011zn}
%\bibitem{dWurm:2011zn}
%  M.~Wurm {\it et al.} (LENA Collaboration),
%  %``The next-generation liquid-scintillator neutrino observatory LENA,''
%  Astropart.\ Phys.\ {\bf 35}, 685 (2012)
%  [arXiv:1104.5620].
%  %%CITATION = ARXIV:1104.5620;%%
%  %75 citations counted in INSPIRE as of 01 May 2014

%%\cite{Keil:2002in}
%\bibitem{dKeil:2002in}
%  M.~T.~Keil, G.~G.~Raffelt and H.-T.~Janka,
%  %``Monte Carlo study of supernova neutrino spectra formation,''
%  Astrophys.\ J.\  {\bf 590}, 971 (2003)
%  [astro-ph/0208035].
%  %%CITATION = ASTRO-PH/0208035;%%
%  %252 citations counted in INSPIRE as of 01 May 2014

%\cite{Tamborra:2012ac}
\bibitem{Tamborra:2012ac}
  I.~Tamborra, B.~M\"uller, L.~H\"udepohl, H.-T.~Janka and G.~Raffelt,
  %``High-resolution supernova neutrino spectra represented by a simple fit,''
  Phys.\ Rev.\ D {\bf 86}, 125031 (2012)
  [arXiv:1211.3920].
  %%CITATION = ARXIV:1211.3920;%%
  %8 citations counted in INSPIRE as of 01 May 2014


%\cite{Mollenberg:2014pwa}
\bibitem{Mollenberg:2014pwa}
  R.~M?llenberg, F.~von Feilitzsch, D.~Hellgartner, L.~Oberauer, M.~Tippmann, V.~Zimmer, J.~Winter and M.~Wurm,
  %``Detecting the Diffuse Supernova Neutrino Background with LENA,''
  Phys.\ Rev.\ D {\bf 91}, no. 3, 032005 (2015)
  [arXiv:1409.2240 [astro-ph.IM]].
  %%CITATION = ARXIV:1409.2240;%%

  \bibitem{jileixu15}
  Jilei Xu,
  %``Relic neutrino background from cosmological supernovae,''
   ``Rock Neutron Background Simulation", JUNO General Meeting, Guangzhou (2015)
  %%CITATION = ASTRO-PH/0410061;%%
  %70 citations counted in INSPIRE as of 30 Apr 2014

%\bibitem{Strumia:2003zx}
%A.~Strumia and F.~Vissani, Phys.\ Lett\ B {\bf 564}, 42-54 (2003) [astro-ph/0302055]

%\cite{Wurm:2007cy}
\bibitem{Wurm:2007cy}
  M.~Wurm, F.~von Feilitzsch, M.~G\"oger-Neff, K.~A.~Hochmuth, T.~Marrod\'an~Undagoitia, L.~Oberauer and W.~Potzel,
  %``Detection potential for the diffuse supernova neutrino background in the large liquid-scintillator %detector LENA,''
  Phys.\ Rev.\ D {\bf 75}, 023007 (2007)
  [astro-ph/0701305].
  %%CITATION = ASTRO-PH/0701305;%%
  %36 citations counted in INSPIRE as of 02 May 2014

%%\cite{Rolke:2004mj}
%\bibitem{Rolke:2004mj}
%  W.~A.~Rolke, A.~M.~L\'opez and J.~Conrad,
%  %``Limits and confidence intervals in the presence of nuisance parameters,''
%  Nucl.\ Instrum.\ Meth.\ A {\bf 551}, 493 (2005)
%  [physics/0403059].
%  %%CITATION = PHYSICS/0403059;%%
%  %167 citations counted in INSPIRE as of 02 May 2014
%
%
%%\cite{Tamborra:2014aua}
%\bibitem{Tamborra:2014aua}
%  I.~Tamborra, F.~Hanke, H.-T.~Janka, B.~M\"uller, G.~G.~Raffelt and A.~Marek,
%  %``Self-sustained asymmetry of lepton-number emission: A new phenomenon during the supernova %shock-accretion phase in three dimensions,''
%  Astrophys.\ J.\  {\bf 792}, 96 (2014)
%  [arXiv:1402.5418].
%  %%CITATION = ARXIV:1402.5418;%%
%  %14 citations counted in INSPIRE as of 03 Jul 2015

%\cite{Tamborra:2014hga}
\bibitem{Tamborra:2014hga}
  I.~Tamborra, G.~Raffelt, F.~Hanke, H.-T.~Janka and B.~M\"uller,
  %``Neutrino emission characteristics and detection opportunities based on three-dimensional %supernova simulations,''
  Phys.\ Rev.\ D {\bf 90}, 045032 (2014)
  [arXiv:1406.0006].
  %%CITATION = ARXIV:1406.0006;%%
  %7 citations counted in INSPIRE as of 03 Jul 2015

%\cite{Kistler:2008us}
\bibitem{Kistler:2008us}
  M.~D.~Kistler, H.~Y\"uksel, S.~Ando, J.~F.~Beacom and Y.~Suzuki,
  %``Core-Collapse Astrophysics with a Five-Megaton Neutrino Detector,''
  Phys.\ Rev.\ D {\bf 83}, 123008 (2011)
  [arXiv:0810.1959].
  %%CITATION = ARXIV:0810.1959;%%
  %50 citations counted in INSPIRE as of 02 May 2014

%\cite{MICA:2013}
\bibitem{MICA:2013}
 A.~Gross (for the IceCube/PINGU Collaboration),
 Talk at the NNN 2013 Conference, Kashiwa, Japan, (2013).
 %see http://indico.ipmu.jp/indico/conferenceTimeTable.py?confId=17



% chap: Solar
%%%%%%%%%%%%%%%%%%%%%%%%%%%%%%%%%%%%%%%%%%%%%%%%%%%%%%%%%%%%%%%%%%%%%%%%%%%%%%%%%%%%%%%%
%%%%%%%%%%%%%%%%%%%%%%%%%%%%%%%%%%%%%%%%%%%%%%%%%%%%%%%%%%%%%%%%%%%%%%%%%%%%%%%%%%%%%%%%
% chap: Solar

%\bibitem{SK-atmospheric} %changed
%\cite{Fukuda:1998mi}
\bibitem{Fukuda:1998mi}
Y.~Fukuda {\it et al.}  [Super-Kamiokande Collaboration],
% Evidence for oscillation of atmospheric neutrinos,
Phys.\ Rev.\ Lett.\  {\bf 81} (1998) 1562.
%%CITATION = HEP-EX/9807003;%%

%\bibitem{K2K} % changed
%\cite{Ahn:2002up}
\bibitem{Ahn:2002up}
M.~H.~Ahn {\it et al.}  [K2K Collaboration],
% Indications of neutrino oscillation in a 250 km long baseline experiment,
Phys.\ Rev.\ Lett.\  {\bf 90} (2003) 041801.
%%CITATION = HEP-EX/0212007;%%

%\bibitem{KamLAND-2002}
%K. Eguchi et al. [KamLAND Collaboration],
%Phys.\ Rev.\ Letts.\ {\bf 90} (2003) 021802.

%\cite{Eguchi:2002dm}
\bibitem{Eguchi:2002dm}
  K.~Eguchi {\it et al.}  [KamLAND Collaboration],
  %``First results from KamLAND: Evidence for reactor anti-neutrino disappearance,''
  Phys.\ Rev.\ Lett.\  {\bf 90}, 021802 (2003)
  [hep-ex/0212021].  %%CITATION = HEP-EX/0212021;%%

%\bibitem{review-Haxton}
%W.~C.~Haxton, R.~G.~Hamish Roberston, and A.~M.~Serenelli,
%% Solar Neutrinos: Status and Prospects,
%Ann.\ Rev.\ Astron.\ Astrophys.\ {\bf 51} (2013) 21.
%%%CITATION = ARXIV:1208.5723;%%

%\cite{Robertson:2012ib}
\bibitem{Robertson:2012ib}
  W.~C.~Haxton, R.~G.~Hamish Robertson and A.~M.~Serenelli,
  %``Solar Neutrinos: Status and Prospects,''
  Ann.\ Rev.\ Astron.\ Astrophys.\  {\bf 51}, 21 (2013)
  [arXiv:1208.5723 [astro-ph.SR]].  %%CITATION = ARXIV:1208.5723;%%

%\bibitem{reviewAM-long}
%V. Antonelli, L. Miramonti, C. Pena-Garay, A. Serenelli,
%% Solar Neutrinos,
%Adv.\ High Energy Phys.\, {\bf 2013} (2013) 351926.
%%%CITATION = ARXIV:1208.1356;%%

%\cite{Antonelli:2012qu}
\bibitem{Antonelli:2012qu}
  V.~Antonelli, L.~Miramonti, C.~Pena Garay and A.~Serenelli,
  %``Solar Neutrinos,''
  Adv.\ High Energy Phys.\  {\bf 2013}, 351926 (2013)
  [arXiv:1208.1356 [hep-ex]].  %%CITATION = ARXIV:1208.1356;%%

  %\cite{Wolfenstein:1977ue}
\bibitem{Wolfenstein:1977ue}
  L.~Wolfenstein,
  %``Neutrino Oscillations in Matter,''
  Phys.\ Rev.\ D {\bf 17}, 2369 (1978).  %%CITATION = PHRVA,D17,2369;%%

%\cite{Mikheev:1986gs}
\bibitem{Mikheev:1986gs}
  S.~P.~Mikheev and A.~Y.~Smirnov,
  %``Resonance Amplification of Oscillations in Matter and Spectroscopy of Solar Neutrinos,''
  Sov.\ J.\ Nucl.\ Phys.\  {\bf 42}, 913 (1985)
  [Yad.\ Fiz.\  {\bf 42}, 1441 (1985)].  %%CITATION = SJNCA,42,913;%%

%%\cite{Alimonti:2008gc}
%\bibitem{Borexino}
%G.~Alimonti {\it et al.}  [Borexino Collaboration],
%% The Borexino detector at the Laboratori Nazionali del Gran Sasso,
%Nucl.\ Instrum.\ and Meth.\ in Phys.\ Res.\ A {\bf Vol. 600, Issue 3} (2009) 568.
%%CITATION = ARXIV:0806.2400;%%

%\cite{Alimonti:2008gc}
\bibitem{Alimonti:2008gc}
  G.~Alimonti {\it et al.}  [Borexino Collaboration],
  %``The Borexino detector at the Laboratori Nazionali del Gran Sasso,''
  Nucl.\ Instrum.\ Meth.\ A {\bf 600}, 568 (2009)
  [arXiv:0806.2400 [physics.ins-det]].  %%CITATION = ARXIV:0806.2400;%%

%\cite{Fukuda:1998fd}
\bibitem{Fukuda:1998fd}
  Y.~Fukuda {\it et al.}  [Super-Kamiokande Collaboration],
  %``Measurements of the solar neutrino flux from Super-Kamiokande's first 300 days,''
  Phys.\ Rev.\ Lett.\  {\bf 81}, 1158 (1998)  [Phys.\ Rev.\ Lett.\  {\bf 81}, 4279 (1998)]
  [hep-ex/9805021].  %%CITATION = HEP-EX/9805021;%%

%\bibitem{Homestake}
%Cleveland, B.T., et al.,
%% Measurement of the Solar Electron Neutrino Flux with the Homestake Chlorine Detector,
% ApJ {\bf 496} (1998) 505;
%R.~Davis, Jr., D.~S.~Harmer and K.~C.~Hoffman,
%% Search for neutrinos from the sun,
%Phys.\ Rev.\ Lett.\  {\bf 20} (1968) 1205.
%%%CITATION = PRLTA,20,1205;%%

%\cite{Cleveland:1998nv}
\bibitem{Cleveland:1998nv}
  B.~T.~Cleveland, T.~Daily, R.~Davis, Jr., J.~R.~Distel, K.~Lande, C.~K.~Lee, P.~S.~Wildenhain and J.~Ullman,
  %``Measurement of the solar electron neutrino flux with the Homestake chlorine detector,''
  Astrophys.\ J.\  {\bf 496}, 505 (1998).  %%CITATION = ASJOA,496,505;%%

%%\cite{Davis:1968cp}
%\bibitem{Davis:1968cp}
%  R.~Davis, Jr., D.~S.~Harmer and K.~C.~Hoffman,
%  %``Search for neutrinos from the sun,''
%  Phys.\ Rev.\ Lett.\  {\bf 20}, 1205 (1968).  %%CITATION = PRLTA,20,1205;%%

%\bibitem{Kamiokande}
%K.S. Hirata et al. [KamiokaNDE Collaboration],
%% Observation of $^8$B solar neutrinos in the Kamiokande-II detector,
%Phys.Rev. Lett. {\bf 63} (1989) 16.

%\cite{Hirata:1989zj}
\bibitem{Hirata:1989zj}
  K.~S.~Hirata {\it et al.}  [Kamiokande-II Collaboration],
  %``Observation of B-8 Solar Neutrinos in the Kamiokande-II Detector,''
  Phys.\ Rev.\ Lett.\  {\bf 63}, 16 (1989).  %%CITATION = PRLTA,63,16;%%

%\bibitem{Gallex}
%W. Hampel et al. [GALLEX Collaboration],
%% GALLEX solar neutrino observations: Results for GALLEX IV,
%Phys. Lett. B {\bf 447} (1999) 127.

%\cite{Hampel:1998xg}
\bibitem{Hampel:1998xg}
  W.~Hampel {\it et al.}  [GALLEX Collaboration],
  %``GALLEX solar neutrino observations: Results for GALLEX IV,''
  Phys.\ Lett.\ B {\bf 447}, 127 (1999).  %%CITATION = PHLTA,B447,127;%%

%\bibitem{GNO}
%M. Altmann et al. [GNO Collaboration],
% Complete results for five years of GNO solar neutrino observations,
%Phys. Lett. B {\bf 616} (2005) 174.

%\cite{Altmann:2005ix}
\bibitem{Altmann:2005ix}
  M.~Altmann {\it et al.}  [GNO Collaboration],
  %``Complete results for five years of GNO solar neutrino observations,''
  Phys.\ Lett.\ B {\bf 616}, 174 (2005)  [hep-ex/0504037].  %%CITATION = HEP-EX/0504037;%%

%\bibitem{SAGE}
%J.N. Abdurashitov et al. [SAGE collaboration],
%% Measurement of the Solar Neutrino Capture Rate by SAGE and Implications for Neutrino Oscillations in Vacuum,
%Phys. Rev. Lett. {\bf 83}  (1999) 4686.

%\cite{Abdurashitov:1999bv}
\bibitem{Abdurashitov:1999bv}
  J.~N.~Abdurashitov {\it et al.}  [SAGE Collaboration],
  %``Measurement of the solar neutrino capture rate by SAGE and implications for neutrino oscillations in vacuum,''
  Phys.\ Rev.\ Lett.\  {\bf 83}, 4686 (1999)  [astro-ph/9907131].

%\bibitem{SNO-review}
%N.~Jelley, A.~B.~McDonald and R.~G.~H.~Robertson,
%% The Sudbury Neutrino Observatory,
%Ann.\ Rev.\ Nucl.\ Part.\ Sci.\  {\bf 59} (2009) 431.
%%%CITATION = ARNUA,59,431;%%

%\cite{Jelley:2009zz}
\bibitem{Jelley:2009zz}
  N.~Jelley, A.~B.~McDonald and R.~G.~H.~Robertson,
  %``The Sudbury Neutrino Observatory,''
  Ann.\ Rev.\ Nucl.\ Part.\ Sci.\  {\bf 59}, 431 (2009).  %%CITATION = ARNUA,59,431;%%

%\bibitem{SNO-1NC}
% Q.~R.~Ahmad {\it et al.}  [SNO Collaboration],
%% Direct evidence for neutrino flavor transformation from neutral current interactions in the Sudbury %Neutrino Observatory,
%Phys.\ Rev.\ Lett.\  {\bf 89} (2002) 011301.
%%%CITATION = NUCL-EX/0204008;%%

%\bibitem{SNO-1BNC}
 %Q.~R.~Ahmad {\it et al.}  [SNO Collaboration],
%% Measurement of day and night neutrino energy spectra at SNO and constraints on neutrino mixing %parameters,
%Phys.\ Rev.\ Lett.\  {\bf 89} (2002) 011302.
%%%CITATION = NUCL-EX/0204009;%%

%\bibitem{Bahcall04}
%J.N.Bahcall and M.H.Pinsonneault,
%% What do we (not) know theoretically about solar neutrino fluxes?,
%Phys. Rev. Lett. {\bf 92} (2004) 121301.

%\cite{Bahcall:2004fg}
\bibitem{Bahcall:2004fg}
  J.~N.~Bahcall and M.~H.~Pinsonneault,
  %``What do we (not) know theoretically about solar neutrino fluxes?,''
  Phys.\ Rev.\ Lett.\  {\bf 92}, 121301 (2004)
  [astro-ph/0402114].  %%CITATION = ASTRO-PH/0402114;%%

%\bibitem{SNO-ESCC}
% Q.~R.~Ahmad {\it et al.}  [SNO Collaboration],
%% Measurement of the rate of $\nu_e+d \to p+p+e^-$ interactions produced by $^8$B solar neutrinos at the Sudbury Neutrino Observatory,
%Phys.\ Rev.\ Lett.\  {\bf 87} (2001) 071301.
%%%CITATION = NUCL-EX/0106015;%%

%\cite{Ahmad:2001an}
\bibitem{Ahmad:2001an}
  Q.~R.~Ahmad {\it et al.}  [SNO Collaboration],
  %``Measurement of the rate of $\nu_e+d \to p+p+e^-$ interactions produced by $^8B$ solar neutrinos at the Sudbury Neutrino Observatory,''
  Phys.\ Rev.\ Lett.\  {\bf 87}, 071301 (2001)
  [nucl-ex/0106015].  %%CITATION = NUCL-EX/0106015;%%

%\bibitem{Arpesella08}
%C. Arpesella et al. [Borexino Collaboration],
%%Direct Measurement of the $^7$Be Solar Neutrino Flux with 192 Days of Borexino Data,
%Phys. Rev. Lett. {\bf 101} (2008)  091302.

%\cite{Arpesella:2008mt}
\bibitem{Arpesella:2008mt}
  C.~Arpesella {\it et al.}  [Borexino Collaboration],
  %``Direct Measurement of the Be-7 Solar Neutrino Flux with 192 Days of Borexino Data,''
  Phys.\ Rev.\ Lett.\  {\bf 101}, 091302 (2008)
  [arXiv:0805.3843 [astro-ph]].  %%CITATION = ARXIV:0805.3843;%%

%\bibitem{KamLAND-Be-7}
% %\cite{Gando:2014wjd}
% A.~Gando {\it et al.}  [KamLAND Collaboration],
% % $^7$Be Solar Neutrino Measurement with KamLAND,
% arXiv:1405.6190 [hep-ex].
% %%CITATION = ARXIV:1405.6190;%%

 %\cite{Gando:2014wjd}
\bibitem{Gando:2014wjd}
  A.~Gando {\it et al.}  [KamLAND Collaboration],
  %``7Be Solar Neutrino Measurement with KamLAND,''
  arXiv:1405.6190 [hep-ex].  %%CITATION = ARXIV:1405.6190;%%

%%\cite{Collaboration:2011nga}
%\bibitem{Borexino-pep}
%G.~Bellini {\it et al.}  [Borexino Collaboration],
%% First evidence of pep solar neutrinos by direct detection in Borexino,
%Phys.\ Rev.\ Lett.\  {\bf 108} (2012) 051302.
%%%CITATION = ARXIV:1110.3230;%%

%\cite{Collaboration:2011nga}
\bibitem{Collaboration:2011nga}
  G.~Bellini {\it et al.}  [Borexino Collaboration],
  %``First evidence of pep solar neutrinos by direct detection in Borexino,''
  Phys.\ Rev.\ Lett.\  {\bf 108}, 051302 (2012)
  [arXiv:1110.3230 [hep-ex]].  %%CITATION = ARXIV:1110.3230;%%

%\bibitem{bxpp}
%G.~Bellini {\it et al.}  [Borexino Collaboration],
%% Neutrinos from the primary proton-proton fusion precess in the Sun,
%Nature {\bf 512}, 383 (2014).

%\cite{Bellini:2014uqa}
\bibitem{Bellini:2014uqa}
  G.~Bellini {\it et al.}  [BOREXINO Collaboration],
  %``Neutrinos from the primary proton¨Cproton fusion process in the Sun,''
  Nature {\bf 512}, 383 (2014).  %%CITATION = NATUA,512,383;%%

%%\cite{Bellini:2013lnn}
%\bibitem{Borexino-fase-I}
%%\cite{Bellini:2013lnn}
%G.~Bellini {\it et al.}  [Borexino Collaboration],
%% Final results of Borexino Phase-I on low energy solar neutrino spectroscopy,
%Phys.\ Rev.\ D {\bf 89} (2014) 112007.
%%%CITATION = ARXIV:1308.0443;%%

%\cite{Bellini:2013lnn}
\bibitem{Bellini:2013lnn}
  G.~Bellini {\it et al.}  [Borexino Collaboration],
  %``Final results of Borexino Phase-I on low energy solar neutrino spectroscopy,''
  Phys.\ Rev.\ D {\bf 89}, 112007 (2014)
  [arXiv:1308.0443 [hep-ex]].  %%CITATION = ARXIV:1308.0443;%%

%\bibitem{Fogli-Lisi-global-2013}
%%\cite{Capozzi:2013csa}
%F.~Capozzi, G.~L.~Fogli, E.~Lisi, A.~Marrone, D.~Montanino and A.~Palazzo,
%% Status of three-neutrino oscillation parameters, circa 2013,
%Phys.\ Rev.\ D {\bf 89} (2014) 093018.
%%%CITATION = ARXIV:1312.2878;%%

%%\cite{Tortola:2012te}
%\bibitem{Tortola:2012te}
%  D.~V.~Forero, M.~Tortola and J.~W.~F.~Valle,
%  %``Global status of neutrino oscillation parameters after Neutrino-2012,''
%  Phys.\ Rev.\ D {\bf 86}, 073012 (2012)
%  [arXiv:1205.4018 [hep-ph]].
%  %%CITATION = ARXIV:1205.4018;%%
%
%  %\cite{Capozzi:2013csa}
%\bibitem{Capozzi:2013csa}
%  F.~Capozzi, G.~L.~Fogli, E.~Lisi, A.~Marrone, D.~Montanino and A.~Palazzo,
%  %``Status of three-neutrino oscillation parameters, circa 2013,''
%  Phys.\ Rev.\ D {\bf 89}, 093018 (2014)
%  [arXiv:1312.2878 [hep-ph]].
%  %%CITATION = ARXIV:1312.2878;%%
%
%  %\cite{Gonzalez-Garcia:2014bfa}
%\bibitem{Gonzalez-Garcia:2014bfa}
%  M.~C.~Gonzalez-Garcia, M.~Maltoni and T.~Schwetz,
%  %``Updated fit to three neutrino mixing: status of leptonic CP violation,''
%  JHEP {\bf 1411}, 052 (2014)
%  [arXiv:1409.5439 [hep-ph]].
%  %%CITATION = ARXIV:1409.5439;%%
%
%  %\cite{GonzalezGarcia:2012sz}
%\bibitem{GonzalezGarcia:2012sz}
%  M.~C.~Gonzalez-Garcia, M.~Maltoni, J.~Salvado and T.~Schwetz,
%  %``Global fit to three neutrino mixing: critical look at present precision,''
%  JHEP {\bf 1212}, 123 (2012)
%  [arXiv:1209.3023 [hep-ph]].
%  %%CITATION = ARXIV:1209.3023;%%

%\bibitem{Valle-global-2014}
%%\cite{Forero:2014bxa}
%D.~V.~Forero, M.~Tortola and J.~W.~F.~Valle,
%% Neutrino oscillations refitted,
%arXiv:1405.7540 [hep-ph].
%%%CITATION = ARXIV:1405.7540;%%

%%\cite{Forero:2014bxa}
%\bibitem{Forero:2014bxa}
%  D.~V.~Forero, M.~Tortola and J.~W.~F.~Valle,
%  %``Neutrino oscillations refitted,''
%  Phys.\ Rev.\ D {\bf 90}, no. 9, 093006 (2014)
%  [arXiv:1405.7540 [hep-ph]].  %%CITATION = ARXIV:1405.7540;%%

%\bibitem{Fogli-Lisi-global-2012}
%%\cite{Fogli:2012ua}
%G.~L.~Fogli, E.~Lisi, A.~Marrone, D.~Montanino, A.~Palazzo and A.~M.~Rotunno,
%% Global analysis of neutrino masses, mixings and phases: entering the era of leptonic CP violation searches,
%Phys.\ Rev.\ D {\bf 86} (2012) 013012.
%%%CITATION = ARXIV:1205.5254;%%

%\bibitem{Valle-global-2012}
%%\cite{Tortola:2012te}
%D.~V.~Forero, M.~Tortola and J.~W.~F.~Valle,
%% Global status of neutrino oscillation parameters after Neutrino-2012,
%Phys.\ Rev.\ D {\bf 86} (2012) 073012.
%%%CITATION = ARXIV:1205.4018;%%
%
%\bibitem{Goncia-Maltoni-global-2012}
%%\cite{GonzalezGarcia:2012sz}
%M.~C.~Gonzalez-Garcia, M.~Maltoni, J.~Salvado and T.~Schwetz,
%% Global fit to three neutrino mixing: critical look at present precision,
%JHEP {\bf 1212} (2012) 123.
%%%CITATION = ARXIV:1209.3023;%%
%An updated version of these results can be found at the website www.nu-fit.org, see the link therein:
%``v1.3: Three-neutrino results after the Neutrino 2014 conference''.

%\bibitem{LBNE}
%C. Adams {\it et al.} [LBNE Collaboration],  arXiv:1307.7335v3, April 2014.

%%\cite{Adams:2013qkq}
%\bibitem{Adams:2013qkq}
%  C.~Adams {\it et al.}  [LBNE Collaboration],
%  %``The Long-Baseline Neutrino Experiment: Exploring Fundamental Symmetries of the Universe,''
%  arXiv:1307.7335 [hep-ex].
%  %%CITATION = ARXIV:1307.7335;%%

%\bibitem{HyperKamiokande}
%K. Abe {\it et al.} [Hyper-Kamiokande Working Group Collaboration], arXiv:1412.4673 [physics.ins-det].

%%\cite{Abe:2011ts}
%\bibitem{Abe:2011ts}
%  K.~Abe, T.~Abe, H.~Aihara, Y.~Fukuda, Y.~Hayato, K.~Huang, A.~K.~Ichikawa and M.~Ikeda {\it et al.},
%  %``Letter of Intent: The Hyper-Kamiokande Experiment --- Detector Design and Physics Potential ---,''
%  arXiv:1109.3262 [hep-ex].
%  %%CITATION = ARXIV:1109.3262;%%
%
%%\cite{Abe:2014oxa}
%\bibitem{Abe:2014oxa}
%  K.~Abe {\it et al.}  [Hyper-Kamiokande Working Group Collaboration],
%  %``A Long Baseline Neutrino Oscillation Experiment Using J-PARC Neutrino Beam and Hyper-Kamiokande,''
%  arXiv:1412.4673 [physics.ins-det].
%  %%CITATION = ARXIV:1412.4673;%%

%\bibitem{NSI}
%T. Ohlsson, Rept. Prog. Phys. {\bf 76} (2013) 044201.

%\cite{Ohlsson:2012kf}
\bibitem{Ohlsson:2012kf}
  T.~Ohlsson,
  %``Status of non-standard neutrino interactions,''
  Rept.\ Prog.\ Phys.\  {\bf 76}, 044201 (2013)
  [arXiv:1209.2710 [hep-ph]].  %%CITATION = ARXIV:1209.2710;%%

%\bibitem{Villante-Serenelli-2014}
%%\cite{Villante:2013mba}
%F.~L.~Villante, A.~M.~Serenelli, F.~Delahaye and M.~H.~Pinsonneault,
%% The chemical composition of the Sun from helioseismic and solar neutrino data,
%Astrophys.\ J.\  {\bf 787} (2014) 13.
%%%CITATION = ARXIV:1312.3885;%%

%\cite{Villante:2013mba}
\bibitem{Villante:2013mba}
  F.~L.~Villante, A.~M.~Serenelli, F.~Delahaye and M.~H.~Pinsonneault,
  %``The chemical composition of the Sun from helioseismic and solar neutrino data,''
  Astrophys.\ J.\  {\bf 787}, 13 (2014)
  [arXiv:1312.3885 [astro-ph.SR]].  %%CITATION = ARXIV:1312.3885;%%

%\bibitem{Serenelli-2014}
%%\cite{Bergemann:2014vaa}
%M.~Bergemann and A.~M.~Serenelli,
%% Solar abundance problem,
%arXiv:1403.3097 [astro-ph.SR].
%%%CITATION = ARXIV:1403.3097;%%

%\cite{Bergemann:2014vaa}
\bibitem{Bergemann:2014vaa}
  M.~Bergemann and A.~Serenelli,
  %``Solar abundance problem,''
  arXiv:1403.3097 [astro-ph.SR].  %%CITATION = ARXIV:1403.3097;%%

%\bibitem{Grevesse98}
%N. Grevesse and A. J. Sauval,
%% Standard Solar Composition,
%Space Science Reviews {\bf 85} (1998) 161.

%\cite{Grevesse:1998bj}
\bibitem{Grevesse:1998bj}
  N.~Grevesse and A.~J.~Sauval,
  %``Standard Solar Composition,''
  Space Sci.\ Rev.\  {\bf 85}, 161 (1998).  %%CITATION = SPSRA,85,161;%%

%\bibitem{Asplund05}
%M. Asplund, N. Grevesse, and J. Sauval,
%% The solar chemical composition,
%Nucl. Phys. A {\bf 77} (2006) 1-4,
%% (Cosmic Abundances as Records of Stellar Evolution and Nucleosynthesis) ASP Conf. Ser. {\bf 336} (2005) 25.

%\cite{Asplund:2004eu}
\bibitem{Asplund:2004eu}
  M.~Asplund, N.~Grevesse and J.~Sauval,
  %``The Solar chemical composition,''
  Nucl.\ Phys.\ A {\bf 777}, 1 (2006)
  [ASP Conf.\ Ser.\  {\bf 336}, 25 (2005)]
  [astro-ph/0410214].  %%CITATION = ASTRO-PH/0410214;%%

%\bibitem{Asplund09}
%Asplund, M., Grevesse, N., Sauval, J., and Scott, P. 2009,
%% The chemical composition of the Sun,
%ARAA {\bf 47} (2009) 481.

%\cite{Asplund:2009fu}
\bibitem{Asplund:2009fu}
  M.~Asplund, N.~Grevesse, A.~J.~Sauval and P.~Scott,
  %``The chemical composition of the Sun,''
  Ann.\ Rev.\ Astron.\ Astrophys.\  {\bf 47}, 481 (2009)
  [arXiv:0909.0948 [astro-ph.SR]].  %%CITATION = ARXIV:0909.0948;%%

%%\cite{Serenelli:2009yc}
%\bibitem{Serenelli09}
%A.~Serenelli, S.~Basu, J.~W.~Ferguson and M.~Asplund,
%% New Solar Composition: The Problem With Solar Models Revisited,
%Astrophys.\ J.\  {\bf 705} (2009) L123.
%%%CITATION = ARXIV:0909.2668;%%

%\cite{Serenelli:2009yc}
\bibitem{Serenelli:2009yc}
  A.~Serenelli, S.~Basu, J.~W.~Ferguson and M.~Asplund,
  %``New Solar Composition: The Problem With Solar Models Revisited,''
  Astrophys.\ J.\  {\bf 705}, L123 (2009)
  [arXiv:0909.2668 [astro-ph.SR]].  %%CITATION = ARXIV:0909.2668;%%

  %\bibitem{SKIV}
%A. Renshaw [Super-Kamiokande Collaboration],
%% Solar Neutrino Results from Super-Kamiokande,
%Physics Procedia {\bf 61} (2015) 345 (arXiv:1403.4575 [hep-ex]).

%\cite{Renshaw:2014awa}
\bibitem{Renshaw:2014awa}
  A.~Renshaw [Super-Kamiokande Collaboration],
  %``Solar Neutrino Results from Super-Kamiokande,''
  Phys.\ Procedia {\bf 61}, 345 (2015)
  [arXiv:1403.4575 [hep-ex]].  %%CITATION = ARXIV:1403.4575;%%


\bibitem{Serenelli-talk}
A.~M.~Serenelli,
% Status of Standard Models,
talk given at ``A special Borexino event - Borexino Mini-Workshop", Gran Sasso, (2014).
%Sept.~5 2014,
%See the talk at https://agenda.infn.it/conferenceDisplay.py?confId=8548

%\bibitem{Zuber}
%%\cite{Zuber:2013uya}
%K.~Zuber,
%% Metallicities in stars - what solar neutrinos can do,
%PoS NICXII {\bf } (2012) 012.
%%%CITATION = POSCI,NICXII,012;%%

%\cite{Zuber:2013uya}
\bibitem{Zuber:2013uya}
  K.~Zuber,
  %``Metallicities in stars - what solar neutrinos can do,''
  PoS NICXII {\bf }, 012 (2012).

%\cite{Serenelli:2009ww}
\bibitem{Serenelli:2009ww}
  A.~M.~Serenelli,
  %``New Results on Standard Solar Models,''
  Astrophys.\ Space Sci.\  {\bf 328}, 13 (2010)
  [arXiv:0910.3690 [astro-ph.SR]].  %%CITATION = ARXIV:0910.3690;%%

%%\cite{Voloshin:2010vm}
%\bibitem{Voloshin}
%M.~B.~Voloshin,
%% Neutrino scattering on atomic electrons in searches for neutrino magnetic moment,
%  Phys.\ Rev.\ Lett.\  {\bf 105} (2010) 201801
%   [Erratum-ibid.\  {\bf 106} (2011) 059901].
%%%CITATION = ARXIV:1008.2171;%%

%\cite{Voloshin:2010vm}
\bibitem{Voloshin:2010vm}
  M.~B.~Voloshin,
  %``Neutrino scattering on atomic electrons in searches for neutrino magnetic moment,''
  Phys.\ Rev.\ Lett.\  {\bf 105}, 201801 (2010)
  [Phys.\ Rev.\ Lett.\  {\bf 106}, 059901 (2011)]
  [arXiv:1008.2171 [hep-ph]].  %%CITATION = ARXIV:1008.2171;%%

%%\cite{Giunti:2014ixa}
%\bibitem{Giunti}
%C.~Giunti and A.~Studenikin,
%% Neutrino electromagnetic interactions: a window to new physics,
%arXiv:1403.6344 [hep-ph].
%%%CITATION = ARXIV:1403.6344;%%

%\cite{Giunti:2014ixa}
\bibitem{Giunti:2014ixa}
  C.~Giunti and A.~Studenikin,
  %``Neutrino electromagnetic interactions: a window to new physics,''
  arXiv:1403.6344 [hep-ph].  %%CITATION = ARXIV:1403.6344;%%

%\cite{Broggini:2012df}
\bibitem{Broggini:2012df}
  C.~Broggini, C.~Giunti and A.~Studenikin,
  %``Electromagnetic Properties of Neutrinos,''
  Adv.\ High Energy Phys.\  {\bf 2012}, 459526 (2012)
  [arXiv:1207.3980 [hep-ph]].  %%CITATION = ARXIV:1207.3980;%%

%%\cite{Beda:2013mta}
%\bibitem{GEMMA}
%A.~G.~Beda, V.~B.~Brudanin, V.~G.~Egorov, D.~V.~Medvedev, V.~S.~Pogosov, E.~A.~Shevchik, M.~V.~Shirchenko and A.~S.~Starostin {\it et al.},
%[Gemma experiment],
%% The results of neutrino magnetic moment search,
%Phys.\ Part.\ Nucl.\ Lett.\  {\bf 10} (2013) 139.
%%%CITATION = 00438,10,139;%%

%\cite{Beda:2013mta}
\bibitem{Beda:2013mta}
  A.~G.~Beda, V.~B.~Brudanin, V.~G.~Egorov, D.~V.~Medvedev, V.~S.~Pogosov, E.~A.~Shevchik, M.~V.~Shirchenko and A.~S.~Starostin {\it et al.},
  %``Gemma experiment: The results of neutrino magnetic moment search,''
  Phys.\ Part.\ Nucl.\ Lett.\  {\bf 10}, 139 (2013).  %%CITATION = 00438,10,139;%%

%%\cite{Wong:2006nx}
%\bibitem{TEXONO}
%H.~T.~Wong {\it et al.}  [TEXONO Collaboration],
%% A Search of Neutrino Magnetic Moments with a High-Purity Germanium Detector at the Kuo-Sheng Nuclear Power Station,
%Phys.\ Rev.\ D {\bf 75} (2007) 012001.
%%%CITATION = HEP-EX/0605006;%%

%\cite{Wong:2006nx}
\bibitem{Wong:2006nx}
  H.~T.~Wong {\it et al.}  [TEXONO Collaboration],
  %``A Search of Neutrino Magnetic Moments with a High-Purity Germanium Detector at the Kuo-Sheng Nuclear Power Station,''
  Phys.\ Rev.\ D {\bf 75}, 012001 (2007)
  [hep-ex/0605006].  %%CITATION = HEP-EX/0605006;%%

%%\cite{Daraktchieva:2005kn}
%\bibitem{MUNU}
%  Z.~Daraktchieva {\it et al.}  [MUNU Collaboration],
%% Final results on the neutrino magnetic moment from the MUNU experiment,
%  Phys.\ Lett.\ B {\bf 615} (2005) 153.
%  %%CITATION = HEP-EX/0502037;%%

  %\cite{Daraktchieva:2005kn}
\bibitem{Daraktchieva:2005kn}
  Z.~Daraktchieva {\it et al.}  [MUNU Collaboration],
  %``Final results on the neutrino magnetic moment from the MUNU experiment,''
  Phys.\ Lett.\ B {\bf 615}, 153 (2005)  [hep-ex/0502037].  %%CITATION = HEP-EX/0502037;%%

%%\cite{Raffelt:1999gv}
%\bibitem{Raffelt}
%G.~G.~Raffelt,
%% Limits on neutrino electromagnetic properties: An update,
%Phys.\ Rept.\  {\bf 320} (1999) 319.
%  %%CITATION = PRPLC,320,319;%%

%  %\cite{Raffelt:1999gv}
%\bibitem{Raffelt:1999gv}
%  G.~G.~Raffelt,
%  %``Limits on neutrino electromagnetic properties: An update,''
%  Phys.\ Rept.\  {\bf 320}, 319 (1999).  %%CITATION = PRPLC,320,319;%%

%%\cite{Viaux:2013lha}
%\bibitem{Viaux}
%N.~Viaux, M.~Catelan, P.~B.~Stetson, G.~Raffelt, J.~Redondo, A.~A.~R.~Valcarce and A.~Weiss,
%% Neutrino and axion bounds from the globular cluster M5 (NGC 5904),
%Phys.\ Rev.\ Lett.\  {\bf 111} (2013) 231301.
%%%CITATION = ARXIV:1311.1669;%%

%\cite{Viaux:2013lha}
\bibitem{Viaux:2013lha}
  N.~Viaux, M.~Catelan, P.~B.~Stetson, G.~Raffelt, J.~Redondo, A.~A.~R.~Valcarce and A.~Weiss,
  %``Neutrino and axion bounds from the globular cluster M5 (NGC 5904),''
  Phys.\ Rev.\ Lett.\  {\bf 111}, 231301 (2013)
  [arXiv:1311.1669 [astro-ph.SR]].  %%CITATION = ARXIV:1311.1669;%%

 %\cite{Beacom:1999wx}
\bibitem{Beacom:1999wx}
  J.~F.~Beacom and P.~Vogel,
  %``Neutrino magnetic moments, flavor mixing, and the Super-Kamiokande solar data,''
  Phys.\ Rev.\ Lett.\  {\bf 83}, 5222 (1999)
  [hep-ph/9907383].  %%CITATION = HEP-PH/9907383;%%

%\cite{Giunti:2015gga}
\bibitem{Giunti:2015gga}
C.~Giunti, K.~A.~Kouzakov, Y.~F.~Li, A.~V.~Lokhov, A.~I.~Studenikin and S.~Zhou,
  %``Electromagnetic neutrinos in terrestrial experiments and astrophysics,''
  arXiv:1506.05387 [hep-ph].

%\bibitem{SNO+}
%%\cite{Kraus:2010zzb}
%C.~Kraus {\it et al.}  [SNO+ Collaboration],
%% The rich neutrino programme of the SNO+ experiment,
%Prog.\ Part.\ Nucl.\ Phys.\  {\bf 64} (2010) 273;
%%%CITATION = PPNPD,64,273;%%
%%\cite{MOTTRAM:2014nna}
%M.~MOTTRAM,
%% SNO+ Experiment Status,
%PoS EPS {\bf HEP2013} (2014) 524.
%%%CITATION = POSCI,EPS-HEP2013,524;%%

%\cite{Kraus:2010zzb}
\bibitem{Kraus:2010zzb}
  C.~Kraus {\it et al.}  [SNO+ Collaboration],
  %``The rich neutrino programme of the SNO+ experiment,''
  Prog.\ Part.\ Nucl.\ Phys.\  {\bf 64}, 273 (2010).  %%CITATION = PPNPD,64,273;%%

%\cite{MOTTRAM:2014nna}
\bibitem{MOTTRAM:2014nna}
  M.~Mottram [SNO+ Collaboration],
  %``SNO+ Experiment Status,''
  PoS EPS {\bf -HEP2013}, 524 (2013).

%\bibitem{SK-IV-bis}
%%\cite{Renshaw:2013dzu}
%A.~Renshaw {\it et al.}  [Super-Kamiokande Collaboration],
%% First Indication of Terrestrial Matter Effects on Solar Neutrino Oscillation,
%Phys.\ Rev.\ Lett.\  {\bf 112} (2014) 091805.
%%%CITATION = ARXIV:1312.5176;%%

%\cite{Renshaw:2013dzu}
\bibitem{Renshaw:2013dzu}
  A.~Renshaw {\it et al.}  [Super-Kamiokande Collaboration],
  %``First Indication of Terrestrial Matter Effects on Solar Neutrino Oscillation,''
  Phys.\ Rev.\ Lett.\  {\bf 112}, 091805 (2014)
  [arXiv:1312.5176 [hep-ex]].  %%CITATION = ARXIV:1312.5176;%%

%\bibitem{bp05} J.~Bahcall, A.~Serenelli and S.~Basu, Astrophys.\ J.\ {\bf 621} 85 (2005).

%\cite{Bahcall:2004pz}
\bibitem{Bahcall:2004pz}
  J.~N.~Bahcall, A.~M.~Serenelli and S.~Basu,
  %``New solar opacities, abundances, helioseismology, and neutrino fluxes,''
  Astrophys.\ J.\  {\bf 621}, L85 (2005)
  [astro-ph/0412440].  %%CITATION = ASTRO-PH/0412440;%%

%\bibitem{KLcosmo}
%S.Abe et al.,
%% Production of radioactive isotopes through cosmic muon spallation in KamLAND
%Phys.Rev.C {\bf 81} (2010).

%%\cite{Abe:2009aa}
%\bibitem{Abe:2009aa}
%  S.~Abe {\it et al.}  [KamLAND Collaboration],
%  %``Production of Radioactive Isotopes through Cosmic Muon Spallation in KamLAND,''
%  Phys.\ Rev.\ C {\bf 81}, 025807 (2010)
%  [arXiv:0907.0066 [hep-ex]].  %%CITATION = ARXIV:0907.0066;%%

%\bibitem{Bxcosmo}
%G.Bellini et al.,
%% Cosmogenic Backgrounds in Borexino at 3800 m water-equivalent depth,
%JCAP {\bf 049} (2013).

%\cite{Bellini:2013pxa}
\bibitem{Bellini:2013pxa}
  G.~Bellini {\it et al.}  [Borexino Collaboration],
  %``Cosmogenic Backgrounds in Borexino at 3800 m water-equivalent depth,''
  JCAP {\bf 1308}, 049 (2013)
  [arXiv:1304.7381 [physics.ins-det]].  %%CITATION = ARXIV:1304.7381;%%

%\bibitem{Moellenberg13} %changed
%%R. Moellenberg, Monte Carlo Study of the Fast Neu-
%%tron Background in LENA, Diploma thesis, Technis-
%% che Universita ̈t Mu ̈nchen, 2009.
%R. M\"ollenberg, PhD thesis, TU M\"unchen (2013).

%%\cite{Mollenberg:2009lka}
%\bibitem{Mollenberg:2009lka}
%  R.~Mollenberg,
%  %``Monte Carlo Study of the Fast Neutron Background in LENA,''
%  %%CITATION = INSPIRE-1320499;%%

% %\cite{Mollenberg:2013qka}
\bibitem{Mollenberg:2013qka}
  R.~Mollenberg,
  ``Monte Carlo Study of Solar $^8$B Neutrinos and the Diffuse Supernova Neutrino Background in LENA,''
  %%CITATION = INSPIRE-1319828;%%
  PhD thesis, TU M\"unchen (2013). see
  http://mediatum.ub.tum.de/node?id=1175550

%\bibitem{sli1} S.~Li, and J.~F.~Beacom, Phys. Rev. C\ {\bf 89} (2014) 045801.
%\bibitem{sli2} S.~Li, and J.~F.~Beacom, Phys. Rev. D\ {\bf 91} (2015) 105005.

%\cite{Li:2014sea}
\bibitem{Li:2014sea}
  S.~W.~Li and J.~F.~Beacom,
  %``First calculation of cosmic-ray muon spallation backgrounds for MeV astrophysical neutrino signals in Super-Kamiokande,''
  Phys.\ Rev.\ C {\bf 89}, 045801 (2014)
  [arXiv:1402.4687 [hep-ph]].  %%CITATION = ARXIV:1402.4687;%%

%\cite{Li:2015kpa}
\bibitem{Li:2015kpa}
  S.~W.~Li and J.~F.~Beacom,
  %``Spallation Backgrounds in Super-Kamiokande Are Made in Muon-Induced Showers,''
  Phys.\ Rev.\ D {\bf 91}, 105005 (2015)
  [arXiv:1503.04823 [hep-ph]].  %%CITATION = ARXIV:1503.04823;%%


% chap: atmospheric
%%%%%%%%%%%%%%%%%%%%%%%%%%%%%%%%%%%%%%%%%%%%%%%%%%%%%%%%%%%%%%%%%%%%%%%%%%%%%%%%%%%%%%%%
%%%%%%%%%%%%%%%%%%%%%%%%%%%%%%%%%%%%%%%%%%%%%%%%%%%%%%%%%%%%%%%%%%%%%%%%%%%%%%%%%%%%%%%%
% chap: atmospheric


%\bibitem{Capozzi:2013csa_ATM}
%  F.~Capozzi, G.~L.~Fogli, E.~Lisi, A.~Marrone, D.~Montanino and A.~Palazzo,
%  %``Status of three-neutrino oscillation parameters, circa 2013,''
%  Phys.\ Rev.\ D {\bf 89}, 093018 (2014)
%  [arXiv:1312.2878 [hep-ph]].
%  %%CITATION = ARXIV:1312.2878;%%
%  %74 citations counted in INSPIRE as of 04 Jul 2014

%  %\cite{Capozzi:2013csa}
%\bibitem{Capozzi:2013csa}
%  F.~Capozzi, G.~L.~Fogli, E.~Lisi, A.~Marrone, D.~Montanino and A.~Palazzo,
%  %``Status of three-neutrino oscillation parameters, circa 2013,''  Phys.\ Rev.\ D {\bf 89}, 093018 (2014)
%  [arXiv:1312.2878 [hep-ph]].  %%CITATION = ARXIV:1312.2878;%%

%\cite{deGouvea:2013onf}
\bibitem{deGouvea:2013onf}
  A.~de Gouvea {\it et al.}  [Intensity Frontier Neutrino Working Group Collaboration],
  %``Working Group Report: Neutrinos,''
  arXiv:1310.4340 [hep-ex].  %%CITATION = ARXIV:1310.4340;%%

%%\cite{Fukuda:1998mi}
%\bibitem{Fukuda:1998mi}
%  Y.~Fukuda {\it et al.}  [Super-Kamiokande Collaboration],
%  %``Evidence for oscillation of atmospheric neutrinos,''
%  Phys.\ Rev.\ Lett.\  {\bf 81}, 1562 (1998)
%  [hep-ex/9807003].
%  %%CITATION = HEP-EX/9807003;%%
%  %4316 citations counted in INSPIRE as of 27 Dec 2014

%\bibitem{Wolfenstein:1977ue}
%  L.~Wolfenstein,
%  %``Neutrino Oscillations in Matter,''
%  Phys.\ Rev.\ D {\bf 17}, 2369 (1978);
%  %%CITATION = PHRVA,D17,2369;%%
%  %3735 citations counted in INSPIRE as of 12 Jul 2014
%  S.~P.~Mikheev and A.~Y.~.Smirnov,
%  %``Resonance Amplification of Oscillations in Matter and Spectroscopy of Solar
%  %Neutrinos,''
%  Sov.\ J.\ Nucl.\ Phys.\  {\bf 42}, 913 (1985)
%  [Yad.\ Fiz.\  {\bf 42}, 1441 (1985)].
%  %%CITATION = SJNCA,42,913;%%
%  %2749 citations counted in INSPIRE as of 12 Jul 2014

%  %\cite{Wolfenstein:1977ue}
%\bibitem{Wolfenstein:1977ue}
%  L.~Wolfenstein,
%  %``Neutrino Oscillations in Matter,''
%  Phys.\ Rev.\ D {\bf 17}, 2369 (1978).  %%CITATION = PHRVA,D17,2369;%%
%
%%\cite{Mikheev:1986gs}
%\bibitem{Mikheev:1986gs}
%  S.~P.~Mikheev and A.~Y.~Smirnov,
%  %``Resonance Amplification of Oscillations in Matter and Spectroscopy of Solar Neutrinos,''
%  Sov.\ J.\ Nucl.\ Phys.\  {\bf 42}, 913 (1985)
%  [Yad.\ Fiz.\  {\bf 42}, 1441 (1985)].  %%CITATION = SJNCA,42,913;%%

%%\cite{Wendell:2014dka}
%\bibitem{Wendell:2014dka}
%  R.~Wendell [Super-Kamiokande Collaboration],
%  %``Atmospheric Results from Super-Kamiokande,''
%  arXiv:1412.5234 [hep-ex].  %%CITATION = ARXIV:1412.5234;%%

%%\cite{Aartsen:2014oha}
%\bibitem{Aartsen:2014oha}
%  M.~G.~Aartsen {\it et al.}  [IceCube PINGU Collaboration],
%  %``Letter of Intent: The Precision IceCube Next Generation Upgrade (PINGU),''
%  arXiv:1401.2046 [physics.ins-det].
%  %%CITATION = ARXIV:1401.2046;%%

%\cite{Katz:2014tta}
\bibitem{Katz:2014tta}
  U.~F.~Katz [KM3NeT Collaboration],
  %``The ORCA Option for KM3NeT,''
  %Submitted to: PoS
  [arXiv:1402.1022 [astro-ph.IM]].
  %%CITATION = ARXIV:1402.1022;%%

%%\cite{Abe:2011ts}
%\bibitem{Abe:2011ts}
%  K.~Abe, T.~Abe, H.~Aihara, Y.~Fukuda, Y.~Hayato, K.~Huang, A.~K.~Ichikawa and M.~Ikeda {\it et al.},
%  %``Letter of Intent: The Hyper-Kamiokande Experiment --- Detector Design and Physics Potential ---,''
%  arXiv:1109.3262 [hep-ex].
%  %%CITATION = ARXIV:1109.3262;%%
%
%%\cite{Abe:2014oxa}
%\bibitem{Abe:2014oxa}
%  K.~Abe {\it et al.}  [Hyper-Kamiokande Working Group Collaboration],
%  %``A Long Baseline Neutrino Oscillation Experiment Using J-PARC Neutrino Beam and Hyper-Kamiokande,''
%  arXiv:1412.4673 [physics.ins-det].
%  %%CITATION = ARXIV:1412.4673;%%


%\bibitem{Thakore:2013xqa} % changed
%  T.~Thakore, A.~Ghosh, S.~Choubey and A.~Dighe,
%  %``The Reach of INO for Atmospheric Neutrino Oscillation Parameters,''
%  JHEP {\bf 1305}, 058 (2013)
%  [arXiv:1303.2534 [hep-ph]].
%  %%CITATION = ARXIV:1303.2534;%%
%  %12 citations counted in INSPIRE as of 27 Dec 2014

%  %\cite{Ahmed:2015jtv}
%\bibitem{Ahmed:2015jtv}
%  S.~Ahmed {\it et al.}  [ICAL Collaboration],
%  %``Physics Potential of the ICAL detector at the India-based Neutrino Observatory (INO),''
%  arXiv:1505.07380 [physics.ins-det].
%  %%CITATION = ARXIV:1505.07380;%%

%\cite{Learned:2009rv}
\bibitem{Learned:2009rv}
  J.~G.~Learned,
  %``High Energy Neutrino Physics with Liquid Scintillation Detectors,''
  arXiv:0902.4009 [hep-ex].  %%CITATION = ARXIV:0902.4009;%%

%\bibitem{Wurm:2011zn} %changed
%  M.~Wurm {\it et al.}  [LENA Collaboration],
%  %``The next-generation liquid-scintillator neutrino observatory LENA,''
%  Astropart.\ Phys.\  {\bf 35}, 685 (2012)
%  [arXiv:1104.5620 [astro-ph.IM]].
%  %%CITATION = ARXIV:1104.5620;%%
%  %100 citations counted in INSPIRE as of 27 Dec 2014

%\cite{Honda:2011nf}
\bibitem{Honda:2011nf}
  M.~Honda, T.~Kajita, K.~Kasahara and S.~Midorikawa,
  %``Improvement of low energy atmospheric neutrino flux calculation using the JAM nuclear interaction model,''
  Phys.\ Rev.\ D {\bf 83}, 123001 (2011)  [arXiv:1102.2688 [astro-ph.HE]].  %%CITATION = ARXIV:1102.2688;%%
%  http://www.icrr.u-tokyo.ac.jp/~mhonda/nflx2014/.
  %%CITATION = ARXIV:1102.2688;%%

%\cite{Lisi:1997yc}
\bibitem{Lisi:1997yc}
  E.~Lisi and D.~Montanino,
  %``Earth regeneration effect in solar neutrino oscillations: An Analytic approach,''
  Phys.\ Rev.\ D {\bf 56}, 1792 (1997)
  [hep-ph/9702343].  %%CITATION = HEP-PH/9702343;%%

%\bibitem{PREM} %changed
%  A. M. Dziewonski and D. L. Anderson, Phys. Earth Planet. Inter. {\bf 25}, 297
%(1981).

%\cite{Dziewonski:1981xy}
\bibitem{Dziewonski:1981xy}
  A.~M.~Dziewonski and D.~L.~Anderson,
  %``Preliminary reference earth model,''
  Phys.\ Earth Planet.\ Interiors {\bf 25}, 297 (1981).  %%CITATION = PEPIA,25,297;%%

%\cite{Wallraff:2014qka}
\bibitem{Wallraff:2014qka}
  M.~Wallraff and C.~Wiebusch,
  %``Calculation of oscillation probabilities of atmospheric neutrinos using nuCraft,''
  arXiv:1409.1387 [astro-ph.IM].
  %%CITATION = ARXIV:1409.1387;%%

%\cite{Cervera:2000kp}
\bibitem{Cervera:2000kp}
  A.~Cervera, A.~Donini, M.~B.~Gavela, J.~J.~Gomez Cadenas, P.~Hernandez, O.~Mena and S.~Rigolin,
  %``Golden measurements at a neutrino factory,''
  Nucl.\ Phys.\ B {\bf 579}, 17 (2000)
  [Nucl.\ Phys.\ B {\bf 593}, 731 (2001)]
  [hep-ph/0002108].  %%CITATION = HEP-PH/0002108;%%

%\cite{Freund:2001pn}
\bibitem{Freund:2001pn}
  M.~Freund,
  %``Analytic approximations for three neutrino oscillation parameters and
  %probabilities in matter,''
  Phys.\ Rev.\ D {\bf 64}, 053003 (2001)
  [hep-ph/0103300].
  %%CITATION = HEP-PH/0103300;%%

%\cite{Akhmedov:2004ny}
\bibitem{Akhmedov:2004ny}
  E.~K.~Akhmedov, R.~Johansson, M.~Lindner, T.~Ohlsson and T.~Schwetz,
  %``Series expansions for three flavor neutrino oscillation probabilities in matter,''
  JHEP {\bf 0404}, 078 (2004)  [hep-ph/0402175].  %%CITATION = HEP-PH/0402175;%%

%\cite{Choubey:2005zy}
\bibitem{Choubey:2005zy}
  S.~Choubey and P.~Roy,
  %``Probing the deviation from maximal mixing of atmospheric neutrinos,''
  Phys.\ Rev.\ D {\bf 73}, 013006 (2006)
  [hep-ph/0509197].  %%CITATION = HEP-PH/0509197;%%


%\bibitem{Andreopoulos:2009rq_ATM} %changed
%  C.~Andreopoulos, A.~Bell, D.~Bhattacharya, F.~Cavanna, J.~Dobson, S.~Dytman,
%H.~Gallagher and P.~Guzowski {\it et al.},
%  %``The GENIE Neutrino Monte Carlo Generator,''
%  Nucl.\ Instrum.\ Meth.\ A {\bf 614}, 87 (2010)
%  [arXiv:0905.2517 [hep-ph]].
%  %%CITATION = ARXIV:0905.2517;%%

  %\cite{Andreopoulos:2009rq}
\bibitem{Andreopoulos:2009rq}
  C.~Andreopoulos, A.~Bell, D.~Bhattacharya, F.~Cavanna, J.~Dobson, S.~Dytman, H.~Gallagher and P.~Guzowski {\it et al.},
  %``The GENIE Neutrino Monte Carlo Generator,''
  Nucl.\ Instrum.\ Meth.\ A {\bf 614}, 87 (2010)
  [arXiv:0905.2517 [hep-ph]].  %%CITATION = ARXIV:0905.2517;%%

%\cite{Agostinelli:2002hh}
\bibitem{Agostinelli:2002hh}
  S.~Agostinelli {\it et al.}  [GEANT4 Collaboration],
  %``GEANT4: A Simulation toolkit,''
  Nucl.\ Instrum.\ Meth.\ A {\bf 506}, 250 (2003).  %%CITATION = NUIMA,A506,250;%%

%%\cite{Peltoniemi:2009xx}
%\bibitem{Peltoniemi:2009xx}
%  J.~Peltoniemi,
%  %``Liquid scintillator as tracking detector for high-energy events,''
%  arXiv:0909.4974 [physics.ins-det]; arXiv:0911.4876 [hep-ex].
%  %%CITATION = ARXIV:0909.4974;%%

%\cite{Peltoniemi:2009xx}
\bibitem{Peltoniemi:2009xx}
  J.~Peltoniemi,
  %``Liquid scintillator as tracking detector for high-energy events,''
  arXiv:0909.4974 [physics.ins-det].
  %%CITATION = ARXIV:0909.4974;%%

%\cite{Peltoniemi:2009hv}
\bibitem{Peltoniemi:2009hv}
  J.~Peltoniemi,
  %``Simulations of neutrino oscillations for a wide band beam from CERN to LENA,''
  arXiv:0911.4876 [hep-ex].
  %%CITATION = ARXIV:0911.4876;%%

%\cite{Huber:2008yx}
\bibitem{Huber:2008yx}
  P.~Huber and T.~Schwetz,
  %``A Low energy neutrino factory with non-magnetic detectors,''
  Phys.\ Lett.\ B {\bf 669}, 294 (2008)
  [arXiv:0805.2019 [hep-ph]].  %%CITATION = ARXIV:0805.2019;%%

%\cite{Abe:2011ph}
\bibitem{Abe:2011ph}
  K.~Abe {\it et al.}  [Super-Kamiokande Collaboration],
  %``Search for Differences in Oscillation Parameters for Atmospheric Neutrinos and Antineutrinos at Super-Kamiokande,''
  Phys.\ Rev.\ Lett.\  {\bf 107}, 241801 (2011)
  [arXiv:1109.1621 [hep-ex]].
  %%CITATION = ARXIV:1109.1621;%%

%\cite{GonzalezGarcia:2004wg}
\bibitem{GonzalezGarcia:2004wg}
  M.~C.~Gonzalez-Garcia and M.~Maltoni,
  %``Atmospheric neutrino oscillations and new physics,''
  Phys.\ Rev.\ D {\bf 70}, 033010 (2004)
  [hep-ph/0404085].
  %%CITATION = HEP-PH/0404085;%%

%%\cite{Adamson:2013whj}
%\bibitem{Adamson:2013whj}
%  P.~Adamson {\it et al.}  [MINOS Collaboration],
%  %``Measurement of Neutrino and Antineutrino Oscillations Using Beam and
%  %Atmospheric Data in MINOS,''
%  Phys.\ Rev.\ Lett.\  {\bf 110}, no. 25, 251801 (2013)
%  [arXiv:1304.6335 [hep-ex]].
%  %%CITATION = ARXIV:1304.6335;%%

%%\cite{Abe:2014ugx}
%\bibitem{Abe:2014ugx}
%  K.~Abe {\it et al.}  [T2K Collaboration],
%  %``Precise Measurement of the Neutrino Mixing Parameter \theta_{23} from Muon
%  %Neutrino Disappearance in an Off-axis Beam,''
%  Phys.\ Rev.\ Lett.\  {\bf 112}, 181801 (2014)
%  [arXiv:1403.1532 [hep-ex]].
%  %%CITATION = ARXIV:1403.1532;%%


% chap: Geoneutrino
%%%%%%%%%%%%%%%%%%%%%%%%%%%%%%%%%%%%%%%%%%%%%%%%%%%%%%%%%%%%%%%%%%%%%%%%%%%%%%%%%%%%%%%%
%%%%%%%%%%%%%%%%%%%%%%%%%%%%%%%%%%%%%%%%%%%%%%%%%%%%%%%%%%%%%%%%%%%%%%%%%%%%%%%%%%%%%%%%
% chap: Geoneutrino


%\bibitem{Jaupart}
%C. Jaupart, S. Labrosse, and J.C. Mareschal (2007) Temperatures, Heat and Energy in the Mantle of the Earth. In: Bercovici, Schubert, eds. Treatise on Geophysics, vol. 7: Mantle dynamics. Elsevier $B-V$. $253-303$, doi:10.1016/B978-044452748-6.00114-0.

\bibitem{Jaupart}
C.~Jaupart, S.~Labrosse, and J.~C.~Mareschal, ``Temperatures, Heat and Energy in the Mantle of the Earth'',
In: Bercovici, Schubert, eds. Treatise on Geophysics, vol. 7: Mantle dynamics. Elsevier {B-V}, 253-303 (2007).
%doi:10.1016/B978-044452748-6.00114-0.

\bibitem{Davies}
J.~H.~Davies and D.~R.~Davies, Solid Earth {\bf 1}, 5 (2010).

%\bibitem{Borexino2013}
%G. Bellini et al.(Borexino Collaboration), Phys. Lett. B {\bf 722} 295 (2013).

%\cite{Bellini:2013nah}
\bibitem{Bellini:2013nah}
  G.~Bellini {\it et al.}  [Borexino Collaboration],
  %``Measurement of geo-neutrinos from 1353 days of Borexino,''
  Phys.\ Lett.\ B {\bf 722}, 295 (2013)
  [arXiv:1303.2571 [hep-ex]].
  %%CITATION = ARXIV:1303.2571;%%

%\bibitem{KLgeo} A. Gando et al.(KamLAND Collaboration), Phys. Rev. D {\bf  88}, 033001 (2013).

%\cite{Gando:2013nba}
\bibitem{Gando:2013nba}
  A.~Gando {\it et al.}  [KamLAND Collaboration],
  %``Reactor On-Off Antineutrino Measurement with KamLAND,''
  Phys.\ Rev.\ D {\bf 88}, 033001 (2013)
  [arXiv:1303.4667 [hep-ex]].
  %%CITATION = ARXIV:1303.4667;%%


%\bibitem{McDonough}
%W.F. McDonough, J.G. Learned, and S.T. Dye, Physics Today {\bf 65}, 46 (2012).

%\cite{McDonough:2012zz}
\bibitem{McDonough:2012zz}
  W.~F.~McDonough, J.~G.~Learned and S.~T.~Dye,
  %``The many uses of electron antineutrinos,''
  Phys.\ Today {\bf 65N3}, 46 (2012).

%\bibitem{Strati2014}
%V.~Strati {\it et al.}, Prog. in Earth and Planet. Sci. {\bf 2}, 1 (2015).

%\cite{Strati:2014kaa}
\bibitem{Strati:2014kaa}
  V.~Strati, M.~Baldoncini, I.~Callegari, F.~Mantovani, W.~F.~McDonough, B.~Ricci and G.~Xhixha,
  %``Expected geoneutrino signal at JUNO,''
Progress in Earth and Planetary Science {\bf 2}, 1 (2015).

\bibitem{Huang13}
Y.~Huang {\it et al.}, Geochemistry Geophysics Geosystems {\bf 14}, 2003 (2013).

\bibitem{Javoy2010}
M.~Javoy {\it et al.}, Earth Planet. Sci. Lett. {\bf 293}, 259 (2010).

%\bibitem{Fiorentini2012}
%G. Fiorentini et al., Phys. Rev. D {\bf 86}, 033004 (2012).

%\cite{Fiorentini:2012yk}
\bibitem{Fiorentini:2012yk}
  G.~Fiorentini, G.~L.~Fogli, E.~Lisi, F.~Mantovani and A.~M.~Rotunno,
  %``Mantle geoneutrinos in KamLAND and Borexino,''
  Phys.\ Rev.\ D {\bf 86}, 033004 (2012)
  [arXiv:1204.1923 [hep-ph]].
  %%CITATION = ARXIV:1204.1923;%%

\bibitem{Turcotte2002}
D.L. Turcotte and G. Schubert, ``Geodynamics, Applications of Continuum
Physics to Geological Problems'', second ed., Cambridge University Press (2002).

%\bibitem{Sramek2013}
%O. \v{S}r\'{a}mek et al., Earth Planet. Sci. Lett. {\bf 361}, 356 (2013).

%\cite{Sramek:2012nk}
\bibitem{Sramek:2012nk}
  O.~Sramek, W.~F.~McDonough, E.~S.~Kite, V.~Lekic, S.~Dye and S.~Zhong,
  %``Geophysical and geochemical constraints on geoneutrino fluxes from Earth's mantle,''
  Earth Planet.\ Sci.\ Lett.\  {\bf 361}, 356 (2013)
  [arXiv:1207.0853 [physics.geo-ph]].
  %%CITATION = ARXIV:1207.0853;%%

\bibitem{Enomoto}
S.~Enomoto {\it et al.}, Earth Planet. Sci. Lett. {\bf 258}, 147 (2007).

%\bibitem{Fiorentini}
%G. Fiorentini et al., Phys. Rev. D {\bf 72}, 1 (2005).

%\cite{Fiorentini:2005cu}
\bibitem{Fiorentini:2005cu}
  G.~Fiorentini, M.~Lissia, F.~Mantovani and R.~Vannucci,
  %``How much Uranium is in the Earth? Predictions for geo-neutrinos at KamLAND,''
  Phys.\ Rev.\ D {\bf 72}, 033017 (2005)
  [hep-ph/0501111].
  %%CITATION = HEP-PH/0501111;%%

%\bibitem{Coltorti}
%M. Coltorti et al. , Geochim. Cosmochim. Acta {\bf 75(9)}, 2271 (2011).

%\cite{Coltorti:2011gr}
\bibitem{Coltorti:2011gr}
  M.~Coltorti, R.~Boraso, F.~Mantovani, M.~Morsilli, G.~Fiorentini, A.~Riva, G.~Rusciadelli and R.~Tassinari {\it et al.},
  %``U and Th content in the Central Apennines continental crust: a contribution to the determination of the geo-neutrinos flux at LNGS,''
  Geochim.\ Cosmochim.\ Acta {\bf 75}, 2271 (2011)
  [arXiv:1102.1335 [astro-ph.EP]].
  %%CITATION = ARXIV:1102.1335;%%

%\bibitem{Huang14}
%Y. Huang et al., Geochem. Geophys. Geosyst., 15 doi:10.1002/2014GC005397(2014).

%\cite{Huang:2014dpa}
\bibitem{Huang:2014dpa}
  Y.~Huang, V.~Strati, F.~Mantovani, S.~B.~Shirey, R.~L.~Rudnick and W.~F.~McDonough,
  %``Regional study of the Archean to Proterozoic crust at the Sudbury Neutrino Observatory (SNO+), Ontario: Predicting the geoneutrino flux,''
  Geochem.\ Geophys.\ Geosyst.\  {\bf 15}, 3925 (2014)
  [arXiv:1404.6692 [physics.geo-ph]].
  %%CITATION = ARXIV:1404.6692;%%

\bibitem{huang2013intraplate}
H-Q.~Huang, {\it et al.}, J. Asian Earth Sci. {\bf 74}, 280 (2013).

%\bibitem{Bellini2013}
%G. Bellini et al. (Borexino Collaboration), Phys Lett. B {\bf 722}, 295 (2013).

%%\cite{Bellini:2013nah}
%\bibitem{Bellini:2013nah}
%  G.~Bellini {\it et al.}  [Borexino Collaboration],
%  %``Measurement of geo-neutrinos from 1353 days of Borexino,''
%  Phys.\ Lett.\ B {\bf 722}, 295 (2013)
%  [arXiv:1303.2571 [hep-ex]].
%  %%CITATION = ARXIV:1303.2571;%%

\bibitem{zhou2000origin}
X.~M.~Zhou, W.~X.~Li., Tectonophysics {\bf 326}, 269 (2000).

\bibitem{gilder1996isotopic}
S.~A.~Gilder {\it et al.}, J. Geophysical Research {\bf 101}, B7, 16137 (1996).

\bibitem{niu2005generation}
Y.~L.~Niu, Geological J. China Universities, {\bf 11}, 9 (2005).

\bibitem{zhou2006petrogenesis}
X.~Zhou {\it et al.}, Episodes, {\bf 29}, 26 (2006).

\bibitem{li2007formation}
Z-X.~Li, X-H.~Li, Geology, {\bf 35}, 179 (2007).

\bibitem{yang2013}
Y-T.~Yang,  Earth Sci. Reviews, {\bf 126}, 96 (2013).

\bibitem{Laske2013}
G.~Laske, G.~Masters, Z.~Ma, and M.~Pasyanos, Geophys. Research Abstracts, {\bf 15}, 2658 (2013).

\bibitem{DeMets1990}
C.~DeMets {\it et al.}, Geophys. J. International, {\bf 101}, 425 (1990).

%\bibitem{Araki}
%T. Araki et al. (KamLAND collaboration), Nature {\bf 436}, 499 (2005).

%\cite{Araki:2005qa}
\bibitem{Araki:2005qa}
  T.~Araki, S.~Enomoto, K.~Furuno, Y.~Gando, K.~Ichimura, H.~Ikeda, K.~Inoue and Y.~Kishimoto {\it et al.},
  %``Experimental investigation of geologically produced antineutrinos with KamLAND,''
  Nature {\bf 436}, 499 (2005).
  %%CITATION = NATUA,436,499;%%

%\bibitem{Bellini}
%G. Bellini et al. (Borexino Collaboration) , Phys. Lett. B {\bf 687}, 299 (2010).

%\cite{Bellini:2010hy}
\bibitem{Bellini:2010hy}
  G.~Bellini {\it et al.}  [Borexino Collaboration],
  %``Observation of Geo-Neutrinos,''
  Phys.\ Lett.\ B {\bf 687}, 299 (2010)
  [arXiv:1003.0284 [hep-ex]].
  %%CITATION = ARXIV:1003.0284;%%

%\bibitem{Chen}
%M.C. Chen,  Earth Moon and Planets {\bf 99}, 221 (2006).

%\cite{Chen:2005zza}
\bibitem{Chen:2005zza}
  M.~C.~Chen,
  %``Geo-neutrinos in SNO+,''
  Earth Moon Planets {\bf 99}, 221 (2006).
  %%CITATION = EMPLD,99,221;%%

\bibitem{IAEA}
J.~Mandula, Nuclear Power Engeneering Section, IAEA-PRIS database 2013.

%\bibitem{Baldoncini2014}
%M. Baldoncini et al., Phys. Rev. D {\bf 91}, 065002 (2015).

%\cite{Baldoncini:2014vda}
\bibitem{Baldoncini:2014vda}
  M.~Baldoncini, I.~Callegari, G.~Fiorentini, F.~Mantovani, B.~Ricci, V.~Strati and G.~Xhixha,
  %``Reference worldwide model for antineutrinos from reactors,''
  Phys.\ Rev.\ D {\bf 91}, 065002 (2015)
  [arXiv:1411.6475 [physics.ins-det]].
  %%CITATION = ARXIV:1411.6475;%%

%\bibitem{Ciuffoli2014}
%E. Ciuffoli et al., Phys. Rev. D {\bf 89}, 7 (2014).

%%\cite{Ciuffoli:2013pla}
%\bibitem{Ciuffoli:2013pla}
%  E.~Ciuffoli, J.~Evslin, Z.~Wang, C.~Yang, X.~Zhang and W.~Zhong,
%  %``Advantages of Multiple Detectors for the Neutrino Mass Hierarchy Determination at Reactor Experiments,''
%  Phys.\ Rev.\ D {\bf 89}, 073006 (2014)
%  [arXiv:1308.0591 [hep-ph]].
%  %%CITATION = ARXIV:1308.0591;%%

%\bibitem{Mueller2011}
%Th. A. Mueller et al., Phys. Rev. C  {\bf 83}, 054615 (2011).

%%\cite{Mueller:2011nm}
%\bibitem{Mueller:2011nm}
%  T.~A.~Mueller, D.~Lhuillier, M.~Fallot, A.~Letourneau, S.~Cormon, M.~Fechner, L.~Giot and T.~Lasserre {\it et al.},
%  %``Improved Predictions of Reactor Antineutrino Spectra,''
%  Phys.\ Rev.\ C {\bf 83}, 054615 (2011)
%  [arXiv:1101.2663 [hep-ex]].
%  %%CITATION = ARXIV:1101.2663;%%

%\bibitem{Ma}
%X.B. Ma et al., Phys. Rev. C {\bf 88}, 014605 (2013).

%\cite{Ma:2012bm}
\bibitem{Ma:2012bm}
  X.~B.~Ma, W.~L.~Zhong, L.~Z.~Wang, Y.~X.~Chen and J.~Cao,
  %``Improved calculation of the energy release in neutron-induced fission,''
  Phys.\ Rev.\ C {\bf 88}, no. 1, 014605 (2013)
  [arXiv:1212.6625 [nucl-ex]].
  %%CITATION = ARXIV:1212.6625;%%

%\bibitem{Strumia}
%A.~Strumia and F.~Vissani, Phys. Lett. B {\bf 564}, 42 (2003).

%%\cite{Strumia:2003zx}
%\bibitem{Strumia:2003zx}
%  A.~Strumia and F.~Vissani,
%  %``Precise quasielastic neutrino/nucleon cross-section,''
%  Phys.\ Lett.\ B {\bf 564}, 42 (2003)
%  [astro-ph/0302055].
%  %%CITATION = ASTRO-PH/0302055;%%

%\bibitem{Capozzi2014}
%F. Capozzi et al., Phys. Rev D. {\bf 89}, 093018 (2014).

%%\cite{Capozzi:2013csa}
%\bibitem{Capozzi:2013csa}
%  F.~Capozzi, G.~L.~Fogli, E.~Lisi, A.~Marrone, D.~Montanino and A.~Palazzo,
%  %``Status of three-neutrino oscillation parameters, circa 2013,''
%  Phys.\ Rev.\ D {\bf 89}, 093018 (2014)
%  [arXiv:1312.2878 [hep-ph]].
%  %%CITATION = ARXIV:1312.2878;%%

%\bibitem{Huber2004}
%P. Huber and T. Schwetz, Phys. Rev. D {\bf 70}, 053011 (2004).

%\cite{Huber:2004xh}
\bibitem{Huber:2004xh}
  P.~Huber and T.~Schwetz,
  %``Precision spectroscopy with reactor anti-neutrinos,''
  Phys.\ Rev.\ D {\bf 70}, 053011 (2004)
  [hep-ph/0407026].
  %%CITATION = HEP-PH/0407026;%%

%\bibitem{Cao}
%J. Cao, Nucl. Phys. B Proc. Suppl. 00, 1 (2011).

%\cite{Cao:2011gb}
\bibitem{Cao:2011gb}
  J.~Cao,
  %``Determining Reactor Neutrino Flux,''
  Nucl.\ Phys.\ Proc.\ Suppl.\  {\bf 229-232}, 205 (2012)
  [arXiv:1101.2266 [hep-ex]].
  %%CITATION = ARXIV:1101.2266;%%

%\bibitem{Djurcic2009}
%Z. Djurcic et al., Journal of Physics G {\bf 36}, 045002 (2009).

%\cite{Djurcic:2008ny}
\bibitem{Djurcic:2008ny}
  Z.~Djurcic, J.~A.~Detwiler, A.~Piepke, V.~R.~Foster, Jr., L.~Miller and G.~Gratta,
  %``Uncertainties in the Anti-neutrino Production at Nuclear Reactors,''
  J.\ Phys.\ G {\bf 36}, 045002 (2009)
  [arXiv:0808.0747 [hep-ex]].
  %%CITATION = ARXIV:0808.0747;%%

%\bibitem{DBb}F. P. An et al., (Daya Bay Collaboration),
%  Phys.\ Rev.\ Lett.\  {\bf 108}, 171803 (2012).

%\bibitem{Bin}
%Z. Bin et al., Chinese Physics C {\bf 36}, 1  (2012).

%\cite{Zhou:2012zzc}
\bibitem{Zhou:2012zzc}
  B.~Zhou, X.~C.~Ruan, Y.~B.~Nie, Z.~Y.~Zhou, F.~P.~An and J.~Cao,
  %``A study of antineutrino spectra from spent nuclear fuel at Daya Bay,''
  Chin.\ Phys.\ C {\bf 36}, 1 (2012).
  %%CITATION = CHPHD,C36,1;%%

%\bibitem{Chooz}
%M. Apollonio, et al. (CHOOZ collaboration), Phys. Rev. D, {\bf 61}, 012001 (2000).

%%\cite{Apollonio:1999jg}
%\bibitem{Apollonio:1999jg}
%  M.~Apollonio {\it et al.}  [CHOOZ Collaboration],
%  %``Determination of neutrino incoming direction in the CHOOZ experiment and supernova explosion location by scintillator detectors,''
%  Phys.\ Rev.\ D {\bf 61}, 012001 (2000)
%  [hep-ex/9906011].
%  %%CITATION = HEP-EX/9906011;%%

\bibitem{Batygov}
M. Batygov, Earth, Moon, and Planets {\bf 183}, 192 (2006).

%\end{thebibliography}

% chap: Sterile
%%%%%%%%%%%%%%%%%%%%%%%%%%%%%%%%%%%%%%%%%%%%%%%%%%%%%%%%%%%%%%%%%%%%%%%%%%%%%%%%%%%%%%%%
%%%%%%%%%%%%%%%%%%%%%%%%%%%%%%%%%%%%%%%%%%%%%%%%%%%%%%%%%%%%%%%%%%%%%%%%%%%%%%%%%%%%%%%%
% chap: Sterile


%\begin{thebibliography}{99}


%\cite{Kusenko:2009up}
\bibitem{Kusenko:2009up}
  A.~Kusenko,
  %``Sterile neutrinos: The Dark side of the light fermions,''
  Phys.\ Rept.\  {\bf 481}, 1 (2009)
  [arXiv:0906.2968 [hep-ph]].
  %%CITATION = ARXIV:0906.2968;%%

%\cite{Gariazzo:2015rra}
\bibitem{Gariazzo:2015rra}
  S.~Gariazzo, C.~Giunti, M.~Laveder, Y.~F.~Li and E.~M.~Zavanin,
  %``Light sterile neutrinos,''
  arXiv:1507.08204 [hep-ph].
  %%CITATION = ARXIV:1507.08204;%%

%\cite{Giunti:2010zu}
\bibitem{Giunti:2010zu}
  C.~Giunti and M.~Laveder,
  %``Statistical Significance of the Gallium Anomaly,''
  Phys.\ Rev.\ C {\bf 83}, 065504 (2011)
  [arXiv:1006.3244 [hep-ph]].
  %%CITATION = ARXIV:1006.3244;%%

%\cite{deHolanda:2003tx}
\bibitem{deHolanda:2003tx}
P.~C.~de Holanda and A.~Y.~Smirnov,
%``Homestake result, sterile neutrinos and low-energy solar neutrino experiments,''
  Phys.\ Rev.\ D {\bf 69}, 113002 (2004)
  [hep-ph/0307266].
  %%CITATION = HEP-PH/0307266;%%
 %86 citations counted in INSPIRE as of 03 Jul 2015

%\cite{deHolanda:2010am}
\bibitem{deHolanda:2010am}
P.~C.~de Holanda and A.~Y.~Smirnov,
%``Solar neutrino spectrum, sterile neutrinos and additional radiation in the Universe,''
  Phys.\ Rev.\ D {\bf 83}, 113011 (2011)
  [arXiv:1012.5627 [hep-ph]].
  %%CITATION = ARXIV:1012.5627;%%
 %54 citations counted in INSPIRE as of 03 Jul 2015

%\cite{Maltoni:2002xd}
\bibitem{Maltoni:2002xd}
  M.~Maltoni, T.~Schwetz, M.~A.~Tortola and J.~W.~F.~Valle,
  %``Ruling out four neutrino oscillation interpretations of the LSND anomaly?,''
  Nucl.\ Phys.\ B {\bf 643}, 321 (2002)
  [hep-ph/0207157].
  %%CITATION = HEP-PH/0207157;%%

%\cite{ALEPH:2005ab}
\bibitem{ALEPH:2005ab}
  S.~Schael {\it et al.}  [ALEPH and DELPHI and L3 and OPAL and SLD and LEP Electroweak Working Group and SLD Electroweak Group and SLD Heavy Flavour Group Collaborations],
  %``Precision electroweak measurements on the $Z$ resonance,''
  Phys.\ Rept.\  {\bf 427}, 257 (2006)
  [hep-ex/0509008].
  %%CITATION = HEP-EX/0509008;%%

%\cite{Sousa:2015bxa}
\bibitem{Sousa:2015bxa}
  A.~Sousa [MINOS and MINOS+ Collaborations],
  %``First MINOS+ Data and New Results from MINOS,''
  arXiv:1502.07715 [hep-ex].  %%CITATION = ARXIV:1502.07715;%%

%\cite{Armbruster:2002mp}
\bibitem{Armbruster:2002mp}
  B.~Armbruster {\it et al.}  [KARMEN Collaboration],
  %``Upper limits for neutrino oscillations muon-anti-neutrino ---> electron-anti-neutrino from muon decay at rest,''
  Phys.\ Rev.\ D {\bf 65}, 112001 (2002)
  [hep-ex/0203021].  %%CITATION = HEP-EX/0203021;%%

  %\cite{Astier:2003gs}
\bibitem{Astier:2003gs}
  P.~Astier {\it et al.}  [NOMAD Collaboration],
  %``Search for nu(mu) ---> nu(e) oscillations in the NOMAD experiment,''
  Phys.\ Lett.\ B {\bf 570}, 19 (2003)
  [hep-ex/0306037].  %%CITATION = HEP-EX/0306037;%%

  %\cite{Antonello:2013gut}
\bibitem{Antonello:2013gut}
  M.~Antonello {\it et al.}  [ICARUS Collaboration],
  %``Search for anomalies in the ${\nu}_e$ appearance from a ${\nu}_{\mu}$ beam,''
  Eur.\ Phys.\ J.\ C {\bf 73}, 2599 (2013)
  [arXiv:1307.4699 [hep-ex]].  %%CITATION = ARXIV:1307.4699;%%

  %\cite{Agafonova:2013xsk}
\bibitem{Agafonova:2013xsk}
  N.~Agafonova {\it et al.}  [OPERA Collaboration],
  %``Search for $\nu_\mu \rightarrow \nu_e$  oscillations with the OPERA experiment in the CNGS beam,''
  JHEP {\bf 1307}, 004 (2013)  [JHEP {\bf 1307}, 085 (2013)]
  [arXiv:1303.3953 [hep-ex]].  %%CITATION = ARXIV:1303.3953;%%

%\cite{Conrad:2012qt}
\bibitem{Conrad:2012qt}
J.~M.~Conrad, C.~M.~Ignarra, G.~Karagiorgi, M.~H.~Shaevitz and J.~Spitz,
%``Sterile Neutrino Fits to Short Baseline Neutrino Oscillation Measurements,''
  Adv.\ High Energy Phys.\ {\bf 2013}, 163897 (2013)
  [arXiv:1207.4765 [hep-ex]].
  %%CITATION = ARXIV:1207.4765;%%
 %82 citations counted in INSPIRE as of 03 Jul 2015

%\cite{Antonello:2015lea}
\bibitem{Antonello:2015lea}
  M.~Antonello {\it et al.}  [MicroBooNE and LAr1-ND and ICARUS-WA104 Collaborations],
  %``A Proposal for a Three Detector Short-Baseline Neutrino Oscillation Program in the Fermilab Booster Neutrino Beam,''
  arXiv:1503.01520 [physics.ins-det].
  %%CITATION = ARXIV:1503.01520;%%

%\cite{Dwyer:2011xs}
\bibitem{Dwyer:2011xs}
D.~A.~Dwyer, K.~M.~Heeger, B.~R.~Littlejohn and P.~Vogel,
%``Search for Sterile Neutrinos with a Radioactive Source at Daya Bay,''
  Phys.\ Rev.\ D {\bf 87}, no. 9, 093002 (2013)
  [arXiv:1109.6036 [hep-ex]].
  %%CITATION = ARXIV:1109.6036;%%
 %21 citations counted in INSPIRE as of 03 Jul 2015

%\cite{Borexino:2013xxa}
\bibitem{Borexino:2013xxa}
  G.~Bellini {\it et al.}  [Borexino Collaboration],
  %``SOX: Short distance neutrino Oscillations with BoreXino,''
  JHEP {\bf 1308}, 038 (2013)
  [arXiv:1304.7721 [physics.ins-det]].
  %%CITATION = ARXIV:1304.7721;%%

%\cite{Gando:2013zoa}
\bibitem{Gando:2013zoa}
  A.~Gando, Y.~Gando, S.~Hayashida, H.~Ikeda, K.~Inoue, K.~Ishidoshiro, H.~Ishikawa and M.~Koga {\it et al.},
  %``CeLAND: search for a 4th light neutrino state with a 3 PBq 144Ce-144Pr electron antineutrino generator in KamLAND,''
  arXiv:1312.0896 [physics.ins-det].
  %%CITATION = ARXIV:1312.0896;%%

%\cite{Riis:2010zm}
\bibitem{Riis:2010zm}
  A.~S.~Riis and S.~Hannestad,
  %``Detecting sterile neutrinos with KATRIN like experiments,''
  JCAP {\bf 1102}, 011 (2011)
  [arXiv:1008.1495 [astro-ph.CO]].
  %%CITATION = ARXIV:1008.1495;%%

%\cite{Kornoukhov:1994zq}
\bibitem{Kornoukhov:1994zq}
V.~N.~Kornoukhov,
%``Some aspects of the creation and application of anti-neutrino artificial sources,''
  ITEP-90-94.
  %%CITATION = ITEP-90-94;%%
 %1 citations counted in INSPIRE as of 03 Jul 2015

%\cite{Cribier:2011fv}
\bibitem{Cribier:2011fv}
M.~Cribier, M.~Fechner, T.~Lasserre, A.~Letourneau, D.~Lhuillier, G.~Mention, D.~Franco and V.~Kornoukhov {\it et al.},
%``A proposed search for a fourth neutrino with a PBq antineutrino source,''
  Phys.\ Rev.\ Lett.\ {\bf 107}, 201801 (2011)
  [arXiv:1107.2335 [hep-ex]].
  %%CITATION = ARXIV:1107.2335;%%
 %44 citations counted in INSPIRE as of 03 Jul 2015

%\cite{Conrad:2013ova}
\bibitem{Conrad:2013ova}
J.~M.~Conrad and M.~H.~Shaevitz,
%``Electron Antineutrino Disappearance at KamLAND and JUNO as Decisive Tests of the Short Baseline $\overline{\nu}_\mu \to \overline{\nu}_e$ Appearance Anomaly,''
  Phys.\ Rev.\ D {\bf 89}, 057301 (2014)
  [arXiv:1310.3857 [hep-ex]].
  %%CITATION = ARXIV:1310.3857;%%
 %3 citations counted in INSPIRE as of 03 Jul 2015

%\cite{Agarwalla:2011qf}
\bibitem{Agarwalla:2011qf}
S.~K.~Agarwalla, J.~M.~Conrad and M.~H.~Shaevitz,
%``Short-baseline Neutrino Oscillation Waves in Ultra-large Liquid Scintillator Detectors,''
  JHEP {\bf 1112}, 085 (2011)
  [arXiv:1105.4984 [hep-ph]].
  %%CITATION = ARXIV:1105.4984;%%
 %18 citations counted in INSPIRE as of 03 Jul 2015

%%\cite{Mention:2011rk}
%\bibitem{Mention:2011rk}
%G.~Mention, M.~Fechner, T.~Lasserre, T.~A.~Mueller, D.~Lhuillier, M.~Cribier and A.~Letourneau,
%%``The Reactor Antineutrino Anomaly,''
%  Phys.\ Rev.\ D {\bf 83}, 073006 (2011)
%  [arXiv:1101.2755 [hep-ex]].
%  %%CITATION = ARXIV:1101.2755;%%
% %453 citations counted in INSPIRE as of 03 Jul 2015

%%\cite{Giunti:2013aea}
%\bibitem{Giunti:2013aea}
%C.~Giunti, M.~Laveder, Y.~F.~Li and H.~W.~Long,
%%``Pragmatic View of Short-Baseline Neutrino Oscillations,''
%  Phys.\ Rev.\ D {\bf 88}, 073008 (2013)
%  [arXiv:1308.5288 [hep-ph]].
%  %%CITATION = ARXIV:1308.5288;%%
% %73 citations counted in INSPIRE as of 03 Jul 2015


%%\cite{Wolfenstein:1977ue}
%\bibitem{Wolfenstein:1977ue}
%L.~Wolfenstein,
%%``Neutrino Oscillations in Matter,''
%  Phys.\ Rev.\ D {\bf 17}, 2369 (1978).
%  %%CITATION = PHRVA,D17,2369;%%
% %3980 citations counted in INSPIRE as of 03 Jul 2015
%
%%\cite{Mikheev:1986gs}
%\bibitem{Mikheev:1986gs}
%S.~P.~Mikheev and A.~Y.~Smirnov,
%%``Resonance Amplification of Oscillations in Matter and Spectroscopy of Solar Neutrinos,''
%  Sov.\ J.\ Nucl.\ Phys.\ {\bf 42}, 913 (1985)
%  [Yad.\ Fiz.\ {\bf 42}, 1441 (1985)].
%  %%CITATION = SJNCA,42,913;%%
% %2897 citations counted in INSPIRE as of 03 Jul 2015

%\cite{Aharmim:2008kc}
\bibitem{Aharmim:2008kc}
B.~Aharmim {\it et al.} [SNO Collaboration],
%``An Independent Measurement of the Total Active B-8 Solar Neutrino Flux Using an Array of He-3 Proportional Counters at the Sudbury Neutrino Observatory,''
  Phys.\ Rev.\ Lett.\ {\bf 101}, 111301 (2008)
  [arXiv:0806.0989 [nucl-ex]].
  %%CITATION = ARXIV:0806.0989;%%
 %304 citations counted in INSPIRE as of 03 Jul 2015

%\cite{Hosaka:2005um}
\bibitem{Hosaka:2005um}
J.~Hosaka {\it et al.} [Super-Kamiokande Collaboration],
%``Solar neutrino measurements in super-Kamiokande-I,''
  Phys.\ Rev.\ D {\bf 73}, 112001 (2006)
  [hep-ex/0508053].
  %%CITATION = HEP-EX/0508053;%%
 %424 citations counted in INSPIRE as of 03 Jul 2015

%\cite{Abe:2008aa}
\bibitem{Abe:2008aa}
S.~Abe {\it et al.} [KamLAND Collaboration],
%``Precision Measurement of Neutrino Oscillation Parameters with KamLAND,''
  Phys.\ Rev.\ Lett.\ {\bf 100}, 221803 (2008)
  [arXiv:0801.4589 [hep-ex]].
  %%CITATION = ARXIV:0801.4589;%%
 %667 citations counted in INSPIRE as of 03 Jul 2015

%\cite{Bakhti:2013ora}
\bibitem{Bakhti:2013ora}
P.~Bakhti and Y.~Farzan,
%``Constraining Super-light Sterile Neutrino Scenario by JUNO and RENO-50,''
  JHEP {\bf 1310}, 200 (2013)
  [arXiv:1308.2823 [hep-ph]].
  %%CITATION = ARXIV:1308.2823;%%
 %9 citations counted in INSPIRE as of 03 Jul 2015

%\cite{Girardi:2014wea}
\bibitem{Girardi:2014wea}
I.~Girardi, D.~Meloni, T.~Ohlsson, H.~Zhang and S.~Zhou,
%``Constraining Sterile Neutrinos Using Reactor Neutrino Experiments,''
  JHEP {\bf 1408}, 057 (2014)
  [arXiv:1405.6540 [hep-ph]].
  %%CITATION = ARXIV:1405.6540;%%
 %5 citations counted in INSPIRE as of 03 Jul 2015

%\cite{Palazzo:2013bsa}
\bibitem{Palazzo:2013bsa}
A.~Palazzo,
%``Constraints on very light sterile neutrinos from $\theta_{13}$-sensitive reactor experiments,''
  JHEP {\bf 1310}, 172 (2013)
  [arXiv:1308.5880 [hep-ph]].
  %%CITATION = ARXIV:1308.5880;%%
 %11 citations counted in INSPIRE as of 03 Jul 2015

%%\cite{An:2013zwz}
%\bibitem{An:2013zwz}
%F.~P.~An {\it et al.} [Daya Bay Collaboration],
%%``Spectral measurement of electron antineutrino oscillation amplitude and frequency at Daya Bay,''
%  Phys.\ Rev.\ Lett.\ {\bf 112}, 061801 (2014)
%  [arXiv:1310.6732 [hep-ex]].
%  %%CITATION = ARXIV:1310.6732;%%
% %124 citations counted in INSPIRE as of 03 Jul 2015

%\cite{An:2014bik}
\bibitem{An:2014bik}
F.~P.~An {\it et al.} [Daya Bay Collaboration],
%``Search for a Light Sterile Neutrino at Daya Bay,''
  Phys.\ Rev.\ Lett.\ {\bf 113}, 141802 (2014)
  [arXiv:1407.7259 [hep-ex]].
  %%CITATION = ARXIV:1407.7259;%%
 %12 citations counted in INSPIRE as of 03 Jul 2015

%%\cite{Huber:2011wv}
%\bibitem{Huber:2011wv}
%P.~Huber,
%%``On the determination of anti-neutrino spectra from nuclear reactors,''
%  Phys.\ Rev.\ C {\bf 84}, 024617 (2011)
%  [Phys.\ Rev.\ C {\bf 85}, 029901 (2012)]
%  [arXiv:1106.0687 [hep-ph]].
%  %%CITATION = ARXIV:1106.0687;%%
% %239 citations counted in INSPIRE as of 03 Jul 2015

%%\cite{Mueller:2011nm}
%\bibitem{Mueller:2011nm}
%T.~A.~Mueller, D.~Lhuillier, M.~Fallot, A.~Letourneau, S.~Cormon, M.~Fechner, L.~Giot and T.~Lasserre {\it et al.},
%%``Improved Predictions of Reactor Antineutrino Spectra,''
%  Phys.\ Rev.\ C {\bf 83}, 054615 (2011)
%  [arXiv:1101.2663 [hep-ex]].
%  %%CITATION = ARXIV:1101.2663;%%
% %312 citations counted in INSPIRE as of 03 Jul 2015

%\cite{Hayes:2013wra}
\bibitem{Hayes:2013wra}
A.~C.~Hayes, J.~L.~Friar, G.~T.~Garvey, G.~Jungman and G.~Jonkmans,
%``Systematic Uncertainties in the Analysis of the Reactor Neutrino Anomaly,''
  Phys.\ Rev.\ Lett.\ {\bf 112}, 202501 (2014)
  [arXiv:1309.4146 [nucl-th]].
  %%CITATION = ARXIV:1309.4146;%%
 %42 citations counted in INSPIRE as of 03 Jul 2015

%%\cite{Adamson:2013whj}
%\bibitem{Adamson:2013whj}
%P.~Adamson {\it et al.} [MINOS Collaboration],
%%``Measurement of Neutrino and Antineutrino Oscillations Using Beam and Atmospheric Data in MINOS,''
%  Phys.\ Rev.\ Lett.\ {\bf 110}, no. 25, 251801 (2013)
%  [arXiv:1304.6335 [hep-ex]].
%  %%CITATION = ARXIV:1304.6335;%%
% %106 citations counted in INSPIRE as of 03 Jul 2015
%\end{thebibliography}

% chap: NucleonDecay
%%%%%%%%%%%%%%%%%%%%%%%%%%%%%%%%%%%%%%%%%%%%%%%%%%%%%%%%%%%%%%%%%%%%%%%%%%%%%%%%%%%%%%%%
%%%%%%%%%%%%%%%%%%%%%%%%%%%%%%%%%%%%%%%%%%%%%%%%%%%%%%%%%%%%%%%%%%%%%%%%%%%%%%%%%%%%%%%%
% chap: NucleonDecay

%\begin{thebibliography}{99}

%\cite{Sakharov:1967rr}
\bibitem{Sakharov:1967rr}
A.~D.~Sakharov,
%``Quark - Muonic Currents And Violation Of Cp Invariance,''
  JETP Lett.\ {\bf 5}, 27 (1967)
  [Pisma Zh.\ Eksp.\ Teor.\ Fiz.\ {\bf 5}, 36 (1967)].
  %%CITATION = JTPLA,5,27;%%
 %18 citations counted in INSPIRE as of 03 Jul 2015

%\cite{Nath:2006ut}
\bibitem{Nath:2006ut}
P.~Nath and P.~Fileviez Perez,
%``Proton stability in grand unified theories, in strings and in branes,''
  Phys.\ Rept.\ {\bf 441}, 191 (2007)
  [hep-ph/0601023].
  %%CITATION = HEP-PH/0601023;%%
 %232 citations counted in INSPIRE as of 03 Jul 2015

%\cite{Reines:1954pg}
\bibitem{Reines:1954pg}
F.~Reines, C.~L.~Cowan and M.~Goldhaber,
%``Conservation of the number of nucleons,''
  Phys.\ Rev.\ {\bf 96}, 1157 (1954).
  %%CITATION = PHRVA,96,1157;%%
 %40 citations counted in INSPIRE as of 03 Jul 2015

%\cite{Krishnaswamy:1981uc}
\bibitem{Krishnaswamy:1981uc}
M.~R.~Krishnaswamy, M.~G.~K.~Menon, N.~K.~Mondal, V.~S.~Narasimham, B.~V.~Sreekantan, N.~Ito, S.~Kawakami and Y.~Hayashi {\it et al.},
%``Candidate Events for Nucleon Decay in the Kolar Gold Field Experiment,''
  Phys.\ Lett.\ B {\bf 106}, 339 (1981).
  %%CITATION = PHLTA,B106,339;%%
 %68 citations counted in INSPIRE as of 03 Jul 2015

%\cite{Battistoni:1985na}
\bibitem{Battistoni:1985na}
G.~Battistoni, E.~Bellotti, C.~Bloise, G.~Bologna, P.~Campana, C.~Castagnoli, V.~Chiarella and O.~Cremonesi {\it et al.},
%``The Nusex Detector,''
  Nucl.\ Instrum.\ Meth.\ A {\bf 245}, 277 (1986).
  %%CITATION = NUIMA,A245,277;%%
 %49 citations counted in INSPIRE as of 03 Jul 2015

%\cite{Berger:1987ke}
\bibitem{Berger:1987ke}
C.~Berger {\it et al.} [FREJUS Collaboration],
%``The Frejus Nucleon Decay Detector,''
  Nucl.\ Instrum.\ Meth.\ A {\bf 262}, 463 (1987).
  %%CITATION = NUIMA,A262,463;%%
 %66 citations counted in INSPIRE as of 03 Jul 2015

%\cite{Thron:1989cd}
\bibitem{Thron:1989cd}
J.~L.~Thron,
%``The Soudan-{II} Proton Decay Experiment,''
  Nucl.\ Instrum.\ Meth.\ A {\bf 283}, 642 (1989).
  %%CITATION = NUIMA,A283,642;%%
 %39 citations counted in INSPIRE as of 03 Jul 2015

%\cite{BeckerSzendy:1992hr}
\bibitem{BeckerSzendy:1992hr}
R.~Becker-Szendy, C.~B.~Bratton, D.~R.~Cady, D.~Casper, R.~Claus, S.~T.~Dye, W.~Gajewski and M.~Goldhaber {\it et al.},
%``IMB-3: A Large water Cherenkov detector for nucleon decay and neutrino interactions,''
  Nucl.\ Instrum.\ Meth.\ A {\bf 324}, 363 (1993).
  %%CITATION = NUIMA,A324,363;%%
 %62 citations counted in INSPIRE as of 03 Jul 2015

%\cite{Hirata:1989kn}
\bibitem{Hirata:1989kn}
K.~S.~Hirata {\it et al.} [Kamiokande-II Collaboration],
%``Experimental Limits on Nucleon Lifetime for Lepton + Meson Decay Modes,''
  Phys.\ Lett.\ B {\bf 220}, 308 (1989).
  %%CITATION = PHLTA,B220,308;%%
 %116 citations counted in INSPIRE as of 03 Jul 2015

\bibitem{kearns:isoup}
E.~Kearns, talk presented at the ISOUP Symposium (2013).

%\cite{Nishino:2012ipa}
\bibitem{Nishino:2012ipa}
H.~Nishino {\it et al.} [Super-Kamiokande Collaboration],
%``Search for Nucleon Decay into Charged Anti-lepton plus Meson in Super-Kamiokande I and II,''
  Phys.\ Rev.\ D {\bf 85}, 112001 (2012)
  [arXiv:1203.4030 [hep-ex]].
  %%CITATION = ARXIV:1203.4030;%%
 %66 citations counted in INSPIRE as of 03 Jul 2015

%\cite{Regis:2012sn}
\bibitem{Regis:2012sn}
C.~Regis {\it et al.} [Super-Kamiokande Collaboration],
%``Search for Proton Decay via $p -> \mu^+ K^0$ in Super-Kamiokande I, II, and III,''
  Phys.\ Rev.\ D {\bf 86}, 012006 (2012)
  [arXiv:1205.6538 [hep-ex]].
  %%CITATION = ARXIV:1205.6538;%%
 %27 citations counted in INSPIRE as of 03 Jul 2015

%\cite{Abe:2013lua}
\bibitem{Abe:2013lua}
K.~Abe {\it et al.} [Super-Kamiokande Collaboration],
%``Search for Nucleon Decay via $n \to \bar{\nu} \pi^{0}$ and $p \to \bar{\nu} \pi^{+}$ in Super-Kamiokande,''
  Phys.\ Rev.\ Lett.\ {\bf 113}, no. 12, 121802 (2014)
  [arXiv:1305.4391 [hep-ex]].
  %%CITATION = ARXIV:1305.4391;%%
 %14 citations counted in INSPIRE as of 03 Jul 2015

%\cite{Undagoitia:2005uu}
\bibitem{Undagoitia:2005uu}
T.~M.~Undagoitia, F.~von Feilitzsch, M.~Goger-Neff, C.~Grieb, K.~A.~Hochmuth, L.~Oberauer, W.~Potzel and M.~Wurm,
%``Search for the proton decay p ---> K+ anti-nu in the large liquid scintillator low energy neutrino astronomy detector LENA,''
  Phys.\ Rev.\ D {\bf 72}, 075014 (2005)
  [hep-ph/0511230].
  %%CITATION = HEP-PH/0511230;%%
 %40 citations counted in INSPIRE as of 03 Jul 2015

%\cite{Stefan:2008zi}
\bibitem{Stefan:2008zi}
D.~Stefan and A.~M.~Ankowski,
%``Nuclear effects in proton decay,''
  Acta Phys.\ Polon.\ B {\bf 40}, 671 (2009)
  [arXiv:0811.1892 [nucl-th]].
  %%CITATION = ARXIV:0811.1892;%%
 %6 citations counted in INSPIRE as of 03 Jul 2015

%\cite{Hayato:1999az}
\bibitem{Hayato:1999az}
Y.~Hayato {\it et al.} [Super-Kamiokande Collaboration],
%``Search for proton decay through p ---> anti-neutrino K+ in a large water Cherenkov detector,''
  Phys.\ Rev.\ Lett.\ {\bf 83}, 1529 (1999)
  [hep-ex/9904020].
  %%CITATION = HEP-EX/9904020;%%
 %149 citations counted in INSPIRE as of 03 Jul 2015

%\cite{Bueno:2007um}
\bibitem{Bueno:2007um}
A.~Bueno, Z.~Dai, Y.~Ge, M.~Laffranchi, A.~J.~Melgarejo, A.~Meregaglia, S.~Navas and A.~Rubbia,
%``Nucleon decay searches with large liquid argon TPC detectors at shallow depths: Atmospheric neutrinos and cosmogenic backgrounds,''
  JHEP {\bf 0704}, 041 (2007)
  [hep-ph/0701101].
  %%CITATION = HEP-PH/0701101;%%
 %56 citations counted in INSPIRE as of 03 Jul 2015

%\cite{Feldman:1997qc}
\bibitem{Feldman:1997qc}
G.~J.~Feldman and R.~D.~Cousins,
%``A Unified approach to the classical statistical analysis of small signals,''
  Phys.\ Rev.\ D {\bf 57}, 3873 (1998)
  [physics/9711021 [physics.data-an]].
  %%CITATION = PHYSICS/9711021;%%
 %2052 citations counted in INSPIRE as of 03 Jul 2015

%\end{thebibliography}

% chap: IndirectDarkMatter
%%%%%%%%%%%%%%%%%%%%%%%%%%%%%%%%%%%%%%%%%%%%%%%%%%%%%%%%%%%%%%%%%%%%%%%%%%%%%%%%%%%%%%%%
%%%%%%%%%%%%%%%%%%%%%%%%%%%%%%%%%%%%%%%%%%%%%%%%%%%%%%%%%%%%%%%%%%%%%%%%%%%%%%%%%%%%%%%%
% chap: IndirectDarkMatter

%\begin{thebibliography}{99}

%\cite{Rubin:1970zza}
\bibitem{Rubin:1970zza}
  V.~C.~Rubin and W.~K.~Ford, Jr.,
  %``Rotation of the Andromeda Nebula from a Spectroscopic Survey of Emission Regions,''
  Astrophys.\ J.\  {\bf 159}, 379 (1970).
  %%CITATION = ASJOA,159,379;%%
  %262 citations counted in INSPIRE as of 27 Jun 2014

%\cite{Zwicky:1933gu}
\bibitem{Zwicky:1933gu}
  F.~Zwicky,
  %``Die Rotverschiebung von extragalaktischen Nebeln,''
  Helv.\ Phys.\ Acta {\bf 6}, 110 (1933).
  %%CITATION = HPACA,6,110;%%
  %672 citations counted in INSPIRE as of 27 Jun 2014

%\cite{Zwicky:1937zza}
\bibitem{Zwicky:1937zza}
  F.~Zwicky,
  %``On the Masses of Nebulae and of Clusters of Nebulae,''
  Astrophys.\ J.\  {\bf 86}, 217 (1937).
  %%CITATION = ASJOA,86,217;%%
  %293 citations counted in INSPIRE as of 27 Jun 2014

%\cite{Hoekstra:2002nf}
\bibitem{Hoekstra:2002nf}
  H.~Hoekstra, H.~Yee and M.~Gladders,
  %``Current status of weak gravitational lensing,''
  New Astron.\ Rev.\  {\bf 46}, 767 (2002)
  [astro-ph/0205205].
  %%CITATION = ASTRO-PH/0205205;%%
  %53 citations counted in INSPIRE as of 27 Jun 2014


%\cite{Clowe:2006eq}
\bibitem{Clowe:2006eq}
  D.~Clowe, M.~Bradac, A.~H.~Gonzalez, M.~Markevitch, S.~W.~Randall, C.~Jones and
  D.~Zaritsky,
  %``A direct empirical proof of the existence of dark matter,''
  Astrophys.\ J.\  {\bf 648}, L109 (2006)
  [astro-ph/0608407].
  %%CITATION = ASTRO-PH/0608407;%%
  %672 citations counted in INSPIRE as of 27 Jun 2014


%\cite{Dietrich:2012mp}
\bibitem{Dietrich:2012mp}
  J.~P.~Dietrich, N.~Werner, D.~Clowe, A.~Finoguenov, T.~Kitching, L.~Miller and
  A.~Simionescu,
  %``A filament of dark matter between two clusters of galaxies,''
  Nature {\bf 487}, 202 (2012)
  [arXiv:1207.0809 [astro-ph.CO]].
  %%CITATION = ARXIV:1207.0809;%%
  %4 citations counted in INSPIRE as of 27 Jun 2014

%%\cite{Ade:2013zuv}
%\bibitem{Ade:2013zuv}
%  P.~A.~R.~Ade {\it et al.}  [Planck Collaboration],
%  %``Planck 2013 results. XVI. Cosmological parameters,''
%  Astron. Astrophys. {\bf 571}, A16 (2014)
%  [arXiv:1303.5076 [astro-ph.CO]].
%  %%CITATION = ARXIV:1303.5076;%%
%  %1995 citations counted in INSPIRE as of 27 Jun 2014

%\bibitem{Feng}
%J. L. Feng, Ann. Rev. Astron. Astrophys. {\bf 48}, 495 (2010)
%[arXiv:1003.0904 [atro-ph.CO]].

%\cite{Feng:2010gw}
\bibitem{Feng:2010gw}
  J.~L.~Feng,
  %``Dark Matter Candidates from Particle Physics and Methods of Detection,''
  Ann.\ Rev.\ Astron.\ Astrophys.\  {\bf 48}, 495 (2010)
  [arXiv:1003.0904 [astro-ph.CO]].
  %%CITATION = ARXIV:1003.0904;%%

%\cite{Bernabei:2010mq}
\bibitem{Bernabei:2010mq}
  R.~Bernabei {\it et al.}  [DAMA and LIBRA Collaborations],
  %``New results from DAMA/LIBRA,''
  Eur.\ Phys.\ J.\ C {\bf 67}, 39 (2010)
  [arXiv:1002.1028 [astro-ph.GA]].
  %%CITATION = ARXIV:1002.1028;%%
  %562 citations counted in INSPIRE as of 27 Jun 2014


%\cite{Aalseth:2010vx}
\bibitem{Aalseth:2010vx}
  C.~E.~Aalseth {\it et al.}  [CoGeNT Collaboration],
  %``Results from a Search for Light-Mass Dark Matter with a P-type Point Contact Germanium Detector,''
  Phys.\ Rev.\ Lett.\  {\bf 106}, 131301 (2011)
  [arXiv:1002.4703 [astro-ph.CO]].
  %%CITATION = ARXIV:1002.4703;%%
  %650 citations counted in INSPIRE as of 27 Jun 2014


%\cite{Aalseth:2011wp}
\bibitem{Aalseth:2011wp}
  C.~E.~Aalseth {\it et al.} [CoGeNT Collaboration],
  %``Search for an Annual Modulation in a P-type Point Contact Germanium Dark Matter Detector,''
  Phys.\ Rev.\ Lett.\  {\bf 107}, 141301 (2011)
  [arXiv:1106.0650 [astro-ph.CO]].
  %%CITATION = ARXIV:1106.0650;%%
  %370 citations counted in INSPIRE as of 27 Jun 2014

%\cite{Aalseth:2014eft}
\bibitem{Aalseth:2014eft}
  C.~E.~Aalseth {\it et al.}  [CoGeNT Collaboration],
  %``Search for An Annual Modulation in Three Years of CoGeNT Dark Matter Detector Data,''
  arXiv:1401.3295 [astro-ph.CO].
  %%CITATION = ARXIV:1401.3295;%%
  %27 citations counted in INSPIRE as of 27 Jun 2014

%\cite{Angloher:2011uu}
\bibitem{Angloher:2011uu}
  G.~Angloher {\it et al.},
  %``Results from 730 kg days of the CRESST-II Dark Matter Search,''
  Eur.\ Phys.\ J.\ C {\bf 72}, 1971 (2012)
  [arXiv:1109.0702 [astro-ph.CO]].
  %%CITATION = ARXIV:1109.0702;%%
  %422 citations counted in INSPIRE as of 27 Jun 2014

%\cite{Agnese:2013rvf}
\bibitem{Agnese:2013rvf}
  R.~Agnese {\it et al.}  [CDMS Collaboration],
  %``Silicon Detector Dark Matter Results from the Final Exposure of CDMS II,''
  Phys.\ Rev.\ Lett.\  {\bf 111}, 251301 (2013)
  [arXiv:1304.4279 [hep-ex]].
  %%CITATION = ARXIV:1304.4279;%%
  %225 citations counted in INSPIRE as of 27 Jun 2014


%\cite{Aprile:2012nq}
\bibitem{Aprile:2012nq}
  E.~Aprile {\it et al.}  [XENON100 Collaboration],
  %``Dark Matter Results from 225 Live Days of XENON100 Data,''
  Phys.\ Rev.\ Lett.\  {\bf 109}, 181301 (2012)
  [arXiv:1207.5988 [astro-ph.CO]].
  %%CITATION = ARXIV:1207.5988;%%
  %613 citations counted in INSPIRE as of 27 Jun 2014


%\cite{Aprile:2013doa}
\bibitem{Aprile:2013doa}
  E.~Aprile {\it et al.}  [XENON100 Collaboration],
  %``Limits on spin-dependent WIMP-nucleon cross sections from 225 live days of XENON100 data,''
  Phys.\ Rev.\ Lett.\  {\bf 111}, 021301 (2013)
  [arXiv:1301.6620 [astro-ph.CO]].
  %%CITATION = ARXIV:1301.6620;%%
  %71 citations counted in INSPIRE as of 27 Jun 2014


%\cite{Akerib:2013tjd}
\bibitem{Akerib:2013tjd}
  D.~S.~Akerib {\it et al.}  [LUX Collaboration],
  %``First results from the LUX dark matter experiment at the Sanford Underground Research Facility,''
  Phys.\ Rev.\ Lett.\  {\bf 112}, 091303 (2014)
  [arXiv:1310.8214 [astro-ph.CO]].
  %%CITATION = ARXIV:1310.8214;%%
  %287 citations counted in INSPIRE as of 27 Jun 2014


%\cite{Agnese:2014aze}
\bibitem{Agnese:2014aze}
  R.~Agnese {\it et al.}  [SuperCDMS Collaboration],
  %``Search for Low-Mass WIMPs with SuperCDMS,''
  Phys.\ Rev.\ Lett.\  {\bf 112}, 241302 (2014)
  [arXiv:1402.7137 [hep-ex]].
  %%CITATION = ARXIV:1402.7137;%%
  %34 citations counted in INSPIRE as of 27 Jun 2014


%\cite{Felizardo:2011uw}
\bibitem{Felizardo:2011uw}
  M.~Felizardo {\it et al.} (SIMPLE Collaboration),
  %``Final Analysis and Results of the Phase II SIMPLE Dark Matter Search,''
  Phys.\ Rev.\ Lett.\  {\bf 108}, 201302 (2012)
  [arXiv:1106.3014 [astro-ph.CO]].
  %%CITATION = ARXIV:1106.3014;%%
  %162 citations counted in INSPIRE as of 27 Jun 2014


%\cite{Yue:2014qdu}
\bibitem{Yue:2014qdu}
  Q.~Yue {\it et al.}  [CDEX Collaboration],
  %``Limits on light WIMPs from the CDEX-1 experiment with a p-type point-contact germanium detector at the China Jingping Underground Laboratory,''
  Phys. Rev. D {\bf 90}, 091701 (2014)
  [arXiv:1404.4946 [hep-ex]].
  %%CITATION = ARXIV:1404.4946;%%
  %2 citations counted in INSPIRE as of 27 Jun 2014


%\cite{Adriani:2008zr}
\bibitem{Adriani:2008zr}
  O.~Adriani {\it et al.}  [PAMELA Collaboration],
  %``An anomalous positron abundance in cosmic rays with energies 1.5-100 GeV,''
  Nature {\bf 458}, 607 (2009)
  [arXiv:0810.4995 [astro-ph]].
  %%CITATION = ARXIV:0810.4995;%%
  %1267 citations counted in INSPIRE as of 27 Jun 2014


%\cite{Chang:2008aa}
\bibitem{Chang:2008aa}
  J.~Chang {\it et al.},
  %``An excess of cosmic ray electrons at energies of 300-800 GeV,''
  Nature {\bf 456}, 362 (2008).
  %%CITATION = NATUA,456,362;%%
  %615 citations counted in INSPIRE as of 27 Jun 2014


%\cite{Abdo:2009zk}
\bibitem{Abdo:2009zk}
  A.~A.~Abdo {\it et al.}  [Fermi LAT Collaboration],
  %``Measurement of the Cosmic Ray e+ plus e- spectrum from 20 GeV to 1 TeV with the Fermi Large Area Telescope,''
  Phys.\ Rev.\ Lett.\  {\bf 102}, 181101 (2009)
  [arXiv:0905.0025 [astro-ph.HE]].
  %%CITATION = ARXIV:0905.0025;%%
  %728 citations counted in INSPIRE as of 27 Jun 2014


%\cite{Aguilar:2013qda}
\bibitem{Aguilar:2013qda}
  M.~Aguilar {\it et al.}  [AMS Collaboration],
  %``First Result from the Alpha Magnetic Spectrometer on the International Space Station: Precision Measurement of the Positron Fraction in Primary Cosmic Rays of 0.5?350 GeV,''
  Phys.\ Rev.\ Lett.\  {\bf 110}, 141102 (2013).
  %%CITATION = PRLTA,110,141102;%%
  %217 citations counted in INSPIRE as of 27 Jun 2014


\bibitem{Aartsen:2012kia}
 M.~G.~Aartsen {\it et al.}  [IceCube Collaboration],
 %``Search for dark matter annihilations in the Sun with the 79-string IceCube detector,''
 Phys.\ Rev.\ Lett.\  {\bf 110}, 131302 (2013).
% [arXiv:1212.4097 [astro-ph.HE]].
 %%CITATION = ARXIV:1212.4097;%%
%
%\bibitem{Capozzi:2013csa}
%  F.~Capozzi, G.~L.~Fogli, E.~Lisi, A.~Marrone, D.~Montanino and A.~Palazzo,
%  %``Status of three-neutrino oscillation parameters, circa 2013,''
%  Phys.\ Rev.\ D {\bf 89}, no. 9, 093018 (2014)
%  [arXiv:1312.2878 [hep-ph]].
%  %%CITATION = ARXIV:1312.2878;%%
%
\bibitem{Krauss:1985aaa}
  L.~M.~Krauss, M.~Srednicki and F.~Wilczek,
  %``Solar System Constraints and Signatures for Dark Matter Candidates,''
  Phys.\ Rev.\ D {\bf 33}, 2079 (1986).
  %%CITATION = PHRVA,D33,2079;%%

\bibitem{Griest:1986yu}
  K.~Griest and D.~Seckel,
  %``Cosmic Asymmetry, Neutrinos and the Sun,''
  Nucl.\ Phys.\ B {\bf 283}, 681 (1987)
  [Erratum-ibid.\ B {\bf 296}, 1034 (1988)].
  %%CITATION = NUPHA,B283,681;%%

 %\cite{Nauenberg:1986em}
\bibitem{Nauenberg:1986em}
  M.~Nauenberg,
  %``Energy Transport And Evaporation Of Weakly Interacting Particles In The Sun,''
  Phys.\ Rev.\ D {\bf 36}, 1080 (1987).
  %%CITATION = PHRVA,D36,1080;%%
%
\bibitem{Gould:1987ju}
  A.~Gould,
  %``{WIMP} Distribution in and Evaporation From the Sun,''
  Astrophys.\ J.\  {\bf 321}, 560 (1987).
  %%CITATION = ASJOA,321,560;%%
  %104 citations counted in INSPIRE as of 20 Apr 2014
 %
\bibitem{Chen:2014oaa}
  C.~S.~Chen, F.~F.~Lee, G.~L.~Lin and Y.~H.~Lin,
  %``Probing Dark Matter Self-Interaction in the Sun with IceCube-PINGU,''
  JCAP {\bf 1410}, 049 (2014)
  [arXiv:1408.5471 [hep-ph]].
  %%CITATION = ARXIV:1408.5471;%%
%
\bibitem{Jungman:1995df}
  G.~Jungman, M.~Kamionkowski and K.~Griest,
  %``Supersymmetric dark matter,''
  Phys.\ Rept.\  {\bf 267}, 195 (1996)
  [hep-ph/9506380].
  %%CITATION = HEP-PH/9506380;%%
  %
\bibitem{Bertone:2004pz}
  G.~Bertone, D.~Hooper and J.~Silk,
  %``Particle dark matter: Evidence, candidates and constraints,''
  Phys.\ Rept.\  {\bf 405}, 279 (2005)
  [hep-ph/0404175].
  %%CITATION = HEP-PH/0404175;%%
  %
\bibitem{Blennow:2007tw}
M.~Blennow, J.~Edsjo and T.~Ohlsson,
%``Neutrinos from WIMP annihilations using a full three-flavor Monte Carlo,''
JCAP {\bf 0801}, 021 (2008).
%[arXiv:0709.3898 [hep-ph]].
%%CITATION = ARXIV:0709.3898;%%
%

%\bibitem{Andreopoulos:2009rq}
%  C.~Andreopoulos, A.~Bell, D.~Bhattacharya, F.~Cavanna, J.~Dobson, S.~Dytman, H.~Gallagher and P.~Guzowski {\it et al.},
%  %``The GENIE Neutrino Monte Carlo Generator,''
%  Nucl.\ Instrum.\ Meth.\ A {\bf 614}, 87 (2010)
%  [arXiv:0905.2517 [hep-ph]].
%  %%CITATION = ARXIV:0905.2517;%%
%
\bibitem{Aartsen:2012uu}
M.~G.~Aartsen {\it et al.} [IceCube Collaboration],
%``Measurement of the Atmospheric $\nu_e$ flux in IceCube,''
Phys.\ Rev.\ Lett.\  {\bf 110}, 151105 (2013).
%[arXiv:1212.4760 [hep-ex]].
%%CITATION = ARXIV:1212.4760;%%
%
\bibitem{Honda:2015fha}
  M.~Honda, M.~S.~Athar, T.~Kajita, K.~Kasahara and S.~Midorikawa,
  %``Atmospheric neutrino flux calculation using the NRLMSISE00 atmospheric model,''
  arXiv:1502.03916 [astro-ph.HE].
  %%CITATION = ARXIV:1502.03916;%%

\bibitem{Behnke:2012ys}
  E.~Behnke {\it et al.}  [COUPP Collaboration],
  %``First Dark Matter Search Results from a 4-kg CF$_3$I Bubble Chamber Operated in a Deep Underground Site,''
  Phys.\ Rev.\ D {\bf 86}, 052001 (2012)
  [arXiv:1204.3094 [astro-ph.CO]].
  %%CITATION = ARXIV:1204.3094;%%

\bibitem{Archambault:2012pm}
  S.~Archambault {\it et al.}  [PICASSO Collaboration],
  %``Constraints on Low-Mass WIMP Interactions on $^{19}F$ from PICASSO,''
  Phys.\ Lett.\ B {\bf 711}, 153 (2012)
  [arXiv:1202.1240 [hep-ex]].
  %%CITATION = ARXIV:1202.1240;%%  %%CITATION = ARXIV:1106.3014;%%

\bibitem{Agnese:2013jaa}
  R.~Agnese {\it et al.}  [SuperCDMS Collaboration],
  %``Search for Low-Mass Weakly Interacting Massive Particles Using Voltage-Assisted Calorimetric Ionization Detection in the SuperCDMS Experiment,''
  Phys.\ Rev.\ Lett.\  {\bf 112}, 041302 (2014)
  [arXiv:1309.3259 [physics.ins-det]].
  %%CITATION = ARXIV:1309.3259;%%
%\end{thebibliography}


% chap: OtherExotic
%%%%%%%%%%%%%%%%%%%%%%%%%%%%%%%%%%%%%%%%%%%%%%%%%%%%%%%%%%%%%%%%%%%%%%%%%%%%%%%%%%%%%%%%
%%%%%%%%%%%%%%%%%%%%%%%%%%%%%%%%%%%%%%%%%%%%%%%%%%%%%%%%%%%%%%%%%%%%%%%%%%%%%%%%%%%%%%%%
% chap: OtherExotic

%
%begin{thebibliography}{99}
%%\cite{Agashe:2014kda}
%\bibitem{Agashe:2014kda}
%K.~A.~Olive {\it et al.} [Particle Data Group Collaboration],
%%``Review of Particle Physics,''
%  Chin.\ Phys.\ C {\bf 38}, 090001 (2014).
%  %%CITATION = CHPHD,C38,090001;%%
% %1371 citations counted in INSPIRE as of 03 Jul 2015

%\cite{Antusch:2008tz}
\bibitem{Antusch:2008tz}
S.~Antusch, J.~P.~Baumann and E.~Fernandez-Martinez,
%``Non-Standard Neutrino Interactions with Matter from Physics Beyond the Standard Model,''
  Nucl.\ Phys.\ B {\bf 810}, 369 (2009)
  [arXiv:0807.1003 [hep-ph]].
  %%CITATION = ARXIV:0807.1003;%%
 %143 citations counted in INSPIRE as of 03 Jul 2015

%%\cite{Ohlsson:2012kf}
%\bibitem{Ohlsson:2012kf}
%T.~Ohlsson,
%%``Status of non-standard neutrino interactions,''
%  Rept.\ Prog.\ Phys.\ {\bf 76}, 044201 (2013)
%  [arXiv:1209.2710 [hep-ph]].
%  %%CITATION = ARXIV:1209.2710;%%
% %39 citations counted in INSPIRE as of 03 Jul 2015

%\cite{Severijns:2006dr}
\bibitem{Severijns:2006dr}
N.~Severijns, M.~Beck and O.~Naviliat-Cuncic,
%``Tests of the standard electroweak model in beta decay,''
  Rev.\ Mod.\ Phys.\ {\bf 78}, 991 (2006)
  [nucl-ex/0605029].
  %%CITATION = NUCL-EX/0605029;%%
 %154 citations counted in INSPIRE as of 03 Jul 2015

%\cite{Ohlsson:2013nna}
\bibitem{Ohlsson:2013nna}
T.~Ohlsson, H.~Zhang and S.~Zhou,
%``Nonstandard interaction effects on neutrino parameters at medium-baseline reactor antineutrino experiments,''
  Phys.\ Lett.\ B {\bf 728}, 148 (2014)
  [arXiv:1310.5917 [hep-ph]].
  %%CITATION = ARXIV:1310.5917;%%
 %10 citations counted in INSPIRE as of 03 Jul 2015

%\cite{Li:2014mlo}
\bibitem{Li:2014mlo}
Y.~F.~Li and Y.~L.~Zhou,
%``Shifts of neutrino oscillation parameters in reactor antineutrino experiments with non-standard interactions,''
  Nucl.\ Phys.\ B {\bf 888}, 137 (2014)
  [arXiv:1408.6301 [hep-ph]].
  %%CITATION = ARXIV:1408.6301;%%
 %3 citations counted in INSPIRE as of 03 Jul 2015

%\cite{Kostelecky:1988zi}
\bibitem{Kostelecky:1988zi}
V.~A.~Kostelecky and S.~Samuel,
%``Spontaneous Breaking of Lorentz Symmetry in String Theory,''
  Phys.\ Rev.\ D {\bf 39}, 683 (1989).
  %%CITATION = PHRVA,D39,683;%%
 %749 citations counted in INSPIRE as of 03 Jul 2015

%\cite{Colladay:1996iz}
\bibitem{Colladay:1996iz}
D.~Colladay and V.~A.~Kostelecky,
%``CPT violation and the standard model,''
  Phys.\ Rev.\ D {\bf 55}, 6760 (1997)
  [hep-ph/9703464].
  %%CITATION = HEP-PH/9703464;%%
 %1024 citations counted in INSPIRE as of 03 Jul 2015

%\cite{Colladay:1998fq}
\bibitem{Colladay:1998fq}
D.~Colladay and V.~A.~Kostelecky,
%``Lorentz violating extension of the standard model,''
  Phys.\ Rev.\ D {\bf 58}, 116002 (1998)
  [hep-ph/9809521].
  %%CITATION = HEP-PH/9809521;%%
 %1272 citations counted in INSPIRE as of 03 Jul 2015

%\cite{Li:2014rya}
\bibitem{Li:2014rya}
Y.~F.~Li and Z.~h.~Zhao,
%``Tests of Lorentz and CPT Violation in the Medium Baseline Reactor Antineutrino Experiment,''
  Phys.\ Rev.\ D {\bf 90}, no. 11, 113014 (2014)
  [arXiv:1409.6970 [hep-ph]].
  %%CITATION = ARXIV:1409.6970;%%
 %1 citations counted in INSPIRE as of 03 Jul 2015

%%\cite{An:2015rpe}
%\bibitem{An:2015rpe}
%F.~P.~An {\it et al.} [Daya Bay Collaboration],
%%``A new measurement of antineutrino oscillation with the full detector configuration at Daya Bay,''
%  arXiv:1505.03456 [hep-ex].
%  %%CITATION = ARXIV:1505.03456;%%
% %3 citations counted in INSPIRE as of 03 Jul 2015

%%\cite{Gando:2013nba}
%\bibitem{Gando:2013nba}
%A.~Gando {\it et al.} [KamLAND Collaboration],
%%``Reactor On-Off Antineutrino Measurement with KamLAND,''
%  Phys.\ Rev.\ D {\bf 88}, no. 3, 033001 (2013)
%  [arXiv:1303.4667 [hep-ex]].
%  %%CITATION = ARXIV:1303.4667;%%
% %57 citations counted in INSPIRE as of 03 Jul 2015

%\cite{Kostelecky:2001mb}
\bibitem{Kostelecky:2001mb}
V.~A.~Kostelecky and M.~Mewes,
%``Cosmological constraints on Lorentz violation in electrodynamics,''
  Phys.\ Rev.\ Lett.\ {\bf 87}, 251304 (2001)
  [hep-ph/0111026].
  %%CITATION = HEP-PH/0111026;%%
 %307 citations counted in INSPIRE as of 03 Jul 2015

%\cite{Kostelecky:2002hh}
\bibitem{Kostelecky:2002hh}
V.~A.~Kostelecky and M.~Mewes,
%``Signals for Lorentz violation in electrodynamics,''
  Phys.\ Rev.\ D {\bf 66}, 056005 (2002)
  [hep-ph/0205211].
  %%CITATION = HEP-PH/0205211;%%
 %428 citations counted in INSPIRE as of 03 Jul 2015

%\cite{Kostelecky:2003cr}
\bibitem{Kostelecky:2003cr}
V.~A.~Kostelecky and M.~Mewes,
%``Lorentz and CPT violation in neutrinos,''
  Phys.\ Rev.\ D {\bf 69}, 016005 (2004)
  [hep-ph/0309025].
  %%CITATION = HEP-PH/0309025;%%
 %270 citations counted in INSPIRE as of 03 Jul 2015

%\cite{Diaz:2009qk}
\bibitem{Diaz:2009qk}
J.~S.~Diaz, V.~A.~Kostelecky and M.~Mewes,
%``Perturbative Lorentz and CPT violation for neutrino and antineutrino oscillations,''
  Phys.\ Rev.\ D {\bf 80}, 076007 (2009)
  [arXiv:0908.1401 [hep-ph]].
  %%CITATION = ARXIV:0908.1401;%%
 %80 citations counted in INSPIRE as of 03 Jul 2015

%%\cite{Giunti:2015gga}
%\bibitem{Giunti:2015gga}
%C.~Giunti, K.~A.~Kouzakov, Y.~F.~Li, A.~V.~Lokhov, A.~I.~Studenikin and S.~Zhou,
%%``Electromagnetic neutrinos in terrestrial experiments and astrophysics,''
%  arXiv:1506.05387 [hep-ph].
%  %%CITATION = ARXIV:1506.05387;%%

%\cite{deGouvea:2012hg}
\bibitem{deGouvea:2012hg}
A.~de Gouvea and S.~Shalgar,
%``Effect of Transition Magnetic Moments on Collective Supernova Neutrino Oscillations,''
  JCAP {\bf 1210}, 027 (2012)
  [arXiv:1207.0516 [astro-ph.HE]].
  %%CITATION = ARXIV:1207.0516;%%
 %20 citations counted in INSPIRE as of 03 Jul 2015

%\cite{deGouvea:2013zp}
\bibitem{deGouvea:2013zp}
A.~de Gouvea and S.~Shalgar,
%``Transition Magnetic Moments and Collective Neutrino Oscillations:Three-Flavor Effects and Detectability,''
  JCAP {\bf 1304}, 018 (2013)
  [arXiv:1301.5637 [astro-ph.HE]].
  %%CITATION = ARXIV:1301.5637;%%
 %14 citations counted in INSPIRE as of 03 Jul 2015

%\end{thebibliography}

% chap: Appendix
%%%%%%%%%%%%%%%%%%%%%%%%%%%%%%%%%%%%%%%%%%%%%%%%%%%%%%%%%%%%%%%%%%%%%%%%%%%%%%%%%%%%%%%%
%%%%%%%%%%%%%%%%%%%%%%%%%%%%%%%%%%%%%%%%%%%%%%%%%%%%%%%%%%%%%%%%%%%%%%%%%%%%%%%%%%%%%%%%
% chap: Appendix

%\begin{thebibliography}{99}

%\bibitem{ILL-F}
%F.~von~Feilitzsch {\it et al.}, Phys.\ Lett.\ B {\bf 118}, 162 (1982)
%
%\bibitem{ILL-S} K.~Schreckenbach {\it et al.}, Phys.\ Lett.\ B {\bf 160}, 325 (1985)
%
%\bibitem{ILL-H} A.~A.~Hahn {\it et al.}, Phys.\ Lett.\ B {\bf 218}, 365 (1989).
%
%\bibitem{Huber}
%P.~Huber, Phys.\ Rev.\ C {\bf 84}, 024617 (2011).
%
%\bibitem{Mueller}
%T.~Mueller {\it et al.}, Phys.\ Rev.\ C {\bf 83}, 054615 (2011).
%
%\bibitem{Vogel}
%P.~Vogel, G. K. Schenter, F. M. Mann and R. E. Schenter, {\it et al.}, Phys.\ Rev.\ C{\bf 24}, 1543 (1981); P. Vogel, Phys. Rev. D{\bf 29}, 1918 (1984).
%
%\bibitem{Dan}
%D. A. ~Dwyer, T. J. ~Langford, arXiv: 1407.1281


%%\cite{VonFeilitzsch:1982jw}
%\bibitem{VonFeilitzsch:1982jw}
%  F.~Von Feilitzsch, A.~A.~Hahn and K.~Schreckenbach,
%  %``Experimental Beta Spectra From Pu-239 And U-235 Thermal Neutron Fission Products And Their Correlated Anti-neutrinos Spectra,''
%  Phys.\ Lett.\ B {\bf 118}, 162 (1982).  %%CITATION = PHLTA,B118,162;%%
%
%  %\cite{Schreckenbach:1985ep}
%\bibitem{Schreckenbach:1985ep}
%  K.~Schreckenbach, G.~Colvin, W.~Gelletly and F.~Von Feilitzsch,
%  %``Determination Of The Anti-neutrino Spectrum From U-235 Thermal Neutron Fission Products Up To 9.5-mev,''
%  Phys.\ Lett.\ B {\bf 160}, 325 (1985).  %%CITATION = PHLTA,B160,325;%%
%
%%\cite{Hahn:1989zr}
%\bibitem{Hahn:1989zr}
%  A.~A.~Hahn, K.~Schreckenbach, G.~Colvin, B.~Krusche, W.~Gelletly and F.~Von Feilitzsch,
%  %``Anti-neutrino Spectra From $^{241}$Pu and $^{239}$Pu Thermal Neutron Fission Products,''
%  Phys.\ Lett.\ B {\bf 218}, 365 (1989).  %%CITATION = PHLTA,B218,365;%%
%
%%\cite{Huber:2011wv}
%\bibitem{Huber:2011wv}
%  P.~Huber,
%  %``On the determination of anti-neutrino spectra from nuclear reactors,''
%  Phys.\ Rev.\ C {\bf 84}, 024617 (2011)
%  [Phys.\ Rev.\ C {\bf 85}, 029901 (2012)]  [arXiv:1106.0687 [hep-ph]].  %%CITATION = ARXIV:1106.0687;%%
%
%
%%\cite{Mueller:2011nm}
%\bibitem{Mueller:2011nm}
%  T.~A.~Mueller, D.~Lhuillier, M.~Fallot, A.~Letourneau, S.~Cormon, M.~Fechner, L.~Giot and T.~Lasserre {\it et al.},
%  %``Improved Predictions of Reactor Antineutrino Spectra,''
%  Phys.\ Rev.\ C {\bf 83}, 054615 (2011)  [arXiv:1101.2663 [hep-ex]].  %%CITATION = ARXIV:1101.2663;%%
%
%
%%\cite{Vogel:1980bk}
%\bibitem{Vogel:1980bk}
%  P.~Vogel, G.~K.~Schenter, F.~M.~Mann and R.~E.~Schenter,
%  %``Reactor Anti-neutrino Spectra and Their Application to Anti-neutrino Induced Reactions. 2.,''
%  Phys.\ Rev.\ C {\bf 24}, 1543 (1981).  %%CITATION = PHRVA,C24,1543;%%
%
%%\cite{Dwyer:2014eka}
%\bibitem{Dwyer:2014eka}
%  D.~A.~Dwyer and T.~J.~Langford,
%  %``Spectral Structure of Electron Antineutrinos from Nuclear Reactors,''
%  Phys.\ Rev.\ Lett.\  {\bf 114}, no. 1, 012502 (2015)
%  [arXiv:1407.1281 [nucl-ex]].  %%CITATION = ARXIV:1407.1281;%%
%

%\bibitem{caojPower}
%J. ~Cao, Nuclear Physics B (Proc. Suppl.) 229-232 (2012) 205-209 [arXiv:11012266]

%%\cite{Cao:2011gb}
%\bibitem{Cao:2011gb}
%  J.~Cao,
%  %``Determining Reactor Neutrino Flux,''
%  Nucl.\ Phys.\ Proc.\ Suppl.\  {\bf 229-232}, 205 (2012)
%  [arXiv:1101.2266 [hep-ex]].
%  %%CITATION = ARXIV:1101.2266;%%

%\bibitem{DYBpub} F. P. An {\it et al.} (Daya Bay Collaboration),
%Phys. Rev. Lett. {\bf 108}, 171803 (2012); Chin. Phys. C{\bf 37},
%011001 (2013); Phys. Rev. Lett. {\bf 112}, 061801 (2014).

%\bibitem{Bugey3}
%B. Achkar {\it et al.}, Phys. Lett. B {bf 374}, 243 (1996).

%\cite{Achkar:1996rd}
\bibitem{Achkar:1996rd}
  B.~Achkar, R.~Aleksan, M.~Avenier, G.~Bagieu, J.~Bouchez, R.~Brissot, J.~F.~Cavaignac and J.~Collot {\it et al.},
  %``Comparison of anti-neutrino reactor spectrum models with the Bugey-3 measurements,''
  Phys.\ Lett.\ B {\bf 374}, 243 (1996).
  %%CITATION = PHLTA,B374,243;%%

%\bibitem{Ranomaly} G. Mention {\it et al.}, Phys. Rev. D{\bf 83}, 073006 (2011).

%\bibitem{IBDCS}
%P.~Vogel and J.~F.~Beacom, Phys.\ Rev.\ D {\bf 60}, 053003 (1999).

\bibitem{kme}
{\it Imrpoving Pressurized Water Reactor Performance Through Instrumentation:
Application Case of Reducing Uncertainties on Thermal Power},
EPRI report prepared by Electricite de France, 2001;
{\it Application of Orifice Plates for Measurement of Feedwater Flow},
EPRI report prepared by Electricite de France, 2001.

\bibitem{apollo2}
R.~Sanchez {\it et al.}, Nuclear Engineering and Technology {\bf 42}, 474 (2010).

\bibitem{dragon}
G.~Marleau, A.~Hebert and R.~Roy, A User Guide for DRAGON. Report {\rm IGE-174} Rev.~5 (2000).

%\bibitem{kopeikin_ei}
%V.~Kopeikin, L.~Mikaelyan, and V.~Sinev, Phys. \ Atom.\ Nucl. {\bf 67}, 1892 (2004).

%\cite{Kopeikin:2004cn}
\bibitem{Kopeikin:2004cn}
  V.~Kopeikin, L.~Mikaelyan and V.~Sinev,
  %``Reactor as a source of antineutrinos: Thermal fission energy,''
  Phys.\ Atom.\ Nucl.\  {\bf 67}, 1892 (2004)
  [Yad.\ Fiz.\  {\bf 67}, 1916 (2004)]
  [hep-ph/0410100].
  %%CITATION = HEP-PH/0410100;%%


%\bibitem{fissfraction}
%X. B.~Ma, L. Z.~Wang, Y. X.~Chen  {\it et al.} arXiv:1405.6807

%\cite{Ma:2014bpa}
\bibitem{Ma:2014bpa}
  X.~B.~Ma, L.~Z.~Wang, Y.~X.~Chen, W.~L.~Zhong and F.~P.~An,
  %``Uncertainties analysis of fission fraction for reactor antineutrino experiments,''
  arXiv:1405.6807 [nucl-ex].
  %%CITATION = ARXIV:1405.6807;%%

%\bibitem{offEfp}
%T.~Mueller {\it et al.}, Phys. Rev. C {\bf 83}, 054615 (2011).

%\bibitem{snfkopeikin}
%V.~I.~Kopeikin, L.~Mikaelyan, V.~Sinev, Phys. Atom. Nucl. {\bf 64}, 849 (2001);
%{\it ibid} {\bf 66}, 500 (2003); {\it ibid} {\bf 69}, 185 (2006).

%\cite{Kopeikin:2001qv}
\bibitem{Kopeikin:2001qv}
  V.~I.~Kopeikin, L.~A.~Mikaelyan and V.~V.~Sinev,
  %``Inverse beta decay in a nonequilibrium anti-neutrino flux from a nuclear reactor,''
  Phys.\ Atom.\ Nucl.\  {\bf 64}, 849 (2001)
  [Yad.\ Fiz.\  {\bf 64}, 914 (2001)]
  [hep-ph/0110290].
  %%CITATION = HEP-PH/0110290;%%

%\cite{Afanasev:2003ci}
\bibitem{Afanasev:2003ci}
  V.~N.~Afanasev, V.~V.~Barmin, A.~A.~Burenkov, V.~I.~Demekhin, A.~G.~Dolgolenko, A.~S.~Gerasimov, V.~S.~Goryachev and Y.~I.~Povarov {\it et al.},
  %``Observation of primary and secondary (proportional) scintillation in two-phase xenon detector,''
  Phys.\ Atom.\ Nucl.\  {\bf 66}, 500 (2003)
  [Yad.\ Fiz.\  {\bf 66}, 527 (2003)].
  %%CITATION = PANUE,66,500;%%

%\cite{Kopeikin:2004fd}
\bibitem{Kopeikin:2004fd}
  V.~Kopeikin, L.~Mikaelyan and V.~Sinev,
  %``Antineutrino background from spent fuel storage in sensitive searches for theta(13) at reactors,''
  Phys.\ Atom.\ Nucl.\  {\bf 69}, 185 (2006)
  [hep-ph/0412044].
  %%CITATION = HEP-PH/0412044;%%
%
%\bibitem{snfihep}
%F. P.~An, X. C.~Tian, L.`Zhan, and J.~Cao, Chinese Phys. C {\bf 33}, 711 (2009).

%\cite{An:2009zz}
\bibitem{An:2009zz}
  F.~P.~An, X.~C.~Tian, L.~Zhan and J.~Cao,
  %``Systematic impact of spent nuclear fuel on Theta(13) sensitivity at reactor neutrino experiment,''
  Chin.\ Phys.\ C {\bf 33}, 711 (2009).
  %%CITATION = CHPHD,C33,711;%%

%\bibitem{snfciae}
%B. Zhou, X. C. Ruan, Y. B. Nie, Z.Y. Zhou, F. P. An, and J.~Cao, Chinese Phys. C {\bf 36}, 1 (2012).

%%\cite{Zhou:2012zzc}
%\bibitem{Zhou:2012zzc}
%  B.~Zhou, X.~C.~Ruan, Y.~B.~Nie, Z.~Y.~Zhou, F.~P.~An and J.~Cao,
%  %``A study of antineutrino spectra from spent nuclear fuel at Daya Bay,''
%  Chin.\ Phys.\ C {\bf 36}, 1 (2012).
%  %%CITATION = CHPHD,C36,1;%%

%\bibitem{ichepdyb}
%W. L.~Zhong, for the Daya Bay collaboration, talk at ICHEP 2014.
%
%\bibitem{DC1412}
%Y. Abe {\it et al.} (Double Chooz Collaboration), JHEP {\bf 1410}, 086 (2014).
%
%\bibitem{RENOtalk}
%S. H.~Seo, for the RENO Collaboration, talk at Neutrino 2014.

%%\cite{DYBbump}
%\bibitem{DYBbump}
%W.~L.~Zhong, for Daya Bay collaboration, talk at ICHEP 2014. See also
%%\cite{Zhan:2015aha}
%%\bibitem{Zhan:2015aha}
%  L.~Zhan [Daya Bay Collaboration],
%  %``Recent Results from Daya Bay,''
%  arXiv:1506.01149 [hep-ex].
%
%%\cite{Seon-HeeSeofortheRENO:2014jza}
%\bibitem{Seon-HeeSeofortheRENO:2014jza}
%  S.~H.~Seo [RENO Collaboration],
%  %``New Results from RENO and The 5 MeV Excess,''
%  arXiv:1410.7987 [hep-ex].  %%CITATION = ARXIV:1410.7987;%%
%
%%\cite{Abe:2014bwa}
%\bibitem{Abe:2014bwa}
%  Y.~Abe {\it et al.}  [Double Chooz Collaboration],
%  %``Improved measurements of the neutrino mixing angle $\theta_{13}$ with the Double Chooz detector,''
%  JHEP {\bf 1410}, 086 (2014)  [JHEP {\bf 1502}, 074 (2015)]
%  [arXiv:1406.7763 [hep-ex]].  %%CITATION = ARXIV:1406.7763;%%

\bibitem{Hayes}
A. C.~Hayes, J. L.~Friar, G. T.~Garvey, G.~Jungman, G.~Jonkmans, Phys. Rev. Lett. {bf 112}, 202501 (2014).

%%\cite{Hayes:2013wra}
%\bibitem{Hayes:2013wra}
%A.~C.~Hayes, J.~L.~Friar, G.~T.~Garvey, G.~Jungman and G.~Jonkmans,
%%``Systematic Uncertainties in the Analysis of the Reactor Neutrino Anomaly,''
%  Phys.\ Rev.\ Lett.\ {\bf 112}, 202501 (2014)
%  [arXiv:1309.4146 [nucl-th]].
%  %%CITATION = ARXIV:1309.4146;%%
% %42 citations counted in INSPIRE as of 03 Jul 2015

%\bibitem{nufact14anfp}
%F. P. ~An, for the Daya Bay Collaboration, talk at NuFact 2014.

%\bibitem{Geant4}
%S. Agostinelli, {\em et al.}, Nucl. Instr. and Meth. A {\bf 506} (2003) 250;
%J. Allison, {\em et al.}, IEEE Trans. Nucl. Sci. NS-53(1) (2006) 270.

%%\cite{Agostinelli:2002hh}
%\bibitem{Agostinelli:2002hh}
%  S.~Agostinelli {\it et al.}  [GEANT4 Collaboration],
%  %``GEANT4: A Simulation toolkit,''
%  Nucl.\ Instrum.\ Meth.\ A {\bf 506}, 250 (2003).  %%CITATION = NUIMA,A506,250;%%

%\cite{Allison:2006ve}
\bibitem{Allison:2006ve}
  J.~Allison, K.~Amako, J.~Apostolakis, H.~Araujo, P.~A.~Dubois, M.~Asai, G.~Barrand and R.~Capra {\it et al.},
  %``Geant4 developments and applications,''
  IEEE Trans.\ Nucl.\ Sci.\  {\bf 53}, 270 (2006).
  %%CITATION = IETNA,53,270;%%

%\bibitem{MUSIC}
%P. Antonioli {\it et al.}, Astro. Phys. 7, 357 (1997).

%\cite{Antonioli:1997qw}
\bibitem{Antonioli:1997qw}
  P.~Antonioli, C.~Ghetti, E.~V.~Korolkova, V.~A.~Kudryavtsev and G.~Sartorelli,
  %``A Three-dimensional code for muon propagation through the rock: Music,''
  Astropart.\ Phys.\  {\bf 7}, 357 (1997)
  [hep-ph/9705408].
  %%CITATION = HEP-PH/9705408;%%

\bibitem{ENDF}
See the website:~http://www-nds.iaea.org/exfor/endf.htm

%\bibitem{soeren}
%W.~Q.~Jiang {\em et al.}, Chinese Physics C {\bf 36}, (2012) 235-240;
%S.~Jetter {\em et al.}, Chinese Physics C {\bf 36}, (2012) 733-741.

%\cite{Jiang:2012zze}
\bibitem{Jiang:2012zze}
  W.~Q.~Jiang, S.~D.~Gu, J.~Joseph, D.~W.~Liu, K.~B.~Luk, H.~Steiner, Z.~Wang and Q.~Wu,
  %``Suppressing ringing caused by large photomultiplier tube signals,''
  Chin.\ Phys.\ C {\bf 36}, 235 (2012).
  %%CITATION = CHPHD,C36,235;%%

%\cite{Jetter:2012xp}
\bibitem{Jetter:2012xp}
  S.~Jetter, D.~Dwyer, W.~Q.~Jiang, D.~W.~Liu, Y.~F.~Wang, Z.~M.~Wang and L.~J.~Wen,
  %``PMT waveform modeling at the Daya Bay experiment,''
  Chin.\ Phys.\ C {\bf 36}, 733 (2012).
  %%CITATION = CHPHD,C36,733;%%

%\bibitem{dybProposal}
%Daya Bay Proposal, arXiv:hep-ex/0701029

%\cite{Guo:2007ug}
\bibitem{Guo:2007ug}
  X.~Guo {\it et al.}  [Daya Bay Collaboration],
  %``A Precision measurement of the neutrino mixing angle $\theta_{13}$ using reactor antineutrinos at Daya-Bay,''
  hep-ex/0701029.
  %%CITATION = HEP-EX/0701029;%%

%\bibitem{mu_para}
%Y. Becherini, A. Margiotta, M. Sioli, M. Spurio, arXiv:hep-ph/0507228.

%\cite{Becherini:2005sr}
\bibitem{Becherini:2005sr}
  Y.~Becherini, A.~Margiotta, M.~Sioli and M.~Spurio,
  %``A Parameterisation of single and multiple muons in the deep water or ice,''
  Astropart.\ Phys.\  {\bf 25}, 1 (2006)
  [hep-ph/0507228].
  %%CITATION = HEP-PH/0507228;%%

%\bibitem{showerMu} FIXIT: Borexino, KamLAND reference

\bibitem{TUNLndg}
Triangular Universities Nuclear Laboratory (TUNL), the Nuclear Data Evaluation Group: http://www.tunl.duke.edu/nucldata/


%\end{thebibliography}


\end{thebibliography}
\end{document}